\documentclass{article}
\usepackage[utf8]{inputenc}
\usepackage{graphicx}
\usepackage{caption}
\usepackage{sidecap}
\sidecaptionvpos{figure}{c}
\usepackage{xcolor}
\usepackage{soul}
\usepackage[letterpaper, margin=1.5 in, headheight=14pt]{geometry}
\usepackage[square,numbers,sort&compress]{natbib}
\usepackage{amsmath}
\usepackage{wrapfig}
\usepackage[colorlinks=true, urlcolor=blue, linkcolor=blue, citecolor=blue]{hyperref}
\usepackage{authblk}
\usepackage{amssymb}
\usepackage{bm}
\usepackage[export]{adjustbox}
\usepackage{multirow}
\def\gtorder{\mathrel{\raise.3ex\hbox{$>$}\mkern-14mu
 \lower0.6ex\hbox{$\sim$}}}
\def\ltorder{\mathrel{\raise.3ex\hbox{$<$}\mkern-14mu
 \lower0.6ex\hbox{$\sim$}}}

\usepackage[shortlabels]{enumitem}

\usepackage{eso-pic}
\newcommand{\reportnum}[2]{
  \AddToShipoutPictureBG*{%
    \AtPageUpperLeft{%
      \hspace{0.75\paperwidth}%
      \raisebox{#1\baselineskip}{%
        \makebox[0pt][l]{\textnormal{#2}}
  }}}%
}

\title{Strong Interaction Physics at the Luminosity Frontier \\ with 22 GeV Electrons at Jefferson Lab}
\author[1]{A.~Accardi}
\author[2]{P.~Achenbach}
\author[3]{D.~Adhikari}
\author[4]{A.~Afanasev}
\author[5]{C.S.~Akondi}
\author[6]{N.~Akopov}
\author[7]{M.~Albaladejo}
\author[8]{H.~Albataineh}
\author[2]{M.~Albrecht}
\author[9]{B.~Almeida-Zamora}
% ...
\affil[1]{\small Hampton University, Hampton, VA, 23669, USA}
\affil[2]{\small Thomas Jefferson National Accelerator Facility, Newport News, VA 23606, USA}
\affil[3]{\small Virginia Tech, Blacksburg, VA 24061 USA}
\affil[4]{\small The George Washington University, Washington, D.C. 20052, USA}
\affil[5]{\small Florida State University, Tallahassee, FL 32306, USA}
\affil[6]{\small A.I. Alikhanyan National Science Laboratory (Yerevan Physics Institute), Yerevan 0036, Armenia}
\affil[7]{\small Instituto de Fisica Corpuscular (IFIC), Centro Mixto CSIC-Universidad de Valencia, E-46071 Valencia, Spain}
\affil[8]{\small Texas A\&M University-Kingsville, Kingsville, TX 78363, USA}
\affil[9]{\small Departamento de Investigaci\'on en F\'isica, Universidad de Sonora, Boulevard Luis Encinas J. y Rosales, Colonia Centro, Hermosillo, Sonora 83000, M\'exico}
% ...

\begin{document}
\reportnum{-7}{JLAB-PHY-23-3840}

\date{}
\maketitle
\clearpage 
\renewcommand{\baselinestretch}{0.75}\normalsize
{
  \hypersetup{linkcolor=black}
  \tableofcontents
}
\renewcommand{\baselinestretch}{1.0}\normalsize

\clearpage \section{Executive Summary}

The purpose of this document is to outline the developing scientific case for pursuing an energy upgrade to 22~GeV of the Continuous Electron Beam Accelerator Facility (CEBAF) at the Thomas Jefferson National Accelerator Facility (TJNAF, or JLab). This document was developed with input from a series of workshops  held in the period between March 2022 and April 2023 that were organized by the JLab user community and staff with guidance from JLab management (see Sec.~\ref{workshops}). The scientific case for the 22~GeV energy upgrade leverages \textit{ existing or already planned Hall equipment and world-wide uniqueness of CEBAF high-luminosity operations}. 

CEBAF delivers the world’s highest intensity and highest precision multi-GeV electron beams and has been do so for more than 25 years. In Fall 2017, with the completion of the 12~GeV upgrade and the start of the 12~GeV science program, a new era at the Laboratory began. The 12~GeV era is now well underway, with many important experimental results already published, and an exciting portfolio Program Advisory Committee approved experiments planned for at least the next 8–10 years \cite{Arrington:2021alx}. At the same time, the CEBAF community is looking toward its future and the science that could be obtained through a future cost-effective upgrade to 22~GeV. The great potential to upgrade CEBAF to higher energies opens a rich and unique experimental nuclear physics program that combines illustrious history with an exciting future, extending the life of the facility well into the 2030s and beyond.

JLab at 22~GeV will provide unique, world-leading science with high-precision, high-luminosity experiments elucidating the properties of quantum chromodynamics (QCD) in the valence regime $(x \geq 0.1)$. JLab at 22~GeV also enables researchers to probe the transition to a region of sea dominance, with access to hadrons of larger mass and different structures. With a fixed-target program at the ``luminosity frontier'', large acceptance detection systems, as well as high-precision spectrometers, CEBAF will continue to offer unique opportunities to shed light on the nature of QCD and the emergence of hadron structure for decades to come. In fact, CEBAF today, and with an energy upgrade, 
will continue to operate with several orders of magnitude higher luminosity than what is planned at the Electron-Ion Collider (EIC). CEBAF’s current and envisioned capabilities enable exciting scientific opportunities that complement the EIC operational reach, thus giving scientists the full suite of tools necessary to comprehensively understand how QCD builds hadronic matter.

The physics program laid out in this document spans a broad range of exciting initiatives that focus on a common theme, namely, investigations that explore different facets of the nonperturbative dynamics that manifest in hadron structure and probe the richness of these strongly interacting systems. The central themes of this program are reviewed in Section~\ref{sec:intro} - Introduction. The main components of the research program are highlighted in Sections~\ref{sec:wg1a} through \ref{sec:wg6}, followed by Section~\ref{sec:acc}, which provides a brief overview of the 22~GeV CEBAF energy-doubling concept. These sections outline the key measurements in different areas of experimental studies possible at a 22~GeV CEBAF accelerator in the existing JLab experimental end stations. They provide details on the key physics outcomes and unique aspects of the programs not possible at other existing or planned facilities.  

The 22~GeV physics program is being developed following three main principles: \textit{a)} identify the flagship measurements that can be done only with 22~GeV and their science impacts (Uniqueness); \textit{b)} 
identify the flagship measurements with 22~GeV that can extend and improve the 12~GeV measurements, helping the physics interpretation through multidimensional bins in extended kinematics (Enrichment); \textit{c)} identify the measurements with 22~GeV that can set the bridge between JLab12 and EIC (Complementarity). Even if a sharp separation among these three categories sometimes is difficult to maintain, we highlight the main points in the following.


\subsection*{\textit{Uniqueness}}

An energy upgrade to CEBAF will dramatically enhance the discovery potential of the existing world-unique hadron physics programs at Jefferson Lab. Several unique thrusts include:
\begin{itemize}
    
    \item In the area of hadron spectroscopy, with real photons in Hall~D and quasi-real photons in Hall~B, a unique production environment of exotic states will be probed providing cross section results, complementary to high-energy facilities. Photoproduction cross sections of exotic states could be decisive in understanding the nature of a subset of the pentaquark and tetraquark candidates that contain charm and anti-charm quarks. Moreover, in Hall~B the high-intensity flux of quasi-real photons at high energy will add the extra capability of studying the $Q^2$ evolution of any new state produced.
    
    \item JLab will be able to explore the proton's gluonic structure by unique precise measurements of the photo and electroproduction cross section near threshold of $J/\psi$ and higher-mass charmonium states, $\chi_c$ and $\psi(2S)$. Moreover, with an increase of the polarization figure-of-merit by an order of magnitude, GlueX will be able to measure polarization observables that are critical to disentangle the reaction mechanism and draw conclusions about the mass properties of the proton.  
    
    \item The JLab 22~GeV upgrade will enable high-precision measurements of the Primakoff production of pseudoscalar mesons with results: to explore the chiral anomaly and the origin and dynamics of chiral symmetry breaking; and to determine the light quark-mass ratio and the $\eta$-$\eta'$ mixing angle model independently. In particular, JLab will be able, for the first time, to perform precision measurements of the radiative decay width of $\pi^0$ off an electron to reach a sub-percent precision on $\Gamma(\pi^0 \rightarrow \gamma \gamma)$, necessary to better understand the discrepancy between the existing experimental results and the high-order QCD predictions, and therefore offering a stringent test of low-energy QCD.
\end{itemize}

\subsection*{\textit{Enrichment} and \textit{Complementarity}}

\begin{itemize}
    \item The 22~GeV upgrade will extend the phase space, in particular in momentum transfer $Q^2$ and hadronic transverse momenta, for studying the momentum space tomography of nucleons and nuclei through the transverse momentum dependent (TMDs) of parton distribution functions, offering a new complementary window between the 12~GeV program and the future EIC. Combined with the high luminosity and precision detecting capabilities of multiparticle final state observables in a multidimensional space, it will make JLab unique to disentangle the genuine intrinsic transverse structure of hadrons encoded in TMDs with controlled systematics. These capabilities are critical for the interpretation of the measurements carried out both at JLab and EIC and for full understanding of the complex nature of nucleon structure properties and hadronization processes. Moreover, JLab has a uniquely fundamental role to play in the EIC era in the realm of precision separation measurements between the longitudinal ($\sigma_L$) and transverse ($\sigma_T$) photon contributions to the cross section, which are critical for studies of both semi-inclusive and exclusive processes. 

    \item The 22~GeV upgrade will be crucial for carrying out elastic and hard-exclusive process experiments. Such measurements require sufficient energy for reaching the scaling and factorization regime, high luminosity for measurements of low-rate processes and multivariable differential analysis, and excellent detector resolution for cross section measurements. Essential physics applications are:
    \begin{enumerate}[a)]
        \item High-quality extraction of the $D$-term form factor of the QCD energy-momentum tensor and the ``pressure'' distribution inside the proton.
        \item Fully differential 3D imaging of the nucleon using novel processes such as Double Deeply	Virtual	Compton	Scattering (DDVCS) and exclusive diphoton production. 
        \item Exploring hadron structure with novel exclusive processes such as $N \rightarrow N^\ast$ transition GPDs	and $N \rightarrow$ meson transition distribution amplitudes.
        \item Extending nucleon, pion, and resonance transition form factor measurements to momentum transfers $Q^2 \sim 30$ GeV$^2$, probing short-range hadron structure, QCD interactions, and the mechanism of the emergence of hadron mass.
    \end{enumerate}

    \item JLab upgrade can offer critical insights for precision studies of partonic structure filling the gap in kinematics of the combined scientific program of JLab 12~GeV and EIC. With its enhanced energy range, the JLab 22~GeV upgrade will allow:
    \begin{enumerate}[a)]
        \item Precision measurements of the nucleon light sea in the intermediate to high-$x$ range, which can help validate novel theoretical predictions for the intrinsic sea components in the nucleon wave function and support Beyond Standard Model searches at colliders.
        \item Precision determination of the helicity structure of the nucleon at large $x$ and of the strong coupling at levels well below one percent in $\Delta \alpha/\alpha $.
        \item Unique opportunities to explore the internal structure of mesons in the intermediate to high-$x$ range.
    \end{enumerate}

    \item The 22~GeV~high intensity beam will create an unprecedented opportunity for Nuclear Sciences to significantly advance our knowledge of QCD dynamics of nuclear forces at core distances. Some highlights of the JLab~22~GeV upgrade program include:
    \begin{enumerate}[a)]
        \item Exploring nuclear forces dominated by nuclear repulsion by carrying out the first-ever direct study of nuclear DIS structure at $x > 1.25$, as well as measuring deuteron structure at sub-Fermi distances in exclusive deuteron break-up reactions with missing momenta above GeV region.
        \item Providing definitive proof of three-nucleon short-range correlations by verifying the existence of the new nuclear scaling at $x > 2$ and $Q^2$=~10~-~15~GeV$^2$ in inclusive $e-A$ scattering;
        \item Extending the reach of medium modification studies to the antishadowing region with unprecedentedly precise measurements using a rich variety of techniques (including tagging) and targets; 
        \item Proving the existence of Color Transparency phenomena in the baryonic sector.
        \item Providing an unprecedented kinematic reach for studies of hadronization in the nuclear medium.
    \end{enumerate}
\end{itemize}

CEBAF energy upgrade will be realized taking advantage of recent novel advances in accelerator technology, which will make it possible to extend the energy reach of the CEBAF accelerator up to 22~GeV within the existing tunnel footprint and using the {\textit existing CEBAF SRF cavity system.} The proposal is to replace the highest-energy arcs with Fixed Field Alternating Gradient (FFA) arcs and increase the number of recirculations through the accelerating cavities. The new pair of arcs configured with an FFA lattice would support simultaneous transport of 6 passes with energies spanning a factor of two. This novel permanent magnet technology will have a big positive impact on JLab operations since it saves energy and lowers operating cost.

CEBAF is a facility in high demand and JLab continues to invest to make optimum use of CEBAF’s capabilities to  produce high-impact science across different areas within Nuclear Physics and beyond. With CEBAF at higher energy, some important thresholds would be crossed and an energy window that sits between JLab at 12~GeV and EIC would be available. This, together with CEBAF capabilities to run electron scattering experiment at the luminosity frontier, can provide unique insight into the nonperturbative QCD dynamics and will place JLab as a unique facility capable of exploring the emergent phenomena of QCD and its associated effective degrees of freedom.

\clearpage \section{Introduction}
\label{sec:intro}

The proposed energy upgrade to the CEBAF accelerator at the Thomas Jefferson National Accelerator Facility would enable the only facility worldwide, planned or foreseen, that can address the complexity at the scientific frontier of emergent hadron structure with its high luminosity and probing precision at the hadronic scale. While high-energy facilities will illuminate the perturbative dynamics and discover the fundamental role of gluons in nucleons and nuclei, a medium energy electron accelerator at the luminosity frontier will be critical to understand the rich and extraordinary variety of non-perturbative effects manifested in hadronic structure. 

\begin{wrapfigure}{r}{0.45\textwidth}
\begin{center}
    \vspace{-0.2in}
    \includegraphics[width=0.43\textwidth]{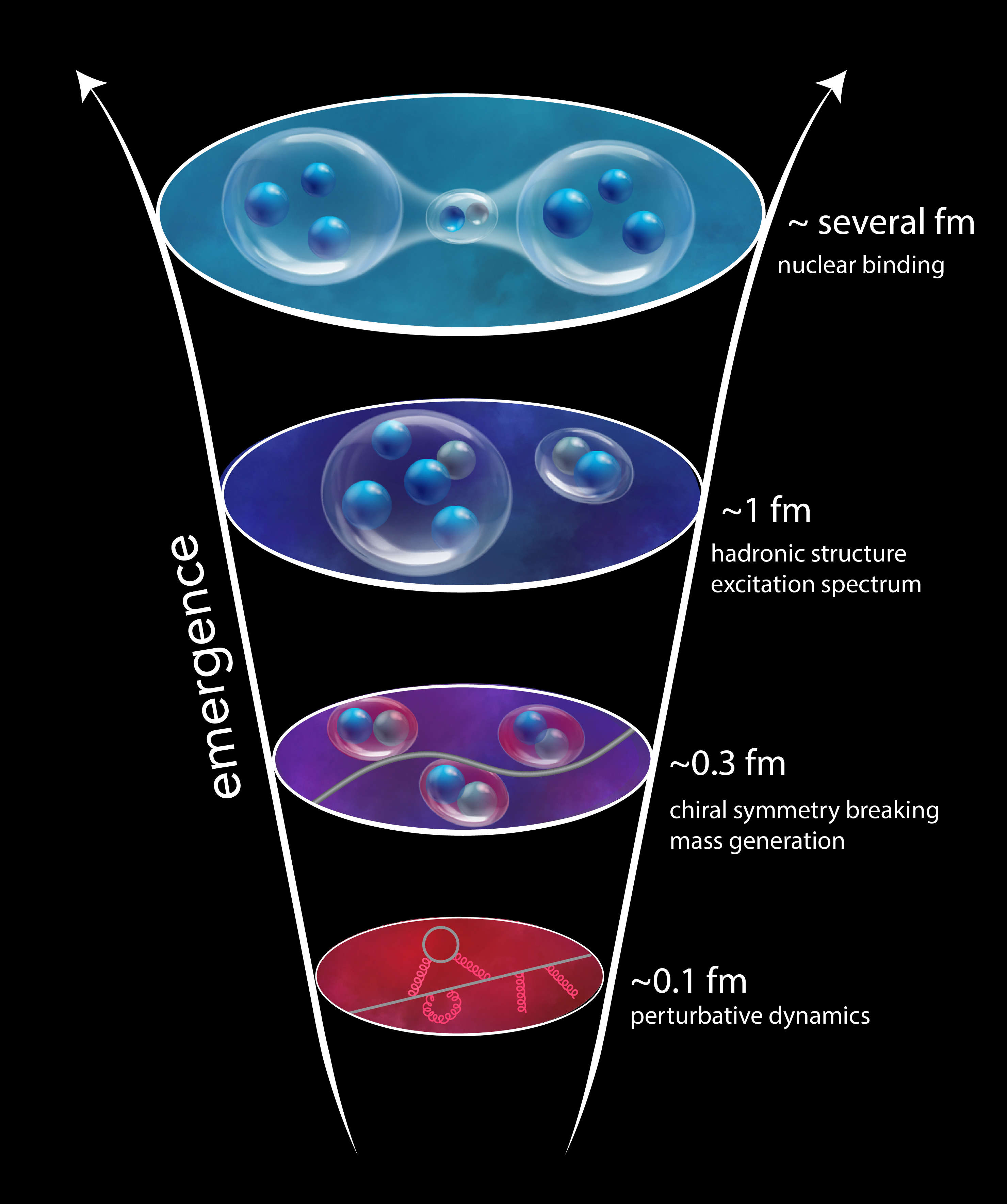}
    \caption{\label{fig:emergence}
        \footnotesize The emergence of structure in QCD from the perturbative regime of quarks and gluons to bound hadrons to hadrons bound in nuclei.}
\end{center}
\vspace{-0.25in}
\end{wrapfigure} 

The Lagrangian of QCD, which we believe governs the dynamics of quarks and gluons, is not easily or directly connected to the complicated observables that we measure in electron-proton and electron-nucleus scattering experiments. At one end of the spectrum, the elementary quark/gluon degrees of freedom are manifested only at distances {$\lesssim 0.1$}~fm, where the quark-gluon interactions can be understood using methods of perturbation theory; however, at hadronic distances $\sim$ 1~fm the dynamics undergo qualitative changes, causing the appearance of effective degrees of freedom expressed in new structure and dynamics.~While these new structures develop in the context of the underlying QCD degrees of freedom, their experimental interpretation remains challenging. This places strong interaction physics in the context of ``emergent phenomena'', a powerful paradigm for the study of complex systems used in other areas of physics, such as condensed matter, as well as biological and social sciences. Here, the behavior of larger and complex aggregates of elementary particles may not be understood in terms of an extrapolation of the properties of a few particles. Instead, at each level of complexity, entirely new properties appear – and the understanding of each new behavior warrants study. This is demonstrated in Fig.~\ref{fig:emergence} where the distance scale incorporates emergence. 

Experimental scattering observables are shaped by certain effects rooted in the quantum and nonlinear nature of QCD. These effects create dynamical scales not present in the original theory (see Fig.~\ref{fig:emergence}). One effect is the breaking of scale invariance by quantum fluctuations at high energies beyond the range of observation, which creates a mass/length scale that acts as the source of all other dynamical scales emerging from the theory (the so-called trace anomaly). Another effect is the spontaneous breaking of chiral symmetry, which generates a dynamical mass of the quarks that provides most of the mass of the light hadrons,  including the nucleon, and is therefore the source of 99\% of the mass of the visible Universe. Yet another effect is confinement, which limits the propagation of QCD color charges over hadronic distances and influences the long-range structure of hadrons and their excitation spectrum. Understanding these nonperturbative effects is the key to understanding the emergence of hadrons and nuclei from QCD.

Many expressions of these nonperturbative effects can be seen already in established hadron spectra, structure, and interactions. Chiral symmetry breaking is expressed in the unnaturally small mass of the pion, which emerges as the Goldstone boson mediating the long-range QCD interactions, and its momentum-dependent coupling to other hadrons; confinement is visible in the spectra of heavy quarkonia. However, in order to truly understand ``how'' the effective dynamics emerges from QCD, it is necessary to study the fields of QCD through scattering processes that probe nonperturbative dynamics hadron structure and spectra. Formulating such processes has been a priority of theoretical and experimental research in recent years.

Since the 2015 Long-Range Plan, several novel processes probing hadron structure and spectra have come into focus, revealing specific aspects of nonperturbative dynamics and providing insight into the emergence of structure from QCD. Studies of nucleon elastic electromagnetic form factors and nucleon resonance electroexcitation amplitudes for many excited states of the nucleon have demonstrated the capability to explore the emergence of hadron mass and the structure of ground and excited nucleon states at a distance scale comparable with the hadron size. In concert, recent advances in accelerator science and technology have made possible a promising, cost-effective extension of the energy reach of the CEBAF accelerator to 22~GeV within the existing tunnel footprint. To map the emergence of hadronic structure from perturbative dynamics, several experimental requirements must be met. One is the need for a large four-momentum transfer, $Q>2$~GeV, to have a well-controlled and localized probe ($ < 0.1$ fm). Importantly, a second momentum scale is simultaneously needed to be sensitive to the emergent regime across the scales shown in Fig.~\ref{fig:emergence}. Such two-scale experimental observables are naturally accessible at a lepton-hadron facility like CEBAF, including exclusive electron-hadron deep virtual Compton scattering (DVCS): $e(\ell)+h(p) \to e(\ell') + h(p') +\gamma$ with the hard scale $Q^2 = -(\ell-\ell')^2$ and the second scale $ t = (p-p')^2$, and semi-inclusive deep inelastic scattering (SIDIS): $e(\ell)+h(p) \to e(\ell') + h'(p') +X$ with the momentum imbalance between $\ell'$ and $p'$ as the second scale. However, once the hadron is broken, larger momentum transfer $Q$ leads to more collision induced radiation, which could significantly shadow the structure information probed at the second (and the soft) scale and reduce our precision to probe the emergent hadron structure. The requisite electron beam energy to probe hadron structure is determined by the need to, for instance, reach the charm threshold in deep-virtual processes and to separate the produced hadronic systems from the target remnants. The studies presented in this document show that the optimal beam energy for performing such two-scale measurements is $\sim 20$~GeV. Another determining constraint to the measurements is the need to precisely measure small cross sections in a multidimensional phase space, needed also for separation of different dynamical mechanisms, which requires high luminosity and multiple devices with differing but complementary experimental capabilities. The fixed-target experiments with the CEBAF accelerator at JLab will achieve luminosities $\sim10^{38} {\rm cm}^{-2} {\rm s}^{-1}$ with the high-resolution spectrometers and SoLID, and $\sim 10^{35} {\rm cm}^{-2} {\rm s}^{-1}$ with the CLAS12 large-acceptance detector. The foreseen Jefferson Lab experimental equipment, including the Solenoidal Large Intensity Device (SoLID) in Hall A, high luminosity CLAS12 in Hall B, precision magnetic spectrometers in Hall C, and polarized, tagged photon beams in Hall D, matches the science need.~It is a major advantage that the measurements can be performed using the existing and well-understood JLab12 detectors, reducing cost and minimizing technical risk to the program. 

The experimental program proposed here is complementary and synergistic with both the current JLab 12~GeV program (including SoLID) and the future EIC. It provides a critical bridge between the two, exploring fascinating and  essential aspects of the emergence of hadrons that are needed for full understanding but are not covered by either JLab $12$~GeV or the EIC. The center-of-mass energies reached in these fixed-target experiments ($\sqrt{s} \sim 6$~GeV) are still substantially below those reached in colliding-beam experiments at EIC ($\sqrt{s} > 20$~GeV), while the luminosity of the fixed target facilities is $\sim 3-4$ orders of magnitude larger. At the same time, there is considerable synergy between the scientific programs pursued with the upgraded CEBAF and the EIC. The experimental requirements for many of the measurements needed to answer the myriad questions posed by the emergence of structure have been assessed in simulations, and some highlights are described in Sections~\ref{sec:wg1a} through \ref{sec:wg6}. The experimental program at 22~GeV is based on an energy-upgraded CEBAF facility that may be considered due to exciting and cost-effective advances in accelerator technology that are highlighted in Section~\ref{sec:acc}.
\clearpage \section{Hadron Spectroscopy}
\label{sec:wg1a}

From the development of the Bohr model of the atom to the
quark model of hadrons, the idea of measuring and organizing spectra of energy
states has proven to be an invaluable tool in gaining insight into the fundamental
theory that generates such states.  A fascinating aspect of quantum chromodynamics (QCD)
is the broad variety of phenomena that emerge from the underlying theory and the 
scales at which this emergence occurs.  In the context of hadron spectroscopy, one
aims to study the spectrum of semi-stable hadrons or hadronic resonances and use
this information to understand how and what types of hadrons are generated by
QCD.

The quark model originally arose from the need to explain the landscape of 
hadrons observed in particle collisions in the mid-twentieth century, and we now
understand that QCD is the fundamental theory underlying the model.  However,
the light quarks of QCD are not the same as those in the quark model, and a detailed
understanding of how QCD generates not only the spectrum predicted by the quark model but also 
perhaps states with additional gluonic degrees of freedom remains an open question.
Until recently, it seemed as if almost all hadrons observed in Nature were
composed of three-quark baryons or quark-antiquark mesons.  While the original
quark model allows the possibility of more complex configurations of quarks
and anti-quarks, Nature appears not to prefer them.  
At the same time, our understanding of gluon self-interactions
in QCD motivated ideas that glueballs, with no quarks, or quark-gluon hybrids
might exist.  These ideas have evolved tremendously into predictions using
lattice QCD techniques about the existence and properties of exotic hybrid
mesons~\cite{Bulava:2022ovd}.  The experimental search for hybrid mesons 
is a key thrust of the
JLab 12 GeV program and complementary experiments around the world,
some of which have reported evidence of such 
states~\cite{Meyer:2015eta,JPAC:2018zyd,BESIII:2022riz}.  Establishing a spectrum
of hybrids would further our understanding of how the unique properties of
gluons in QCD affect the emergence of the hadron spectrum.

Theoretical techniques for connecting QCD to experimental data have advanced 
significantly in recent decades but new discoveries indicate our understanding
of QCD dynamics is far from complete.  In the last twenty years, 
high-energy and high-intensity experiments have produced
a mountain of discoveries in the spectroscopy of hadrons containing heavy quarks
(charm and bottom)~\cite{Olsen:2017bmm,Lebed:2016hpi,Briceno:2015rlt}.  
For example, observation of peaks in the invariant mass of $J/\psi\,\pi^-$ 
around 4~GeV~\cite{BESIII:2013ris,Belle:2013yex}
suggest new tetraquark classes of particles: being heavy, 
such states must have a $c\bar{c}$ but the presence of charge requires at least an
additional $d\bar{u}$. There are numerous similar states, in both the bottom~\cite{Belle:2011aa}
as well as charm spectra~\cite{BESIII:2013ouc}, which have masses at level where 
light-quark meson interactions become relevant.  If these states are in fact
hadron resonances, then one would like to know their nature.  For example, are these
systems compact four-quark objects or more like a meson-meson molecule?  These
hadrons are all instances of confinement in QCD, and it is valuable to
understand their place in the hadron landscape.  Just as probing the nuclear
landscape has led to a better understanding of nuclear structure and nucleon
interactions, the hadron landscape provides a path to explore interesting features
of QCD.  While the 12~GeV program at JLab
is able to explore light-quark systems in isolation and can produce
the lowest-mass $c\bar{c}$ systems, an energy upgrade is essential for
JLab to contribute unique information on photoproduction of systems
with light and heavy degrees of freedom that appear to exhibit exotic properties.
Throughout this section we consider not only the final upgrade target electron
energy of 22~GeV but also demonstrate that significant new results can be obtained
by interim operations at 17~GeV if a phased upgrade strategy is adopted.

\subsection{Photoproduction as Tool for Spectroscopy}

An energy upgrade to CEBAF would dramatically enhance the discovery potential
of the existing world-unique hadron spectroscopy experimental programs at JLab, 
namely the CLAS12 experiment in Hall B~\cite{Burkert:2020akg} and the GlueX experiment in Hall D~\cite{GlueX:2020idb}.  Both
experiments feature high-acceptance, multi-particle spectrometers that are designed
to detect the decays of hadronic resonances and enable studies of properties and
production mechanisms of hadrons.  The CLAS12 experiment uses polarized virtual photoproduction
at low $Q^2$ via electron-proton collisions, while the Hall D facility at JLab provides a
real photon beam that is partially linearly polarized through coherent bremsstrahlung 
scattering of the CEBAF electron beam off of a diamond radiator.  With Hall D, the
recoil electron momentum is measured, which provides an energy determination of each
photon incident on the proton target at the center of the GlueX spectrometer, while 
for CLAS12, detection of the scattered electron provides energy and (linear) 
polarization of the virtual photon.  The future high-energy and high-luminosity 
spectroscopy program leverages existing experience with these two facilities and 
their complementary photoproduction mechanisms.  

\begin{figure}
    \centering
    \includegraphics[width=\linewidth]{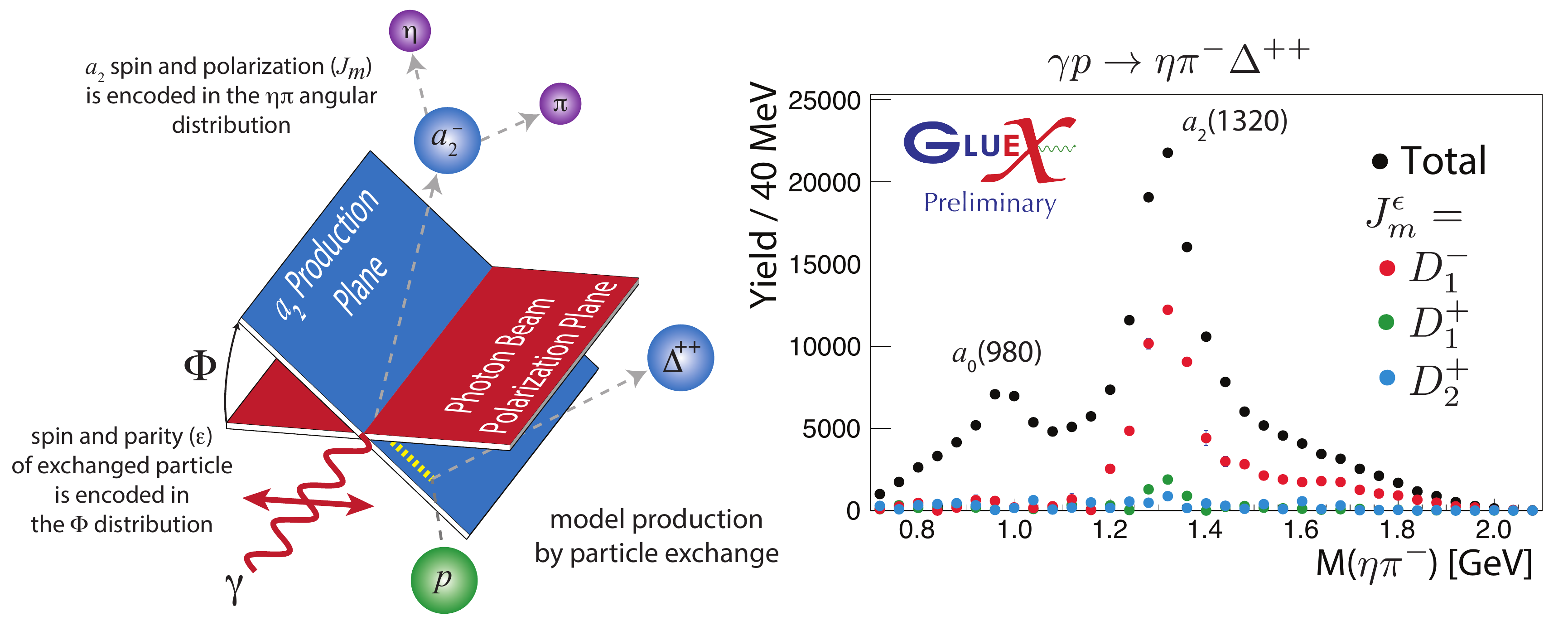}
    \caption{A sketch of the polarized photoproduction of $a_2^-(1320)$ via $t$-channel interaction with the target.  Preliminary data from GlueX
    indicates that the dominant production mechanism of the spin-2 ($D-$wave) peak consistent with the $a_2$ in the $\eta\pi^-$
    spectrum is by exchange of an unnatural parity particle ($\epsilon=-$).}
    \label{fig:eta_pi_prod}
\end{figure}

Photoproduction of mesons by linearly polarized photons provides an opportunity to extract
information about the production mechanisms and how these mechanisms vary with kinematics.
Figure~\ref{fig:eta_pi_prod} illustrates a typical model for production of $\eta\pi^-$ 
by linearly polarized photons where the beam photon interacts with a particle emitted
by the target, {\it i.e.,} a $t$-channel production.
Preliminary results from the GlueX experiment illustrate
that $a_2^-(1320)$ can be identified in the $\eta\pi^-$ mass spectrum using the angular
distribution of the $\eta\pi^-$.  By analyzing the angle between the production and
polarization planes, one learns that the dominant production mechanism of $a_2^-$ is
by exchange of a particle with unnatural parity ($J^P=0^-,1^+,\ldots$) like a pion ($J^P=0^-$).  
This is consistent with expectations: the photon beam can be considered a virtual $\rho$ meson
that scatters off of a $\pi^-$ emitted by the target to produce the $a_2^-$ via its
well-known $\rho\pi$ coupling.  The use of linearly polarized photons enables this
additional angular analysis that sheds light on the production coupling of the hadronic
resonance in addition to the decay coupling.

An advantage of using photoproduction to study hadronic resonances is the variety
of different production mechanisms that are available -- many virtual particles can
be exchanged between the beam and the target over a broad kinematic range.  In contrast,
when studying resonances in $B$ meson decays or $e^+e^-$ collisions, the quantum numbers
and kinematics of the initial state are fixed.  These well-known initial conditions in
the latter case simplifies the analysis and interpretation of experimental
data, but can also constrain the opportunities for exploring resonance production.
The use of linearly polarized real or virtual photoproduction relaxes production 
constraints at a cost of increasing analysis complexity, and provides a unique and 
complementary tool to study hadronic resonances.  In addition, the GlueX and CLAS12
experiments offer the opportunity to cross-check results over a wide range of kinematics
and final states using two similar but complementary photoproduction mechanisms.

\subsection{Spectroscopy of Exotic States with $c\bar{c}$}

The last two decades have produced numerous discoveries of new particles in
the charm and bottom sectors by experiments like BaBar, BESIII, and Belle 
at $e^+e^-$ machines, as well as LHCb at the LHC.  All of these experiments
have pushed the luminosity frontier and as a result have the ability to
discover new hadrons that are rarely produced.  Some of these new discoveries,
like the observation of excited states of the $\Omega_b$ ($bss$)~\cite{LHCb:2021ptx}, 
extend our knowledge of 
conventional hadrons containing heavy quarks, while many others, like the 
charged $Z_c$ tetraquark candidates~\cite{BESIII:2013ris,Belle:2013yex,BESIII:2013ouc,LHCb:2014zfx,Belle:2014nuw}, have forced a reconsideration of
long-standing ideas about the valence quark content of hadrons generated 
by QCD.  Extensive reviews of these new particles can be found 
in Refs.~\cite{Lebed:2016hpi,Briceno:2015rlt, Olsen:2017bmm,Brambilla:2019esw}.
While these particles are colloquially referred to as the $XYZ$ states, the 
community has yet to agree on a naming scheme, let alone an underlying 
theoretical interpretation, for the numerous new additions to the hadron landscape.

The $XYZ$ states are exotic because they have properties that are
inconsistent with the well-understood heavy $q\bar{q}$ mesons.  Some exotic
features are clear:  a meson with non-zero electric charge cannot be a $c\bar{c}$ state.
Other states are unusual because they have masses, quantum numbers, or decay 
properties that do not align with expectation based on our understanding
of heavy-quark systems.  A common feature to all of the $XYZ$ states is that
they have masses where both heavy and light quarks play a key role in their
structure and decays, that is, the path to understanding these particles
involves QCD in the strongly interacting regime.

\begin{SCfigure}
    \centering
    \includegraphics[width=0.6\linewidth]{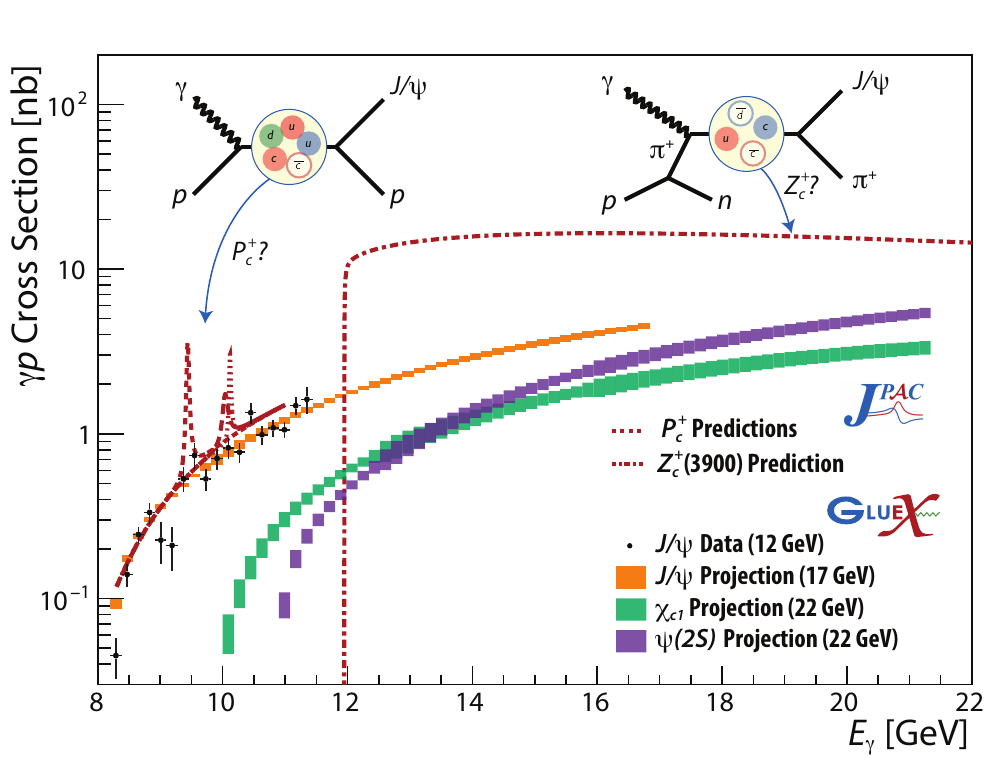}
    \caption{Photoproduction cross sections of states containing $c\bar{c}$ as a function of photon beam energy.  The points are GlueX data~\cite{Adhikari:2023fcr}  The colored boxes are projections of statistical precision using the GlueX detector with different assumptions about the electron energy.  The collection of dashed and dotted curves indicate how pentaquark $P_c$~\cite{HillerBlin:2016odx} or tetraquark $Z_c$~\cite{Winney:2022tky} candidates might appear.}
    \label{fig:cc_xc}
\end{SCfigure}

A peculiar feature of the $XYZ$ states is that, with the exception of the
$X(3872)$, none so far have been observed in multiple production mechanisms~\cite{Workman:2022ynf}.  
Most observations of $XYZ$ states come from analysis of $e^+e^-$ collisions
or the decay of hadrons containing $b$ quarks, but these two production
mechanisms seem to have generated a non-overlapping set of exotic candidates.
The reason why some states appear in certain production environments but
not others is not understood.  A feature of both production mechanisms is
that they require exotic candidates to be produced in conjunction with other
hadrons. For example, if one wants to produce a charged tetraquark candidate
$Z_c$ in an $e^+e^-$ collision, another charged hadron must be present in the
final state.  The fact that most $XYZ$ states appear as a peak in the mass
spectrum of two particles in a three-body system has led to the suggestion that
some of the states are not hadronic resonances but kinematic effects known as
triangle singularities~\cite{Guo:2016bkl,Bayar:2016ftu,Nakamura:2021qvy,Guo:2019twa}.  
These effects arise when relatively long-lived particles,
like $D$ mesons, are produced with kinematic conditions that are favorable
for rescattering into some final state of interest.  A peak in the invariant
mass of the final state is then an indicator of meeting the criteria for
rescattering and not the signal of a resonance.  In these three-body decays, 
separating the signature of a new type of hadron from a kinematic effect, 
a question of utmost importance to understanding the spectrum of hadrons 
generated by QCD, requires precise measurement and theoretical understanding 
of the lineshape~\cite{Pilloni:2016obd}.

With an energy upgrade, JLab is capable of providing unique and
complementary information that could be decisive in understanding the nature
of a subset of the $XYZ$ states.  Those states that are particularly
well suited for exploration at JLab are the ones that are candidates 
for resonances in $J/\psi\,p$, $J/\psi\,\pi$, and $\psi(2S)\,\pi$ systems.  
Since the photon has the same quantum numbers as the $J/\psi$ or $\psi(2S)$,
photoproduction can be viewed as a mechanism to scatter a virtual $J/\psi$
directly off of a proton or off the charged pion cloud around a proton.  In such
production mechanisms, sketched at the top of Fig.~\ref{fig:cc_xc}, the
production vertex is directly related to the decay vertex.  These clean
two-to-two scattering processes are free from triangle singularities that
can complicate the interpretation of existing data from $e^+e^-$ collisions and
$B$ decay.

\subsubsection{The $X(3872)$ and Conventional $c\bar{c}$}

The discovery of the $X(3872)$ by Belle in 2003~\cite{Belle:2003nnu} marked the start of what continues
today as a very exciting investigation into the spectroscopy of systems containing
a heavy $q\bar{q}$ component. The $X(3872)$ is the most robust and most extensively
studied of all of the $XYZ$ states.  It has been observed by numerous experiments
and in numerous production environments. Its quantum numbers are determined:  
$J^{PC}=1^{++}$~\cite{LHCb:2013kgk} and as such the PDG has designated the state $\chi_{c1}(3872)$.  The
state has a very narrow width of about 1~MeV, a mass that is consistent with the 
$D^0\overline{D^{*0}}$ threshold, and exhibits large isospin violation in its
decays.  Explanations of its underlying structure include conventional
$\chi_{c1}(2P)$, $D\overline{D^*}$ molecule, and compact tetraquark.  Because of
its robustness, it serves as an interesting standard candle in photoproduction
investigations with an upgraded CEBAF.  Virtual photoproduction of the 
$X(3872)$ with a muon beam was recently explored by COMPASS~\cite{COMPASS:2017wql}, 
but interestingly, the state observed, while having a mass consistent 
with the $X(3872)$, exhibited different decay properties than established by other 
experiments.  The COMPASS observations are based on a total of $13.2\pm5.2$
events.  A follow-up investigation in electroproduction or photoproduction with
a high-luminosity CEBAF is warranted.

\begin{SCfigure}
    \centering
    \includegraphics[width=0.4\linewidth]{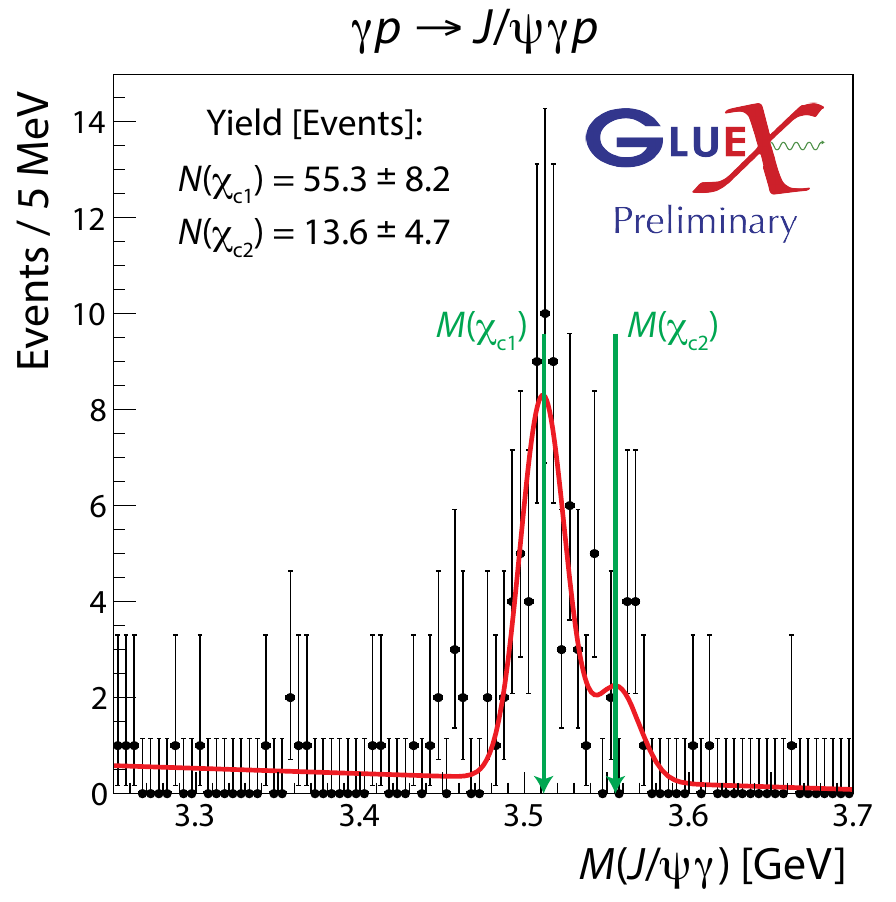}
    \caption{Preliminary results from the GlueX Collaboration showing evidence for exclusive photoproduction of $\chi_{c1}$.  The $\chi_{c1}$ candidates are reconstructed in the $\gamma J/\psi$ decay mode with $J/\psi\to e^+e^-$.  These preliminary results use the entire photon beam energy range available to GlueX and come from a subset of about $3\times 10^{11}$ events collected over 100 days of beam on target.  The PDG~\cite{ParticleDataGroup:2022pth} values for the masses of the $\chi_{c1}$ and $\chi_{c2}$ are given by the green arrows.}
    \label{fig:gluex_chic12}
\end{SCfigure}

By looking to the edge of the capability of the current machine, one can
see potential for new explorations of charmonium production.
The CEBAF configuration is such that the GlueX experiment receives one
additional pass through the north accelerating LINAC than the other halls.  
Therefore, the highest energy photons in the JLab 12~GeV machine are 
impinging on the GlueX target.  Figure~\ref{fig:gluex_chic12} shows the 
$J/\psi\,\gamma$ invariant mass in the reaction $\gamma p\to J/\psi\,\gamma p$
where a preliminary signal of about 50 events consistent with 
$\gamma p\to \chi_{c1} p$ is observed.  This result, which uses 100 days
of beam on the GlueX target, will yield the first measurement of the
photoproduction cross section of the $\chi_{c1}(1P)$.  An upgrade of the electron
energy to 22~GeV is projected to increase the $\chi_{c1}$ yield by two orders
of magnitude, which would enable a measurement of the cross section dependence
on energy.  The $X(3872)$ has some similarities to the $\chi_{c1}(2P)$ state
of the $c\bar{c}$ system.  If the $X(3872)$ is observed in photoproduction
at an energy-upgraded CEBAF, then JLab can contribute unique information 
that may provide insight to the nature of the $X(3872)$ by conducting a 
comparison of photoproduction mechanisms of the $\chi_{c1}$ 
and $X(3872)$.

\subsubsection{Pentaquark $P_c$ Candidates}

In 2015, the LHCb Collaboration reported the observation of two pentaquark candidates in the
decay of $\Lambda_b\to J\psi\,p\,K^-$~\cite{LHCb:2015yax}.  
The states appear as peaks in the $J/\psi\,p$
mass spectrum around 4.4~GeV and have minimum quark content of $c\bar{c}uud$.  In
2019, using a significantly larger dataset, LHCb was able to further resolve three
narrow peaks in the mass spectrum known as the $P_c(4312)^+$, $P_c(4440)^+$, 
and $P_c(4457)^+$~\cite{LHCb:2019kea}.  Like some other $XYZ$ candidates, 
the three-body final state, 
as well as the presence of charm baryon and meson mass thresholds, invites
an explanation for some of the peaks as triangle singularities.
If the states are true $J/\psi\,p$ resonances, then they should
be produced in photon-proton collisions, with the photon acting
like a virtual $J/\psi$, as pictured at the top of Fig.~\ref{fig:cc_xc}.  One would expect
the $J/\psi$ production cross section to peak at photon beam energies that excite the
pentaquark resonance~\cite{HillerBlin:2016odx}.  

\begin{SCfigure}
    \centering    
    \includegraphics[width=0.45\linewidth]{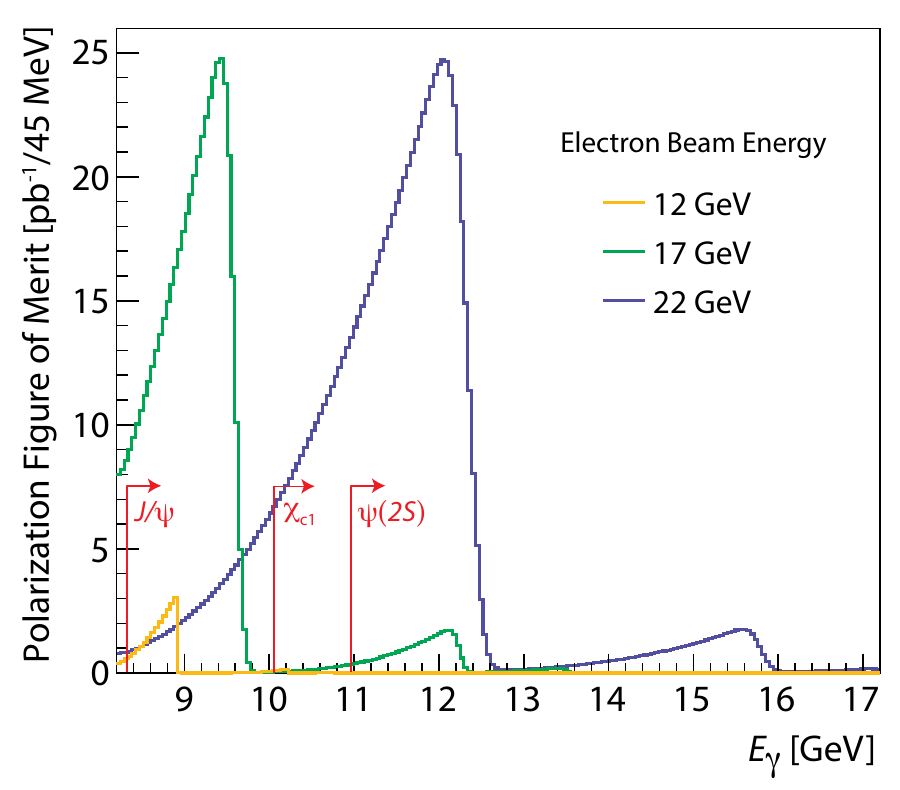}
    \caption{The polarization figure of merit ($P^2(dN_\gamma/dE)$) as a function of photon beam energy 
    $E_\gamma$ for the existing 12~GeV GlueX configuration assuming 100 days of beam on target (yellow).   
    Figures of merit assuming equal beam time are shown for 17~GeV and 22~GeV electrons, both of which
    are drawn for the same diamond orientation.  Various $c\bar{c}$ production thresholds are shown.}
    \label{fig:pol_fom}
\end{SCfigure}

Figure~\ref{fig:cc_xc} shows data from the GlueX
experiment for the $J/\psi$ cross section as a function of photon beam energy.  While 
the cross section shows some structure, discussed extensively in Refs.~\cite{Adhikari:2023fcr,Winney:2023oqc},
it is not evident that this structure is a result of a pentaquark resonance.  The shape
of the cross section at threshold is thought to be linked to the gluonic structure
of the proton (see Section~\ref{sec:wg4}). The orange boxes 
in Fig.~\ref{fig:cc_xc} show the projections for the statistical precision on the $J/\psi$ 
cross section assuming a similar amount of integrated luminosity on the GlueX target
but with an electron beam energy of 17~GeV.  One can define a polarization figure-of-merit
as $P^2I$, where $P$ is the beam polarization and $I$ is the beam intensity.  (The 
statistical uncertainty on polarized observables scales like the inverse square-root of
the figure of merit.)  Figure~\ref{fig:pol_fom} shows the polarization figure-of-merit for
a typical 12~GeV configuration of the GlueX beamline and a configuration that uses
17~GeV electrons incident on the GlueX radiator.  The 17~GeV beam increases the
polarization figure of merit by an order of magnitude near $J/\psi$ production threshold.
A precision measurement of $J/\psi$ polarization would inform our understanding of the
production mechanism and potentially validate the use of such data to draw 
conclusions about the proton structure. 
An upgrade to 22~GeV would permit measurements of the $\chi_{c1}$ and $\psi(2S)$ 
cross sections with similar precision.  To probe the $c\bar{c}$ threshold region, these
investigations must be conducted with photon beam energies in the 8-20~GeV region, 
making them well-suited for the energy and luminosity planned for the upgrade.

If a positive signal for any of the $P_c$ states can be established in photoproduction, then
this eliminates the possibility that the corresponding peak observed in $\Lambda_b$ decay 
is due to a kinematic effect and solidifies the interpretation as a resonance.
Signals in photoproduction open the door for new measurements, for example, an 
analysis of the $J/\psi\,p$ angular distributions would provide 
information about the quantum numbers of the corresponding $P_c$ state.  
A stringent upper-limit on the photoproduction cross section further constrains 
the interpretation of $\Lambda_b$ decay results.  In the near future, combined analyses 
of $J/\psi$, $\chi_{cJ}$, and open charm final states using data from existing facilities
are expected to put severe constraints on models that describe the LHCb signals as a consequence of 
kinematic effects. This is important, as no rescattering model is 
able to quantitatively reproduce 
the pentaquark signals so far, and the information of other channels 
is needed to improve this description. If an independent observation of any of 
the $P_c$ states becomes available in a different production mechanism, then 
the null result in photoproduction is informative of the internal structure of 
the pentaquark, as one must explain why the underlying structure  causes a 
suppression in photoproduction.

\subsubsection{Tetraquark $Z_c$ Candidates} 

Studies of $e^+e^-$ collisions at center-of-mass energies at both the $c\bar{c}$ and
$b\bar{b}$ scales, as well as studies of $B$ meson decay, have uncovered a large
collection of tetraquark candidates, often labeled $Z_c$ or $Z_b$.  The exotic signature
of many of these states is a peak in the invariant mass spectrum of a charged pion and
a $c\bar{c}$ ($J/\psi$, $h_c$, or $\psi(2S)$) or $b\bar{b}$ ($\Upsilon(nS)$ or $h_b$) 
hadron.  The mass and charge imply a minimum $qq\bar{q}\bar{q}$ content of these states. 
The pattern of exotic $c\bar c$ and $b \bar b$ has similarities, in particular 
for these charged states~\cite{BESIII:2013ouc,BESIII:2013ris,Belle:2011aa}.

Let us consider as an example the $Z_c(3900)$, which has been observed by BESIII~\cite{BESIII:2013ris}
and Belle~\cite{Belle:2013yex} in the $e^+e^- \to J/\psi \pi^+\pi^-$ reaction as an unambiguous signal 
in the $J/\psi\,\pi$ invariant mass. The vicinity to the $\bar D D^*$ thresholds favors an 
interpretation in terms of a molecule of the two open-charm mesons, but a compact tetraquark 
hypothesis is not ruled out. Nontrivial rescattering of the three-body final state can 
also generate a signal that mimics a resonance. The biggest obstacle to the understanding 
of the nature of the $Z_c(3900)$ comes from the fact that the state has been observed in one production 
channel only, and most notably it does not appear in the high-statistics $B$ decay 
datasets available from LHCb. This casts doubts on its very existence as a QCD resonance.
In this respect, photoproduction offers an ideal setup to study the $Z_c$. Indeed, the 
coupling to photons can be related to the $Z_c \to J/\psi \pi$ decay, as in both 
cases the $c$ and $\bar c$ quarks must tunnel from the respective clusters (mesons 
for a molecule and diquarks for a tetraquark) before forming the charmonium or 
annihilate into a photon. If the state actually exists, we thus expect to see it 
in photoproduction, with the fine details depending on its internal structure. 

\begin{SCfigure}
    \centering
    \includegraphics[width=0.52\linewidth]{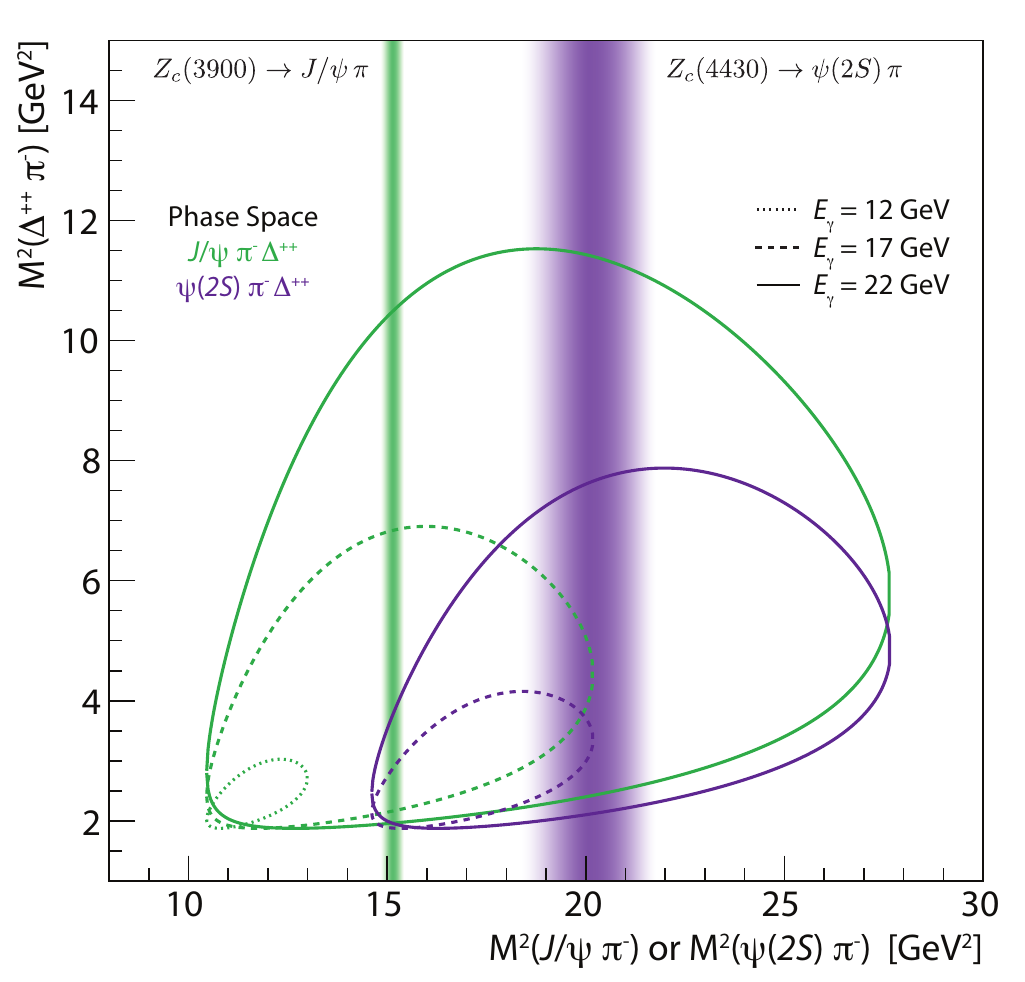}
    \caption{Sketches of the available phase space (the Dalitz plot boundary) for the $J/\psi\,\pi\,\Delta$ (green) and $\psi(2S)\,\pi\,\Delta$ (purple) systems produced in $\gamma p$ collisions under different assumptions about the incident photon energy (indicated by line styles).  The regions of phase space that would be populated by decays of $Z_c(3900)\to J/\psi\,\pi$ and $Z_c(4430)\to \psi(2S)\,\pi$ are shaded.}
    \label{fig:dalitz}
\end{SCfigure}

As previously noted, the $Z_c$ states that are observed to couple to a pion and a vector meson 
are ideal for exploration with polarized photon beams at an upgraded JLab.  As has
been demonstrated with GlueX data (Fig.~\ref{fig:eta_pi_prod}), analysis of the production
angles provides the signature for pion exchange.  This, coupled with the upgrade in energy,
allows one to explore $J/\psi\,\pi^\pm$ and $\psi(2S)\,\pi^\pm$ scattering, which provides
a unique production environment that is free from rescattering triangle singularities.  
Figure~\ref{fig:dalitz} shows the phase space available for $J/\psi\,\pi^-\Delta^{++}$ and
$\psi(2S),\pi^-\Delta^{++}$ under a variety of assumptions about the photon beam energy in
photon-proton collisions.  At 17~GeV one can search for the $Z_c(3900) \to J/\psi\,\pi$ in 
a region of phase space that is kinematically separated from $\Delta\pi$ resonances.  With
22~GeV photons the phase space opens significantly and searches for states like the
$Z_c(4430)\to\psi(2S)\pi$, a state observed in $B$ decay but not in $e^+e^-$ collisions, 
are permitted.  Both searches use the demonstrated capability
of the GlueX and CLAS12 detectors for reconstructing $J/\psi\to e^+e^-$ and pions.  In addition,
the pion-exchange process is strongest at threshold~\cite{Winney:2022tky}, making the 
upgraded CEBAF an ideal machine for these studies.

In parallel to this, further developments in lattice QCD will also allow us to study 
the state from a different perspective. In these numerical calculations it is 
indeed possible to simulate the elastic scattering of $J/\psi\,p$, which one cannot 
achieve in experiments because of the short lifetime of the $J/\psi$. Exploratory 
studies performed in the past with unphysically heavy pion masses showed no evidence 
for a $Z_c(3900)$~\cite{Cheung:2017tnt}. However, the recent calculations of doubly charm channels 
highlighted a strong dependence on the pion mass, which affects the previous studies 
and calls for new ones at the physical point in the future~\cite{Lyu:2023xro,
Padmanath:2022cvl, Francis:2018jyb}. If the $Z_c$ emerges from 
these calculations, its status as a QCD resonance be strengthened. 

We expect that the same conclusions will be reached by the photoproduction 
searches and the lattice results. If the state is found, that would be the final 
confirmation of a four-quark state, and opens a long list of new measurements 
related to its internal structure, for example, the $Q^2$ dependence in electroproduction.
If both lattice and experiments agree on the nonexistence of a $Z_c(3900)$, this will teach 
us more on the hadron final state interactions that generate the signal in $e^+e^-$
collisions and could have implications on the interpretation of other
members of the family of $Z_c$ and $Z_b$ states. Even more interesting, if lattice 
QCD and photoproduction data find opposite conclusions, it will change our 
understanding of final state interactions and of hadron structure in order to 
justify such an unexpected result.

\subsection{Light Meson Spectroscopy with 22 GeV Electrons}
\vspace{-0.2cm}
An increase in electron beam energy provides enhanced capabilities to explore the
spectrum of light hadrons, thereby extending the existing 12~GeV program in
hadron spectroscopy.  Currently, the real photon beam used in the GlueX experiment
that is generated from the 12~GeV electrons has a peak polarization of about
35\% at an energy of about 9~GeV.  This configuration is obtained by a choice
of orientation of the diamond lattice with respect to the photon beam.  Higher
polarization can be obtained but it comes with a cost of lowering the energy of
the peak intensity.  Therefore, an energy upgrade allows 
not only the obvious increase in photon beam energy
but also an option of producing similar energies to the current 12~GeV 
configuration but with dramatic enhancements in degree of polarization and flux.  The
secondary coherent peak, visible in the example in Fig.~\ref{fig:pol_fom}, 
can also be used.  For example, using a 22~GeV electron beam it is possible
to configure the beamline such that one has 12~GeV photons with about 70\%
linear polarization, while at the same time, having 15~GeV photons with about
50\% linear polarization.  This capability allows for exploration of energy-dependent
effects.  

\begin{SCfigure}
    \centering    
    \includegraphics[width=0.45\linewidth]{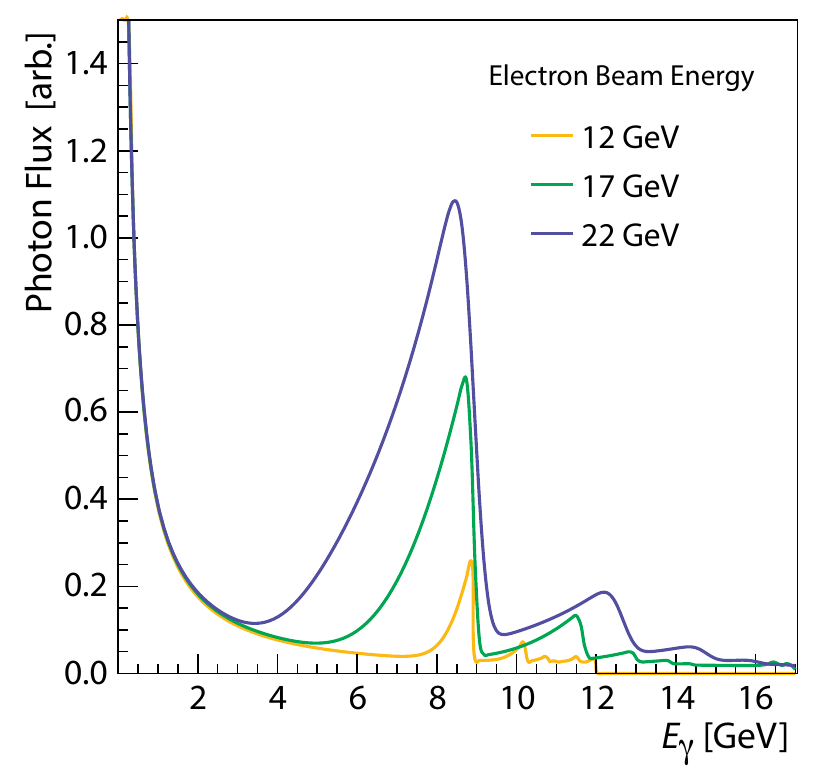}
    \caption{The coherent bremsstrahlung photon spectrum shown for three different choices
    of electron energy and three different choices of diamond radiator orientation such that
    the primary peak is the same position.  All three curves are normalized such that they have
    equal area above a common low energy cutoff.  Note that the vertical scale is truncated.}
    \label{fig:brem_spec}
\end{SCfigure}

The increase in rate at the high-energy end of the
photon spectrum is not anticipated to cause complications with increased
noise or electromagnetic background in the Hall D photon
beam line.  Figure~\ref{fig:brem_spec} shows photon energy spectra for three 
different electron beam energies.  The dominant low energy portion of the 
spectrum is responsible for detector backgrounds that ultimately limit the 
usable beam current.  The spectra in Fig.~\ref{fig:brem_spec} are normalized
such that all curves have a common area above the low-energy cutoff.
Therefore one can see that increasing the electron beam energy while adjusting 
the diamond to keep the coherent photon flux peak at a fixed energy results 
in a larger fraction of useful high-energy photons with respect to the 
background-generating low energy portion of the spectrum.

The increased capabilities of an upgraded machine can be used in a variety
of ways.  By increasing the polarization of real photons with GlueX,
one enhances the dependence of the production amplitude 
on the orientation of the decay plane with respect
to production plane, the angle $\Phi$ depicted in Fig.~\ref{fig:eta_pi_prod}.
This provides enhanced capability to discriminate between production mechanisms.
Conducting meson spectroscopy studies at higher energy also enhances the
kinematic separation between the decay products of produced mesons and
excited baryons, {\it i.e.,} one has much better distinction between beam 
fragmentation and target excitation regimes.  Finally, one has the ability
to study production mechanism dependence on energy.  The CLAS12 light hadron 
spectroscopy program will also greatly benefit from the energy upgrade, 
providing a high intensity flux of quasi-real photons at high energy and 
the extra capability of studying the $Q^2$ evolution of any new state produced.

In summary, an upgraded electron beam at 22 GeV will allow the hadron spectroscopy program at Jefferson Lab to cross the critical threshold into the region where $c\bar{c}$ states can be produced in large quantities, and with additional light quark degrees of freedom. This opens a new opportunity to study production of exotic states and contribute with potentially decisive information about their internal structure, which would us to understand their place in the landscape of hadrons generated by QCD. Moreover, with a 22 GeV electron beam it will be possible to extend a large class of spectroscopic studies conducted with the 12 GeV program.
 
\clearpage \section{Partonic Structure and Spin}
\label{sec-wg2}

The 1970's marked a significant turning point in the field of particle physics with the development of the parton model. This revolutionary concept transformed our understanding of the structure of hadrons, revealing them to be composed of quarks and gluons that interact as described by the theory of quantum chromodynamics (QCD). Since then, there has been remarkable progress in both experimental and theoretical research on the subject, providing us with tantalizing insights into the internal structure of protons and neutrons. While our understanding still remains incomplete, these advancements have paved the way for exciting new discoveries in particle physics.

The parton model introduced the fundamental quantities known as parton distribution functions (PDFs). These functions enable quantification of the number densities of quarks and gluons within hadrons as a function of their momentum fraction relative to the parent hadron. Precise determination of PDFs from experimental data has been a challenging, active area of research that requires a variety of experimental data, ranging from low-energy reactions, such as those at JLab, to high-energy experiments at the LHC, where PDFs also play an important role in searches for physics beyond the standard model (BSM). This concept has proven to be a cornerstone in the field, providing crucial insights into the nature of hadronic matter. 

The field of parton physics is on the cusp of a new era of exploration, with the planned experiments of the JLab 12 GeV program and those at the future EIC with dedicated beam polarization capabilities. These experiments are expected to provide high-definition maps of the internal structure of hadrons, using a range of techniques such as reactions involving polarized hadrons to access helicity-dependent PDFs, and dedicated tagged experiments to probe the structure of mesons, among others. These cutting-edge techniques promise to deepen our understanding of the fundamental constituents of matter and their interactions, and represent a major step forward in our quest to unravel the emergent phenomena of QCD.  However, the combined scientific program of JLab 12 GeV and EIC has a gap in kinematics, which is where the JLab 22~GeV upgrade can offer critical insights for precision studies of partonic structure. With its enhanced energy range, the JLab 22 GeV upgrade can provide access to a wider range of kinematic regimes, enabling the investigation of specific partonic processes and their properties with greater precision. 

In the upcoming sections, we will explore the importance of upgrading the energy capacity of the CEBAF accelerator for significantly improving precision and phase space coverage of experimental data. We will discuss key measurements and physics outcomes, including:
\begin{itemize}
    \item Precision measurements of the nucleon light sea in the intermediate to high-$x$ range, which can help validate novel theoretical predictions for the intrinsic sea components in the nucleon wave function and support BSM searches at colliders;
    \item Precision determination of the strong coupling and helicity structure of the nucleon;
    \item Unique opportunities to explore the internal structure of mesons.
\end{itemize}
An energy upgrade to 22 GeV will provide unprecedented and complementary opportunities to both existing and planned experiments, laying the foundation for groundbreaking discoveries that will allow us to understand the emergent phenomena of QCD.

\subsection{Nucleon Light Sea in the Intermediate-$x$ Range}
\label{sec:lightsea}

There is an extensive worldwide effort to determine parton distribution functions (PDFs) of the nucleon, led by QCD analysis collaborations such as ABM, CTEQ-JLab (CJ), CTEQ-TEA (CT), JAM, MSHT, and NNPDF. These groups focus on extracting PDFs from high-energy data, comprised of legacy experiments as well as more recent data sets measured at the LHC and other facilities. To ensure the validity of the purely leading-twist interpretation of theoretical predictions for all datasets, most groups employ kinematical cuts in Deep Inelastic Scattering (DIS) processes. For instance, commonly used cuts include virtuality of the exchanged photon $Q^2 > 4~{\rm GeV}^2$ and the invariant mass of the unobserved system in the final state $W^2 > 12.25~{\rm GeV}^2$. 

\begin{figure}[t!]
    \begin{center}  
        \includegraphics[width=0.52\textwidth]{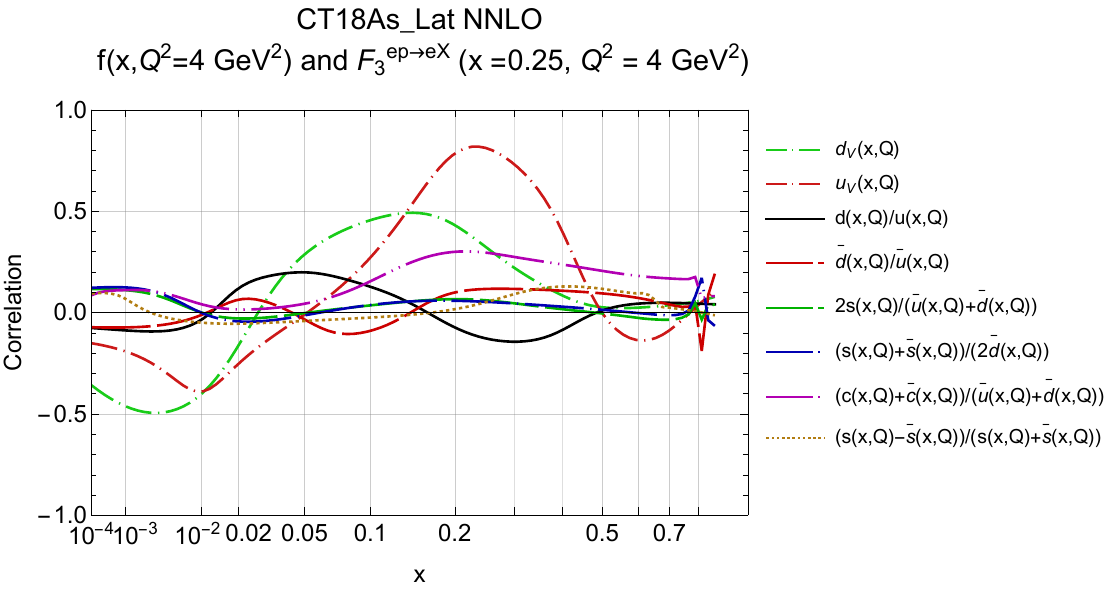}
        \includegraphics[trim={0 8cm 0 4.5cm},clip,width=0.47\textwidth]{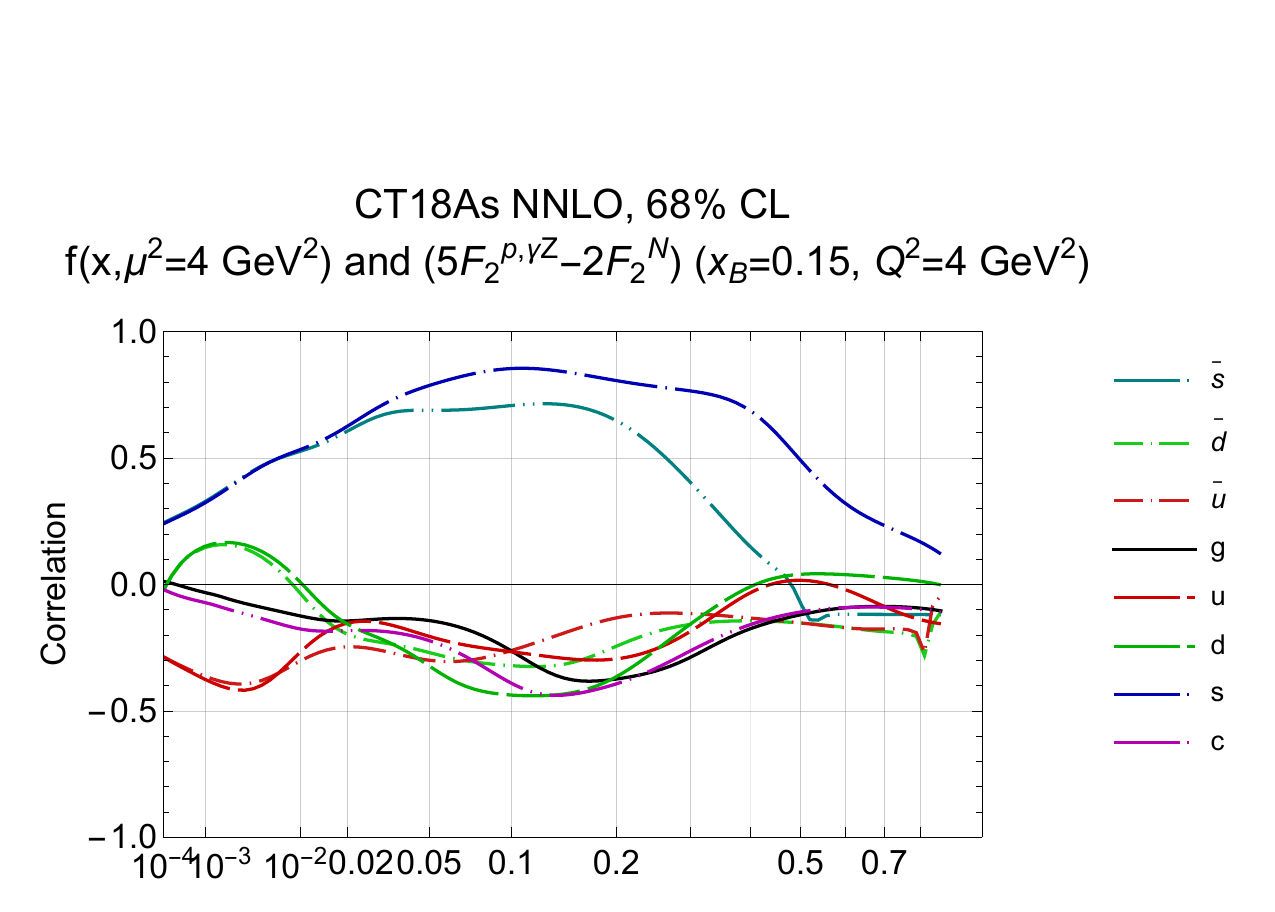}
    \end{center}
    \caption{Hessian correlation~\cite{Pumplin:2001ct,Nadolsky:2001yg,Nadolsky:2008zw} obtained for two CT18 variant fits, CT18AS\_Lat and CT18As NNLO, between various PDFs and the neutral-current $F_3$ structure function (left) and a structure function combination $\mathcal{F} =(5F_2^{\gamma Zp}) - 2 F_2^{\gamma N}$ (right) in a specific kinematic region accessible to the 22 GeV program.}
    \label{fig:CTcorr}
\end{figure}

However, our understanding of the sea-quark PDFs remains limited, especially for large $x$, such as $x\! >\! 0.3$. In this region, the valence PDFs dominate and are much larger than the sea-quark PDFs. Moreover, even the valence PDFs are poorly known in the large-$x$ extrapolation region ($x > 0.7$). Practitioners have proposed different scenarios to handle the light-quark sea at $x > 0.3$. For instance, the ``low-sea'' scenario adopted by CT and MSHT assumes that the sea PDFs are considerably smaller than the valence PDFs at high $x$. On the other hand, the ``high-sea'' scenario for instance, NNPDF4.0 permits the sea-quark PDFs to be SU(3)-symmetric at the highest $x$ or even comparable in size to the $d$-quark PDF at $x\! >\! 0.7$. The detailed flavor separation among the sea PDFs, especially at high $x$, constitutes a significant challenge: existing data do not possess sufficient discriminating power to invalidate either scenario, a fact which underscores the need for further constraints on the sea PDFs. To address this, a potential energy upgrade of JLab to 22 GeV would yield additional experimental data at higher energies and enable access to larger values of $Q^2$, thereby enhancing our exploration of the nucleon's  sea-quark sector. One way to quantify the potential sensitivities within the CT framework is to examine the Hessian correlation \cite{Pumplin:2001ct,Nadolsky:2001yg,Nadolsky:2008zw} between specific combinations of PDFs and observables accesible at JLab 22, particularly those related to parity-violating $\gamma Z$-interference structure functions. In Fig.~\ref{fig:CTcorr} (left), we provide an illustrative example, depicting the PDF correlations of $F^{\gamma Z}_3$. The latter exhibits notably significant effects on the high-$x$ light-quark $q-\bar{q}$ PDF combinations, thereby improving control over these would establish a more robust foundation for unraveling the flavor dependence of the high-$x$ sea.

Mapping out the light-sea nucleon sector is an essential endeavor in the study of hadron structure. Sea asymmetries like $\bar{d}-\bar{u}$ and $\bar u(x) + \bar d(x) - s(x) - \bar s(x)$, exhibit maximal decorrelation from the gluonic field configurations of the nucleon, providing unique insights into QCD's non-perturbative flavor dynamics. For instance, the authors in \cite{ChangPRL}, utilizing a QCD-inspired model developed by Brodsky, Hoyer, Peterson, and Sakai (BHPS)~\cite{BHPS}, have shown intriguing agreement between the BHPS model and the prevailing experimental constraints on sea asymmetries. This concurrence suggests the potential existence of the so-called ``intrinsic'' nucleon sea, despite significant systematic uncertainties on $s+\bar{s}$ that arises from incomplete knowledge of kaon fragmentation functions to analyze semi-inclusive DIS (SIDIS) data with kaons in the final state. 

Similarly, the recently completed SeaQuest experiment at Fermilab has placed constraints on $\bar{d}-\bar{u}$ using lepton-pair production in $pp$ and $pd$ collisions across a broad Bjorken-$x$ range~\cite{SeaQuest:2021zxb}, achieving higher statistical precision compared to the earlier measurements by the NuSea experiment~\cite{PhysRevD.64.052002}. These measurements exhibit relatively good agreement with similar findings reported by the HERMES Collaboration~\cite{HERMES:1998uvc} using SIDIS data, albeit in a different region of the exchange photon virtuality $Q^2$ and with lower statistics. To establish the universality of these observations, new SIDIS measurements are essential. The 12 GeV program and its upgrade will offer a clear pathway to studying nucleon asymmetries and validating their universal nature within the next generation of QCD global analyses of PDFs.

\noindent
\underline{\emph{Nucleon Strangeness}}. 
Emblematic of the challenge of disentangling the flavor dependence of the nucleon sea, the size and high-$x$ shape of the strange-quark
PDF remain poorly determined.
Thus, a quantitative understanding of the nonperturbative strange quark sea remains elusive despite numerous investigative attempts to clarify its structure. Uncertainties in nucleon strangeness limit a variety of other extractions based on collider data --- for example, determinations of the CKM matrix element, $V_{cs}$, or the $W$-boson mass, $m_W$, both of which rely on precise knowledge of the strange quark PDF.

As photons interact with $d$- and $s$-quarks with equal strength, it is challenging to isolate the behavior of the strange PDF solely using inclusive Deep Inelastic Scattering (DIS) observables. This is the case even when using proton and neutron targets, without also resorting to weak probes which may access independent flavor currents within the nucleon~\cite{Cooper-Sarkar:2015boa}. Information to constrain the strange-quark PDF has therefore conventionally involved charged-current neutrino DIS, typically involving inclusive charm-meson production off heavy nuclei to obtain sufficient event rates. Assessments of the CCFR~\cite{CCFR:1994ikl} and NuTeV~\cite{NuTeV:2007uwm} neutrino and antineutrino cross-sections from the Tevatron, along with more recent data from the CHORUS \cite{Kayis-Topaksu:2011ols} and NOMAD \cite{NOMAD:2013hbk} experiments at CERN, have led to a strange to light-antiquark ratio \mbox{$R_s = (s+\bar s)/(\bar u + \bar d)$} of approximately $R_s\! \approx\! 0.5$. However, interpreting the neutrino-nucleus data is complicated by uncertainties in nuclear effects both in the initial and final states. These uncertainties arise from the challenges in connecting nuclear structure functions to those of free nucleons \cite{Kalantarians:2017mkj} and dealing with the charm quark energy loss and $D$ meson-nucleon interactions during hadronization within the nucleus \cite{Accardi:2009qv, Majumder:2010qh}.

A complementary avenue to information on nucleon strangeness leverages inclusive $W^\pm$ and $Z$ boson production in $pp$ collisions, which involve weak without nuclear structure complications. At the same time, hadroproduction loses the comparative simplicity of the DIS production mechanism discussed above. Recent ATLAS Collaboration data at the LHC suggested a larger strange quark sea than traditionally obtained from neutrino scattering, with $R_s \approx 1.13$ at $x\! =\! 0.023$ and $Q^2\! =\! 1.9$~GeV$^2$ \cite{ATLAS:2012sjl, ATLAS:2016nqi}. The latest analysis combined HERA and ATLAS data and found results consistent with the earlier enhancement, although this result exhibited some tension with the $\bar d > \bar u$ behavior typically preferred by the Fermilab E866 Drell-Yan (DY) experiment \cite{NuSea:1998kqi, NuSea:2001idv}, as well as the more recent SeaQuest data~\cite{SeaQuest:2021zxb}. Alekhin 
{\it et al.} argued that the strange quark enhancement was due to the suppression of the $\bar d$ sea at small $x$ \cite{Alekhin:2008mb, Alekhin:2014sya, Alekhin:2017olj}, again emphasizing the difficulty of isolating the strange PDF from the rest of the light-quark sea. The ATLAS $Z \to \ell \ell$ data were found to be at odds with CMS results, which align with the ABMP16 global QCD analysis \cite{Alekhin:2017kpj}. A recent analysis by Cooper-Sarkar and Wichmann (CSKK) found no significant tension between HERA, ATLAS, and CMS data and supported an unsuppressed strange PDF at low $x$ \cite{Cooper-Sarkar:2018ufj}. However, their standard fit again appears to be in tension with the E866 DY data, although forcing $\bar d > \bar u$ only reduces $R_s$ by around 10\% \cite{Cooper-Sarkar:2018ufj}.

Yet another approach to obtain information on the strange-quark PDF at lower energies involves semi-inclusive deep-inelastic scattering (SIDIS). In this method, the detection of charged pions or kaons in the final state serves as an indicator of the initial state PDFs; whereas the experimental approaches noted above are limited by nuclear uncertainties (for neutrino DIS) or additional uncertainties in hadroproduction (for $pp$ scattering), SIDIS is limited by uncertainties in the fragmentation functions governing hadronization of the strange quark. Previously, the HERMES Collaboration \cite{HERMES:2008pug} examined the $K^+ + K^-$ production data from deuterons and discovered a significant increase in the extracted strange PDF for $x \lesssim 0.1$ when using leading-order (LO) hard coefficients, but a notable suppression for $x \gtrsim 0.1$. A later analysis \cite{HERMES:2013ztj}, utilizing new $\pi$ and $K$ multiplicity data, observed a less marked rise at small $x$, but virtually zero strangeness for $x > 0.1$. The analysis in Ref.~\cite{HERMES:2013ztj}, like others, operates under the strong assumption that the nonstrange PDFs and fragmentation functions (FFs) are well-understood, disregarding potential correlations. However, previous analyses of polarized SIDIS data revealed that FF assumptions can significantly influence the extracted helicity PDFs~\cite{Leader:2010rb, Leader:2011tm}, necessitating a concurrent analysis of PDFs and FFs for conclusive results \cite{Sato:2016wqj}. Aschenauer {\it et al.}~\cite{Aschenauer:2015rna} highlighted the importance of an LO extraction as an initial step towards a next-to-leading-order (NLO) analysis of semi-inclusive DIS data, given its current unavailability. Borsa {\it et al.} \cite{Borsa:2017vwy} later explored how SIDIS data can constrain unpolarized proton PDFs through an iterative reweighting procedure, advancing towards a comprehensive global analysis of PDFs and FFs. More recently, the JAM Collaboration has carried out a combined analysis to {\it simultaneously} determine unpolarized PDFs and FFs using DIS, SIDIS, and hadron production in $e^+e^-$ reactions. The analysis found mild trends for the kaon SIDIS data to suppress the strange quark PDF around $x \gtrsim 0.1$ \cite{Sato:2019yez}.

Parity-Violating Deep Inelastic Scattering (PVDIS)~\cite{Brady:2011uy, Hobbs:2008mm} provides another independent approach to disentangle the flavor structure of the nucleon sea, including the strange PDF. This can be attributed to the distinct electroweak currents --- especially the interference of electromagnetic and weak exchanges --- which probe distinct parton-level flavor currents in the nucleon. The associated low-energy observables like the parity-violating asymmetry, $A^\mathrm{PV}$, and interference structure functions, $F^{\gamma Z}_i$, are sensitive to unique PDF combinations, such that information on the nucleon's light sea and strange content may be obtained through the joint analysis of certain electroweak observables.
%
%
For illustration, certain combinations of electroweak structure functions on the proton and isoscalar nucleon like ${\cal F}\equiv (5F_2^{\gamma Zp}) - 2 F_2^{\gamma N}$ are proportional to $x(s+\bar{s})$ at LO in pQCD; as such, parity-violating scattering from the proton and high-luminosity electromagnetic scattering from the deuteron could offer new sensitivity to the symmetric strange sea. This can be seen in the right panel of Fig.~\ref{fig:CTcorr}.
While higher-order perturbative corrections, 2-body nuclear effects in the deuteron, and other considerations may diminish the resulting correlation between
this structure function combination and the strange PDF, it suggests the possibility of using a mixture
of electroweak observables to enhance sensitivity to nucleon strangeness and the
nucleon sea.
It also highlights the unique importance of parity-violating observables like the structure function $F_2^{\gamma Zp}$ to future PDF analyses. However, PVDIS measurements pose a challenge due to their high-luminosity requirements compared to other reactions, and currently, there are no PVDIS data available for QCD global analysis. This situation is anticipated to improve in the upcoming years, thanks to the high-luminosity capabilities of the SoLID 12 GeV program at JLab and the potential upgrade to the 22 GeV energies.  
The anticipated distinct correlations between given  PDFs and an observable $\cal F$ (Fig.~\ref{fig:CTcorr}) were obtained using the latest PDF fits from the CT collaboration, CT18As NNLO~\cite{Hou:2022onq}. We stress that  these conclusions are independent of any consideration of projected experimental and theoretical systematic uncertainties but instead they rely only on the inherent correlation between these PVDIS observables and the underlying PDFs from which they are computed.

\begin{figure}[t!]
    \begin{center}  
        \includegraphics[width=0.4\textwidth]{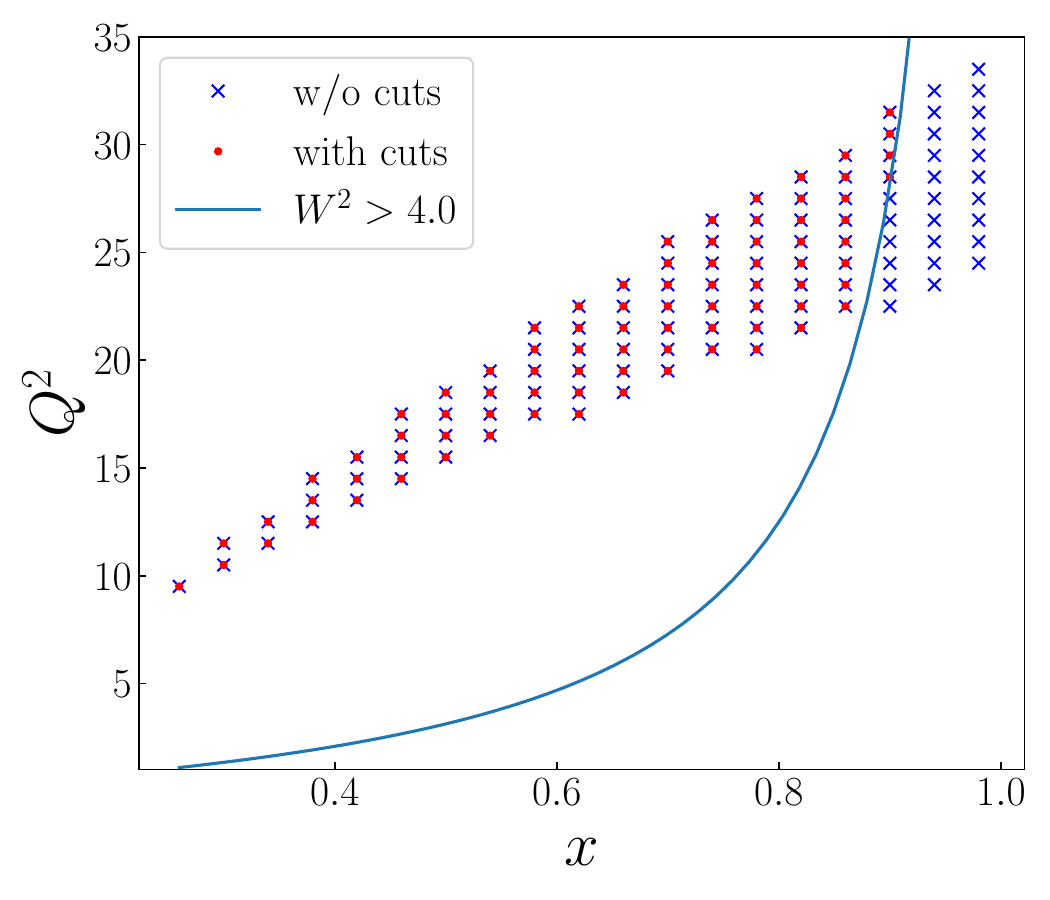}
        \includegraphics[width=0.4\textwidth]{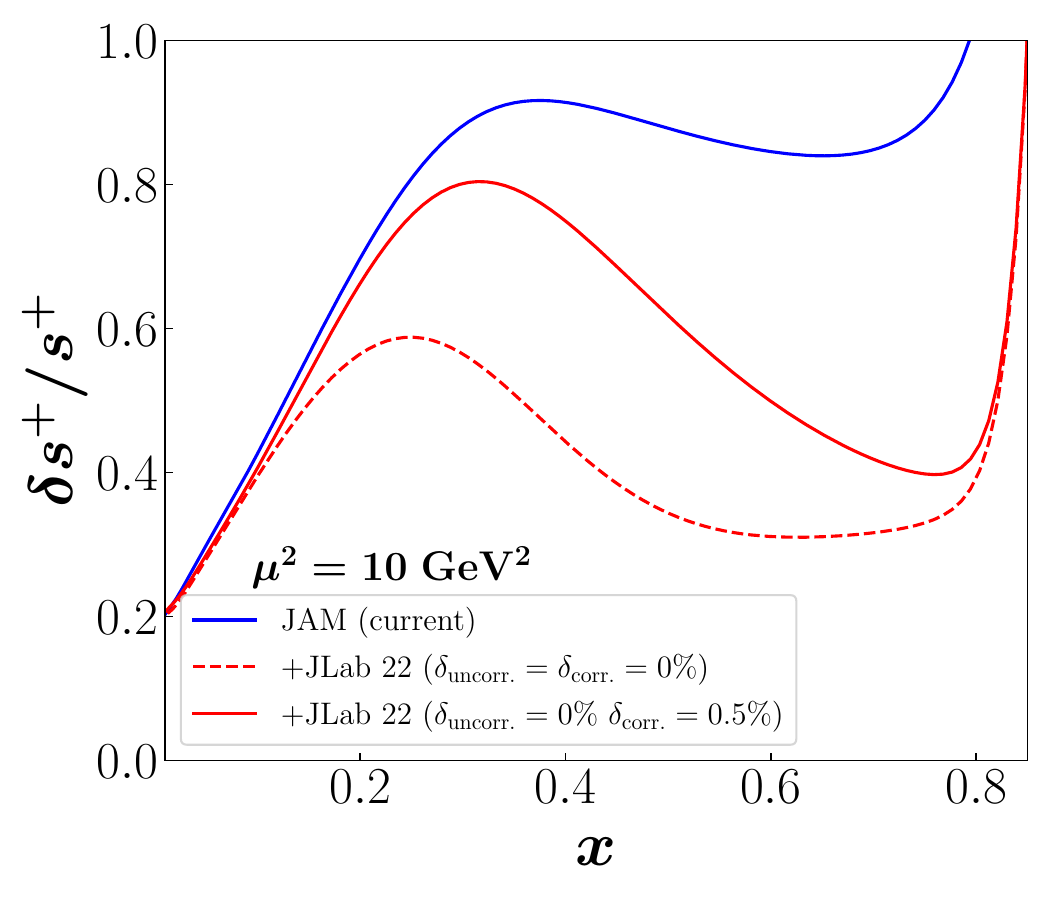}
    \end{center}
    \caption{(Left) The kinematic coverage available from a 22~GeV beam and the unmodified acceptance of the SoLID spectrometer. (Right) The impact of simulated 22~GeV PVDIS proton and deuteron data on the strange quark PDF $s^+=s+\bar{s}$ in the JAM framework~\cite{Cocuzza:2022jye}.  A high statistics measurement of $A_{PV}$ with realistic normalization uncertainty measured with 22~GeV beam, the SoLID spectrometer, and the luminosity described in the text is simulated.}
    \label{fig:kinematics}
\end{figure}

For a pseudodata-based impact study analysis, we carry out an NLO global analysis within the JAM PDF analysis framework. We consider simulated $A^{\rm PV}$ proton and deuteron data at the expected kinematics from the unmodified SoLID spectrometer, as shown in Fig.~\ref{fig:kinematics} (left). This setup provides access to PVDIS asymmetries approximately in the range of 0.07\% to 0.21\%. A comparable experiment at 11~GeV would cover a similar $x$ range. 
However, operating at 22~GeV offers several advantages: the higher $Q^2$ suppresses power corrections and expands the $x$ region suitable for analysis within a joint QED+QCD factorization framework \cite{Liu:2020rvc}. Consequently, a significant fraction of the higher-energy data could be incorporated into various global PDF fits. Furthermore, at the smaller $x$ values accessible at 22 GeV, PVDIS will demonstrate enhanced sensitivity to $F_3^{\gamma Z}$, which can may probe the matter-antimatter asymmetry in the light sea sector, potentially including the strange sector {\it i.e.}, 
$s-\bar{s}$. This sensitivity would represent another unique aspect of the JLab22 PVDIS program.

\noindent
\underline{\emph{Charm Content of the Proton}}. Employing a 22 GeV electron beam opens a sufficient phase space for the creation of charm-anticharm pairs, significantly boosting charm production rates. If we can identify the charm quark in the final state, charm structure function measurements at JLab22 could offer valuable insights into charm production, especially within the intermediate to large-$x$ range. Such data could shed light on any possible nonperturbative charm component to the proton, an area recently explored by the NNPDF Collaboration \cite{Ball:2016neh,Ball:2022qks,NNPDF:2021njg} with findings qualitatively in line with two model predictions based on intrinsic charm~\cite{BHPS,Hobbs:2013bia}. In contrast, however, the CT Collaboration~\cite{Guzzi:2022rca} also recently explored the hypothesis of fitted charm, finding no clear evidence based on its default sets of high-energy data. These distinct findings suggest a need for additional data to resolve existing tensions among various PDF analysis efforts. While the EIC plans to measure the charm structure function to glean information on intrinsic charm in the range of $0.001 < x < 0.6$ at relatively low energies \cite{Kelsey:2021gpk,AbdulKhalek:2021gbh}, the high-luminosity capabilities of the proposed JLab 22~GeV upgrade could provide unprecedentedly precise data for probing the charm structure function in the mid-to-high-$x$ region. Precise data on charm production are important given the necessity of understanding the dynamics of heavy quarks, which are relevant for accurate perturbative QCD calculations~\cite{Gao:2021fle} as well as astrophysical processes. For instance, prompt neutrinos originating from charm decays are a dominant background for cosmic neutrinos at IceCube and KM3Net \cite{Gauld:2015yia}. Therefore, precise measurements of the charm structure function, as obtainable through the JLab 22 GeV upgrade, are vital for advancing our understanding of prompt neutrino production and reducing background uncertainties in astroparticle physics. 

\begin{figure}[t]
    \begin{center}
        \includegraphics[width=0.335\textwidth,angle=-90]{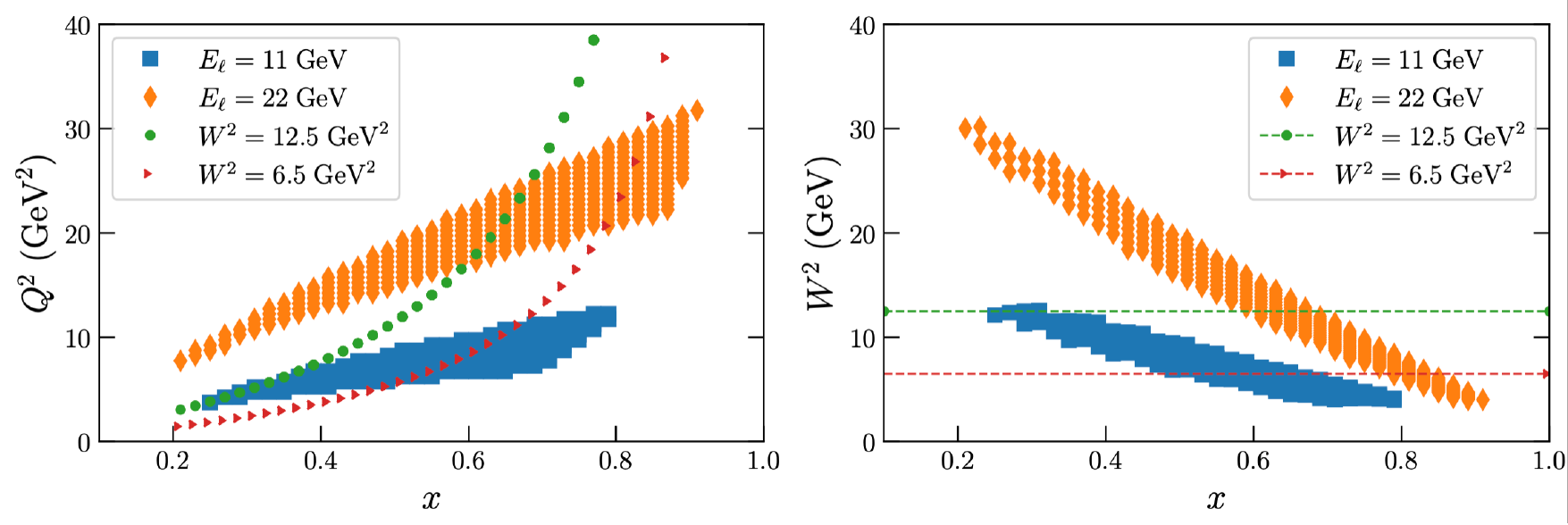}
    \end{center}
    \caption{The kinematic coverage in the $(x,Q^2)$ plane (left) and in the $(x,W^2)$ plane (right) of the projected JLab measurements with a lepton beam energy of 22~GeV, compared with the corresponding coverage of the current data taken with an 11~GeV beam. We also indicate $W^2=12.5$ GeV$^2$, the usual kinematic cut in most global PDF determinations, as well as $W^2=6.5$ GeV$^2$.}
    \label{fig:1-DIS}
\end{figure}

\noindent
\underline{\emph{High-$x$ PDFs and Synergies with HEP}}. 
Boosting the JLab lepton-beam energy to $E_e\! =\! 22$ GeV offers the prospect of extending experimental coverage of the large-$x$ region to an approximate value of $x\simeq\! 0.65$ as shown in Fig.~\ref{fig:1-DIS}; this in turn has implications for high-energy physics (HEP), aiding in the quest for a comprehensive understanding of the unpolarized quark and gluon structure of the proton~\cite{Gao:2017yyd,PDF4LHCWorkingGroup:2022cjn}. An enhanced energy will keep measurements within the perturbatively calculable DIS domain, wherein higher-twist and target-mass effects are comparatively suppressed and may be better controlled. To prevent contamination from non-leading twist when extracting twist-2 PDFs, QCD analyses~\cite{Hou:2019efy,NNPDF:2021njg,Bailey:2020ooq} commonly apply kinematical cuts, {\it e.g.}, $W^2=(p+q)^2 \ge 12.5$ GeV$^2$ and $Q^2=-q^2\!>\!4$\ GeV$^2$, where $p$ and $q$ denote the standard DIS kinematics of the target and exchanged virtual photon, respectively. With $E_e\! =\! 22$ GeV, many more DIS events at JLab would traverse these kinematical cuts as shown in Fig.~\ref{fig:1-DIS} in a fashion which could permit more detailed separation of the twist-2 PDFs at high $x$. This sensitivity to the high-$x$ PDFs is further reflected in the PDF-mediated correlations computed by the CT Collaboration as shown and discussed in Fig.~\ref{fig:CTcorr}. Ultimately, the flavor separation provided by JLab 22 GeV measurements would complement future LHC and EIC measurements in exploring various PDF combinations --- for instance, differences between the high-sea and low-sea scenarios for proton PDFs and provide controls over higher-twist effects to constrain subleading contributions relevant for antiquark PDFs at large $x$. 

\begin{figure}[t]
    \begin{center}
        \includegraphics[width=0.8\textwidth]{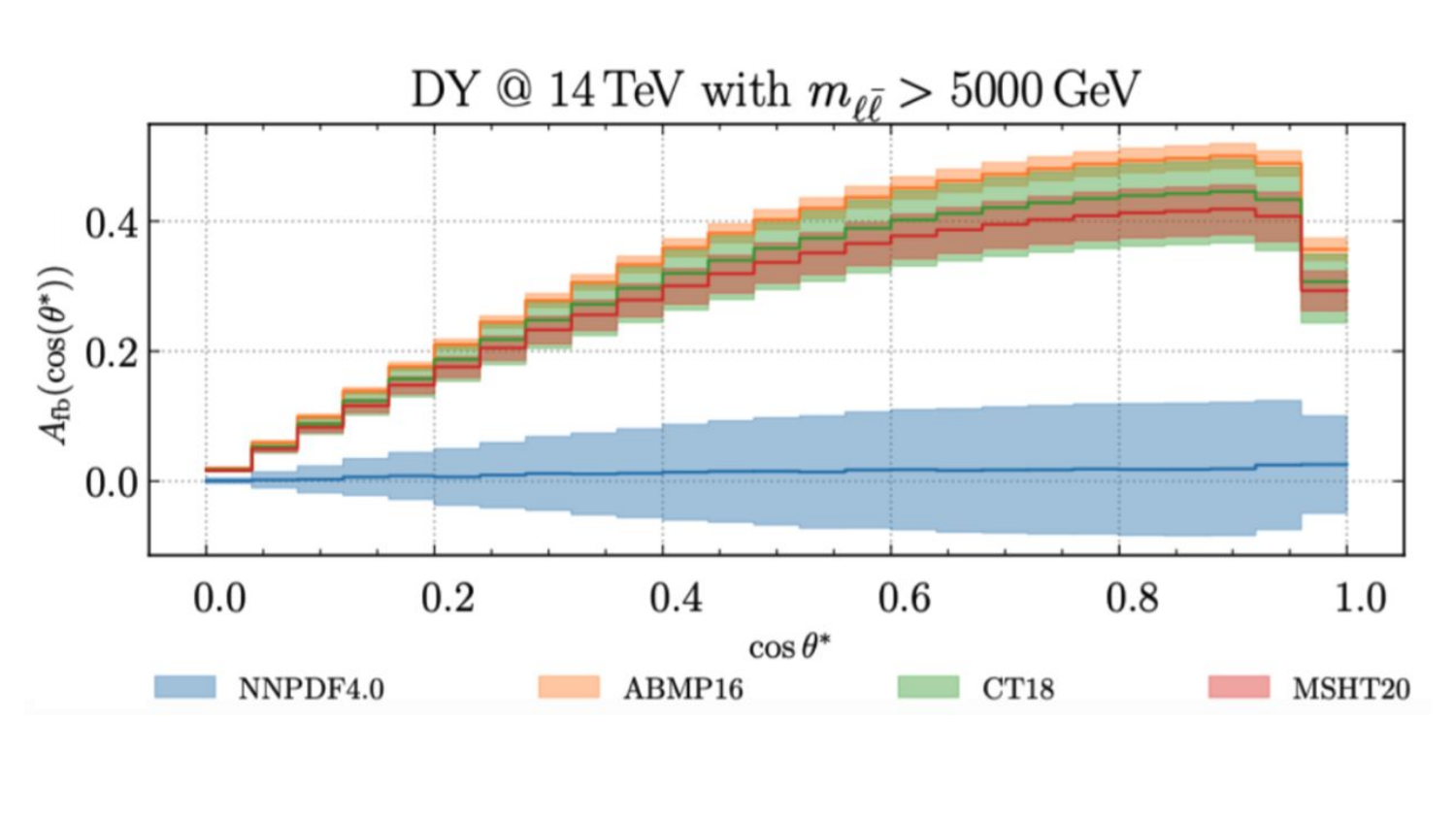}
    \end{center}
    \vspace{-1cm}
    \caption{The forward-backward asymmetry as a function of Collins-Soper angle \cite{Collins:1977iv} in high-mass DY production at the LHC~\cite{Ball:2022qtp} provides enhanced sensitive to the behavior of the quark and antiquark PDFs in the large-$x$ extrapolation region.}
    \label{fig:2}
\end{figure}

In light of this, data procured from the JLab 22~GeV upgrade will uniquely augment our understanding of the large-$x$ proton structure, thereby influencing other physics analyses that are sensitive to this structure. These include searches for BSM signatures in the tails of high-mass Drell-Yan (DY) distributions or measurements at forward rapidities at the LHC and high-energy astroparticle physics at neutrino telescopes. To underscore this assertion, the region accessible up to $x\simeq 0.65$ is instrumental in generating reliable predictions for Beyond the Standard Model (BSM) searches at the LHC.
To illustrate this, Fig.~\ref{fig:2} presents predictions made by the NNPDF Collaboration for the forward-backward asymmetry in high-mass DY production as a function of Collins-Soper angle $\cos(\theta^*)$~\cite{Collins:1977iv} at the LHC \cite{Ball:2022qtp}. This observable has an enhanced sensitivity to the slope of quark and anti-quark PDF in the large-$x$ region such that, for instance, if $q=\bar{q}$, the asymmetry $A_{FB}$ is zero.  At present, theory predictions for $A_{FB}$ give both (positive and zero) scenarios, which prevents the use of the observable to discriminate BSM physics. In this context, measurements such as PVDIS at the 22 GeV upgrade will provide the necessary constraints on the sea PDFs at intermediate to large $x$, and those will help to identify potential BSM signs in the DY rapidity and invariant mass distributions in the context of EFT-based analyses, including simultaneous fits of BSM contributions alongside PDFs as developed in, {\it e.g.}, Refs.~\cite{Greljo:2021kvv,Gao:2022srd}.
In addition to high-energy scattering, efforts to
ascertain $\delta_\mathrm{CP}$ in the neutrino sector will depend on a next generation of long-baseline experiments like
DUNE; for these, theoretical accuracy~\cite{Ruso:2022qes} requires knowledge of the high-$x$ PDFs for which JLab 22 data could serve as a valuable baseline relevant to extrapolations to lower $W^2$.


\subsection{Polarized PDFs and Strong Coupling}

The JLab accelerator, operating at 22 GeV with a high level of polarized luminosity, offers a unique opportunity to conduct precise determination of the nucleon spin structure functions. Specifically, this machine is well-suited to investigate the deep valence quark (high-$x$) region and to explore the polarized sea in the middle-$x$ region. Additionally, the data obtained from these experiments will be crucial for achieving a more accurate determination of the strong coupling constant.

\begin{figure}[htb!]
    \begin{center}
        \includegraphics[width=0.6\textwidth]{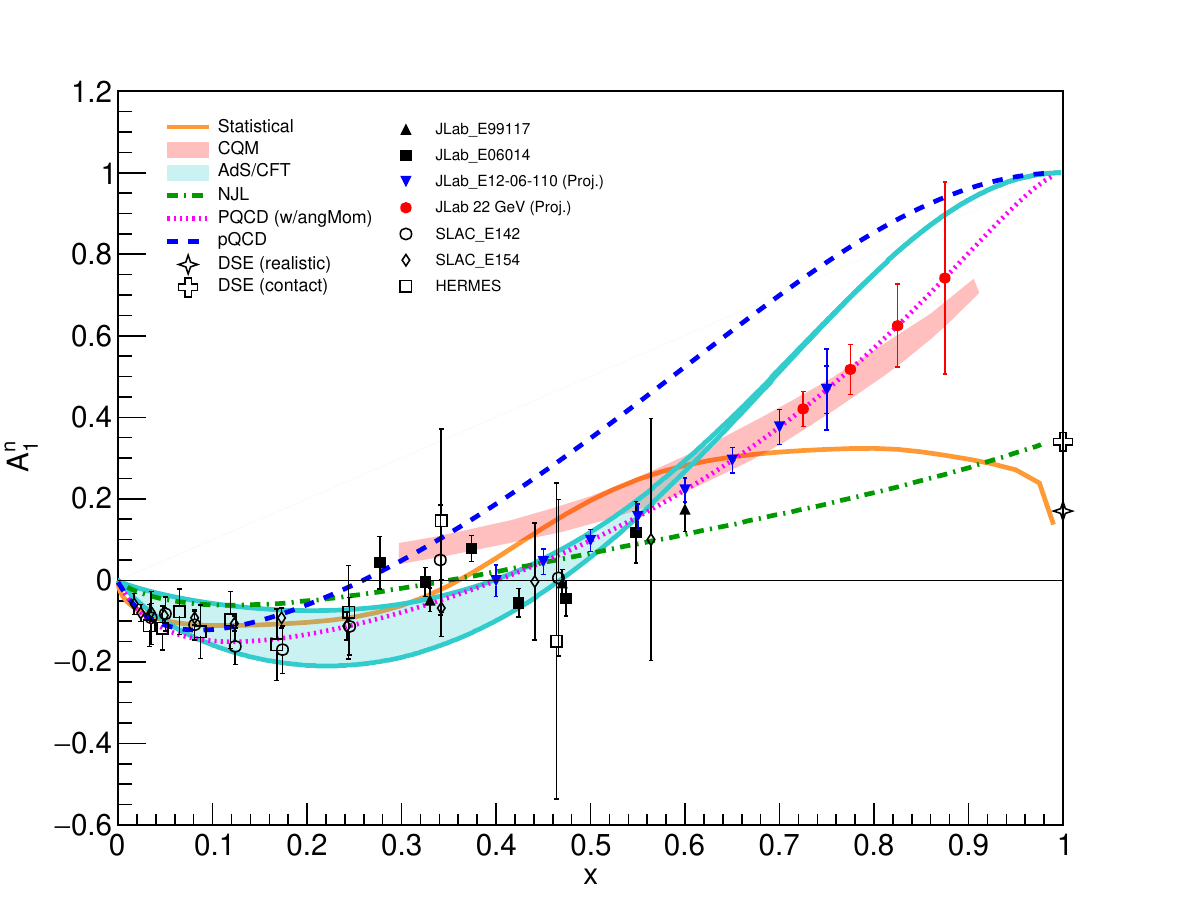}
    \end{center}
    \caption{Neutron spin asymmetry $A_1^n$ from 6 GeV experiments (black symbols) and expected precision from the completed 11 GeV experiment E12-06-110 (blue down triangles). A possible follow-on experiment with 30 days of 22 GeV beam and standard Hall C equipment would extend the precision and kinematic reach of the existing data significantly (red circles). Various previous models for the $x$-dependence and the expected asymptotic value at $x \rightarrow 1$ are shown in addition to a new prediction based on  light-front holographic QCD (AdS/CFT)~\cite{AdSCFT}.}
 \label{A1n}
\end{figure}

\noindent
\underline{\emph{Polarized PDFs from JLab at 22 GeV}}.  Inclusive structure functions  of polarized nucleons have been relatively well measured across a broad of DIS kinematic ranges. Particularly in the valence region, where $x > 0.5$, data from JLab at 6 GeV, and 12 GeV, have yielded and will continue to provide unparalleled insights into the nucleon's quark helicity and flavor structure. By increasing the beam energy to 22 GeV, we can eliminate the remaining gap in determining the asymptotic valence quark structure at the highest achievable $x$, effectively cutting in half the unexplored region $x = 0.8 ... 1$ inaccessible at JLab at 11 GeV (see Fig.~\ref{A1n}). Some of the latest predictions indicate a significant shift in the spin carried by $d$-quarks from negative values below $x = 0.8$ to full polarization of $+1$ at $x = 1$~\cite{AdSCFT}. A higher beam energy would enable access to significantly higher momentum transfer $Q^2$ and final state mass $W$, thereby reducing model uncertainties stemming from resonance contributions, higher twist, and target mass effects. A pleasant side effect of a higher beam energy is an increase in count rates for the same kinematics, enhancing statistical precision. Beyond the extreme $x \rightarrow 1$ limit, an extended $Q^2$ range could offer opportunities to examine the $Q^2$ evolution of PDFs, as well as production of hadrons with high transverse momentum in the moderate $x$ region ($x = 0.1...0.6$), thereby indirectly revealing the largely enigmatic ``valence gluon PDFs'' in this region. By juxtaposing different beam energies at the same $x$ and $Q^2$, subleading structure functions such as $R$, $g_2$, and $A_2$ can be obtained, offering insights into quark-gluon correlations within the nucleon. Accessing a higher final-state invariant mass considerably widens the interpretable range for flavor tagging through semi-inclusive production of pions, kaons, and other mesons and baryons. This opens up possibilities for in-depth studies of sea quark PDFs in this intermediate to high-$x$ region, believed to be dominated by the nucleon meson cloud.

\begin{figure}[htb!]
    \begin{center}
    \includegraphics[width=0.4\textwidth]{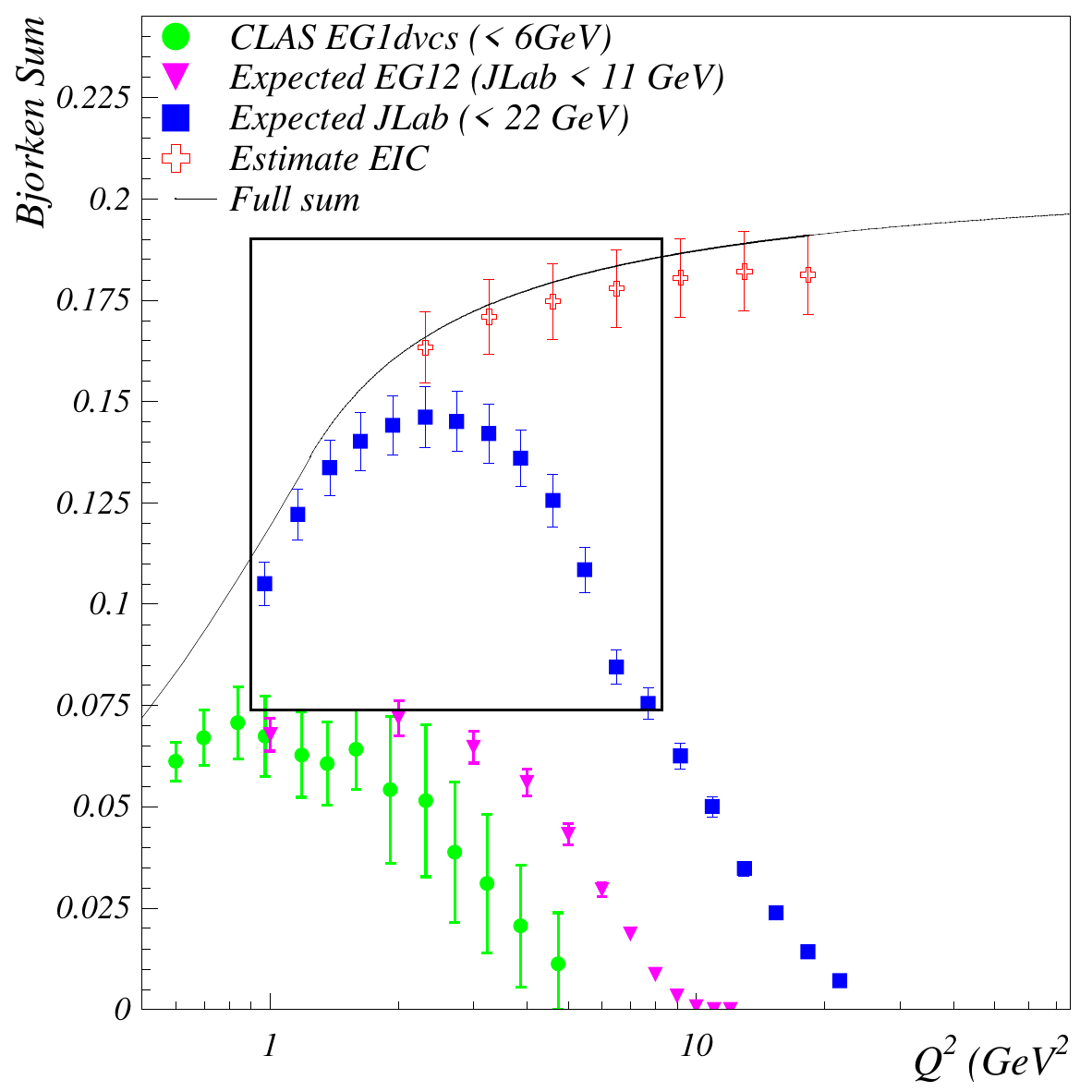}
    \includegraphics[width=0.4\textwidth]{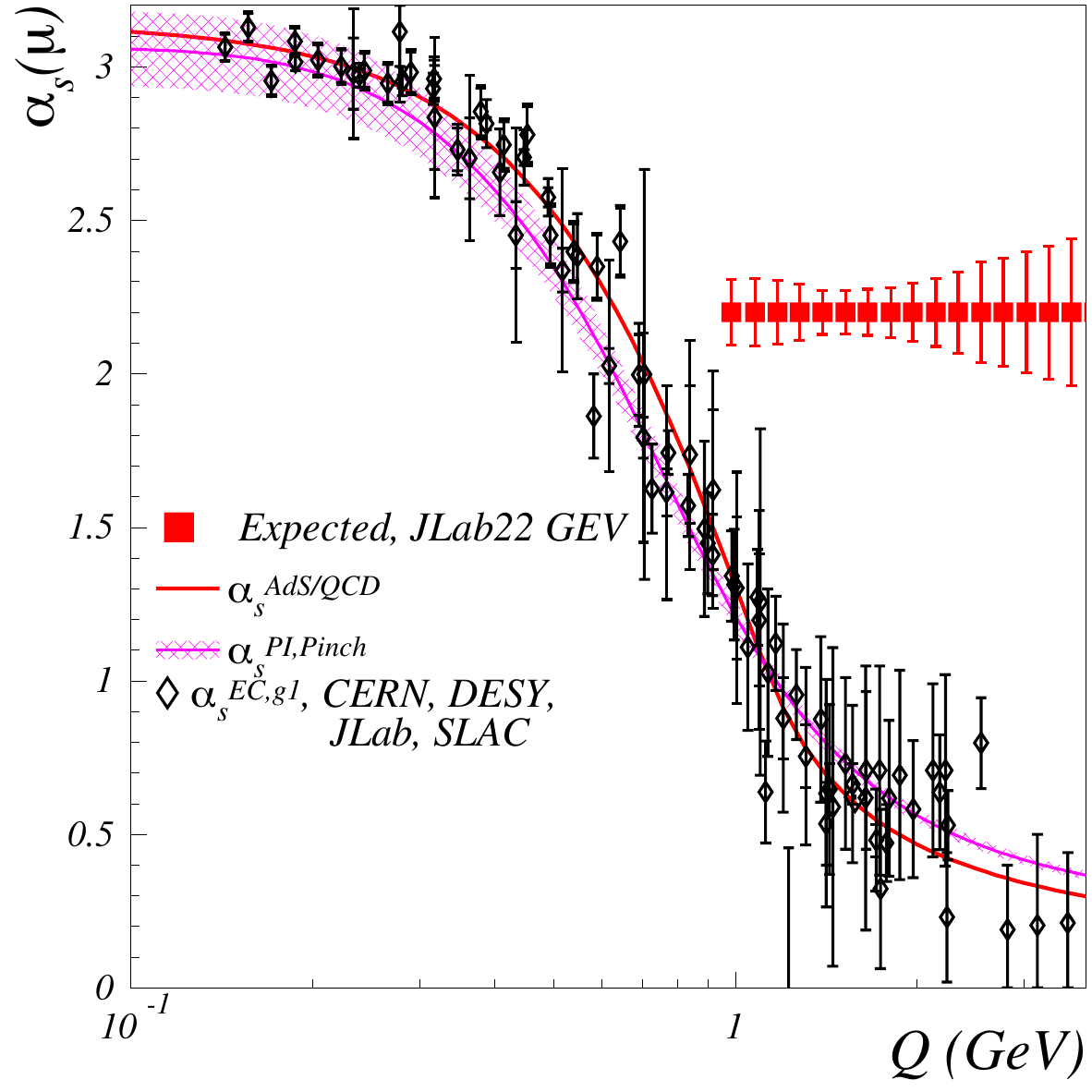}
    \end{center}
     \caption{
        (Left) Expected $ \Gamma_1^{p-n}$ from 22 GeV (squares), 11 GeV (triangles), and EIC (crosses).  6 GeV 
        data (circles) and theory expectation (plain line) are also shown. The rectangle shows the optimal range to extract $\alpha_s$.
        (Right) Expected accuracy on mapping $\alpha_s(Q^2)$ (squares) compared to world data~\cite{Deur:2022msf} (rhombi) and predictions~\cite{Brodsky:2010ur, Cui:2019dwv}.
        }
 \label{fig:bj-22}
\end{figure}

\noindent
\underline{\emph{Prospects for High-Precision Determination of $\alpha_s$}}. The strong coupling constant, denoted as $\alpha_s$, is a crucial quantity in QCD and a key parameter of the Standard Model~\cite{Deur:2016tte}. However, its current uncertainty of $\Delta \alpha_s/\alpha_s = 0.8\%$~\cite{ParticleDataGroup:2020ssz}  is the least precise among all fundamental couplings.  The QCD community is currently investing significant efforts in reducing the uncertainty of $\alpha_s$, with active research being conducted in this area~\cite{dEnterria:2022hzv}. A high-luminosity 22 GeV electron accelerator is an ideal tool for obtaining a more accurate value of $\alpha_s$ using the Bjorken sum rule~\cite{Bjorken:1966jh} defined as $\Gamma_1^{p-n}(Q^2) \equiv \int \left(g_1^p(x,Q^2)-g_1^n(x,Q^2)\right)\, dx$. The expected uncertainty, $\Delta \alpha_s/ \alpha_s \simeq 0.6\%$, is smaller than the current world average and significantly smaller than any single measurement. 
This level of precision is achievable because 22 GeV strikes a balance between high sensitivity to $\alpha_s$ and a small perturbative QCD truncation uncertainty, which typically dominates high-precision extractions of $\alpha_s$. It is important to note that this level of accuracy cannot be reached at 11 GeV, where the missing low-$x$ part of $\Gamma_1^{p-n}$ is expected to become sizable as illustrated in Fig.~\ref{fig:bj-22} (left) where the difference between measurements and theory is attributed to the missing low-$x$ part. While studying the Bjorken sum role  is part of the JLab 12 GeV program~\cite{EG12}, precision extraction of $\alpha_s$ is not. Similarly, the focus of the EIC is not on accurate extraction of $\alpha_s$, although it can provide constraints on $\alpha_s$ at the $1.5-2\%$ level. However, the EIC would be important to access low-$x$ data and, therefore, a 22 GeV upgrade should be seen as a synergistic effort between the EIC and JLab.

Due to the isovector nature, Bjorken sum role has a simpler $Q^2$ evolution and is well-known to higher orders in perturbative QCD ~\cite{Kataev:1994gd,Bjorken_a5}, which helps to limit the inaccuracies in the extraction of $\alpha_s$ in general. Furthermore, the extraction of $\alpha_s$ from the Bjorken sum rule has the advantage of having only a few non-perturbative inputs, the most important being the precisely measured axial charge $g_A=1.2762(5)$~\cite{ParticleDataGroup:2020ssz}. Higher-twist contributions are also known to be small for $\Gamma_1^{p-n}$~\cite{Deur:2014vea}.  One can expect negligible statistical uncertainties on $\Gamma_1^{p-n}$ due to the high-luminosity of JLab at 22 GeV, which has a polarized DVCS and TMD program that can produce sufficient statistics for an inclusive and integrated observable. In the 6 GeV EG1-DVCS experiment, the statistical uncertainties on $ \Gamma_1^{p-n}$ from the EG1-DVCS experiment were below $0.1\%$ at all $Q^2$~\cite{Deur:2014vea}. Therefore, for the 22 GeV experiment, we can safely project a statistical uncertainty of $0.1\%$ for each $Q^2$ point, with bin sizes increasing exponentially to account for the decrease in cross section with $Q^2$. We estimate the experimental systematic uncertainty (excluding the inaccessible low-$x$ region) to be around $5\%$ from several sources, including 3\% for polarimetry (beam and target), 3\% for the target dilution/purity (NH$_3$ and $^3$He), 2\% for the nuclear corrections (due to the extraction of the neutron data from polarized $^3$He, which has an uncertainty of 5\% and contributes approximately one-third of $\Gamma_1^{p-n}$), 2\% for the structure function $F_1$ (required to form $g_1$ from the measured $A_1$ asymmetry), and 1\% for radiative corrections. We assign a 10\% uncertainty for the low-$x$ region that will be covered by the EIC and a 100\% uncertainty beyond that coverage. For the $Q^2$ values relevant to extracting $\alpha_s$ from the Bjorken sum rule, this unmeasured region is that of $x \lesssim 10^{-4}$. 
This approach allows us to account for the variability in uncertainty due to the availability of PDF fits, and to provide a comprehensive estimate for the uncertainty in the low-$x$ part of the project measurements at the 22 GeV.

The resulting $\Gamma_1^{p-n}$ is shown in Fig.~\ref{fig:bj-22} (left), along with the best 6 GeV JLab DIS data\ \cite{Deur:2014vea} and the expected results for 11 GeV and the EIC. The statistical uncertainties on $\Gamma_1^{p-n}$ were found to be below 0.1\% for all $Q^2$ in the 6 GeV EG1-DVCS experiment~\cite{Deur:2014vea}. To account for the exponential increase in bin size as $Q^2$ increases, we use 0.1\% for each $Q^2$ point. The experimental systematic uncertainty (excluding the missing low-$x$ part) is expected to be around 5\%. The improved precision of the 22 GeV measurement over previous 6 and 11 GeV measurements is evident in Fig.~\ref{fig:bj-22} (left), demonstrating the ideal complementarity with the expected measurements from the EIC. To determine the value of $\alpha_s(M^2_{z^0})$, we fit the simulated data using the $\Gamma_1^{p-n}$ approximation at N$^4$LO+twist-4. The fit involves determining the twist-4 coefficient and $\Lambda_s$ as free parameters, where $\Lambda_s$ is the non-perturbative scale of QCD (corresponding to the scale where the perturbatively-defined coupling would diverge). The pQCD approximation for $\alpha_s$ is also at N$^4$LO. To estimate the uncertainty arising from pQCD truncation, we use N$^5$LO+twist-4 with $\alpha_s$ at N$^5$LO \cite{Kniehl:2006bg} and take half of the difference between N$^4$LO and N$^5$LO as the truncation error. In order to minimize the total uncertainty, we carefully select the number of low-$Q^2$ points and high-$Q^2$ points used in the fit. Low-$Q^2$ points have higher $\alpha_s$ sensitivity and smaller systematics, but they also contribute more to the truncation error. On the other hand, high-$Q^2$ points have lower $\alpha_s$ sensitivity and larger systematics, but they contribute less to the truncation error. After optimization, we find that the optimal fit range is 1$<Q^2<$8~GeV$^2$, which leads to a value of $\bm{\Delta \alpha_s/\alpha_s \simeq 6.1\times10^{-3}}$. 

\begin{figure} 
    \centering
    \includegraphics[width=0.6\textwidth]{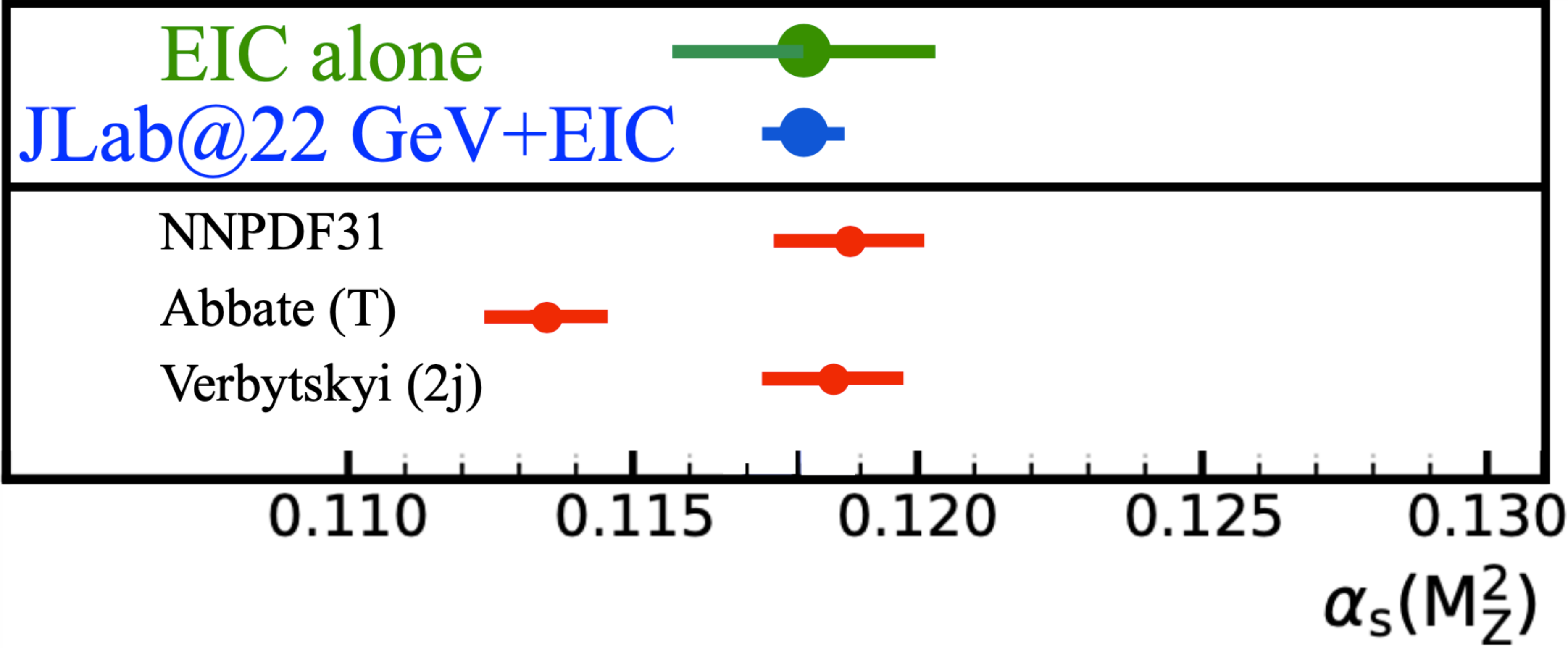}
    \caption{\small{Expected accuracy on $\alpha_s(M^2_{z^0})$ from JLab at 22 GeV (blue), compared the EIC expectation (green) and the three most precise world data~\cite{ParticleDataGroup:2020ssz}.}} 
    \label{fig:alphas-22}
\end{figure}

In summary, utilizing the Bjorken sum rule method at JLab with a beam energy of 22~GeV can enable the measurement of $\Delta \alpha/\alpha$ at levels well below one percent as shown in Fig.~\ref{fig:alphas-22}. The already small missing low-$x$ contribution at moderate $Q^2$ is further reduced with the addition of EIC data. In addition, the pQCD truncation error that typically limits $\alpha_s(M^2_{z^0})$ extractions is reduced since the pQCD series for $\Gamma_1^{p-n}$ and $\alpha_s$ have been estimated up to N$^5$LO. The steep $Q^2$-dependence of the Bjorken sum rule method at JLab 22~GeV, which is approximately 50 times steeper than that of the EIC, provides a high level of sensitivity. Furthermore, the extraction at moderate $Q^2$  reduces uncertainty by a factor of 5 compared to extractions near the $Z$ boson mass, $M^2_{z^0}$. Overall, the Bjorken sum rule method at JLab 22~GeV can deliver a precise measurement of the strong coupling with high sensitivity and have the potential to reduce the its current uncertainty.
%

\subsection{Meson Structure}

Tagged deep inelastic scattering (TDIS) provides a mechanism to access meson structure via the Sullivan process \cite{Barry:2023qqh,Cao:2021aci}.  These measurements will be among only a few to study the essentially unknown and yet fundamental structure of mesons with planned experiments at JLab 12 GeV and EIC. The TDIS program at JLab a 11 GeV is expected to provide new information on pion and kaon structure in the valence regime~\cite{PionTDISProposal, KaonTDISProposal} specially for kaons, which has essentially no existing data for PDF analysis. In addition, it is possible to carry out semi-inclusive measurements, by measuring low momentum final state hadrons in coincidence with scattered electrons  from hydrogen and deuterium targets. The reactions for pion TDIS will be $H(e, e'p)X$ and $D(e, e'pp)X$, and for kaon TDIS it will be $H(e, e'\pi^{-}p)X$. The hadrons will be measured in a multiple time projection chamber (mTPC) surrounding the target, and the electrons will be measured by the Super Bigbite Spectrometer. The mTPC must be a high-rate capable detector and its development is one of the driving forces of streaming readout developments at JLab. The experimental conditions to realize the TDIS program are extremely challenging and  the mTPC detector under development is expected to be capable of tracking at one of the highest rates to date. The 11\,GeV program will therefore be pivotal in establishing the technology and experimental technique, as well as analysis methods and model development, making future experiments at JLab 22~GeV and the Electron Ion Collider possible.

Comparing TDIS Sullivan process measurements directly with existing data from DY experiments at CERN~\cite{NA10:1985ibr} and Fermilab~\cite{Conway:1989fs} and upcoming DY measurements from AMBER at CERN will provide important tests of universality of the meson structures, particularly the valence quark distributions at large $x_\pi$. However, unlike the DY experiments, the TDIS data will be almost entirely free of nuclear corrections. These measurements will complement the low-$x_\pi$ collider data taken from HERA~\cite{H1:2010hym,ZEUS:2002gig} and future EIC~\cite{AbdulKhalek:2021gbh,Arrington:2021biu}. TDIS at 11\,GeV benefits from the 
high-luminosity capabilities of JLab and will offer much better handle on uncertainties in  the valence region, precisely  where the EIC's reach is statistically limited~\cite{Arrington:2021biu}.  To perform a reliable QCD extraction of pion (and kaon) PDFs, the observed final state invariant mass $W_{\pi}$ must be large enough to avoid the expected resonances.  The 11\,GeV facility will be able to map out the previously unmeasured resonance regions of the pion at low-$W_\pi^2$ to high precision, whereas the 22\,GeV experiment will provide a larger range of $W_\pi^2$ for available PDF analysis. To assess the kinematic range where a meson PDF analysis is realizable, we have calculated the contribution of the $\rho$ meson (lowest-lying resonance) to the exclusive $F_2^\pi$ structure function and find a non-negligible signal of about $20-40\%$ of the inclusive structure function in the extrapolated kinematic regions of the 11\,GeV experiment. Because of the width of the $\rho$ decay, an estimate for a minimum $W_\pi^2$ for a safe PDF analysis is $W_\pi^{2} > 1.04~{\rm GeV}^2$. 
\begin{figure}
    \centering
    \includegraphics[width=0.9\textwidth]{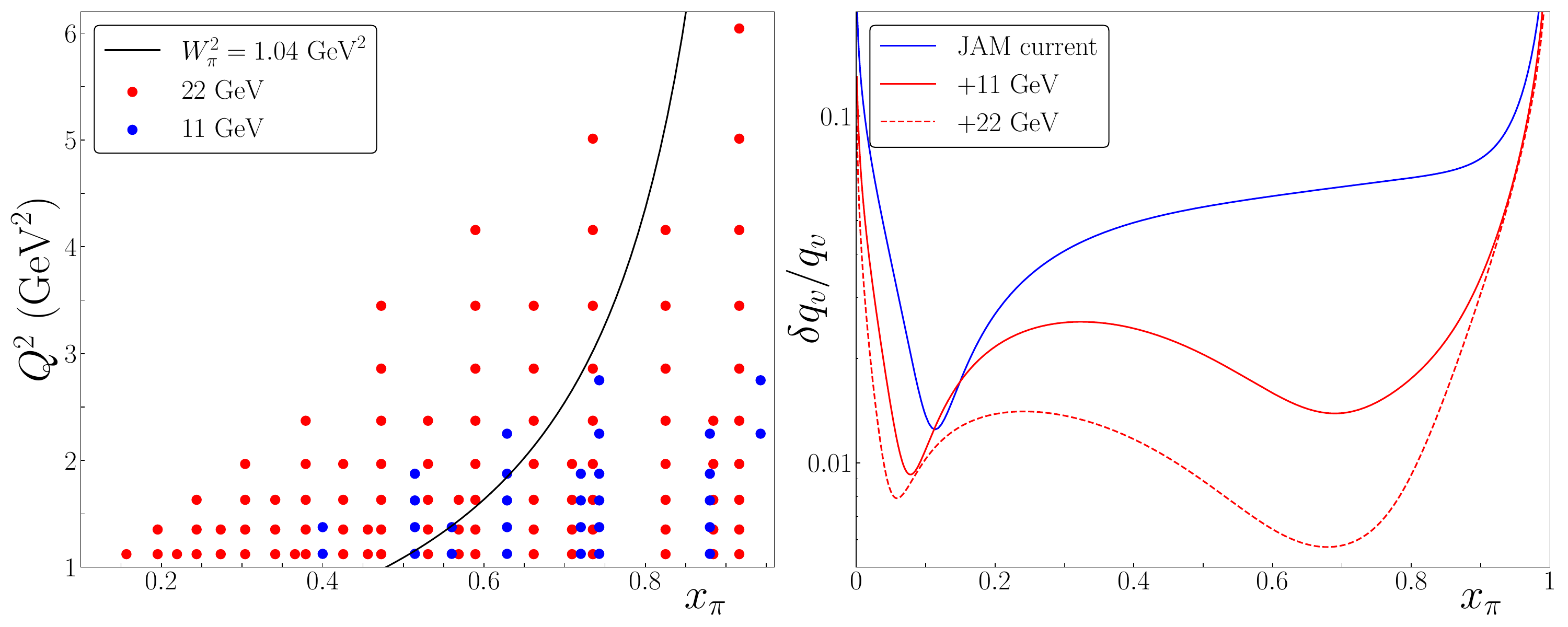}
    \caption{(Left) The kinematics of the 11\,GeV (blue) and 22\,GeV (red) points in $Q^2$ versus $x_\pi$ along with the line of $W_\pi^2=1.04~{\rm GeV}^2$. Multiple bins in $t$ are on each red point. (Right) The impact on the valence quark distribution from the JLab TDIS experiments.}
    \label{f.impact}
\end{figure}
In the left panel of Fig.\ \ref{f.impact}, we illustrate the phase space available for PDF analysis in a bin of $t=-0.05$~GeV$^2$, the virtuality of the scattered meson, for the 11~GeV and 22~GeV experiments using blue and red points, respectively. The black curve represents a contour of fixed $W_\pi^2$ value at 1.04~GeV$^2$. Points located to the right of  the curve will be eliminated, resulting in a significant reduction of available phase space for the 11~GeV experiment. In contrast, the 22~GeV experiment offers a much larger phase space, allowing for more comprehensive PDF analysis. In addition, due to the cut on $W_\pi^2$, the range of $x_\pi$ suitable for PDF analysis is greatly restricted to $0.4 < x_\pi < 0.6$ in the case of the 11~GeV experiment, while the $x_\pi$ coverage is expected to be enhanced in the 22~GeV kinematics. We also perform an impact study on the pion PDFs with the inclusion of the 11~GeV and 22~GeV TDIS experiments. We assume a 1.2\% systematic uncertainty, with an integrated luminosity of $8.64\times 10^{6}~{\rm fb}^{-1}$ with 100\% acceptance corresponding to 200 days of data taking at $d{\cal L}/dt-5\times10^{38}$/cm$^2$/s. After the cut of $W_{\pi}^{2} < 1.04~{\rm GeV}^2$, only 26 pseudodata points remained from the 11\,GeV experiment, while 231 data points were permitted from the 22\,GeV experiment. In the right panel of Fig.~\ref{f.impact}, we show the relative uncertainty of the valence quark PDF in the current state (blue), and with the inclusion of the 11\,GeV pseudodata (solid red) and 22\,GeV pseudodata (dashed red) as a function of $x_\pi$. Notably, with the inclusion of the 22~GeV data, we see a marked improvement in the knowledge of pion PDFs across the large available $x_\pi$ range.

To summarize, the TDIS program at the 11 and 22 GeV JLab will play a pivotal role in elucidating the internal structure of mesons, including access to their TMDs. This exploration of the meson sector is a significant undertaking in hadronic physics, offering a deeper understanding of QCD emergent phenomena. As mesons provide a crucial link between fundamental quarks and the observable world, enhancing our knowledge of their structure and dynamics promises profound insights into the fundamental principles of the strong interaction. This includes the understanding of confinement, the dynamics of quark-gluon interactions and chiral symmetry, which are central aspects of QCD. Furthermore, the high-precision data expected from the TDIS program can lead to a refinement of existing theoretical models and potentially inform the development of new ones.

\clearpage 
\section{Hadronization and Transverse Momentum}
\label{sec:wg3}

Semi-inclusive deep inelastic scattering (SIDIS) is a powerful tool that enables us to study the momentum space tomography of nucleons and nuclei through a range of quantum correlation functions in QCD such as transverse momentum dependent PDFs (TMDs).  Thanks to dedicated experimental programs at HERMES (DESY), COMPASS (CERN), and the 12-GeV program at JLab, significant progress has been made in recent years in understanding SIDIS reactions. These experiments have provided us with intriguing glimpses into the 3D structure of hadrons in momentum space, revealing the complex interplay between quarks and gluons. With continued progress in this field, we can expect to gain even deeper insights into the structure of hadrons and the nature of the strong force, with implications for both particle physics and nuclear physics.

In general, in the one-photon-exchange approximation, SIDIS reactions can be decomposed in terms of 18 structure functions (SFs) \cite{Bacchetta:2006tn} stemming from multiple degrees of freedom, such as beam and target polarizations. These objects contain various convolutions of twist-2 or higher-twist PDFs and fragmentation functions that are multiplied by specific kinematic pre-factors~\cite{Bacchetta:2006tn} and they offer unique information about quark-gluon dynamics in the nucleon. In addition to standard DIS kinematic variables $x$ and $Q^2$, the SFs responsible for different azimuthal modulations in $\phi_h$ (azimuthal angle between hadronic and leptonic planes), and $\phi_S$ (azimuthal angle of the transverse spin),  depend also on the fraction of the virtual photon energy carried by the final state hadron, $z$, and its transverse momentum with respect to the virtual photon, $P_T$. 

The complexity of the SIDIS reaction poses significant experimental challenges to isolate each SFs from cross sections/asymmetries since SFs have intricate kinematic dependencies, such as $x$, $Q^2$, and $P_T$. In particular, measuring each of these requires the full $\phi$ dependence of the reaction and, in some cases, the $\epsilon$ dependence, which defines the relative cross section contributions from longitudinal ($\sigma_L$) and transverse photons ($\sigma_T$).  Moreover, their determination becomes increasingly difficult in the high-energy valence region where certain SFs, such as helicity-dependent SFs sensitive to longitudinal spin-dependent TMDs, are suppressed due to kinematic factors. 

Most of the current SIDIS programs have mainly focused on studying SFs related to transversely polarized virtual photons. Unfortunately, longitudinal SFs have not received much attention, and their contribution to TMD phenomenology remains largely unexplored. This lack of understanding of longitudinal photon contributions introduces systematic uncertainties that can only be evaluated through direct measurements. Therefore, it is crucial to expand our program by measuring the wide range of SFs across an enhanced multidimensional phase space. This will help to validate and improve ultimately our understanding of parton dynamics in SIDIS reactions. 

In addition, the interpretation of SIDIS data in terms of TMDs has been a significant challenge in recent years, as it involves multiple physical mechanisms that contribute to the production of hadrons in the final state \cite{Gonzalez-Hernandez:2018ipj, Wang:2019bvb, Boglione:2016bph, Collins:2016hqq} in addition to the complexity of the reactions in terms of structure functions. The connection between SIDIS data and TMDs is only established within the TMD \emph{current region}, which overlaps with other mechanisms such as target, central, and hard collinear fragmentation regions depending on the overall collision energy \cite{Boglione:2016bph, Boglione:2019nwk}. 

The 22 GeV upgrade, with its extended $Q^2$ coverage, offers a new complementary window between the 12 GeV program and the future Electron Ion Collider (EIC). In addition, this new energy range makes JLab unique to disentangle the genuine intrinsic transverse structure of hadrons encoded in TMDs with controlled systematics. The availability of two energies or in-between energy ranges also allows us to identify the scaling properties of the SIDIS reaction, validate the measurements of leading contributions, and explore sub-leading contributions associated with multi-parton dynamics of QCD. A combined 11~GeV and 22 GeV SIDIS program is therefore needed to address these issues.

The importance of separating the structure functions cannot be overstated and a potential JLab 22~GeV upgrade could provide a significant boost to the $Q^2$ and $P_T$ range, enabling us to access more accurate measurements of these structure functions with the following benefits:
\begin{itemize}
    \item High-luminosity measurements over an enhanced multidimensional phase space without the need of averaging or marginalization of the SIDIS phase space, {\it e.g.}, will allow access to the $Q^2$ dependence of structure functions at fixed $x$ or fixed $P_T$ and validate the expectations from theoretical frameworks in QCD;
    
    \item Large acceptance for existing JLab spectrometers that allows precise determination of the $\phi$ dependence of the cross section, allowing unambiguous identification of the relevant structure functions;
    
    \item Finally, the high energy and luminosity, combined with well-understood magnetic focusing spectrometers, will provide the ability to make measurements of the $\epsilon$-dependent terms over a large region of $(x, Q^2)$ phase space, allowing the measurement of $R=\sigma_L/\sigma_T$ in SIDIS.
\end{itemize}

In the following we will discuss how the CEBAF upgrade will be essential to boost the scientific reach of the SIDIS program, enabling us to make new discoveries about the fundamental nature of hadronic matter. With this upgrade, we will be able to explore new frontiers in the study of quarks and gluons, and to gain a deeper understanding of QCD's emergent phenomena.

\subsection{Importance of Multi-Dimensional SIDIS Measurements}

SIDIS cross sections, hadron multiplicities, and polarization independent azimuthal asymmetries are multi-differential in nature. Therefore a Multi-Dimensional (Multi-D) analysis is mandatory to disentangle the intricate dependencies on the kinematical variables $x$, $Q^2$, $P_T$, $z$. Comparing results obtained by different SIDIS experiments operating with different beam energies and phase-space coverage is often impractical and error-prone if the comparisons are done on a one-dimensional basis. Looking at single-dimensional kinematic dependencies of cross sections or asymmetries obtained from different experiments while integrating over other dimensions of non-equal phase space contours, may result in significant discrepancies. Precision measurements in Multi-D for all variables are also critical to understand effects induced by phase space limitations. It was suggested that even at COMPASS energies the phase space available for single-hadron production in deep-inelastic scattering should be taken into account to describe data in the standard pQCD formalism. 

Another class of effects that Multi-D measurements can help is to understand the systematics associated with initial and final state hadron mass corrections in SIDIS. The HERMES experiments provided pioneering measurements at a similar energy as the proposed 22 GeV upgrade but with limited integrated luminosity. JLab22, in contrast, will enable detailed Multi-D measurements that will answer open questions from the HERMES program, confront the factorized description of hadron production in SIDIS with a wealth of precision data bridging the sub-asymptotic and Bjorken regimes. Increasing JLab's beam energy from 6-12~GeV to 22 GeV will allow one to measure SIDIS events at higher values of $Q^2$ than previously possible, and further allowing to study subleading power corrections originating from higher-twist parton correlations. 

Multi-D measurements play a crucial role in the investigation of helicity-dependent TMD PDFs, specifically the relatively unknown $g_1(x,k_T)$, where $k_T$ is the transverse momentum of the quark. However, obtaining measurements of helicity TMDs at higher energies is challenging due to the suppression of the kinematic factor $\sqrt{1-\epsilon^2}$. In the valence region and at high energies, this factor becomes relatively small, making it difficult to extract the double spin asymmetries needed for the determination of $g_1(x,k_T)$ in the multidimensional space. Recent measurements of the $P_T$-dependence in double spin asymmetries (DSAs), conducted for the first time across different $x$-bins, have revealed interesting insights. These measurements suggest the existence of different average transverse momenta for quarks aligned or anti-aligned with the nucleon spin~\cite{CLAS:2010fns,CLAS:2017yrm}, consistent with findings from LQCD simulations~\cite{Musch:2010ka}. The extended range of $P_T$ accessible at the JLab 22~GeV kinematics will allow for exploration of the $P_T$-range where contributions from vector mesons are expected to be negligible~\cite{Avakian:2019uzf} and shed light on the nature of $g_1(x,k_T)$.

\begin{figure}[t]
  \begin{center}
    \includegraphics[width=0.4\textwidth]{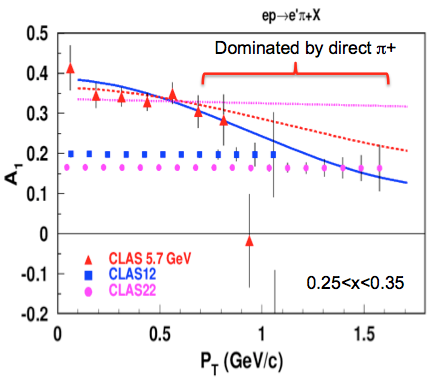}
    \includegraphics[width=0.43\textwidth]{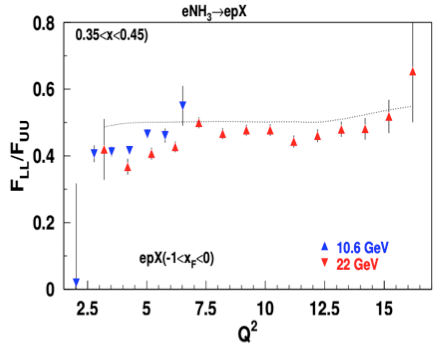}
 \end{center}
  \caption{The double spin asymmetry as a function of $P_T$ in $ep\rightarrow e^\prime\pi^+X$ in a given bin in $x$ (left) and the $Q^2$-dependence of the double spin asymmetry in a given bin in $x$ for $ep\rightarrow e^\prime pX$ (right). The projections for 100 days, use the existing simulation and reconstruction chain, and the luminosity currently used for the CLAS12 detector (see Fig.~\ref{fig:JLAB4D}). The curves correspond to different widths in $k_T$ of $g_1(x,k_T)$ compared to $f_1(x,k_T)$.}
 \label{all}
\end{figure}

In order to understand the systematics involved in extracting helicity TMD-PDFs from DSAs, it is necessary to conduct thorough investigations into the $P_T$ and $Q^2$ dependencies. In addition, it is crucial to examine the potential backgrounds arising from other SFs that contribute to various azimuthal modulations. Figure~\ref{all} illustrates projected measurements of the kinematic dependencies of DSAs for a 22 GeV beam utilizing the existing CLAS12 detector. Expanding the range of $Q^2$ would enable precision tests of the evolution properties of $g_1(x,k_T)$, thereby facilitating the validation of different phenomenological approaches.

\begin{figure}[h!]
    \begin{center}
        \includegraphics[width=\textwidth]{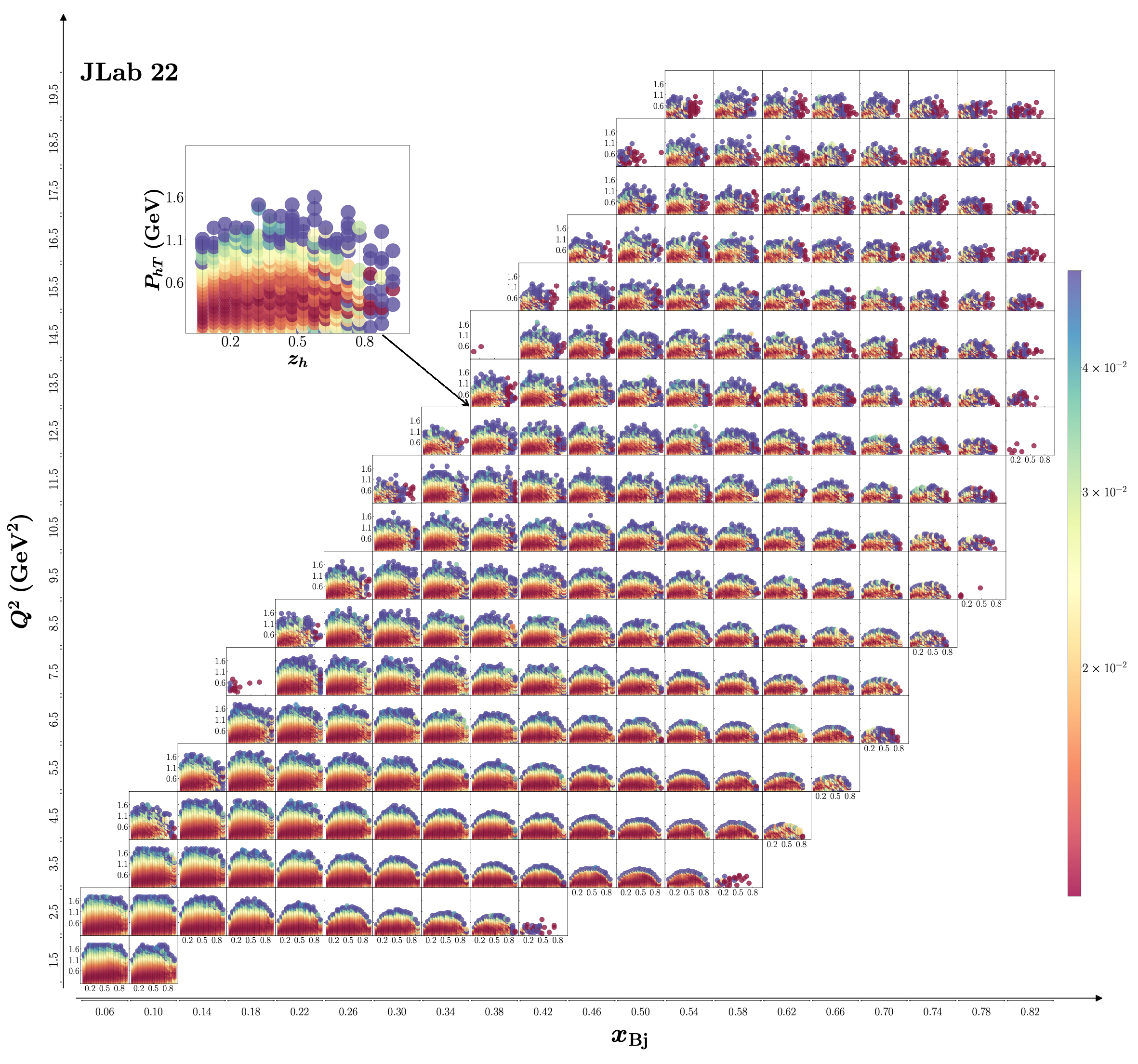}
    \end{center}
    \caption{Multi-D phase space of SIDIS at 22 GeV kinematics. The color range shows the expected relative uncertainties for SIDIS cross sections in 4D bins, using the existing CLAS12 simulation/reconstruction chain for 100 days of running with $10^{35}$~cm$^{-2}$s$^{-1}$ luminosity.}
    \label{fig:JLAB4D}
\end{figure}

In summary, the utilization of multi-dimensional analysis approaches carries numerous implications and benefits. The intricate nature of nucleon structure properties and hadronization processes necessitates precise multi-dimensional measurements for a comprehensive understanding of SIDIS reactions. Such measurements will be delivered by the JLab 22 GeV program in fine 4D bins as shown in Fig.~\ref{fig:JLAB4D} as projected by simulations of the existing  CLAS12 detector. The combined measurements at the upgraded machine with high luminosity and extended phase space coverage across all JLab Halls involved in the SIDIS program, will provide unprecedented measurements of SIDIS reactions for the hadronic physics community.  

\subsection{Role of Longitudinal Photon and SIDIS Structure Functions}

One of the key features to understand hard reactions with virtual photons like in DIS and SIDIS reactions is the distinction between the    longitudinal ($\sigma_L$) and transverse ($\sigma_T$) photons contributions. Such distinctions at $Q^2$ above 10 GeV will be only possible at JLab with an upgraded 22 GeV beam, delivering high-luminosity data. With well-understood magnetic focusing spectrometers, the new experiments will be able to measure the most precise ratios of the longitudinal to transverse cross sections $R=\sigma_L/\sigma_T$. While moderately accurate measurements of this ratio have been made for inclusive deep inelastic scattering ~\cite{Bebek:1975pn, Bebek:1976wv, Bebek:1977pf}, there are hardly any measurements of $R_{\rm SIDIS}$ for the SIDIS process. Therefore, it is imperative to address this limitation, as modern measurements of SIDIS at HERMES, COMPASS, and JLab have relied on $R_{\rm SIDIS}=R_{DIS}$, which is necessarily independent of $z$, $P_T$, $\phi$, and hadron and target nucleon types. 

\begin{figure}[h!]
  \begin{center}
    \includegraphics[width=0.65\textwidth]{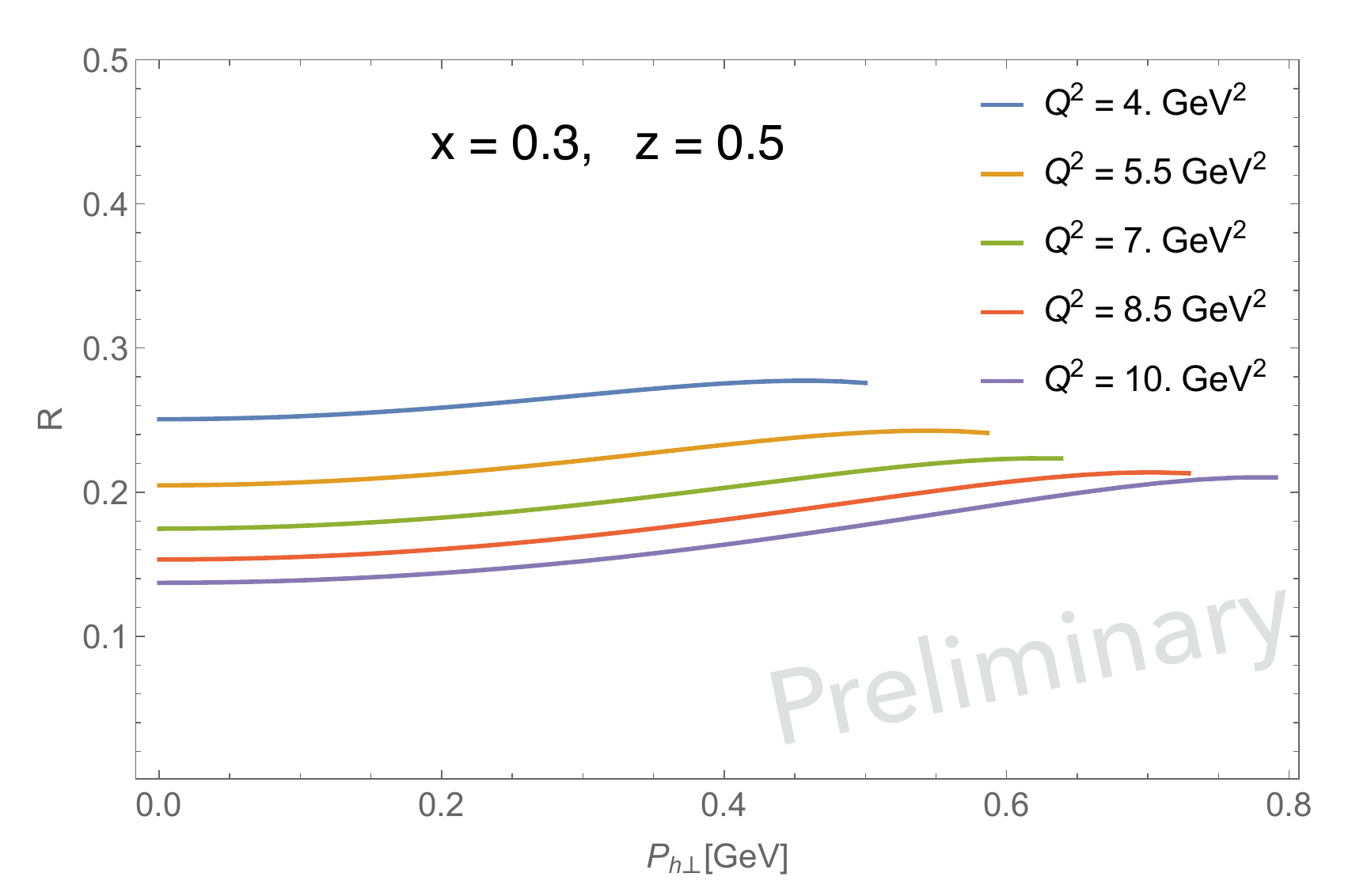}
 \end{center}
  \caption{Estimate of $R_{\rm SIDIS} = F_{UU,L}/F_{UU,T}$ versus the hadron transverse momentum $P_T (P_{hT})$ at fixed values of $x$ and $z$ and for different values of $Q^2$, compatible with JLab22 kinematics, using MAP22 TMD analysis~\cite{Bacchetta:2022awv}.}
 \label{fig:Restimate}
\end{figure}

In TMD phenomenology it is often assumed that $F_{UU,L}$ is negligible at low transverse momentum. However, recent investigations from Refs.~\cite{Bacchetta:2008xw,Anselmino:2005nn}  have indicated that such assumptions might not be valid as also indicated in Fig.~\ref{fig:Restimate} where predictions for the ratio $R=F_{UU,L}/F_{UU,T}$, 
display sizable contributions reaching up to 30\% ~\cite{Bacchetta:2022awv}. These findings demonstrate that the contribution of $F_{UU,L}$ cannot be disregarded, as it can be substantial and essential for an accurate interpretation of $F_{UU,T}$, which is associated with standard leading-twist TMDs.
The empirical separation of the transverse  and longitudinal components of cross sections is typically achieved through the measurement of the photo-hadron cross section under various kinematic conditions. These measurements correspond to the same photon 4-momentum $Q^2$ and $x$ values but differ in the virtual photon polarization parameter $\epsilon$. In order to acquire these essential measurements, experiments must be conducted at different kinematic combinations, \textbf{including varying incident electron energies}. To facilitate such experiments and fully realize the next generation of SIDIS  measurements, the CEBAF accelerator is therefore required to be running higher energies beyond the existing 12 GeV program.

As $z$ approaches 1 ({\it i.e.}, exclusive scattering), the $Q^2$ dependence of $R_{\rm SIDIS}=F_{UU,L}/F_{UU,T}$ is expected to change from $1/Q^2$ to $Q^2$. Experimental measurements at COMPASS on the deuteron~\cite{COMPASS:2014kcy}  and the proton~\cite{Moretti:2021naj}, at HERMES~\cite{Airapetian:2012yg} and CLAS/CLAS12~\cite{Osipenko:2008aa, Diehl:2021rnj} have shown that the $F_{UU}^{\cos 2 \phi_h}$ related in the perturbative limit to $F_{UU,L}$~\cite{Bacchetta:2008xw}, and the $F_{UU}^{\cos \phi_h}$ arising from the interference between longitudinal and transverse photons 
can be very significant, with $\cos \phi_h$-modulations as high as 30\%~\cite{COMPASS:2014kcy,Moretti:2021naj,Airapetian:2012yg,Osipenko:2008aa, Diehl:2021rnj}, and unexpectedly getting higher at large $Q^2$, which can potentially indicate the dominant role of longitudinal photons in certain kinematics. Similarly, a strong signal for the SF $F_{UT}^{\sin\phi_S}$ at large $z$ has been observed by both the HERMES and COMPASS Collaborations, which can be attributed to the unsuppressed nature of longitudinal photons cross sections. The  longitudinal cross sections and its associated structures functions play a prominent role in SIDIS reactions, in particular for unpolarized and transversely polarized targets~\cite{Bacchetta:2006tn}.

\begin{figure}[h!]
    \begin{center}
        \includegraphics[width=0.95\textwidth]{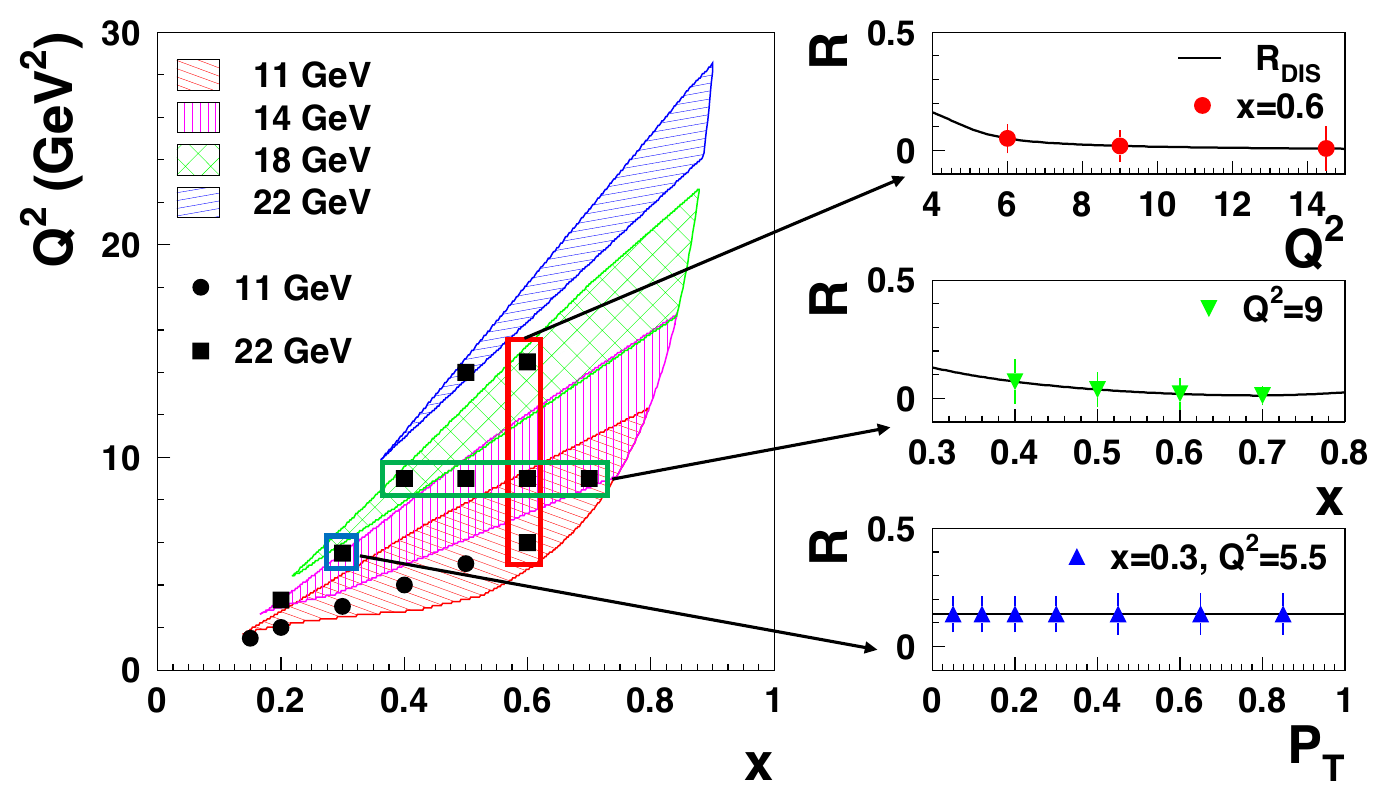}
    \end{center}
    \caption{Projections for measurements of $R_{\rm SIDIS} = \sigma_{L,{\rm SIDIS}}/\sigma_{T,{\rm SIDIS}}$ with electron beam energies up to 22~GeV. The left panel shows the available kinematic space in Hall C using the existing HMS and SHMS spectrometers. The right three panels demonstrate the accuracy achievable with 3~months of nominal 40~$\mu$A current on LH2 and LD2 targets, collecting both $\pi^+$ and $\pi^-$ SIDIS data for a measurement of the $Q^2$ dependence, the $x$ dependence, and the $P_{hT}$ dependence. The black curves indicate the measured value of $R_{DIS}$.}
\label{fig:R1}
\end{figure}

The existing uncertainty surrounding the experimental knowledge of $R_{\rm SIDIS}$ raises doubts about the reliability of using current SIDIS data to infer quark flavor and spin distributions in hadrons. While, a new experiment in Hall C is expected to provide precise measurements of $R_{\rm SIDIS}$ using the Rosenbluth technique in the next few years, these measurements will have limited kinematic space due to the 11 GeV maximum beam energy, but will enable the extraction of the transverse SIDIS cross section without any uncontrolled assumptions about $R$. Extension of the beam energy to 22 GeV would significantly expand the kinematic phase space that is critical to accurately interpret the data in QCD. In Fig.~\ref{fig:R1} we present projections for measuring $R_{\rm SIDIS}$ by combining multiple energies from 11 GeV up to 22 GeV  using simulated SIDIS data with pions in the final state. The results were  obtained with the existing Hall C spectrometers, assuming 3 months of nominal beam time, a beam current of 40~$\mu$A, and measurement of both $\pi^+$ and $\pi^-$ for LH2 and LD2 targets. The projections assume a difference in $\epsilon$ values of at least 0.2, and point-to-point systematic uncertainties of 1.4\%, consistent with other Hall C experiments. Note that the projections for the $P_{T}$ dependence assume that information about the $\phi$-dependent interference terms in the cross section will be constrained by other experiments ({\it e.g.} Hall B), since the Hall C spectrometers provide full $\phi$ coverage only up to $P_{T}=$0.4 to 0.5~GeV. 

In summary, 
with high energy and luminosity, along with well-characterized magnetic focusing spectrometers, it becomes feasible to make measurements of the $\epsilon$-dependent terms over a wide range of the $(x, Q^2, z, P_T)$ phase space enabling the most precise measurements of $R = \sigma_L/\sigma_T$ in SIDIS.

\subsection{Physics Opportunities}


\noindent
\underline{\emph{Quark-Spin Dependence of Hadronization and Correlations of Hadrons in CFR}}. The measurement of the Collins asymmetries \cite{Collins:1992kk} in SIDIS off a transversely polarized target is a unique opportunity to access simultaneously the partonic transverse spin structure of the nucleons and the spin-dependence of the hadronization process. These asymmetries  can be written~\cite{Bacchetta:2006tn} as a convolution of the chiral-odd transversity TMD PDF $h_1^q$ and the chiral-odd and T-odd spin-dependent TMD fragmentation function (FF) $H_{1q}^{\perp\,h}$ \cite{Collins:1992kk}, referred to as the Collins function. While the transversity TMD PDF $h_1^q$ describes the transverse polarization of a quark with flavor $q$ in a transversely polarized nucleon, the spin-dependent FF describes the fragmentation of a transversely polarized quark $q$ into the unpolarized hadron $h$. The structure function can depend on the Bjorken $x$, $Q^2$, the fractional energy $z$ carried by $h$, and on the transverse momentum $P_{\rm T}$ of $h$. 

The Collins FF offers a unique opportunity of studying the quark-spin dependence of the hadronization process, which is a still poorly understood non-perturbative phenomenon in QCD. A wider knowledge of this function would bring very useful input information to build the models of the polarized hadronization, {\it e.g.} models inspired to the string fragmentation model \cite{Kerbizi:2018qpp} or field-theoretical models \cite{Matevosyan:2016fwi}, which in turn will help to understand hadronization process that correlates with quark-spin.
\begin{figure}[!t]
    \includegraphics[width=\textwidth]{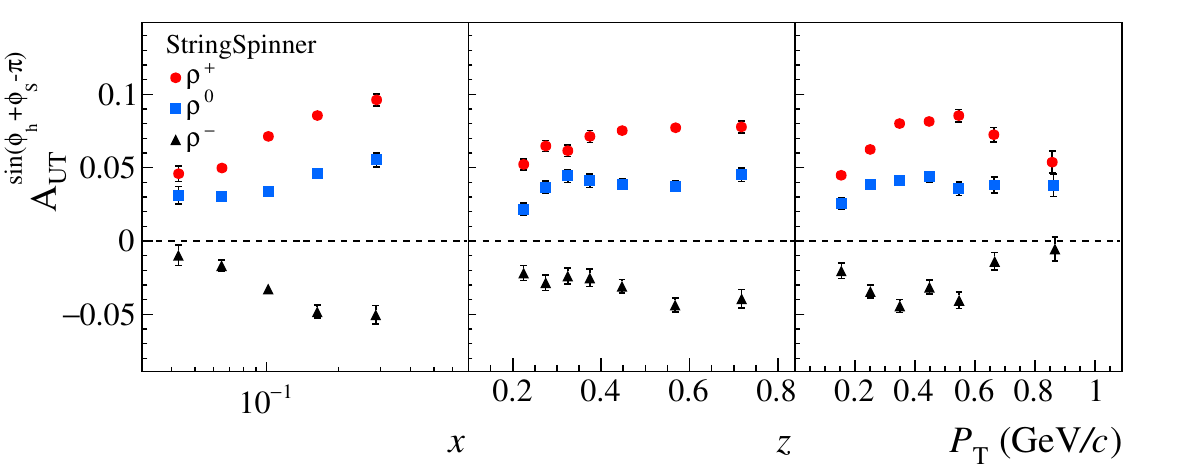}
    \includegraphics[width=\textwidth]{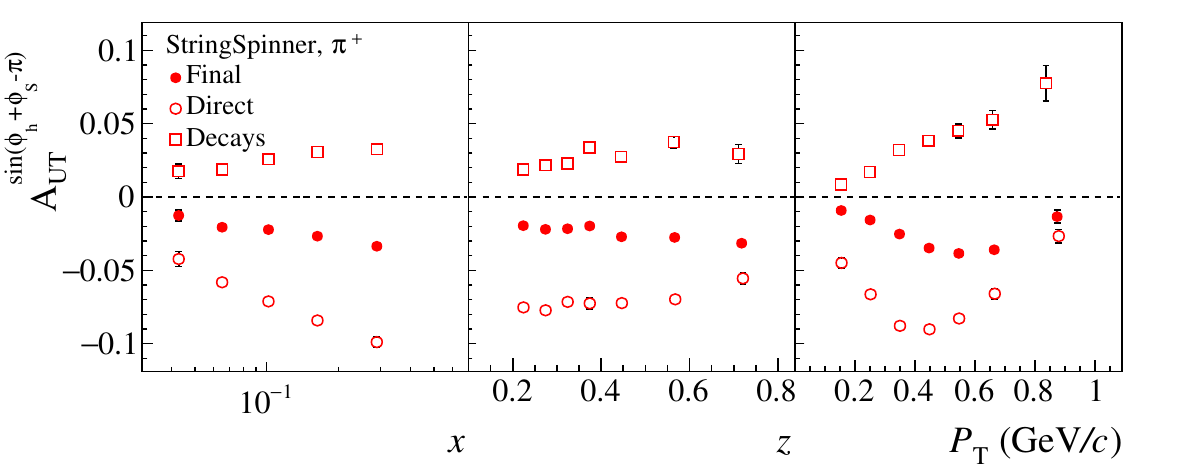}
    \caption{Prediction for the Collins asymmetries for $\rho^+$ (circles), $\rho^0$ (squares) and $\rho^-$ (triangles) in SIDIS off    transversely polarized protons in the JLab22 kinematics and the impact of VMs on the inclusive pion Collins SSA. The simulations are carried using the StringSpinner package \cite{Kerbizi:2021pzn} and the PYTHIA 8.2 \cite{Sjostrand:2014zea} event generator.}
    \label{fig:Collins-rho}
\end{figure}

\begin{figure}[t]
    \includegraphics[width=\textwidth]{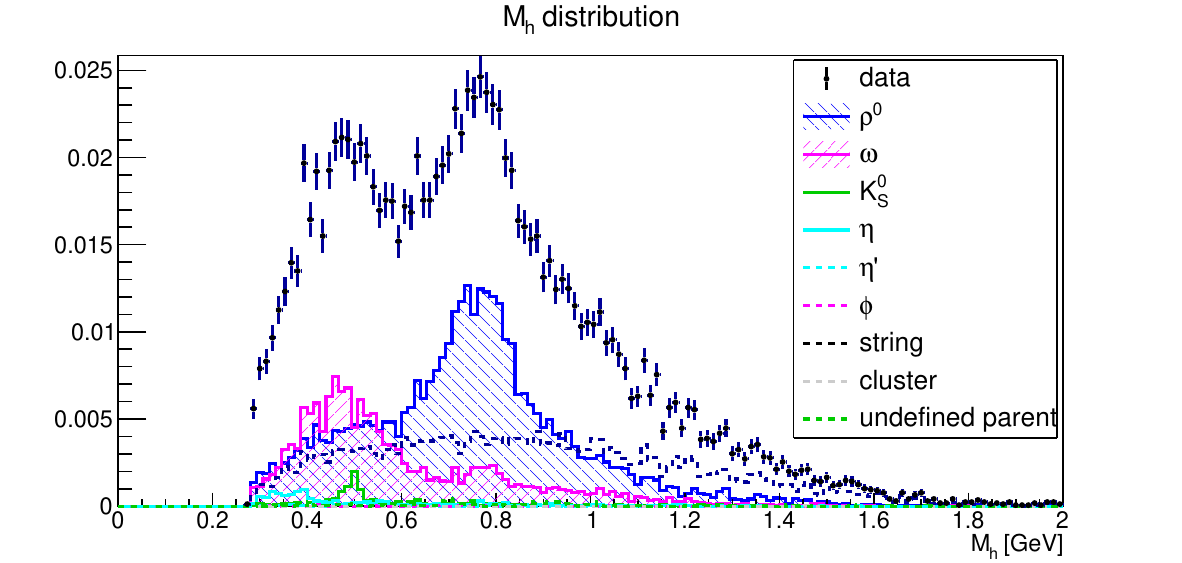}
    \caption{Invariant mass $M_h$ distribution for $\pi^+\pi^-$ dihadrons from CLAS22 Monte Carlo data (black points). The colored distributions represent dihadrons where one or both of the hadrons are produced from the indicated parent. The dominant contributions are from $\rho^0$ and $\omega$ decays. One histogram entry is filled for each single pion. }
 \label{fig:mh}
\end{figure}

To date the Collins FF has been extracted for the production of pseudoscalar (PS) mesons, but not for heavier hadron species. Contributions from vector mesons (VMs) to the Collins asymmetry due to correlation of hadrons produced in the current fragmentation region (CFR), is one of the important sources of systematics involved in TMD extraction, so far completely ignored in phenomenology. 
Dihadron production in the Current Fragmentation (CFR) Region, in general, plays a crucial role in this regard, as it provides access to intricate details of QCD dynamics that are not readily accessible through single-hadron SIDIS measurements. Furthermore, dihadron production becomes essential in understanding the systematics arising from various simplifying assumptions (such as independent fragmentation and isospin symmetry) used in the extraction of TMD PDFs from single-hadron SIDIS. The data from polarized SIDIS experiments, such as HERMES, COMPASS, and more recently CLAS, have enabled access to multiparton correlations. Given the current state-of-the-art in extracting PDFs within the realm of pQCD, it is crucial to undertake a comprehensive analysis of the $Q^2$ behavior of different relevant observables needed for validation of underlying frameworks for analysis of single hadron SIDIS, neglecting hadronic correlations, and separation of different contributions to relevant SFs, such as the SF describing the hadronization of transversely polarized quarks.  For dihadron observables, the spectra in $(z, M_h)$, where $z$ is the fragmentation variable and $M_h$ is the invariant mass of the hadron pair, could provide insights into contributions beyond the expected two-pion mass distribution as illustrated in Fig.~\ref{fig:mh}. Measured Single-Spin Asymmetries (SSAs) \cite{Hayward:2021psm} clearly demonstrate a dependence on the invariant mass of the pion pair, indicating significant correlation effects in hadron production in the Current Fragmentation Region (CFR).

In the context of the recently developed ``string+$^3P_0$ model" \cite{Kerbizi:2021gos} of polarized hadronization \cite{Kerbizi:2021gos}, it was shown that a deeper insight into the spin dependence of hadronization is encoded in the Collins FF associated with the  production of vector mesons (VMs), more in particular for the case of $\rho$ mesons (see also  Ref.~\cite{Bacchetta:2000jk}).  Since the contamination of the $\rho$ meson sample from decays of heavier resonances is  expected to be negligible according to simulations, the produced VMs are mostly sensitive to the direct mechanisms of quark fragmentation. Therefore, measurements of the Collins asymmetries for VMs is relevant to constrain the free parameters of the hadronization models, which in turn will shed new light on the mechanisms of quark fragmentation. However, the experimental information on the Collins asymmetries for VMs is presently limited because \textit{a}) the high combinatorial background that must be subtracted when constructing the VM candidates in experimental data, and \textit{b}) the low statistics of VMs as compared to the final state mesons. The only existing measurement is the asymmetry for $\rho^0$ mesons measured by the COMPASS experiment in SIDIS off transversely polarized protons \cite{COMPASS:2022jth}. This pioneering work shows that the measurement of Collins asymmetries for $\rho$ mesons is feasible and can be done with higher precision at future facilities. 

An upgraded CEBAF accelerator running at 22 GeV provides a unique opportunity to study Collins asymmetries with VMs since it is expected to lead to a lower combinatorial background from non-resonant hadron pairs in the invariant mass regions of the $\rho$ mesons, {\it e.g.} as compared to the EIC operating at much higher energies. The high luminosity design of JLab22 favors the collection of a sizeable number of $\rho$ meson-candidates and of pion pairs in the invariant mass regions before and after the $\rho$ region, needed for a reliable background estimation to cleanly extract the Collins asymmetries with VMs in the $\rho$ region. 

Using the simulation framework of the ``string+${}^3P_0$ model" in the PYTHIA 8.2 Monte Carlo event generator \cite{Sjostrand:2014zea} via the StringSpinner package \cite{Kerbizi:2021pzn}, sizable Collins asymmetries for VMs up to 10$\%$ for the case of $\rho^+$ have been projected in the kinematic regions available at the  JLab22 GeV as shown in Fig.~\ref{fig:Collins-rho}. The large values of the asymmetries stems from the behavior of the transversity PDF in the valence region which (see {\it e.g.} Ref.~\cite{COMPASS:2014bze}). The intriguing dependencies of the asymmetry as a function of $z$ and $P_{\rm T}$ are a genuine prediction of the model which can be confronted with measurements at the JLab 22 GeV. Similar studies performed for kaons may give a hint on the origin of kaon SSAs observed in SIDIS which are typically higher that SSAs on pions. While the expected relative fractions of vector mesons vs. scalar mesons in the strange sector may be higher, the relative fractions of decay particles in the kaon samples are less. The measurement of the Collins asymmetries for $\rho$ and $K^*$ mesons at JLab22 will thus serve as a benchmark for the ``string+${}^3P_0$ model" of hadronization, and in general for the models of hadronization aiming at explaining the experimentally observed spin effects. The developed models can then be used for the systematic inclusion of spin effects in present Monte Carlo event generators, which are needed for the future experiments such as the EIC.

\clearpage
\noindent
\begin{figure}[!t]
    \centering
    \includegraphics[trim={0 1cm 0 2cm},clip,width=\textwidth]{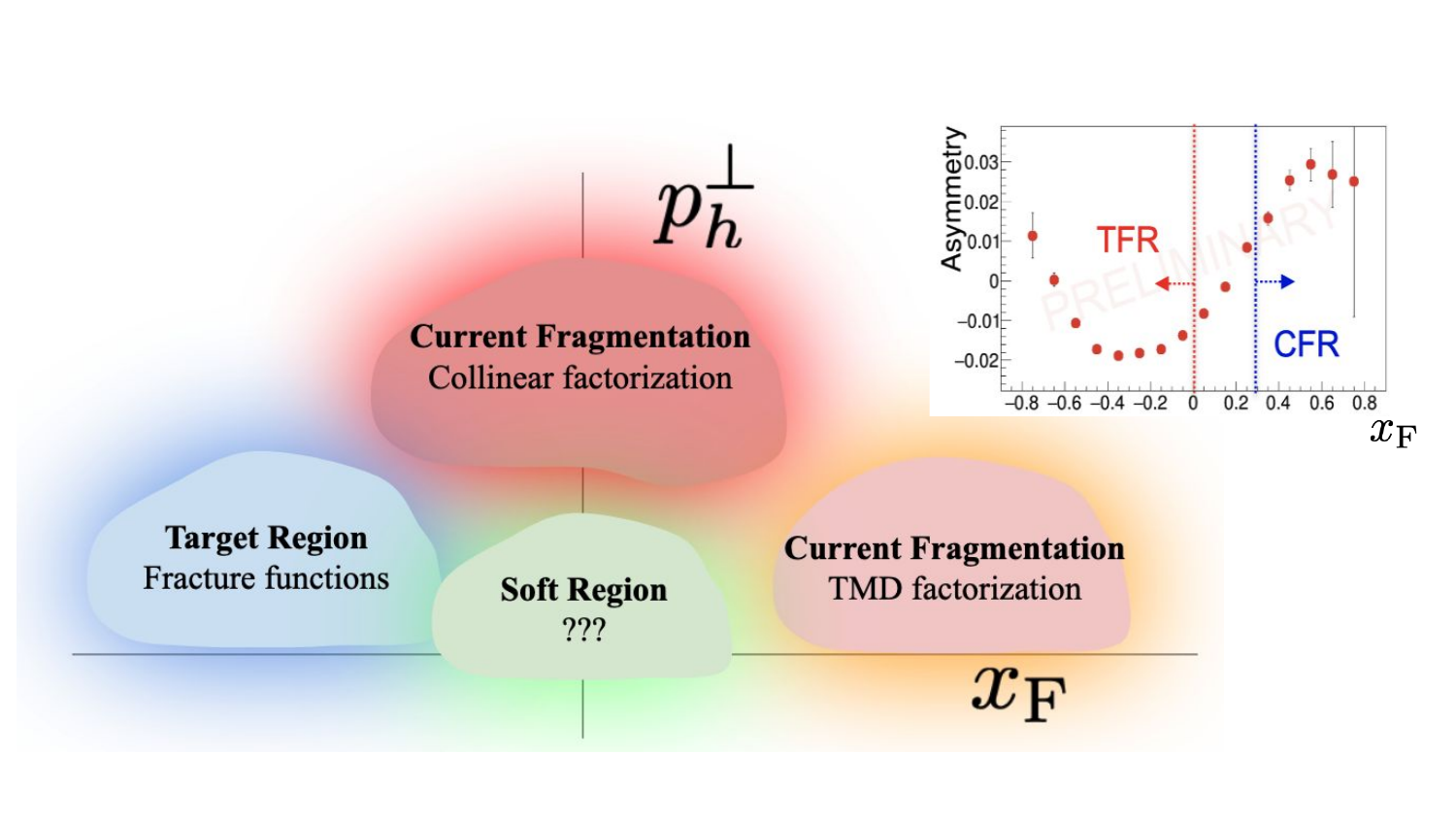}
    \caption{Conceptual separation of SIDIS regions  and preliminary CLAS12 beam-spin asymmetry for the inclusive $ep \rightarrow e'pX$ sample with a 10.6~GeV longitudinally polarized electron beam incident on a liquid-hydrogen target as a function of $x_F$, and the missing mass of the $epX$ system (left)}
    \label{fig:tfr_vs_cfr_epX}
\end{figure}
\underline{\emph{Correlations of Hadrons in the Current and Target Fragmentation Regions}}. In order to conduct detailed studies of the correlations in hadron production, it is necessary to detect additional hadrons produced either in the  CFR, or in the Target Fragmentation Region (TFR), 
where observed hadrons are generated from the hadronization of spectator partons that did not interact with the virtual photon. In the TFR, hadrons are not described by factorization into PDFs and FFs. Instead, the theoretical framework to study these reactions is based on the concept of Fracture Functions (FrFs), originally established in Ref.~\cite{Trentadue:1993ka} and later extended to the spin- and transverse-momentum-dependent case \cite{Anselmino:2011ss}. Similar to PDFs and FFs, FrFs describe the conditional probability of forming a specific final state hadron after the ejection of a particular quark. Studies of the TFR \cite{CLAS:2022sqt} are not only interesting in their own right but are also critical for properly interpreting many CFR measurements, which have been the driving force behind numerous experiments over the past few decades. For instance, while it is sometimes possible to kinematically separate the TFR and CFR 
({\it e.g.}, in high-energy Drell-Yan processes), it is not always clear where this demarcation occurs, particularly in fixed-target experiments. 

In the absence of a comprehensive understanding of the signals anticipated from target fragmentation, there exists a potential risk of misinterpreting results that are erroneously attributed to current fragmentation.
Therefore, studying the TFR is crucial for a comprehensive interpretation of SIDIS measurements and to avoid potential misattributions in the analysis of experimental data.

In the literature, various methods have been proposed to differentiate between contributions from CFR and TFR in SIDIS experiments. The commonly used variables for this purposes are the hadron rapidity, defined as $\eta = \frac{1}{2} \log \left[ \frac{E_h + p_z}{E_h - p_z}\right]$, and the $x$-Feynman variable, denoted as $x_F = \frac{2 p \cdot q}{|q| W}$. Recently, new phenomenological ideas have emerged \cite{Boglione:2016bph}, offering additional tools to quantify the likelihood of a given kinematic bin in SIDIS to be controlled by a particular physical production mechanisms as illustrated in  Fig.~\ref{fig:tfr_vs_cfr_epX} (left). However, it is important to note that no exact experimental observable exists that can precisely separate the TFR from the CFR. One possible approach to discriminate between the two regions is by leveraging the fact that although the form of the inclusive cross sections remains the same in both regions \cite{Anselmino:2011ss}, the structure functions governing the kinematic dependence of individual modulations have distinct origins. For instance, in Fig.~\ref{fig:tfr_vs_cfr_epX}, the CLAS12 beam-spin asymmetry from inclusive proton production is presented as a function of $x_F$. In the negative $x_F$ region, where the reconstructed proton moves opposite to the virtual photon, a negative 2\% asymmetry is observed. As the asymmetry transitions to the positive $x_F$ region, the sign flips, and it levels out at around positive 2\%. The magnitude of the asymmetry in the intermediate transition region, when compared to both edge cases, can provide indications about the relative composition of CFR and TFR events. Accurately tracking this transition and appropriately accounting for background events originating from the opposite kinematic region to the one of interest are crucial components in fixed-target experiments, such as JLab22. Moreover, such considerations could even impact the interpretation of TMD studies at the EIC.

\begin{figure}[ht]
    \centering
    \includegraphics[width=0.5\textwidth]{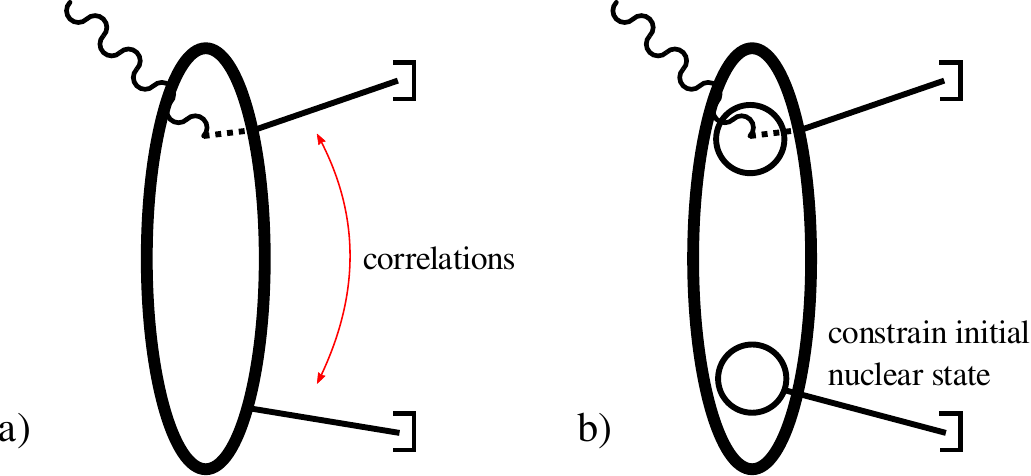}
    \caption{Schematic diagram of tagged SIDIS processes where we illustrate the interaction between the target and the virtual photon.  a) Detection of a hadron in both the current fragmentation region (top line) and target fragmentation region (bottom line) gives access to partonic interactions and correlations.  b) Tagged measurements on light nuclei with tagged nuclear fragments constrain the initial nuclear configuration.}
    \label{fig:tagged}
\end{figure}

\noindent
\underline{\emph{Tagged SIDIS Measurements}}. Tagged measurements refer to processes where a hadron is detected with momenta of the order of zero up to several 100 MeV (relative to the target center-of-mass) in the target fragmentation region.  For SIDIS a tagged measurement means a hadron detected both in the current and fragmentation region, giving access to parton interactions and correlations~\cite{Schweitzer:2012hh}, see Fig.~\ref{fig:tagged}a.  Both correlations in kinematic variables (longitudinal momentum, transverse momentum $P_T$, azimuthal angles) as well as in quantum numbers (flavor, spin, charge) could be studied.  The $P_T$ correlations in particular could shed light on the origin of the intrinsic quark transverse momentum and discriminate that from other possible sources such as final-state interactions and soft radiation~\cite{Schweitzer:2012hh}.  Even without detection of a current fragmentation hadron, the tagged DIS measurement would yield information on the dynamics of target fragmentation~\cite{Trentadue:1993ka} and hadronization, which is an area with few available measurements.

On light nuclei such as $^2$H, $^3$He and $^4$He, nuclear fragments (so-called spectator nucleon(s) or an $A-1$ nucleus) can be tagged.  The momenta of the detected fragments then result in an additional handle on the configuration of the initial nuclear state (see Fig.~\ref{fig:tagged}b), to be contrasted with non-tagged measurements where one averages over all possible nuclear configurations. For the deuteron with proton spectator tagging, measurements at small spectator momenta ($\sim$100 MeV) can be used to perform on-shell extrapolation, which probes free neutron structure~\cite{Sargsian:2005rm,Cosyn:2020kwu}. In the case of tagged SIDIS, this would allow the extraction of neutron TMDs free from nuclear effects and corrections, an essential ingredient to perform TMD flavor decompositions. At larger spectator momenta, a differential study of medium modifications or non-nucleonic components of the nuclear wave function belongs to the possibilities.

Tagged measurements are in general challenging in fixed-target experiments, since they need dedicated detectors to detect the low-momentum tagged particles. JLab will have the necessary equipment and experience from the 12 GeV era measurements with the BONUS12~\cite{Bonus12:2006} and ALERT~\cite{Armstrong:2017wfw} detectors. Moreover, JLab will have a monopoly on these data in the very high Bjorken-$x$ region and the tagged measurements -- which invite highly differential studies in the measured variables -- will benefit from the very high luminosity of the proposed upgrade.

\noindent
\underline{\emph{Independent Fragmentation and Role of Charge Symmetry}}.  At leading order in perturbative QCD, the assumption of independent fragmentation allows us to utilize the ratios of semi-inclusive $\pi^+$ and $\pi^-$ production to investigate potential charge symmetry violation (CSV) effects in quark PDFs. Recently, Experiment E12-09-002 was conducted in Hall C, which collected SIDIS data for $\pi^+$ and $\pi^-$ production from deuterium. The aim of this experiment was to explore CSV effects using the formalism proposed in Ref.~\cite{Londergan:1996vf}. Additional data were also obtained from hydrogen to assess the validity of the assumption of independent fragmentation. By studying the $z$ dependence and magnitude of various charge and target ratios, such as the ``difference ratio'' $H(\pi^+-\pi^-)/D(\pi^+-\pi^-)$, it is possible to examine the dependence on the valence quark distributions. Any deviation from the expected behavior of these ratios would indicate violations of charge symmetry or inconsistencies in the independent fragmentation assumption.

The extraction of CSV from SIDIS pion production requires knowledge of the ratio of unfavored to favored fragmentation functions, $D_{unfav}/D_{fav}$.  This ratio can either come from existing parameterizations or can be fit to the data from the experiment.  A multiparameter fit performed on the Hall C data revealed a determination of the charge symmetry violating quark Parton Distribution Functions (PDFs) consistent with the upper limit derived from a previous global fit of quark PDFs \cite{Martin:2003sk}. Unlike most PDF fits that do not consider quark charge symmetry violation as a free parameter, this earlier fit allowed for the inclusion of quark CSV as a degree of freedom. However, the quality of the fit of the Hall C data is not ideal indicating some tension between the experimental data and the assumed leading order form. On the other hand, if fragmentation functions that incorporate some degree of charge symmetry violation are assumed (for example, using the fragmentation function fit from Ref.~\cite{deFlorian:2007aj}), the quality of the fit is much improved. 

These preliminary results are intriguing, and it is difficult to disentangle charge symmetry violating effects in the quark PDFs and fragmentation functions from the possible lack of validity of the leading order fragmentation assumption.  Better understanding of the $P_T$-dependence of fragmentation functions, and the impact of vector mesons discussed above, will certainly help in quantifying the systematics of the CSV and the assumption of independent fragmentation in general. The large $Q^2$ range allowed by 22 GeV would be of enormous benefit in that the expected $Q^2$ behavior of the SIDIS cross sections as well as the ratios discussed above could be explored with high precision.  For example, the unexpected $Q^2$ dependence would suggest that the leading-order assumption and  CSV extractions from SIDIS data will require proper evaluation of the systematics.  Conversely, if the expected $Q^2$ dependence is observed, one would have greater confidence that any observed CSV effects are valid.  

\noindent
\underline{\emph{Precision TMD Studies}}. From the perspective of phenomenological applications, the JLab 22~GeV upgrade is expected to provide unprecedented accuracy in the measurement of SIDIS cross sections in the large-$x$ region, which will help to determine the TMD nucleon structure with greater precision than ever before. The impact of the JLab22 data on reducing uncertainties is estimated to be about two orders of magnitude for $x=0.1$, as demonstrated in Fig.~\ref{fig:MAP22TMD impact}. This estimation is based on the MAP22TMD~\cite{Bacchetta:2022awv} extraction as the baseline. A similar impact is expected when using the SV19 extraction~\cite{Scimemi:2019cmh}.

It is crucial to acknowledge that the impact studies conducted with the current generation of TMD fits have limitations. The main factor behind this is that the available data does not have the resolution required to capture the finer details of the TMD distributions, which leads to the use of simplified ansatzes in the extractions that may be biased. Notably, none of the current fits account for flavor dependence or PDF uncertainty. The unaccounted uncertainties from these factors could potentially be significant \cite{Bury:2022czx}. The energy range of JLab22 measurements, $Q^2<20$~GeV$^2$, is also a critical aspect to consider. At these energies, power corrections to the factorization theorems are not negligible. The effects of power corrections have been shown to be significant in existing SIDIS measurements, such as those at COMPASS or HERMES, but are often neglected due to insufficient precision and resolution. This is particularly true for polarized measurements. With the high precision of JLab22, the impact of power correction effects can be explored with great accuracy, providing valuable input for theoretical studies.

The precision and resolution expected from JLab22 present an opportunity to directly study TMD physics in position space. One promising avenue is the direct determination of the Collins-Soper (CS) kernel, a critical element of the theory which relates the TMD-PDFs at different energy scales, by combining SIDIS measurements at different $Q$, thus avoiding parametric bias \cite{BermudezMartinez:2022ctj}. The high precision of JLab22 will allow for a very accurate exploration of the CS kernel, including power corrections, and provide valuable input for theoretical studies. Similar studies are also possible at the EIC. Figure~\ref{fig:CS_JLabvsEIC} shows the estimated uncertainties for JLab22 and EIC, with the baseline being the SV19 value of the CS kernel \cite{Scimemi:2019cmh}. The results demonstrate the complementary nature of EIC and JLab22 for such studies, with JLab22 capable of accessing larger values of $b$ due to finer resolution at small $P_T$, while EIC provides more accurate values at small $b$ due to wider coverage in $P_T$ and higher $Q$.

As previously mentioned, the measurements of SIDIS at JLab22 will be affected by power corrections. Consequently, the extraction of the CS kernel will take the form illustrated in Fig.~\ref{fig:CS_JLabvsEIC} (right). At small $b$, the value of the CS kernel is purely perturbative, and the difference between the left and right panels in Fig.~\ref{fig:CS_JLabvsEIC} demonstrates the effect of power corrections. By comparing extractions at different $Q$, $x$, and $z$, it will be possible to reconstruct the shape of power corrections with great precision and without any modeling. This presents a unique opportunity that can only be realized with high-luminosity $ep$ machines such as JLab22 and EIC.

%
\begin{figure}[t]
  \begin{center}
    \includegraphics[width=0.40\textwidth]{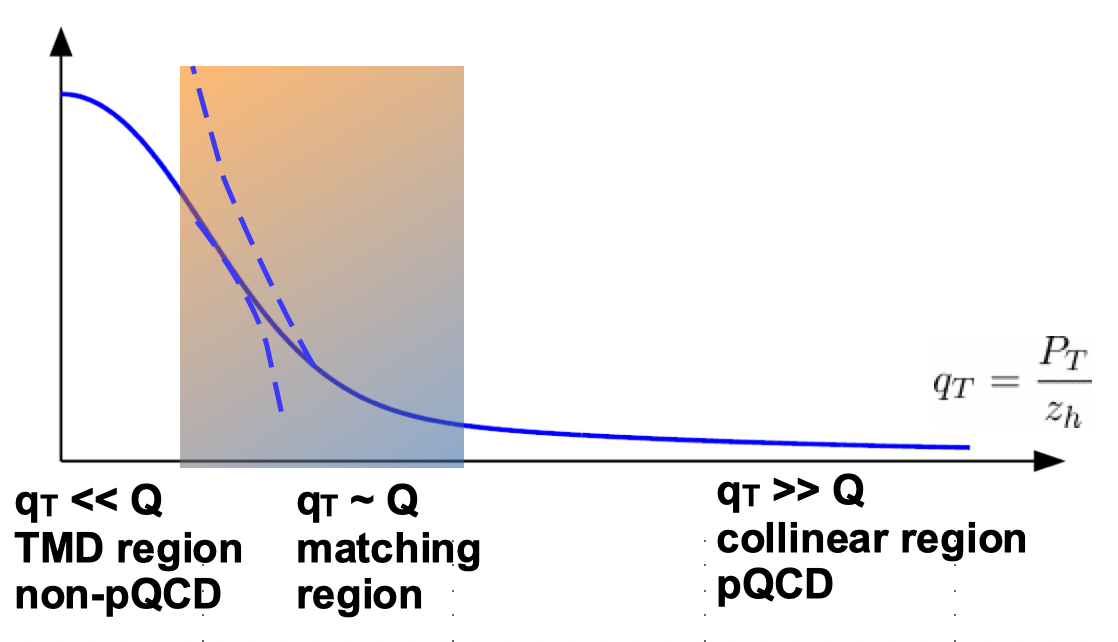}
    \hspace{0.5cm}
    \includegraphics[width=0.55\textwidth]{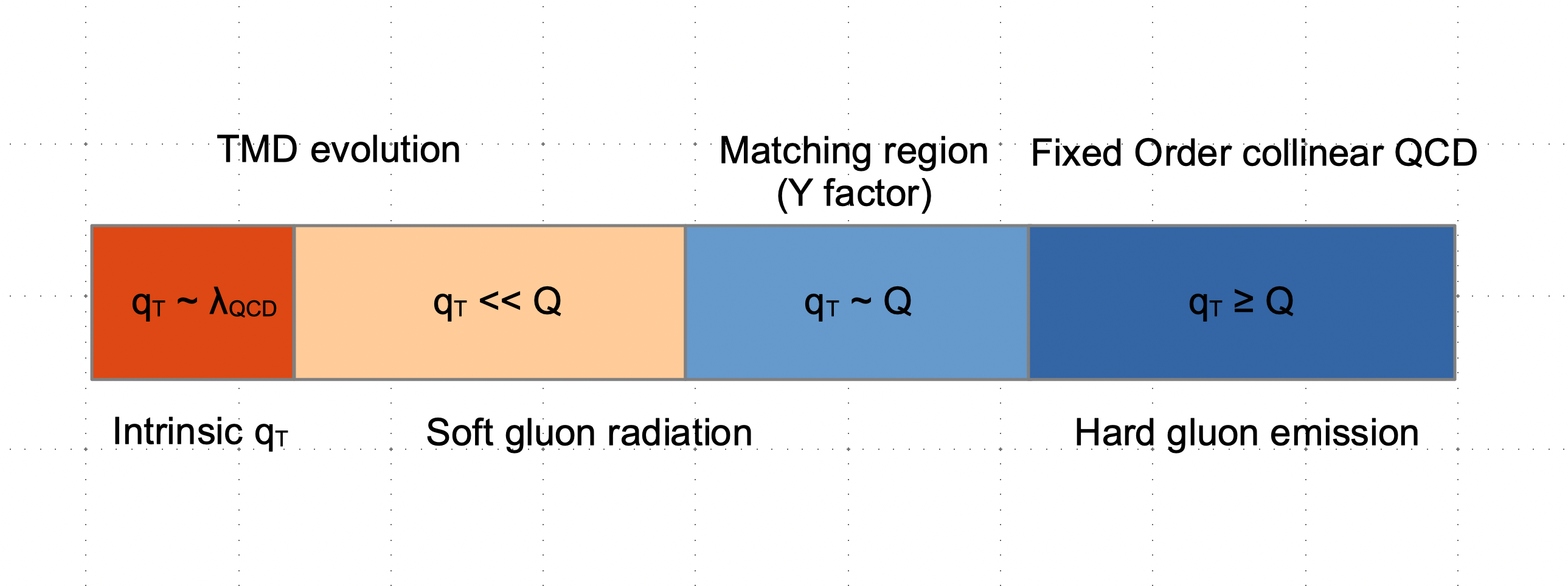}
 \end{center}
  \caption{The $q_T$ distribution of the SIDIS cross section can be divided in different regions according to the size of $q_T$ with respect to $Q$. Large $q_T$s correspond to the collinear region, where we expect pQCD to be at work. Small $q_T$s correspond to the TMD region, where non-perturbative effects become dominant. A smooth matching is supposed to happen in the intermediate region, the so-called matching region. }
 \label{secfig:1}
\end{figure}

\noindent
\underline{\emph{The Matching Region in SIDIS}}. The theoretical study of SIDIS is based on factorization theorems which, in principle, allow for the description of the $q_T$ ($q_T=P_T/z$) distribution of the SIDIS cross section over the full $q_T$ range. More specifically, TMD factorization allows for the description of the small $q_T$ region, where $q_T \ll Q$, but fails at larger values of $q_T$. In turn, collinear factorization describes the cross section at large $q_T$, where $q_T \gg Q$, but becomes divergent at small $q_T$. As shown in the left panel of Fig.~\ref{secfig:1}, the region where the two schemes are supposed to match is called ``matching region''. Usually the matching procedure is devised to work on the basis of the $Y$ term contribution, which corrects for the misbehavior of the non-perturbative contribution ($W$) as $q_T$ becomes large, providing a consistent (and positive) $q_T$ differential cross section,
$\sigma = W+Y$. The $Y$-term should provide an effective smooth transition to large $q_T$, where fixed-order perturbative calculations are expected to apply. For this scheme to work four distinct kinematic regions, large enough and well separated between one another, have to be clearly identified. A pictorial representation of these regions is shown in the right panel of Fig.~\ref{secfig:1}. Given the major part of the polarized SIDIS data is in the range of $0.2<z<0.8$ the ``matching region'' will be for $P_T\sim 0.5-1.5$ GeV.

Serious issues, however, affect the practical implementation of this scheme. In fact, comparison with existing SIDIS experiments like HERMES, COMPASS, and JLab12 have shown large discrepancies with the cross section computed in the collinear formalism, which is expected to be valid at large $q_T$, as shown in Ref.~\cite{Gonzalez-Hernandez:2018ipj}. Moreover, matching through the $Y$ term often fails as $Y$ is very large (as large as the cross section itself) even at low $q_T$ and it is affected by very large theoretical uncertainties, see for example the study of Ref.~\cite{Boglione:2014oea}. The issues described above lead to some fundamental questions, which urgently need  answering: How does QCD manifest itself in the matching region? Is it just a transition region or does it need a new theoretical approach of its own?  

In order to gain a thorough understanding of the transition region between different kinematic regimes in SIDIS, it is crucial to have high-statistics data that precisely cover the relevant kinematics. This transition region represents intervals where competing processes, production mechanisms, and even exclusive diffractive processes contribute to the SIDIS reaction. These intervals can be referred to as ``matching regions'' where different mechanisms coexist and complement each other. The relative contribution of TMD and collinear factorization regions in these matching regions strongly depends on the specific kinematic sector being probed in the multi-dimensional phase space. Similarly, the correlations between hadrons produced in the CFR and TFR, as well as other competing processes, are intricately linked to the dynamics of these transitions. To effectively separate and study these transitions, it becomes crucial to have precise and high-quality experimental measurements in the multi-dimensional phase space. Such measurements will provide valuable information on the dependencies of various observables on $Q^2$, which are expected to differ for different contributing mechanisms. This enables the validation and proper separation of these distinct mechanisms. 

A unique feature of JLab22 is that it will offer an unprecedented insight into the matching region, a region that cannot be explored with similar resolution in any other SIDIS experiment. The upgraded JLab22, with its amazing statistics, will be a magnifying glass on the central region and will allow us to explore an energy and transverse momentum range that is crucial to improve our current understanding of QCD in terms of factorization theorems. With its fine resolution and largely extended reach in $x$, JLab22 will acquire an unprecedented ability to perform multiple binning analyses as shown in Fig.~\ref{fig:JLAB4D}, which will  provide high precision information on the TMD region.

\begin{figure}[t]
    \begin{center}
    \includegraphics[width=0.45\textwidth]{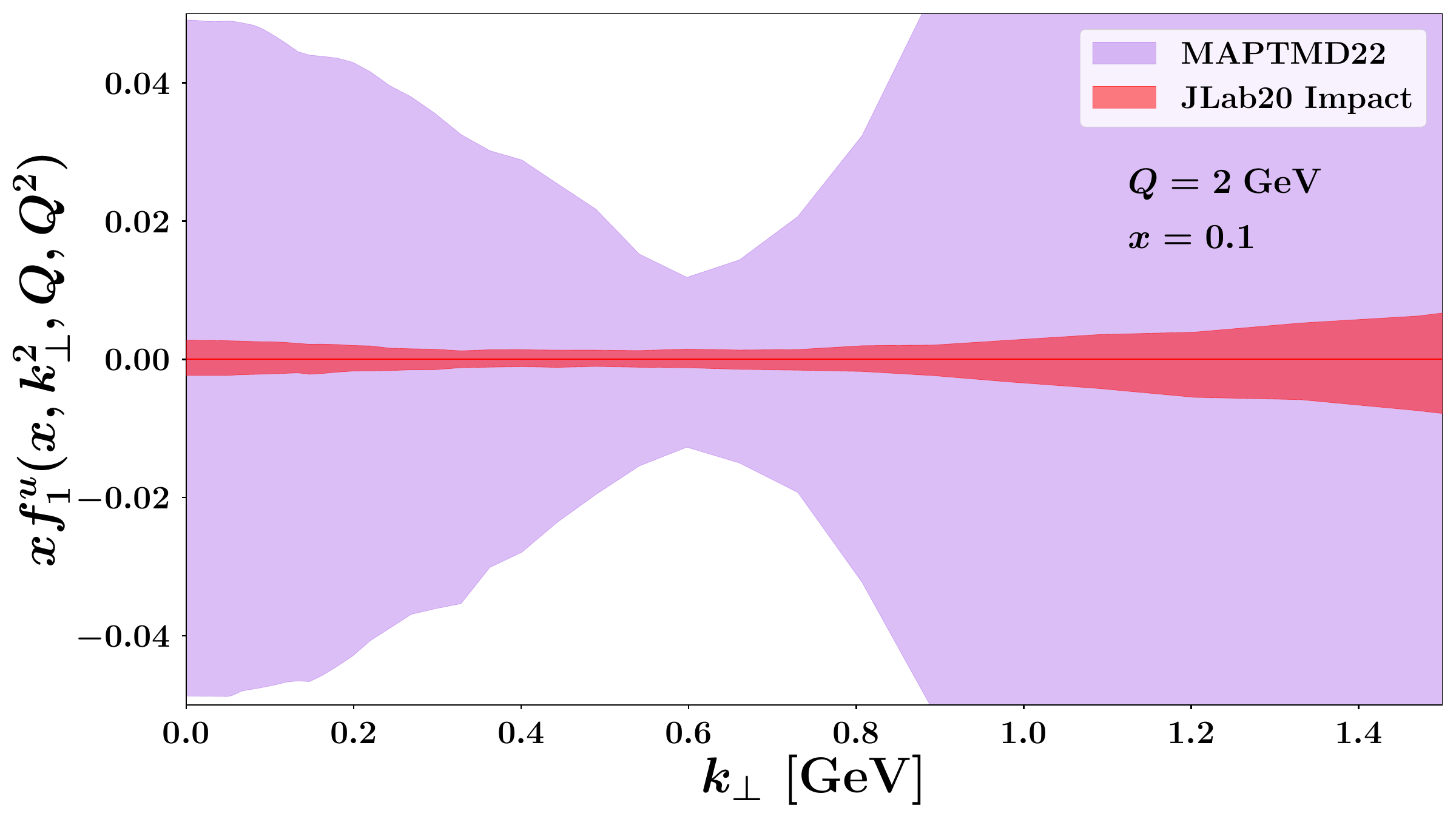}
    \includegraphics[width=0.45\textwidth]{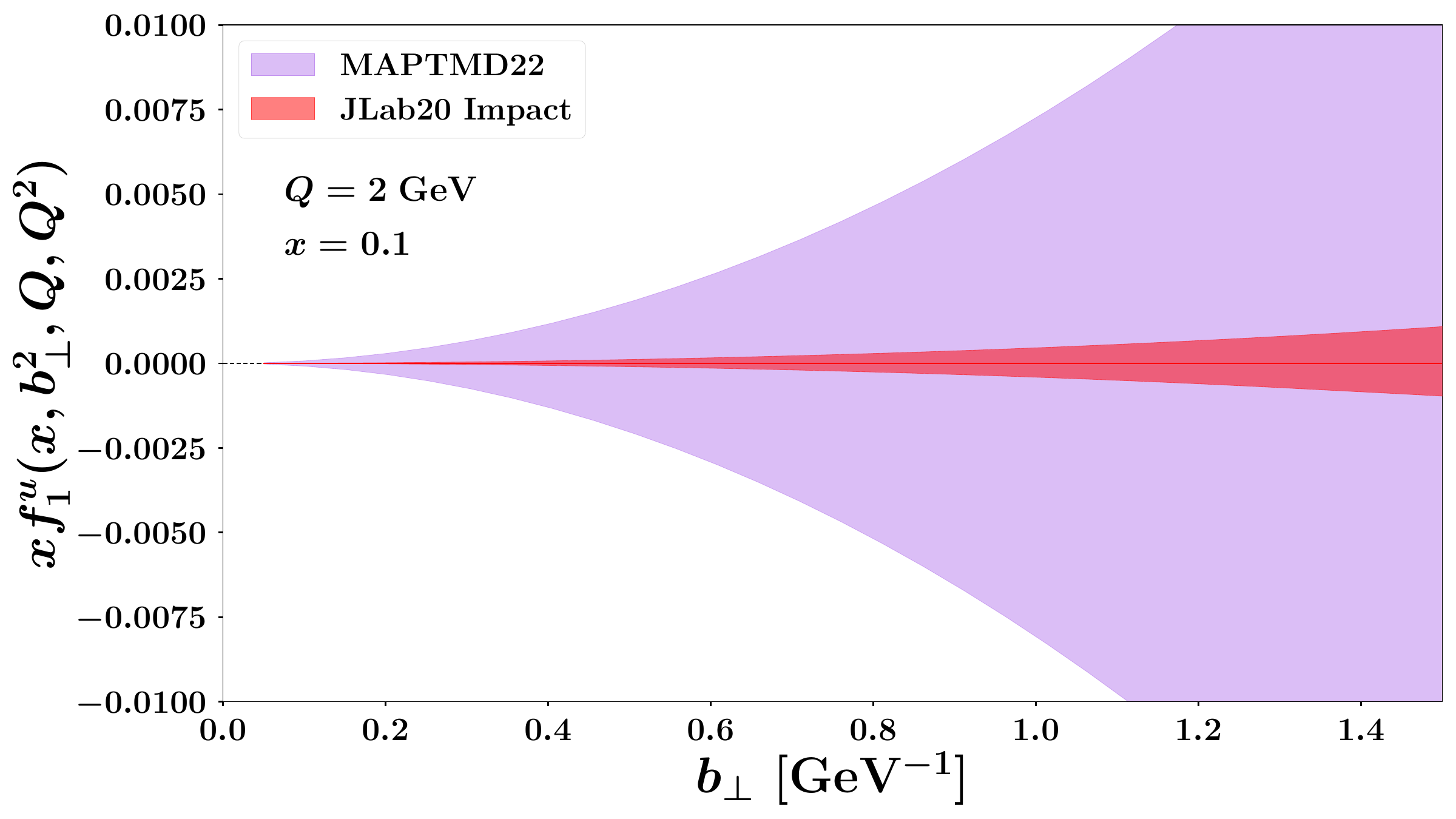}
    \end{center}
    \caption{Impact on the error bands of the TMD in $k_{\perp}$ space (left) and its Fourier-conjugate  $b_{\perp}$ (right) at two values of $x$ and at $Q= 2$ GeV, based on the MAP22TMD analysis~\cite{Bacchetta:2022awv}. Purple bands: current situation. Red bands: after the inclusion of JLab22 data.}
    \label{fig:MAP22TMD impact}
\end{figure}
\begin{figure}[t]
    \begin{center}
    \includegraphics[width=0.45\textwidth]{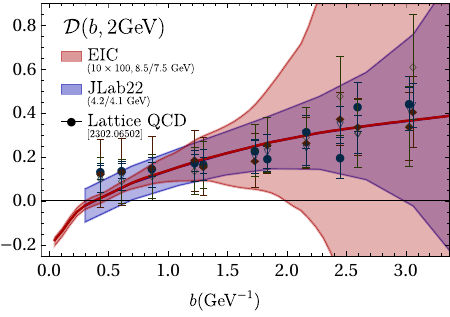}
    ~~
    \includegraphics[width=0.45\textwidth]{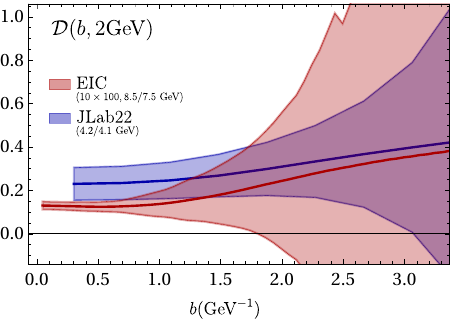}
    \end{center}
    \caption{Comparison of uncertainty bands for the Collins-Soper (CS) kernel versus $b_\perp$ ($b=b\perp$), directly extracted from the data using the method proposed in \cite{BermudezMartinez:2022ctj}, for both EIC and JLab22. The extractions consider only a single ratio of two bins in $Q$, integrated over $x$ and $z$. In the left figure, the CS kernels are normalized to the SV19 value \cite{Scimemi:2019cmh} for better visibility of the uncertainty bands. The right figure shows a more realistic picture of the extracted CS kernel, including the effects of power corrections.}
    \label{fig:CS_JLabvsEIC}
\end{figure}

\noindent
\underline{\emph{Role of LQCD}}. 
In addition to experimental efforts in SIDIS, significant progress has been made in recent years to calculate TMD-PDFs inside a nucleon from lattice QCD. One direction is the computation of ratios of TMD $x$-moments in the $b_\perp$ space~\cite{Yoon:2017qzo}, and the other is large-momentum effective theory (LaMET)~\cite{Ji:2013dva,Ji:2020ect} that allows for the extraction of the Collins-Soper kernel~\cite{Ebert:2018gzl} as well as the full $(x,b_\perp)$ dependence of the TMD-PDFs~\cite{Ji:2019ewn}. So far, there have been several lattice calculations of the CS kernel at unphysical quark masses~\cite{Shanahan:2020zxr,LatticeParton:2020uhz,Schlemmer:2021aij,LPC:2022ibr,Shu:2023cot}, where the systematic uncertainties are gradually being understood and improved. Besides, the first exploratory calculation of the unpolarized proton TMD-PDF with lattice renormalization and one-loop perturbative matching has also been carried out~\cite{LPC:2022zci}. The lattice results are currently in qualitative agreement with phenomenological results and exhibit similar behaviors in the Fourier-conjugate position space $b_\perp$, with uncertainties increasing gradually as $b_\perp$ increases. This highlights the importance of experimental studies at large $b_\perp$, which requires fine binning of experimental data in the $P_T$ of hadrons. On the lattice side, with the improvement of statistical and systematic errors in the large $b_\perp$ region, there will be a more precise comparison between theory and experiment on these TMD observables, and JLab data can be critical to test the theory.

%
\begin{figure}[t]
   \includegraphics[height=.3\textwidth,width=0.5\textwidth]{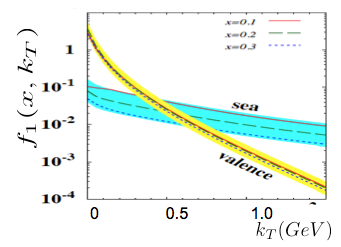}
   \includegraphics[height=.3\textwidth,width=0.45\textwidth]{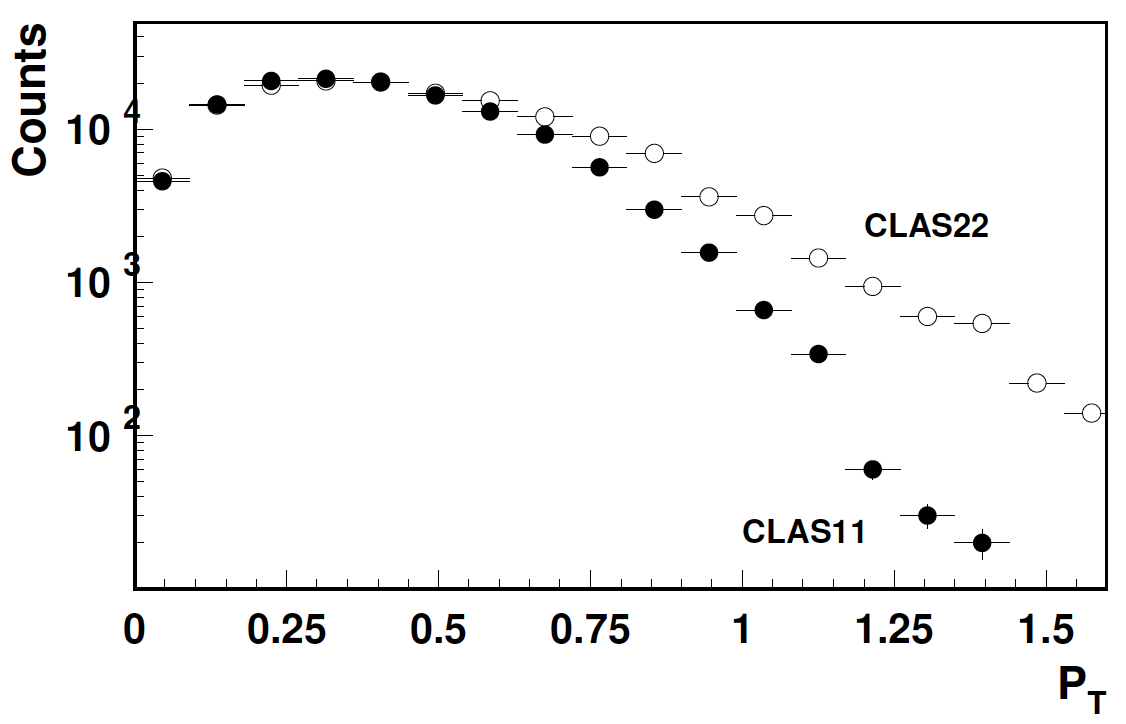}
\caption{Transverse momentum dependence of sea and valence quarks \cite{Schweitzer:2012hh} (left) and extension of the transverse momentum coverage with JLab22 (open circles) for a given bin in $x$ and $z$ ($0.25<x<0.3, 0.35<z<0.45$) at $Q^2>3$ GeV$^2$.}
 \label{fig:TMDcase}
\end{figure}

\begin{figure}[t]
\centering
\includegraphics[width=0.48\textwidth]{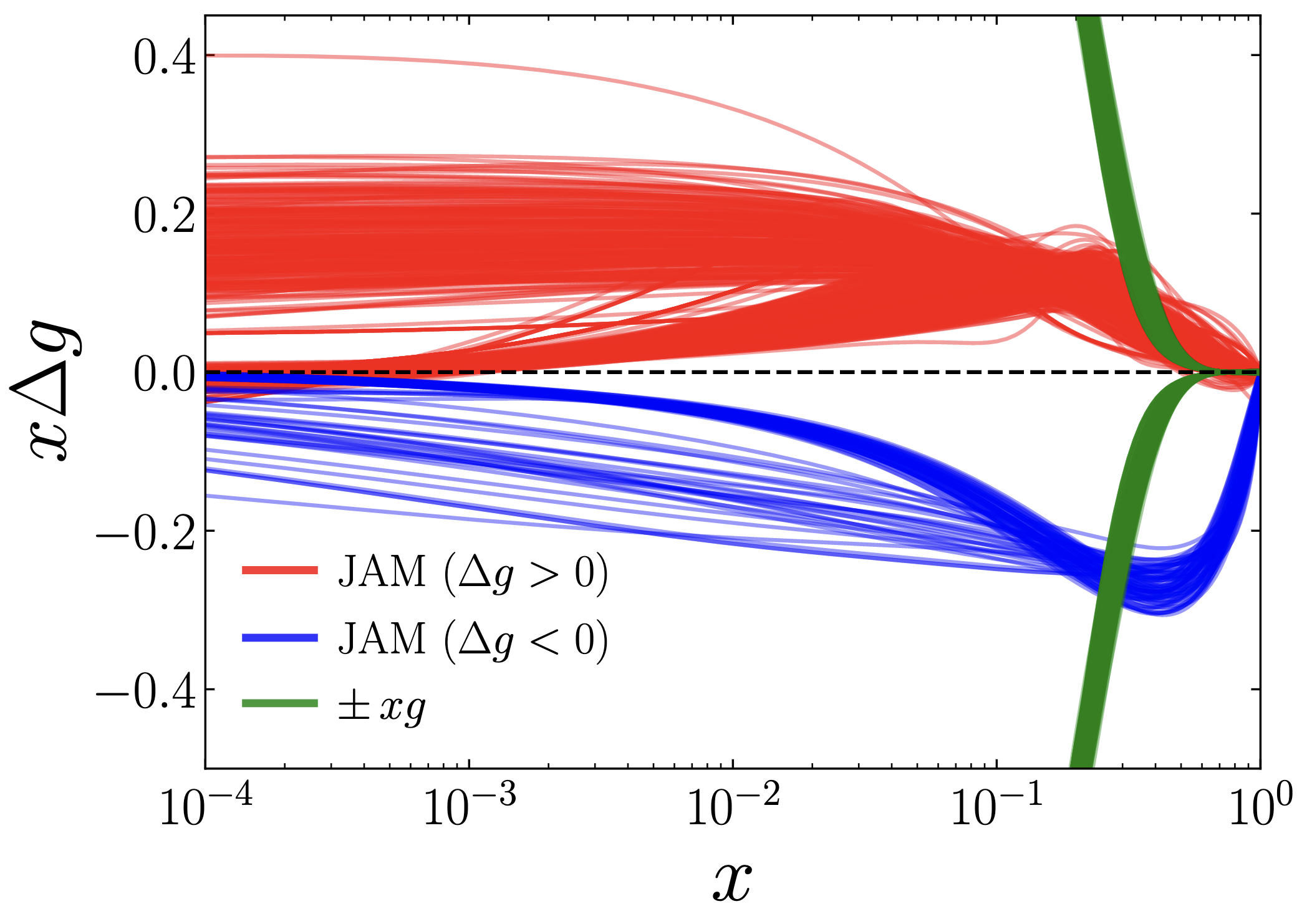}
\includegraphics[width=0.5\textwidth]{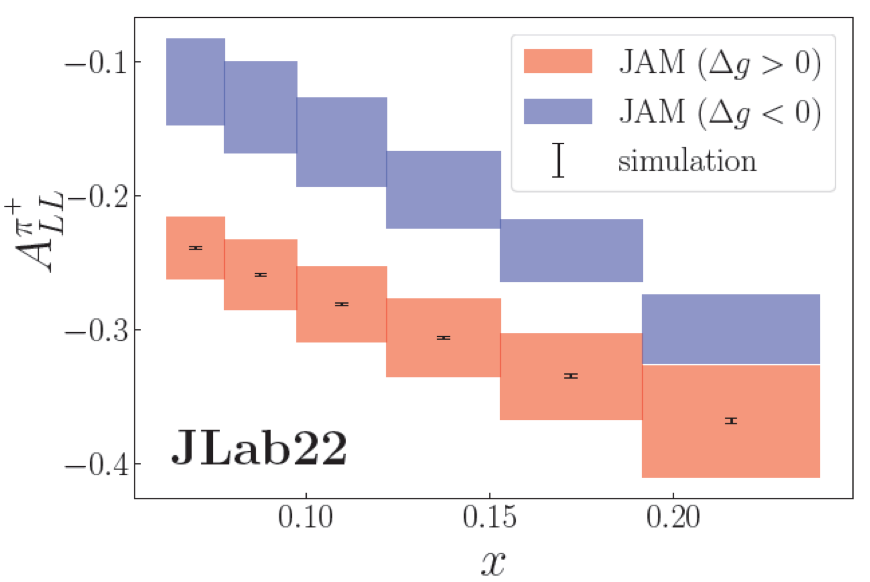}
\caption{Left panel: Polarized gluon distribution $x \Delta g(x)$ at $Q^2 = 10$~GeV$^2$ from JAM~\cite{Zhou:2022wzm}, showing separately  $\Delta g > 0$ (red lines) and $\Delta g < 0$ (blue lines) and contrasted to $\pm$ the unpolarized gluon distribution, $x|g(x)|$ (green lines). Right panel: double longitudinal spin asymmetry $A_{LL}^{\pi^+}$ for semi-inclusive $\pi^+$ production from a proton, for a selected kinematics at JLab with 22~GeV electron beam. Note that the heights of the colored boxes give a $1\sigma$ uncertainty in the asymmetry from the PDF replicas, while the error bars give the expected statistical uncertainty with 100\% acceptance. }
\label{fig:gluon-ppdfs}
\end{figure}

\noindent
\underline{\emph{Mapping out the Large-$x$ Sea}}.  Understanding the dynamics of partons, including the non-perturbative sea quarks, is crucial for gaining insights into strong interactions. The correlations between the spin of the target and/or the momentum and spin of quarks, along with final state interactions, determine the azimuthal distributions of produced particles. Measurements of flavor asymmetries in sea quark distributions, carried out in Drell-Yan experiments, have revealed substantial non-perturbative effects at large Bjorken-$x$, where valence quarks dominate \cite{Alberg:2017ijg}. Earlier measurements by the NMC experiment indicated that the integrated $\bar{d}$ distribution is larger than the integrated $\bar{u}$ distribution \cite{NewMuon:1991hlj}. The E866 and SeaQuest Collaborations have provided more recent measurements suggesting a significantly larger $\bar{d}$ distribution compared to $\bar{u}$ across the accessible range of $x$ \cite{Garvey:2001yq, Nagai:2017dhp}. Non-perturbative $q\bar{q}$ pairs, which are also correlated with spins, play a crucial role in spin-orbit correlations and the measurement of single-spin asymmetries observed in various experiments over the past few decades.

Predictions indicate that the distribution of unpolarized sea quarks exhibits a power-like tail, approximately proportional to $1/P_T^2$, extending up to the chiral symmetry-breaking scale \cite{Schweitzer:2012hh}. A similar behavior is observed in the flavor-nonsinglet polarized sea. The transverse momentum distributions of valence and sea quarks are predicted to have distinct shapes, particularly at large values of $P_T$. The effect of dynamical chiral symmetry breaking on the partonic structure of nucleons has important implications for the transverse momentum distributions of particles produced in hard scattering processes. With the significant increase in phase space provided by 22 GeV experiments, a much wider range of transverse momenta can be accessed, which is critical for studying TMDPDFs.

\noindent
\underline{\emph{Gluon Polarization}}. For the past thirty years~\cite{Aidala:2012mv}, the nuclear physics community has been driven by the pursuit of understanding the proton spin puzzle - the breakdown of the proton's spin into its quark and gluon helicity and orbital angular momentum components. Experimental programs around the world have been dedicated to this effort, and we now have a relatively comprehensive understanding of the total fraction of helicity carried by quarks. However, questions remain about the specific flavor decomposition of the sea quark contributions. A significant breakthrough was achieved when double spin asymmetries were observed in inclusive jet production in polarized proton-proton collisions at RHIC~\cite{STAR:2014wox}, allowing for the first detection of a polarized gluon distribution. Follow-up data from the STAR~\cite{STAR:2019yqm, STAR:2021mfd, STAR:2021mqa} and PHENIX~\cite{PHENIX:2010aru} Collaborations have reinforced these findings, giving us greater confidence in our understanding of both the quark and gluon helicity content of the proton.

The JAM Collaboration~\cite{Zhou:2022wzm} recently conducted a review of the analysis of jet data to assess the degree to which the results rely on the theoretical assumptions made in the analysis, such as SU(3) flavor symmetry for the axial vector charges that govern non-singlet combinations of spin-dependent PDFs~\cite{Bass:2009ed, Ethier:2017zbq}, and the positivity constraints for unpolarized PDFs. Their analysis revealed that a second set of solutions could be possible without the PDF positivity constraints~\cite{Candido:2020yat, Collins:2021vke}, which are not technically necessary based on theoretical grounds. As illustrated in Fig.~\ref{fig:gluon-ppdfs}, this set of solutions could lead to $\Delta g < 0$, contrary to the traditional small and positive $\Delta g$ and positive quark polarization $\Delta q$ that combine to produce a positive asymmetry, as observed in the STAR data. The data suggest that negative $\Delta g$ could also be feasible, with a greater magnitude, which, when paired with positive $\Delta q$, can generate a cancellation between the positive contribution from the gluon-gluon channel and the negative contribution from the quark-gluon channel, producing equally valid explanations for the inclusive jet data.

In a previous study, J\"{a}ger {\it et al.}~\cite{Jager:2003ch} explored potential constraints on the sign of $\Delta g$ by examining inclusive pion production in polarized $pp$ collisions using the PHENIX data~\cite{PHENIX:2015fxo} for neutral pions.  They were able to derive a small but negative lower limit for the double spin asymmetry that corresponds to negative gluon helicity PDF comparable to those observed in the JAM analysis~\cite{Zhou:2022wzm}. Furthermore, the sign of $\Delta g$ has been investigated through a comparison of PHENIX data on inclusive charged pion production~\cite{PHENIX:2014axc, PHENIX:2020trf}. 

The comparison of predictions based on recent JAM analysis~\cite{Zhou:2022wzm} for $\pi^+$ and $\pi^-$ asymmetries at $pp$ center of mass energies $\sqrt{s} = 200$~GeV and 510~GeV as a function of $x_T = 2 p_T/\sqrt{s}$, where $p_T$ is the transverse momentum of the final state pion in the laboratory frame, indicate the uncertainties on these data are still too large to exclude either a positive or negative value of $\Delta g$, even though the $\pi^+$ asymmetry has the potential to differentiate between the different solutions for $\Delta g$.

Although EIC might provide the definitive resolution on $\Delta g$ and its sign via scaling violation in DIS, an alternative possibility to resolve this problem on a comparable time scale can be found in the context of double spin asymmetries (DSAs) in SIDIS. Specifically, longitudinally polarized lepton-nucleon reactions with large transverse momentum hadrons produced in the final state have direct sensitivity on polarized gluons inside the initial state nucleons because the corresponding hard scattering matrix elements are at the same order of magnitude in the strong coupling constant $\alpha_s$ as the quark scattering contributions. The feasibility to discriminate the sign of gluon polarization with this process was recently studied at Ref.~\cite{Whitehill:2022mpq} with an analysis that compared the discriminatory capabilities at JLab 12~GeV and the potential upgrade to the 22 GeV with the projected $A_{LL}^{\pi^+}$ asymmetries. The statistical uncertainties for the JLab projections were based on a luminosity of $d \mathcal{L}/d t = 10^{-35}~{\rm cm}^{-2}{\rm s}^{-1}$. The asymmetries at JLab 12 GeV have relatively large values and small statistical uncertainties. However, for most kinematics, the asymmetry bands with positive and negative polarized gluons overlap significantly, making it difficult to differentiate between the positive and negative $\Delta g$ solutions. The upgraded 22 GeV electron beam allows access to a larger portion of the intermediate and low-$x$ region and provides better discrimination between the two possible scenarios for $\Delta g$ (see Fig.~\ref{fig:gluon-ppdfs}). We stress that this analysis did not included acceptance effects nor systematic effects stemming from depolarization.  The analysis in Ref.~\cite{Whitehill:2022mpq} concludes that a high luminosity JLab with a $\approx 20$~GeV beam is well-suited for differentiating between positive and negative solutions due to the asymmetry's scaling behavior with $\sqrt{s}$. 

\subsection{Summary}

The energy upgrade at JLab presents a unique opportunity to extend measurements to a wider range in $x$, $Q^2$, and $P_T$. This will be crucial for advancing our understanding of QCD dynamics, including the evolution properties and transverse momentum dependences of PDFs. To achieve a detailed understanding of the contributions to measured cross sections and asymmetries in SIDIS with controlled systematics, it is necessary to consider all involved kinematical variables ($x$, $Q^2$, $z$, $P_T$, and $\phi$). JLab is the only facility capable of separating different structure functions involved in polarized SIDIS, including longitudinal photon contributions. By performing precision Multi-D measurements of single and dihadron SIDIS with an upgraded CEBAF accelerator, and by studying the $Q^2$ dependences of observables, we can test the impact of several theoretical assumptions used in TMD phenomenology. This will also provide validation of the extraction frameworks, which is critical for proper evaluation of systematic uncertainties. Additionally, the detection of multiparticle final states and the study of multiplicities and asymmetries of dihadrons and vector mesons will offer crucial insights into the source of single spin asymmetries and the dynamics of the polarized quark hadronization process.

\clearpage \section{Spatial Structure, Mechanical Properties, and Emergent Hadron Mass}
\label{sec:wg4}

\subsection{Introduction}

The extended spatial structure of hadrons is one of the basic expressions of their emergence from QCD. It attests to their composite nature and reveals the dynamical scales created by the non-perturbative phenomena of chiral symmetry breaking and confinement (see Sec.~\ref{sec:intro}). It also reveals the mechanical properties (internal motion, forces) and allows one to discuss hadron structure in terms similar to those used
for nonrelativistic systems such as atoms or nuclei. The study of the spatial structure of hadrons is
a rapidly expanding field of science, with experimental programs ongoing at JLab12 and planned at EIC,
and many theoretical and experimental developments and opportunities reaching further into the future.

One source of information on the spatial structure are the hadron form factors of operators measuring
local physical quantities. Originally, the concept of form factors was developed for the electromagnetic
currents operators, and extensive efforts have been devoted mapping the distributions of charge
and magnetization in hadrons and nuclei. Recently, the concept of form factors has been extended to a
much larger class of local QCD operators composed from quark and gluon fields. The form
factors of the QCD energy-momentum tensor (spin-2 quark and gluon operators) describe the spatial
distributions of momentum, angular momentum, and forces in the nucleon and quantify the mechanical
properties of the dynamical system \cite{Polyakov:2018zvc,Lorce:2018egm}. The form factor of the
trace anomaly (spin-0 gluon operator) describes the spatial distribution of the gluonic fields
involved in scale symmetry breaking and plays an important role in the proton mass decomposition
\cite{Lorce:2017xzd,Hatta:2018sqd,Metz:2020vxd}.

Another source of information on the spatial structure of hadrons are the generalized
parton distributions (GPDs) \cite{Goeke:2001tz,Diehl:2003ny,Belitsky:2005qn}, which describe the
spatial distributions of quarks and gluons in the transverse plane seen by a high-energy probe
sampling field components with given longitudinal momentum.  They allow one to create ``tomographic images''
of the hadron in terms of quark/gluon degrees of freedom and bring them to life as 3D objects in space.
Extensive efforts are under way to extract the GPDs from experimental data and lattice QCD calculations
and construct the tomographic images. While essential progress will be made with the data from JLab12
and EIC, there remain significant challenges to hadron imaging that can be overcome with new
theoretical and experimental developments beyond these programs.

Form factors and GPDs are measured in exclusive electro/photoproduction processes, where the initial hadron
emerges intact in the final state, and the momentum transfer is conjugate to the spatial structure investigated.
Such measurements generally require high luminosity because of low rates and the need for differential
measurements. At the same time, they require collision energies allowing for energy and momentum transfers
significantly above the hadronic scale $\sim$ 1 GeV (high-$Q^2$ electroproduction, heavy quarkonium production).
The proposed high-intensity 22 GeV facility would provide the necessary combination of both capabilities
and substantially
expand the possibilities for exploring the spatial structure of hadrons in both gluon
and quark degrees of freedom. Qualitatively new applications are the measurement of gluonic form factors
of hadrons through exclusive charmonium production, and fully differential 3D imaging of the nucleon
using dilepton/diphoton production. In addition, the new facility would offer essential quantitative advances
in the study of electromagnetic form factors, GPDs with exclusive photon/meson production, and the study of
meson form factors and GPDs.

The emergence of hadronic mass from the massless theory of QCD is certainly the most fundamental phenomenon
of strong interaction physics. It gives rise to more than 90\% of the visible mass of the Universe
residing in the protons and neutrons in atomic nuclei. The question ``how'' this happens is essential
for understanding the structure	of matter. Hadronic	mass is	generated by nonperturbative interactions
between	quarks and gluons associated with mass scales significantly larger than the QCD scale parameter
$\Lambda_{\rm QCD} \approx$ 200 MeV (or distance scales significantly smaller than 1 fm). For quarks the mass generation
is associated with the well-known phenomenon of dynamical chiral symmetry breaking, which converts the nearly
massless QCD quarks into constituent quarks with a dynamical mass. The same nonperturbative interactions
mediate	high-momentum scattering processes on hadrons such as $N \rightarrow N$ elastic form factors
or $N \rightarrow N^*$ transition form factors at momentum transfers $Q^2 \sim$ few GeV$^2$.
This connection	can be made explicit in	approaches such	as continuum Schwinger methods, which describe
quark/gluon mass generation	and	baryon three-quark structure in a unified framework.
In this context only measurements at JLab - after a 22~GeV energy upgrade - of electromagnetic ground state and $N \rightarrow N^*$ transition form factors would allow us to continuously map out the transition from the strongly coupled to the perturbative QCD regime and hence to explore the full range of distances where the dominant part of hadron mass and bound three-quark structures emerge from QCD \cite{Carman:2023zke}.

\subsection{QCD Energy-Momentum Tensor}

\subsubsection{Gluonic Mass and Momentum Distributions from Charmonium Production}

Explaining the origin of the nucleon mass is essential for understand the structure of all visible
matter in the Universe. The $u$ and $d$ quark masses in the QCD Lagrangian account only for a tiny
fraction of the nucleon mass, and most of it is generated by gluon fields through the effect of
dynamical chiral symmetry breaking. The mass distribution in the nucleon is encoded in the form factors
of the gluonic energy-momentum tensor (so-called gravitational form factors) and can be quantified
in this way \cite{Lorce:2017xzd,Hatta:2018sqd,Metz:2020vxd}.
They include the spin-0 gluon operator describing the trace anomaly, and the
spin-2 gluon operator measuring the gluon momentum distribution. Theoretical studies have shown that
these form factors can be extracted from measurements of exclusive $J/\psi$ photo/electroproduction
near the threshold ($W - W_{thr} \sim$ 2-4 GeV), by analyzing the combined $W$ and $t$ dependence
of the differential cross
section \cite{Kharzeev:1998bz,Gryniuk:2016mpk,Mamo:2019mka,Mamo:2022eui,Guo:2021ibg,Duran:2022xag}.
This opens the prospect of exploring the mass, pressure, and force distributions of gluons in the proton.
Such measurements are complementary to $J/\psi$ production at EIC energies ($W \gtrsim 10$~GeV), which
probe the gluon GPDs at $x \lesssim 0.1$ \cite{AbdulKhalek:2021gbh}.

%
%
\begin{figure}[ht]
\begin{center}
\includegraphics[width=0.4\textwidth]{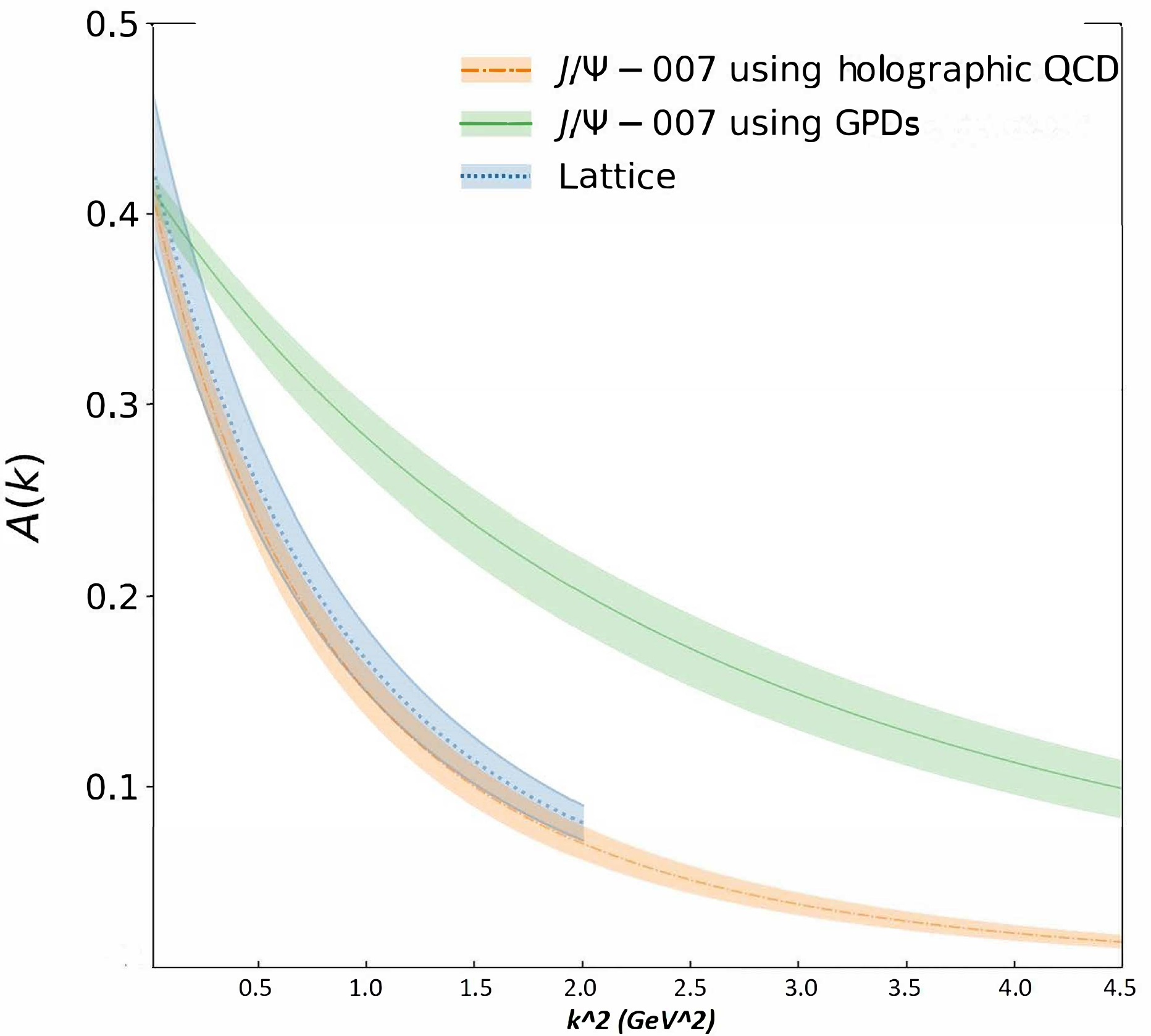}
\includegraphics[width=0.4\textwidth]{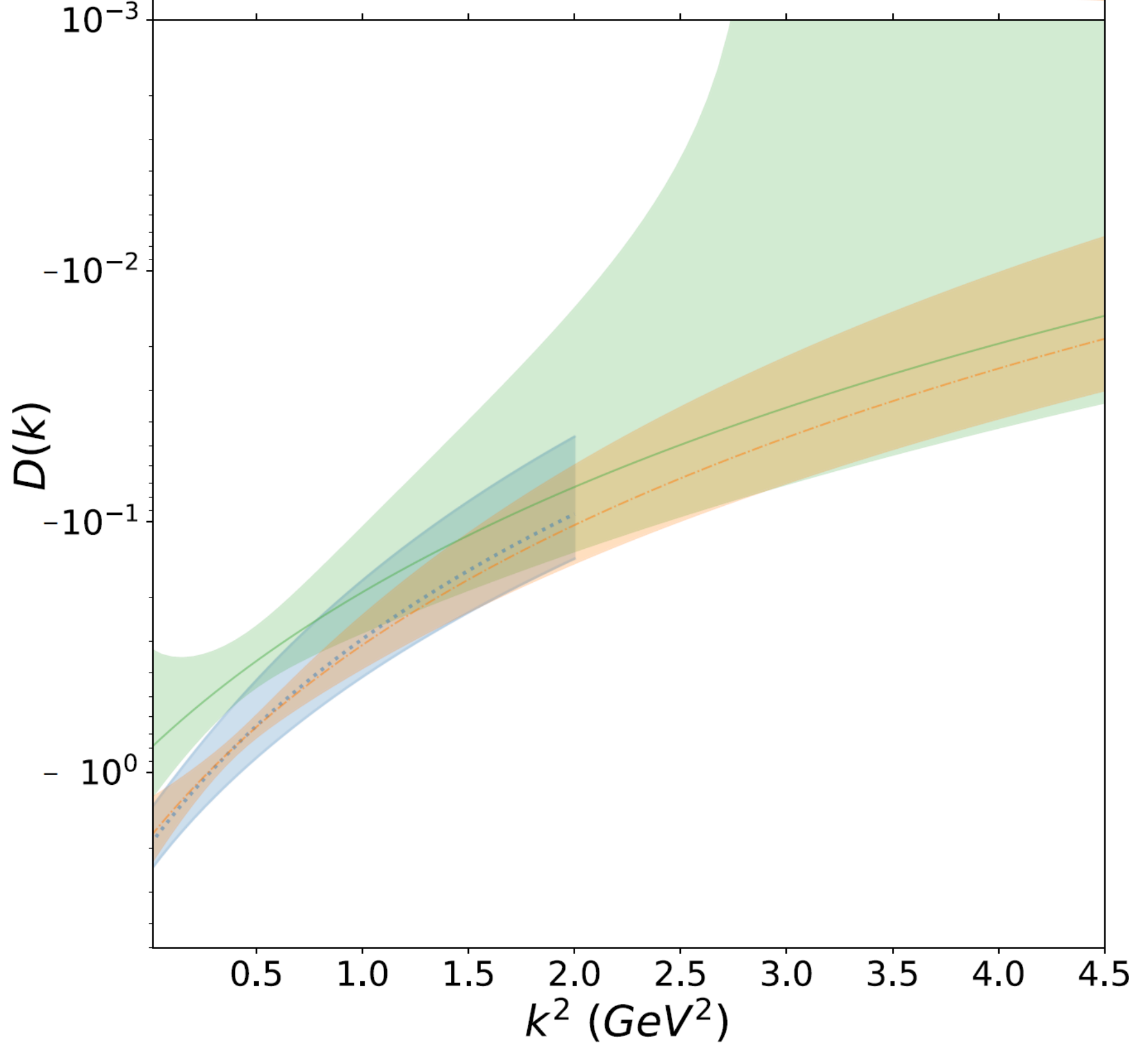}
\end{center}
\caption{The gluonic form factors $A(k^2=-t)$ (left) and $D(k^2)=4C(k^2)$ (right) extracted from the two-dimensional cross 
section data of $J/\psi-007$ Collaboration \cite{Duran:2022xag} using the holographic QCD approach \cite{Mamo:2019mka,Mamo:2022eui} 
(dash-dot curve) and the GPD approach \cite{Guo:2021ibg} (green solid curve), compared to the recent lattice calculation 
\cite{Pefkou:2021fni} (blue dotted curve).}
\label{AK_DK}
\end{figure}
Recent experimental results from JLab12 show the feasibility of extracting gluonic structure from
near-threshold $J/\psi$ production \cite{Duran:2022xag}.
Figure~\ref{AK_DK} shows the gluonic form factors $A(t)$ and $D(t)$ of the proton, extracted from the exclusive $J/\psi$ photoproduction differential cross section measured in the JLab Hall C experiment E12-16-007 \cite{Duran:2022xag} using two different theoretical descriptions of the reaction mechanism: a GPD-based model implementing collinear factorization \cite{Guo:2021ibg}, and a holographic QCD model based on gauge-string duality \cite{Mamo:2019mka,Mamo:2022eui}. 
The analyses have also extracted the gluonic radii of the nucleon, which can be compared with 
the electric charge radius. The results suggest that the gluonic mass distribution of the proton resides within the electric charge distribution. 
Further measurements of $J/\psi$ photo/electroproduction 
at 11 GeV are planned with the future SoLID detector at JLab \cite{Chen:2014psa}.
The theoretical interpretation of these measurements is a matter 
of on-going research and raises several questions which cannot be definitively answered with the present 
12 GeV data and require a broader kinematic range.

The JLab 22 GeV upgrade will be crucial to realizing the potential of this program.
The extended energy range will make it possible to measure the interplay of $W$ and $t$ dependence
over a range sufficient for separating the spin-0 and 2 contributions and testing/improving the proposed
models of the reaction mechanism. The 22 GeV fixed-target energy covers exactly the region where the differences
between different reaction models are maximal and can be distinguished by the data. The luminosity $\sim
10^{37}$ cm$^{-2}$ s$^{-1}$ will be critical for these low-rate differential measurements. For near-threshold
$J/\psi$ production the JLab 22 GeV facility would be unique. Complementary measurements of near-threshold $\Upsilon$
production would be possible with the EIC at 100 fb$^{-1}$ luminosity with suitable
detectors \cite{AbdulKhalek:2021gbh}. This field of research is evolving rapidly,
and significant progress in theory is expected over the next 10 years.

\subsubsection{Quark Pressure Distribution from Deeply Virtual Compton Scattering}

\begin{figure}[t]
\begin{center}
\includegraphics[width=0.405\textwidth]{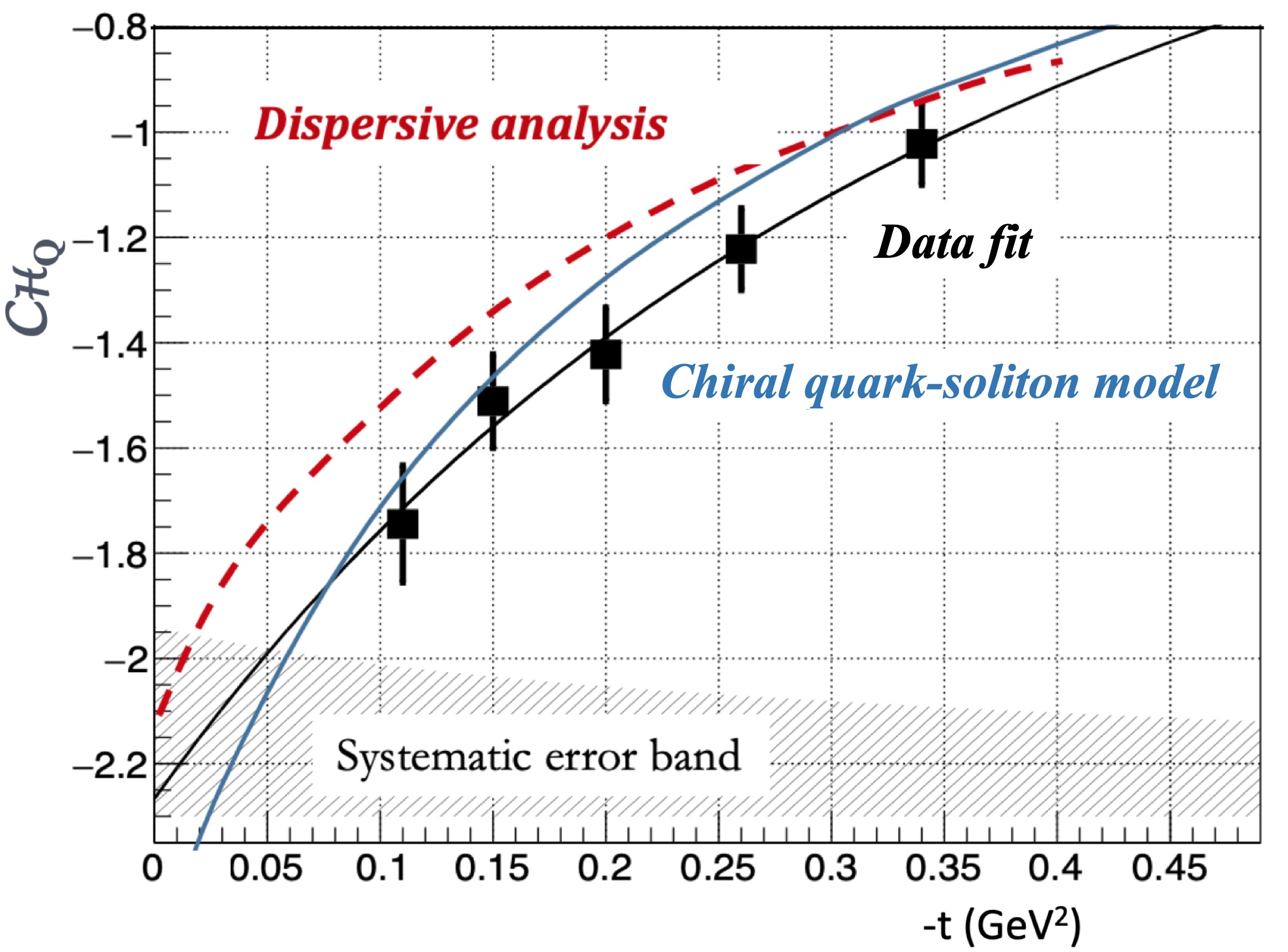}
\includegraphics[width=0.48\textwidth]{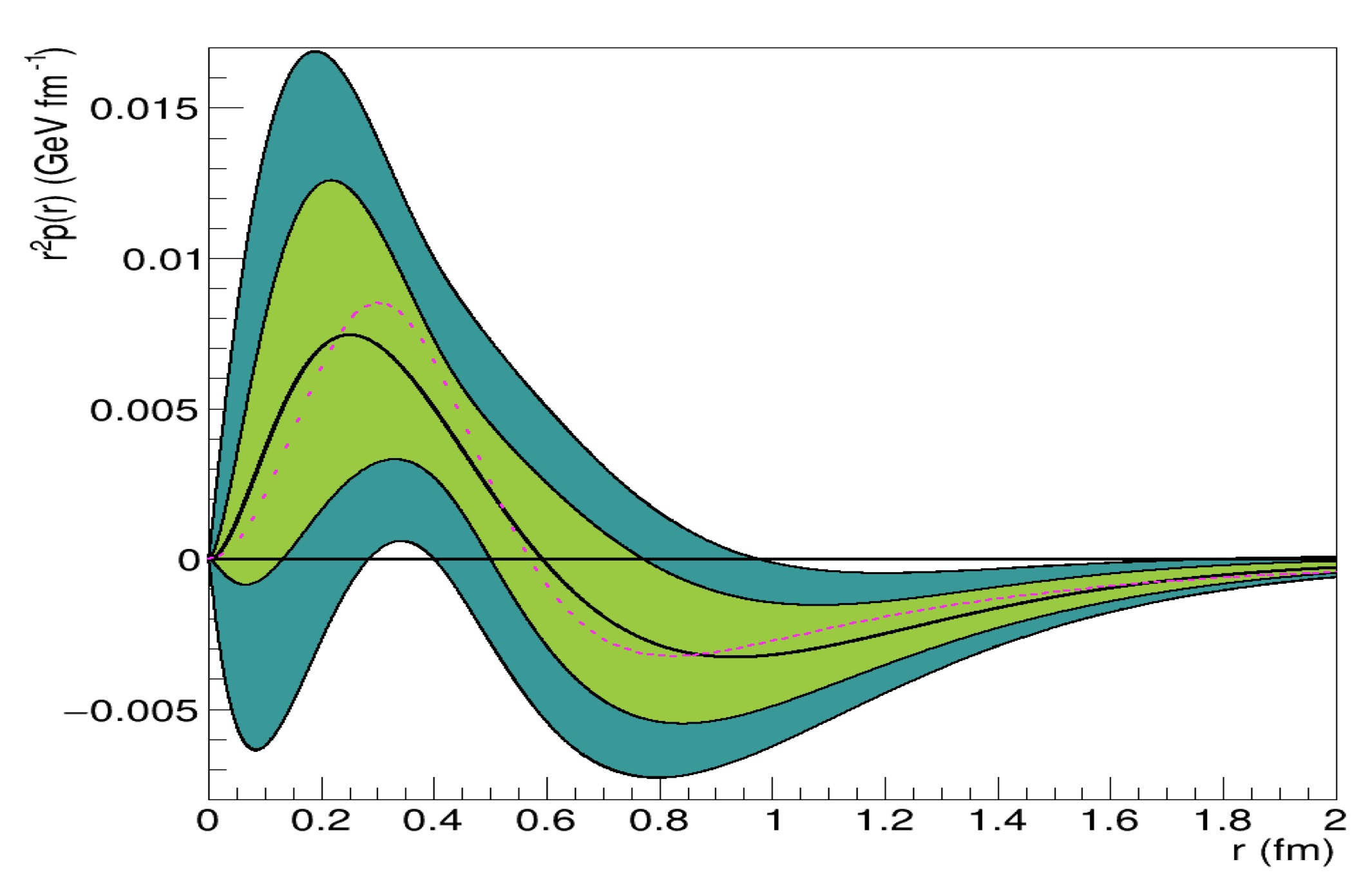}
\end{center}
\caption{Left: $D_q(t)$ vs. $-t$ from 6 GeV DVCS data compared to theoretical predictions. Right: Black line is the pressure 
distribution versus distance from the proton center employing a Fourier transform of $D_q(t)$ in $t$. The light-green band shows 
the estimated systematic uncertainty of the fit.}
 \label{fig:dterm_6gev}
\end{figure}

The form factors of the energy-momentum tensor are at the center of modern nucleon structure
physics. Of particular interest is the $D$-term \cite{Polyakov:1999gs}, which describes aspects of the
distribution of QCD forces on the quarks in the nucleon (``pressure'') \cite{Polyakov:2002yz} and has
become the subject of numerous theoretical studies of the ``mechanical properties'' of the nucleon
\cite{Polyakov:2018zvc,Lorce:2018egm}. The $D$-term form factor appears as the subtraction constant in
a dispersion relation of the amplitude of the deeply-virtual Compton scattering (DVCS) process and be
extracted from the experimental data on $ep \rightarrow e' \gamma p$ with minimal model dependence.
First empirical extractions of the $D$-term form factor and the ``pressure'' distribution have been performed with
the JLab 6 GeV data (see Fig.~\ref{fig:dterm_6gev}) \cite{Burkert:2018bqq,Burkert:2023wzr}; see also discussion and results
in Refs.~\cite{Kumericki:2019ddg,Moutarde:2019tqa}.
The restricted kinematic range covered by the 6 GeV data resulted in very large uncertainties in
extracting the distribution of pressure and shear stress. The main limitations are the small range of
energies (or longitudinal momentum fraction $\xi$) available for evaluating the dispersion integral,
small range of $t$ (limited by a condition on $t/Q^2$) available for computing the Fourier transform
of the form factor.
%
%

%
%
\begin{figure}[t]
  \begin{center}
    \includegraphics[width=0.95\textwidth]{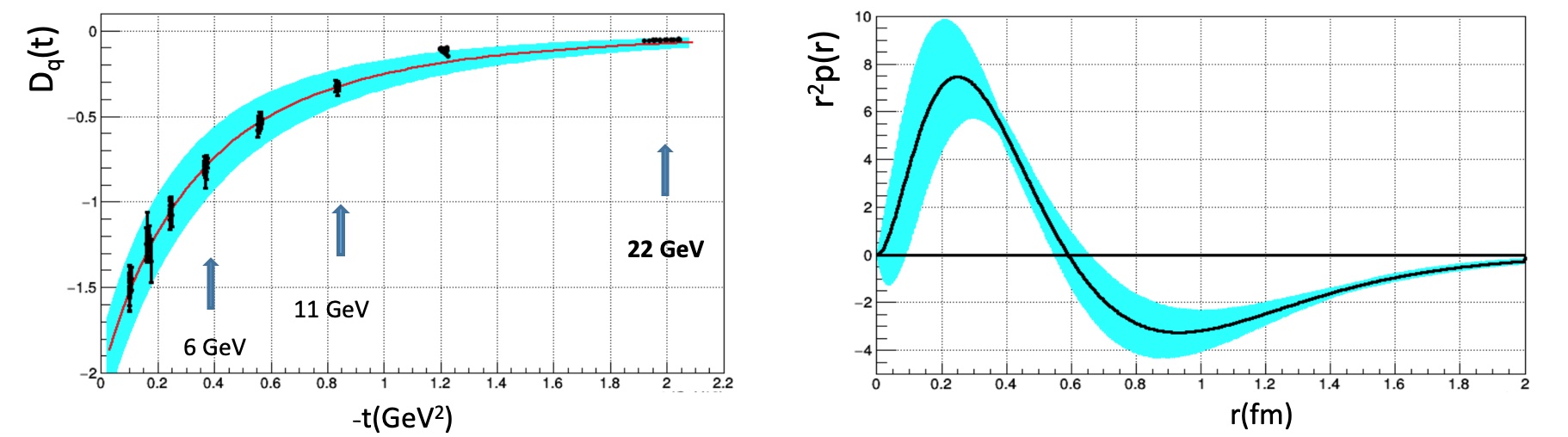}
 \end{center}
 \vspace{-5mm}
  \caption{Left: The $D$-term form factor of the QCD energy-momentum tensor, $D_q(t)$,
  as a function of the momentum transfer $-t$, as extracted from DVCS experiments. The arrows
  indicate the $-t$ ranges covered at different beam energies with the constraint $-t/Q^2 < 0.2$.
  Right: Quark pressure distribution in the proton, $p(r)$, as a function of the distance from the
  proton center, $r$, obtained as the Fourier transform of $D_q(t)$.}
 \label{fig:dterm_projection}
\end{figure}
A high quality
extraction of the $D(t)$ form factor would be possible with the proposed 22 GeV facility.
The available energy would improve the convergence of dispersion integral and permit tests of the stability
of the subtraction. It would also allow one to extend the measurements to significantly higher values of $t$,
while keeping $t/Q^2$ at values such that power corrections are under control. Projections of the
form factor measurements and extraction of the pressure distribution at various energies are shown in
Fig.~\ref{fig:dterm_projection}.

\begin{figure}[t!]
\begin{center}
\includegraphics[width=1.0\textwidth]{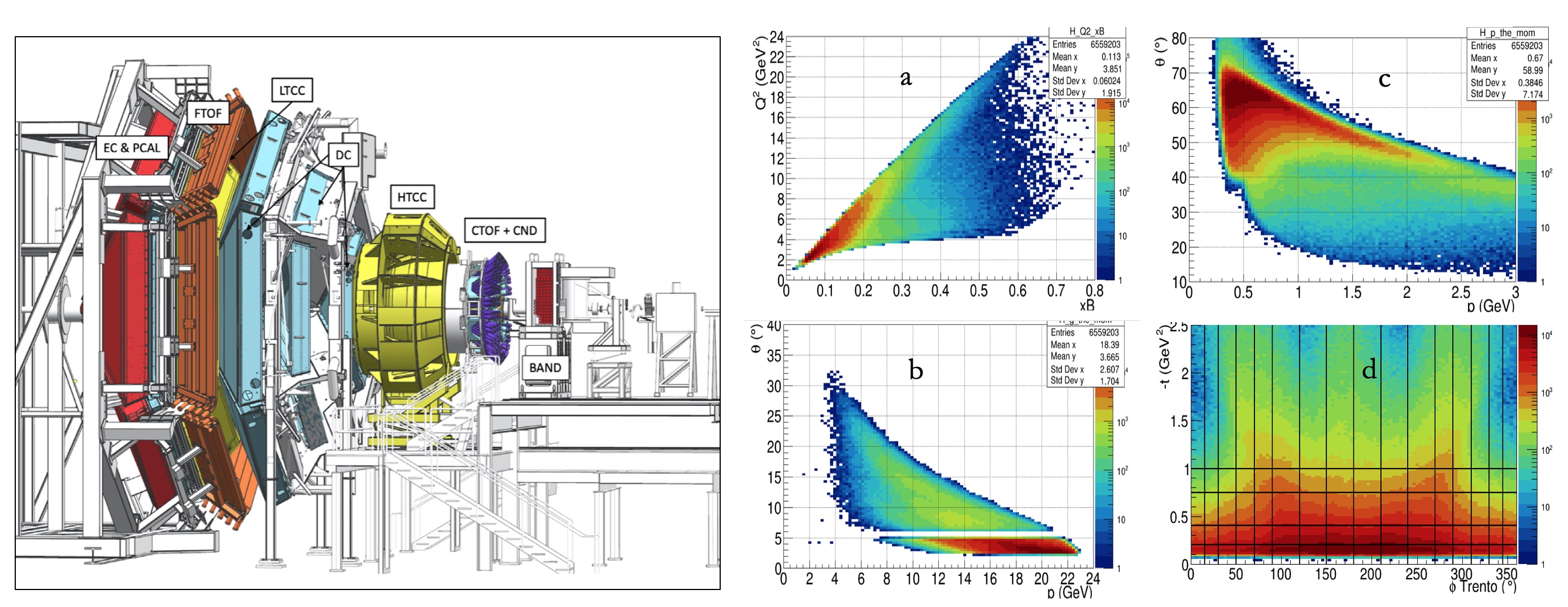}
\end{center}
\caption{Left: The CLAS12 detector system. Right: Response to exclusive DVCS + Bethe-Heitler events at 22 GeV beam energy: 
(a) Scattered electron kinematics in $Q^2$ vs. $x_B$. (b) DVCS-BH photon kinematics in polar angle vs. momentum, (c) proton 
kinematics in polar angle vs. momentum, (d) event distribution in $-t$ vs. azimuthal angle $\phi$.}
 \label{CLAS12}
\end{figure}

The CLAS12 detector~\cite{Burkert:2020akg} shown in Fig.~\ref{CLAS12} has been designed with
measurements of GPD-related processes in mind. It has demonstrated with data published
recently~\cite{CLAS:2022syx} that CLAS12 is well suited for measurements of the DVCS-BH process in
large connected kinematic domains in $Q^2, x_B, -t$ and azimuthal angle~$\phi$ to measure cross
sections and polarized and polarized target processes simultaneously at the currently available
maximum beam energy of 10.6~GeV. Simulations of the same processes at 22 GeV shown in
Fig.~\ref{CLAS12} demonstrate that CLAS12 is also well matched to measure the response to DVCS and
BH events at an upgraded JLab beam energy of 22 GeV. Connected kinematic ranges in $1.5 < Q^2 < 20$~GeV$^2$, 
$0.05 < x_B < 0.6$, $0.1 < -t < 2.5$~GeV$^2$, and azimuthal angle $0 < \phi < 360^\circ$ will be simultaneously 
measured. The large ranges in $x_B$ and $-t$ are essential as applying the dispersion relation requires the full 
integration in $\xi = x_B/(2 - x_B)$.
%
%

\subsection{3D Imaging with GPDs}

\subsubsection{Longitudinal/Transverse Separation in Exclusive Processes}

The past few decades have been ripe with progress in our understanding of the QCD structure of
strongly interacting systems, from unraveling the spin structure of the proton in terms of the quark
and gluon spin and orbital angular momentum, to developing a microscopic definition of pressure and
mass while giving a detailed description of the spatial distributions of quarks and gluons.  The key
objects that made it possible to perform quantitative detailed studies, and that connect all of
these physical properties are the GPDs (\cite{Ji:1996ek,Radyushkin:1997ki} for reviews, see
Refs.~\cite{Goeke:2001tz,Diehl:2003ny,Belitsky:2005qn,Kumericki:2016ehc}). GPDs are measurable through
deeply virtual exclusive scattering (DVES) processes, in particular DVCS, where they appear embedded in the Compton Form Factors 
(CFFs), which are
convolutions over the longitudinal partonic momentum fraction, $x$, with known perturbative
kernels. In DVES the electron scatters off the proton or nucleus with momentum, $p$, at high
four-momentum transfer squared, $Q^2$. A high momentum photon or a meson is produced, leaving a
proton with momentum $p'=p-\Delta$, in the final state.  By Fourier transforming the GPDs in the
variable $\Delta_T$, the transverse momentum transfer between the initial and final proton, one can
access the impact parameter dependent parton distribution functions, $\rho_{q,g}(x,b)$
\cite{Soper:1976jc,Burkardt:2000za}, which simultaneously define the distributions in longitudinal
momentum fraction, $x$, of a quark/gluon positioned at a transverse distance, $b$, from the hadron's
center of mass (CoM). Measuring CFFs and GPDs allows us, therefore, to extend the grasp on the physics information
contained in the nucleon elastic form factors, providing a unique probe of QCD at the amplitude
level.

Extracting 3D images of the proton's interior from experimental data while pinning down the origin
of it's mass and spin are defining goals of the nuclear and particle physics experimental programs
at JLab and the upcoming EIC. While the EIC is mostly focused at low Bjorken $x_B$, where
gluonic components are dominant, JLab above 12 GeV allows us to explore in detail the valence quark
region. By expanding the $x_B$ and $Q^2$ domain, a JLab upgrade to 22 GeV would allow us to
explore the emergence of antiquarks in exclusive reactions.  Despite two decades long efforts of
DVES measurements pioneered by experiments at the HERA collider, and subsequently carried out in
dedicated programs at JLab and at the COMPASS experiment at CERN, CFFs and GPDs remain
elusive quantities to extract from data. A major obstacle affecting all analyses has been posed, so
far, by the rather involved expressions for the cross section in both the unpolarized and polarized
configurations, which do not allow us to associate a given GPD in a specific polarization
configuration with a polarization measurement for the corresponding configuration
\cite{Belitsky:2001ns,Belitsky:2010jw}. While, on one side, this prevents us from using previous
experience on inclusive and semi-inclusive experiments, which are characterized by a more
transparent structure of the cross section for various polarization configurations (see {\it e.g.}
the analysis in Ref.~\cite{Bacchetta:2006tn} and references therein), it has now become clear that
the DVCS cross section tracks the one for electron-nucleon elastic scattering experiments
determining the nucleon electromagnetic and weak form factors.

The analyses presented in Refs.~\cite{Kriesten:2020apm,Kriesten:2020wcx,Kriesten:2019jep} performed along
these lines show, in fact, that a Rosenbluth separation of data from an unpolarized target allow us
to simultaneously determine two of the GPDs entering the cross section for the interference term
between DVCS and the Bethe-Heitler (BH) process. Together with the more efficient reformulation of
the cross section for the $ep \rightarrow e' p' \gamma (M)$ process presented in
Refs.~\cite{Kriesten:2019jep,Kriesten:2020apm,Kriesten:2021sqc}, refined statistical analyses such as
the ones afforded by machine learning (ML) techniques can decisively improve the extraction of
physical observables form data. Several analyses using ML tools to extract CFFs from data were
already presented in Refs.~\cite{Kumericki:2019mgk,Moutarde:2019tqa,Grigsby:2020auv,Cuic:2020iwt,Almaeen:2022imx}.

A program for extracting meaningful information from data can be continued with more precise data
taking with dedicated experiments beyond JLab12 GeV. An extensive analysis using precision data in the
valence and emerging sea quarks region would allow us, to separate the leading twist component from
the ${\cal O}(1/Q)$/power corrections dependent effects arising from both dynamical higher twists,
and kinematic corrections. Furthermore, the onset of the scaling regime of QCD will be ultimately
confirmed by the comparison of DVCS and Timelike Compton Scattering (TCS) data over a sufficiently
large range of four-momentum transfer, $Q^2$. Other outstanding questions to be addressed in this
program will be: 1) separating the twist two and three components through meaningful Rosenbluth
separations; 2) separating out the various flavor GPDs; 3) zooming into the onset of gluon
components.

%
%
\begin{figure}[!]
\centering
\includegraphics[width=8.0cm]{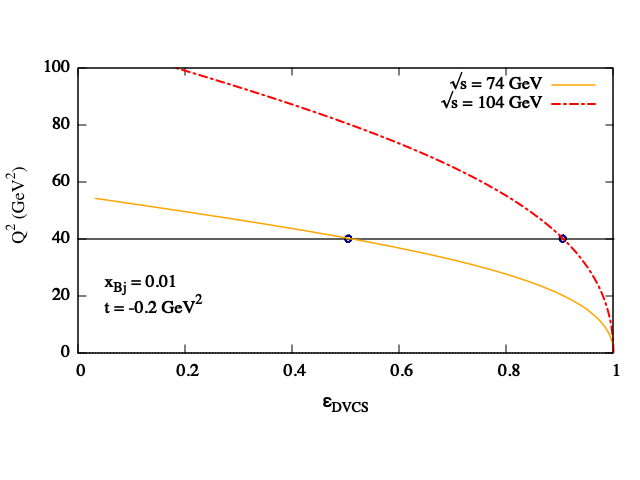}
\hspace{0.25cm}
\includegraphics[width=8.0cm]{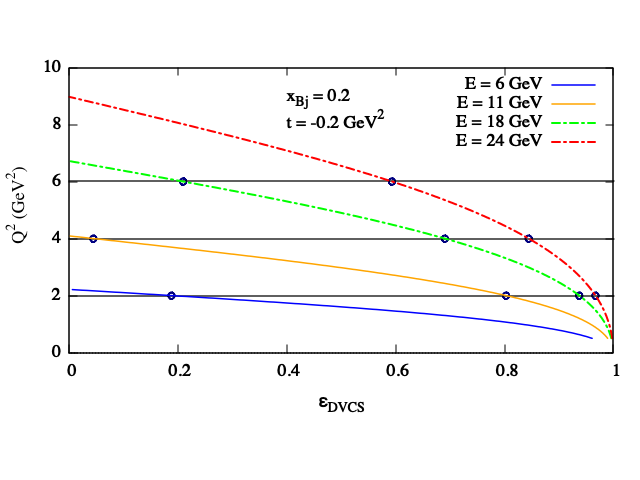}
\caption{Longitudinal to transverse virtual photon polarization parameter for the DVCS process,
$\epsilon_{DVCS}$. The L/T separation of the DVCS term can be accomplished at the kinematic points
shown on (left) $E_{e} = 6,\, 11,\, 18,\, 24$~GeV and fixed $x_{B} = 0.2$, $t = -0.2$~GeV$^2$; (right) EIC configurations are given as $\sqrt{s} = 74$ GeV corresponds to $E_{e} = 5$~GeV, $E_p = 275$~GeV; and 
$\sqrt{s} = 104$~GeV corresponds to $E_{e} = 10$~GeV, $E_{p} = 275$~GeV at kinematics $x_{B} = 0.01$, $t = -0.2$~GeV$^2$ (work in progress with Brandon Kriesten)}
\label{fig:Rosenbluth}
\end{figure}
%
%
\begin{figure}[!]
\centering
\includegraphics[width=8.0cm]{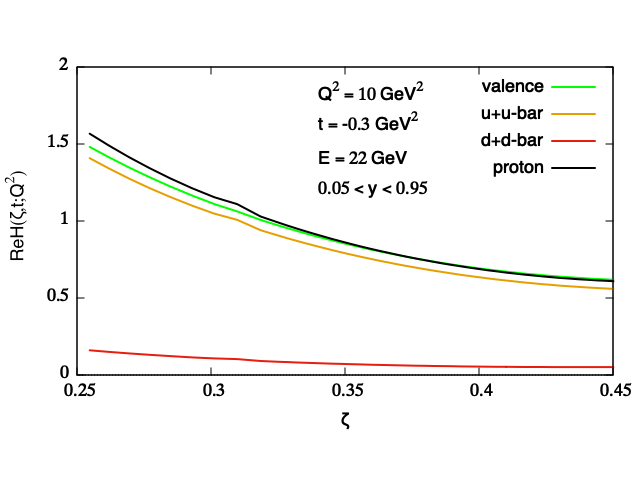}
\hspace{0.25cm}
\includegraphics[width=8.0cm]{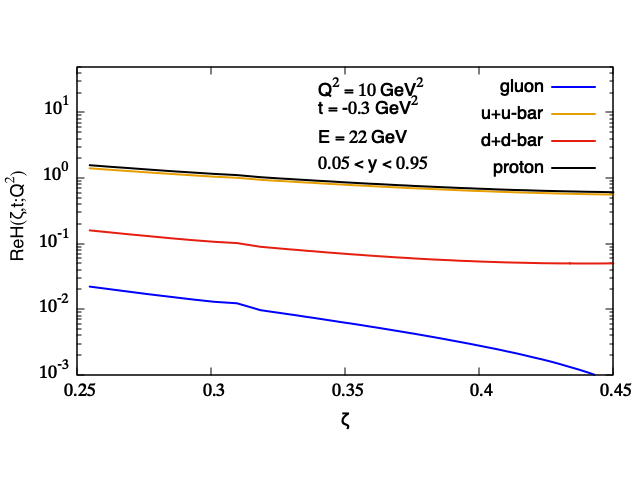}
\caption{Flavor separated and gluon contributions to the CFF plotted as a function of $x_{B}$ at
JLab kinematics typical of a 22 GeV upgrade. Notice the role of the valence component in the
evaluation of the CFF given that $q + \bar{q} = q_{v} + 2\bar{q}$. (work in progress with Brandon
Kriesten)}
    \label{fig:flavor}
\end{figure}
Figure~\ref{fig:Rosenbluth} shows the longitudinal to transverse virtual photon polarization
parameter for the DVCS process at JLab and EIC kinematics, respectively. The advantage of the JLab
setting can be clearly seen. Figure~\ref{fig:flavor} addresses the issue of the separation of
various quark flavors contributions to the $e p \rightarrow e' p' \gamma$ cross section. On the
right we also present the gluon contribution through its effect on perturbative QCD evolution at
NLO. All curves were calculated in the model of Ref.~\cite{Kriesten:2021sqc}.

In conclusion, the JLab higher energy upgrade will be uniquely important to perform the
Rosenbluth separations that allow us to separate out twist-2 from twist-3 components
proportional to OAM \cite{Kriesten:2020apm}, and to perform in depth studies of the scale dependence
of GPDs centered on the valence contribution and the emergence of sea quark effects.

\subsubsection{Differential Imaging with Double Deeply Virtual Compton Scattering}
\label{subsubsec:double_dvcs}
The program of ``tomographic imaging'' of the nucleon with GPDs requires the full information on
their dependence on the longitudinal momentum variables --- the parton momentum fraction $x$, and
the longitudinal momentum transfer $\xi$. Conventional exclusive processes such as DVCS cannot fully
disentangle the $x$ and $\xi$ dependence, because the observables sample the GPDs only in the
special kinematics of $x = \xi$ or as integrals over $x$. Novel processes such as dilepton
production
\begin{align}
e + p \rightarrow e' + (l^+l^-) + p, \hspace{2em} l = e \;\; \textrm{or} \;\; \mu
\end{align}
(Double Deeply Virtual Compton Scattering, or DDVCS)
can disentangle the longitudinal momentum variables in the GPDs by varying the dilepton
mass in the process \cite{Guidal:2002kt,Belitsky:2002tf}. By measuring the $t$-dependence in this
configuration, they could provide fully differential tomographic images of nucleon structure.

%
%
\begin{figure}[!]
\begin{center}
\includegraphics[width=.49\textwidth]{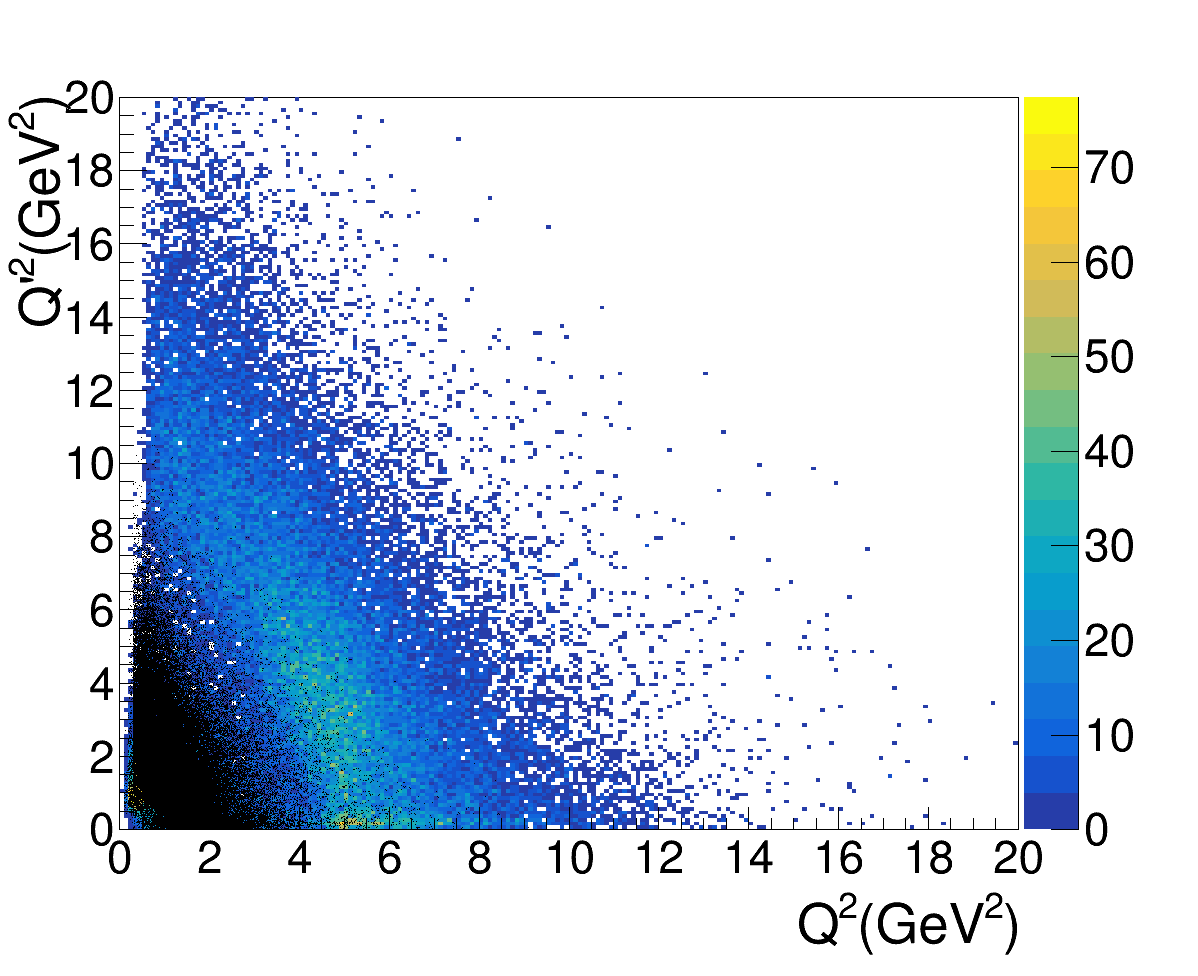}
\includegraphics[width=.49\textwidth]{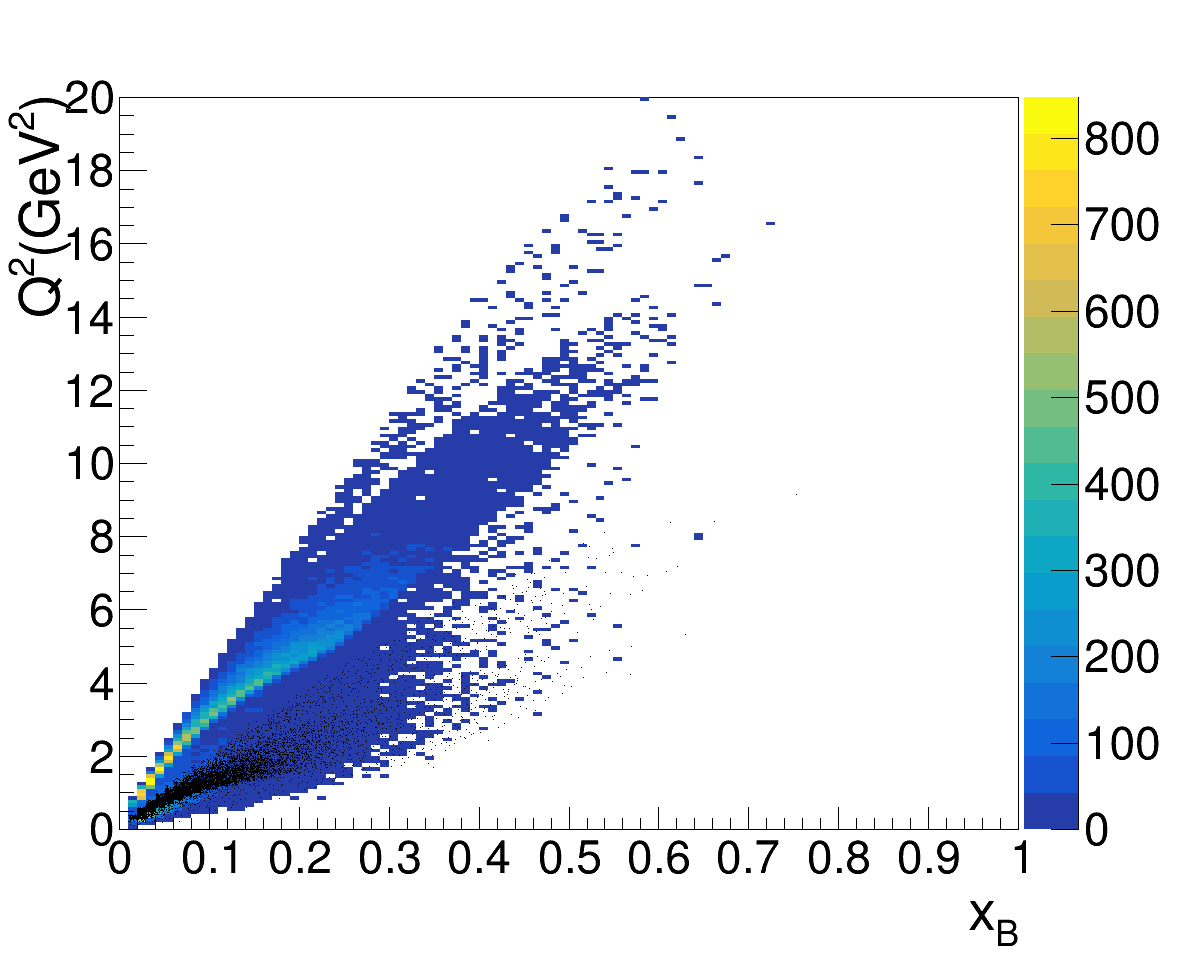}
\end{center}
\caption{
Comparison of SoLID DDVCS kinematic coverage between 11 GeV (black) and 22 GeV (color) electron beams. They both use the same running conditions and detect the scattered electron and decay muon pair at forward and large angle. $x_B$ is Bjorken-$x$, $Q^2$ is the negative four momentum transfer squared, and $Q'^2$ is the invariant mass of muon pair squared. 22 GeV electron beam will allow access to substantially higher $x_B$, $Q^2$, and $Q'^2$.}
\label{fig:DDVCS_Q2_x}
\end{figure}
These next-generation measurements are challenging, and exploratory studies with JLab 12 GeV are under way
\cite{Zhao:2021zsm}. The focus is on the production of muon pairs, which eliminates the additional complexity
of mixing electron from the pair with the scattered electron. Muons also go through larger amount of
materials allowing to improve signal to noise by using large amounts of shielding. Letters of Intent
for exploratory measurements of DDVCS at 11 GeV have been submitted to the JLab PAC, one using the
the SoLID spectrometer and another the CLAS12 setup.

The proposed 22 GeV facility would be ideally suited for this program. The high luminosity is essential
because the DDVCS cross section is suppressed by a factor $\alpha_{\rm em} \sim 10^{-2}$ compared to DVCS.
The energy is needed for reaching dilepton masses $M(l^+l^-) \sim$ few GeV in electroproduction with
$Q^2 \sim$ few GeV$^2$, where one can perform scaling studies and apply the GPD-based reaction mechanism
based on QCD factorization. Figure~\ref{fig:DDVCS_Q2_x} shows the kinematical coverage of the
SoLID and CLAS setups at 22 GeV. For both, the range in $Q^2$ and $Q'^2 \equiv M^2(\mu^+\mu^-)$
is much enlarged compared to 11 GeV. The cross section estimates were obtained using the GRAPE and VGG
models, and indicate a reduction of cross section at 22 GeV by about a factor of 3 compared to 11 GeV.
The beam spin asymmetries expected in 22 GeV kinematics are sizable, of the order of several per cent, as in the 11 GeV kinematics, and can be measured
reliably with the proposed setups.

\subsubsection{Novel GPD Probes with Exclusive Diphoton Production}
%
%
\begin{figure}
\begin{center}
\includegraphics[width=0.65\textwidth]{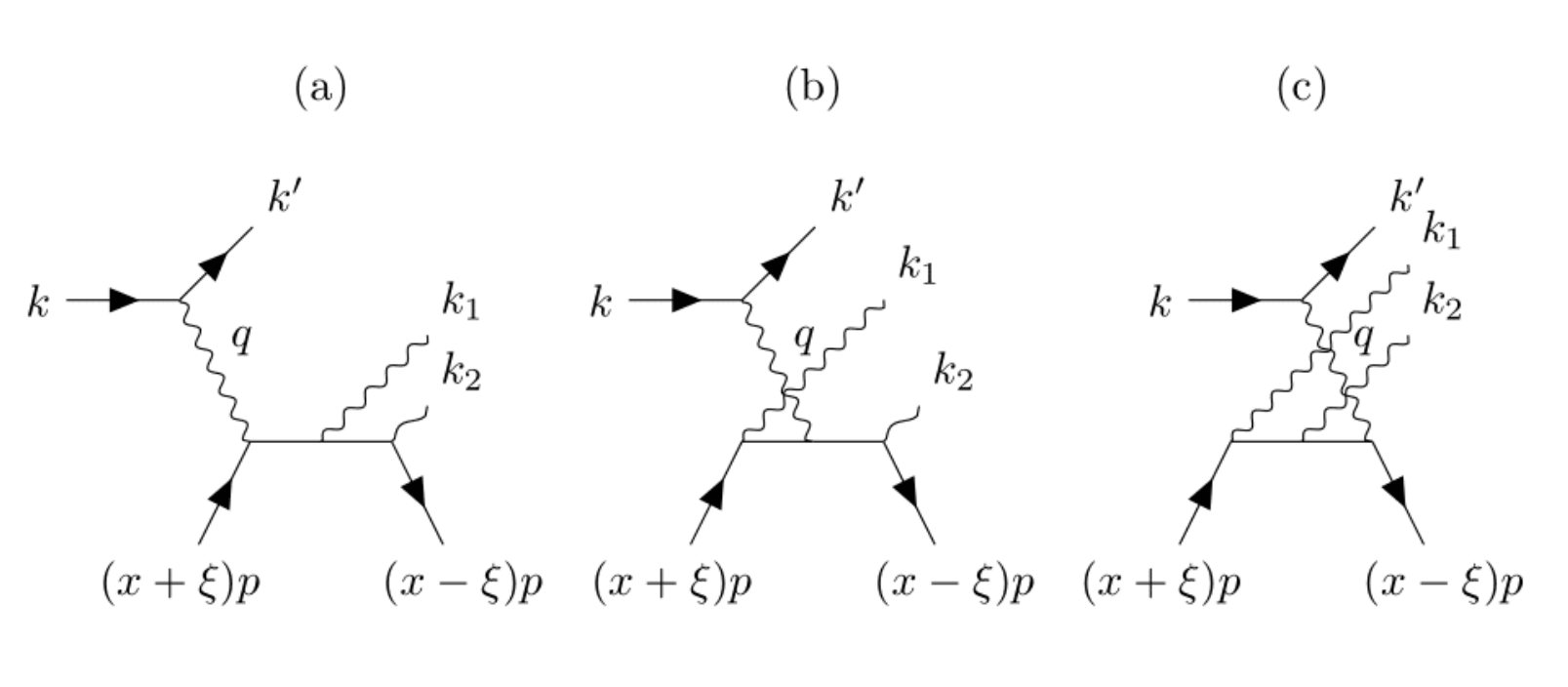}
\end{center}
\caption{The lowest order hard amplitude for the photoproduction of a large mass diphoton
(complementary diagrams with $k_1 \leftrightarrow k_2$ are not shown).  }
\label{fig.Born}
\end{figure}

\begin{SCfigure}
    \includegraphics[width=0.5\linewidth]{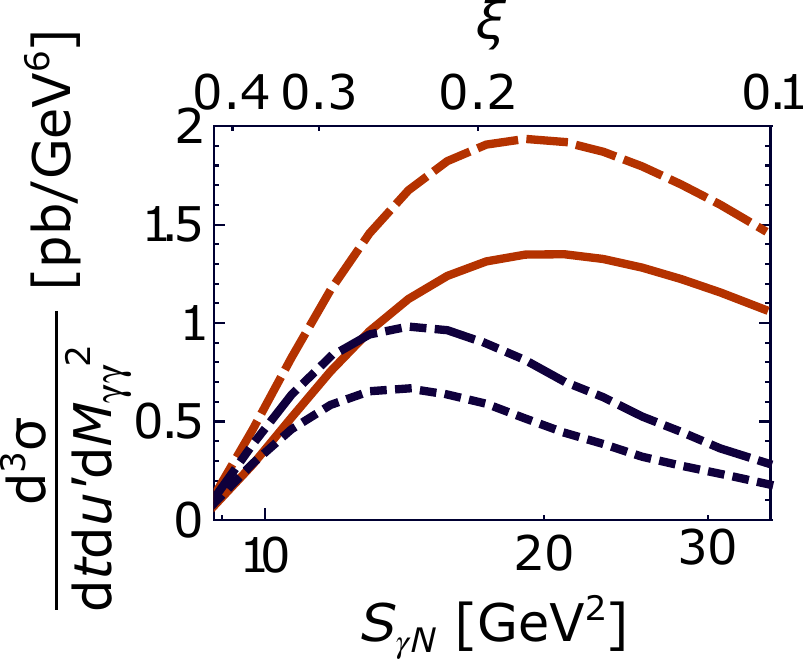}
   \caption{Differential cross section as a function of $S_{\gamma N}$ (bottom axis) and the corresponding $\xi$ (top axis) for $M_{\gamma\gamma}^2 = 4~\mathrm{GeV}^2$, $t=t_{min}$, and $u' = -1~\mathrm{GeV}^2$. The leading order result is denoted by the solid (dashed) red line, while the next-to-leading order one by the dotted (dash-dotted) blue line for the GK (MMS) GPD model.}
     \label{cutplotgaga}
\end{SCfigure}

The quest for nucleon tomography in terms of GPDs necessitates the study of as many exclusive processes as possible. Processes with $2\to 3$ hard amplitudes have been shown to provide many examples of interesting reactions~\cite{Ivanov:2002jj, Boussarie:2016qop, Duplancic:2023kwe}. The quasi-real photoproduction of a large invariant mass photon pair~\cite{Pedrak:2017cpp, Grocholski:2022rqj}: 
\begin{equation}
\gamma(k,\epsilon) + N(p_1,s_1) \rightarrow \gamma(k_1,\epsilon_1) +  \gamma(k_2,\epsilon_2)+ N(p_2,s_2)\,,
\label{process}
\end{equation}
is the simplest among them since it is purely electromagnetic at the Born order level, as shown in Fig.~\ref{fig.Born}. It has been proven to factorize~\cite{Grocholski:2021man, Qiu:2022pla} into a hard amplitude and quark GPDs. Because of the symmetries of the process, the contributing GPDs  are the charge conjugation odd parts of quark GPDs,
\begin{eqnarray}
  H^+(x,\xi,t) &=& H(x,\xi,t) - H(-x,\xi,t)\,,\\
  \tilde H^+(x,\xi,t) &=& \tilde H(x,\xi,t) + \tilde H(-x,\xi,t)\,,
\end{eqnarray}
and similar equations for $E^+$ and $\tilde E^+$,  
which are decoupled from the DVCS, TCS, and DDVCS reactions. Gluonic GPDs do not contribute for the same reason.

 The hard scale of this reaction is the diphoton invariant squared mass, $M_{\gamma\gamma}^2 = (k_1+k_2)^2$, while the skewness variable, $\xi$, similarly as in the TCS case, is related to $\tau = (M_{\gamma\gamma}^2)/(S_{\gamma N}-M^2)$ with $S_{\gamma N} = (k+p_1)^2$ through $\xi \approx \tau/(2-\tau)$. The estimated cross section is shown in Fig.~\ref{cutplotgaga} as a function of $S_{\gamma N}$ (bottom axis) and the corresponding $\xi$ (top axis) for $M_{\gamma\gamma}^2 = 4~\mathrm{GeV}^2$, $-t=-(p_2-p_1)^2 = -t_{min}$, and $u' = (k-k_2)^2 = -1~\mathrm{GeV}^2$. This energy range for a quasi-real photon beam will be easily accessible with the JLab22 facility. We show the leading-order and next-to-leading order results for two GPD models \cite{Goloskokov_2014, Mezrag:2013mya}. The electroproduction reaction, which receives contributions from two Bethe-Heitler like processes is discussed in detail in Ref.~\cite{Pedrak:2020mfm}.

The presented estimation of cross sections shows that the experiment is feasible with a high-luminosity facility in the JLab22 energy range. The CLAS12 detector looks adequate for observing and measuring this process. Dedicated experiments may be prepared in Hall A or Hall C.  The difference between our results with two different GPD models shows that this process is discriminative and will induce severe constraints on our understanding of the quite badly known charge conjugation even part of quark GPDs. Note that the process is available in both the PARTONS framework \cite{Berthou:2015oaw} and EpIC Monte Carlo generator \cite{Aschenauer:2022aeb}, making impact and measurability studies convenient for the experimental community. 
Since the diphoton process is insensitive to the gluon and sea quark GPDs, there is no enhancement of the amplitude at small $\xi$. This is the reason why in Fig.~\ref{cutplotgaga} the cross section drops in the sea region. Therefore, contrarily to the DVCS or TCS cases, the large photon energy reached by the future EIC does not help to have larger scattering amplitudes.

\subsubsection{$x$-dependent GPDs with Back-to-Back Photon-Hadron Production}
The $x$-moments of GPDs, uniquely sensitive to GPDs' $x$-dependence, are responsible for many
emergent hadronic properties such as the hadron's mass~\cite{Ji:1994av,Ji:1995sv,Lorce:2017xzd,Metz:2020vxd}
and spin~\cite{Ji:1996ek}, as well as its internal pressure and shear
force~\cite{Polyakov:2018zvc,Burkert:2018bqq}.  However, as noted in Sec.~\ref{subsubsec:double_dvcs}, the well-studied
DVCS and DVMP processes cannot provide the full information on the $x$-dependence of GPDs.  In
addition to the DDVCS, which requires the luminosity that a collider may not be able to deliver,
several {\it single-diffractive hard exclusive processes (SDHEP)}, along with criteria to enhance
the sensitivity on $x$-dependence of GPDs, were introduced and corresponding QCD factorization was
justified~\cite{Qiu:2022bpq,Qiu:2022pla}.  In particular, the exclusive photoproduction of a
photon-pion pair with high back-to-back transverse momentum~\cite{Qiu:2023mrm},
\begin{align}
	N(p_1) + \gamma(p_2) \to N'(p'_1) + \pi(q_1) +  \gamma(q_2), 
\label{eq:sdhep}
\end{align}
can be studied at JLab in Hall D by GlueX collaboration with controllable photon beam
polarization, as well as in Hall A/C with quasi-real photon beams, and in Hall B with polarized quasi-real photon beams.
Unlike DVCS and DVMP, the relative momentum of two active partons (quark-antiquark for quark GPDs or
two gluons for gluon GPDs), which defines the $x$-dependance of GPDs, is entangled with the
transverse momentum flow between the observed photon and the pion.  Consequently, the transverse
momentum $q_{1T}$ distribution of the observed pion (or $q_{2T}=-q_{1T}$ of the photon), or
equivalently, its polar angular $\theta$ distribution, is directly sensitive to the $x$-shape of the
GPDs.
%
%
\begin{SCfigure}
\sidecaptionvpos{figure}{htbp}
\includegraphics[width=0.48\textwidth]{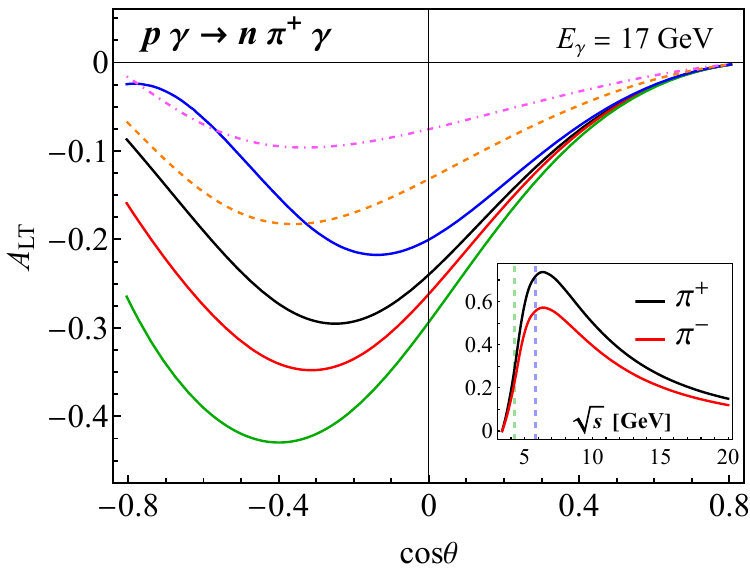} 
{\hskip 0.1in}
\caption{Double polarization asymmetry $A_{LT}$ for the exclusive photoproduction of
  $\gamma$-$\pi^+$ pair as a function of the pion's polar angle $\theta$ with a linearly polarized
  photon beam of energy $E_{\gamma} = 17~{\rm GeV}$ on a longitudinally polarized nucleon target, which
  is accessible at JLab Hall D with the upgraded JLab $22~{\rm GeV}$ setup. The black curve represents
  the asymmetry calculated with the GK model for GPDs.  The other solid lines correspond to adding
  different shadow GPDs to unpolarized GPD $H$, while the dashed lines are generated by adding
  shadow GPDs to the polarized GPD $\tilde{H}$.  In the inset figure, the total production cross
  section with $q_T \geq 1~{\rm GeV}$ and $|t| \leq 0.2~{\rm GeV}^2$ is shown as a function of the
  center-of-mass collision energy for producing $\gamma$-$\pi^+$ (black) and a $\gamma$-$\pi^-$
  (red).  The vertical green and blue dashed lines refer to the photon beam energies $E_{\gamma} =
  9$ and $17~{\rm GeV}$, respectively.  }
	\label{fig:ALT}
\end{SCfigure}

The double polarization asymmetry $A_{LT}$ of the photoproduction in Eq.~(\ref{eq:sdhep}) are
calculated by using the GK model of GPDs~\cite{Goloskokov:2005sd, Goloskokov:2007nt,
  Goloskokov:2009ia, Kroll:2012sm}, plus various polynomially parametrized shadow GPDs $S(x, \xi)$~\cite{Bertone:2021yyz, Moffat:2023svr} with different $x$-distributions, for $t = -0.2~{\rm GeV}^2$, $\xi
= 0.2$ and $E_{\gamma} = 17~{\rm GeV}$, and presented in Fig.~\ref{fig:ALT}.  As constructed, adding the
shadow GPDs to the GK model does not impact the theory predictions to DVCS-like
processes (at least to the leading order), but, it clearly alters theoretical predictions for the
plotted asymmetries in Fig.~\ref{fig:ALT}.  That is, the type of exclusive process in
Eq.~(\ref{eq:sdhep}) is advantageous in determining the $x$-dependence of GPDs~\cite{Qiu:2023mrm}.
In the inset of Fig.~\ref{fig:ALT}, the total production cross section with $q_T \geq 1~{\rm GeV}$ and $|t|
\leq 0.2~{\rm GeV}^2$ is shown as a function of the center-of-mass collision energy for producing a pair
of $\gamma$-$\pi^+$ (black) and $\gamma$-$\pi^-$ (red).  Exclusive production of a pair of high
transverse momentum particles requires a high luminosity and a sufficient collision energy.  The
$22~{\rm GeV}$ CEBAF energy upgrade (the blue dashed line) hits the production rate at the sweet spot.
In addition, with the polarized photon beam and hadron target at JLab, various asymmetries of
polarized cross sections lead to different azimuthal angle $\phi$ distribution of the observed pion,
helping to distinguish contributions from different GPDs~\cite{Qiu:2023mrm}.

\subsubsection{Resonance Structure with $N \rightarrow N^*$ Transition GPDs}

While measurements of exclusive processes have already provided much insight into the 3D structure
of the ground state nucleon, little is known about the 3D structure of resonances so far.
This information is encoded in so-called transition GPDs \cite{Goeke:2001tz,Belitsky:2005qn}, which can be accessed
in exclusive processes with a $N \rightarrow N^{*}$
transition \cite{Kroll:2022roq,Guichon:2003ah,Semenov-Tian-Shansky:2023bsy}.
The simplest such reaction is the $N \rightarrow N^{*}$ DVCS process,
\begin{equation}
\gamma^{*} + N \; \longrightarrow \; \gamma + N^{*} \; \longrightarrow \; \gamma + (N\; \textrm{meson}),
\end{equation}
where a real photon is produced in addition to a nucleon resonance, which then decays into a ground state nucleon
and a meson. Another reaction is $N \rightarrow N^{*}$ pseudoscalar meson electroproduction,
\begin{equation}
\gamma^{*} + N \; \longrightarrow \; M + N^{*} \longrightarrow \; M + (N \; \textrm{meson}).
\end{equation}
For both reactions, a QCD factorization theorem holds in the Bjorken limit: $-t/Q^{2} \ll 1$ and $x_{B}$ fixed,
with an additional condition on $Q^{2}$ in relation to the resonance mass:
$Q^{2} \gg m^{2}_{N^{*}}$ \cite{Kroll:2022roq,Guichon:2003ah}.

The theory and interpretation of transition GPDs has made substantial progress in recent years
and has become a field of study in its own right. For the simplest case of the $N\to\Delta$ transition,
the structural decomposition of the matrix elements has been studied for the chiral-even
(quark helicity-conserving) \cite{Goeke:2001tz} and chiral-odd (quark helicity-flipping,
or transversity) GPDs \cite{Kroll:2022roq}. The first moments of the chiral-even $N\to\Delta$ GPDs are
related to the Jones-Scadron electromagnetic form factors \cite{Jones:1972ky,Goeke:2001tz}
and the Adler axial form factors \cite{Adler:1968tw,Adler:1975mt,Goeke:2001tz}
of the $N \to \Delta$ transition. The second moments of the chiral-even GPDs give access to the
$N\to\Delta$ transition matrix elements of the QCD energy-momentum tensor \cite{Kim:2022bwn},
including the QCD angular momentum of the $N \to \Delta$ transition \cite{Kim:2023xvw}, and possess a
rich mechanical structure. Of particular interest is the possibility of connecting the $N \rightarrow N$ and
$N \rightarrow \Delta$ GPDs through the $1/N_c$ expansion of QCD, using the emergent spin-flavor symmetry
in the large-$N_c$ limit, which enables a unified analysis of ground state and transition GPDs
\cite{Goeke:2001tz,Pascalutsa:2006ne,Schweitzer:2016jmd,Kroll:2022roq}.

%
%
\begin{SCfigure}
\includegraphics[width=0.5\textwidth]{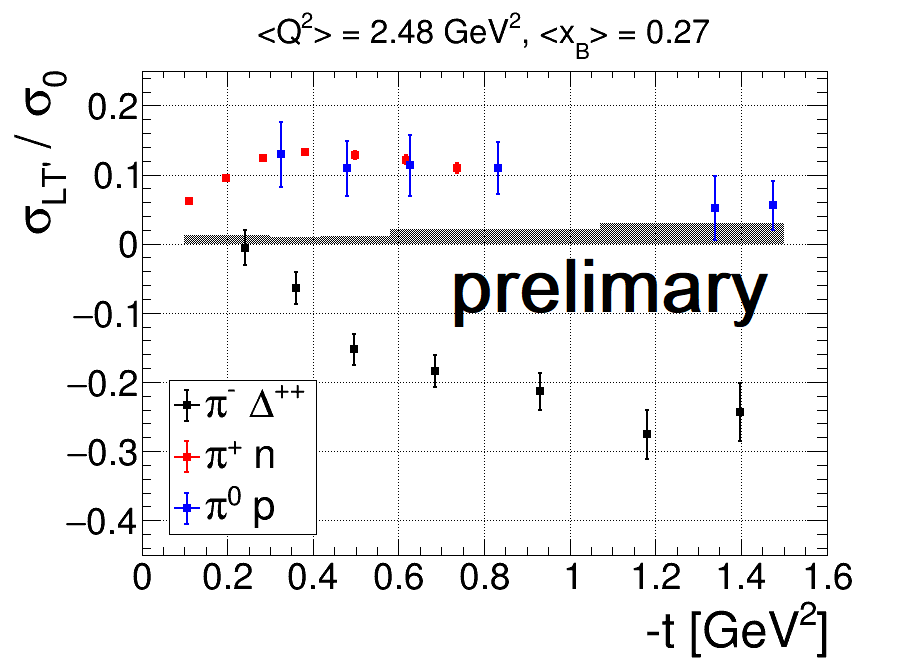}
\caption{Preliminary results for the structure function ratio $\sigma_{LT'}/\sigma_{0}$ for
	$\pi^{-}\Delta^{++}$ (black) \cite{CLAS:2023akb} in comparison to results from $\pi^{+}n$
	(red)~\cite{CLAS:2022iqy} and $\pi^{0}p$ (blue)~\cite{Kim23}. The gray histogram shows the
	systematic uncertainty of the $\pi^{-}\Delta^{++}$ measurement.}
	\label{fig:pions_comparison}
\end{SCfigure}

The chiral-even transition GPDs are probed in DVCS with $N \rightarrow \Delta$ transitions.
The process has been described theoretically in detail in Ref.~\cite{Guichon:2003ah}, and with a special
focus on CLAS12 kinematics in Ref. \cite{Semenov-Tian-Shansky:2023bsy}. The first experimental study of
$p \rightarrow \Delta^{+}$ DVCS beam spin asymmetries and cross sections
is currently ongoing based on CLAS12 data. The chiral-odd transition GPDs are probed
in hard exclusive pion electroproduction with $N \rightarrow \Delta$ transitions.
Such processes have been studied theoretically in Ref.~\cite{Kroll:2022roq}, using predictions
for the transition GPDs based on the large-$N_c$ limit of QCD.
On the experimental side, no publications of observables sensitive to transition GPDs
were available for a long time, since either statistics or the beam energy were not sufficient to
study the high $Q^{2}$ regime of such $N \rightarrow N^{*}$ reactions and to have enough phase space
to suppress the dominant backgrounds. With the structure function ratio $\sigma_{LT'}/\sigma_{0}$ of
the hard exclusive $\pi^{-}\Delta^{++}$ electroproduction process, the first observable sensitive to
transition GPDs was recently submitted for publication by the CLAS Collaboration
\cite{CLAS:2023akb}. Figure~\ref{fig:pions_comparison} shows a comparison of preliminary results for
the structure function ratio $\sigma_{LT'}/\sigma_{0}$ for $\pi^{-}\Delta^{++}$ in comparison to
results from $\pi^{+}n$ \cite{CLAS:2022iqy} and $\pi^{0}p$ \cite{Kim23}.  The large absolute magnitude
for $\pi^{-}\Delta^{++}$, compared to $\pi^{+}n$ can be seen as a clear effect of the excitation
process \cite{CLAS:2023akb}.

%
%
\begin{figure}[!]
	\centering
		\includegraphics[width=0.53\textwidth]{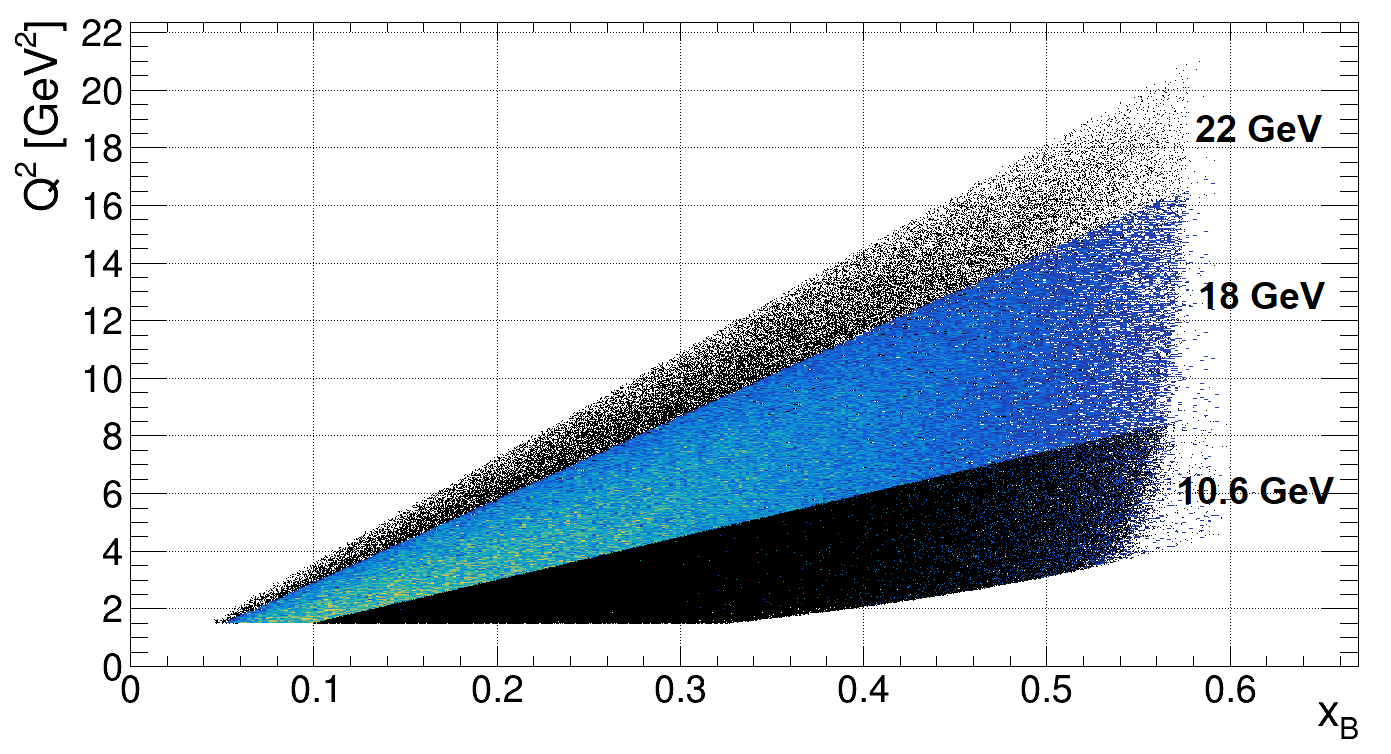}
		\includegraphics[width=0.43\textwidth]{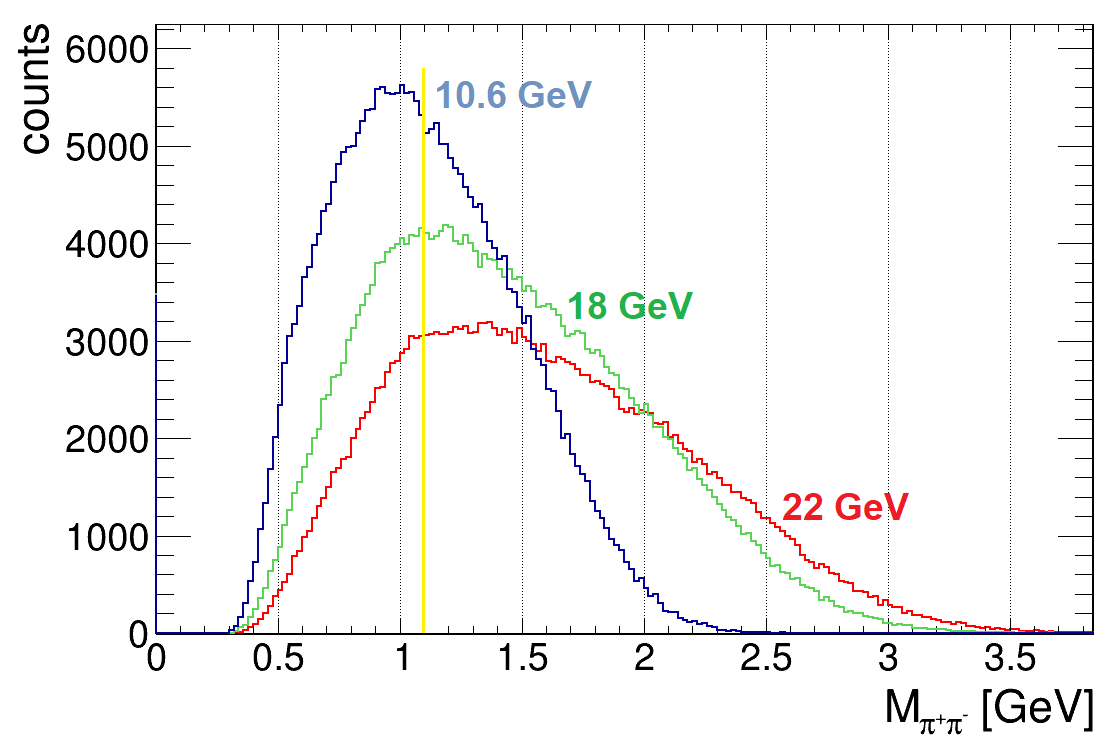}
	\caption{Comparison of the available phase space, accessible with the present CLAS12 setup, in $Q^{2}-x_{B}$ or the $\pi^{-}\Delta^{++}$ process under forward kinematics ($-t <$~1.5~GeV$^{2}$) (left) and for the $\pi^{+}\pi^{-}$ invariant mass of the same process, which is used to suppress the dominant $\rho$ production background by the cut on $M(\pi^{+}\pi^{-}) >$~1.1~GeV, indicated by the yellow line (right) for a 10.6~GeV, 18~GeV and 22~GeV electron beam.}
	\label{fig:q2_xB_coverage}
\end{figure}

The $N \rightarrow N^{*}$ DVCS, as well as the $N \rightarrow N^{*}$ DVMP processes, will both strongly profit from an energy and luminosity upgrade of JLab/CLAS12.
From the statistics point of view, the low efficiency for the detection of the multi-particle final states, in combination with the background suppression cuts, strongly limit the available statistics of the final sample. From the beam energy point of view, the currently available beam energy of 10.6~GeV allows a study of the lower lying nucleon and $\Delta$ resonances in a limited $Q^{2}$ range. However, especially for higher mass resonances, the factorization requirement $Q^{2} \gg m^{2}_{N^{*}}$ strongly limits the option based on a 10.6 GeV electron beam. Here, a 22 GeV upgrade of JLab will enable the investigation of higher-mass resonances and extend the accessible $Q^{2}$ range for the lower-mass resonances. Based on this extended range, a detailed study of the scaling behavior of the different observables will become possible.  

Figure~\ref{fig:q2_xB_coverage} shows the available phase space, accessible with the present CLAS12
setup, in $Q^{2}-x_{B}$ for the $\pi^{-}\Delta^{++}$ process under forward kinematics and the
$\pi^{+}\pi^{-}$ invariant mass of the same process for a 10.6~GeV, 18~GeV, and 22~GeV electron
beam. The distributions of the $N \rightarrow \Delta$ DVCS process show similar characteristics.
It can be seen that a 22 GeV upgrade of JLab will provide a significantly increased $Q^{2}$ range
for a fixed value of $x_{B}$. This will provide a big advantage for the study of these processes,
since the factorization of the $N \rightarrow N^{*}$ DVCS and DVMP processes requires a high
virtuality $Q^{2}$ to be above the resonance mass squared. While this condition can be already
fulfilled with a 10.6 GeV electron beam for lower-mass resonances, such as the $\Delta(1232)$, an
energy upgrade to 22 GeV will be essential ensure the factorization of the process for higher-mass
resonances and to study the scaling behavior of the observables.  As shown in the right part of
Fig.~\ref{fig:q2_xB_coverage}, the increase of the phase space for the different invariant mass
combinations will allow a more efficient suppression of non-resonant background from exclusive meson
production and also from other (differently charged) nucleon resonance production channels, which is
mostly expected at lower masses. Higher beam energies will, therefore, also provide a more efficient
event selection and a better suppression of the non-resonant background.
The 22~GeV upgrade of JLab, in combination with a luminosity upgrade of CLAS12, will thus
provide ideal conditions for the study of the 3D structure of nucleon resonances via transition GPDs.

\subsubsection{Transition Distribution Amplitudes in Backward-Angle Processes
\label{sec:tda}}
%
%
\begin{figure}[hb]
\begin{center}
\includegraphics[width=0.3\textwidth]{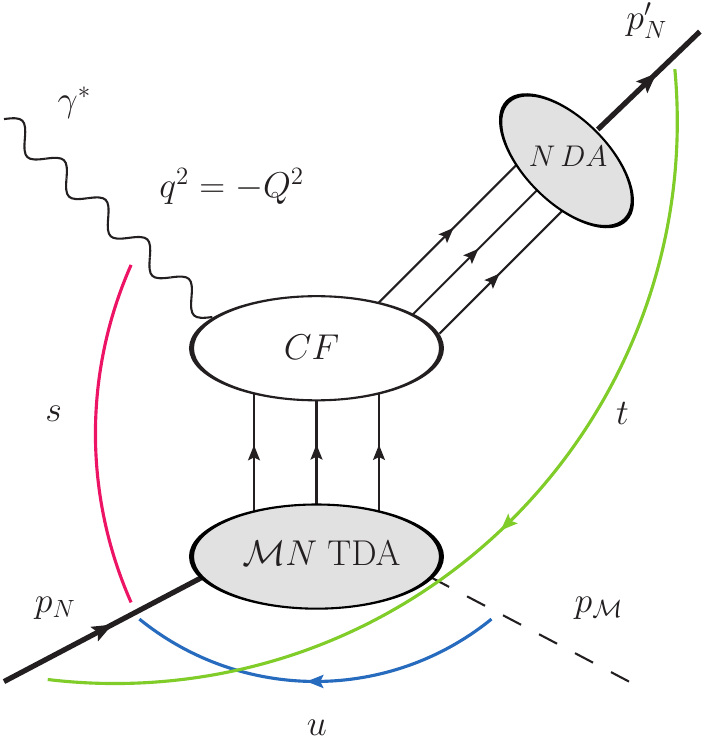}
\hspace{.2\textwidth}
\includegraphics[width=0.3\textwidth]{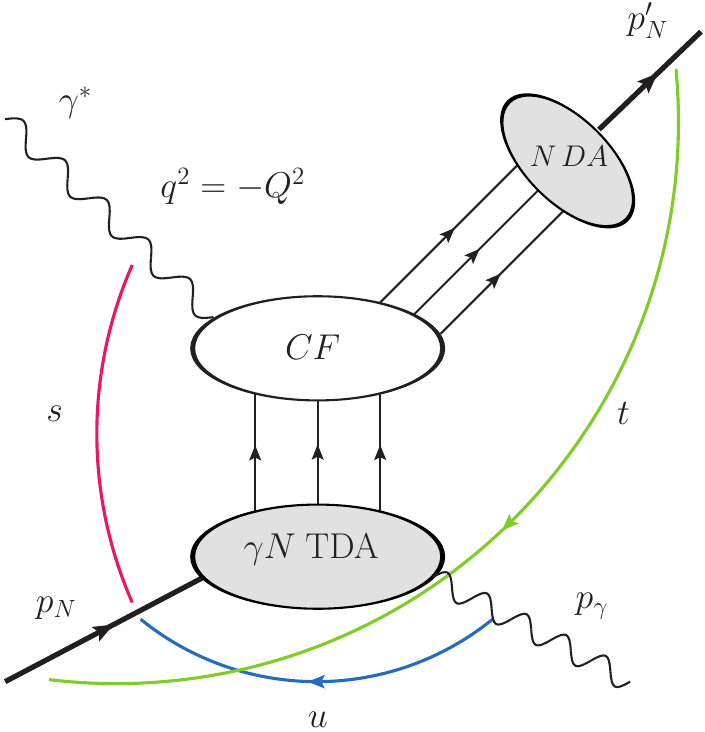}
\end{center}
\caption{ The factorized amplitude for backward electroproduction of a meson (left) or a
photon (right). CF denotes the perturbatively calculable coefficient function.}
\label{fig.process}
\end{figure}

While the pronounced forward peak of exclusive electroproduction cross sections has been investigated extensively in the last two decades, revealing the backward-angle peak was quite delayed
(for a review, see Ref.~\cite{Gayoso:2021rzj}), 
although several theoretical predictions 
\cite{Frankfurt:1999fp,Pire:2004ie} 
advocated for its study in the framework of the collinear factorization approach of perturbative QCD. 
In a nutshell, the argument relies on the fact that a deeply virtual photon is able to resolve the partonic structure of the nucleon in a quite similar way in the backward as in the forward regimes. It  thus seems legitimate to assume the extension of the validity of collinear factorization in near-backward kinematics. The scattering amplitude is then presented as a convolution of non-perturbative hadronic matrix elements
of the light-cone three quark operators, with a hard subprocess amplitude describing the interaction of partons with the hard electromagnetic probe. 

The reaction mechanism for electroproduction of a backward meson off a nucleon, and backward electroproduction of a real photon off a nucleon  
(backward DVCS, bDVCS) is presented, respectively, in Figs.~5 and~11 of Ref.~\cite{Pire:2021hbl}. Apart from familiar nucleon distribution amplitudes (DAs), this description involves a new class of non-perturbative non-diagonal objects: nucleon-to-meson (${\cal M} N$) 
and nucleon-to-photon ($\gamma N$) transition distribution amplitudes (TDAs).
The concept of transition distribution amplitudes (TDAs) 
 \cite{Pire:2021hbl} 
naturally extends both the concept of nucleon DAs and nucleon GPDs.
To leading twist-$3$ accuracy, TDAs are defined as matrix elements of the same three quark light cone operator occurring in the definition of nucleon DAs, with color structure $\varepsilon_{c_1 c_2 c_3}  q^{c_1} (z_1) q^{c_2} (z_2) q^{c_3} (z_3)$. 
However, these
matrix elements are taken between two states of different baryonic charges (a nucleon and a meson, or a nucleon and a photon). Moreover,  similarly to the case of GPDs, the non-zero transfer of longitudinal momenta is characterized by a skewness variable $\xi$ defined with respect to the $u$-channel momentum transfer $p_{{\cal M}, \gamma}-p_N$.  As in the forward case, $\xi$ is related to
the Bjorken variable $x_B$ as $\xi \approx \frac{x_B}{2-x_B}$.
Nucleon-to-meson and nucleon-to-photon TDAs quantify partonic correlations inside hadrons; they give access to the baryon charge distribution in the transverse plane and provide new tools to study the shape of nucleon's mesonic and electromagnetic clouds.

The first experimental studies of near-backward hard exclusive reactions at JLab have been recently presented in
Refs.~\cite{CLAS:2017rgp, JeffersonLabFp:2019gpp,CLAS:2020yqf}.
A dedicated study of the exclusive backward electroproduction of a $\pi^0$  above the resonance region has recently been approved with JLab Hall C~\cite{Li:2020nsk}.
The goal of this experiment is to perform cross section measurements at several different $Q^2$ values
with complete $\sigma_L$, $\sigma_T$, $\sigma_{LT}$, and $\sigma_{TT}$
separation to verify the $\sigma_T$ dominance, revealed in Ref.~\cite{JeffersonLabFp:2019gpp}, for
near-backward $\omega$-meson electroproduction.

Challenging the validity of the collinear factorized
description of hard backward meson and photon electroproduction reactions is of primary importance to elaborate a unified and consistent approach for physics of hard exclusive reactions both in the forward 
and in the backward regions with non-trivial cross channel baryon number exchange.
The improved luminosity and perfect angular coverage after the suggested $22$ GeV
upgrade makes JLab a unique experimental facility to probe hadron dynamics in the vicinity of the backward peak, to confirm or disprove the validity of factorized description, discriminate between different TDA models, and recover the hadronic structural information encoded in TDAs.

%
%
\begin{figure}
\begin{center}
\includegraphics[width=0.45\textwidth]{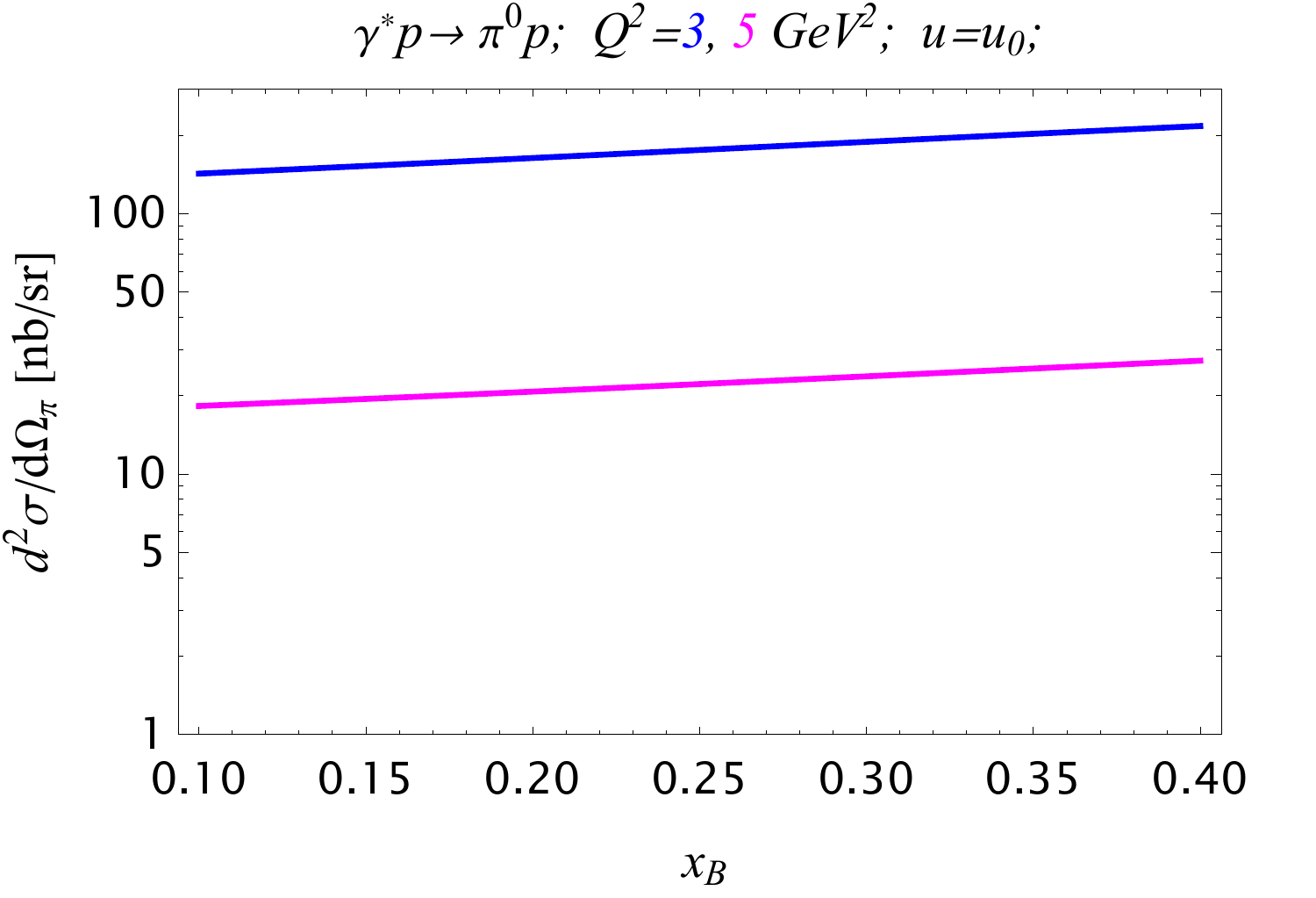}
\hspace{0.03\textwidth}
\includegraphics[width=0.45\textwidth]{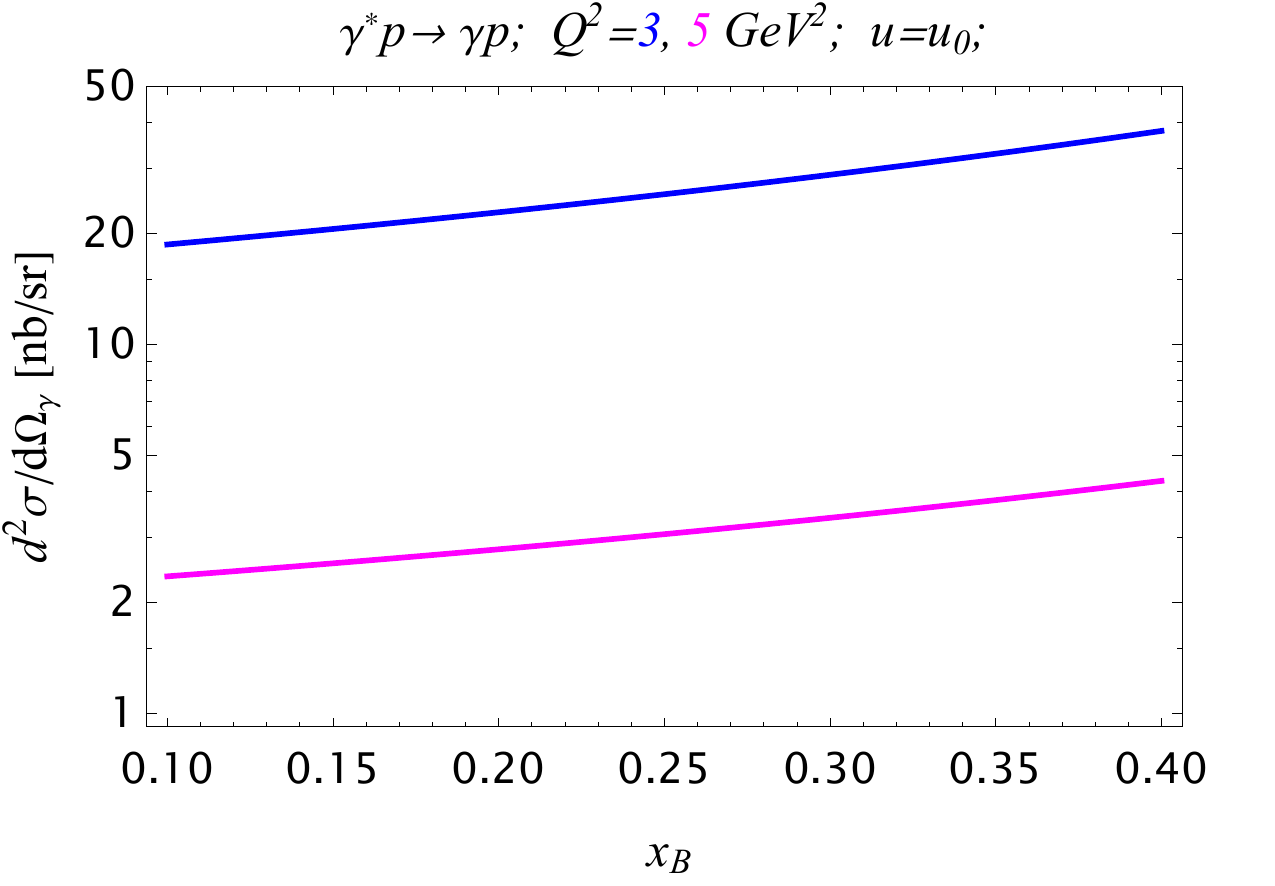} 
\end{center}
\caption{An estimate of $\frac{d^2 \sigma}{d \Omega}$ cross sections for (left) backward electroproduction of a $\pi^0$ meson, 
(right) or of a photon, as a function of $x_B$ for $Q^2=3$, $5$ GeV$^2$. The $x_B$ range is representative of the kinematics that 
may be accessed by a JLab22 experiment.}
\label{fig.estimates}
\end{figure}

In order to examine the prospects of experimental studies of backward reactions at the kinematical conditions of JLab22, we present theoretical predictions for the cross section of hard backward electroproduction of pions and photons in the $(Q^2, x_B)$ range appropriate to this facility.
In Fig.~\ref{fig.estimates}
are shown the differential cross sections $\frac{d^2 \sigma}{d \Omega}$ of
$\gamma^* N \to \pi N'$ (left)
and
$\gamma^* N \to \gamma N'$ (right) as a function of the Bjorken variable $x_B$ for the two values of $Q^2=3$, $5$ GeV$^2$. For the case of backward pion production these cross section estimates are based on
the cross-channel nucleon exchange model for $\pi N$ TDAs suggested in Ref.~\cite{Pire:2011xv}.
For the case of backward DVCS \cite{Lansberg:2006uh}, we rely on the nucleon-to-photon TDA model
\cite{Pire:2022fbi,Pire:2022kwu}
with adjustable normalization  devised in  to account for the recent backward $J/\psi$ photoproduction data presented by the Hall D Collaboration \cite{GlueX:2023pev}.
The current model TDAs are still quite primitive and the results must be taken only as  very rough order of magnitude estimates.
However, taken at face value, these predictions clearly show that
the corresponding cross sections are accessible with JLab22.
The higher electron energy will allow to extend in a crucial way the results obtained up to now at JLab \cite{CLAS:2017rgp,JeffersonLabFp:2019gpp}, as well as those expected in the near future \cite{Li:2020nsk}
and help to get a detailed understanding of hadron dynamics in the vicinity of the backward peak. 
Note that, contrarily to the forward electroproduction (DVCS, DVMP, or TCS) processes, backward amplitudes do not benefit from small $\xi$ enhancement since TDAs mostly probe the valence quark content of nucleons. Very high energy processes  at EIC are thus not the favored channels for their study.

As a final comment, let us stress that experiments with nuclear targets can be used to explore the phenomenon of color transparency \cite{Jain:2022xzo} for backward hard reactions \cite{Huber:2022wns}, which is a crucial prediction of the short distance nature of the underlying mechanism.

\subsection{Short-Range Electromagnetic Structure}
\subsubsection{Pion and Kaon Form Factors}

Measurement of the $\pi^+$ electromagnetic form factor for $Q^2>0.3$ GeV$^2$ can be accomplished at by the detection of the exclusive reaction $p(e,e'\pi^+)n$ at low $-t$. 
This is best described as quasi-elastic ($t$-channel) scattering of the electron from the
virtual $\pi^+$ cloud of the proton, where $t=(p_{p}-p_{n})^2$ is the
Mandelstam momentum transfer to the target nucleon.
Scattering from the $\pi^+$ cloud dominates the longitudinal
photon cross section ($d\sigma_L/dt$),  when $|t|\ll m_p^2$.  To reduce background contributions, one preferably separates the
components of the cross section due to longitudinal (L) and transverse (T)
virtual photons (and the LT, TT interference contributions), via a Rosenbluth
separation.  The value of $F_{\pi}(Q^2)$ is determined by
comparing the measured $\sigma_L$ values at small $-t$ to the best
available electroproduction model.  The obtained $F_{\pi}$ values are in
principle dependent upon the model used, but one anticipates this dependence to
be reduced at sufficiently small $-t$.

Hall C has a uniquely important role to play in the EIC era, particularly in
the realm of precision L/T-separation measurements.  Conventional Rosenbluth
separations are impractical at the EIC, because statistical and random
systematic uncertainties in $\sigma_L$ are magnified by $1/\delta\epsilon$,
where $\delta\epsilon$ is the difference in the virtual photon polarization
parameters at high and low beam energies.  To keep the uncertainties in
$\sigma_L$ to an acceptable level, $\delta\epsilon >0.2$ is typically required,
{\it i.e.} an uncertainty magnification no more than 5.  This is not feasible at the EIC,
so physicists will need to rely on an extrapolation of L/T-separated data from
Hall C for the interpretation of EIC data.  

We consider a two phase program.  In phase 1, only measurements with the existing 
HMS+SHMS instrumentation were explored, to see what can be accomplished in a
``cost-effective phased-approach''.  In this phase, a higher energy JLab electron beam,
in concert with the existing HMS+SHMS spectrometers in Hall C will enable
important Deep Exclusive Meson Production (DEMP) measurements which
significantly extend the kinematic range of the 11 GeV physics measurements and
improve the range of overlap between JLab L/T-separated measurements and the
unseparated measurements of the EIC.  This improved overlap will greatly ease
the interpretability of the higher $Q^2$ EIC data and be a significant
scientific contribution.  Several programs of measurement are significantly 
enhanced.  The pion form
factor is a key observable to study in our understanding the physics of color
confinement, {\it i.e.} understanding the transition of the behavior of QCD from long
distance scales (low $Q^2$, where confinement dominates and the interaction is
very strong) to short distance scales (high $Q^2$, where the quarks act as if
they are free).  The pion is one of the simplest QCD systems available for
study, and the measurement of its elastic form factor is the best hope for
seeing this transition experimentally.  This is possible via high resolution
measurements of the $p(e,e'\pi^+)n$ reaction.  18 GeV electron beam will allow
the $\pi^+$ form factor to determined to $Q^2=10$~GeV$^2$ with small uncertainties, and up to
11.5 GeV$^2$ with somewhat larger model uncertainties, a significant advance
over the 12 GeV data.  Similarly, assuming the extracted $\sigma_L$
are sufficiently sensitive to the $K+$ pole, high resolution measurements of
the $p(e,e'K^+)\Lambda$ reaction may allow the $K^+$ form factor to be
determined up to $Q^2=7.0$ GeV$^2$ with small uncertainties, and to 9.0 GeV$^2$ with
larger uncertainties.

A separate program with broad significance is the study of hard-soft
factorization in exclusive meson production.  To access the physics contained
in GPDs, one is limited to the kinematic regime where the hard-soft
factorization theorem applies.  There is no single criterion for the kinematic
region of applicability, but tests of necessary conditions can provide evidence
that the factorization regime has been reached.  One of the most stringent
tests is the $Q^2$-dependence of the $\pi$, $K$ exclusive electroproduction
cross sections, {\it i.e.} $\sigma_L$ scales to leading order as $Q^{-6}$;
$\sigma_T$ does not, with the expectation of $Q^{-8}$ scaling; and $\sigma_L
\gg \sigma_T$.  The experimental validation of the onset of the hard scattering
regime is essential for the reliable interpretation of JLab GPD program
results.  One question that can be addressed is whether the onset of scaling is
different for kaons than pions.  Furthermore, by studying both $K^+$ and
$\pi^+$, a quasi-model-independent study of hard-soft factorization can be
performed.  Such a program is already underway with E12-09-011 \cite{E12-09-011} and E12-19-006 \cite{E12-19-006}.
Electron beams up to 18 GeV with the HMS+SHMS nearly double the $Q^{-n}$
scaling test range of these experiments, greatly reducing uncertainties and extending
the range of these measurements over a wider $x$ range.  
A related issue is the
study of hard-soft factorization in backward angle DEMP reactions, which can be
described in Sec. \ref{sec:tda}.
An 18 GeV beam will enable a significant improvement in the $Q^{-n}$ scaling test in these reactions as well.

\begin{SCfigure}
 \includegraphics[width=0.65\textwidth]{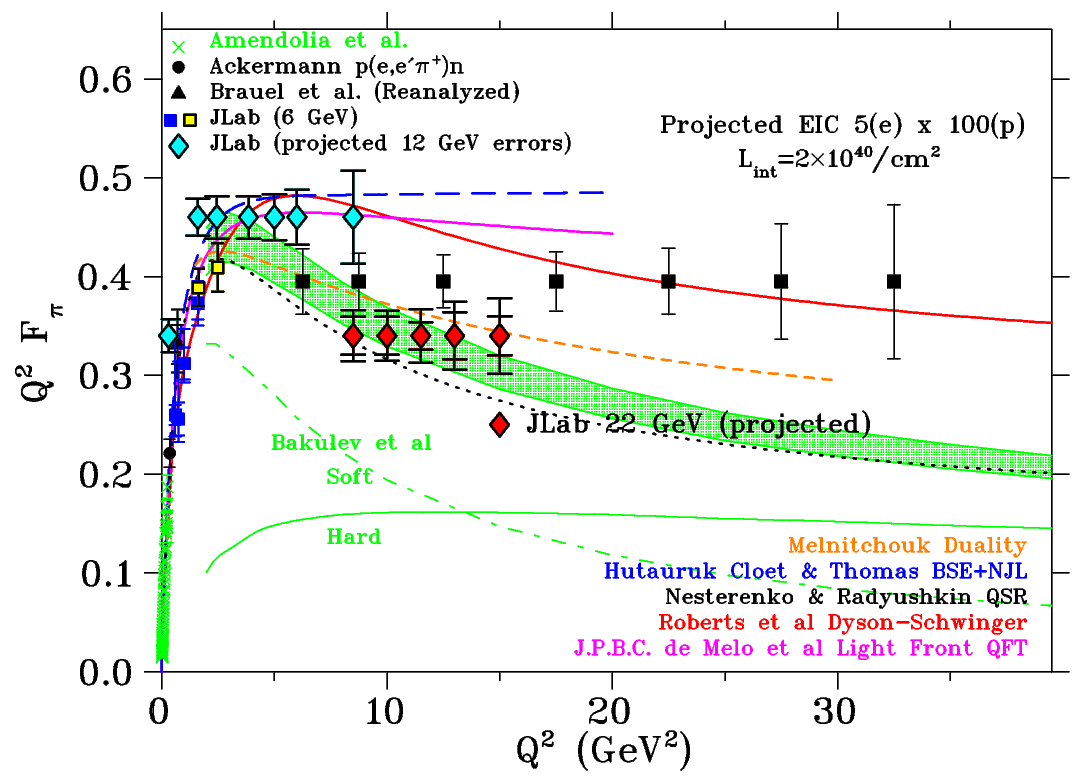} 
  \caption{\label{fig:Fpi}
Existing data (blue, black, yellow, green) and projected uncertainties for
future data on the pion form factor from JLab (PionLT: cyan; 22 GeV VHMS+SHMS:
red) and EIC (black), in comparison to a variety of hadronic structure models.
JLab 22 GeV with an upgraded VHMS will dramatically improve the overlap between
the $F_{\pi}$ from true L/T-separations at JLab and non-L/T-separated data from
the EIC.}
\end{SCfigure}

A phase 2 set of measurements replaces the HMS with a new spectrometer we dub
VHMS, to enable measurements utilizing the full 22 GeV electron beam energy.
For pion form factor measurements, the scattered electron would be detected in
the SHMS and the high-momentum, forward-going $\pi^+$ in the upgraded VHMS. 
In this scenario, assuming similar
$p(e,e'\pi^+)n$ statistics to the recently completed PionLT experiment \cite{E12-19-006}, we
project 22 GeV electron beam and an upgraded VHMS will extend the region of
high quality $F+{\pi}$ values from $Q^2=$6.0~GeV$^2$ (PionLT) to $Q ^2$=13.0 GeV$^2$,
and with somewhat larger errors to $Q^2$=15 GeV$^2$ (see Fig.~\ref{fig:Fpi}.  Here
the error bars are calculated, but $y$-positions of the projected
data are arbitrary.  The 22 GeV upgrade will provide greatly improved overlap
between the $F_{\pi}$ JLab and EIC datasets.  As the interpretation of some EIC
data ({\it e.g.} GPD extraction from DEMP data) will depend on an extrapolation of
Hall C L/T-separated data, maximizing the overlap between the Hall C and EIC
datasets are a high priority.

\subsubsection{Nucleon Electromagnetic Form Factors at High Momentum Transfer}

The elastic electromagnetic form factors of hadrons are among the simplest measurable quantities for
probing the spatial distributions and interactions of their elementary quark-gluon constituents.
The precise polarization transfer measurements of the proton form factor ratio $\mu_p G_E^p/G_M^p$
from JLab's Halls A~\cite{JeffersonLabHallA:1999epl,JeffersonLabHallA:2001qqe} and
C~\cite{Puckett:2010ac}, that revealed the unexpected decrease of this ratio for momentum transfers
$Q^2 \gtrsim 1$ GeV$^2$, are among the best-known and most widely cited experimental results from
JLab \cite{perdrisat_bonner}.
Measurements of hadron elastic form factors at large momentum transfers
$Q^2$ are sensitive to the interesting and theoretically challenging region of transition in QCD
between the non-perturbative regime of strong coupling and confinement and the perturbative regime
of weak coupling and asymptotic freedom~\cite{Barabanov:2020jvn,Gross:2022hyw}.

The nucleon form factors are accessible experimentally through measurements of
differential cross sections and double-polarization observables. A summary of the current state of knowledge of the nucleon
electromagnetic form factors can be found in Ref.~\cite{Gross:2022hyw}.
\begin{figure}[t!]
\begin{center}
\includegraphics[width=0.75\textwidth]{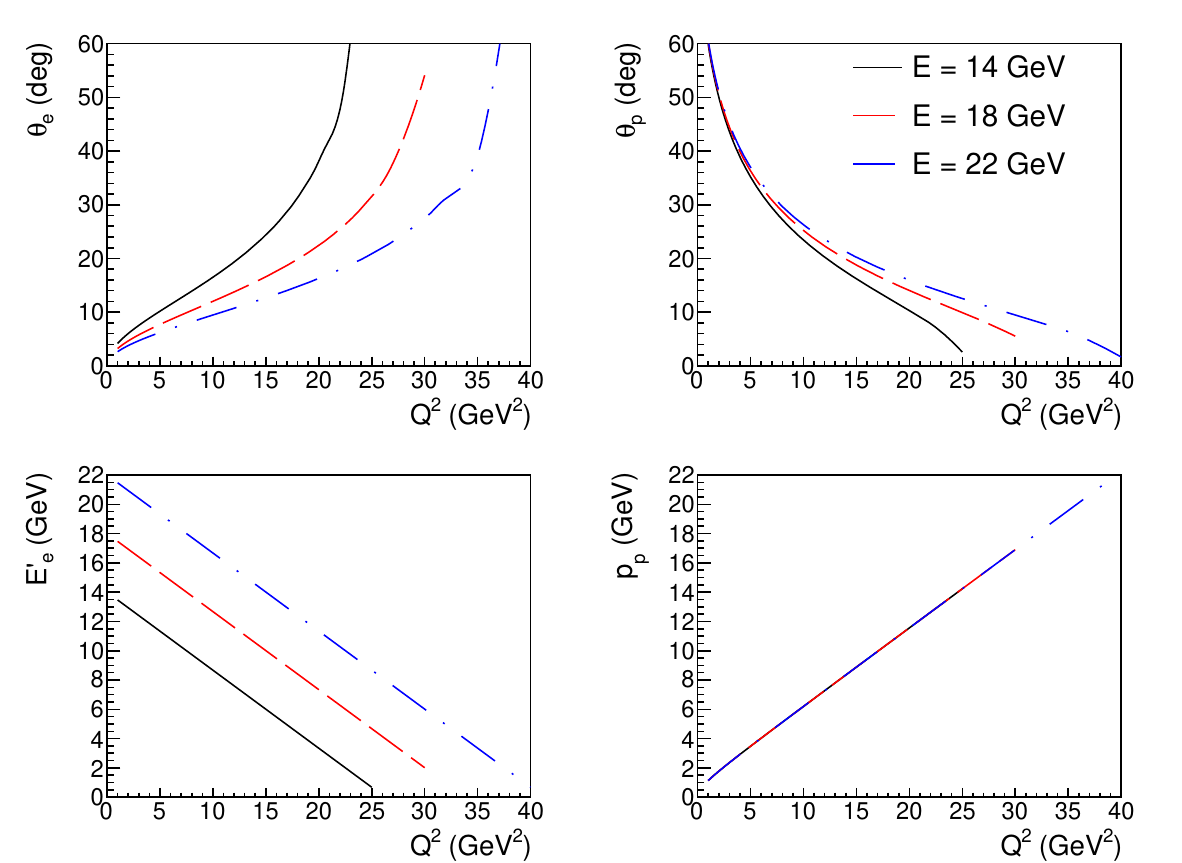} 
\end{center}
\vglue -0.2in
\caption{Fixed target elastic $eN$ scattering kinematics for beam energies of 14, 18, and 22~GeV. From top left to bottom right: $Q^2$ dependence of electron and proton scattering angles $\theta_e$, $\theta_p$, scattered electron energy $E'_e$, and scattered proton momentum $p_p$. }
\label{fig:KinePlots1}
\end{figure}
%
%
\begin{figure}[b!]
    \centering
    \includegraphics[width=0.49\textwidth]{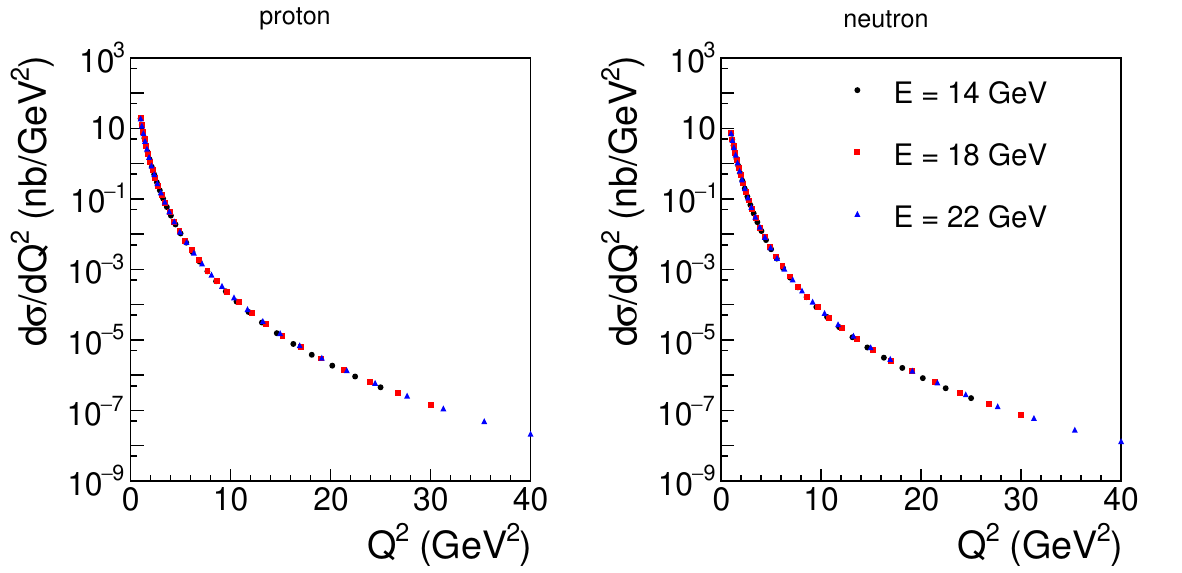}
    \includegraphics[width=0.49\textwidth]{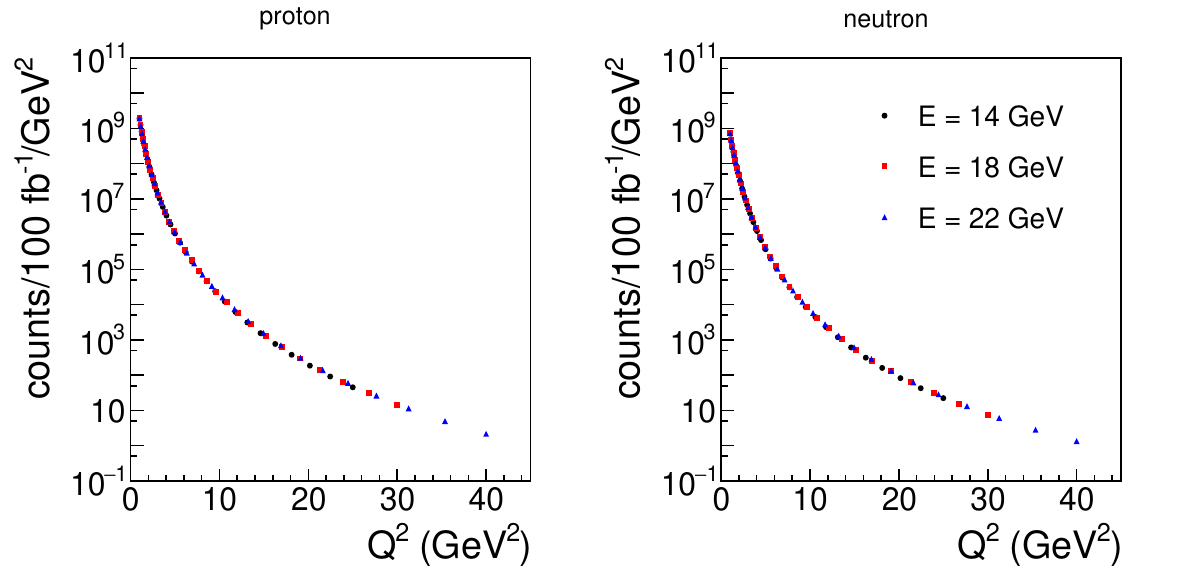}
    \vglue -0.05in
    \caption{Left: Differential cross section $d\sigma/dQ^2$ for different beam energies for proton and neutron. 
    Right: Same as left, expressed in terms of an event rate per unit integrated luminosity per $Q^2$ interval (assuming 2$\pi$ azimuthal angle acceptance).}
    \label{fig:rates}
\end{figure}
A future energy upgrade of CEBAF would facilitate extending these measurements to $Q^2$ of at least 20-30 GeV$^2$ for the magnetic form factors, and at least 15-20 GeV$^2$ for the electric form factors.
Due to the approximate $Q^{-12}$ dependence of the elastic $eN$ scattering cross section at large
$Q^2$, these measurements require very high luminosities that are only achievable in fixed-target
experiments, combined with large-acceptance detectors. Moreover, measurements of polarization
observables require not merely large $Q^2$, but also virtual photon polarization $\epsilon$
meaningfully different from one, as the transverse asymmetry/recoil polarization $A_t = P_t$ that is
sensitive to the form factor ratio vanishes in both the forward and backward-angle limits ($\epsilon
= 1$ and $\epsilon = 0$), and is maximum at $\epsilon = 0.5$ for any given $Q^2$.

Figure~\ref{fig:KinePlots1} shows the $Q^2$ dependence of the scattering angles and momenta of the
outgoing particles in two-body $eN \rightarrow eN$ scattering for beam energies of 14, 18, and 22
GeV, representative of different numbers of passes through an upgraded CEBAF. In the range of 10-30
GeV$^2$, the particle angles are well-matched to the acceptances of existing or planned
large-acceptance spectrometers such as CLAS12, SBS+BigBite, and SoLID. The outgoing particle
energies are rather high, which would pose some challenges in terms of acceptance for precision
focusing spectrometers such as those in Hall C, and would be challenging in terms of momentum
resolution for medium and large-acceptance spectrometers with moderate field integral, such as
SBS+BB and/or SoLID in Hall A and CLAS12 in Hall B. Nonetheless, the measurements appear feasible
over a wide range of $Q^2$ without requiring major new detector
construction. 

Figure~\ref{fig:rates} shows the cross section $d\sigma/dQ^2$ for proton and neutron, expressed in
nb/GeV$^2$ and in terms of counts per 100 fb$^{-1}$ per GeV$^2$. Note that the cross section
differential in $Q^2$ is independent of the beam energy at a given $Q^2$ (in contrast to the cross
section differential in electron solid angle $d\sigma/d\Omega_e$). The cross section $d\sigma/dQ^2$
assumes $2\pi$ azimuthal acceptance. While the planned Electron-Ion Collider (EIC) at BNL should be
capable of measuring elastic $ep$ cross sections to fairly large $Q^2$ values at $\epsilon \approx
1$~\cite{Schmookler:2022gxw}, the EIC operating at its design luminosity will produce $\approx 100$
fb$^{-1}$ per \textit{year} in the best-case scenario. In contrast, a typical CEBAF fixed-target
luminosity of $\approx 10^{38}$ cm$^{-2}$ s$^{-1}$ (liquid hydrogen/deuterium) produces $\approx
10,000$ fb$^{-1}$ per \textit{day}, allowing for precision measurements of cross sections and
polarization observables over a wide range of $Q^2$ and $\epsilon$. As such, high-$Q^2$ elastic form
factor measurements are a unique worldwide capability of CEBAF, and will remain so even in the EIC
era.

\subsection{Bound Three-Quark Structure of Excited Nucleons and Emergence of Hadron Mass}

\subsubsection{The Emergent Hadron Mass Paradigm}

\begin{figure}[ht]

\includegraphics[clip, width=0.52\textwidth]{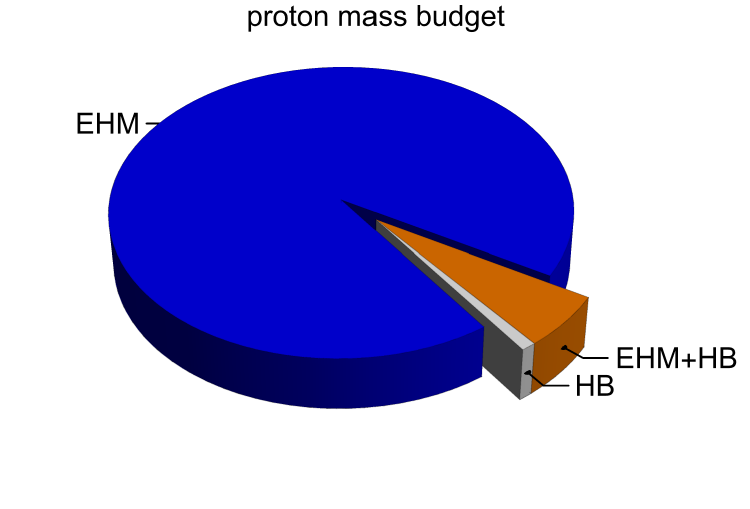}

\vspace*{-38ex}

\rightline{\parbox[t][10em][c]{0.44\textwidth}{
\caption{\label{Fmassbudget}
Proton mass budget, drawn using a Poincar\'e-invariant decomposition: emergent hadron mass (EHM) $= 94$\%; Higgs boson (HB) contribution
$= 1$\%; and EHM+HB interference $= 5$\%. (Separation at renormalization scale $\zeta = 2\,$GeV, calculated using information from
Refs.\,\cite{Flambaum:2005kc, RuizdeElvira:2017stg, Aoki:2019cca, Workman:2022ynf}).}}}

\vspace*{5ex}

\end{figure}

The Standard Model of Particle Physics has one well-known mass-generating mechanism for the most elementary constituents of Nature,
\textit{viz}.\ the Higgs boson \cite{Englert:2014zpa, Higgs:2014aqa}, which is critical to the evolution of the Universe. Yet, alone,
the Higgs is responsible for just 1\% of the visible mass in the Universe.  Visible matter is constituted from nuclei found on Earth
and the mass of each such nucleus is largely the sum of the masses of the nucleons they contain.  However, only 9\,MeV of a nucleon's
mass, $m_N=940$~MeV, is directly generated by Higgs boson couplings into quantum chromodynamics (QCD). Evidently, as highlighted by
Fig.\,\ref{Fmassbudget}, Nature has another, very effective, mass-generating mechanism.  Often called emergent hadron mass (EHM)
\cite{Binosi:2022djx, Ferreira:2023fva, Ding:2022ows, Carman:2023zke}, it is responsible for 94\% of $m_N$, with the remaining 5\% 
generated by constructive interference between EHM and the Higgs boson. This makes studies of the structure of ground and excited 
nucleon states in experiments with electromagnetic probes a most promising avenue to gain insight into the strong interaction dynamics
that underlie the  emergence of the dominant part of the visible mass in the Universe \cite{Barabanov_2021, Proceedings:2020fyd,
Burkert:2017djo, Carman:2023zke, Deur:2022msf}.

These experiments are key to address still open questions in contemporary hadron physics. What is the dynamical origin of EHM; what are
its connections with gluon and quark confinement, themselves seemingly characterized by a single nonperturbative mass scale; and are
these phenomena linked or even the underlying cause of dynamical chiral symmetry breaking (DCSB)? DCSB has long been argued to provide
the key to understanding the pion, Nature's most fundamental Nambu-Goldstone boson, with its unusually low mass and structural
peculiarities \cite{Aguilar:2019teb, Anderle:2021wcy}.
 
\begin{figure}[t]
\centering
\includegraphics[width=0.8\textwidth]{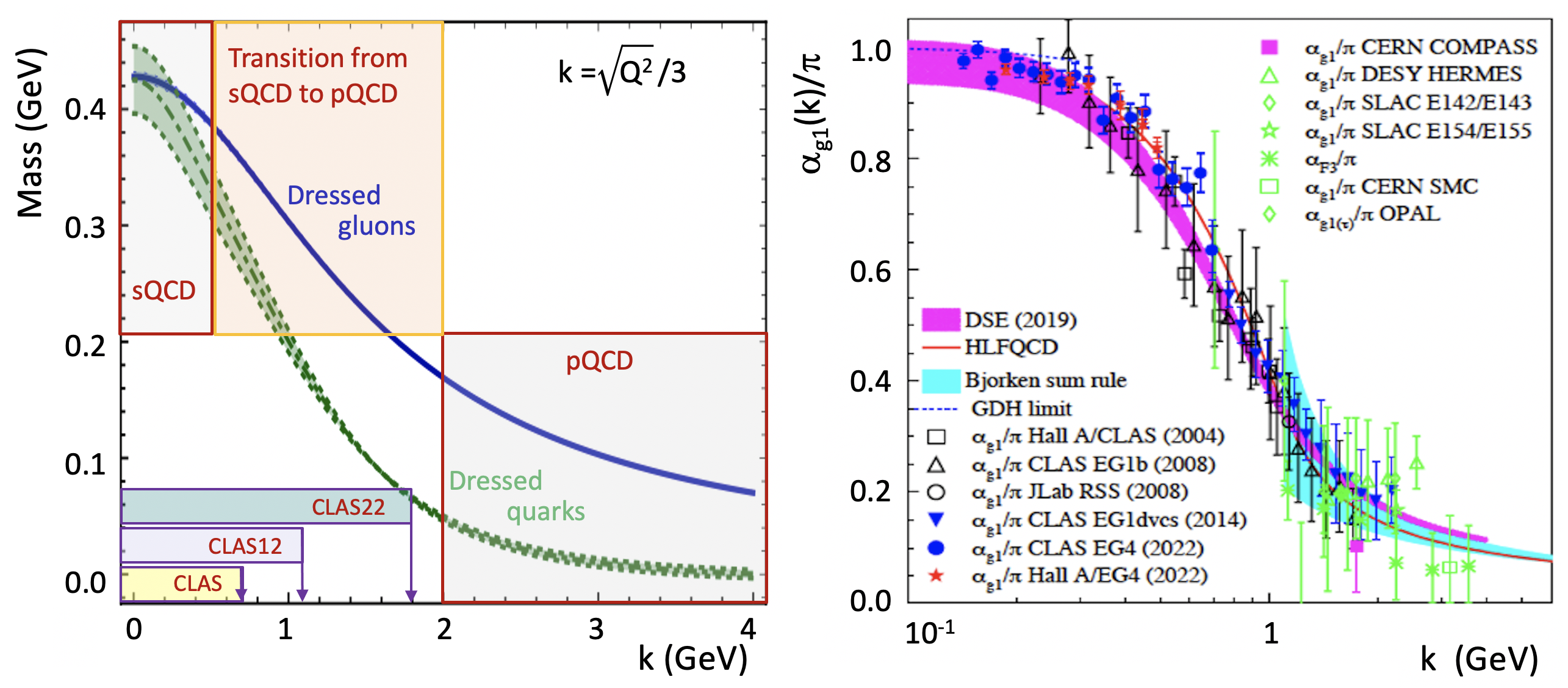}
\vspace{-3mm}
\caption{Momentum-dependent dressed quark and gluon masses (left) \cite{Roberts:2021nhw} and the QCD running coupling (right)
\cite{Deur:2022msf} deduced using CSMs from the QCD Lagrangian as a solution of the equations of motion for the quark and gluon 
fields. On the left the ranges of momenta accessible for mapping the dressed quark mass function from the results on the evolution 
of the $\gamma_vpN^*$ electrocouplings with $Q^2$ from the measurements of the 6-GeV/12-GeV eras with the CLAS/CLAS12 detectors are 
shown, as well as the corresponding momentum range that will be accessible after an increase of the CEBAF energy up to 22~GeV from the
anticipated results on the $\gamma_vpN^*$ electrocouplings at $Q^2$ from 10-30~GeV$^2$ from the measurements with the CLAS22 detector.}
\label{running_mass_reach}
\end{figure}

After the pioneering work of Schwinger in the early sixties studying nonperturbative gauge-sector dynamics in Poincar\'e-invariant 
quantum field theories \cite{Schwinger:1962tp, Cornwall:1981zr, Mandula:1987rh}, treatments of QCD using continuum Schwinger function
methods (CSMs) have delivered self-consistent calculations of the base ingredients capable of explaining EHM. The results are drawn in
Fig.\,\ref{running_mass_reach} and demonstrate that a Schwinger mechanism is active in QCD \cite{Binosi:2022djx, Ferreira:2023fva}. The
gluon vacuum polarization tensor remains four-transverse. Owing to the gluon self-interactions encoded in the QCD Lagrangian, the three
four-transverse modes of the gluon acquire a momentum-dependent mass. Existence of a gluon mass-scale enables the definition and
calculation of a unique QCD analog of the Gell-Mann-Low effective charge, well-known from quantum electrodynamics (QED). This charge
saturates on the infrared domain and is practically identical to the process-dependent charge extracted in experimental studies of the
Bjorken sum rule \cite{Deur:2022msf}, as shown in the right part in Fig.~\ref{running_mass_reach}. The massive gluon propagator and
effective charge are the principal elements in the quark gap equation and, together, they ensure that light (even massless) quarks 
acquire a running mass, $M_q(k)$, whose value at infrared momenta matches that which is typically identified with a constituent quark 
mass \cite[Sec.\,2C]{Roberts:2021nhw}. The QCD running coupling and the momentum-dependent dressed quark and gluon masses constitute 
the three pillars of EHM, about which more will be explained below. Contemporary theory is now engaged in elucidating the huge array 
of their observable consequences and paths to measuring them.  The challenge for experiment is to test the predictions so that the
boundaries of the Standard Model can finally be drawn.

Testing these predictions requires a paradigm shift going beyond the studies of the only stable ground state of hadron, {\it i.e.}, 
the proton. Just as studying the ground state of the hydrogen atom did not reveal the need for and intricacies of QED, focusing on 
the ground state of only one form of hadron matter will not elucidate the full complexity of the strong interaction dynamics in the 
regime where the QCD running coupling $\alpha_s/\pi$ is comparable with unity, referred to as the strong QCD (sQCD) regime. A new 
era is dawning, with science poised to construct and begin operating high-luminosity, high-energy, and large acceptance facilities 
that will enable precision studies of new types and excited states of hadron matter. For instance, one may anticipate a wealth of 
highly precise data that will reveal the inner workings of Nature's most fundamental Nambu-Goldstone bosons, the $\pi$ and $K$ mesons
\cite{Aguilar:2019teb, Anderle:2021wcy}, and if the future is planned well, critical empirical information on the structure of nucleon
excited states ($N^\ast$s) will be extended over the full range of distances where the dominant part of their masses and structure are
anticipated to emerge from QCD \cite{Carman:2023zke}.

\subsubsection{Experimentally Driven Studies on $N^*$ Structure, Emergent Hadron Mass, and Strong QCD}

During the last decade, crucial progress has been achieved in the exploration of the $N^*$ electroexcitation amplitudes, the so-called
$\gamma_vpN^*$ electrocouplings, stimulating research efforts with an emphasis on how the masses and properties of $N^*$ states emerge
from QCD \cite{Burkert:2017djo, Proceedings:2020fyd, Barabanov_2021, Ding:2022ows, Carman:2023zke}. High-quality meson electroproduction
data of the 6-GeV era at Jefferson Lab (JLab) from the CLAS detector have enabled reliable extraction of the electrocouplings of the 
most prominent $N^\ast$s in the mass range up to 1.8~GeV. The data for different $N^*$ states expresses many facets of the strong 
interaction that generate $N^*$ structure and mass in the $Q^2$ range up to 5~GeV$^2$ \cite{Barabanov_2021, Carman:2023zke,
Proceedings:2020fyd, Burkert:2017djo}. This places heavy pressure on theory to deliver interpretations and explanations.

Synergistically engaging with the JLab experimental program, a diverse collection of theoretical models and methods have been employed 
in attempts to address the questions raised by the data and connect them with QCD \cite{Suzuki:2009nj,Giannini:2015zia,Burkert:2017djo}. 
In the past decade, notable successes have been achieved using CSMs \cite{Qin:2020rad}, which have delivered numerous predictions for 
hadron structure observables in the meson and baryon sectors, both for ground and excited states \cite{Barabanov_2021, 
Proceedings:2020fyd, Burkert:2017djo, Gao:2017mmp, Xu:2018cor, Wang:2018kto, Chen:2018rwz,Binosi:2018rht,
Chen:2019fzn, Qin:2019hgk, Lu:2019bjs, Souza:2019ylx, Raya:2021zrz, Cui:2021mom, Ding:2022ows, Liu:2022nku}.

CSM analyses are characterized by the use of dressed (quasiparticle) quarks and gluons as active degrees of freedom with momentum- and,
hence, distance-dependent masses. The framework's predictions for these momentum-dependent mass functions (see 
Fig.~\ref{running_mass_reach} left) have been confirmed in numerical simulations of lattice-regularized QCD \cite{Gao:2017uox,
Oliveira:2018lln, Binosi:2019ecz, Boito:2022rad}. They have also been used to define and calculate the momentum evolution of the QCD 
analog of the Gell-Mann-Low running coupling in QCD \cite{Deur:2022msf} (see Fig.~\ref{running_mass_reach} right). These three pillars 
of the CSM paradigm for explaining EHM -- momentum dependence of the dressed quark and gluon mass functions and the QCD running coupling
-- have become available from the solution of the QCD equations of motion for the quark and gluon fields with the results shown in
Fig.~\ref{running_mass_reach} \cite{Binosi:2022djx, Ding:2022ows, Ferreira:2023fva}. The observed ground state nucleon and $N^*$ 
masses emerge mostly from the running masses of the three dressed quarks that approach the hadron mass scale in the infrared at quark
momenta $k < 0.5$~GeV. Consequently, the electromagnetic elastic nucleon form factors and $\gamma_vpN^*$ electrocouplings exhibit
particular sensitivity to the emergent part of hadron mass. Dressed gluons acquire running masses owing to the gluon self-interaction
encoded in the QCD Lagrangian \cite{Binosi:2022djx, Ferreira:2023fva}. At the distances/quark momenta where the transition from the
perturbative QCD (pQCD) to sQCD regimes is anticipated and as the QCD running coupling $\alpha_s/\pi$ becomes comparable with unity, 
the energy stored in the gluon field is absorbed into the running mass of the dressed quark. A particular strength of CSMs is their 
ability to simultaneously treat and unify the physics of mesons and baryons.

\begin{figure}[t]
\centering
\includegraphics[width=0.95\textwidth]{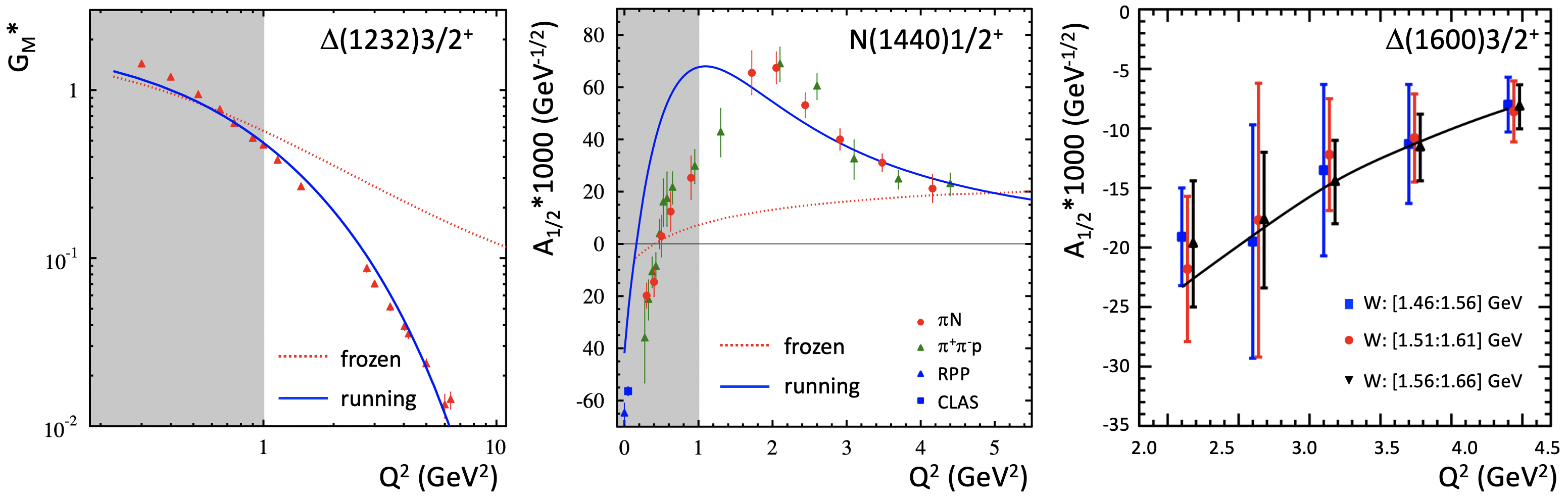}
\vspace{-3mm}
\caption{Results for the $p \to \Delta(1232)3/2^+$ magnetic transition form factor (left) and the $N(1440)1/2^+$ $A_{1/2}(Q^2)$ electrocoupling (middle) \cite{Burkert:2022ioj, Aznauryan:2011qj,Mokeev:2022xfo} from studies of $\pi N$ and $\pi^+\pi^-p$ electroproduction in measurements of the JLab 6-GeV era. 
CSM predictions with the running dressed quark mass deduced from the QCD Lagrangian, see Fig.~\ref{running_mass_reach} (left), are shown as blue solid lines \cite{Segovia:2015hra, Burkert:2017djo} and by employing a simplified contact $qq$-interaction resulting in a momentum-independent (frozen) quark mass of $\approx 0.36$~GeV as red dotted lines \cite{Wilson:2011aa}. 
Comparisons between the CSM prediction (solid line) on the $A_{1/2}(Q^2)$ $\Delta(1600)3/2^+$ electrocoupling~\cite{Lu:2019bjs} and preliminary results from the studies of $\pi^+\pi^-p$ electroproduction with CLAS are shown on the right. The data points with error bars have become available from independent analyses of the cross sections in overlapping $W$-intervals with substantial contributions from the $\Delta(1600)3/2^+$ as labeled for $Q^2$ from 2 to 5~GeV$^2$.}
\label{data_vs_prediction}
\end{figure}

Acquiring insight into the dressed quark mass function from data on hadron structure represents a challenge for experimental hadron 
physics. The amplitude of the virtual photon interaction with a dressed quark in the process of $N^*$ electroexcitation is sensitive 
to the dressed quark mass, making it possible to map out the momentum dependence of the dressed quark mass from the results on the
evolution of the $\gamma_vpN^*$ electrocouplings with $Q^2$. Analyses of the JLab 6-GeV era results on the $N^*$ electroexcitation 
amplitudes have vastly improved our understanding of the momentum dependence of the dressed quark mass function, while the running 
gluon mass and QCD running coupling are constrained by the results on $N^*$ electroexcitation \cite{Carman:2023zke, Mokeev:2022xfo,
Proceedings:2020fyd, Burkert:2017djo}. A good description of the $\Delta(1232)3/2^+$ and $N(1440)1/2^+$ electrocouplings at 
$Q^2 < 5$~GeV$^2$ for these resonances of different structure (see Fig.~\ref{data_vs_prediction}, left and middle), achieved using 
CSMs with the same dressed quark mass function deduced from the QCD Lagrangian and employed elsewhere in the successful description 
of experimental results on nucleon electroweak elastic and transition form factors \cite{Cui:2020rmu, Chen:2022odn}, offers sound 
evidence for providing insight into the momentum dependence of the dressed quark mass. This link is strengthened by the fact that 
such a mass function is also a key element in the prediction of meson electromagnetic elastic and transition form factors 
\cite{Gao:2017mmp, Ding:2018xwy}. 

The CSM predictions made in 2019 \cite{Lu:2019bjs} on the $Q^2$-evolution of the $\Delta(1600)3/2^+$ electrocouplings, achieved 
without modifying or introducing any new parameters, were confirmed in 2022 by the preliminary results from analysis of $\pi^+\pi^-p$
electroproduction measured with the CLAS detector (see Fig.~\ref{data_vs_prediction} right). This success solidifies evidence for 
understanding the dressed quark mass function and, consequently, EHM from studies of the $\gamma_vpN^*$ electrocouplings.

Most results on the $\gamma_vpN^*$ electrocouplings from the JLab experiments of the 6-GeV era are available for $Q^2 < 5$~GeV$^2$,
allowing for the exploration of the momentum dependence of the dressed quark mass within the range of quark momentum $k < 0.75$~GeV, 
where $<$30\% of hadron mass is anticipated to be generated (see Fig.~\ref{running_mass_reach} left). CLAS12 is the only facility 
capable of extending these results on the $\gamma_vpN^*$ electrocouplings into the unexplored $Q^2$ range from 5 to 10~GeV$^2$ based 
on measurements of $\pi N$, $\pi^+\pi^-p$, $K\Lambda$, and $K\Sigma$ electroproduction~\cite{Carman:2023zke, Proceedings:2020fyd},
spanning the domain of quark momenta up to 1.1~GeV where $\approx\!50$\% of hadron mass is generated (see
Fig.~\ref{running_mass_reach} left).

The already available results on the $\gamma_vpN^*$ electrocouplings from CLAS and those foreseen from the extension with CLAS12 
will offer the information needed to facilitate the development of approaches for the description of the structure of bound 
three-quark baryons based on quantities computed from the QCD Lagrangian. These approaches will ultimately be capable of making 
predictions for ground state nucleon structure observables and $\gamma_vpN^*$ electrocouplings for $Q^2 < 10$~GeV$^2$, as was 
discussed at the JLab Workshop {\it ``Strong QCD from Hadron Structure Experiments"} held in 2019~\cite{Proceedings:2020fyd}.

\begin{figure}[t]
\centering
\includegraphics[width=0.9\textwidth]{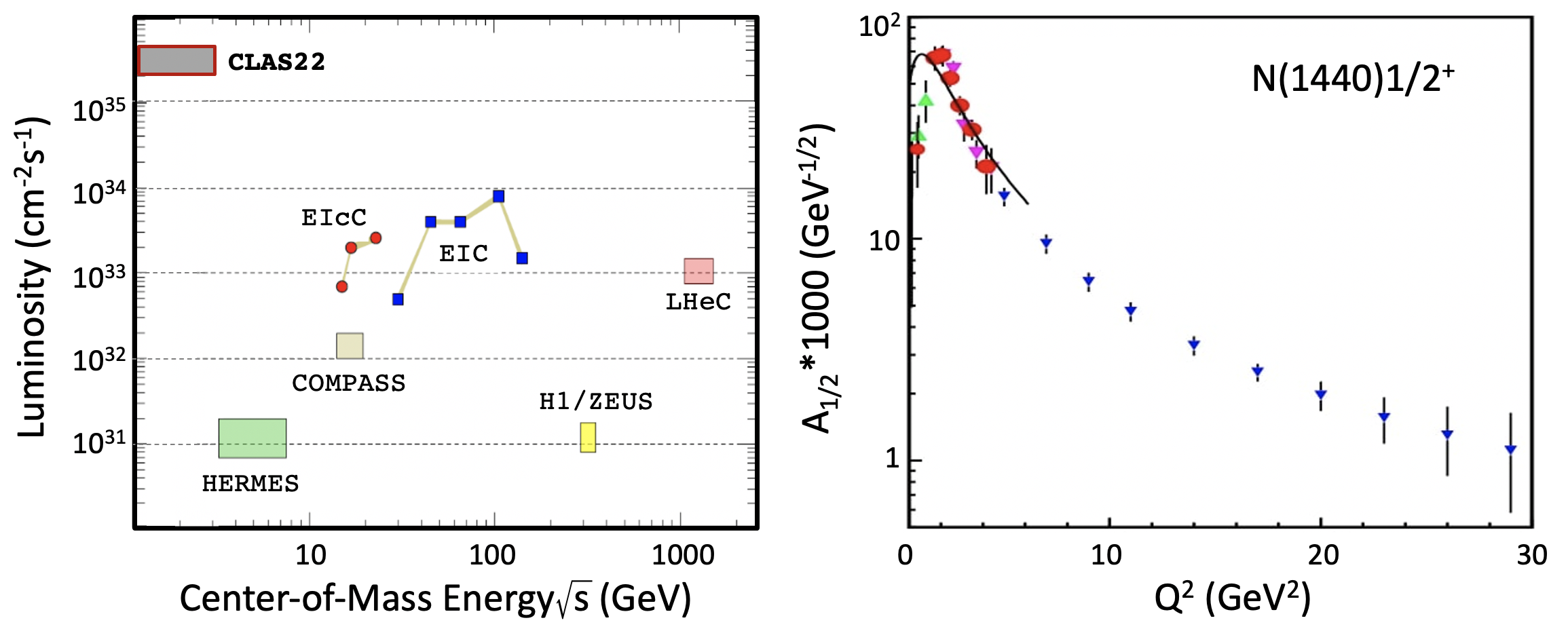}
\vspace{-2mm}
\caption{(Left) Luminosity versus invariant mass coverage of the available and foreseen facilities to explore hadron structure in
experiments with electromagnetic probes. CLAS22 would be the only facility with sufficient luminosity to determine the $\gamma_vpN^*$
electrocouplings at $Q^2$ from 10-30~GeV$^2$ that can map out the dressed quark mass function within the range of quark momenta 
$k < 2$~GeV where the dominant part of hadron mass and the bound three-quark structure of $N^*$s emerge from QCD. (Right) Available
results at $Q^2$ up to 5~GeV$^2$ and those projected for the $Q^2$ evolution of the $N(1440)1/2^+$ $A_{1/2}$ electrocoupling for $Q^2$ 
up to 30~GeV$^2$ for a luminosity of $5 \times 10^{35}$~cm$^{-2}$s$^{-1}$ and six months of data collection time.}
\label{projections}
\end{figure}

Ultimately, pushing the momentum transfer squared to $N^*$s up to 30~GeV$^2$ will extend the coverage of quark momenta over the domain
where $\approx\!90$\% of hadron mass emerges (see Fig.~\ref{running_mass_reach} left). At the $Q^2$ limit of this domain, where the 
QCD running coupling becomes smaller, direct comparisons between nonperturbative and perturbative QCD concepts on how hadron structure 
emerges from QCD can be attempted. To resolve the challenging problem of understanding the underpinnings of EHM, $M_q(k)$ will be 
mapped out over the entire range of quark momenta from zero up to $\approx\!2$~GeV. As one progresses through this momentum scale, the 
transition from strongly coupled to perturbative QCD takes place and the dressed quarks and gluons, which emerge on the domain for 
which $\alpha_s/\pi \to 1$ (see Fig.~\ref{running_mass_reach}), begin to reveal their inner parton-like origin. This unique endeavor,
probing QCD through detailed studies of bound three-quark states, is fully complementary with, \textit{e.g}., hard scattering off 
single quarks, as in deep inelastic scattering, as well as studies of pseudoscalar and hybrid mesons, and paves the way for further
extensions of the exploration of $N^*$ structure in three dimensions \cite{Semenov-Tian-Shansky:2023bsy}.

Simulations of $\pi N$, $\pi^+\pi^-p$, $K\Lambda$, and $K\Sigma$ electroproduction channels with an increased CEBAF beam energy of 
22~GeV show that the $\gamma_vpN^*$ electrocouplings can indeed be extracted up to $Q^2 \approx 30$~GeV$^2$ utilizing the large 
acceptance CLAS12 spectrometer at luminosities ${\cal L} \approx$ 2-5$ \times 10^{35}$~cm$^{-2}$s$^{-1}$ (a configuration referred 
to as CLAS22) as exemplified in Fig.~\ref{projections} (right). A comparison of the parameters for the available and anticipated 
facilities for studies of hadron structure with electromagnetic probes in this regime (see Fig.~\ref{projections} left) demonstrates 
that, after the CEBAF energy increase, CLAS22 would be the only facility capable of delivering results on the $\gamma_vpN^*$
electrocouplings for $Q^2$ up to 30~GeV$^2$. A representative example for the anticipated accuracy in the resonance electrocoupling
extraction is shown in Fig.~\ref{projections} (right) for the $A_{1/2}$ electrocoupling of the $N(1440)1/2^+$.

Baryons are the most fundamental three-body systems in Nature. If we do not understand how QCD -- a Poincar\'e-invariant quantum field
theory -- generates these bound states, then our understanding of Nature is incomplete. Moreover, EHM is not immutable: its 
manifestations are manifold and growing experience is revealing that each hadron manifests different facets. One piece -- the proton --
does not complete a puzzle. Completing the QCD picture requires far more, and precise data relating to the structure of nucleon excited
states will add essential pieces. Extending the results on the $\gamma_vpN^*$ electrocouplings into the $Q^2$ range up to 30~GeV$^2$,
after the increase of the CEBAF energy and pushing the CLAS12 detector capabilities to measure exclusive electroproduction to the 
highest possible luminosity, will offer the only foreseen opportunity to explore how the dominant part of hadron mass and the bound 
three-quark structure of $N^*$s emerge from QCD. These things will make CEBAF at 22~GeV a unique QCD-facility at the luminosity 
frontier.

\clearpage \section{Hadron–Quark~Transition~and~Nuclear~Dynamics~at~Extreme~Conditions}
\label{sec:wg5}

\subsection{Theoretical Overview}\label{sec:theory}

One of the outstanding issues of the strong interactions physics is understanding the dynamics of the transition between hadronic and partonic (quarks and gluons) phases of matter.~At~{\em high temperatures}, such transitions are relevant to the evolution of matter after the Big Bang, which have been studied experimentally in heavy ion collisions. At {\em low (near zero) temperatures and high densities (``cold dense" states)}, such transitions are relevant to understanding the stability of atomic nuclei as well as dynamics of cold dense nuclear matter that exists at the core of neutron stars and defines the limiting density for structured matter.

Two main directions of exploring ``cold dense" transitions are associated with probing the dynamics of nuclear forces at short space-time separation in nuclei (referred to hereafter as ``nuclear dynamics at extreme conditions"), and investigating nuclear medium modifications of hadronic structure for both bound nucleons and hadrons produced in nuclear medium (referred to hereafter as ``hadron-quark transition" in nuclear medium). Overall, the research program of 22~GeV energy upgrade will be aimed at discovering the fundamental QCD basis of short-range nuclear physics phenomena.

\subsubsection{Nuclear Dynamics at Extreme Conditions}

Last two decades have seen a significant progress in our understanding of the nuclear structure at short distances down to internucleon separations of $\sim$ 0.8~fm. JLab experiments at 6~GeV~\cite{Egiyan_2003,Egiyan_2006} and 12~GeV~\cite{Fomin_2012} energies have confirmed early predictions~\cite{Frankfurt:1993sp} of the onset of scaling in ratios of inclusive cross sections at large $Q^2$ and $x$ indicating the dominance of two-nucleon short-range correlations (2N SRCs) in bound nucleon momentum range of $300-650$~MeV. Several significant advances have been made in studies of 2N SRCs affirming that (i) the latter consists of nucleonic components only~\cite{FRANKFURT:2008}; (ii) the factor of 20 dominance of proton-neutron components to that of proton-proton and neutron-neutron components~\cite{Piazetzky_2006,Subedi_2008,Duer_2019}, which is due to the dominance of tensor interactions~\cite{Sargsian:2005,Schiavilla:2006xx} as an indication of probing 2N~SRCs at distances as small as 0.8~fm; (iii) the prediction of momentum sharing rule~\cite{Sargsian:2012sm} according to which the minority component in asymmetric nuclei per nucleon has a larger share of high momentum component in the nuclear ground state wave function - the effect that was confirmed experimentally at JLab~\cite{Hen_2014,Duer_2018}.

The future of exploring the short-distance structure of nuclei is to reach the distances dominated by practically unknown dynamics of nuclear core. The latter is one of the most fascinating subjects of the modern\begin{wrapfigure}[18]{r}{0.55\textwidth}
\vspace{-0.65cm}
    \begin{center}  
        \includegraphics[clip=true, trim= 0.2cm 0.45cm 0 0.4cm,width=0.55\textwidth]{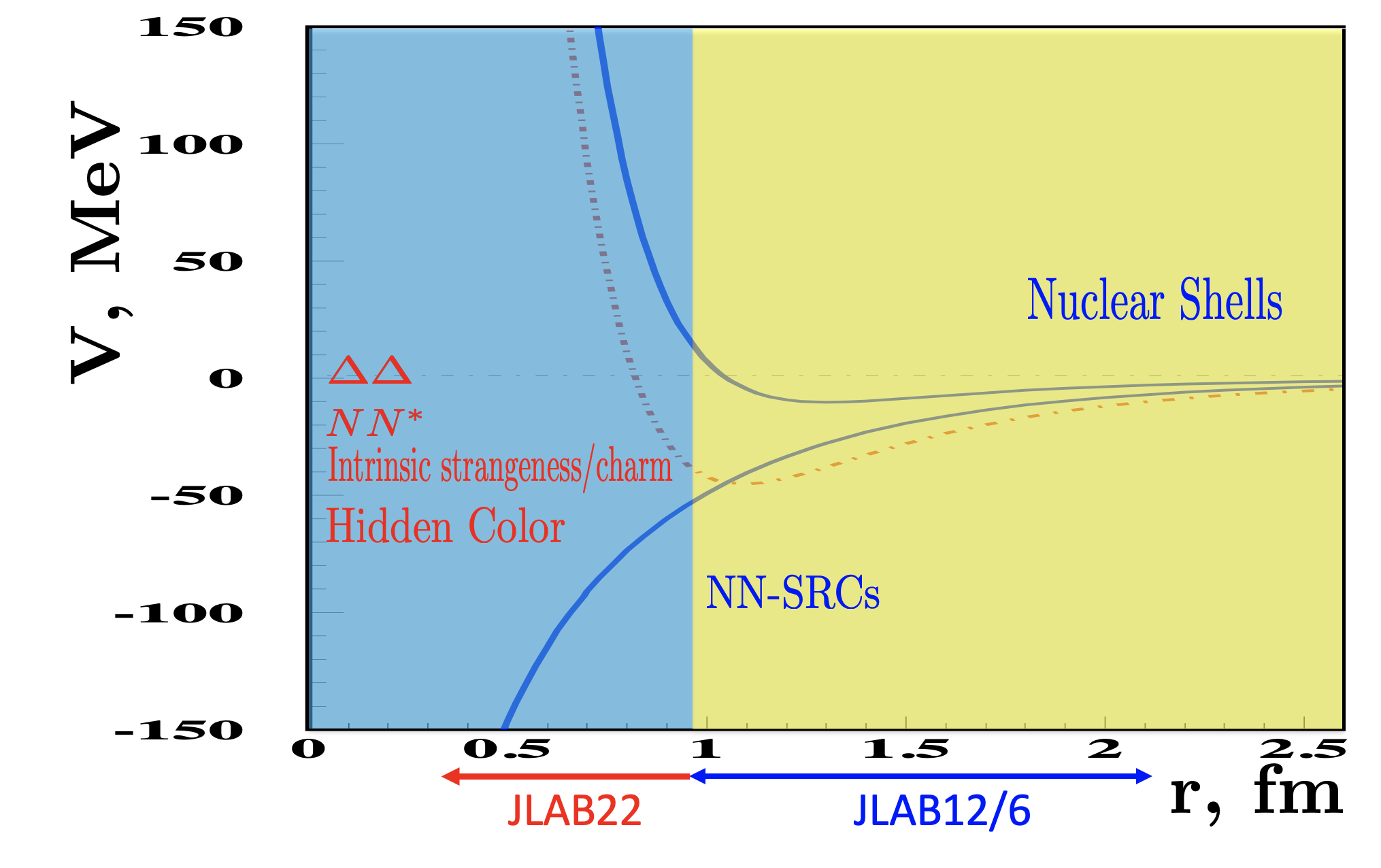}
        \end{center}
            \vglue -0.23in
            \caption{Internucleon force reach at 22~GeV (at distances $\ge 1.2$~fm, $NN$ potentials contribute to the mean field Hartree-Fock potentials, resulting in nuclear shell structure).} 
        \vspace{-0.15cm}
    \label{figcore}        
\end{wrapfigure} nuclear physics. Its existence is essential for the stability of atomic nuclei~\cite{Jastrow:1951} and thus the stability of the structure in the universe. Yet, it is very little known about dynamical origin of nucleon-nucleon ($NN$) repulsive core. The modern phenomenological $NN$ potentials use the Wood-Saxon type ansatz of 1960s (see {\it e.g.}~Ref.~\cite{Wiringa:1994wb}), while in~effective theories the short-range interaction is parameterized by contact terms (see {\it e.g.}~Ref.~\cite{Epelbaum:2008ga}).~QCD gives new perspective to dynamical origin of the nuclear core predicting possibility of sizable contribution from non-nucleonic components including the hidden color~\cite{Harvey:1981,Brodsky:1986,Frankfurt:1981,Miller:2014}. Hidden color states in nuclear wave functions follow from being restricted only by six-quark color singlets in the two-baryon systems at very short ($\le 0.5$~fm) distances. Current constraints on such a six-quark admixture is less than few percent in overall normalization of nuclear ground state wave function and their exploration is essential for understanding the hadron-quark transition in super-dense nuclear matter.

Since the expected excitation energies relevant to nuclear core are in the order of several GeV (see Fig.~\ref{figcore}), it will require probing substantially large internal momenta in the nucleus ($\gtrsim 1$~GeV) to be able to probe the $NN$ core. The repulsive nature of the interaction also indicates that the measured cross section in many cases will be very small. 

As it will be discussed in the next paragraphs, the upgraded energy and high intensity electron beam will provide unprecedented conditions for probing very large internal momenta in nuclei.~Discovering the fundamental QCD basis for short-range nuclear physics phenomena is a primary goal for the 22~GeV energy upgrade.~This contribution builds upon two discovery proposals covered in Subsubsecs.~\ref{sub:fastq} and~\ref{sub:BNS-TDIS} about probing superfast quarks and bound nucleon structure with tagged deep-inelastic scattering~(TDIS). Hidden color states in nuclear wave functions and a quark-gluon basis for the European Muon Collaboration (EMC) effect and SRCs in nuclei are examples of accessible fundamental QCD physics for JLab22.~As mentioned in the Subsubsec.~\ref{sub:fastq}, JLab22 probes of quark-gluon degrees of freedom in $NN$ interactions are described with an emphasis on the six-quark color singlets of two-nucleon system. Hidden color singlets, in contrast to the multiple color singlet nucleons that dominate the nuclear wave function, have been suggested as a solution to the EMC effect in $A > 3$ nuclei~\cite{West:2020tyo}. Such fundamental QCD states can be built with diquark constituents, and a diquark bond formed with valence quarks from two different nucleons has been proposed as the cause of SRCs in nuclei~\cite{West:2020rlk}.

\begin{figure}[!t]
\vspace{-0.70cm}
    \centering   
        \includegraphics[clip=true, trim= 0.01cm 0 0 0.015cm, 
        width=0.45\textwidth,height=0.3\textwidth]{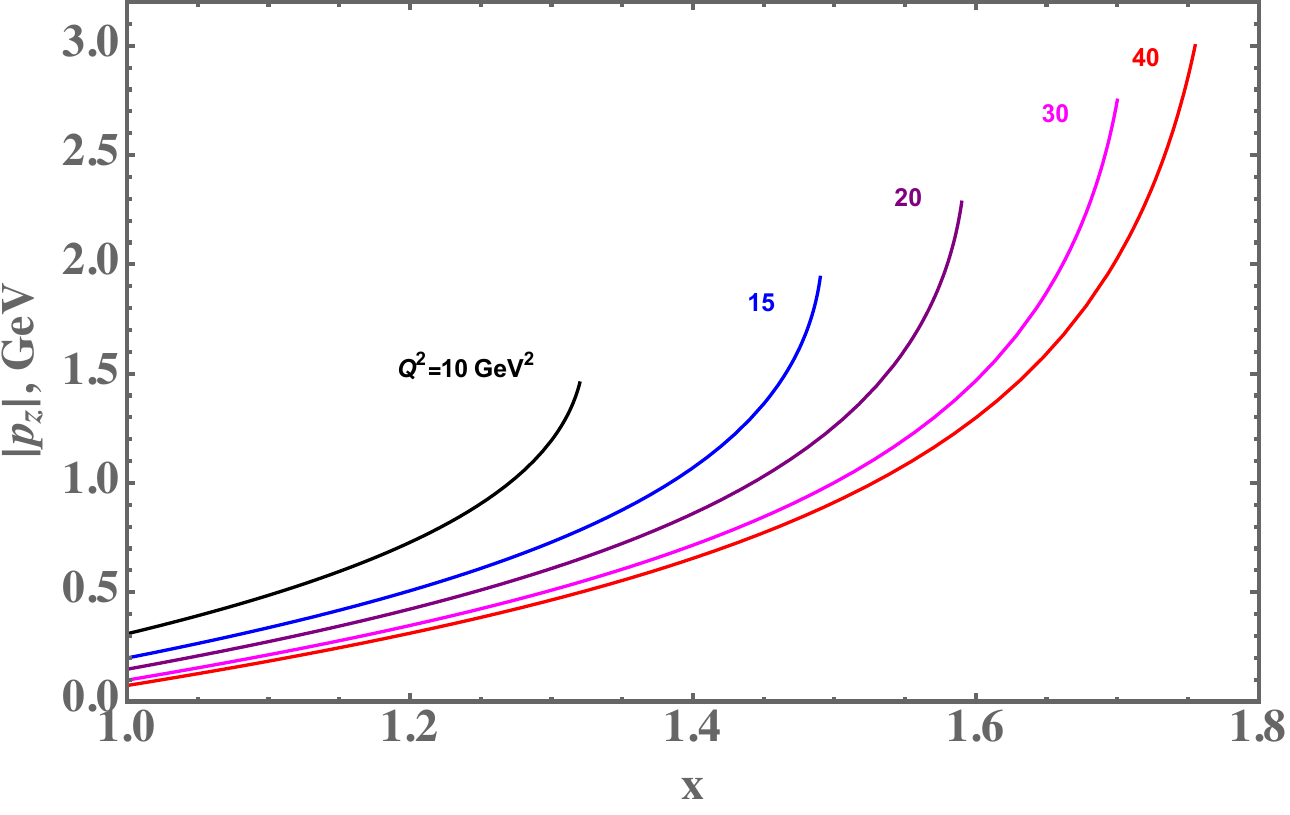}
         \includegraphics[clip=true, trim= 0.25cm 0.785cm 0.75cm 0,width=0.5\textwidth,height=0.355\textwidth]{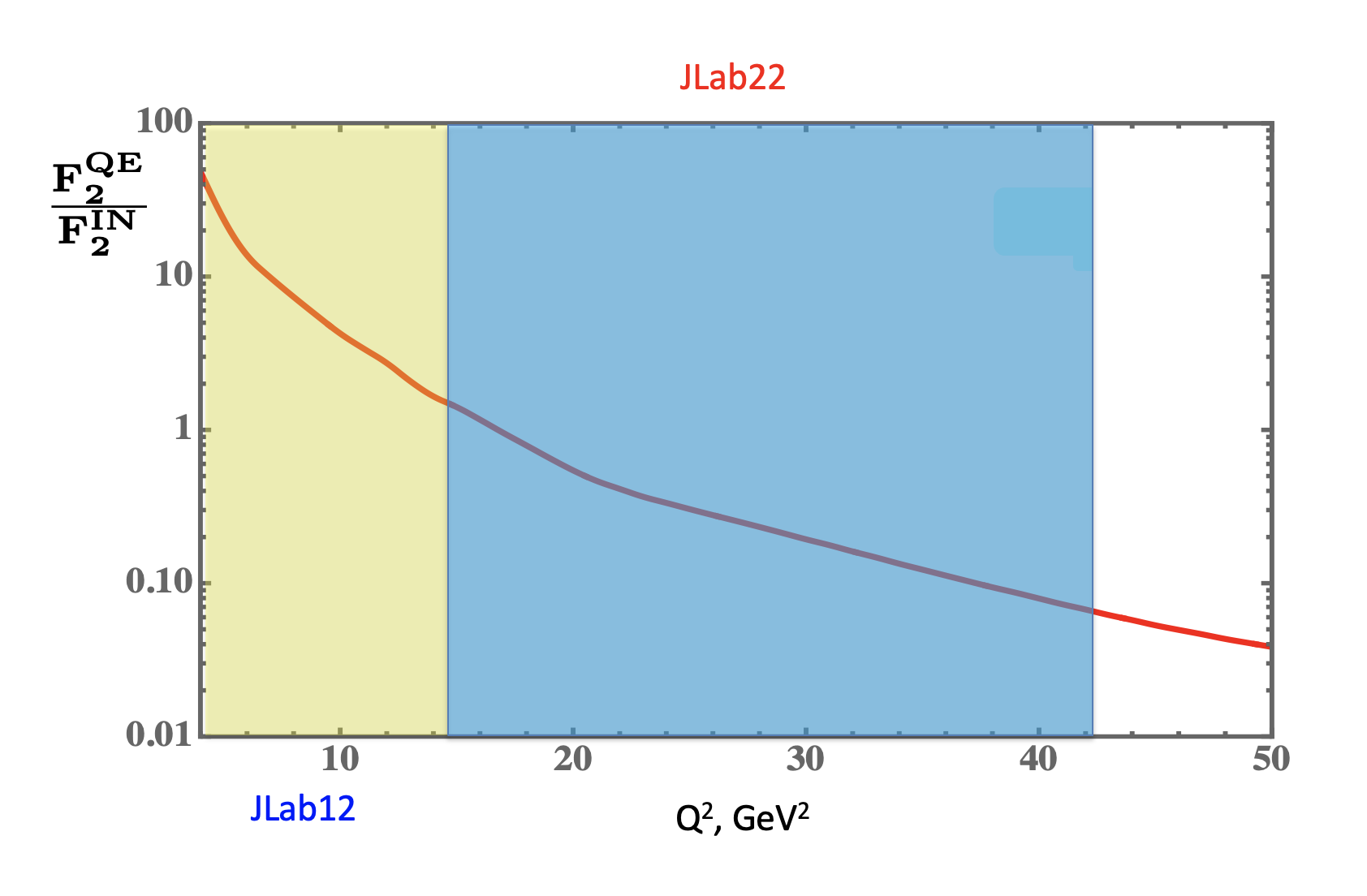}
            \vglue -0.1in
            \caption{Left: Absolute value of internal longitudinal momenta for DIS from bound nucleon with produced final mass of $W=2$~GeV at different $Q^2$. Right: Ratio of quasielastic to DIS contribution of the nuclear structure function $F_2$ for $x=1.5$ and different $Q^2$. No EMC effects are taken into account in this estimation.}
    \label{fig:DISkin}        
\end{figure}

\paragraph*{Probing Superfast Quarks in Nuclei.} It is known that isolating DIS processes in nuclei at Bjorken scaling variable $x>1$ is associated with probing a bound nucleon at very large internal momenta~\cite{Frankfurt:1988,Sargsian:2007gd,Freese:2015,Freese:2019}, which can originate from two or more nucleons being at very close proximity. For the case of deuteron target, it corresponds to measuring preexisting state with baryonic number two that has very large internal momenta, which could be related to very small separations. In Fig.~\ref{fig:DISkin}~left, the kinematics of nuclear DIS is considered responsible for producing final state mass $W=2$~GeV from the bound nucleon in the deuteron.
 
As depicted in Fig.~\ref{fig:DISkin} left, the combination of high $Q^2$ and $x>1$ will allow reaching internal momenta never before measured.~DIS processes in this case will proceed from scattering off nuclear quarks that carry more momentum fraction than quarks from isolated stationary nucleons - these quarks are refereed to as ``superfast quarks"~\cite{Frankfurt:1988}. As it will be shown in Subsubsec.~\ref{sub:fastq}, the generation of superfast quarks in nuclear medium is very sensitive to the dynamical origin of the nuclear core. The experimental exploration of superfast quarks started already at 6~GeV energies~\cite{Fomin:2010} at JLab and currently is underway with 12~GeV beam energies~\cite{E12-06-105}. At current energies, however, inclusive cross section at $x>1$ is dominated by quasielastic (QE) scattering which has to be taken into account to evaluate the pure DIS contribution. The 22~GeV JLab energy upgrade will allow a significant increase of the momentum transfer squared $Q^2$ at $x>1$ in which case it will allow to suppress the QE contribution (see Fig.~\ref{fig:DISkin} right) providing a direct access to DIS processes in the nucleus with $x>1$.

\paragraph*{Deuteron Structure at Sub-Fermi Distances.} Another rather direct way of accessing $NN$ interactions at core distances is to probe deuteron electrodisintegration at $Q^2\ge 4$~GeV$^2$ and at very large missing momenta, approximately above 800~MeV, for wide angular range of recoil nucleon production. In this case, the lower limit of $Q^2$ is defined from the condition that one can clearly distinguish between struck and recoil nucleons originating from the deuteron. For these reactions, the existence of angular anisotropy in light-front momentum distribution will indicate the existence of non-nucleonic component in the core of the $NN$ interaction. The first attempt of measuring such a large internal momenta has already resulted in an unexpected momentum distribution~\cite{Yero:2020}. The increase of scattered electron energy will allow performing systematic studies of these processes with realistic counting rates. The further extension of this program will include the SIDIS processes tagged by baryonic resonances. The upgraded energy will significantly increase the phase space of backward production kinematics allowing access to the preexisting non-nucleonic states in the deuteron that produce baryonic resonances in backward kinematics.

\paragraph*{Probing Three-Nucleon Short-Range Correlations and 3N Forces.} One of the groundbreaking results in nuclear physics with the upgraded 22~GeV energy will be the direct proof of the existence of short-range three-nucleon (3N) SRCs in the ground state of the nuclear wave function. One of such results could be the observation of another scaling (similar to 2N SRCs~\cite{Frankfurt:1993sp,Egiyan_2003,Egiyan_2006,Fomin_2012}) in the ratios of inclusive cross sections for $A>$~3 and $A=$~3 nuclei in the domain of 3N SRCs. As discussed in Subsubsec.~\ref{sub:3NSRCs}, the analysis of existing inclusive data indicates a tantalizing signature for such a scaling~\cite{Sargsian:2019,Day:2023}.

Finally, the program of exploring nuclear forces at extreme kinematics includes the systematic study of 3N forces, irreducible to the sequence of $NN$ interactions. Three-nucleon forces are essential in the dynamics of high-density nuclear matter, which have predicted the existence of supermassive neutron stars (exceeding two solar masses) that was observed in recent years. However, the dynamical origin of these forces are poorly known. The energy upgrade will allow a systematic study of 3N forces by including both electrodisintegration of $A= 3$ nuclei and probing 3N~SRCs in high $Q^2$ reactions off nuclei.

\subsubsection{Hadron-Quark Transition}
Nuclear medium represents a unique environment to explore the QCD dynamics of hadron-quark transition. Several phenomena are related to this transition including the ones associated with medium modifications of quark-gluon substructure of bound nucleons. Such a modification was discovered rather accidentally for the valence quark structure of bound nucleons by the European Muon Collaboration while studying inclusive DIS from nuclei. It was observed that the valence parton distribution functions (PDFs) of bound nucleons are depleted in the region of $0.3 < x < 0.6$ beyond the level expected from the Fermi motion of bound nucleons in nuclei. Follow up experiments have demonstrated an opposite phenomenon in which the enhancement was observed for bound nucleon PDFs in the region of $x\sim 0.1$, the so-called antishadowing region, on the scale of 2\% for inclusive DIS. The latter has indicated possible different dynamics for medium modifications of valence and sea quarks, including gluons, distributions. 

The energy upgrade will allow systematic studies of medium modification effects both in the region of PDFs suppression as well as antishadowing region. Such a program can be realized by mapping significantly larger $Q^2$ and wider $x$ kinematic ranges as well as considering semi-inclusive tagged DIS processes.

The second group of studies relevant to hadron-quark transitions is aimed at understanding the hadronization process by considering the production of hadrons in nuclear medium. For the quasielastic channel, such studies include experiments that probe possible color transparency (CT) phenomena while for inelastic kinematics, they study effects of confinement and dynamics of quark-hadron transition in the process of producing final hadronic states.

\paragraph*{Nuclear Medium Modifications (EMC Effect).} Basic models of nuclear physics describe the nucleus as a collection of unmodified nucleons moving non-relativistically under the influence of two- and three-nucleon forces, treated approximately as a mean field. In such a picture, considering very different scale of nuclear~({\it tens} of MeV) and baryon~({\it hundreds} of MeV) excitation energies, the partonic structure functions of bound and free nucleons should be identical. Therefore, it was generally expected that, except for nucleon motion effects, DIS experiments would give the same result for all nuclei.

Instead, the DIS measurements from nuclear targets~\cite{Aubert:1983} have observed in the valence quark region a reduction of the structure function of nucleons bound in heavier nuclei compared to deuterium, beyond what is expected from simple Fermi motion effects in models in which baryonic and momentum sum rules are satisfied (the effect generally referred to as the EMC effect). Since its initial discovery, a large experimental and theoretical efforts have been put into understanding its origin. The followup experimental advances include the observation of the local nuclear density dependence of the EMC effects~\cite{Seely:2009}, as well as the important role of SRCs in enhancing the strength of the EMC effect for $A\ge 4$ nuclei~\cite{Gomez:2011,Schmookler:2019}. Currently, there is no generally accepted theoretical interpretation of the observed EMC effects. 

In general, the question of understanding the EMC effect in the valence quark domain is coupled to understanding the very nature of confinement.~The parton model interpretation of the EMC effects is that the medium reduces the nuclear structure functions for large $x$ values so that there are fewer high-momentum quarks in a nucleus than in free space.~This momentum reduction leads, via the uncertainty principle, to the notion that quarks in nuclei are confined in a larger volume than that of a free nucleon. 

Overall, the medium modification in the valence quark region is expected to be proportional to the virtuality of bound nucleon.~For example, a recent work uses the idea of holographic duality~\cite{Brodsky:2014yha} to motivate functional forms of free and medium-modified quark distribution functions~\cite{Kim:2022lng}. This work finds that large values of the virtuality are needed to explain the nuclear DIS data. The nuclear presence of such large values can be be tested using the expanded tagging techniques that would be available at JLab22.

\begin{wrapfigure}[16]{r}{0.374\textwidth}
    \vspace{-0.55cm}
  \centering
 \includegraphics[width=0.374\textwidth]{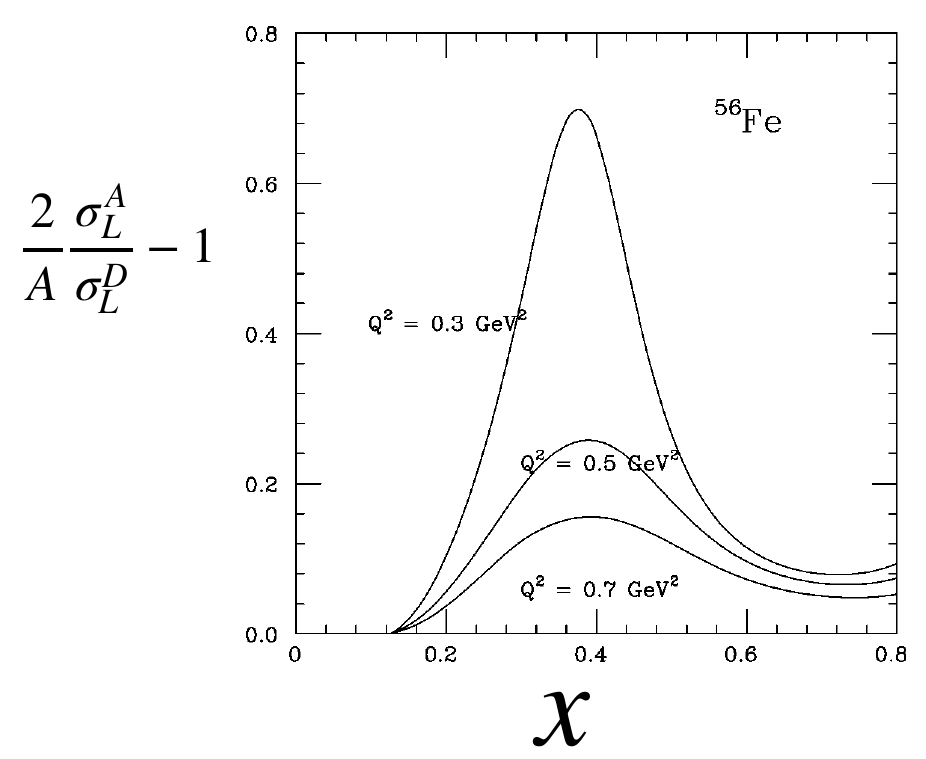}
  \hspace{0.1cm}
  \vglue -0.265in
  \caption{Enhancement of longitudinal cross sections as a function of $Q^2$ and $x=Q^2/2 M_N \nu$, where $M_N$ is the nucleon mass and $\nu$ is the virtual photon energy.}
\label{predict} 
 \end{wrapfigure}
The above discussion has been concerned with the valence regime of large $x$ $(\ge 0.3)$. At smaller $x$ values, the phenomena of shadowing and antishadowing are present. The comprehensive review of Ref.~\cite{Frankfurt:2011cs} shows that restricted amount of data is available for the antishadowing region and only simple expressions based on baryonic sum rules describe qualitatively the data. Widening the $Q^2$ coverage as well as extending it to SIDIS processes will set a new stage in comprehensive investigations of dynamical origin of antishadowing.

An important extension of medium modification studies are the exploration of the behavior of sea quarks and gluons in the nuclear medium. One possible method for studies of sea quark modifications that will allow to isolate pionic degrees of freedom is to use longitudinal/transverse (L/T) separation in DIS processes.~It was shown in Ref.~\cite{Miller:2001yf} that pionic effects with a sea small enough to be consistent with measured nuclear dimuon production data~\cite{{Alde:1990im}} could be large enough to predict substantial nuclear enhancement of the cross section for longitudinally polarized virtual photons for the kinematics accessible at JLab22. Predictions are shown in Fig.~\ref{predict}.

In regards to possible medium modifications of sea quarks, the analysis of the SeaQuest Collaboration~\cite{SeaQuest:2021zxb} data, which studied the Drell-Yan process off nuclei, suggests the presence of an EMC-like modification of nuclear PDFs~\cite{Alvioli:2022tij}. Also, the forward dijet production data in $pA$ collisions at the Large Hadron Collider are consistent with the models which assume gluon PDFs modifications analogous EMC effects~\cite{Kotko:2017oxg}. At 22~GeV beam energy, the one unique feature of high-intensity beam will be the possibility to study gluon modification at large $x$ ($\ge 0.5$). This can be achieved by considering open charm production channels such as $\gamma + N\to \Lambda_c +D+X$ at $W^2 > (M_\Lambda+M_d+ M_N)^2$ and $J/\psi$ production in $\gamma + N\to J/\psi +X$ reactions.

To this end, larger energies of JLab22 would allow to perform a detailed analysis of the EMC ratio for a broad range of $x$, including the ranges in which the EMC ratio is not a linear function of $x$. Additionally, it would extend the experience gained in current pioneering experiments exploring EMC effects in tagged DIS processes to the high energy domain and thus emphasize their important signature for such an investigation.

\paragraph*{Color Transparency Phenomena.}  CT is a fundamental prediction of QCD stating that one can observe reduced initial or/and final state interactions~(FSIs) in coherent production of hadrons in the nuclear medium at high-momentum transfer~\cite{Frankfurt:1994hf}. The basic idea follows from just two points: 1) high-momentum transfer reactions may make point-like color singlet states known as point-like configurations (PLC)\footnote{In the literature, they are also called small size configurations (SSC), and both terms are used interchangeably.}; 2) small color neutral objects have small cross sections for strong interactions. CT experiments need to have well controlled kinematics, such as the QE knockout of protons from nuclei.

CT effects have been observed~\cite{E791:2000kym} in the 500~GeV reaction of $\pi+A\to A+ jj$, where the notation $jj$ refers to two jets at high relative momenta. Tantalizing indications of the onset of CT have been reported at JLab energies in the $A(e,e')\pi^+$~\cite{Clasie:2007aa} and $A(e,e')\rho$~\cite{CLAS:2012tlh,Elfassi:2022} reactions. The putative signature is a rise of the nuclear transparency (defined as a ratio of a measured cross section to a cross section expected in the absence of final state interactions) with increasing values of $Q^2$ that is proportional to $p$. However, the present range is not quite large enough to provide an utterly convincing evidence for CT effects. Higher energy measurements would add a great value.

The lingering concern for the baryonic sector is that the observation of CT effects is as elusive as ever. The results of the recent JLab experiment~\cite{HallC:2020ijh} probing CT effects in the QE $^{12}$C$(e,e'p)$ reaction up to $Q^2$ of 14~GeV$^2$ have claimed no reduction of FSIs which essentially suggests a flat dependence of nuclear transparency with increasing $Q^2$. There are two possible explanations of this observation: i) it is very difficult to form PLC in color single three-quark systems that require significantly large $Q^2$, and/or ii) the expansion of the three-quark PLC is so fast and thus it quickly hadronizes before escaping the nucleus. 

The above two scenarios can be cross-checked and verified at JLab22 for CT studies with baryon production by either increasing the $Q^2$ and thus slowing down the QCD expansion~\cite{Caplow-Munro:2021xwi}, or employing a complementary logic in which case one measures only events that are products of a struck nucleon rescattering off the spectator nucleon in the nucleus on the so-called double-scattering processes~\cite{Egiian:1994ey,Frankfurt:1994kt,Frankfurt:1994kk}. The uniqueness in searching for CT effects in these processes is that one can use lightest nuclei such as deuteron and reach considerable sensitivity to possible modifications of the cross section of PLC interactions with the spectator nucleon and thus keeping expansion effects essentially under control (see Subsubsec.~\ref{sub:CTstudies}).

\paragraph*{Hadronization in Nuclei.}   The confinement of quarks inside hadrons is conceivably one of the most remarkable feature of QCD. The quest to quantitatively understand the confinement dynamics in terms of experimentally measured quantities is an essential goal of modern nuclear physics.~Much experimental attention has been focused on understanding confinement through hadron spectroscopy. Alternatively, the subject is often introduced through the string-breaking mechanism.~This picture is confirmed by lattice calculations using static quarks depicting the gluon field concentrated in a flux-tube (or string)~\cite{GrossWilczek73, Dokshitzer:2003bt}, which extends over a space-time region.~The string has a ``tension" $\kappa$ of a magnitude in the order of 1~GeV/fm that is predicted by the Lund string model to be the rate of the propagating quark's energy loss in the analyzing nuclei~\cite{Andersson:1979ue, Andersson:1983ia}. 

Due to the great success of DIS studies in probing the internal structure of the nucleon since early 1970s at SLAC~\cite{Osborne:1978ai}, DIS off nuclei has been considered the pioneering process in investigating quark propagation, hadron formation, and medium modifications of observable characterizing these transitions~\cite{Accardi:2009qv,Osborne:1978ai, EuropeanMuon:1991jmx, EuropeanMuon:1984inj}. The description of the hadronization process is denoted then by two space-time scales categorizing its two stages. In the first stage following the virtual photon hard scattering, the struck quark propagates in the target nucleus, during the production time~($\tau_{p}$), and undergoes medium-stimulated gluon bremsstrahlung prior to becoming a color-neutral object, known as prehadron. The latter evolves in the second stage into a fully dressed hadron with its own gluonic field within the formation time~($\tau_f$). The hadronization studies are thus performed to provide information on the dynamics scales of the process, and to constrain the existing models with various predictions of its time characteristics either in vacuum or in nuclei~\cite{Andersson:1979ue, Artru:1974hr, Shuryak:1978ij, Wang:2003aw, Kopeliovich:2003py,BROOKS2021136171}.

The fundamental focus of the broad JLab program of 6~GeV and 12~GeV era, which is extended here to 22~GeV (see Subsubsec.~\ref{sub:SIDIS}), is to determine the mechanisms of confinement in forming hadrons. The essential experimental technique that enables these studies is to employ nuclei as space analyzers of hadronization processes. In this approach, hadrons are formed from energetic quarks over distance scales ranging from 0-10~fm, which are perfectly matching the dimensions of atomic nuclei. For example, a recent simple geometrical model has found a strong dependence behavior on observed hadron energy fraction, $z$, for the partonic phase, ranging from 2~fm at high $z$ to 8~fm at smaller $z$ for HERMES data~\cite{BROOKS2021136171}, in quantitative agreement with existing predictions of the Lund string model~\cite{Andersson:1979ue,Andersson:1983ia}. 

\subsubsection{Summary of Flagship Experiments at 22 GeV}

The list of flagship experiments are chosen based on the current progress of nuclear physics and nuclear QCD studies at JLab and elsewhere, emphasizing the kinematic reach that 22~GeV energy will achieve. The uniqueness of the suggested upgrade is that, in addition to the increased energy, the high intensity of the beam will allow to perform measurements of increasingly small cross sections with unprecedented accuracy. Additionally, many impeding effects that currently need to be accounted or subtracted theoretically become corrections and thus grant access to unique phenomena relevant to nuclear structure at small distances and hadronization effects.

For nuclear dynamics at extreme conditions, the experiments highlighting the study of superfast quarks in nuclei, repulsive core in the deuteron, and 3N~SRCs in nuclei are discussed, respectively, in Subsubsecs.~\ref{sub:fastq}, \ref{sub:edepn}, and \ref{sub:3NSRCs}. While the highlighted experiments for hadron-quark transition in nuclear medium are i) probing bound nucleon and partonic structures via tagged processes (see Subsubsecs.~\ref{sub:BNS-TDIS} and \ref{sub:PDFtagging}); ii) investigating unpolarized and polarized EMC effects as well as antishadowing and shadowing regions (see Subsubsecs.~\ref{sub:EMC} and \ref{sub:spinEMC}); iii) CT studies (see Subsubsec.~\ref{sub:CTstudies}); iv) hadronization studies in nuclei (see Subsubsec.~\ref{sub:SIDIS}), and v) coherent nuclear $J/\Psi$ photoproduction (see Subsubsec.~\ref{sub:JPsi}).

All aforementioned experiments have established research groups who will perform similar measurements at 12~GeV. The progress achieved in conducting these experiments at 12~GeV will be significant for further refining and extending the scope of the measurements that can be accessed only with upgraded 22~GeV beam energy.

\subsection{Nuclear Dynamics at Extreme Conditions}\label{sec:Core-Extreme}

\subsubsection{Superfast Quarks}\label{sub:fastq}

The origin of the nuclear repulsive core is one of biggest unknowns in nuclear physics.~The phenomenology of $NN$ interaction indicates that the repulsive core dominates at internucleon distances $\ltorder 0.5$~fm. These are also the distances where one expects the onset of quark-gluon degrees of freedom in hadronic interaction. Therefore, it is very likely that the solution of the nuclear core problem lies in understanding QCD dynamics of the nuclear forces at short distances. QCD introduces additional intrigue in grasping the repulsive core of $NN$ interactions. For example, when considering the six-quark color-singlet cluster as the limiting case of $NN$ system, one expects as much as 80\% of the components of the $NN$ wave function to consist of hidden color states such as two-color octet ``nucleons"~\cite{Brodsky:1986}. In such a picture, hidden colors may play a significant role in the dynamics of the core. 

One way of exploring the dynamics of the $NN$ core is the consideration of exclusive $NN$ scattering at large momentum transfer $-t$. The current observations have indicated a qualitative change of the dynamics of hadronic interaction once relevant distances ($\sim \frac{1}{\sqrt{-t}}$) become comparable to the range of the $NN$ core~\cite{Brodsky:1987xw}, but the complexity of theoretical interpretations of hadronic processes limited the ability to interpret these results in terms of hidden color or other quark-level structure at short distances.

Currently, the most promising direction in exploring the physics of the core is high energy electro-nuclear processes in which the virtual photon scatters from highly correlated bound nucleonic systems at small space-time separations. That such correlations can be isolated and investigated in high energy nuclear processes was one of the achievements of the experimental program of the investigation of high energy electro-nuclear processes~\cite{Sargsian:2003}. The scaling observed in the ratios of inclusive $e-A$ to $e-^2$H cross sections in QE kinematics at Bjorken $x>1$~\cite{Frankfurt:1993es,Fomin_2012} has indicated the possibility of isolating 2N~SRCs~\cite{Sargsian:2003, ARRINGTON:2012}. Subsequent experiments comparing the strengths of the proton-proton and proton-neutron SRCs~\cite{Subedi_2008} have demonstrated the tensor nature of 2N~SRCs and provided input on the momentum structure of SRCs~\cite{Arrington:2022sov}.

\begin{figure}[b!]
    \centering
        \includegraphics[width=0.49\textwidth,height=0.44\textwidth]{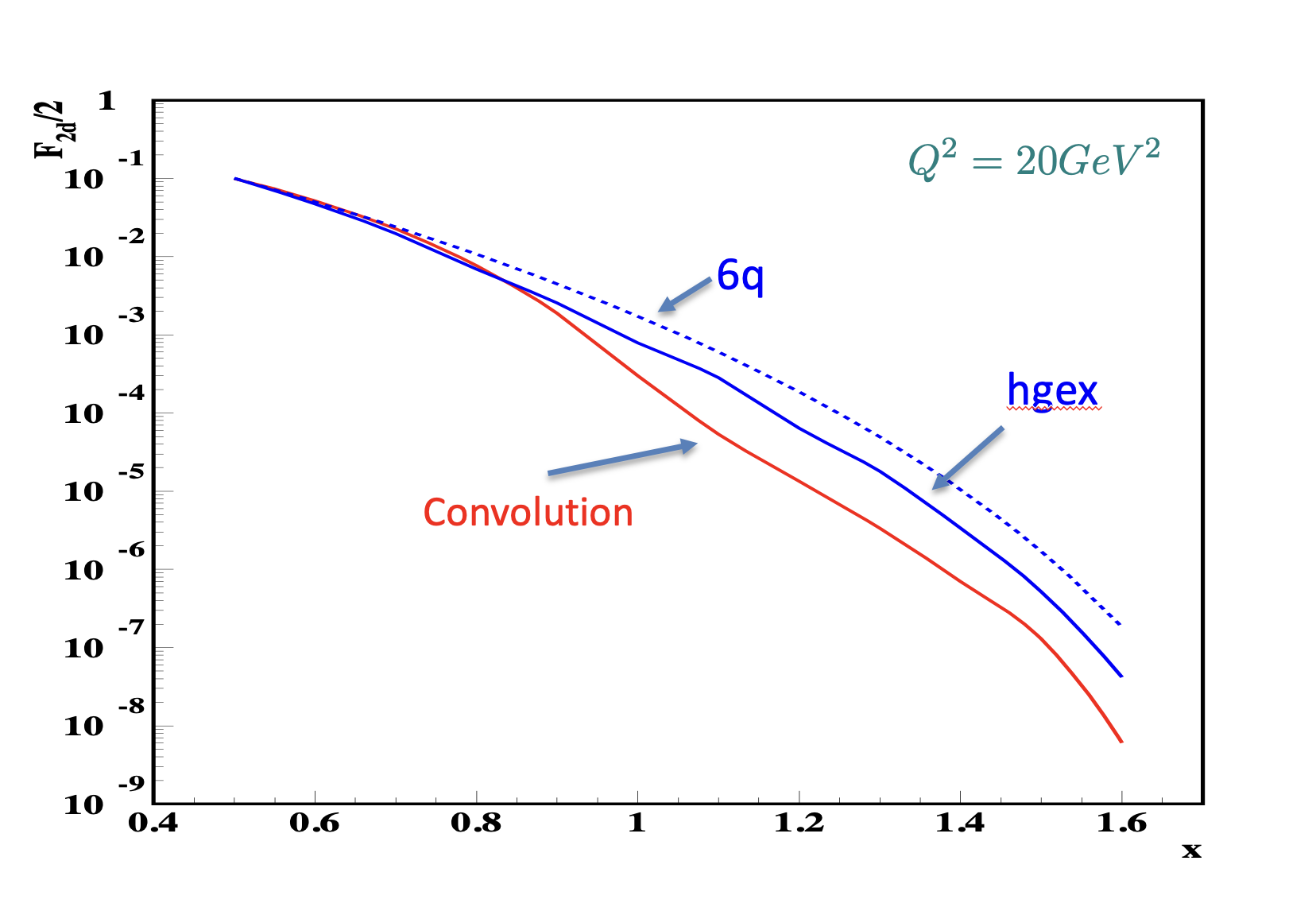}
    \includegraphics[width=0.49\textwidth,height=0.4\textwidth]{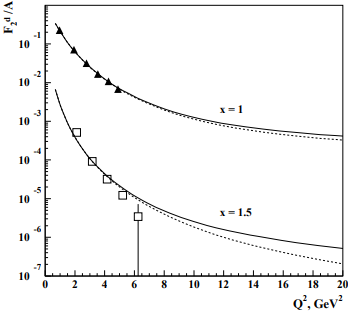}
    \vglue -0.15in
    \caption{Left: Deuteron valence quark distribution based on a simple convolution model (dashed red line~\cite{Arrington:2003qt}), compared to a deuteron with a 5\% component based on the 6$q$-bag model of Ref.~\cite{Mulders:1983au}. In the EMC effect region, the impact is extremely small, while the six-quark bag contribution enhances the PDF for $x>1$, dominating the PDFs above $x=1.1$. Right: Deuteron structure function for unmodified deuteron and modified deuteron based on the color screening model~\cite{Sargsian:2003}. In this case, the deuteron PDF is suppressed at large $x$ and $Q^2$, rather than enhanced. The left figure is adapted from Ref.~\cite{Arrington:2003qt}, and the right figure is reproduced from Ref.~\cite{Sargsian:2003}.}
    \label{fig:qofx}
\end{figure}
While these measurements focused on QE scattering and did not probe the partonic structure of these short-distance nucleons, extension of inclusive studies to high $Q^2$ offer the possibility to combine the kinematic isolation of short-range structures in nuclei at $x>1$ with the extraction of parton distributions via DIS processes. The extraction of the distribution of these superfast quarks, which carry more longitudinal momentum than is possible in a single, stationary nucleon, is sensitive to the partonic structure of the SRCs that dominate scattering in this regime. In a simple convolution model, the PDFs at $x>1$ arise from the convolution of the nucleon PDFs and the nuclear momentum distributions, with the superfast quarks coming from this highest-$x$ quarks in the highest-momentum nucleons. This leads to a distribution which falls off rapidly with $x$. However, modifications of the SRC internal structure can significantly change this picture. In pictures where the overlap of the nucleons in the SRC allows for direct quark exchange between the nucleons, there can be a dramatic enhancement in the distribution of these superfast quarks, as illustrated by two models shown in Fig.~\ref{fig:qofx} left. The models that are based only on nucleonic degrees of freedom predict larger suppression of the distribution of superfast quarks~\cite{Sargsian:2003,Hen:2016}, as illustrated in Fig.~\ref{fig:qofx} right.

There are two major challenges to making DIS measurements at $x>1$. First, reaching the DIS regime for $x>1$ requires extremely high $Q^2$ values, and it is not clear exactly what $Q^2$ is required to cleanly isolate the parton distributions. On top of this, the inclusive cross section is very low due to the simultaneous requirement of reaching very large $x$ values and extremely high $Q^2$. Because the typical cuts used to define DIS are not appropriate for the $x>1$ region, the goal is to aim for $Q^2$ values where the underlying $e-N$ scattering is dominated by DIS. Momenta of bound nucleon that can kinematically generate a quark with $x \ge 1$ in deep-inelastic scattering can be evaluated from the relation:
\begin{equation}
(q + p_{N})= W_N^2,
\label{pNz}
\end{equation}
where $q$ and $p_N$ are four-momenta of virtual photon and bound nucleon and $W_N$ is the final mass produced on the bound nucleon. For the case of the deuteron target $p_N = p_d-p_s$, where $p_d$ and $p_s$ are on-shell momenta of 
deuteron target and spectator nucleon, respectively. As demonstrated in Fig.~\ref{fig:DISkin}~left, DIS processes considering $W_N=2$~GeV results in very large initial momenta for bound nucleon at $x>1$. 
 
In taking the convolution of the $e-N$ cross section with the distribution of bound nucleons, the fraction of scattering for which $W^2_N$, the invariant mass of the $e-N$ system, is in the DIS region and taken to be $W^2_N > 4$~GeV$^2$ for large $Q^2$, can be determined.

\begin{figure}[b!]
    \centering
    \includegraphics[clip=true,trim=0.75cm 0.5cm 0.9cm 1.9cm, angle=90,width=0.69\textwidth]{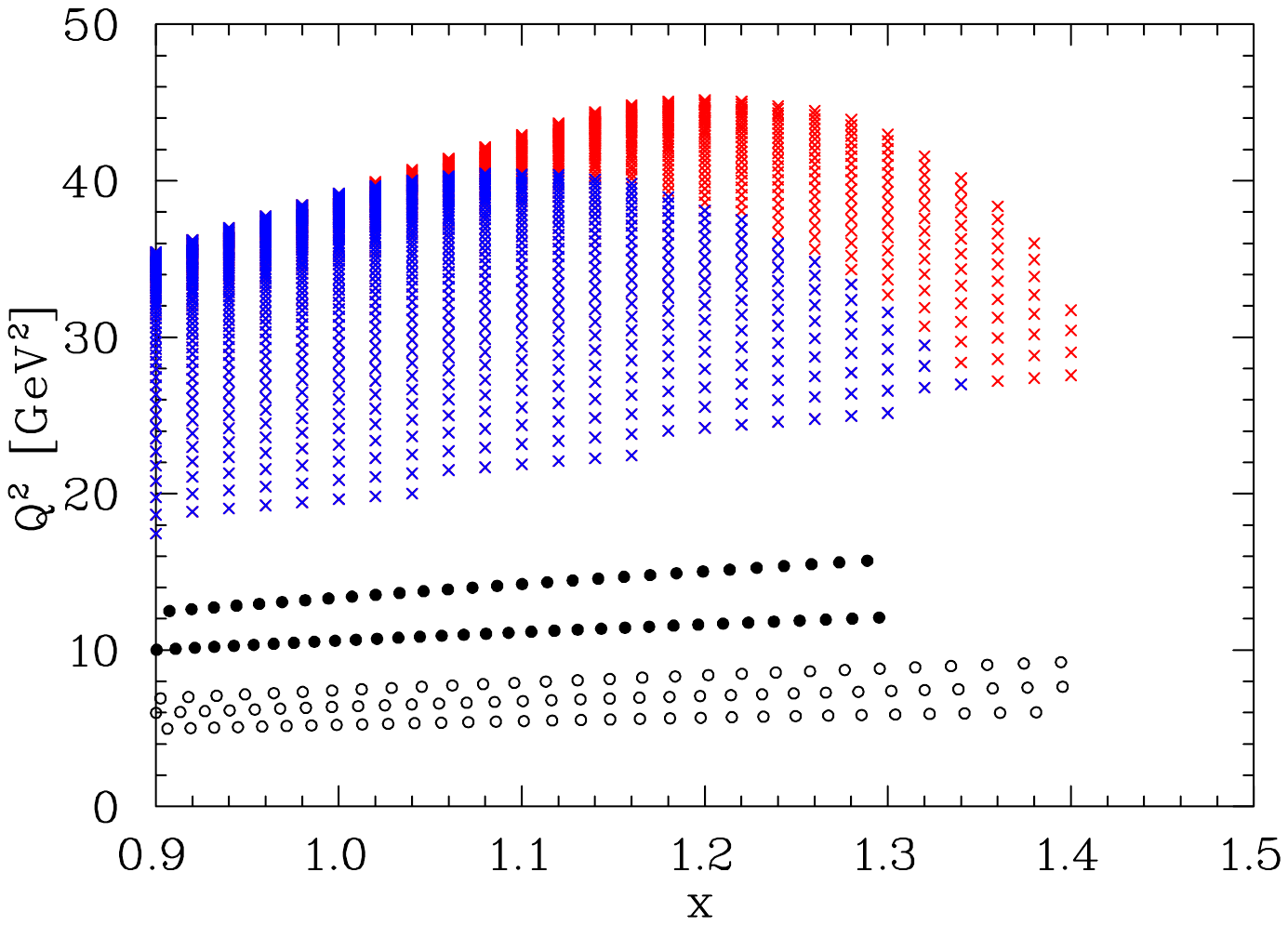}
    \vglue -0.25in
    \caption{Kinematic domain accessible for 22~GeV electron beam. Colorful points are the
    22~GeV projections while the open (solid) black circles are the 6 (12)~GeV measurements.}
    \label{fig:sfq_kine}
\end{figure}
Calculations suggest that the cross section is DIS dominated over a range in $x$ ($x>1$) for scattering with 12~GeV beam energies~\cite{Sargsian:2003}, but with significant contributions from the resonance region as well. Duality of the proton and neutron PDFs~\cite{Niculescu:2000tk,Niculescu:2015wka} implies that the average cross section in the resonance region closely follows that predicated from the DIS structure function, with resonances yielding additional structure in $W^2$ (and thus $x$) on top of this. In nuclei, the Fermi smearing leads to this structure being washed out, yielding scaling behavior even at modest $Q^2$ over most of the resonance region~\cite{Arrington:2003nt}. As such, the resonance contributions are expected to yield modest deviations from the DIS expectation. Data taken at 6~GeV already demonstrate that the $x>1$ structure functions are consistent with expected scaling behavior and can be well reproduced using a QCD scaling inspired fit~\cite{Fomin:2010}, even where the cross section is dominated by resonance contributions. While data recently taken at 12~GeV~\cite{E12-06-105} at $x>1$ and high $Q^2$ will not be free of resonance region contributions, the greatly enhance DIS contribution will allow for a much more precise study of scaling in this region, allowing us to make conclusions about the underlying superfast quark distribution, with modest uncertainties associated with non-DIS contributions. Given the size of the effects predicted by some models, this could be sufficient to have a first indication, if not quantitative measurement, of what sort of PDF modification occurs in SRC-dominated scattering. However, the quantitative conclusions will be limited by these non-DIS contributions, and we will have only the $x$ dependence and a limited $A$ dependence to differentiate between different effects.

Figure~\ref{fig:sfq_kine} shows the kinematics of existing measurements at JLab6 (open) and JLab12 (solid) black circles, along with projections for 22~GeV. The red (blue) points indicate the $x$-$Q^2$ region that yields $>$ 1~count/hr ($>$ 10~counts/hr) for a 50~$\mu$A beam on a $\sim$~2\% carbon target. This will provide high-statistics tests of scaling for $x$ values up to $x$=~1.1-1.2, allowing the quantification of non-DIS contributions. In addition, it will allow to map out both $Q^2$ and $A$ dependencies up to $x=1.3$ for $Q^2~\ge~30$~GeV$^2$, while allowing for measurements $Q^2>40$~GeV$^2$ and pushing to $x>$~1.4 for a limited subset of targets. 

\subsubsection{Probing Deuteron Repulsive Core}\label{sub:edepn}

\begin{figure}[!b]
\centering
\includegraphics[scale=0.465]{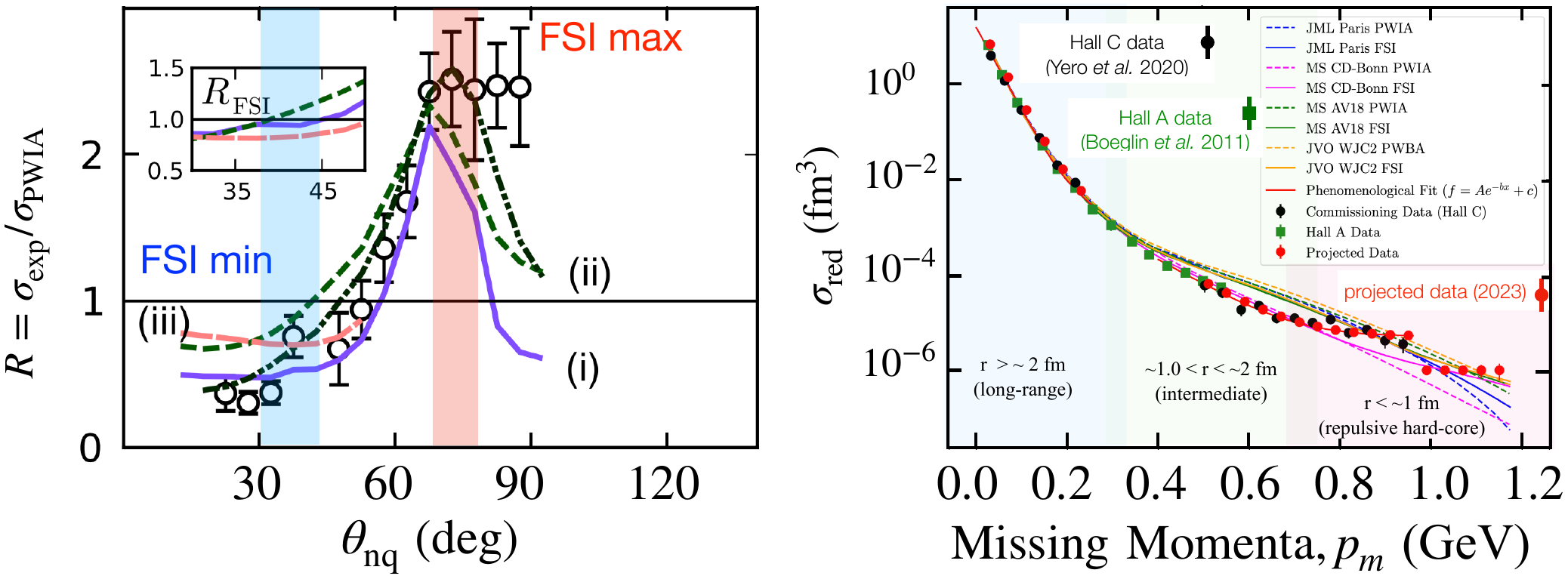}
\vglue -0.1 cm
\caption{Left: Angular distribution ratio $R(\theta_{\mathrm{nq}})= \sigma_{\mathrm{exp}} / \sigma_{PWIA}$ for $p_{m}=0.5$~GeV~\cite{Boeglin_2011}, where PWIA stands for the plane wave impulse approximation. Right: Reduced cross sections for the neutron recoil angle $\theta_{\mathrm{nq}}=35\pm5^{\circ}$~\cite{Yero:2020}.} 
\label{fig:deep_exp}
\end{figure}

Exclusive quasielastic electrodisintegration of the deuteron~$d(e,e^\prime N_f)N_r$
at high $Q^2$, in which $N_r$ can be identified as a recoil nucleon, is expected to provide most direct access to the short-range nuclear structure of the deuteron at very large missing momenta~\cite{Boeglin:2015}. The $d(e,e^\prime p)n$ measurements up to missing momenta of $\sim 550$~MeV and $Q^2=3.5$~GeV$^2$, which were carried out in Hall~A~\cite{Boeglin_2011}, have verified that FSIs are highly anisotropic with respect to the neutron recoil angle, $\theta_{nq}$, as theoretically predicted. The data were reproduced very well by the theoretical calculations of Refs.~\cite{Sargsian_2010, Laget_2005, Orden_2014}, where the kinematic window at $\theta_{nq}\sim$ 35$^{\circ}$ to 45$^{\circ}$ was found to have reduced FSIs and therefore providing an access to the ground state deuteron wave function at internal momenta up to~$\sim$~550~MeV, see Fig.~\ref{fig:deep_exp} left. 

The existence of a kinematic window of the ground sate wave function at large internal momenta was exploited further in recent Hall~C~\cite{Yero:2020} study, which has extended the previous measurement up to missing momenta of $950$~MeV while selecting kinematics where FSIs are reduced. The results are in good agreement with theoretical calculations of M.~Sargsian~\cite{Sargsian_2010} up to missing momenta of $\sim$~700~MeV, however, none of the  existing theoretical calculations were able to describe the data above these momenta, see Fig.~\ref{fig:deep_exp} right. These are very unexpected and surprising results indicating the possible onset of a new regime in the dynamics of the $p-n$ state~\cite{Vera_2021,Sargsian_2022}. 

Taking into account the fact that starting at missing momenta of $\sim 750$~MeV $\Delta\Delta$ excitation threshold is crossed in the $p-n$ system, one expects that reaching such large missing momenta will open up new venue in probing non-nucleonic components in the deuteron, including possible hidden-color states. At JLab22, it will be possible to carry out systematic studies of $d(e,e'N_f)N_r$ processes for up to unprecedented high values of recoil nucleon momenta ($\sim 1.5$~GeV). These measurements can be performed in Hall~C by measuring scattered electron and struck proton in coincidence~\cite{Yero2023}, or in Hall~B by measuring the recoil proton in coincidence with scattered electron. 

\subsubsection{Probing 3N SRCs in Nuclei}\label{sub:3NSRCs}

Three nucleon short-range correlations, in which three nucleons come close together, are unique arrangements in the strong interaction physics.~Unlike 2N~SRCs, 3N~SRCs have never been directly probed.\begin{wrapfigure}[14]{r}{0.45\textwidth}
    \vspace{-0.85cm}
  \centering
 \includegraphics[scale=0.45]{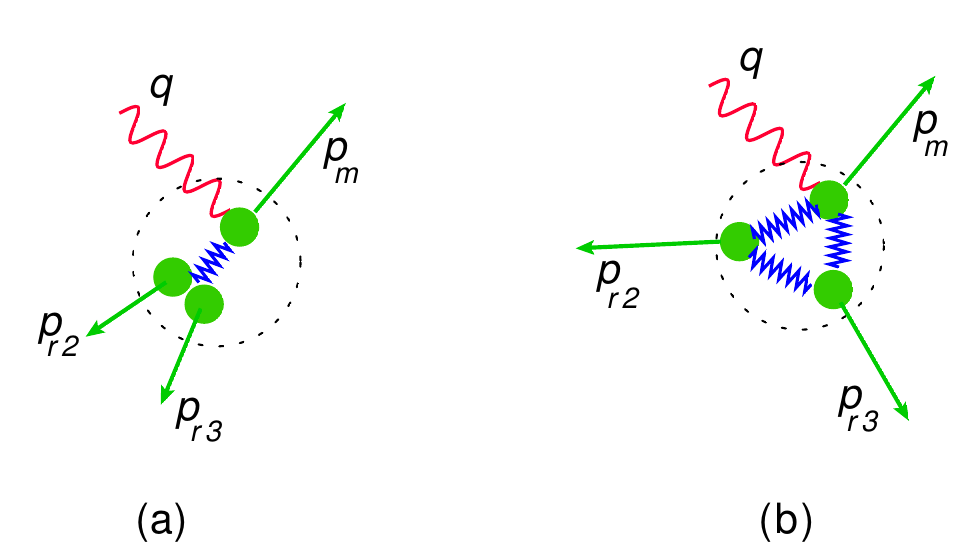}
  \vglue -0.05in
\caption{(a) Type-I 3N~SRCs in which the fast probed nucleon is balanced by two recoil nucleons.~(b) Type-II 3N~SRCs in which all tree nucleons have equal momenta with relative angles of $\sim 120^\circ$.} \label{fig:3Ngrams}
 \end{wrapfigure} It is expected that 3N-SRCs, which dominate the high momentum component of nuclear wave function at internal momenta of $\gtrsim 700$~MeV~\cite{Frankfurt:1981,FRANKFURT:2008}, being almost universal up to a scale factor~(see {\it e.g.}~Refs.~\cite{Frankfurt:1981,CiofidegliAtti:2015lcu}).

The dynamics of three-nucleon short-range configurations reside at the borderline of our knowledge of nuclear forces making their exploration a testing ground for ``beyond the standard nuclear physics" phenomena such as irreducible three-nucleon forces, inelastic transitions in 3N systems as well as the transition from hadronic to quark degrees of freedom. Their strength is expected to grow faster with the local nuclear density than the strength of 2N~SRCs~\cite{Frankfurt:1981,FRANKFURT:2008}. As a result, their contribution  will be essential for the understanding of the dynamics of super-dense nuclear matter (see {\it e.g.} Ref.~\cite{Heiselberg:2000dn}).

Theoretical studies of electrodisintegration of $A=3$ system~\cite{Sargsian:2005,Sargsian:2005ASF} indicate the dominance of two types of 3N~SRCs: type-I in which the high momentum of probed nucleon is balanced by two spectator nucleons each carrying approximately half of the probed nucleon momentum, see Fig.~\ref{fig:3Ngrams}~(a), and type-II in which all three nucleons have equal momenta with relative angles of $\sim 120^\circ$, see Fig.~\ref{fig:3Ngrams}~(b). These configurations dominate at different missing energy values of the reaction indicating different electroproduction processes that can probe 3N~SRCs.

\begin{figure}[!b]
\centering
  \vspace{-0.5cm}
\includegraphics[clip=true, trim= 0.5cm 0.85cm 0 0, width=0.48\textwidth]{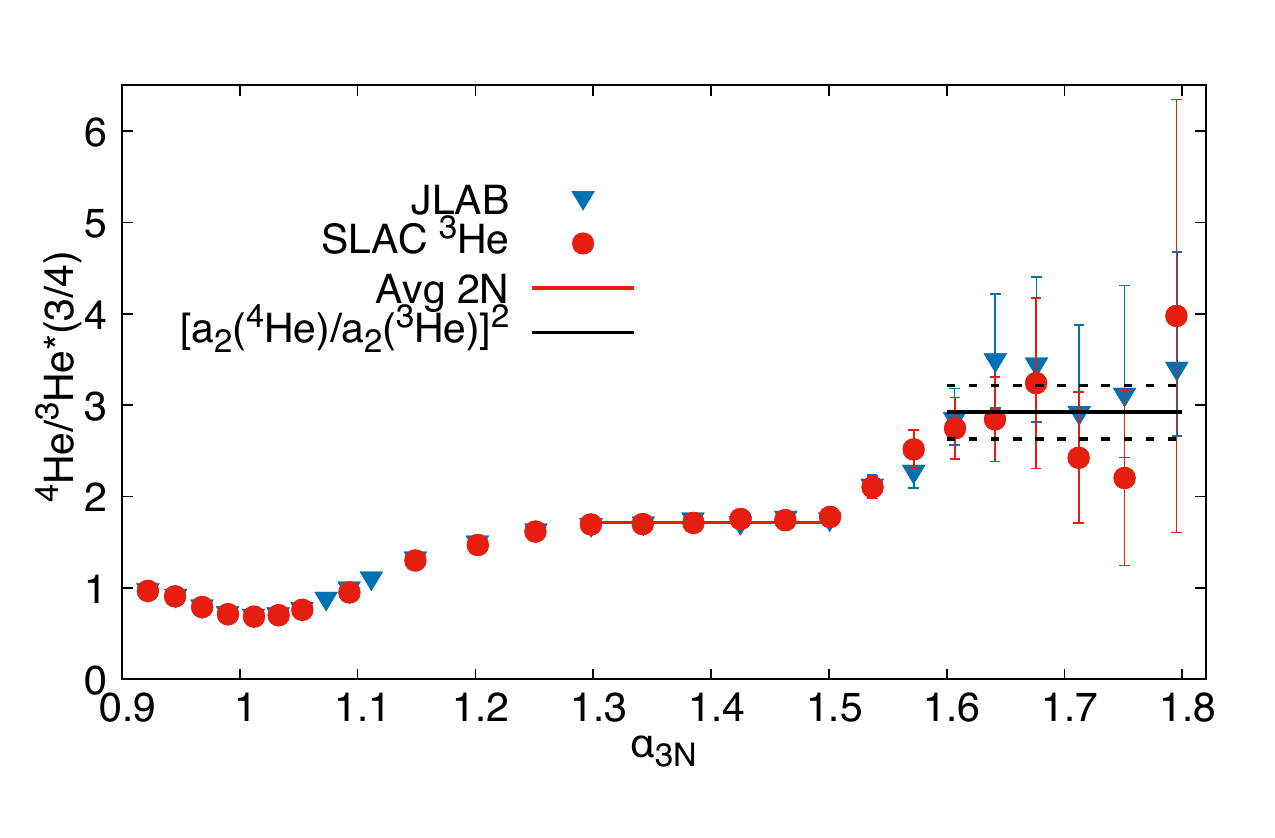}
\vspace{0.1cm}
\includegraphics[clip=true, trim= 0.5cm 0 0 0.0cm, height=4.7cm,width=0.48\textwidth]
{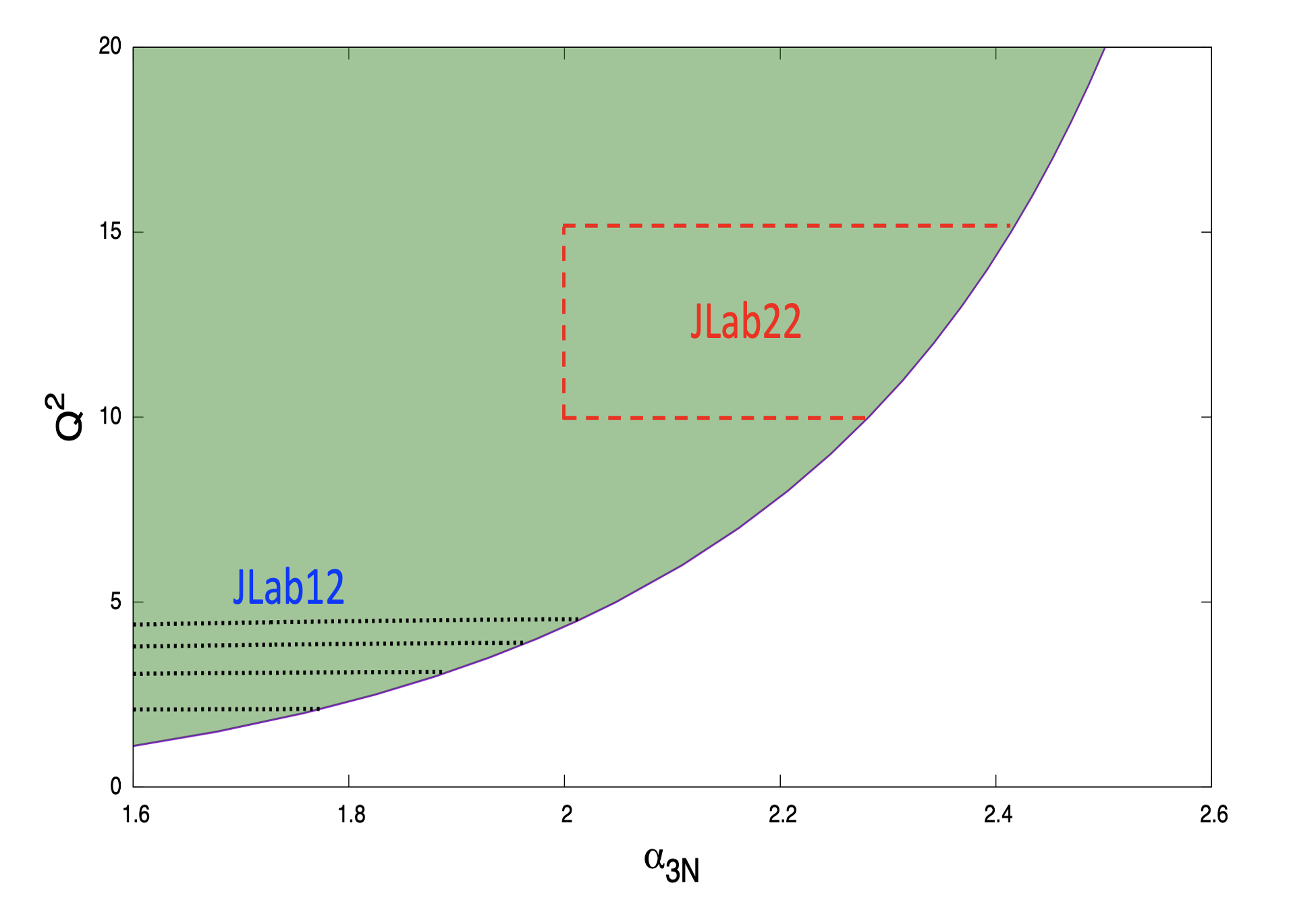}
\vglue -0.125in
\caption{Left: The $\alpha_{3N}$ dependence of inclusive cross section ratios for $^4$He to $^3$He.~The data are from JLab~\cite{Fomin_2012} and SLAC~\cite{Day:1979bx,Rock:1981aa} experiments. The horizontal line at $1.3\le \alpha_{3N}<1.5$ identifies the magnitude of the 2N~SRC plateau~\cite{Sargsian:2019,Day:2023}. Right: The $Q^2$ range necessary to isolate 3N~SRCs for JLab 22~GeV. Also shown are the ranges that will be accessed in the 12~GeV experiments.} 
\label{fig:tantalscaling}
\end{figure}

Recent studies~\cite{Sargsian:2019,Day:2023} have demonstrated that the type-I 3N-SRCs can be probed unambiguously in inclusive scattering at $\alpha_{3N}>2$, where $\alpha_{3N}$ is the light front momentum fraction of 3N~SRCs carried by the interacting nucleon. A spectacular signature of the onset of 3N~SRCs in this case will be the appearance of a new scaling in the ratio of inclusive cross sections of nuclei with $A>3$ to $A=3$ nucleus (similar to what was observed in JLab for 2N~SRCs in the region of $1.3 \le \alpha_{2N} \le 1.5$~\cite{ARRINGTON:2012,Fomin:2017}). Currently, there is no data for $\alpha_{3N}>2$.~Based on the phenomenological point-of-view that 2N~SRCs should be suppressed already at internal momenta above 700~MeV, it was predicted in Refs.~\cite{Sargsian:2019,Day:2023} that 3N~SRC scaling could be set at that values and thus observed at $\alpha_{3N}\ge 1.6$. Currently, there is a very restricted number of data satisfying this condition which demonstrates tantalizing signatures for the onset of nuclear scaling relevant for 3N~SRCs, see Fig.~\ref{fig:tantalscaling} left.
 An unambiguous verification of type-I 3N~SRCs require kinematic conditions that will cover sufficiently a wide range of $\alpha_{3N}>2$ for a great variety of nuclei. As depicted in Fig.~\ref{fig:tantalscaling} right, this will require reaching the range of $Q^2\ge 10-15$~GeV$^2$ which is accessible only at JLab22.

The main concerns for experimental studies of type-II 3N~SRCs is that they require measurements with large removal energies of $E_m\ge 300$~MeV. Such reactions can be probed in a new generation of high $Q^2$ exclusive disintegration of $A=3$ nuclei in which a large missing energy $E_m$ and at least one large recoil momentum of spectator nucleons can be measured simultaneously. As it is demonstrated in Refs.~\cite{Sargsian:2005,Sargsian:2005ASF}, these processes are very sensitive to the presence of irreducible 3N forces resulting in predictions of cross section differing by one order of magnitude.

\subsection{Hadron-Quark Transition in Nuclear Medium}\label{sec:HQ-NM}

\subsubsection{Bound Nucleon Structure from Tagged DIS}\label{sub:BNS-TDIS}
\label{BNSTDIS}

Nucleons in the nucleus are densely packed, strongly interacting, and composite objects. It would be surprising if nucleons in the nucleus did not distort or modify the structure of their neighboring nucleons. However, there is little evidence for this modification beyond the neutron lifetime and the $\sim 1\%$ binding energy. The EMC Effect is one of the few pieces of evidence for bound nucleon modifications~\cite{Geesaman:1995,Norton:2003cb,Malace:2014uea,Hen:2016}.

Inclusive measurements detect only the scattered lepton. In order to select DIS scattering from individual nucleons, we need to ``tag" them, by detecting the spectator nucleons. For example, if an electron scatters from a proton in $^4$He, we would need to detect the scattered electron and the recoil $^3$H. If the recoil\begin{wrapfigure}[26]{r}{0.4\textwidth}
\vspace{-0.95 cm}
  \begin{center}
    \includegraphics[width=0.4\textwidth]{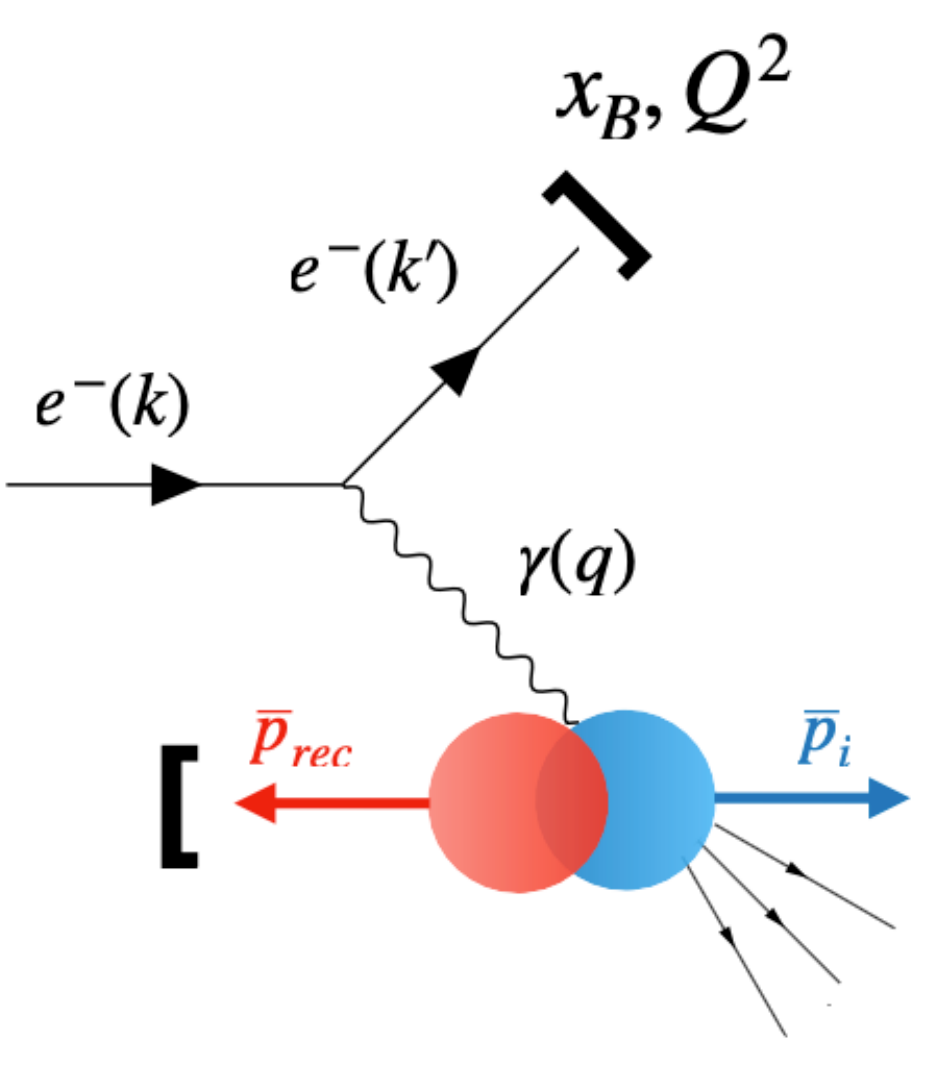}
 \end{center}
 \vglue -0.3in
  \caption[Tagged DIS kinematics]{The TDIS reaction. An electron with four-momentum $k$ exchanges a virtual photon with a nucleon in a deuteron with momentum $\vec p_i$.~The scattered electron has four-momentum $k'$.~The spectator backward nucleon is detected with momentum $\vec p_{rec}$, thus tagging the struck nucleon.~The four-momentum transfer is $Q^2= -(k-k')^2$ and the Bjorken scaling variable is $x_B$=~$x$= $Q^2/2M_N\nu$.}
 \label{TDIS1}
\end{wrapfigure}
$A-1$ nucleus is a spectator to the reaction ({\it i.e.}, if it does not rescatter), then it recoils with momentum equal to and opposite the initial momentum of the struck nucleon, $\vec p_{rec}\approx -\vec p_i$, see Fig.~\ref{TDIS1}. In fixed-target experiments, this is best done with very light nuclei, such as the deuteron, to enable the detection of the relatively low-momentum recoil nucleus.

These experiments need to boost their kinematics into the rest frame of the struck nucleon, using $x'$ and $W'$ rather than $x$ and $W$, the invariant mass of final-state hadrons. They minimize rescattering of the spectator nucleon by detecting back-angle spectators.

By measuring nucleon structure over a range of $p_{rec}$, TDIS measurements can distinguish between slight modification of all nucleons and significant modification of SRC nucleons. Slight modification of all nucleons implies a small $p_i$-independent modification. Large modification of SRC nucleons implies a strongly $p_i$-dependent modification, with large modification at large $p_i$.

Previous tagged DIS measurements at 6~GeV have either focused on almost-free nucleon structure by measuring low-momentum recoil protons from deuterium using the barely off shell nucleon structure (BONuS) detector plus CLAS~\cite{CLAS:2011qvj,Tkachenko:2014,Griffioen:2015hxa} or they have measured higher momentum recoils $300 \le p_{rec}\le 600$~MeV and suffered from a lack of statistics and kinematic range~\cite{CLAS:2005ekq}. Current and near-future 12~GeV measurements include low-momentum recoils measured with a low energy radial tracker (ALERT) or BONuS and higher-momentum recoils measured with backward angle neutron detector (BAND) or Large Angle Detector.  

However, the existence of a possible high-intensity 22~GeV electron beam at JLab, even with the existing CLAS12 and BAND detectors, could dramatically increase the statistical and kinematic reach of the experiments.
\begin{figure}[t!]
  \begin{center}
    \includegraphics[width=0.85\textwidth]{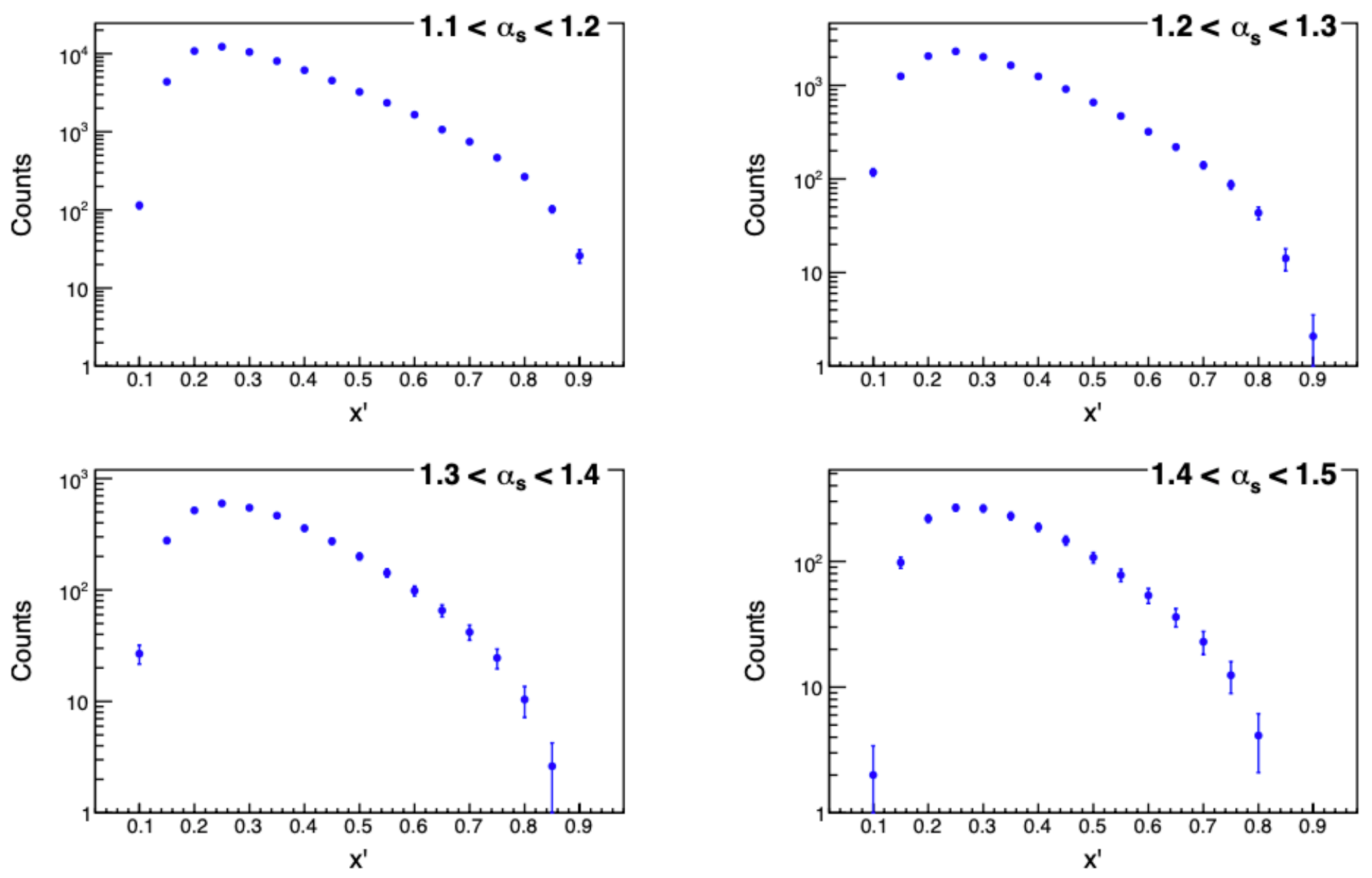}
 \end{center}
 \vglue -0.25 in
  \caption[Tagged DIS statistics]{Expected tagged DIS $d(e,e'n_s)$ statistics for 22~GeV, the CLAS12 and BAND detectors and 180~fb$^{-1}$ of luminosity for four bins in $\alpha_s$ (the light cone momentum fraction) as a function of $x'$.}
 \label{TDIS2}
\end{figure}
As can be seen in Fig.~\ref{TDIS2}, the expected statistical precision over a wide range of $\alpha_s$ and $x'$ is remarkable. This is a dramatic increase over the statistics and kinematic range of existing and planned 12~GeV experiments.

\subsubsection{Probing Partonic Structure with Spectator Tagging}\label{sub:PDFtagging}

The origin of the EMC effect continues to evade a clear explanation, which is often hampered by theoretical or experimental complications that frequently cloud the interpretation. In quasielastic electron scattering, quenching of the Coulomb sum rule has been difficult to interpret for different models~\cite{Lovato:2016gkq,Cloet:2015tha}, and experiments with a recoil polarimeter to measure polarization transfer or induced polarization observables need model calculations to describe the data, but provide little insight into the substructure of the nucleon~\cite{Rvachev_2005,Hu:2006fy,Malace:2010ft,Ford:2014lra}. Unlike the EMC effect, these quasielastic measurements lack a direct partonic interpretation. On the other hand, spectator-tagged deeply virtual Compton scattering (DVCS) provides a partonic interpretation, and through a fully exclusive measurement, yields a separate handle for studying final state interactions much like induced polarization measurements. 

Recent DVCS results on light nuclei from the CLAS Collaboration have sparked interest in the so-called generalized EMC effect~\cite{Dupre:2021}. Predictions for this off-forward beam spin asymmetry ratio of the $\sin\phi$ harmonic in nuclei over the free nucleon are in general agreement with the data~\cite{Guzey:2008fe,Liuti:2005qj}.
The harmonic of the beam spin asymmetry (BSA) is
\begin{equation}
  A_{LU}^{\sin\phi} = \frac{1}{ \pi} \int_{-\pi}^{\pi} d\phi \sin\phi A_{LU}(\phi),
\end{equation}
where $A_{LU}(\phi)$ is the measured DVCS beam spin asymmetry binned in $x$, the momentum transfer square $t$, and $Q^2$. This harmonic is proportional to the following combination of Compton form factors (CFFs)~\cite{Guidal:2013rya}
\begin{equation}
  A_{LU}^{ \sin\phi} \propto \operatorname{Im}( F_1 \mathcal{H}- \frac{t}{4M^2} 
   F_2 \mathcal{E}+ \frac{x_B}{2}(F_1+F_2)\tilde{\mathcal{H}}),
\end{equation}
which, in the case of the proton, is dominated by 
$\operatorname{Im}(\mathcal{H})$, and for the neutron, it is mostly sensitive to $\operatorname{Im}(\mathcal{E})$ and $\operatorname{Im}(\tilde{\mathcal{H}})$.

\begin{wrapfigure}[13]{r}{0.55\textwidth}
  \vspace{-0.5 cm}
  \begin{center}
    \includegraphics[width=0.55\textwidth,clip,trim=50mm 60mm 50mm 65mm]{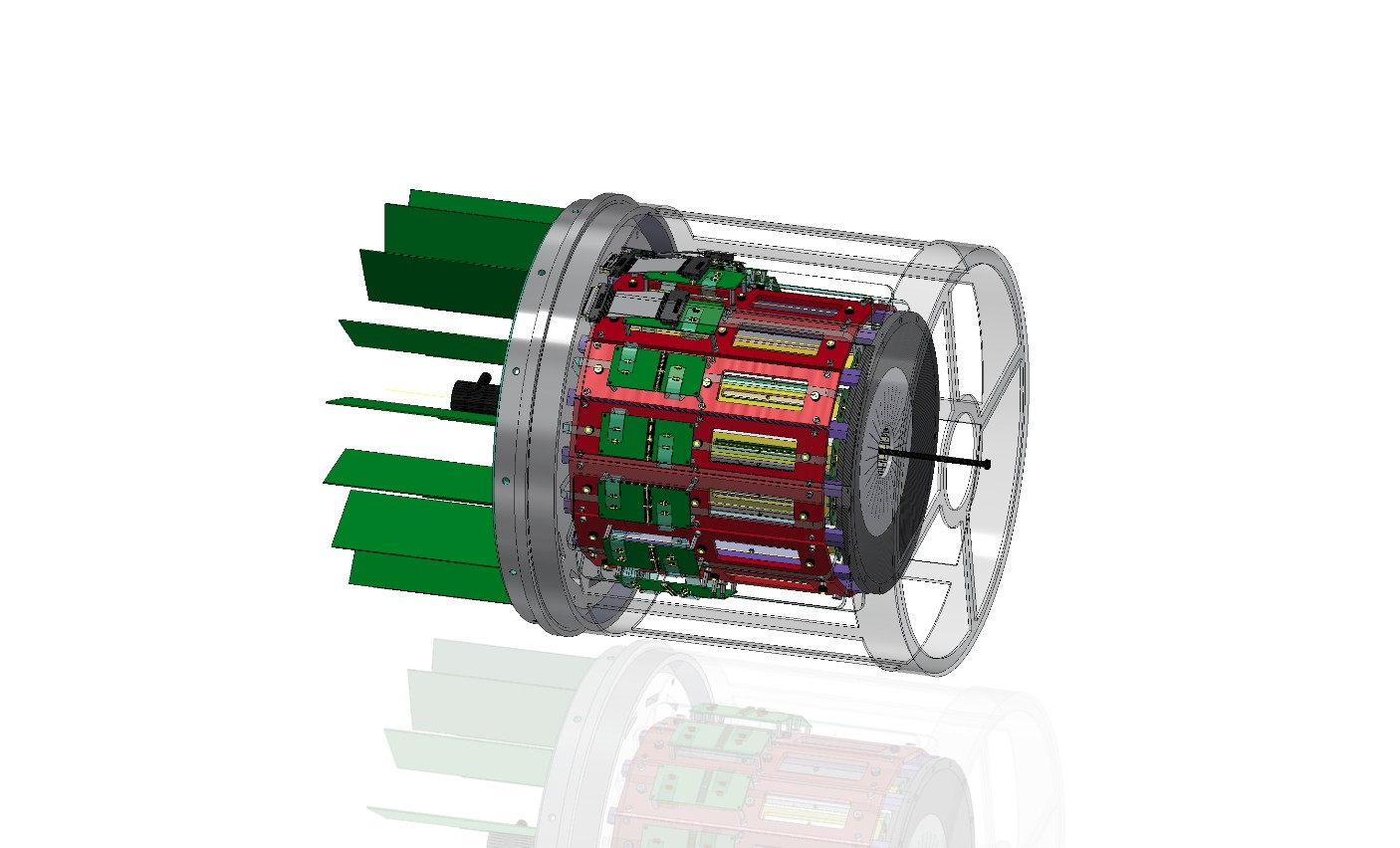}
  \end{center}
  \vglue -0.2in
 \caption{The ALERT detector.\label{fig:alertDet}}
\end{wrapfigure}

There are two tagged DVCS generalized EMC ratios planned for the upcoming 11~GeV CLAS12 ALERT experiment, which will use both $^4$He and $^2$H targets~\cite{Armstrong:2017}. The off-forward proton and neutron ratios are $R_p = A_{LU}^{p^{*}}/A_{LU}^{p}$ and $R_n = A_{LU}^{n^{*}}/A_{LU}^{n}$, where the ``$*$" indicates the in-medium ($^4$He) BSA, the denominators will use the free proton BSA from the accumulated CLAS12 liquid-hydrogen data sets and a quasi-free neutron from $^2$H with ALERT detector depicted in Fig.~\ref{fig:alertDet}.~The latter is built around a 30-cm-long gaseous $^4$He target straw and consists of a small drift chamber surrounded by a time-of-flight hodoscope, which is roughly 20~cm in diameter and 30~cm in length.

Additionally, the measurement of the beam spin asymmetry (BSA) in coherent DVCS is part of the broad ALERT scientific program~\cite{Armstrong:2017wfw}. BSA offers a way to explore partonic spatial distributions leading then to 3-D tomography of nuclei. Combining this coherent nuclear BSA with the free proton DVCS $A_{LU}$ results will allow for discerning among the several competing explanations of the nuclear medium effects.~The $^4$He nucleus is a spin-0 object and therefore at twist-2 its partonic structure can be parameterized by only one chiral even GPD [$H_{A}(x,\xi,t)$]. Thus, proposed asymmetry measurements allow for a significantly simplified extraction of the real and imaginary parts of the Compton form factor $\mathcal{H}_{A}$ in a model independent way. This azimuthal asymmetry arises from the interference between the DVCS and the Bethe-Heitler (BH) amplitudes. The BH process, in which the real photon is emitted by the scattering electron rather than the hadron, and DVCS have identical final states. The BH amplitude depends on the electromagnetic form factors, which is well known. The resulting asymmetry expression is given by
\begin{equation}
  A_{LU} = \frac{\alpha_{0}(\phi_{h})\Im_{A}} { \alpha_{1}(\phi_{h}) + \alpha_{2}(\phi_{h})\Re_{A} + \alpha_{3}(\phi_{h}) ( \Re_{A}+\Im_{A} ) },
\end{equation}
where $\Re_{A}=\mathfrak{Re}\{\mathcal{H}_{A}\}$ and $\Im_{A}=\mathfrak{Im}\{\mathcal{H}_{A}\}$ are the real and imaginary parts of the desired CFF, respectively, $\phi_{h}$ is the azimuthal angle between leptonic and hadronic planes, and $\alpha_{i}(\phi_{h})$'s are $\phi_{h}$-dependent kinematical terms. 
 The experimentally observed asymmetries are calculated in each kinematic bin using DVCS event yields for positive ($^+$) and negative ($^-$) helicity states as
\begin{equation}
  A_{LU} = \frac{1}{P_{b}} \frac{N^{+}-N^{-}}{N^{+}+N^{-}},
\end{equation}
where $P_{b}$ is the polarization of the incident electron beam, and $N^+$ ($N^-$) denotes the DVCS yield for positive (negative) helicity. For the coherent channel, the target $^4$He remains intact and recoils as a whole. The exclusivity is achieved by detecting the scattered electron and real photon in the forward CLAS12 detector and tag the back-scattered low-momentum $^4$He nucleus in ALERT. 
 
\begin{figure}[tp!]
  \begin{center}
  \includegraphics[width=0.9\textwidth]{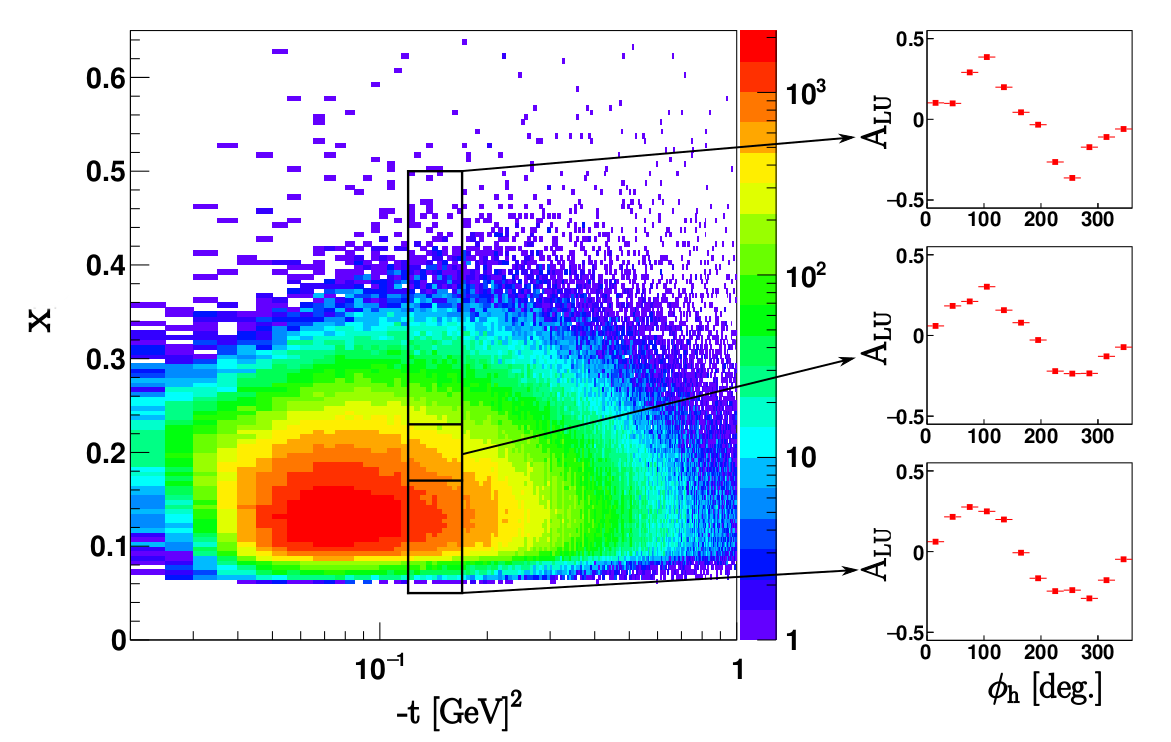}
 \end{center}
 \vglue -0.25 in
  \caption{The 22~GeV kinematic coverage of $x$ vs.~$-t$ phase-space along with the expected $A_{LU}$ for selected set of bins. The shown asymmetry bins correspond to a fixed $0.12<-t<0.17~\text{GeV}^2$ range, and to $0.05<x<0.17$, $0.17<x<0.23$, and $0.23<x_{B}<0.50$ from top to bottom in the right panel.}
 \label{alert_22kin}
\end{figure}

The 22~GeV simulation of nuclear DVCS reactions off $^4$He has been performed using the Monte~Carlo (MC) event generator TOPEG (The Orsay Perugia Event Generator)~\cite{PhysRevC.98.015203}, developed by R.~Dupr\'e {\it et al.}, and the CLAS12 GEANT4 MC (GEMC) package, which includes the full geometry and material specifications of ALERT detector. The CLAS12 Forward Tagger (FT) has been considered in this simulation to improve the acceptance of very forward real photons produced at much lower angles up to about 2 degrees.

The extracted BSA statistical precision from the 22~GeV simulation was scaled to correspond to a luminosity of 10$^{35}$ cm$^{-2}$s$^{-1}$ for 55 PAC days. The analyzed bins for $x$ and $-t$ were chosen as the ones planned for the 11~GeV ALERT experiment~\cite{ Armstrong:2017wfw}. For the BSA, $A_{LU}$, the simulated data were integrated over the full range of $Q^2$. The assumed beam polarization was taken in the range of $80\%$. Figure~\ref{alert_22kin} shows $A_{LU}$ projections for selected set of bins along with the anticipated 22~GeV kinematic coverage. The 22~GeV JLab upgrade will significantly extend the $Q^2$ reach of the measurements (up to 12~GeV) and elevates the statistics of the lower $x$ region, $0.08 <x<0.15$, which would allow for more detailed $x$ dependence studies with optimized bins. The FT inclusion leads to a more than four-fold increase of the DVCS acceptance due to the improved lower angular coverage, which significantly improves the expected BSA statistical precision, as depicted in Fig.~\ref{alert_22kin} right panel. 

\subsubsection{Unpolarized EMC and Antishadowing Regions}\label{sub:EMC}

While tremendous experimentally and theoretically efforts have been put into understanding the EMC effect, much fewer efforts have been invested in studying the antishadowing $x$ region ($x\sim$ 0.1) in the medium. In the inclusive DIS measurement, the DIS cross section of a heavy nucleus indicates enhancements near $x\sim$ 0.1 compared with one of a deuteron target. On the other hand, the Drell-Yan experiments reveal no such enhancement, indicating that the sea quarks may not suffer from the antishadowing effect~\cite{Alde:1990im}. 

SIDIS is a powerful tool for studying the quark distributions in nucleons and nuclei. Using high-energy electrons scattering off a nucleon, the virtual photon knocks out a quark in the nucleons. Then the quark has to undergo a complicated hadronization process due to the color confinement. Out of many colorless final state hadrons generated in the hadronization process, the leading hadron will be measured in coincidence with the scattered electron. Under the factorization framework, the SIDIS structure functions of electron-nucleus scattering can be factorized into the convolution of the colinear PDFs and the colinear fragmentation functions (FFs) for all quark flavors and gluons, in the colinear framework. When the transverse momentum of the leading hadron is measured, the structure function becomes the convolution of TMDs and the 3-D FFs. One of the biggest advantages of using SIDIS to measure PDFs is the ``flavor-tagging" feature where the different detected leading hadrons are uniquely sensitive to the knocked-out quarks. 

One can measure the SIDIS via electron-nucleus scattering and get access to the nuclear~PDF (nPDF) of individual quarks by tagging different hadrons and directly study the flavor-dependence of the EMC effect and antishadowing effect in nuclei. However, the SIDIS cross sections measure not only the medium modification of PDFs but also the nuclear effect of the nuclear FFs (nFFs) which are small for light nuclei but significant for heavy nuclei~\cite{Zurita:2021kli}. A systematic global analysis of high-precision nuclear-SIDIS data with extensive kinematic coverage and multiple hadron productions is essential to decouple the nPDFs and nFFs for different quark flavors. 
\begin{figure}[t!]
  \begin{center}
    \includegraphics[width=\textwidth]{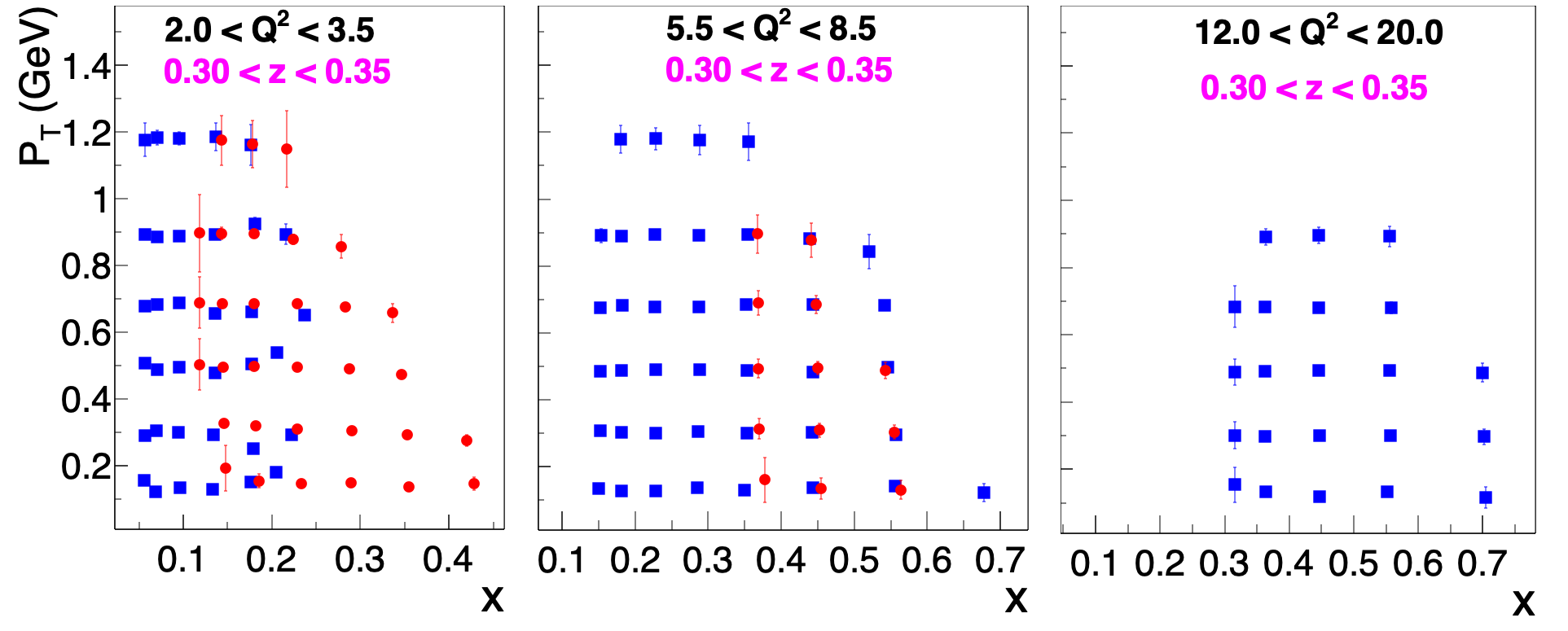}
 \end{center}
 \vspace{-0.6cm}
  \caption{Kaon-SIDIS projection in 4-D binning ($Q^2$, $z$, $x$, $p_T$) for $^3$He at $10^{35}$~cm$^{-2}$ s$^{-1}$ luminosity and beam time of 100 days. Out of totally 38 ($Q^2$, $z$) bins, only three bins are shown here to illustrate the comparison of statistical uncertainties and coverage of $p_T$ and $x$ between 11~GeV (red circles) and 22~GeV (blue squares) beam energies for three different $Q^2$ bins and one fixed $z$ value.}
  \label{nuclear_sidis_kaon_4d}
 \label{fig:1}
\end{figure}

With a 22~GeV electron beam, one can enforce the SIDIS reaction in the current fragmentation region where the cross sections can be factorized as the convolution of nPDFs and nFFs. Theory suggests that at 11~GeV only 70\% of pion-SIDIS data and 20\% of kaon-SIDIS data are in the current fragmentation region~\cite{Boglione:2016bph}. The higher energy also allows for the measurement of heavy mesons such as kaons, protons/antiproton, and lambda which are impossible to be measured with an 11~GeV beam. Broader $Q^2$ and $p_T$ distributions also enable theoretical corrections. Figure~\ref{nuclear_sidis_kaon_4d} show the projection of the $k^+$-SIDIS data in 4-D~($Q^2$, $z$, $x$, $p_T$) with $^3$He at $10^{35}$~cm$^{-2}$ s$^{-1}$ luminosity and 100 days of running. The same projections at 11~GeV are also given as a comparison. Doubling the beam energy not only largely extends the $Q^2$ and $p_T$ coverage but also pushes the $x$ down to the medium region with great precision. 

 \begin{figure}[t!]
  \begin{center}
   \includegraphics[width=0.75\textwidth]{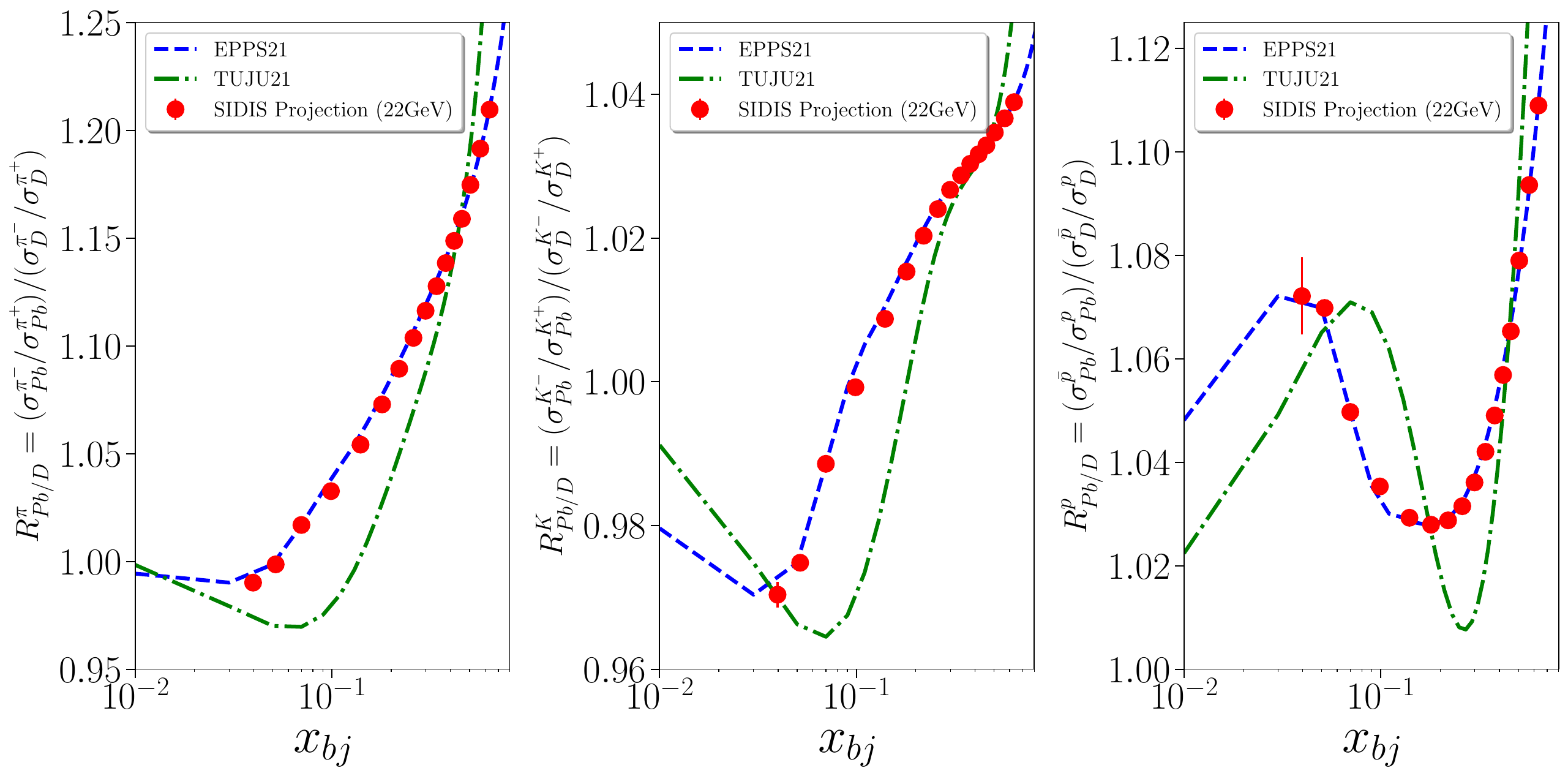}
 \end{center}
   \vspace{-0.75cm}
\caption{SIDIS super-ratios, $R=\frac{\sigma_A^{h+}/\sigma_A^{h-}}{\sigma_D^{h+}/\sigma_D^{h-}}$, between lead and deuteron for pion (left panel), kaon (middle panel), and proton (right panel) production for 3 $<Q^2<$ 4 GeV$^2$ and 0.3 $<z<$ 0.35 after integrating over $p_T$. The curves are EPPS21~\cite{Eskola:2021nhw} global nPDF fits (assuming flavor independence) and TUJU21~\cite{PhysRevD.100.096015} nPDF fits (allowing flavor dependence).}
 \label{fig:pb_sidis_ratios}
\end{figure}

Without going through the complicated global analysis, one fixed ($Q^2$, $z$) range can be picked; {\it e.g.}, 3~$< Q^2<$~4~GeV$^2$ and 0.3~$< z<$~0.35, to compare the variation of SIDIS cross section with $x$ between a heavy nucleus and deuteron.~Figure~\ref{fig:pb_sidis_ratios} shows the sensitivities of super-ratios $R$ ($\frac{\sigma_A^{h+}/\sigma_A^{h-}}{\sigma_D^{h+}/\sigma_D^{h-}}$) with different hadron final states to different nPDF global fits for lead, where EPPS21~\cite{Eskola:2021nhw} nPDFs assumes flavor-independence and TUJU21~\cite{PhysRevD.100.096015} nPDFs allows flavor-dependence. Note that in the collinear framework, the $p_T$ distribution is integrated when extracting the cross sections. The distributions near x$\sim$~0.1 have great statistical precision to determine if there is any indication of flavor-dependence of the EMC and antishadowing effects. There are a total of 36 similar ($Q^2$, $z$) bins allowing for sophisticated global extraction of individual nPDF for different quark flavors in light to heavy nuclei to systematically study their EMC and antishadowing effects in the valance and sea regions. 

\subsubsection{Spin Structure Functions in EMC, Antishadowing, and Shadowing Regions}\label{sub:spinEMC}

\paragraph*{Medium-Modified Spin Structure.} As detailed in Subsubsec.~\ref{BNSTDIS}, the experimental study of structure modifications of bound nucleons has been carried out for decades. Yet, despite much theoretical work, there is not consensus on what causes them. There is an approved 11~GeV JLab experiment to measure~\textit{spin} structure function modifications for the first time, primarily in the EMC and antishadowing regions, for a bound polarized proton embedded in a polarized $\mathrm{^7}$Li nucleus~\cite{pEMC_proposal,pEMC_update}.~The same technique, if used at 22~GeV, would push for the first time into the shadowing region, and with sufficient reach in four momentum transfer $\mathrm{Q^2}$ to give confidence in the validity of theoretical assumptions. The JLab22 \textit{uniquely} gives access to the shadowing region for spin structure functions, measured at high luminosities enabled by fixed target experiments.

This is very interesting physics because the shadowing and antishadowing regions are characterized at least partially by the onset of multi-step diffractive processes which allow constructive and destructive interferences. These are thought in some models to cause enhancement and suppression in the antishadowing and shadowing region, respectively~\cite{PhysRevD.70.116003}, while other models have different approaches~\cite{PhysRevC.61.014002,PhysRevC.95.055208,Cloet:2006,Miller:2005,Fanchiotti:2014,cloet:2005}. Theoretical work that includes these regions for polarized $\mathrm{^7}$Li range from predicting a 10\%  suppression to a 50\% enhancement in spin structure function ratios. It is very clear this is \textit{terra incognita} and further progress in understanding these crucially important kinematic regimes of $\mathrm{x<0.3}$ without new data is improbable. 
\begin{figure}[t!]
  \begin{center}
    \includegraphics[clip=true, trim=0.9cm 0 0.5cm 3cm, width=0.75\textwidth]{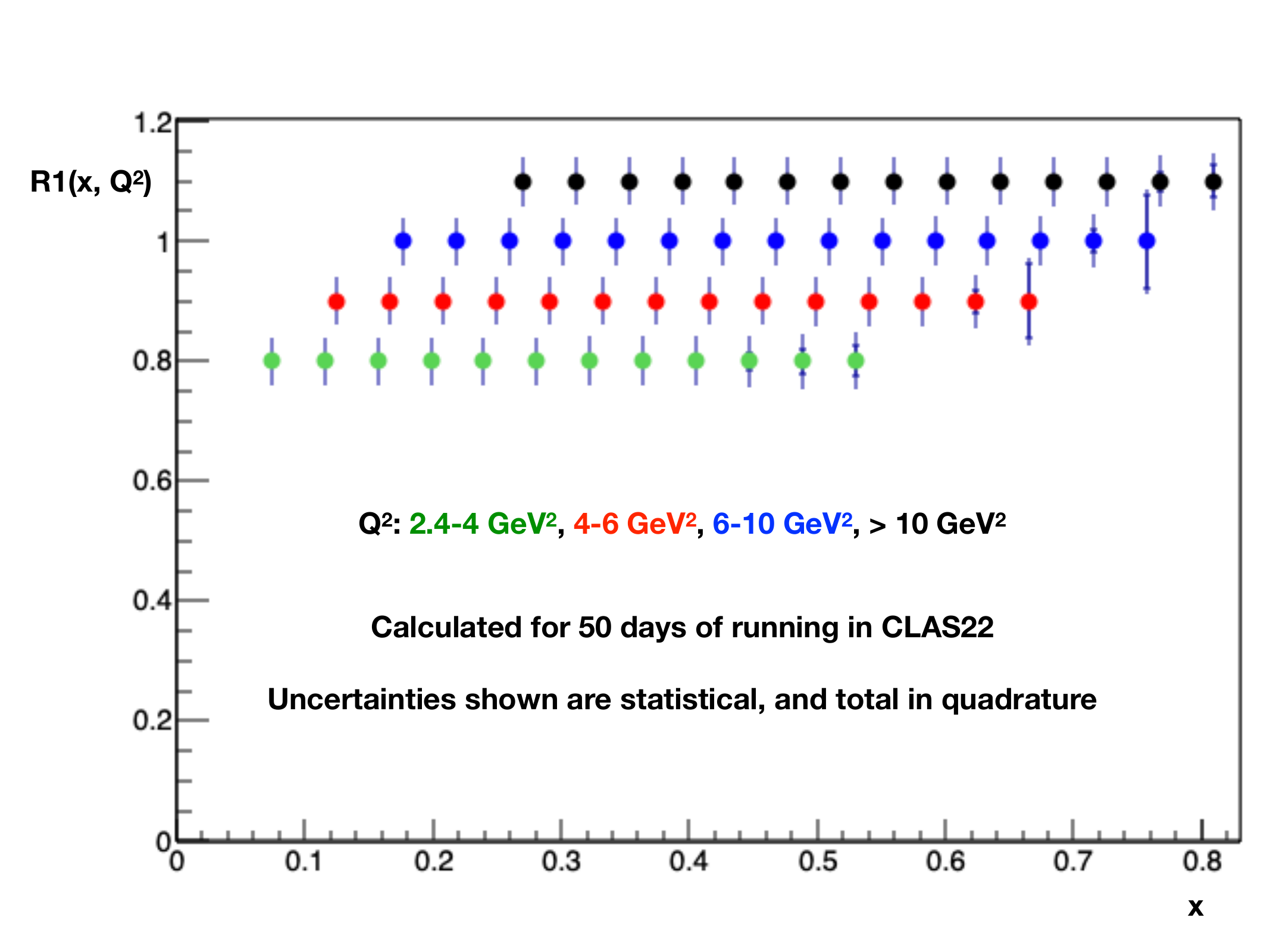}
 \end{center}
 \vglue -0.35in
  \caption[pEMC]{Anticipated results for observable $R_1$ (see text). The assumed luminosity for this prediction is $\text{2x10}^{35} ~\mathrm{cm^{-2}s^{-1}}$, which has been nearly demonstrated in present-day CLAS for light nuclear targets.}
 \label{pEMC}
\end{figure}

\paragraph*{Anticipated Results and Coverage.} Following the above discussion of the approved 12~GeV JLab experiment, the goal now is to produce the 22~GeV projections for one of the main observables, denominated as R$\mathrm{_1}$, which is defined as   
\begin{equation}
R_1 = \frac {\mathrm{[d\sigma_+ - d\sigma_-]_{\mathrm{^7}Li}}} {\mathrm{[d\sigma_+ - d\sigma_-]{_p}}},
\label{pEMC-R1}
\end{equation}
where the positive and negative signs indicate the relative beam and target polarization. In Fig.~\ref{pEMC}, predicted R$\mathrm{_1}$ results are shown for fifty days of beam time in the upgraded CLAS22. For the assumed luminosity of $\text{2x10}^{35} ~\mathrm{cm^{-2}s^{-1}}$, the uncertainties are dominated by systematic uncertainties except for a few bins at high $x$. This luminosity has been nearly achieved in present-day CLAS for light nuclear targets such as deuterium. 

In comparison to the 11~GeV experiment, the 22~GeV measurement will feature much higher four-momentum transfer for each bin in $x$, assuring that we can have full confidence in the interpretation of the results. For example, for a minimum $x$ of 0.08, the momentum transfer will be 3.2 GeV$^2$ at 22~GeV compared to 1.2 GeV$^2$ at 11~GeV. Furthermore, for 22~GeV, we can reach Q$^2$ of 10 GeV$^2$ at $x$=~0.3 compared to $x$=~0.75 at 11~GeV, where the statistical information will be much poorer and the Fermi momentum effects begin to encroach. While the $Q^2$ range will be superior for the 22~GeV study, there is also very much value in intercomparison of the results from the two energies and thus being able to study several aspects such as scaling behavior, higher twist, contribution from the unmeasured $A_2$, and target mass effects. As a result, the two beam energy measurements will complement each other in important ways.

\subsubsection{Color Transparency Studies}
\label{sub:CTstudies}

QCD uniquely predicts the existence of hadrons with their constituent quarks in a small-sized color singlet, thus suppressing interactions between the singlet and the surrounding color field in the nuclear medium~\cite{Brodsky:1988xz,Brodsky:1994kf,Dutta:2012ii,Brodsky:2022bum,Elfassi:2022}. This QCD phenomenon, dubbed as CT, can be observed experimentally in exclusive processes with sufficiently high momentum transfer leading to a significant reduction in FSIs~\cite{Brodsky:1988xz}.  

The experimental observable commonly used to search for CT effects is the nuclear transparency, $T$, taken as a ratio of the cross section per nucleon for a process on a bound nucleon to that of a free nucleon. Thus, the signature of CT is measured as an increase in the nuclear transparency with increasing momentum transfer squared, $Q^2$ (or momentum of the final state hadron). In complete CT, FSIs vanish and $T$ plateaus. In the absence of CT, the nuclear transparency is not expected to change, following the same relative energy independence of the $NN$ cross section. 

Meson experiments in both 6 and 12~GeV era of JLab have explored the onset of CT through the hard exclusive electroproduction of $\rho$ and $\pi$ mesons off nuclei. Pion production measurements in Hall C measured the transparency for $e + A \rightarrow e' + \pi^+ + A^{*}$ using the HMS and SOS spectrometers. The results have indicated both an energy and $A$ dependence of the nuclear transparency consistent with models inclusive of CT effects~\cite{Clasie:2007} for $Q^2$ from 1.1 to 4.7~GeV$^2$. The CLAS Collaboration experiment measured $\rho$-meson production on carbon and iron targets relative to $^2$H in the range of 0.8 to 2.2~GeV$^2$~\cite{ElFassi:2012,Elfassi:2022}. The extracted nuclear transparencies showed a $Q^2$ and $A$ dependence consistent with the very same models of CT and at a lower onset than that measured for the pion~\cite{Frankfurt:2008pz,Gallmeister:2010wn,Cosyn:2013qe}. 

While the onset of CT is anticipated to be at a lower energy regime in mesons than baryons, the precise kinematic regime for protons is not known. Intriguing results from large angle $A(p,2p)$ scattering at Brookhaven National Laboratory (BNL)~\cite{Carroll:1988,Mardor:1998zf,Leksanov:2001,Aclander:2004zm} indicated a region of interest for $A(e,e’p)$ experiments at JLab. The measured $A(e,e’p)$ cross section was compared to calculations in the plane wave impulse approximation which excludes FSIs. Experiments were conducted at SLAC~\cite{Makins:1994mm,ONeill:1994znv}, and JLab~\cite{Abbott:1997,Garrow:2001,Dutta:2021} measuring $Q^2$ up to 14.2~GeV$^2$ and the highest proton momentum of $\sim$~8.5~GeV ruling out the observation of the onset of CT in this regime for protons. In light of these most recent proton results, new ideas for exploring the onset of CT in different kinematics have become increasingly significant.

An increase of the JLab beam energy is crucial for fully evaluating the QCD signature of CT in nuclei. The increased beam energy enables measurements of the entire range, from nearly the onset to full CT, for mesons such as the $\rho$ and pion.~The extended reach in $Q^2$ and the improved rates that accompany the increase of beam energy enable a robust program for exploring the onset of CT in protons in various kinematics to be detailed in the next section. 

\begin{figure}[!b]
\centering
\vspace{-0.5 cm}
\begin{minipage}{0.52\textwidth}
     \includegraphics[clip=true, trim=0.45cm 0.25cm 1.25cm 0.9cm, width=\textwidth]{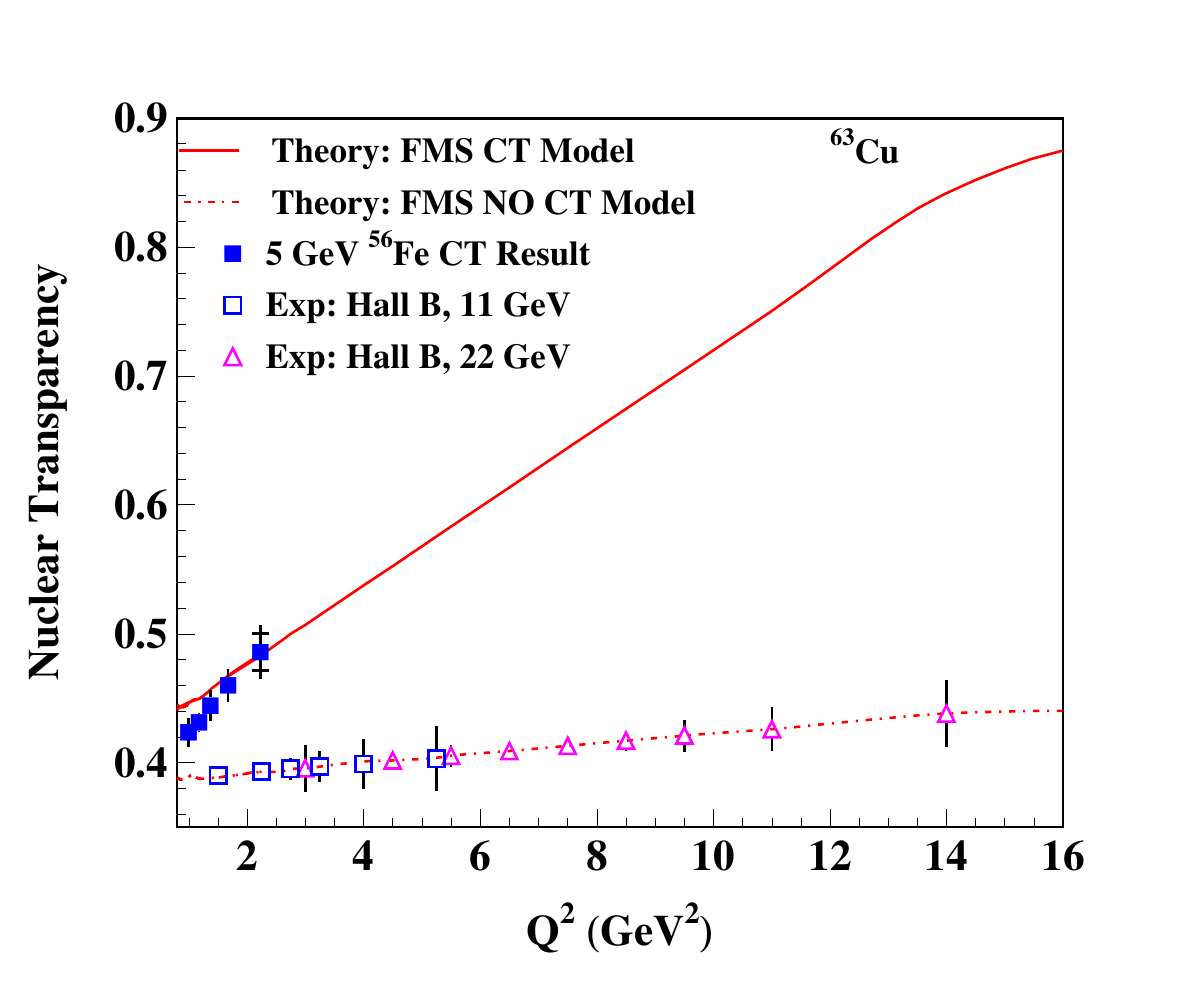}
     \end{minipage}
     \begin{minipage}{0.47\textwidth}
     \includegraphics[clip=true, trim=0.1cm 0 0.9cm 1.5cm,width=\textwidth]{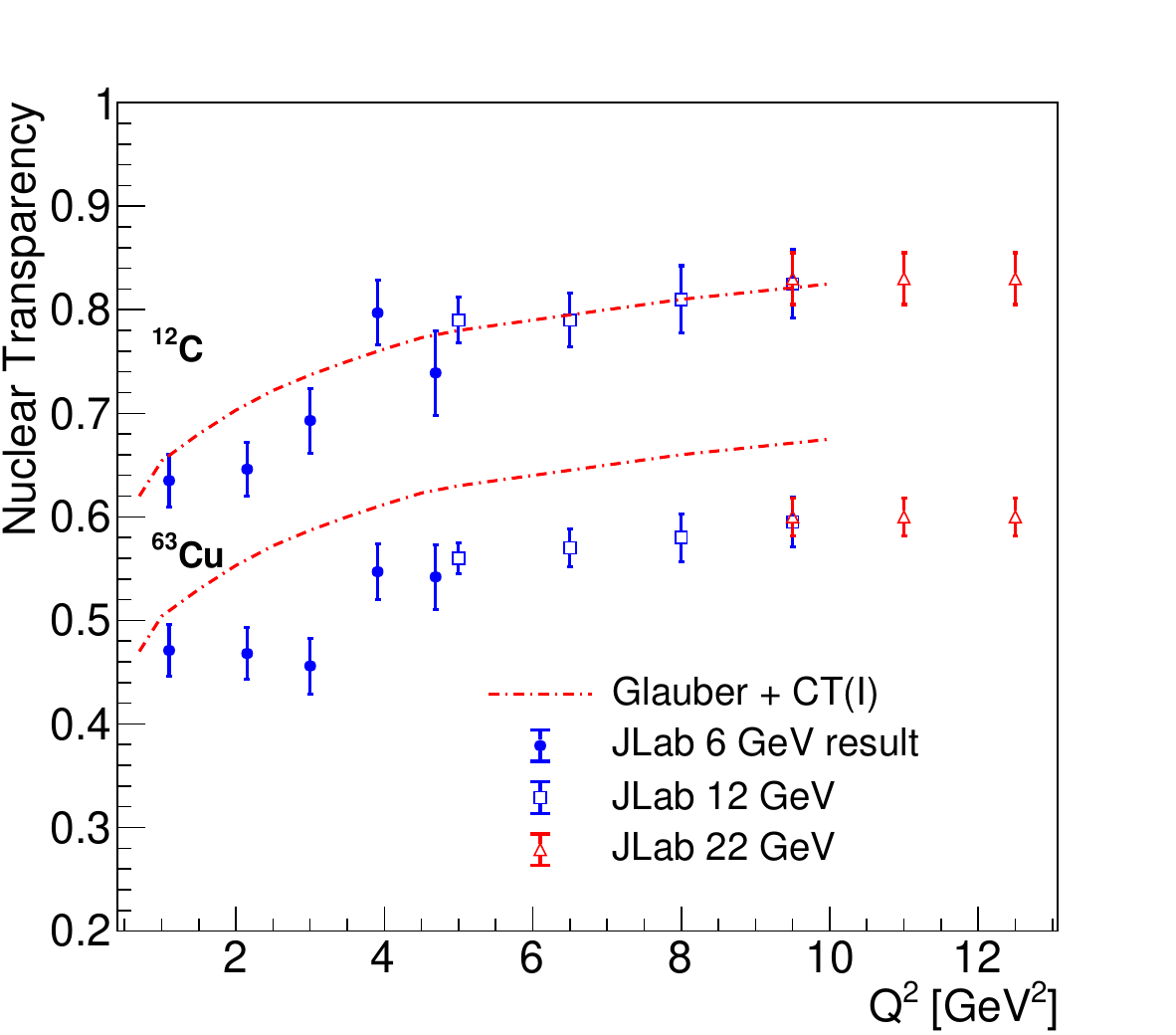}
     \end{minipage}
     \vglue -0.15in
  \caption{Left: the nuclear transparency for the $\rho$-meson experiment in Hall~B, where the solid circles correspond to the already published 5~GeV CLAS6 results on $^{56}$Fe~\cite{ElFassi:2012}, and the open squares (triangles) correspond to the projections with 11~GeV~(22~GeV) beam energy on $^{63}$Cu.~The Frankfurt–Miller–Strikman~(FMS)~\cite{Frankfurt:2008pz} curve is a linear extrapolation of the 11~GeV predictions. Right: the nuclear transparency for the $\pi^+$-meson in Hall~C on carbon and copper targets, where the solid points correspond to results already published~\cite{Dutta:2021,Abbott:1997,Garrow:2001}, and the open points correspond to the projections with 11~GeV and 17.6~GeV beam energies.}
 \label{fig:mesonCTprojections}
\end{figure}

The 22~GeV beam energy upgrade at JLab increases the maximum attainable $Q^2$ and improves the rates at otherwise slow-counting kinematics. The possibilities of extending meson measurements for $\rho$ in Hall~B and pion in Hall~C are presented here as a proof of capability. 

For the $\rho$-meson exclusive diffractive measurement using the CLAS12 spectrometer in Hall~B, the $Q^2$ can be extended up to 14~GeV$^2$ (with the highest bin upper limit is 16~GeV$^2$). The 22~GeV simulation is performed by assuming the same CLAS12 nominal per-nucleon luminosity of 10$^{35}$cm$^{-2}$s$^{-1}$, 7 PAC days as the 11~GeV projections, and the Hall~B flag assembly where the solid-target foils are mounted in series (see Fig.~{\color{blue}4} in Ref.~\cite{Elfassi:2022}). The obtained projections for copper are shown in Fig.~\ref{fig:mesonCTprojections} left for a fixed coherence length representing the lifetime of the $q\Bar{q}$ fluctuation of the virtual photon. In the region where the 11 and 22~GeV beam energy projections overlap, the statistical uncertainties are reduced by almost a factor of 3, except for the 3~GeV$^2$ case at 22~GeV because it is at the lower edge of the CLAS12 acceptance. 

Similarly, one can extend the $\pi^+$ measurements in Hall~C with a higher beam energy, but the maximum $Q^2$ is approximately 12.5~GeV$^2$ due to spectrometer limitations and to maintain $t<1$~GeV$^2$ to reduce FSIs. Assuming a similar experimental setup as that in the 12~GeV era with electrons in the HMS and $\pi^+$ in the SHMS, one can take measurements at $Q^2=9.5, 11, \text{and }12.5$~GeV$^2$ with a 17.6~GeV beam energy on targets of H, $^2$H, $^{12}$C and $^{63}$Cu. Assuming 80~$\mu$A beam current on hydrogen and a 3\% statistical uncertainty for each kinematic, the full experiment could be completed with 200 hours of beam on target with the projections shown in Fig.~\ref{fig:mesonCTprojections} right. Note that the assumption made here is the beam dump power limitations would be increased with the increased beam energy.

Likewise, one can extend the measurements for the proton in parallel kinematics as in~\cite{Dutta:2021} in Hall~C. While the maximum momenta of the SHMS and HMS spectrometers and minimum angles limit the extended reach in $Q^2$, the increase in beam energy significantly improves the rates at the high $Q^2$.~An increase in the beam energy to 13~GeV improves the rates by a factor of three at the previously highest $Q^2=14.2$~GeV$^2$.~Assuming a beam energy of 13~GeV, one can measure additional $Q^2= 14.2, 15.8, \text{and }17.4$~GeV$^2$ at 2.2\% statistical uncertainty on $^{12}$C in 160 hours of 80~$\mu$A beam on target. An increase in the beam energy can provide a strong test for light-front holographic QCD (LFHQCD) predictions (see Ref.~\cite{Brodsky:2022bum}). Eventual upgrades to the spectrometer momenta would improve the reach in attainable $Q^2$ in Hall~C. 

\begin{figure}[!t]
\centering
     \includegraphics[width=0.6\textwidth]{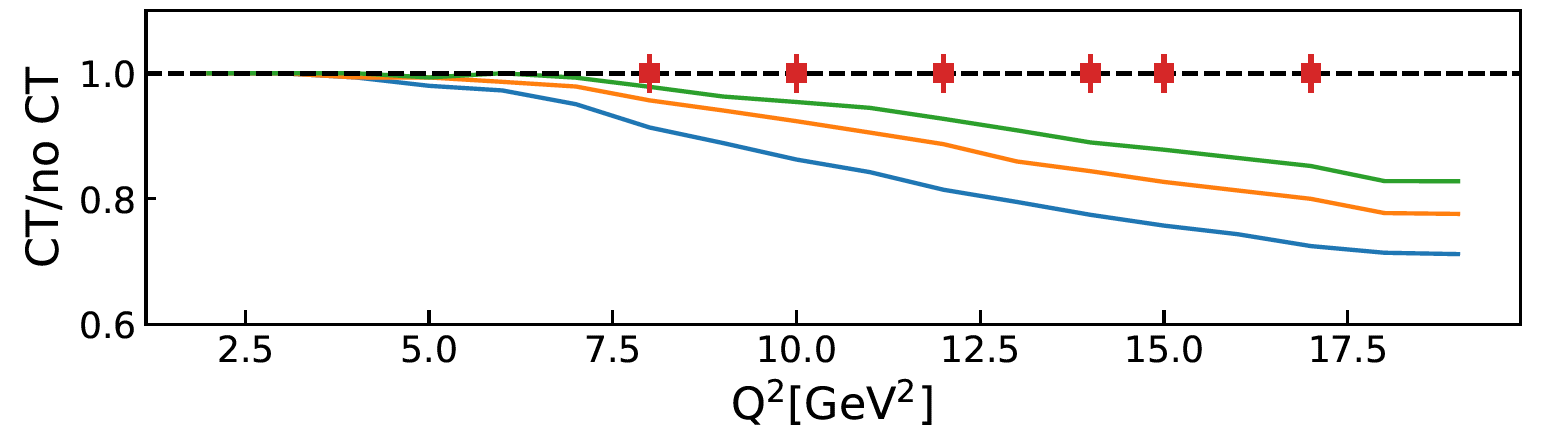}
     \vglue -0.1in
  \caption{Projections for D$(e,e'p)$ in rescattering kinematics. The curves are from Ref.~\cite{physics4040092} corresponding to 3 possible values of the CT model parameter, $\Delta M^2$.}
 \label{fig:deepct_proj}
\end{figure}

New measurements are accessible with the increase in beam energy for exploring the onset of CT in the proton in non-traditional kinematics with high FSIs. This would make it possible to explore the loophole left from previous measurements if the apparent lack of observed CT is due to limitations in parallel kinematics. For example, D$(e,e'p)n$ has well-known FSI contributions from double scattering that is strongly determined by the recoiling neutron angle and momentum. In this way, it is feasible to measure the D$(e,e'p)n$ reaction in the Hall~C spectrometers accessing high $Q^2$ up to 17~GeV$^2$ (see discussion in Ref.~\cite{physics4040092}). A scan in $Q^2= 8,10,12,14,15, \text{and }17$~GeV$^2$ using the HMS and SHMS in coincidence with a 13~GeV beam could be accomplished with a 3\% statistical uncertainty and 3~months of beam on target, as shown in Fig.~\ref{fig:deepct_proj}.

\subsubsection{Hadronization Studies in Nuclei}\label{sub:SIDIS}

With the emergence of the hadronization concept in the late 1970s~\cite{Field:1976ve} to explain the limited transverse momenta of hadrons in jets produced in $e^+e^-$ and $pp$ collisions, many theoretical models were developed with the objective to offer an explanation to the experimental data as well as a theoretical framework for the dynamics of the hadronization process.

Similarly to the EMC effect~\cite{aubert1983ratio}, it is expected that the hadron production from hard reactions, in the presence of hot dense QCD matter or cold nuclear medium, would be different from its equivalent in vacuum. This was confirmed by the PHENIX~\cite{adcox2005formation} and STAR~\cite{STAR:2005gfr} experiments at RHIC, as well as SLAC~\cite{Osborne:1978ai}, HERMES and EMC Collaborations~\cite{HERMES:2010,HERMES:2011,HERMES:2007plz,EuropeanMuon:1991jmx}. The observed hadron attenuation could be attributed to many effects such as: quark energy loss due to induced gluon radiation and quark multiple rescattering with the surrounding medium, and to FSIs of the produced hadron (or prehadron) with the nucleus. Those effects were widely studied in many models such as: Giessen Boltzmann-Uehling-Uhlenbeck (GiBUU) transport model, Lund string model, Rescaling model, Quark energy loss model, Quantitative model, Higher-twist pQCD model, etc. Despite the existence of a large number of phenomenological models, that in general reproduce qualitatively the global features of the data, a clear understanding of the hadronization mechanism is still a challenge and a moderate success in describing the data was achieved. Inputs from experimental data are crucial to test and calibrate these models and also to check the validity of many theoretical calculations.

To establish a detailed picture of the time-space development of the hadronization mechanism in SIDIS production, the two time-distance scales $\tau_{p}$ and $\tau_f$ need to be studied. Simple calculations based on pQCD and the Lund string model showed that the length of the first stage, $L_p \approx \nu(1-z_h)/\kappa$, depends on the energy of the virtual photon transferred to the struck quark(s), $\nu$, the fraction $z=E_h/\nu$, where $E_h$ is the energy transferred to the final hadron, and on the string tension $\kappa$. In the hadron rest frame, the length of the second stage, $L_f$, is approximated by the hadron radius, $0.5-0.8$~fm. Taking into account time dilation, $L_f$, in the lab frame, can range up to distances that exceed the size of the nucleus. The above estimations demonstrate that the hadronization occurs at small time and length scales. Consequently, the hadrons detected at very large distances from the origin of the reaction do not provide detailed information that is sensitive to the production and formation times. These facts clarify the importance of using a nuclear media in the study of hadronization. The scattering centers (nucleons) inside the nuclei act like miniscule detectors placed within distances comparable to the length scales associated with the hadronization. Therefore, measurements of the effects induced by these scattering centers on the propagating quark, prehadron and hadron would provide a unique opportunity to study the hadronization mechanism at its early stages. 

Another important topic in hadronization is the study of two hadron production from lepton-nucleus deep-inelastic scattering. Depending on the evolution of the fragmentation process with time and space, the observed signal might either be dominated by in-medium prehadronic interactions or by partonic energy loss. This could be investigated through the two hadron production in SIDIS off nuclei. If quark energy loss were the dominant mechanism, then one would expect that the hadron attenuation would not depend significantly on the number of hadrons involved, and the double-hadron to single-hadron ratio for a nuclear target should not depend strongly on the atomic mass number $A$. If, on the other hand, hadron absorption were the dominant process, then requiring an additional slow hadron would suppress the two-hadron yield from heavier nuclei. Therefore, the double-hadron to single hadron ratio would decrease with the size of the nuclear medium. Measurements from HERMES~\cite{PhysRevLett.96.162301} showed a slight dependence on the atomic mass number. The data were confronted to three models based on different assumptions and, interestingly, none of them was able to describe the observed double-to single hadron yields. 

In the end, the study of hadronization using SIDIS processes off nuclei offers the possibility to address some challenging questions, such as i) what are the dynamics leading to color confinement?; ii) what are the effects of the nuclear medium on the fragmentation functions?; iii) how long does it take to form the colorless object (prehadron)?; and iv) how long does it take to form the color field of a fully dressed hadron? Answers to those questions would enhance our understanding of hadronization which is one of the exciting frontier subjects in QCD.

\paragraph*{Proposed Measurements and Results.} The SIDIS experimental observables needed to explore the two time-scales associated with the hadronization process are:
\begin{itemize}
\item [i)] The transverse momentum, $p_{T}$, broadening related to the production time of the struck color-neutralized objects, which is defined as \\
\begin{equation}
\Delta \langle p^2_T\rangle^{A}_h(Q^2,\nu, z)= \left[\langle p_T^2\rangle^{A}_h-\langle p_T^2\rangle^{D}_h\right](Q^2, \nu, z), 
\label{eq:Dpt1}
\end{equation}
where $\langle p_T^2\rangle^{A}_h$ is the mean $p_{T}$ squared obtained for a nucleus $A$ and hadron $h$ while $Q^2$ is the four-momentum transfer squared, and 
\item [ii)] The hadron multiplicity ratio related to the hadron formation time and defined as:\\
\begin{equation}
R^A_h(Q^2,z, \nu, p_T)=\frac{N^A_{h}(Q^2,z, \nu, p_T)/N^A_e(\nu, Q^2)}{(N^D_{h}(Q^2,z,\nu, p_T)/N^D_e(\nu, Q^2)},
\label{eq:RAh1}
\end{equation}
where, $N^A_{e}$ and $N^A_h$ are, respectively, the yield of DIS electrons and SIDIS hadrons produced on a nucleus $A$ for a given kinematic bin.
\end{itemize}

Additionally, the double-to-single hadron multiplicity ratio is defined as:\\
\begin{equation}
 R_{2h}(x,Q^2,z_2,p_{T,2}^2, \Delta\phi)=\frac{(N_{h_1,h_2}(x,Q^2,z_1,p_{T,1}^2, \phi_1, z_2, p_{T,2}^2,\Delta\phi)/N_{h_1}(x,Q^2,z_1,p_{T,1}^2, \phi_1))^A_{z_1>0.5}}{(N_{h_1,h_2}(x,Q^2,z_1,p_{T,1}^2, \phi_1, z_2, p_{T,2}^2,\Delta\phi)/N_{h_1}(x,Q^2,z_1,p_{T,1}^2, \phi_1))^D_{z_1>0.5}},
\end{equation}
where $N_{h_1,h_2}^{A,z_1 > 0.5}$ is the number of events containing at least two hadrons, $z_1$ is the energy fraction of the first (leading) hadron, and $z_2$ is the energy fraction of the second (subleading) hadron. $N_{h_1}^{A}$ is the yield of one leading hadron, and $\phi$ is the angle between the leptonic and hadronic planes. Instead of binning in $\phi_2$, the convention of Ref.~\cite{Paul:2022} is adopted by binning on $\Delta \phi \equiv \phi_{1}-\phi_{2}$.

\begin{figure}[b!]
    \centering
    \includegraphics[clip=true,trim=0.25cm 0.1cm 0.1cm 0,width=0.325\textwidth]{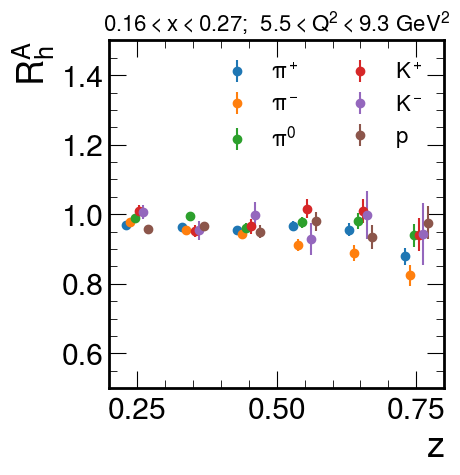}
    \includegraphics[clip=true,trim=0.15cm 0.1cm 0.1cm 0,width=0.33\textwidth]{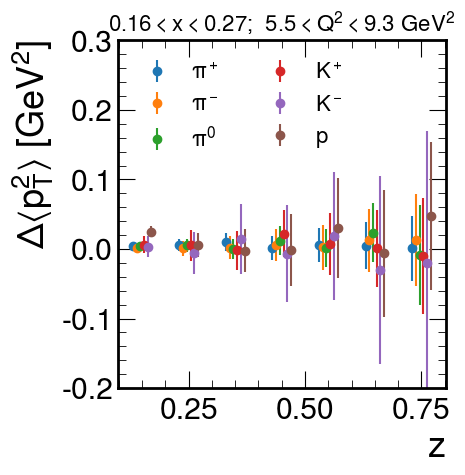}
    \includegraphics[clip=true,trim=0.125cm 0.1cm 0.2cm 0,width=0.33\textwidth]{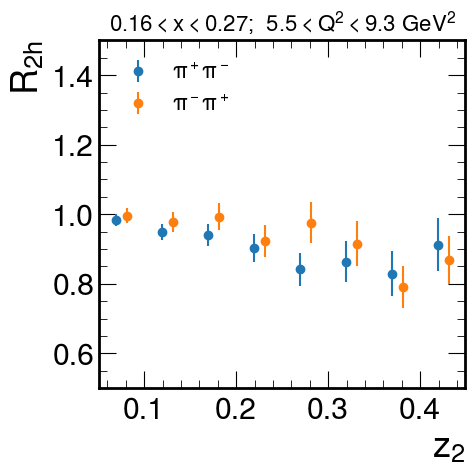}
     \includegraphics[clip=true,trim=0.25cm 0.1cm 0.1cm 0,width=0.325\textwidth]{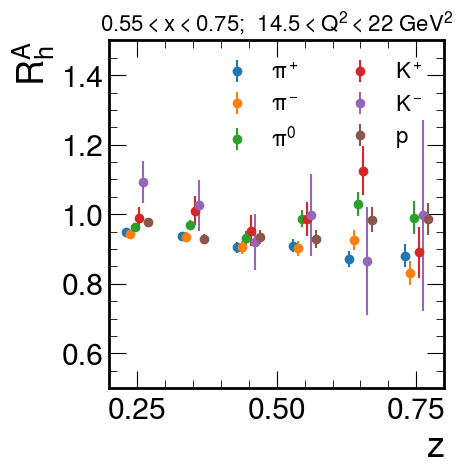}
    \includegraphics[clip=true,trim=0.15cm 0.1cm 0.1cm 0,width=0.33\textwidth]{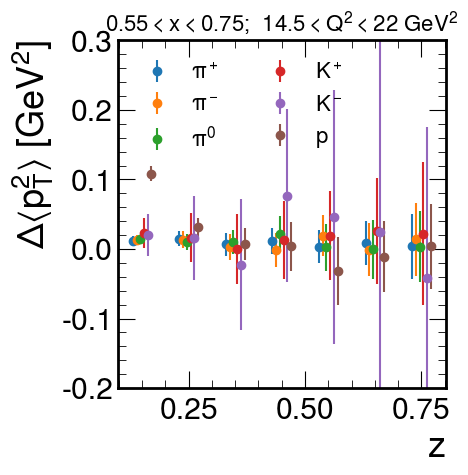}
    \includegraphics[clip=true,trim=0.125cm 0.1cm 0.2cm 0,width=0.33\textwidth]{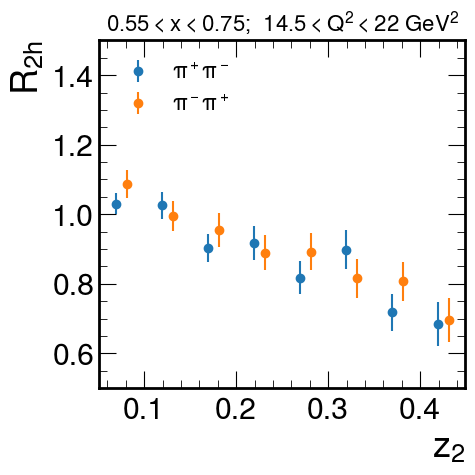}
    \vglue -0.15in
    \caption{Three-fold $z$ ($z_2$)-binned projections of $R^A_h$ (left column), $\Delta\langle p_T^2\rangle$ (middle column), and $R_{2h}$ (right column) for various hadrons production off carbon target represented with different colors and for a combination of $x$ and $Q^2$ bins that are outside the 11~GeV coverage on both rows. Error bars represent the statistical precision of the simulated sample from GiBUU assuming a per-nucleon luminosity of $10^{35}$~cm$^{-2}$s$^{-1}$ and 15~PAC days.}
    \label{fig:nuclear_sidis}
\end{figure}
The $p_T$-broadening is related to the path travelled by the struck quark in the nucleus. Its measurement is very important because it reveals whether or not the hadronization is occurring inside or outside the nuclear medium. Therefore, the hadron multiplicity ratio is the appropriate observable to experimentally measure the hadron formation time. It is normalized by DIS electrons to cancel out initial state nuclear effects, such as nuclear modified partonic distributions, and to isolate final state nuclear modifications. A decrease of the ratio with the mass number $A$ and an increase with the energy $\nu$ is expected because the attenuation has to be larger for large nuclei and the nuclear effects have to weaken with increasing struck quark(s) energy.

Those observables were previously studied by the HERMES and CLAS Collaborations for various hadron channels such as pions~\cite{HERMES:2001,Moran:2022,Paul:2022, PhysRevLett.96.162301,HERMES:2010}, kaons~\cite{HERMES:2003}, proton~\cite{HERMES:2007,HERMES:2011}, and lambda~\cite{PhysRevLett.130.142301}. The proposal here is to measure these observables using CLAS12 at 22~GeV for many reasons. First, it will allow to cover a wide phase-space including both the valence and sea quark regions. The upgraded detection capability as well as luminosity will allow the study of hadronization for a large variety of hadron species. It is expected to perform these measurements with an unprecedented precision, which will help understand some of the HERMES data taken with limited statistics. Furthermore, the combination of CLAS6 and CLAS12 datasets taken at 5~GeV and 11~GeV then 22~GeV will make CLAS the unique place to study hadronization using a wider kinematical coverage.

In Fig.~\ref{fig:nuclear_sidis}, the three-fold differential projections of all three observables are shown for various hadrons production off carbon target. Since systematic uncertainties for similar measurements at CLAS6 and HERMES have had more than 1\% systematic uncertainties, the study of channels listed here will be also limited by systematic uncertainties.~By running for 60 PAC days with a deuterium target in series with one of the heavy nuclei such as carbon, copper, tin, and lead, and assuming the nominal CLAS12 per-nucleon luminosity of $10^{35}$~cm$^{-2}$s$^{-1}$, one can make a variety of measurements of several hadron species to much higher precision than ever achieved. A larger run period may be needed for precise measurements for rare hadron channels, particularly those with charm quarks. 

\subsubsection{Coherent Nuclear J/$\psi$ Photoproduction}\label{sub:JPsi}

Coherent production of heavy vector mesons (VM) from nuclei is considered a critical measurement for understanding the gluon distribution in nuclei~\cite{EIC_whitepaper}, allowing access both to the $x$ dependence and the spatial distribution of the gluons.~The $J/\psi$ meson is particularly promising for such a measurement; as the lightest heavy vector meson, higher-twist and sea-quark effects are suppressed in its production in favor of two-gluon exchange, while its relatively low mass allows phase space for its production at much lower energies than heavier mesons such as the $\Upsilon$.

Photoproduction of VM in Heavy-Ion ultraperipheral collisions (UPC) has been used at the Large Hadron Collider to study the gluon distributions in heavy nuclei such as $^{208}$Pb, using the electromagnetic field of the ions as a source of photons for photon-nucleus interactions~\cite{2017489,20131273,2021136280,ALICE_jpsi_EPJC,PhysRevC.105.L032201}. One study at BNL used UPC production of $J/\psi$ to examine the gluon structure of the deuteron~\cite{PhysRevLett.128.122303}, including the first observation of coherent $J/\psi$ photoproduction from such a light nucleus. However, UPC interactions are limited for a few reasons, including ambiguity in which nucleus is being probed and the lack of exclusivity in measuring the event. Experiments with electron beams and real photon beams, in the case of Hall~D, are necessary to provide both complementarity and more detailed measurements of these processes.

\begin{wrapfigure}[15]{r}{0.45\textwidth}
 \centering
  \vspace{-0.275cm}
   \includegraphics[clip=true, trim= 0 0.6cm 0 0.45cm, width=0.45\textwidth]{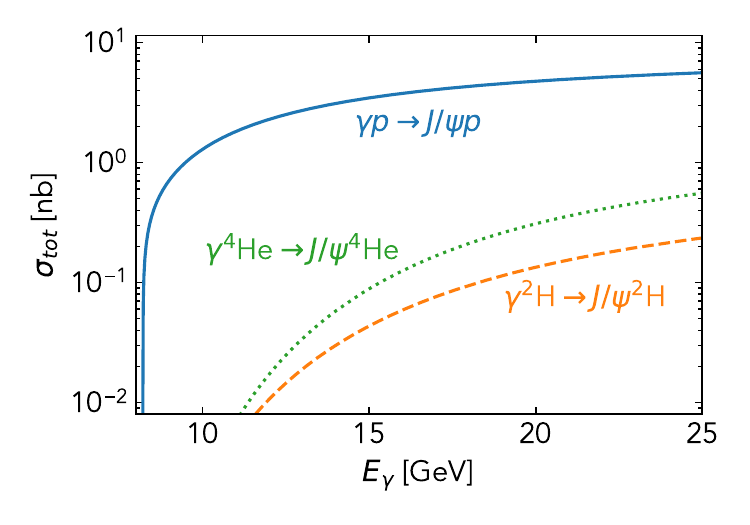}
   \vglue -0.1in \caption{Coherent $J/\psi$ photoproduction cross section from different nuclei, including free proton, deuterium, and $^4$He, as a function of the photon energy.}
\label{fig:coherent_cs}
\end{wrapfigure}

Photoproduction of $J/\psi$ from proton targets has previously been measured in Hall~D~\cite{PhysRevLett.123.072001} at 12~GeV, providing the first detailed measurements of the proton’s gluon distribution in the threshold region of high $x=\frac{m_{J/\psi}^2}{2m_pE_\gamma}$. Similar measurements in nuclei, however, are limited by the sharply falling nuclear form factor. The large mass of the $J/\psi$ requires a high four-momentum transfer $-t=-(p_\gamma-p_{J/\psi})^2$ near the production threshold, heavily suppressing the coherent production of $J/\psi$ with low-energy photons. The minimum momentum transfer decreases with increasing energy of the incident photon, allowing access to the phase space dominated by coherent production. A consequence of this is that while the photoproduction cross section of $J/\psi$ from protons increases slowly with the photon energy, nuclear coherent production cross sections increase much more dramatically as the kinematics grow more favorable, as shown in Fig.~\ref{fig:coherent_cs}. For this reason, while coherent $J/\psi$ photoproduction off nuclei remains challenging at current 12~GeV energies, the tagged photon energies enabled by a 22~GeV electron beam would enable such a measurement. This would provide the first measurement of the nuclear gluon distributions at $x > 0.23$ and would be complementary to UPC measurements in heavy ion collisions and diffractive VM production at the EIC, each of which has limited ability to resolve nuclear gluon dynamics in this near-threshold region.

Two observables are of particular interest in measuring exclusive VM production. The first is the beam photon energy $E_\gamma$, which controls the longitudinal momentum $x$ of the gluons being probed. The second is the momentum transfer , which is conjugate to the impact parameter; performing a Fourier transform of the measured nuclear form factor gives access to the transverse position distribution of the gluons~\cite{PhysRevC.87.024913}. One relation this provides us is the extraction of the gluonic RMS radius of the nucleus, in a manner equivalent to the extraction of the charge radius from the form factor slope:
\begin{equation}
 \langle r^2_g\rangle= \frac{6}{F_A(0)}\frac{dF_A}{dt}\bigg|_{t=0}. 
\end{equation} This observable would be directly comparable to current UPC measurements, allowing validation of the extracted deuteron properties.

The 22~GeV electron beam would enable a much higher-energy flux of real photons in Hall D as compared with the current 12~GeV beam. We perform the following projections assuming an 
electron-tagged photon luminosity of 200 pb$^{-1}$ in the energy range spanning $87.5-97.5\%$ of the beam energy, or $19.25<E_\gamma<20.9$~GeV. This corresponds to a kinematic region of $x\sim 0.25$. We similarly examine the case for a 17~GeV beam, corresponding to $x\sim 0.33$. In each of these cases, lower photon energies, or higher values of $x$, can be reached by reducing the coherent peak of the photon energy spectrum. We note for completeness that the GlueX spectrometer was recently used successfully to measure photonuclear interactions off nuclei for the study of short-range correlations, supporting the ability to perform measurements with nuclear targets such as deuterium, helium and even lead~\cite{GlueX:2020dvv}.

The projections shown here were calculated by fitting previous measurements of the photoproduction cross section from the proton~\cite{PhysRevLett.123.072001, PhysRevLett.35.483}, assuming the dipole form factor for the proton extracted from the GlueX measurements in order to extrapolate the forward $t=0$ photoproduction cross section as a function of $x$, scaling this forward cross section by the appropriate factor $A^2$ for each nucleus, and replacing the proton form factor with nuclear form factors obtained by Fourier transforming single-nucleon densities~\cite{PhysRevC.89.024305} calculated using the AV18~\cite{Wiringa:1994wb} interaction. This provides a data-driven model of the differential cross section for coherent photoproduction as a function of photon energy $E_\gamma$ and momentum transfer $t$. We have considered here the light nuclei $^2$H and $^4$He; heavier nuclei can also be measured, but would result in events concentrated closer to $t=0$, decreasing our ability to resolve the transverse structure.
 
\begin{figure}[!b]
 \begin{center}
   \includegraphics[clip=true, trim= 0.5cm 0 0.2cm 0.45cm,width=0.475\textwidth]{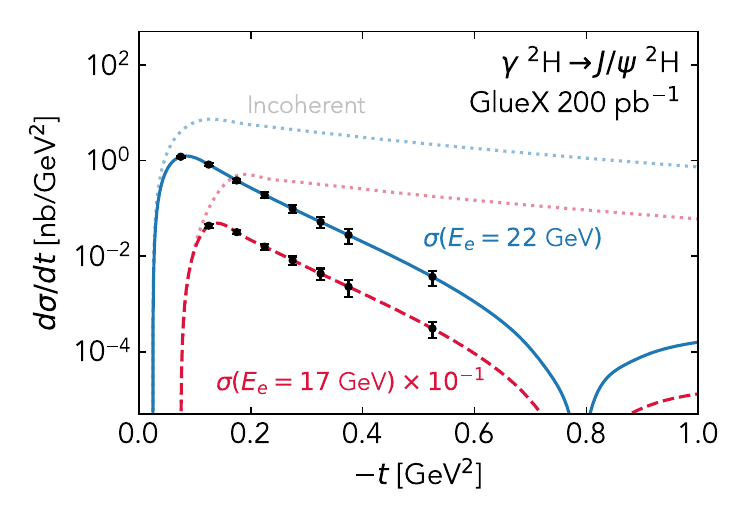}
   \includegraphics[clip=true, trim= 0.5cm 0 0.2cm 0.45cm,width=0.475\textwidth]{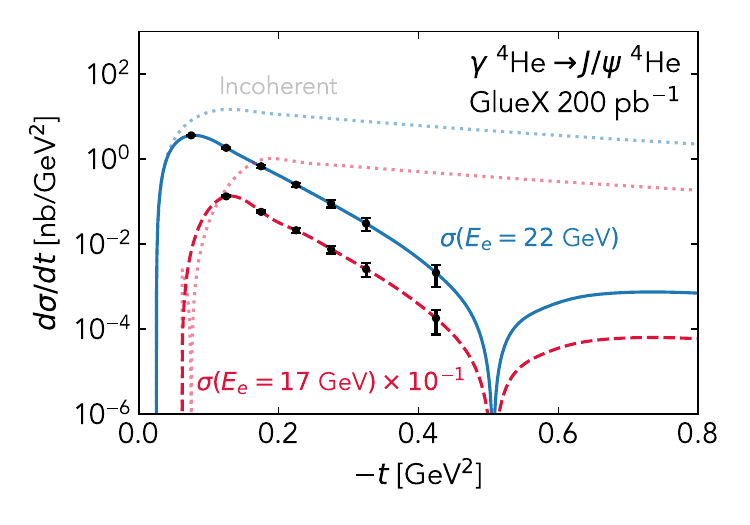}
\end{center}
\vglue -0.3in
 \caption{Expected nuclear coherent $J/\psi$ measurement statistics for deuterium (left) and Helium-4 (right). Calculations were performed at electron energies of 17~GeV (scaled down by a factor of 10) and 22~GeV, assuming an integrated luminosity of 200 pb$^{-1}$ for photons carrying between $87.5-97.5\%$ of the beam energy. Also included are projections for the incoherent background (dotted line), which dominate over the coherent signal at these kinematics.}
\label{fig:coherent_projections}
\end{figure}

Figure~\ref{fig:coherent_projections} shows the projected statistical precision that can be achieved in the measurement of this differential cross section. Here we have considered only the decay channel $J/\psi\rightarrow e^+e^-$, with a branching ratio of 5.97\%, and conservatively assume a 50\% efficiency for detecting the $J/\psi$ and selecting the event. We find that a substantial number of coherent events can be measured in the region of lower momentum-transfer, but statistics do not allow mapping the form factor beyond $|t| > 0.5$ GeV$^2$; more complex features of the form factor such as the node structure at higher $|t|$ would require much higher statistics or beam energy.

We note that a major complication comes from the substantial incoherent background $\gamma A\rightarrow J/\psi X$, in which the nucleus does not remain intact but breaks up into constituent nucleons. This process, the cross sections for which were estimated by scaling the cross section from a free proton, is increasingly dominant over coherent production at higher $|t|$, likely limiting the feasible measurement region to low-momentum-transfer kinematics, and does not give insight to the average gluon structure of the nucleus; rather, it is sensitive to fluctuations in the nuclear state. It is possible to reject this background by detecting final-state particles resulting from nuclear breakup, such as relatively low-momentum protons or neutrons. Such event rejection would likely require target engineering to ensure that incoherent nucleons are able to reliably exit the target. Similar targets have previously been used in JLab experiments requiring low-momentum particle tagging~\cite{CLAS:2011qvj}.
\clearpage \section{QCD Confinement and Fundamental Symmetries}
\label{sec:wg6}

The JLab 22 GeV upgrade will enable high-precision measurements of the Primakoff production of pseudoscalar mesons with results to explore the chiral anomaly and the origin and dynamics of chiral symmetry breaking, allow model-independent determinations of the light quark mass ratio and the $\eta$-$\eta'$ mixing angle, and provide critical input to the hadronic light-by-light corrections to the anomalous magnetic moment of the muon. The higher beam energy will also greatly improve the reach of direct searches for light (sub-GeV) dark matter scalars and pseudoscalers through Primakoff production, add to the reach and robustness of Standard Model tests using parity-violation in deep inelastic scattering (PVDIS), and expand opportunities for studies with secondary beams. 

\begin{figure}[h!]
\centering
\includegraphics*[width=0.6\textwidth]{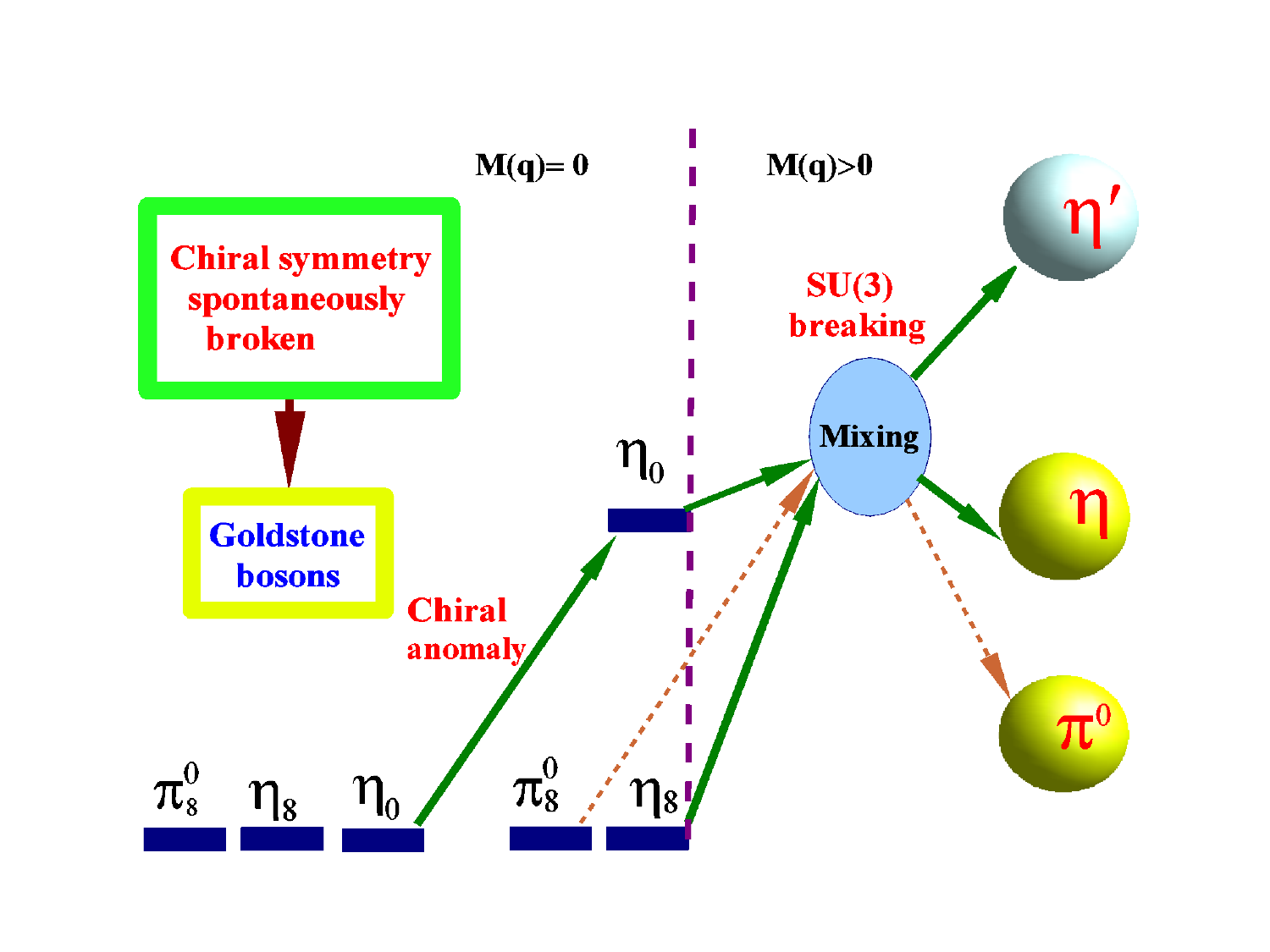}
\caption{The QCD symmetries at low-energy and the properties of light pseudoscalar meson $\pi^0$, $\eta$, and $\eta^{\prime}$.}
\label{primex-physics}
\end{figure}

\subsection{Precision Measurements of $\pi^0$, $\eta$, and $\eta'$ Decays}

Three lightest neutral self-conjugated pseudoscalar mesons, $\pi^0$, $\eta$, and $\eta'$, are manifestation of fundamental symmetries of the nonperturbative QCD ground state. They provide a rich laboratory for tests of the fundamental symmetries in the Standard Model and beyond~\cite{Gan:2020aco}.

In the chiral limit where the quark masses are set to zero, as shown in the left-hand side of Fig.~\ref{primex-physics}, the QCD Lagrangian $\mathcal{L}_{\rm QCD}$ is invariant under the global symmetry group of $SU(3)_L \times SU(3)_R \times U(1)_A \times U(1)_B$. 
 These symmetries, however, appear in nature differently. 
The condensation of quark--antiquark pairs in the QCD vacuum spontaneously breaks the chiral symmetry $SU(3)_L \times SU(3)_R$ down to the flavor $SU(3)_V$ symmetry. Each broken generator results in a massless Nambu--Goldstone boson, corresponding to the octet of pseudoscalar mesons ($\pi^0$, $\pi^{\pm}$, $K^{\pm}$, $K^0$, $\bar{K}^0$, and $\eta_8$).
The $U(1)_A$ symmetry is explicitly broken by the quantum fluctuations of 
the quarks coupling to the gauge fields, known as the
chiral anomaly~\cite{Adler:1969gk,Bell:1969ts}. Such anomalous symmetry breaking has a purely quantum-mechanical origin, representing one of the most profound symmetry breaking phenomena. 
The anomaly  associated with quarks coupling to the gluon fields prevents  $\eta_0$ from being a Goldstone boson; the same anomaly is also  related to so called ``the $\theta$ term" in the strong PC problem. Consequently, the singlet $\eta_0$ acquires a nonvanishing mass in the chiral limit~\cite{tHooft:1976rip}. This axial anomaly is proportional to $1/N_c$,  where $N_c$ is the number of colors. Therefore  $\eta_0$ does become a Goldstone boson~\cite{Witten:1979vv} at the large $N_c$ limit. On the other hand, the  anomaly associated with the coupling of quarks to the electromagnetic field, is primarily responsible for the two-photon decays of $\pi^0$, $\eta$, and $\eta^{\prime}$.  

If the quark masses are turned on (which are small compared to the chiral symmetry breaking scale $\sim 1$~GeV), as shown in the right-hand side of Fig.~\ref{primex-physics}, the chiral symmetry is explicitly broken and thereby generates masses for the Nambu--Goldstone bosons, following the mechanism discovered by Gell-Mann, Oakes, and Renner~\cite{Gell-Mann:1968hlm}.
Furthermore, the unequal quark masses break the $SU(3)_V$ flavor symmetry (and, to a lesser extent, isospin), leading to mixing between the chiral limit states.   The mixing of $\pi^0$ and  $\eta$ is proportional to   $m_u-m_d$, and   the mixing of $\eta$ and $\eta'$  is proportional to $m_s-\hat m$ ($\hat m=(m_u+m_d)/2$). 
In addition, the isospin symmetry breaking gives rise to  hadronic decays of $\eta\to\pi^+\pi^-\pi^0$ and $\eta\to3\pi^0$, where their partial decay widths are normalized to the partial decay width of $\eta\to \gamma\gamma$ experimentally. These decays constitute one of the relatively rare isospin-breaking hadronic observables that the electromagnetic effects are strongly suppressed~\cite{Bell:1968wta,Sutherland:1966zz}, offering clean experimental access to the light quark mass difference $m_u-m_d$.  
Lastly, $U(1)_B$ baryon number symmetry is also broken explicitly by the axial anomaly associated with electroweak gauge fields, however,  this effect is  negligible except in the very early Universe~\cite{Kuzmin:1985mm}.

\vspace{0.3cm}
\begin{wrapfigure}[14]{r}{0.4\textwidth}
   \vspace{-1cm}
    \begin{center}  
        \includegraphics[angle=0, width=0.25\textwidth]{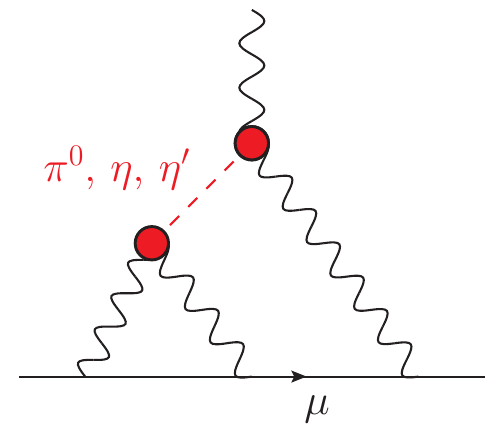}
    \end{center}
     \vspace{-0.8cm}
     \caption{Pseudoscalar pole contributions to hadronic light-by-light scattering in the anomalous magnetic moment of the muon; crossed diagrams are not shown.  The red blobs denote the pseudoscalar transition form factors.  Figure taken from Ref.~\cite{Gan:2020aco}.}
      \label{fig:HLbL}
\end{wrapfigure}

A study of $\pi^0$, $\eta$, and $\eta^{\prime}$ also plays a critical role in searching for new physics beyond the Standard Model. The uncertainty in the Standard Model prediction of the anomalous magnetic moment of the muon ($a_\mu$) is dominated by hadronic effects~\cite{Aoyama:2020ynm}.~Next to hadronic vacuum polarization, the second-most-important contribution is given by a loop topology dubbed hadronic light-by-light scattering (HLbL), which in turn is dominated by pole contributions of the lightest flavor-neutral pseudoscalars $\pi^0$, $\eta$, and $\eta'$, as shown in Fig.~\ref{fig:HLbL}. The strength of these contributions is determined by the singly and doubly virtual transition form factors (TFFs).

For the largest individual HLbL contribution, the $\pi^0$ pole term, a data-driven, dispersion-theoretical
determination of the TFF has been performed~\cite{Hoferichter:2018dmo,Hoferichter:2018kwz}.  It is based on the incorporation of the lowest-lying singularities due to $2\pi$ and $3\pi$ intermediate states, information
on the asymptotic behavior in QCD, and experimental data for the $\pi^0\to\gamma\gamma$ decay width as well as the spacelike singly
virtual TFF available at high energies.
The result allows for a precise assessment of the different sources of uncertainty,
\begin{equation}\label{eq:g-2result}
 a_\mu^{\pi^0}=63.0(0.9)_{F_{\pi\gamma\gamma}}(1.1)_\text{disp}\big({}^{2.2}_{1.4}\big)_\text{BL}(0.6)_\text{asym}\times 10^{-11}
 =63.0\big({}^{2.7}_{2.1}\big)\times 10^{-11} \,,
\end{equation}
where the individual uncertainties refer to the form factor normalization, dispersive input, experimental 
uncertainty in the singly virtual data, and the onset of the asymptotic contribution, in order.
Here, the normalization uncertainty has already been reduced from the original publications~\cite{Hoferichter:2018dmo,Hoferichter:2018kwz} to the value in Eq.~\eqref{eq:g-2result} given in the White Paper~\cite{Aoyama:2020ynm}, thanks to the improved value for the $\pi^0$ radiative width obtained by the PrimEx Collaboration~\cite{PrimEx-II:2020jwd} (shown as a  solid blue point in Fig.~\ref{projection24}).  
The result in Eq.~\eqref{eq:g-2result} is in good agreement with the lattice QCD calculation~\cite{Gerardin:2019vio}, however, mainly after readjusting the TFF normalization
to the PrimEx experimental value~\cite{PrimEx-II:2020jwd}. Remarkably, while the chiral corrections~\cite{Goity:2002nn,Ananthanarayan:2002kj,Kampf:2009tk}
increase the $\pi^0$ width about 4\% compared to the prediction based on the chiral anomaly alone (see Fig.~\ref{projection24}) that is  in slight tension with the PrimEx result, the lattice calculation~\cite{Gerardin:2019vio} points toward to  a form factor normalization that is on the small side.
Further independent studies of the $\pi^0$ TFF in lattice QCD are in progress~\cite{Burri:2022gdg,Verplanke:2022eto}. An experimental opportunity enabled by an energy upgrade to 22~GeV to produce $\pi^0$ off an atomic electron target (described below) to further improve the experimental precision for both decay width and TFF of $\pi^0$, will be clearly important.

Beyond the normalization, a precise measurement of the slope of the $\pi^0$ TFF provides an important constraint on the calculation of $a_\mu$.
 The sum rule~\cite{Hoferichter:2018dmo,Hoferichter:2018kwz}
\begin{equation}
\frac{m_{\pi^0}^2}{F_{\pi\gamma\gamma}}\frac{\partial}{\partial q^2}F_{\pi^0\gamma^*\gamma^*}(q^2,0)\bigg|_{q^2=0}
=31.5(2)_{F_{\pi\gamma\gamma}}(8)_\text{disp}(3)_\text{BL}\times 10^{-3} 
=31.5(9)\times 10^{-3} \,
\end{equation}
is more sensitive to the dispersive input and allows for an important cross-check on the matching to the high-energy asymptotics~\cite{Hoferichter:2014vra}.

While significant work on dispersion-theoretical analyses of the $\eta$ and $\eta'$
TFFs has already been performed~\cite{Hanhart:2013vba,Kubis:2015sga,Holz:2015tcg,Gan:2020aco,Holz:2022hwz},
a fully data-driven determination of the corresponding $a_\mu$ contributions in analogy to $\pi^0$ has not yet been completed. The numbers in Ref.~\cite{Aoyama:2020ynm} are based on a phenomenological data-driven approach using rational approximants~\cite{Masjuan:2017tvw},
\begin{equation}
    a_\mu^{\eta} =  16.3(1.4)\times10^{-11}\,, \qquad
    a_\mu^{\eta'} = 14.5(1.9)\times10^{-11}\,, \label{eq:eta-PA}
\end{equation}
where the uncertainties could be further improved and  be decomposed in terms of individual
sources of input in the manner of Eq.~\eqref{eq:g-2result} by the future  experimental
data. Both $\eta$ and $\eta'$ TFF normalizations and (singly virtual) high-energy asymptotics therein are closely linked to $\eta$--$\eta'$ mixing~\cite{Escribano:2015yup}. For $\eta$, a recent lattice-QCD calculation of $a_\mu^{\eta}$~\cite{Alexandrou:2022qyf} (with a further one in progress~\cite{Verplanke:2022eto}) suggests fair agreement with Eq.~\eqref{eq:eta-PA}, but again demonstrates quite some tension with the phenomenological low-energy parameters, normalization and slope, of the corresponding TFF.~Interestingly, once again the radiative width for $\eta\to\gamma\gamma$ is relatively small compared to the Particle Data Group average~\cite{ParticleDataGroup:2022pth} that includes only the results from the $e^+e^-$ collision measurements, and much closer to a Primakoff result by the Cornell Collaboration~\cite{Browman:1974sj} published in 1974. Clearly, high-precision experimental determinations of the decay widths for $\pi^0$,\,
$\eta,\,\eta'\to\gamma\gamma$ and their space-like singly virtual TFF's at small $Q^2$  within the modern JLab Primakoff program would be most valuable in this respect.

In summary, a study of $\pi^0$, $\eta$, and $\eta^{\prime}$ will have great potential to shed light on some fundamental questions in the Standard Model and beyond: testing the chiral anomaly and probing the origin and dynamics of chiral symmetry breaking; offering a clean path for model independent determinations of the light quark-mass ratio and the $\eta$-$\eta^{\prime}$ mixing angle; and  providing critical inputs to the theoretical calculations of the hadronic light-by-light contributions to the anomalous magnetic moment of the muon~\cite{Gan:2020aco}.

In the past two decades, the PrimEx Collaboration has successfully developed a comprehensive Primakoff experimental program at JLab 6 and 12~GeV with nuclear targets.
The Primakoff effect~\cite{Primakoff:1951iae} is a process of high-energy  photo- or electroproduction of mesons  in the Coulomb field of a target.
This program includes high-precision measurements of the two-photon decay widths $\Gamma(P\to\gamma\gamma)$ and the spacelike transition form factors $F_{P\gamma^*\gamma^*}(-Q^2,0)$ at four-momentum transfers $Q^2 = 0.001 \ldots 0.3$~GeV$^2$,
where $P$ represents $\pi^0$, $\eta$, and $\eta^{\prime}$ ~\cite{Gan:2014pna}. 
The JLab 22 GeV upgrade will enable such measurements with experimental sensitivities not previously achievable.

\subsubsection{Primakoff Production of $\pi^0$ From Atomic Electrons}

\begin{figure}
\centering
\includegraphics*[width=0.6\textwidth]{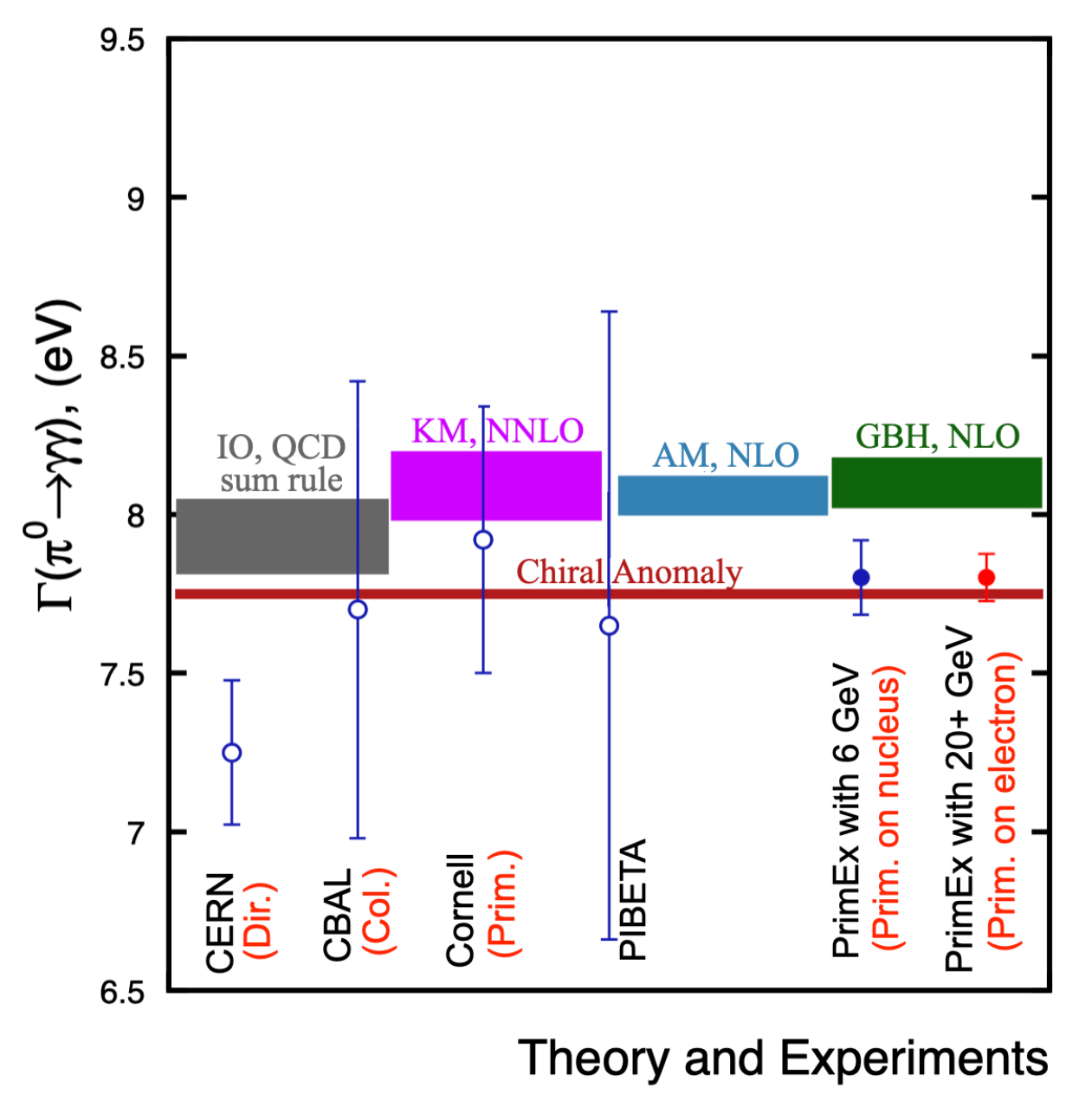}
\vglue -0.2in
\caption{The projected precision on $\Gamma(\pi^0\rightarrow \gamma\gamma)$ with an atomic electron target (the red point) and the previous published results (the blue points) listed in PDG~\cite{ParticleDataGroup:2022pth}.
Theoretical predictions are: chiral anomaly~\cite{Adler:1969gk,Bell:1969ts} (dark red~band); IO, QCD sum rule~\cite{Ioffe:2007eg} (gray~band); KM, ChPT NNLO~\cite{Kampf:2009tk}  (magenta~band); AM, ChPT NLO~\cite{Ananthanarayan:2002kj} (blue~band); GBH, ChPT  NLO~\cite{Goity:2002nn} (green~band). 
}
\label{projection24}
\end{figure}

As the lightest hadron, $\pi^0$ plays a special role in our understand of QCD confinement.
Its radiative decay width is one of very rare parameters in low-energy QCD that can be predicted at $\approx 1$\% precision, offering an important test of QCD confinement. 
 The chiral anomaly drives the decay of the $\pi^0$ meson into two photons with no adjustable parameters in the predicted decay width~\cite{Adler:1969gk,Bell:1969ts,Bernstein:2011bx}: 
\begin{equation}
\Gamma(\pi^0\rightarrow\gamma\gamma) = \frac{m_{\pi^0}^3\alpha^2N_{c}^2}{576\pi^3F_{\pi^0}^2} = 7.750\pm0.016~\mbox{eV}, 
\end{equation}
where $\alpha$ is the fine-structure constant, 
$m_{\pi^0}$ is the $\pi^0$ mass, $N_{c}=3$ is the number of colors in QCD. This prediction, shown as a dark red band in Fig.~\ref{projection24}, is exact in the chiral limit (except for an experimental uncertainty contributed from the pion decay constant, $F_{\pi^0}=92.277\pm 0.095\,$MeV, extracted from the charged pion weak decay~\cite{ParticleDataGroup:2018ovx}).
Due to the small mass of the $\pi^0$,  higher order corrections to this prediction induced by the non-vanishing quark masses are small ($\sim$4\%) and can be calculated with percent accuracy.
 Several independent theoretical calculations  are shown as color bands in 
Fig.~\ref{projection24}. These calculations are performed either in the framework of chiral perturbation theory (ChPT) up to ${\cal{O}}(p^6)$ (NLO)~\cite{Goity:2002nn,Ananthanarayan:2002kj} (NNLO corrections are considered in Ref.~\cite{Kampf:2009tk}) or based on QCD sum rules~\cite{Ioffe:2007eg}.

The two most recent experiments (PrimEx-I and PrimEx-II) on $\pi^0$ were carried out with nuclear targets ($^{12}$C, $^{28}$Si, and $^{208}$Pb) during the JLab 6~GeV era. The weighted average of PrimEx I and II results is $\Gamma (\pi^0\to\gamma\gamma)=7.802 (052)_{\rm stat}(105)_{\rm syst}$ eV~\cite{PrimEx-II:2020jwd}. 
This result with a 1.50\% total uncertainty, shown as a solid blue point in Fig.~\ref{projection24},   represents the most accurate measurement of this fundamental parameter to date. 
Its central value agrees to the leading-order chiral anomaly prediction
~\cite{Bernstein:2011bx}
 and is $2\sigma$ below the theoretical calculations
 ~\cite{Goity:2002nn,Ananthanarayan:2002kj,Kampf:2009tk,Ioffe:2007eg} 
 based on higher-order corrections to the anomaly. 
 This is clearly a significant result calling for further investigations. An experimental opportunity enabled by a JLab 22~GeV upgrade to produce $\pi^0$ off an atomic electron target to reach a sub-percent precision on $\Gamma (\pi^0\to\gamma\gamma)$, as shown in Fig.~\ref{projection24}, is  necessary to better understand the discrepancy between the existing experimental result and the high-order QCD predictions. 
 
\begin{figure}
\centering
\includegraphics*[width=0.5\textwidth]{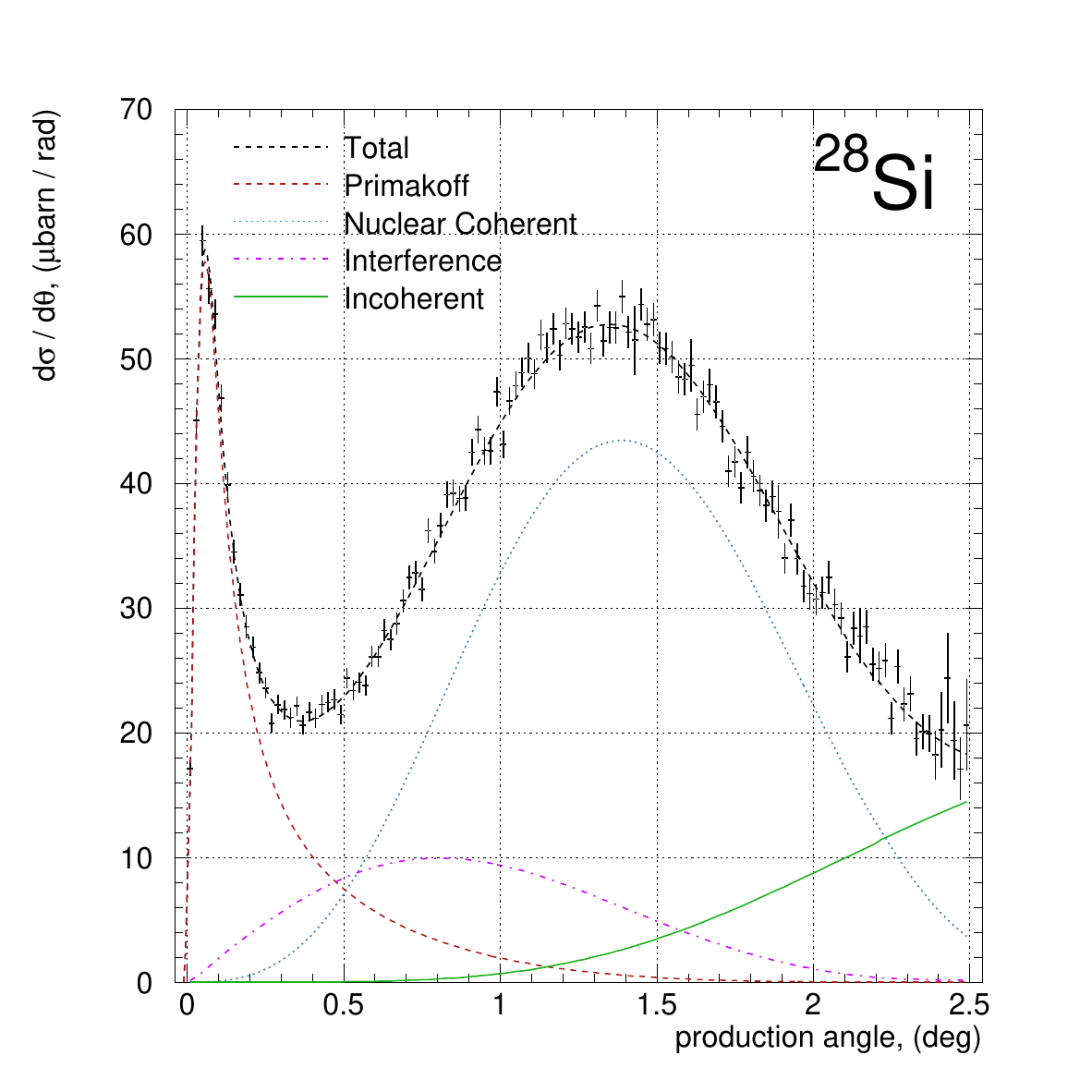}
\includegraphics*[width=0.4\textwidth]{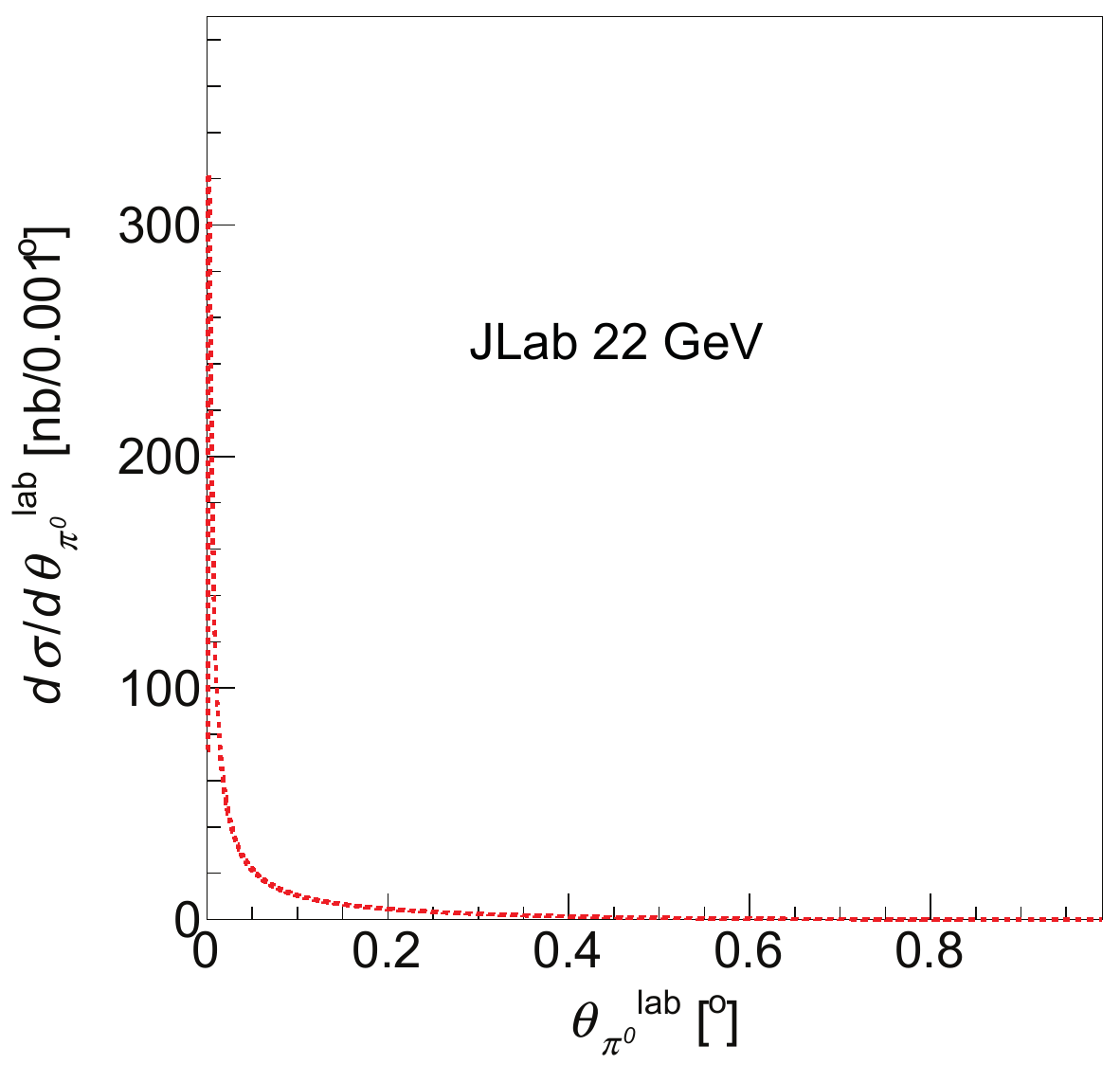}
\vglue -0.15in
\caption{Left: The measured cross section of $\gamma + $Si$\rightarrow \pi^0 + $Si from the PrimEx II experiment
~\cite{PrimEx-II:2020jwd} 
(with $E_\gamma$ of 4.45--5.30 GeV). Right: The projected cross section of $\gamma + e\rightarrow \pi^0 + e$ at JLab 22 GeV (with $E_\gamma$ of 20--22~GeV and without smearing of the experimental resolutions).}
\label{primex-II}
\end{figure}

The biggest challenge for using a nuclear target is that the Primakoff effect is not the only mechanism for the production of mesons, as shown Fig.~\ref{primex-II} (left). There is nuclear coherent background from strong production, an interference between the strong and Primakoff production amplitudes, and the incoherent nuclear process. The classical method of extracting the Primakoff amplitude is to fit the measured total differential cross section in the forward direction based on the different characteristic behaviors of the production mechanisms with respect to the production angles. If using an electron target, all these nuclear backgrounds can be eliminated.
 
The threshold for photo- or electroproduction of $\pi^0$ off an electron target is 18 GeV. An energy upgrade of the electron beam at JLab to 22 GeV will thus enable precision measurements of radiative decay width (using photo-production) and transition form factor (using electroproduction)  of $\pi^0$ off an electron for the first time. The Primakoff production off an atomic electron has significant advantages over a nucleus in the following areas:
\begin{itemize}
\item {\em Elimination of all nuclear backgrounds}. The largest systematic uncertainty in the previous PrimEx I and II experiments
~\cite{PrimEx-II:2020jwd} 
with nuclear targets ($^{12}$C, $^{28}$Si, and $^{208}$Pb) is due to yield extraction ($\sim 1$~\%) to separate Primakoff events from the nuclear backgrounds, as shown in Fig.~\ref{primex-II} (left). With an electron target, all nuclear backgrounds will be eliminated, see Fig.~\ref{primex-II} (right).
\item {\em Elimination of uncertainties due to nuclear effects}. 
Extracting $\Gamma(\pi^0\rightarrow\gamma\gamma)$ from the measured Primakoff cross section off a nucleus requires knowing the nuclear charge form factor, corrected for the initial-state interaction of the incoming photon (or electron in the case of TFF measurement) and the final-state interaction of the outgoing mesons in the nuclear medium. Using an electron target (a point-like particle) eliminates all uncertainties related to these nuclear effects.
\item {\em Enabling recoil detection}. For the case of a nuclear target, the momentum of the recoil nucleus in the Primakoff process is too small to measure. Primakoff production off an electron target will enable detection of the recoil electron to suppress  backgrounds, such as those from the beam line, offering a cleaner Primakoff signal. 

\end{itemize}

\begin{figure}
\centering
\includegraphics*[width=12.cm, clip]{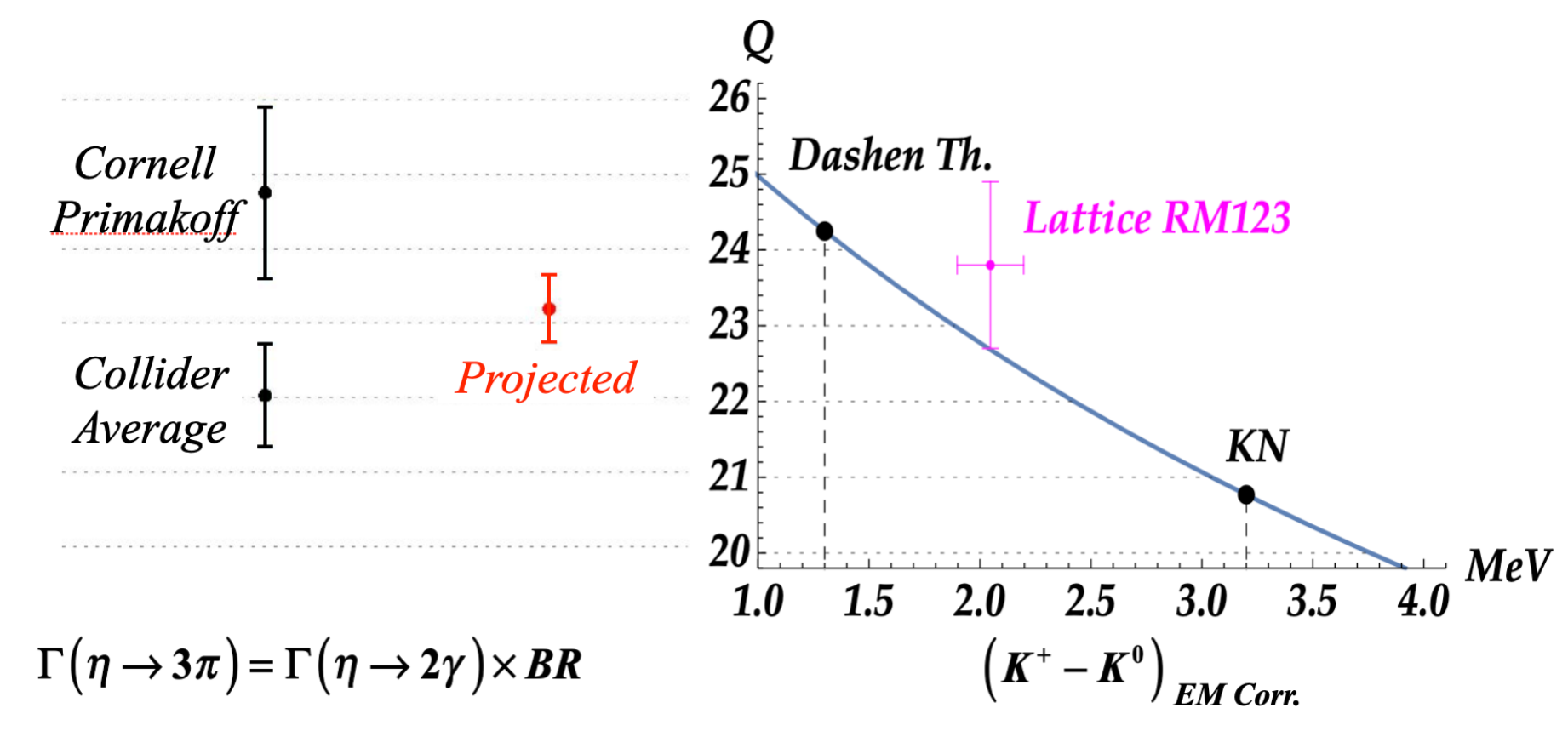}
\vglue -0.15in
\caption{Light quark mass ratio ${\cal Q}$ determined by two different methods.
The left-hand side are calculated from the $\eta\to 3\pi$ decay determined by using the Cornell Primakoff~\cite{Browman:1974sj}, the collider average~\cite{ParticleDataGroup:2018ovx} experimental results, and the projected Primakoff measurement at JLab 22 GeV for $\Gamma(\eta\to \gamma\gamma)$ as input. The right-hand side shows the results of  ${\cal Q}$ obtained from the kaon mass difference with different theoretical estimates for the electromagnetic corrections based on Dashen's theorem, Ref.~\cite{Kastner:2008ch} (KN), and the lattice~\cite{Giusti:2017dmp}. This figure is taken from Ref.~\cite{Gan:2020aco} with modifications.}
\label{fig:quark_mass_ratio}
\end{figure}

The projected precision on $\Gamma(\pi^0\rightarrow \gamma\gamma)$ with an atomic electron target is $\approx 0.95$\% (the red point in Fig.~\ref{projection24}), a one-third reduction of uncertainty from the previous PrimEx I and II  (the blue point in Fig.~\ref{projection24})~\cite{PrimEx-II:2020jwd}. 
It will independently verify the observed discrepancy
between the previous PrimEx result~\cite{PrimEx-II:2020jwd}  and the high-order QCD predictions, offering a stringent test of low-energy QCD.

\subsubsection{Primakoff Productions of $\eta$ and $\eta'$ From Nuclear Targets}
\begin{figure}[t]
\begin{center}
\includegraphics[width=0.45\linewidth,angle=0]{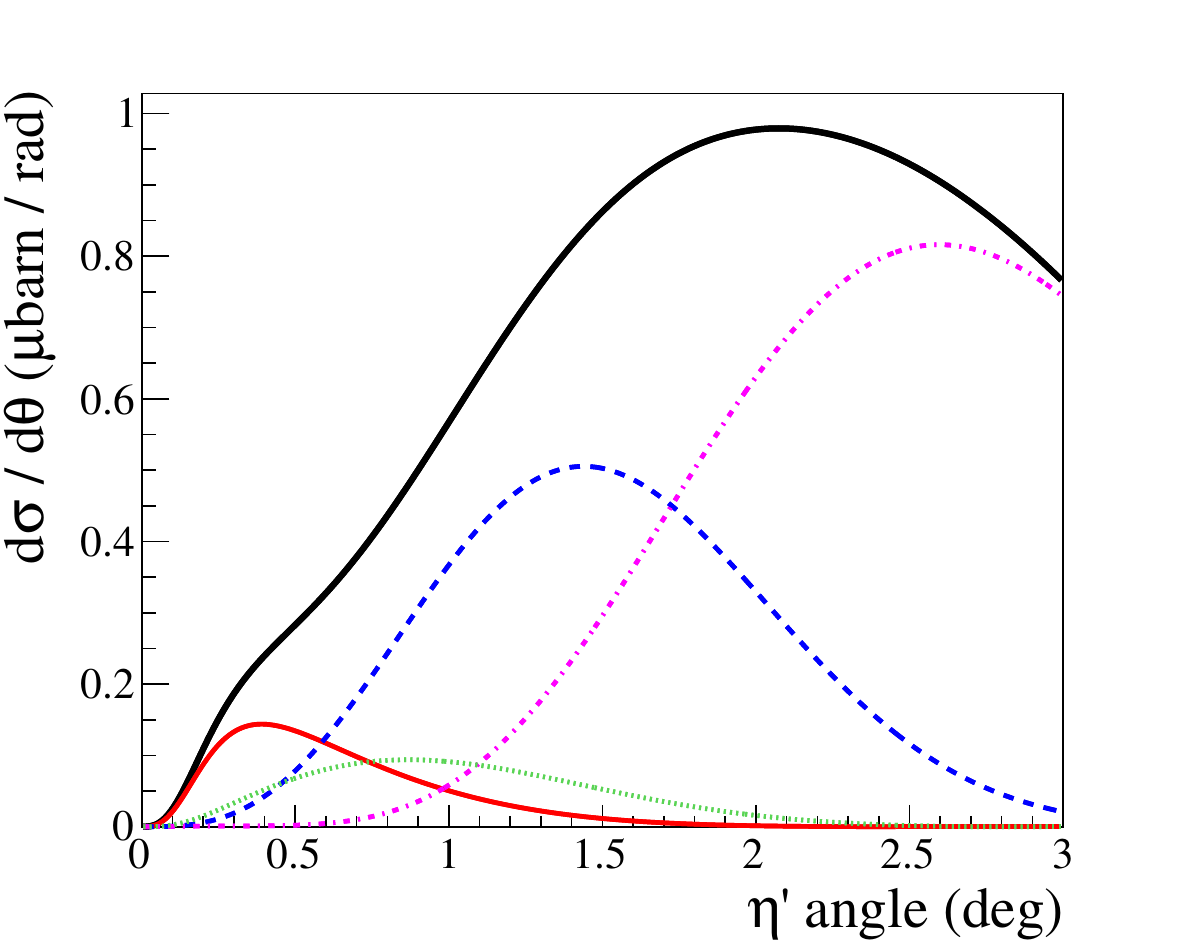} 
\includegraphics[width=0.45\linewidth,angle=0]{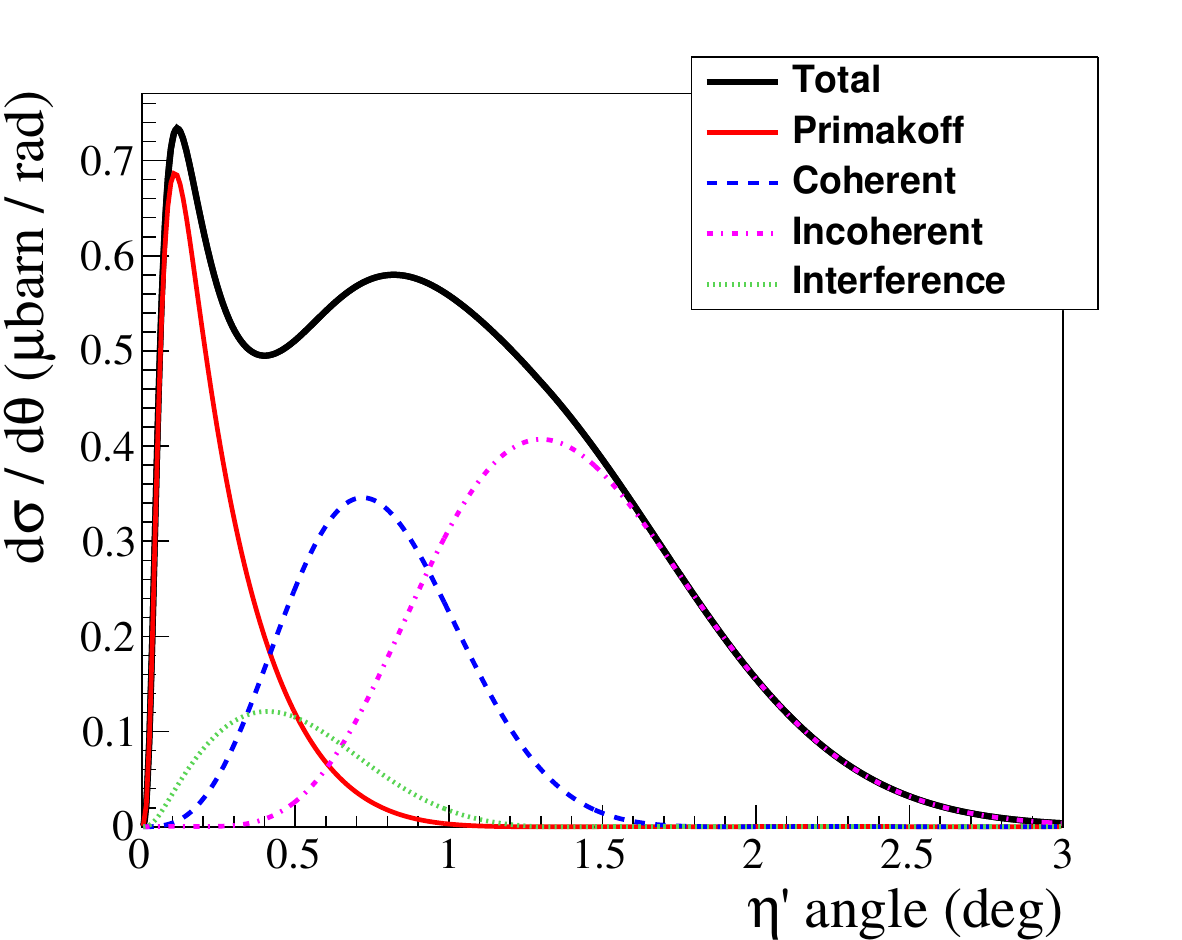}
\end{center}
\vglue -0.15in
\caption{Differential cross sections of  $\gamma + ^4\!He\rightarrow\eta^{\prime}+ ^4\!He$  as a function of the $\eta^{\prime}$ production angles for 
the beam energy of 10 GeV (left) and of 20 GeV (right).}
\label{fig:etap}
\end{figure}

In addition, the 22 GeV upgrade will greatly enhance the  Primakoff measurements of the two-photon decay widths and the transition form factors  of $\eta$ and $\eta'$ off nuclear targets. As shown in Fig.~\ref{primex-II}
(left), the Primakoff cross section is peaked at a small polar angle, $\theta_{Pr} \sim \frac{m^2}{2E^2}$, and increases with the beam energy, 
${[\frac{d\sigma_{Pr}}{d\Omega}]}_{max}\sim \frac{Z^2E^4}{m^3}$,  where  $E$, $m$, and  $Z$ are the beam energy, the meson mass and the  charge of target, respectively. The nuclear coherent cross section has a broader distribution  peaked at a relatively larger angle, $\theta_{NC} \sim\frac {2}{EA^{1/3}}$. A higher beam energy will help better separating  the Primakoff from the nuclear  backgrounds, as well as increasing the Primakoff cross section, which is more important for massive mesons, such as $\eta^{\prime}$, as demonstrated in Fig~\ref{fig:etap}.

The PrimEx-eta experiment~\cite{PrimExeta:2010} on the $\eta$ radiative decay width, $\Gamma(\eta \rightarrow \gamma\gamma)$, was recently completed for data collection in 2022 with JLab 12~GeV. The data analysis is in progress with an anticipated precision of 4-6\% (based on the current preliminary assessment). With a JLab 22~GeV upgrade, the projected precision for $\Gamma(\eta \rightarrow \gamma\gamma)$ will be at level of 2\%. The existing published results on this parameter were performed using two photon interactions either  via the Primakoff effect or through $e^+e^-$ collisions ($e^+e^-\rightarrow\gamma^*\gamma^*e^+e^-\rightarrow\eta e^+e^-$). The collider results listed in PDG~\cite{ParticleDataGroup:2022pth} have individual experimental uncertainty ranging  from 4.6\% to 25\%. They are consistent within the experimental uncertainties, however,  their average value is about $\sim 4\sigma$ larger than the previous Primakoff result~\cite{Browman:1974sj} (with 14\% precision) published by the Cornell Collaboration in 1974.  New precision Primakoff result from JLab at 22~GeV (or even at the current 12~GeV potentially) will help resolving this long-standing puzzle. 

All other $\eta$ partial decay widths listed in the PDG~\cite{ParticleDataGroup:2022pth} are normalized to the two-photon decay width. A precision measurement of $\Gamma(\eta \rightarrow \gamma\gamma)$ will improve all other partial decay widths in the $\eta$ sector, thus  offering a broader impact. One of such examples is to determine the light quark mass ratio, ${\cal Q} \equiv (m_s^2-\hat{m}^2)/(m_d^2-m_u^2)$, by improving the accuracy of the $\eta\rightarrow 3\pi$ decay width~\cite{Leutwyler:1996np}. The fundamental parameter ${\cal Q}$ drives isospin violation in the Standard Model. In most cases, however, the isospin-violating observables are also affected by electromagnetic effects. In order to extract information on ${\cal Q}$, one must first calculate and disentangle the contribution due to electromagnetic interactions. For example, in the case of $K^+$--$K^0$ mass difference as shown in  Fig.~\ref{fig:quark_mass_ratio} (right), the extracted ${\cal Q}$ from such observable is sensitive to the theoretical calculations of the electromagnetic correction.
By contrast, the $\eta\rightarrow 3\pi$ decay is caused almost exclusively by the isospin symmetry 
breaking part of the Hamiltonian $\sim (m_u-m_d)(u{\bar u}-d{\bar d})/2$. Moreover, Sutherland's theorem~\cite{Bell:1968wta,Sutherland:1966zz} forbids electromagnetic contributions in the chiral limit; and contributions of order $\alpha$ are also suppressed by $(m_u+m_d)/\Lambda_\mathrm{QCD}$.
These single out $\eta\rightarrow 3\pi$ to be the best path for an accurate determination of  ${\cal Q}$~\cite{Leutwyler:1996np, Gan:2020aco}. As shown in Fig.~\ref{fig:quark_mass_ratio} (left), the largest systematic uncertainty for the current value of ${\cal Q}$ determined from $\eta\rightarrow 3\pi$ is dominated by the experimental discrepancy between the Cornell Primakoff result~\cite{Browman:1974sj} and the collider average~\cite{ParticleDataGroup:2022pth} for $\Gamma(\eta \rightarrow \gamma\gamma)$. The projected Primakoff measurement of $\Gamma(\eta\to \gamma\gamma)$ at JLab 22~GeV upgrade will offer a more precise determination of ${\cal Q}$ by resolving this discrepancy,  shown as a red point in Fig.~\ref{fig:quark_mass_ratio} (left).

\begin{figure}[h]
\begin{center}
\includegraphics[width=0.75\linewidth,angle=0]{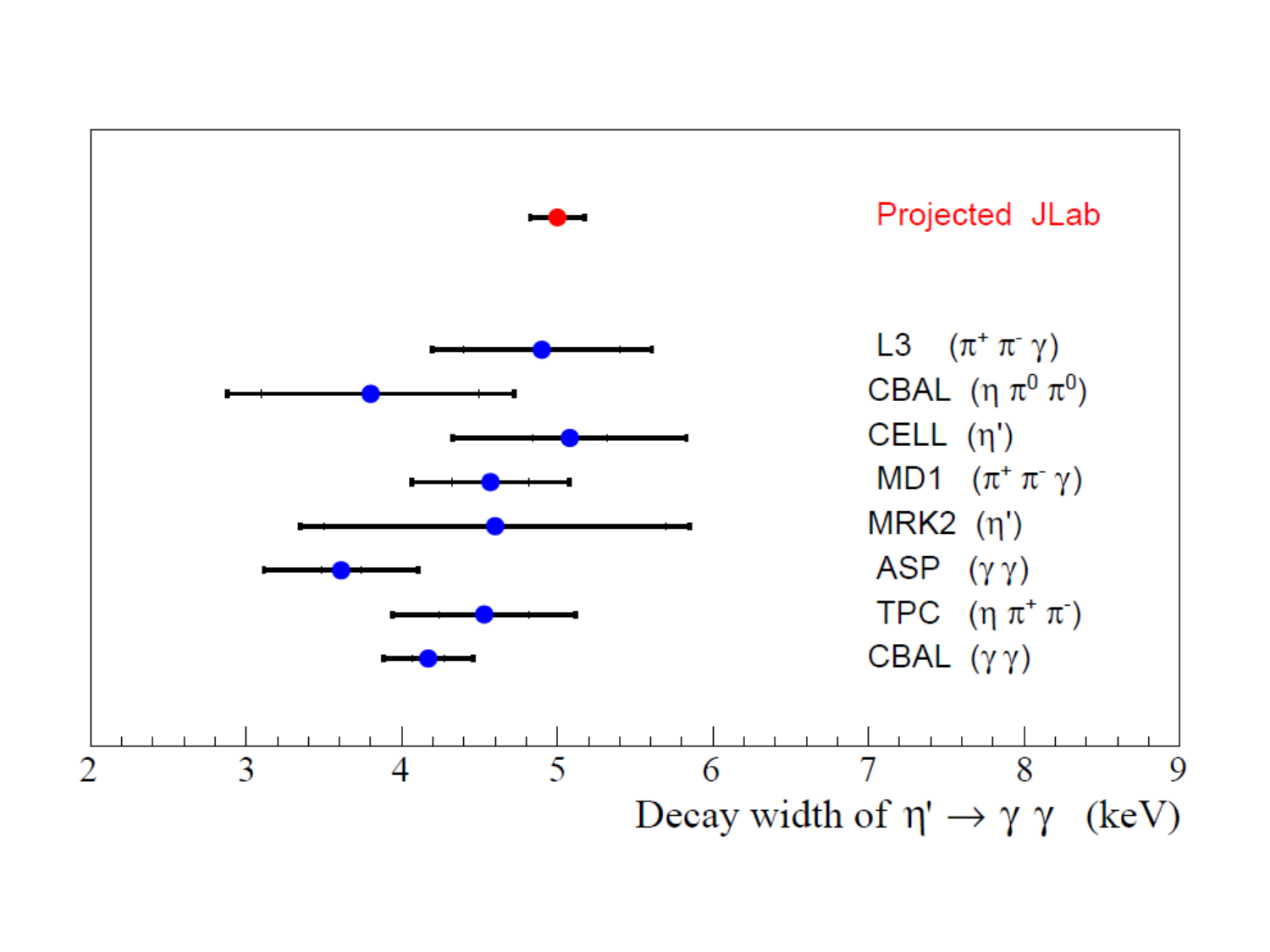} 
\end{center}
\vspace{-1.55cm}
\caption{The existing experimental results (the blue points) of the $\eta^{\prime} \to \gamma \gamma$ decay width by the collider experiments~\cite{ParticleDataGroup:2022pth} and the projected measurement at JLab 22 GeV (the red point) via the Primakoff effect.
} 

\label{fig:projetap}
\end{figure}

All existing measurements of $\Gamma(\eta'\to\gamma\gamma)$ were carried out by using $e^+e^-$ collisions, shown as the blue points in Fig.~\ref{fig:projetap}, with experimental uncertainty for each individual experiment in the range of 7.3\%--27\%~\cite{ParticleDataGroup:2022pth}. The JLab energy upgrade will be essential to perform the first Primakoff measurement on $\Gamma(\eta'\to\gamma\gamma)$ by helping a clean separation of the Primakoff signal from the nuclear backgrounds, as demonstrated in Fig~\ref{fig:etap}. The projected precision of $\sim$3.5\% for $\Gamma(\eta'\to\gamma\gamma)$ at JLab 22~GeV, the red point in Fig.~\ref{fig:projetap}, will help our understanding of  the $U(1)_A$  anomaly coupling to the gluon field. The precision measurements of $\eta$ and $\eta'$, coupled with theory, will provide further input for global analyses of the $\eta$--$\eta^\prime$ system to determine their mixing angles and decay constants.
Moreover, it will further pin down the contributions of $\eta$ and  $\eta^\prime$ to the light-by-light scattering in $(g-2)_\mu$.

\subsection{Search for sub-GeV Dark Scalars and Pseudoscalars via the Primakoff Effect}

The Primakoff cross section, $\sigma_{Pr}\sim \frac{Z^2}{m^3}\log(E)$,  will increase with a higher beam energy and a higher $Z$ target.  The proposed high-energy and high-luminosity upgrade at JLab will offer opportunities to directly search for sub-GeV dark scalars and pseudoscalars via the Primakoff production off a heavy target (such as Pb), probing two out of four the most motivated portals coupling the Standard Model sector to the dark sector. The distinguishable characteristics of Primakoff  mechanism will serve as filters to suppress the QCD backgrounds. The candidates of scalar and pseudoscalar can be explored by hunting for resonant peaks of $\gamma\gamma$, $e^+e^-$, $\mu^+\mu^-$, $\pi\pi$, and $\pi\pi\pi$ in the forward angles where the Primakoff production dominates.

The Dark Matter (DM) constitutes about 85\% of the matter in the Universe. Very little knowledge about the nature of DM is known, except its gravitational property. There is a strong consensus among the physics community about the vital importance of broadening the scope of new physics searches~\cite{Essig:2013lka,Alexander:2016aln,Battaglieri:2017aum}, both in  parameter space and in 
experimental approaches. Recently, sub-GeV dark matter or mediators have gained strong motivation, driven partly by several observed anomalies. The reported excesses in high-energy cosmic rays could be explained by dark matter annihilation~\cite{Arkani-Hamed:2008hhe,Pospelov:2008jd}. 
The muonic anomaly~\cite{Liu:2018qgl, Fayet:2007ua,Pospelov:2008zw} and an anomalous $e^+ e^-$ resonance 
observed in $^8$Be decay~\cite{Krasznahorkay:2015iga,Feng:2016jff} can be resolved with new gauge bosons. In addition, the scalar and pseudoscalar-mediated dark forces can solve small scale structure anomalies in dwarf galaxies and subhalos, while satisfying constraints on larger galaxy and cluster scales~\cite{Tulin:2017ara,Tulin:2012wi, Tulin:2013teo}. If these phenomena are interpreted in terms of new physics, all point toward DM or mediators in the MeV--GeV mass range. A 22~GeV upgrade will greatly advance  searches for such scalars and pseudoscalars via the Primakoff effect. 

\begin{figure}[t]
\centering
\includegraphics*[width=0.6\textwidth]{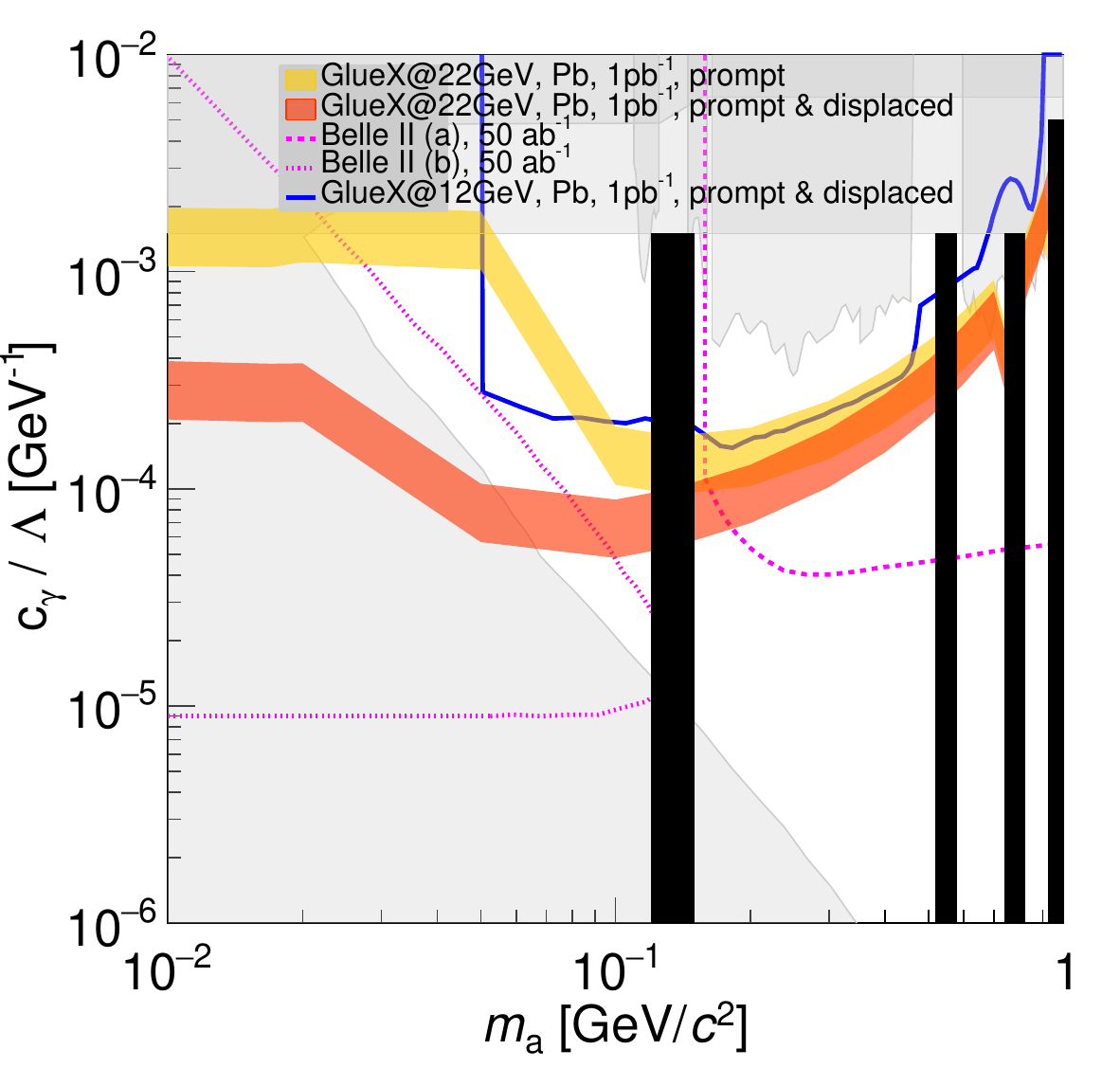}
\vglue -0.2in
\caption{The experimental reaches for the ALP-photon coupling vs. the ALP mass. The projected reaches for GlueX at JLab 22~GeV (in yellow and orange) are estimated for a $Pb$ target with 1~pb$^{-1}$ integrated luminosity. The reach at GlueX 12~GeV (in blue) is from~\cite{Aloni:2019ruo}. The projected Belle II~\cite{Dolan:2017osp} (a: prompt decay and b: displaced vertex) reaches (in pink) are  for 50~ab$^{-1}$ integrated luminosity. The existing limits~\cite{Bjorken:1988as,OPAL:2002vhf,Knapen:2016moh,Blumlein:1991xh} are shown in gray.
}
\label{ALP-reach}
\end{figure}

As an example, we considered 
a hypothetical Axion-Like Particle (ALP)~\cite{Aloni:2018vki,Aloni:2019ruo}, $a$, 
produced from the Primakoff process, $\gamma + Pb\to a +Pb$ with $a \to \gamma\gamma$,  using the GlueX apparatus with an upgraded forward calorimeter (FCAL-II) that is currently underinstallation.
The cross section for this process is $\frac{d\sigma}{d\Omega} = \Gamma_{a\to\gamma\gamma}\frac{8\alpha Z^2}{m^3_a}\frac{\beta^3E^4}{Q^4}\left|F_{e.m.}\left(Q\right)\right|^2\sin^2\left(\mathrm{\theta_a^{lab}}\right)$~\cite{Aloni:2019ruo}. The  ALP radiative decay width is $\Gamma_{a\to\gamma\gamma} = \frac{c_\gamma^2 m_a^3}{64\pi\Lambda^2}$, where $\frac{c_\gamma}{\Lambda}$ is the coupling of axion to the photon as described in Ref.~\cite{Aloni:2018vki,Aloni:2019ruo}. Assuming the space between the $Pb$ target ($\sim 5$\% R.L.) and  FCAL-II has a distance of 10.5~m and is filled with the $He$ gas only, one searches for the ALP that decays either inside of the target (prompt) or after the target (displaced). As shown in Fig.~\ref{ALP-reach}, the projected GlueX reach at JLab 22~GeV for a 2$\sigma$ significance (see orange band) with 1~pb$^{-1}$ integrated luminosity (corresponding to $\sim$120 days of beam time for a photon flux of $\sim 10^{8}\gamma/s$) is  competitive to the reaches by  the Belle II  with 50~ab$^{-1}$ integrated luminosity~\cite{Dolan:2017osp} that is equivalent to about 7 years of running.

\subsection{Electroweak Studies with SoLID}

Parity-violation in deep inelastic scattering (PVDIS) provides a sensitivity to Beyond Standard Model (BSM) couplings. The parity-violating asymmetry $A_{PV}$ is generated from the interference of electromagnetic and weak neutral currents and measured experimentally by scattering a longitudinally polarized electron beam on an unpolarized target. Measuring the the parity-violating asymmetry from electron-deuteron scattering allows for the measurement the effective electron-quark coupling constants (2$C_{iu} - C_{id}$) with $i=1,2$ which, under the Standard Model, can be expressed as a function of the weak mixing angle ${\sin^2\theta_W}$.

An experimental program for measurements of $A_{PV}$ is planned to run at JLab, utilizing the planned Solenoidal Large Intensity Device (SoLID)~\cite{PVDIS_PAC} spectrometer. The large acceptance and high luminosity capability of SoLID makes it ideal for a precision measurement of the parity-violating asymmetry. Upgrading CEBAF to 22~GeV and using SoLID will extend the $Q^{2}$ and $x$ range beyond what is possible at 11~GeV. As described in Section~\ref{sec:lightsea}, a program of measurements with SoLID at 22~GeV will significantly improve knowledge of quark parton distribution functions over a broad range of $x$. Additionally, measurements at 22~GeV of  $A_{PV}^{D}$ from the deuteron will also provide an improvement in the overall uncertainty on the effective electron-quark coupling constants (2$C_{iu} - C_{id}$) of about 15\%, as compared to the expected uncertainty from the 11 GeV experiment.

\subsection{Secondary Beams}

The stopping of the high-energy beam will produce a shower of radiation, most of which will be contained in the thick beam dump, while deeply penetrating muons and neutrinos will continue to propagate, producing high-intensity secondary beams. Simulations have shown that the neutrino flux above the dump is characterized by a decay-at-rest (DAR) spectrum. It is expected that the intense neutrino flux 
of $\sim$9$\cdot$\,10$^{17}$ expected to be available at 10~GeV, a flux that is comparable to flagship DAR-neutrino facilities such as the Spallation Neutron Source at Oak Ridge National Lab, would be doubled by the increase in beam energy to 22~GeV. Such a neutrino facility would enable studies of coherent elastic neutrino-nucleus scattering (CE$\nu$NS), with count rates up to 40 times more than the recently published COHERENT measurement~\cite{COHERENT:2021xmm}. As the CE$\nu$NS cross section is a clean, tree-level prediction of the Standard Model, such measurements would be provide a means to search for BSM signals that might arise from non-standard interactions such as dark matter, new mediators, or a large neutrino magnetic moment. In addition, CE$\nu$NS provides a different and complementary way to measure Standard Model parameters such as the neutron radius of nuclei and weak mixing angle ${\sin^2\theta_W}$.

\begin{figure}[h!]
 \begin{center}
    \includegraphics[width=0.95\textwidth]{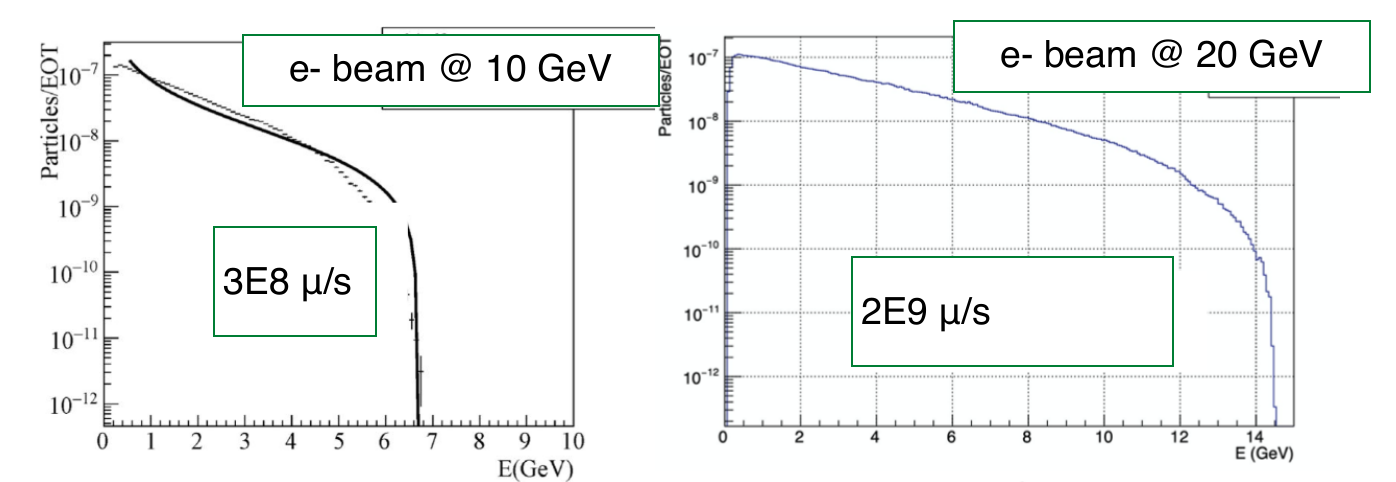}
 \end{center}
 \vglue -0.1in
  \caption{Muon energy distribution produced by interactions of a 10~GeV (20~GeV) electron beam  with the beam dump in Hall A. }
 \label{fig:muonspectrum}
\end{figure}

\begin{figure}[h!]
 \begin{center}
    \includegraphics[width=0.80\textwidth]{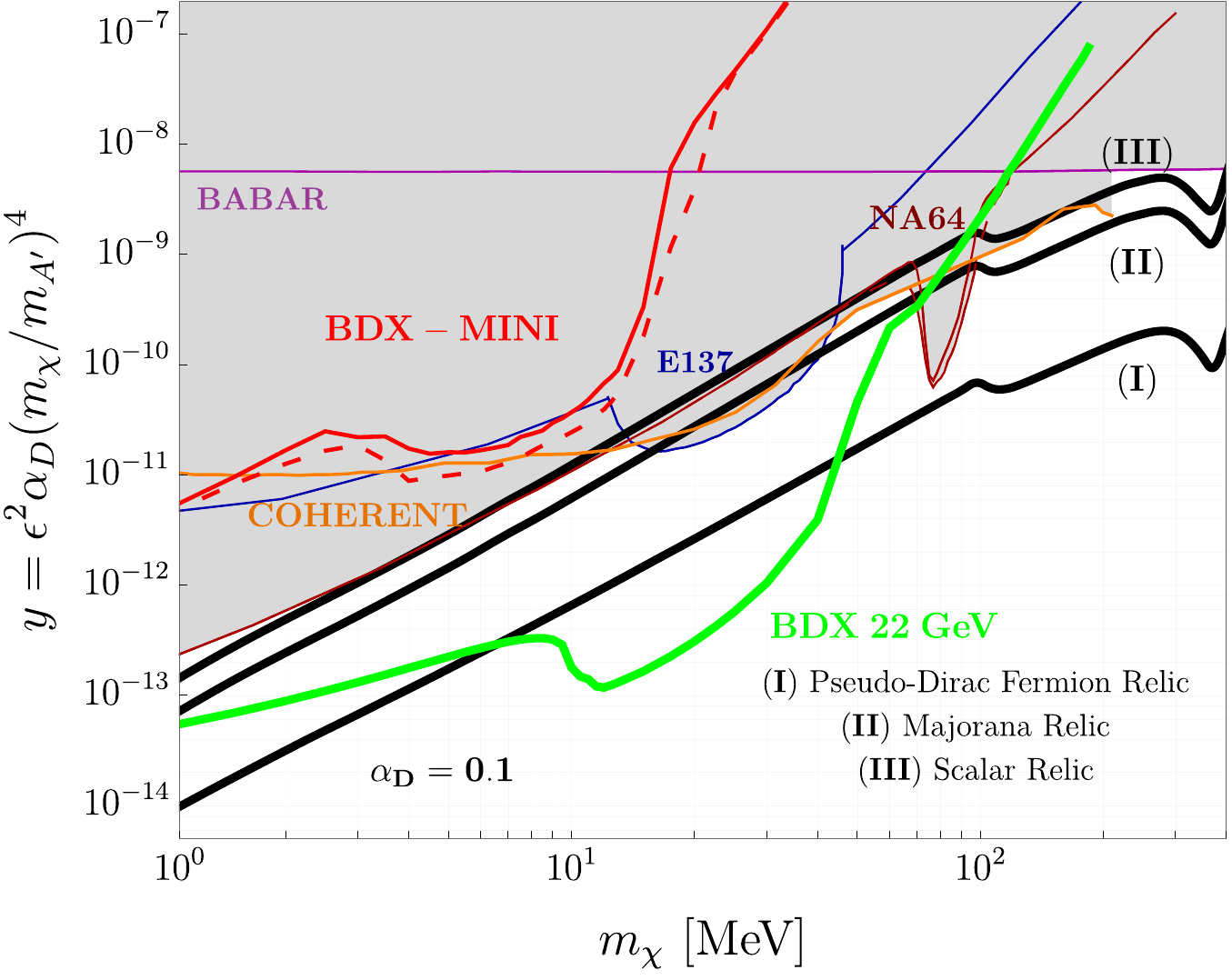}
 \end{center}
 \vglue -0.15in
  \caption{The BDX sensitivity  at 90$\%$ CL (green curve) by using a 22\,GeV  electron beam. The limit is given for the scaling parameter $y$, proportional to the LDM-SM interaction cross section, as a function of the LDM mass m$_{\chi}$. The curve refers to the ideal case of a zero-background measurement, assuming a 300\,MeV energy threshold and an overall 20$\%$ signal efficiency.}
 \label{fig:BDXreach}
\end{figure}

Further opportunities would be gained with the addition of a secondary beamline, such as is envisioned for the approved BDX  experiment~\cite{BDX:2016akw}.  Muons are produced in the electron beamdump primarily by Bethe-Heitler radiation. A high-intensity flux of $\sim3\cdot$\,10$^{8}$\,$\mu$/s ($\sim2\cdot$\,10$^{9}$\,$\mu$/s) is expected at the exit of the concrete vault, produced by and mainly collinear with a a 10\,GeV (20\,GeV) primary e-beam.~Figure~\ref{fig:muonspectrum} compares the simulated Bremsstrahlung-like energy distribution for 10\,GeV and 20~GeV primary e-beams. Such a muon beam would offer the possibility to search for muon-coupling light dark scalars that may explain the $(g-2)_{\mu}$ anomaly \cite{Marsicano:2018vin}. A similar enhancement due to higher energy is also be expected for the production of light dark matter (through $A'$-sstralhung, resonant and non-resonant annihilation) that will be searched for by the BDX experiment~\cite{BDX:2016akw,Battaglieri:2022dcy}, as shown in Fig.~\ref{fig:BDXreach}.

\clearpage \section{CEBAF Energy `Doubling' - Accelerator Concept}
\label{sec:acc}

\begin{figure}[h!]
\centering
\includegraphics*[width=0.6\textwidth]{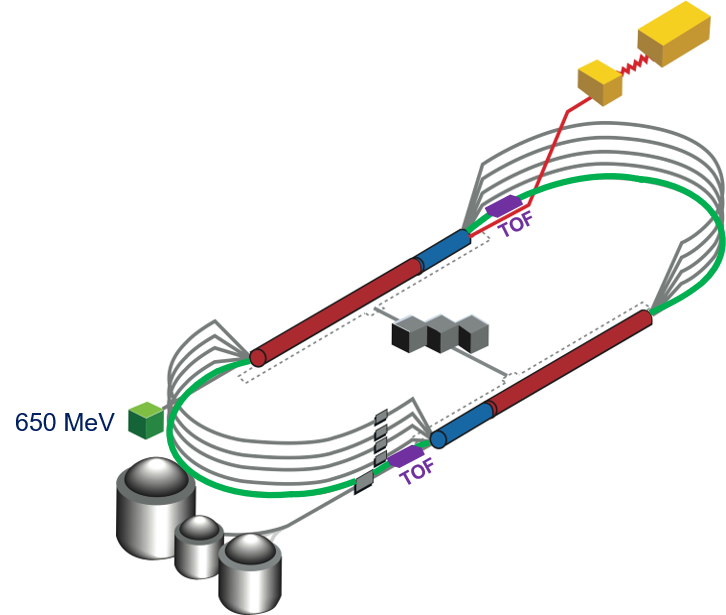}
\caption{Sketch of the CEBAF accelerator with the two highest energy arcs, Arc 9 and Arc A, replaced with a pair of FFA arcs (green).}
\label{CEBAF}
\end{figure}

The previous energy upgrade of CEBAF, from 6 to 12 GeV,
was achieved by installing additional SRF cavities in the
North and South LINACs, increasing the energy gain per pass,
while leaving the maximum number of passes unchanged. Recent advances in accelerator technology have made it possible to further extend the energy reach of the CEBAF accelerator up to 22 GeV within the existing tunnel footprint. In the proposed energy upgrade, the
energy gain per pass remains unchanged, while the number of recirculations through the accelerating cavities is nearly doubled. 
Encouraged by the recent success of the CBETA project (Cornell Brookhaven Electron Test Accelerator)~\cite{Bartnik:2020pos}), 
a proposal was formulated to increase the CEBAF energy from the present 12 GeV to about 22 GeV by replacing the highest-energy arcs with Fixed Field Alternating Gradient (FFA) arcs ~\cite{Bogacz:2021}, as illustrated schematically in Fig.~\ref{CEBAF}). 

\begin{figure}[h!]
\centering
\includegraphics*[width=0.7\textwidth]{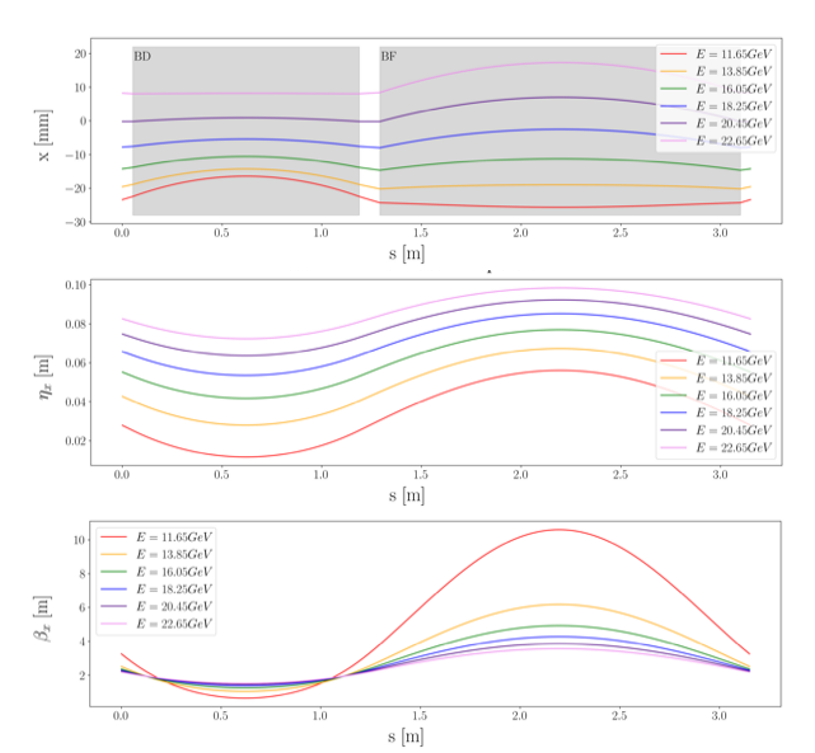}
\caption{Compact FODO cell configured with two combined function magnets featuring closely spaced orbits and small Twiss functions for six different energy beams.}
\label{FFA_cell_W}
\end{figure}

The design is based on an exciting new approach to accelerate electrons efficiently with multiple LINAC passes and transporting them through a single FFA beamline, as was successfully demonstrated by CBETA project. The Non-Scaling FFA approach allows beam acceleration within a small beam pipe as in synchrotrons, but without varying the magnetic field. These recirculating 180$^\circ$ FFA arcs are made up of 86 repeating cells. The arc's building block is a compact, 3.15-m-long, FODO cell composed with two magnets and two drifts. Each of the magnets is a multi-function Halbach magnet ~\cite{Brooks:2022}, ~\cite{Brooks:2023} with dominant dipole and quadrupole fields. One magnet per cell bends and focuses the electron beam, and the other bends and de-focuses in the same plane. As illustrated in Fig.~\ref{FFA_cell_W}, different energy beams may be transported through a narrow beam pipe, since the transverse orbit offsets are confined to small aperture of about 4~cm. Closely spaced orbits and low betas (a few meters) result from very strong focusing, reducing the horizontal dispersion function from meters in conventional separate functions arcs down to a few cm in the FFA arc. 

\begin{figure}[t!]
\centering
\includegraphics*[width=0.8\textwidth]{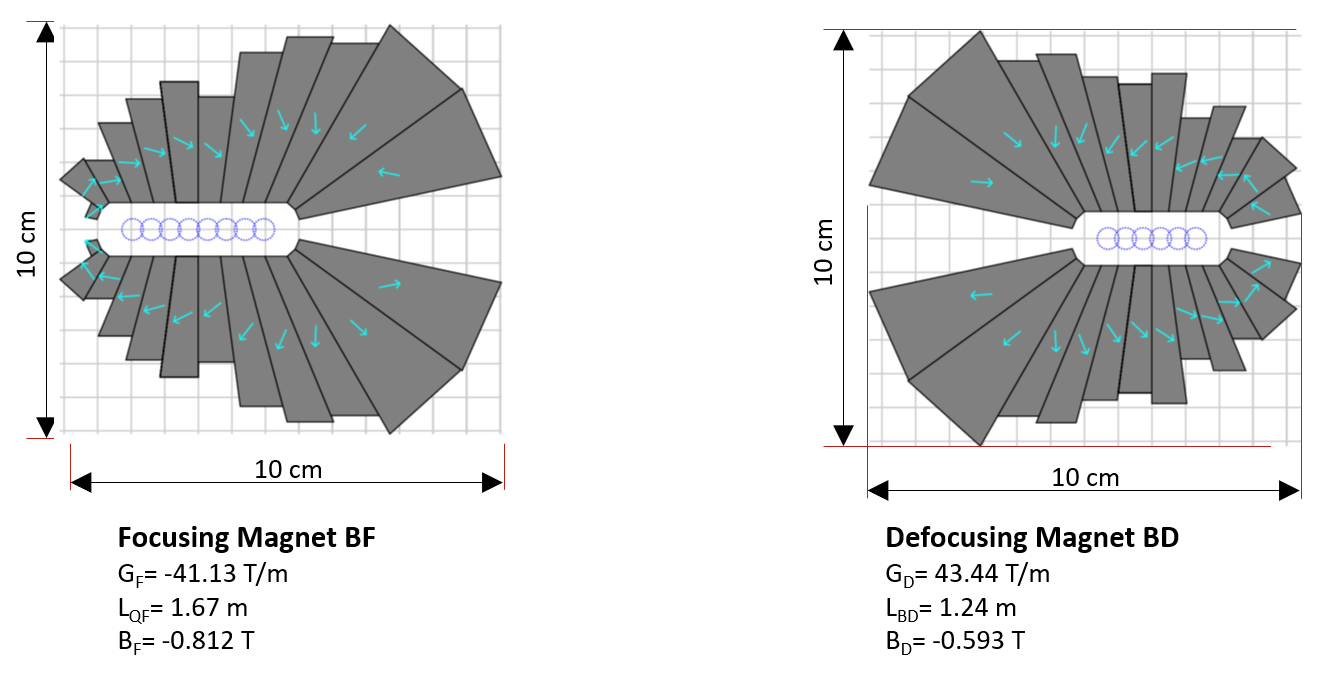}
\caption{The cross section and field specs of the open mid-plane magnets consisting of 24 wedge-shaped pieces of NdFeB. The outer wedges are symmetrical, while the top and bottom wedges have two edges parallel to the horizontal axis.}
\label{Magnets}
\end{figure}

The new pair of arcs configured with an FFA lattice would support simultaneous transport of 6 passes with energies spanning a factor of two. This wide energy bandwidth could be achieved using the non-scaling FFA principle implemented with Halbach-style permanent magnets. As illustrated in Fig.~\ref{Magnets}, the magnet design features an open mid-plane geometry, in order for the synchrotron radiation to pass through the magnets, while minimizing radiation damage to the permanent magnet material. This novel magnet technology saves energy and lowers operating costs. 
  
In addition to the spreaders, one must design the time-of-flight horizontal `Splitters' for each of the energies that pass through the FFA arcs. These will be located along the new FFA arcs, downstream of the spreaders (shown as purple boxes in in Fig.~\ref{CEBAF})  Conceptually, they will be similar to those at CBETA. They will need to fit in the space currently occupied by the highest-energy passes in the CEBAF recirculating arcs. This would necessitate a pair of time-of-flight splitters, which are capable of adjusting the momentum compaction, $M_{56}$, at both East and West FFA arc. 

One of the challenges of the multi-pass LINAC optics is to provide uniform focusing in a vast range of energies, using fixed field lattice. The current CEBAF is configured with a 123~MeV injector feeding into a racetrack Recirculating Linear Accelerator (RLA) with a 1.1~GeV LINAC on each side.  Increasing number of LINAC passes to 10+ makes optical matching virtually impossible due to extremely high energy span ratio (1:175).

\begin{figure}[h!]
\centering
\includegraphics*[width=0.95\textwidth]{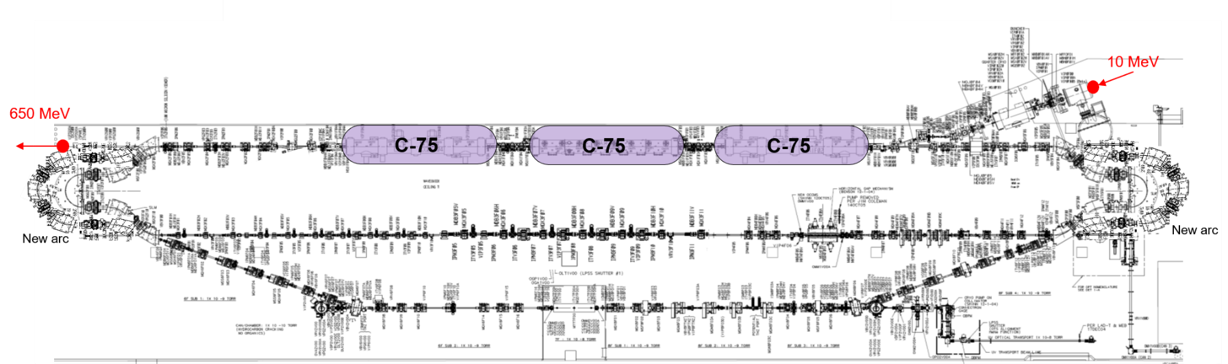}
\caption{Schematic view of $650$~MeV recirculating injector (3-pass) based on LERF.}
\label{Injector}
\end{figure}

 The proposed new building block of LINAC optics is configured as a sequence of triplet cells flanking two cryomodules. Initial triplets, based on 45 Tesla/m quads, are scaled with increasing momentum along the LINAC. This style LINAC focusing provides a stable multi-pass optics compatible with much smaller beta functions in the FFA arcs and it is  capable of covering energy ratio of 1:33. This sets the minimum injection energy at $650$~MeV. In the current concept, it is proposed to replace old 123~MeV injector with a $650$~MeV 3-pass recirculating injector based on the existing LERF facility augmented by three C-70 cryomodules. The upgraded $650$~MeV injector is schematically illustrated in Fig.~\ref{Injector}. The beam is then transferred from the LERF vault through a dedicated fixed energy $650$~MeV transport line and injected into the North LINAC 

Staying within the CEBAF footprint, while transporting high energy beams (10-22 GeV) calls for special mitigation of synchrotron radiation effects. One of them is to increase the bend radius at the arc dipoles (packing factor of  the FFA arcs increased to about 92\%). 
Arc optics was designed to ease individual adjustment of momentum compaction and the horizontal emittance  dispersion, $H$, in each arc to suppress adverse effects of the synchrotron radiation on beam quality: dilution of the transverse and longitudinal emittance due to quantum excitations
Table~\ref{tab:SREmittance} lists arc-by-arc cumulative dilution of the transverse, $\Delta \epsilon_N$, and longitudinal, $\Delta \sigma_{\frac{\Delta E}{E}}$, emittances due to quantum excitations calculated using the following analytic formulas:  
\begin{equation}
  \Delta \epsilon_N = \frac{2 \pi}{3} C_q r_0 <H> \frac{\gamma^6}{\rho^2}\,,
  \label{eq:Emit_dil_2}
\end{equation}
\begin{equation}
  \frac{\Delta \epsilon_E^2}{E^2} = \frac{2 \pi}{3} C_q r_0~ \frac{\gamma^5}{\rho^2}\,,
  \label{eq:Emit_dil_3}
\end{equation}

Here, $\Delta \epsilon^2_E$ is an increment of energy square variance, $r_0$ is the classical electron radius, $\gamma$ is the Lorentz boost and $C_q = \frac{55}{32 \sqrt{3}} \frac{\hbar}{m c} \approx 3.832 \cdot 10^{-13}$~m for electrons (or positrons).
The horizontal emittance dispersion in Eq.~\ref{eq:Emit_dil_2},  is given by the following formula: $H = (1+\alpha^2)/\beta \cdot D^2 + 2 \alpha \ D D' + \beta \cdot D'^2$ where $D, D'$ are the bending plane dispersion and its derivative, with averaging over bends defined as: $<...>~=~\frac{1}{\pi}\int_{bends}...~d\theta$.
\begin{table}[!h]
  \centering
  \small
  \begin{tabular}{cccc} 
  \hline
  Pass number & Beam Energy & $\epsilon^x_N$ & $\sigma_{\frac{\Delta E}{E}}$  \\
    & [GeV] & [mm~mrad] &  [\%]\\
  \hline\hline
  1 & 2.8 & 1.0 & 0.01\\
  2 & 5.0 & 2 & 0.02\\
  3 & 7.2 & 4 & 0.02\\
  4 & 9.4 & 12 & 0.03\\
  5 & 11.5 & 20 & 0.03\\
  6 & 13.7 & 21 & 0.04\\
  7 & 15.8 & 23 & 0.05\\
  8 & 17.9 & 26 & 0.06\\
  9 & 19.9 & 34 & 0.08\\
  10 & 21.9 & 49 & 0.11\\
  10.5 & 22.9 & 61 & 0.12\\
  \hline
  \end{tabular}
  \caption{The horizontal and longitudinal emittances diluted by synchrotron radiation (arcs contribution) after respective passes is summarized. Additional net contribution from the horizontal 'splitters' is estimated as: an emittance dilution of $20$~{mm$\cdot$mrad} with a relative energy spread of $0.3 \cdot 10^{-3}$. Here, $ \sigma_{\frac{\Delta E}{E}} = \sqrt{\frac{ \Delta\epsilon_E^2}{E^2}}$.}
  \label{tab:SREmittance}
\end{table}
\\
In summary, the proposed 22~GeV, 10-pass, design would promise to deliver a normalized emittance of $81$~{mm$\cdot$mrad} with a relative energy spread of $1.5 \cdot 10^{-3}$. Further recirculation beyond 22 GeV is limited by large, 0.9 GeV per electron, energy loss due to synchrotron radiation, which depends on energy to the fourth power. The net energy loss is comparable to the energy gain per LINAC, which clearly sets the limit of reasonable number of recirculations.

Finally, given the greater total energies expected with this upgrade, we are also investigating the impact this will have on our extraction system and beam delivery to the experimental halls. For the extraction system, this will depend partly on the needs of the experimental program, and partly on how we choose to extract the beam. For the beam delivery to the halls, the hall beamlines are currently under investigation. Improvements to the magnetic septa are expected to be required, and the dipoles to the hall lines will need to be strengthened and improved as well. The overall optics will require some adjustments, but should be manageable.

One of the more challenging aspects of this design is the method of beam extraction. Multiple methods are under
consideration, each with associated limitations on the flexibility of beam delivery – the resulting scheme has to balance
the needs of the users, technical feasibility, and cost. 

To conclude, significant progress has been made in the design of the energy upgrade for CEBAF using FFA transport. Over the last
year, we have settled on a design concept, developed more detailed designs of various machine sections, and iterated
some sections as simulations were performed. While the full design is not yet completed, we are working toward that
goal as we begin to consider other aspects of this upgrade concept.

\clearpage\section{Workshops}
\label{workshops}
This White Paper is a culmination of several dedicated workshops conducted since the spring of 2022. These workshops led up to the final event, the mini-symposium held at the annual APS April meeting 2023. We are pleased to provide access to the presentations from these workshops through the following links:

\begin{itemize}
    
    \item[] \href{https://indico.jlab.org/event/520/}{J-FUTURE}, March 28–30, 2022 Jefferson Lab and Messina University (Italy). Organizers: Marco Battaglieri, Alessandro Pilloni, Adam Szczepaniak, Eric Voutier.

    \item[] \href{https://www.jlab.org/conference/hews22}{High Energy Workshop Series 2022}, Jefferson Lab
    \begin{itemize}
        
        \item \href{https://indico.jlab.org/event/528/}{Hadron Spectroscopy with a CEBAF Energy Upgrade}, June 16-17, 2022. Organizers: Marco Battaglieri, Sean Dobbs, Derek Glazier, Alessandro Pilloni, Justin Stevens, Adam Szczepaniak, Alaina Vaughn.
        
        \item \href{https://indico.jlab.org/event/539/}{The Next Generation of 3D Imaging}, July 7-8, 2022. Organizers: 
        Harut Avagyan, Carlos Munoz Camacho, Jian-Ping Chen, Xiangdong Ji, Jianwei Qiu, Patrizia Rossi.

        \item \href{https://indico.jlab.org/event/557/}{Science at Mid-$x$: Anti-shadowing and the Role of the Sea}, July 22-23,  2022. Organizers: John Arrington, Mark Dalton, Cynthia Keppel, Wally Melnitchouk, Jianwei Qiu.
        
        \item \href{https://indico.jlab.org/event/540/}{Physics Beyond the Standard Model}, Aug. 1, 2022. Organizers: 
        Marco Battaglieri, Bob McKeown, Xiaochao Zheng.
        
        \item \href{https://indico.jlab.org/event/543/}{J/Psi and Beyond}, Aug. 16-17, 2022. Organizers:  
        Ed Brash, Ian~Clo\"et, Zein-Eddine Meziani, Jianwei Qiu, Patrizia Rossi. 
    \end{itemize}
    
    \item[] \href{https://indico.knu.ac.kr/event/566/}{APCTP Focus Program on Nuclear Physics 2022: Hadron Physics Opportunities with JLab Energy and Luminosity Upgrade}, July 18-23, 2022, Korea. Organizers: 
    Harut Avagyan, Chueng-Ryong Ji, Kyungseon Joo, Victor Mokeev, Yongseok Oh, A. Vladimirov.

    \item[] \href{https://indico.ectstar.eu/event/152/}{ECT*: Opportunities with JLab Energy and Luminosity Upgrade}, Sep. 26-30, 2022, Italy. Organizers: Moskov Amaryan, John Arrington, Harut Avagyan, Alessandro Bacchetta, Marco Battaglieri, Lamiaa El Fassi, Ralf Gothe, Or Hen, Xiangdong Ji, Kyungseon Joo, Xiaochao Zheng.

    \item[] \href{https://www.jlab.org/conference/luminosity22gev}{Science at the Luminosity Frontier: Jefferson Lab at 22 GeV},  Jan. 23-25, 2023, Jefferson Lab. Organizers:
        ``\emph{Spectra and Structure of Heavy and Light Hadrons as Probes of QCD}'': Ralf Gothe, Matt Shepherd; 
        ``\emph{Sea and Valence Partonic Structure and Spin}'': Jian-Ping Chen, Ioana Niculescu, Nobuo Sato;
        ``\emph{Form Factors, Generalized Parton Distributions and Energy-Momentum Tensor}'': Latifa Elouadrhiri, Garth Huber, Christian Weiss;
        ``\emph{Fragmentation, Transverse Momentum and Parton Correlations}'': Harut Avagyan, Dave Gaskell, Nobuo Sato;
        ``\emph{Hadron-Quark Transition and Nuclear Dynamics at Extreme Conditions}'': Lamiaa El Fassi, Misak Sargasian;
        ``\emph{Low-Energy Tests of the Standard Model and Fundamental Symmetries}'': Liping Gan, Kent Paschke.    

    \item[] APS April Meeting 2023 Mini-Symposium: Opportunies with JLab Upgrades in Energy, Luminosity, and a Positron Beam - [\href{https://meetings.aps.org/Meeting/APR23/Session/B15?utm_campaign=AM23&utm_medium=email&utm_source=unit+events}{Session I},
    \href{https://meetings.aps.org/Meeting/APR23/Session/K16?utm_campaign=AM23&utm_medium=email&utm_source=unit+events}{Session II},
    \href{https://meetings.aps.org/Meeting/APR23/Session/AA02}
    {Session III}], 
    April 15 and 24, 2023. Organizers: Harut Avagyan, Jianping Chen, Liping Gan, Ashot Gasparian.

\end{itemize}

\clearpage\section{Acknowledgements}

The authors would like to express their gratitude to the following colleagues, whose critical and valuable comments contributed to the preparation of this document: Patrick Achenbach, Marco Battaglieri, Daniel Carman, Eugene Chudakov,  Bob McKeown, Mark Jones, and  Viktor Mokeev. 

Furthermore, the authors would also like to acknowledge the support received from the following institutions/agencies: 
    US Department of Energy, Office of Science, Office of Nuclear Physics (contract numbers
        DE-FG02-05ER41374, 
        DE‐FG02‐07ER41522, 
        DE-FG02-01ER41172, 
        DE-FG02-07ER41528, 
        DE-AC05-06OR23177) 
        and the Early Career Program; 
    US National Science Foundation (contract numbers
        PHY-10011349, 
        PHY-1812396, 
        PHY-2111181, 
        PHY 2209421); 
    Natural Sciences and Engineering Research Council of Canada (NSERC). 



\clearpage
\bibliographystyle{unsrtnat}
\bibliography{
references/preface.bib,
references/wg1a.bib,
references/wg1b.bib,
references/wg2.bib,
references/wg3.bib,
references/wg4.bib,
references/wg5.bib,
references/wg6.bib,
references/accelerator.bib}

\begin{thebibliography}{100}
\expandafter\ifx\csname url\endcsname\relax
  \def\url#1{\texttt{#1}}\fi
\expandafter\ifx\csname urlprefix\endcsname\relax\def\urlprefix{URL }\fi
\expandafter\ifx\csname href\endcsname\relax
  \def\href#1#2{#2} \def\path#1{#1}\fi

\bibitem{Arrington:2021alx}
J.~Arrington, et~al., {Physics with CEBAF at 12 GeV and future opportunities},
  Prog. Part. Nucl. Phys. 127 (2022) 103985.
\newblock \href {http://arxiv.org/abs/2112.00060} {\path{arXiv:2112.00060}},
  \href {https://doi.org/10.1016/j.ppnp.2022.103985}
  {\path{doi:10.1016/j.ppnp.2022.103985}}.

\bibitem{Bulava:2022ovd}
J.~Bulava, et~al., {Hadron Spectroscopy with Lattice QCD}, in: {Snowmass 2021},
  2022.
\newblock \href {http://arxiv.org/abs/2203.03230} {\path{arXiv:2203.03230}}.

\bibitem{Meyer:2015eta}
C.~A. Meyer, E.~S. Swanson, {Hybrid Mesons}, Prog. Part. Nucl. Phys. 82 (2015)
  21--58.
\newblock \href {http://arxiv.org/abs/1502.07276} {\path{arXiv:1502.07276}},
  \href {https://doi.org/10.1016/j.ppnp.2015.03.001}
  {\path{doi:10.1016/j.ppnp.2015.03.001}}.

\bibitem{JPAC:2018zyd}
A.~Rodas, et~al., {Determination of the pole position of the lightest hybrid
  meson candidate}, Phys. Rev. Lett. 122~(4) (2019) 042002.
\newblock \href {http://arxiv.org/abs/1810.04171} {\path{arXiv:1810.04171}},
  \href {https://doi.org/10.1103/PhysRevLett.122.042002}
  {\path{doi:10.1103/PhysRevLett.122.042002}}.

\bibitem{BESIII:2022riz}
M.~Ablikim, et~al., {Observation of an Isoscalar Resonance with Exotic JPC=1-+
  Quantum Numbers in
  J/\ensuremath{\psi}\textrightarrow{}\ensuremath{\gamma}\ensuremath{\eta}\ensuremath{\eta}'},
  Phys. Rev. Lett. 129~(19) (2022) 192002, [Erratum: Phys.Rev.Lett. 130, 159901
  (2023)].
\newblock \href {http://arxiv.org/abs/2202.00621} {\path{arXiv:2202.00621}},
  \href {https://doi.org/10.1103/PhysRevLett.129.192002}
  {\path{doi:10.1103/PhysRevLett.129.192002}}.

\bibitem{Olsen:2017bmm}
S.~L. Olsen, T.~Skwarnicki, D.~Zieminska, {Nonstandard heavy mesons and
  baryons: Experimental evidence}, Rev. Mod. Phys. 90~(1) (2018) 015003.
\newblock \href {http://arxiv.org/abs/1708.04012} {\path{arXiv:1708.04012}},
  \href {https://doi.org/10.1103/RevModPhys.90.015003}
  {\path{doi:10.1103/RevModPhys.90.015003}}.

\bibitem{Lebed:2016hpi}
R.~F. Lebed, R.~E. Mitchell, E.~S. Swanson, {Heavy-Quark QCD Exotica}, Prog.
  Part. Nucl. Phys. 93 (2017) 143--194.
\newblock \href {http://arxiv.org/abs/1610.04528} {\path{arXiv:1610.04528}},
  \href {https://doi.org/10.1016/j.ppnp.2016.11.003}
  {\path{doi:10.1016/j.ppnp.2016.11.003}}.

\bibitem{Briceno:2015rlt}
R.~A. Briceno, et~al., {Issues and Opportunities in Exotic Hadrons}, Chin.
  Phys. C 40~(4) (2016) 042001.
\newblock \href {http://arxiv.org/abs/1511.06779} {\path{arXiv:1511.06779}},
  \href {https://doi.org/10.1088/1674-1137/40/4/042001}
  {\path{doi:10.1088/1674-1137/40/4/042001}}.

\bibitem{BESIII:2013ris}
M.~Ablikim, et~al., {Observation of a Charged Charmoniumlike Structure in
  $e^+e^- \to \pi^+\pi^- J/\psi$ at $\sqrt{s}$ =4.26 GeV}, Phys. Rev. Lett. 110
  (2013) 252001.
\newblock \href {http://arxiv.org/abs/1303.5949} {\path{arXiv:1303.5949}},
  \href {https://doi.org/10.1103/PhysRevLett.110.252001}
  {\path{doi:10.1103/PhysRevLett.110.252001}}.

\bibitem{Belle:2013yex}
Z.~Q. Liu, et~al., {Study of $e^+e^- \to \pi^+ \pi^- J/\psi$ and Observation of
  a Charged Charmoniumlike State at Belle}, Phys. Rev. Lett. 110 (2013) 252002,
  [Erratum: Phys.Rev.Lett. 111, 019901 (2013)].
\newblock \href {http://arxiv.org/abs/1304.0121} {\path{arXiv:1304.0121}},
  \href {https://doi.org/10.1103/PhysRevLett.110.252002}
  {\path{doi:10.1103/PhysRevLett.110.252002}}.

\bibitem{Belle:2011aa}
A.~Bondar, et~al., {Observation of two charged bottomonium-like resonances in
  Y(5S) decays}, Phys. Rev. Lett. 108 (2012) 122001.
\newblock \href {http://arxiv.org/abs/1110.2251} {\path{arXiv:1110.2251}},
  \href {https://doi.org/10.1103/PhysRevLett.108.122001}
  {\path{doi:10.1103/PhysRevLett.108.122001}}.

\bibitem{BESIII:2013ouc}
M.~Ablikim, et~al., {Observation of a Charged Charmoniumlike Structure
  $Z_c$(4020) and Search for the $Z_c$(3900) in $e^+e^- \to \pi^+\pi^- h_c$},
  Phys. Rev. Lett. 111~(24) (2013) 242001.
\newblock \href {http://arxiv.org/abs/1309.1896} {\path{arXiv:1309.1896}},
  \href {https://doi.org/10.1103/PhysRevLett.111.242001}
  {\path{doi:10.1103/PhysRevLett.111.242001}}.

\bibitem{Burkert:2020akg}
V.~D. Burkert, et~al., {The CLAS12 Spectrometer at Jefferson Laboratory}, Nucl.
  Instrum. Meth. A 959 (2020) 163419.
\newblock \href {https://doi.org/10.1016/j.nima.2020.163419}
  {\path{doi:10.1016/j.nima.2020.163419}}.

\bibitem{GlueX:2020idb}
S.~Adhikari, et~al., {The GLUEX beamline and detector}, Nucl. Instrum. Meth. A
  987 (2021) 164807.
\newblock \href {http://arxiv.org/abs/2005.14272} {\path{arXiv:2005.14272}},
  \href {https://doi.org/10.1016/j.nima.2020.164807}
  {\path{doi:10.1016/j.nima.2020.164807}}.

\bibitem{LHCb:2021ptx}
R.~Aaij, et~al., {Observation of excited $\Omega_c^0$ baryons in $\Omega_b^-
  \to \Xi_c^+ K^-\pi^-$decays}, Phys. Rev. D 104~(9) (2021) L091102.
\newblock \href {http://arxiv.org/abs/2107.03419} {\path{arXiv:2107.03419}},
  \href {https://doi.org/10.1103/PhysRevD.104.L091102}
  {\path{doi:10.1103/PhysRevD.104.L091102}}.

\bibitem{LHCb:2014zfx}
R.~Aaij, et~al., {Observation of the resonant character of the $Z(4430)^-$
  state}, Phys. Rev. Lett. 112~(22) (2014) 222002.
\newblock \href {http://arxiv.org/abs/1404.1903} {\path{arXiv:1404.1903}},
  \href {https://doi.org/10.1103/PhysRevLett.112.222002}
  {\path{doi:10.1103/PhysRevLett.112.222002}}.

\bibitem{Belle:2014nuw}
K.~Chilikin, et~al., {Observation of a new charged charmoniumlike state in
  $\bar{B}^0 \to J/\psi K^- \pi^+$ decays}, Phys. Rev. D 90~(11) (2014) 112009.
\newblock \href {http://arxiv.org/abs/1408.6457} {\path{arXiv:1408.6457}},
  \href {https://doi.org/10.1103/PhysRevD.90.112009}
  {\path{doi:10.1103/PhysRevD.90.112009}}.

\bibitem{Brambilla:2019esw}
N.~Brambilla, S.~Eidelman, C.~Hanhart, A.~Nefediev, C.-P. Shen, C.~E. Thomas,
  A.~Vairo, C.-Z. Yuan, {The $XYZ$ states: experimental and theoretical status
  and perspectives}, Phys. Rept. 873 (2020) 1--154.
\newblock \href {http://arxiv.org/abs/1907.07583} {\path{arXiv:1907.07583}},
  \href {https://doi.org/10.1016/j.physrep.2020.05.001}
  {\path{doi:10.1016/j.physrep.2020.05.001}}.

\bibitem{Adhikari:2023fcr}
S.~Adhikari, et~al., {Measurement of the J/$\psi $ photoproduction cross
  section over the full near-threshold kinematic region} (4 2023).
\newblock \href {http://arxiv.org/abs/2304.03845} {\path{arXiv:2304.03845}}.

\bibitem{HillerBlin:2016odx}
A.~N. Hiller~Blin, C.~Fern\'andez-Ram\'\i{}rez, A.~Jackura, V.~Mathieu, V.~I.
  Mokeev, A.~Pilloni, A.~P. Szczepaniak, {Studying the P$_c$(4450) resonance in
  J/$\psi$ photoproduction off protons}, Phys. Rev. D 94~(3) (2016) 034002.
\newblock \href {http://arxiv.org/abs/1606.08912} {\path{arXiv:1606.08912}},
  \href {https://doi.org/10.1103/PhysRevD.94.034002}
  {\path{doi:10.1103/PhysRevD.94.034002}}.

\bibitem{Winney:2022tky}
D.~Winney, A.~Pilloni, V.~Mathieu, A.~N. Hiller~Blin, M.~Albaladejo, W.~A.
  Smith, A.~Szczepaniak, {XYZ spectroscopy at electron-hadron facilities. II.
  Semi-inclusive processes with pion exchange}, Phys. Rev. D 106~(9) (2022)
  094009.
\newblock \href {http://arxiv.org/abs/2209.05882} {\path{arXiv:2209.05882}},
  \href {https://doi.org/10.1103/PhysRevD.106.094009}
  {\path{doi:10.1103/PhysRevD.106.094009}}.

\bibitem{Workman:2022ynf}
R.~L. Workman, et~al., {Review of Particle Physics}, PTEP 2022 (2022) 083C01.

\bibitem{Guo:2016bkl}
F.-K. Guo, U.~G. Mei\ss{}ner, J.~Nieves, Z.~Yang, {Remarks on the $P_c$
  structures and triangle singularities}, Eur. Phys. J. A 52~(10) (2016) 318.
\newblock \href {http://arxiv.org/abs/1605.05113} {\path{arXiv:1605.05113}},
  \href {https://doi.org/10.1140/epja/i2016-16318-4}
  {\path{doi:10.1140/epja/i2016-16318-4}}.

\bibitem{Bayar:2016ftu}
M.~Bayar, F.~Aceti, F.-K. Guo, E.~Oset, {A Discussion on Triangle Singularities
  in the $\Lambda_b \to J/\psi K^{-} p$ Reaction}, Phys. Rev. D 94~(7) (2016)
  074039.
\newblock \href {http://arxiv.org/abs/1609.04133} {\path{arXiv:1609.04133}},
  \href {https://doi.org/10.1103/PhysRevD.94.074039}
  {\path{doi:10.1103/PhysRevD.94.074039}}.

\bibitem{Nakamura:2021qvy}
S.~X. Nakamura, {$P_c(4312)^+$, $P_c(4380)^+$, and $P_c(4457)^+$ as double
  triangle cusps}, Phys. Rev. D 103 (2021) 111503.
\newblock \href {http://arxiv.org/abs/2103.06817} {\path{arXiv:2103.06817}},
  \href {https://doi.org/10.1103/PhysRevD.103.L111503}
  {\path{doi:10.1103/PhysRevD.103.L111503}}.

\bibitem{Guo:2019twa}
F.-K. Guo, X.-H. Liu, S.~Sakai, {Threshold cusps and triangle singularities in
  hadronic reactions}, Prog. Part. Nucl. Phys. 112 (2020) 103757.
\newblock \href {http://arxiv.org/abs/1912.07030} {\path{arXiv:1912.07030}},
  \href {https://doi.org/10.1016/j.ppnp.2020.103757}
  {\path{doi:10.1016/j.ppnp.2020.103757}}.

\bibitem{Pilloni:2016obd}
A.~Pilloni, C.~Fernandez-Ramirez, A.~Jackura, V.~Mathieu, M.~Mikhasenko,
  J.~Nys, A.~P. Szczepaniak, {Amplitude analysis and the nature of the
  Z$_c$(3900)}, Phys. Lett. B 772 (2017) 200--209.
\newblock \href {http://arxiv.org/abs/1612.06490} {\path{arXiv:1612.06490}},
  \href {https://doi.org/10.1016/j.physletb.2017.06.030}
  {\path{doi:10.1016/j.physletb.2017.06.030}}.

\bibitem{Belle:2003nnu}
S.~K. Choi, et~al., {Observation of a narrow charmonium-like state in exclusive
  $B^\pm \to K^\pm \pi^+ \pi^- J/\psi$ decays}, Phys. Rev. Lett. 91 (2003)
  262001.
\newblock \href {http://arxiv.org/abs/hep-ex/0309032}
  {\path{arXiv:hep-ex/0309032}}, \href
  {https://doi.org/10.1103/PhysRevLett.91.262001}
  {\path{doi:10.1103/PhysRevLett.91.262001}}.

\bibitem{LHCb:2013kgk}
R.~Aaij, et~al., {Determination of the X(3872) meson quantum numbers}, Phys.
  Rev. Lett. 110 (2013) 222001.
\newblock \href {http://arxiv.org/abs/1302.6269} {\path{arXiv:1302.6269}},
  \href {https://doi.org/10.1103/PhysRevLett.110.222001}
  {\path{doi:10.1103/PhysRevLett.110.222001}}.

\bibitem{COMPASS:2017wql}
M.~Aghasyan, et~al., {Search for muoproduction of $X (3872)$ at COMPASS and
  indication of a new state $\widetilde{X}(3872)$}, Phys. Lett. B 783 (2018)
  334--340.
\newblock \href {http://arxiv.org/abs/1707.01796} {\path{arXiv:1707.01796}},
  \href {https://doi.org/10.1016/j.physletb.2018.07.008}
  {\path{doi:10.1016/j.physletb.2018.07.008}}.

\bibitem{ParticleDataGroup:2022pth}
R.~L. Workman, et~al., {Review of Particle Physics}, PTEP 2022 (2022) 083C01.
\newblock \href {https://doi.org/10.1093/ptep/ptac097}
  {\path{doi:10.1093/ptep/ptac097}}.

\bibitem{LHCb:2015yax}
R.~Aaij, et~al., {Observation of $J/\psi p$ Resonances Consistent with
  Pentaquark States in $\Lambda_b^0 \to J/\psi K^- p$ Decays}, Phys. Rev. Lett.
  115 (2015) 072001.
\newblock \href {http://arxiv.org/abs/1507.03414} {\path{arXiv:1507.03414}},
  \href {https://doi.org/10.1103/PhysRevLett.115.072001}
  {\path{doi:10.1103/PhysRevLett.115.072001}}.

\bibitem{LHCb:2019kea}
R.~Aaij, et~al., {Observation of a narrow pentaquark state, $P_c(4312)^+$, and
  of two-peak structure of the $P_c(4450)^+$}, Phys. Rev. Lett. 122~(22) (2019)
  222001.
\newblock \href {http://arxiv.org/abs/1904.03947} {\path{arXiv:1904.03947}},
  \href {https://doi.org/10.1103/PhysRevLett.122.222001}
  {\path{doi:10.1103/PhysRevLett.122.222001}}.

\bibitem{Winney:2023oqc}
D.~Winney, et~al., {Dynamics in near-threshold $J/\psi$ photoproduction} (5
  2023).
\newblock \href {http://arxiv.org/abs/2305.01449} {\path{arXiv:2305.01449}}.

\bibitem{Cheung:2017tnt}
G.~K.~C. Cheung, C.~E. Thomas, J.~J. Dudek, R.~G. Edwards, {Tetraquark
  operators in lattice QCD and exotic flavour states in the charm sector}, JHEP
  11 (2017) 033.
\newblock \href {http://arxiv.org/abs/1709.01417} {\path{arXiv:1709.01417}},
  \href {https://doi.org/10.1007/JHEP11(2017)033}
  {\path{doi:10.1007/JHEP11(2017)033}}.

\bibitem{Lyu:2023xro}
Y.~Lyu, S.~Aoki, T.~Doi, T.~Hatsuda, Y.~Ikeda, J.~Meng, {Doubly charmed
  tetraquark $T^+_{cc}$ from Lattice QCD near Physical Point} (2 2023).
\newblock \href {http://arxiv.org/abs/2302.04505} {\path{arXiv:2302.04505}}.

\bibitem{Padmanath:2022cvl}
M.~Padmanath, S.~Prelovsek, {Signature of a Doubly Charm Tetraquark Pole in DD*
  Scattering on the Lattice}, Phys. Rev. Lett. 129~(3) (2022) 032002.
\newblock \href {http://arxiv.org/abs/2202.10110} {\path{arXiv:2202.10110}},
  \href {https://doi.org/10.1103/PhysRevLett.129.032002}
  {\path{doi:10.1103/PhysRevLett.129.032002}}.

\bibitem{Francis:2018jyb}
A.~Francis, R.~J. Hudspith, R.~Lewis, K.~Maltman, {Evidence for charm-bottom
  tetraquarks and the mass dependence of heavy-light tetraquark states from
  lattice QCD}, Phys. Rev. D 99~(5) (2019) 054505.
\newblock \href {http://arxiv.org/abs/1810.10550} {\path{arXiv:1810.10550}},
  \href {https://doi.org/10.1103/PhysRevD.99.054505}
  {\path{doi:10.1103/PhysRevD.99.054505}}.

\bibitem{Pumplin:2001ct}
J.~Pumplin, D.~Stump, R.~Brock, D.~Casey, J.~Huston, J.~Kalk, H.~L. Lai, W.~K.
  Tung, {Uncertainties of predictions from parton distribution functions. 2.
  The Hessian method}, Phys. Rev. D 65 (2001) 014013.
\newblock \href {http://arxiv.org/abs/hep-ph/0101032}
  {\path{arXiv:hep-ph/0101032}}, \href
  {https://doi.org/10.1103/PhysRevD.65.014013}
  {\path{doi:10.1103/PhysRevD.65.014013}}.

\bibitem{Nadolsky:2001yg}
P.~M. Nadolsky, Z.~Sullivan, {PDF Uncertainties in WH Production at Tevatron},
  eConf C010630 (2001) P510.
\newblock \href {http://arxiv.org/abs/hep-ph/0110378}
  {\path{arXiv:hep-ph/0110378}}.

\bibitem{Nadolsky:2008zw}
P.~M. Nadolsky, H.-L. Lai, Q.-H. Cao, J.~Huston, J.~Pumplin, D.~Stump, W.-K.
  Tung, C.~P. Yuan, {Implications of CTEQ global analysis for collider
  observables}, Phys. Rev. D 78 (2008) 013004.
\newblock \href {http://arxiv.org/abs/0802.0007} {\path{arXiv:0802.0007}},
  \href {https://doi.org/10.1103/PhysRevD.78.013004}
  {\path{doi:10.1103/PhysRevD.78.013004}}.

\bibitem{ChangPRL}
W.-C. Chang, J.-C. Peng, {Flavor Asymmetry of the Nucleon Sea and the
  Five-Quark Components of the Nucleons}, Phys. Rev. Lett. 106 (2011) 252002.
\newblock \href {http://arxiv.org/abs/1102.5631} {\path{arXiv:1102.5631}},
  \href {https://doi.org/10.1103/PhysRevLett.106.252002}
  {\path{doi:10.1103/PhysRevLett.106.252002}}.

\bibitem{BHPS}
S.~J. Brodsky, P.~Hoyer, C.~Peterson, N.~Sakai, {The Intrinsic Charm of the
  Proton}, Phys. Lett. B 93 (1980) 451--455.
\newblock \href {https://doi.org/10.1016/0370-2693(80)90364-0}
  {\path{doi:10.1016/0370-2693(80)90364-0}}.

\bibitem{SeaQuest:2021zxb}
J.~Dove, et~al., {Publisher Correction: The asymmetry of antimatter in the
  proton [doi: 10.1038/s41586-021-03282-z]}, Nature 590~(7847) (2021) 561--565.
\newblock \href {http://arxiv.org/abs/2103.04024} {\path{arXiv:2103.04024}},
  \href {https://doi.org/10.1038/s41586-022-04707-z}
  {\path{doi:10.1038/s41586-022-04707-z}}.

\bibitem{PhysRevD.64.052002}
R.~S. Towell, et~al.,
  \href{https://link.aps.org/doi/10.1103/PhysRevD.64.052002}{Improved
  measurement of the dbar/ubar asymmetry in the nucleon sea}, Phys. Rev. D 64
  (2001) 052002.
\newblock \href {https://doi.org/10.1103/PhysRevD.64.052002}
  {\path{doi:10.1103/PhysRevD.64.052002}}.
\newline\urlprefix\url{https://link.aps.org/doi/10.1103/PhysRevD.64.052002}

\bibitem{HERMES:1998uvc}
K.~Ackerstaff, et~al., {The Flavor asymmetry of the light quark sea from
  semiinclusive deep inelastic scattering}, Phys. Rev. Lett. 81 (1998)
  5519--5523.
\newblock \href {http://arxiv.org/abs/hep-ex/9807013}
  {\path{arXiv:hep-ex/9807013}}, \href
  {https://doi.org/10.1103/PhysRevLett.81.5519}
  {\path{doi:10.1103/PhysRevLett.81.5519}}.

\bibitem{Cooper-Sarkar:2015boa}
A.~M. Cooper-Sarkar, {HERA Collider Results}, PoS DIS2015 (2015) 005.
\newblock \href {http://arxiv.org/abs/1507.03849} {\path{arXiv:1507.03849}},
  \href {https://doi.org/10.22323/1.247.0005} {\path{doi:10.22323/1.247.0005}}.

\bibitem{CCFR:1994ikl}
A.~O. Bazarko, et~al., {Determination of the strange quark content of the
  nucleon from a next-to-leading order QCD analysis of neutrino charm
  production}, Z. Phys. C 65 (1995) 189--198.
\newblock \href {http://arxiv.org/abs/hep-ex/9406007}
  {\path{arXiv:hep-ex/9406007}}, \href {https://doi.org/10.1007/BF01571875}
  {\path{doi:10.1007/BF01571875}}.

\bibitem{NuTeV:2007uwm}
D.~Mason, et~al., {Measurement of the Nucleon Strange-Antistrange Asymmetry at
  Next-to-Leading Order in QCD from NuTeV Dimuon Data}, Phys. Rev. Lett. 99
  (2007) 192001.
\newblock \href {https://doi.org/10.1103/PhysRevLett.99.192001}
  {\path{doi:10.1103/PhysRevLett.99.192001}}.

\bibitem{Kayis-Topaksu:2011ols}
A.~Kayis-Topaksu, et~al., {Measurement of charm production in neutrino
  charged-current interactions}, New J. Phys. 13 (2011) 093002.
\newblock \href {http://arxiv.org/abs/1107.0613} {\path{arXiv:1107.0613}},
  \href {https://doi.org/10.1088/1367-2630/13/9/093002}
  {\path{doi:10.1088/1367-2630/13/9/093002}}.

\bibitem{NOMAD:2013hbk}
O.~Samoylov, et~al., {A Precision Measurement of Charm Dimuon Production in
  Neutrino Interactions from the NOMAD Experiment}, Nucl. Phys. B 876 (2013)
  339--375.
\newblock \href {http://arxiv.org/abs/1308.4750} {\path{arXiv:1308.4750}},
  \href {https://doi.org/10.1016/j.nuclphysb.2013.08.021}
  {\path{doi:10.1016/j.nuclphysb.2013.08.021}}.

\bibitem{Kalantarians:2017mkj}
N.~Kalantarians, C.~Keppel, M.~E. Christy, {Comparison of the Structure
  Function F2 as Measured by Charged Lepton and Neutrino Scattering from Iron
  Targets}, Phys. Rev. C 96~(3) (2017) 032201.
\newblock \href {http://arxiv.org/abs/1706.02002} {\path{arXiv:1706.02002}},
  \href {https://doi.org/10.1103/PhysRevC.96.032201}
  {\path{doi:10.1103/PhysRevC.96.032201}}.

\bibitem{Accardi:2009qv}
A.~Accardi, F.~Arleo, W.~K. Brooks, D.~D'Enterria, V.~Muccifora, {Parton
  Propagation and Fragmentation in QCD Matter}, Riv. Nuovo Cim. 32~(9-10)
  (2009) 439--554.
\newblock \href {http://arxiv.org/abs/0907.3534} {\path{arXiv:0907.3534}},
  \href {https://doi.org/10.1393/ncr/i2009-10048-0}
  {\path{doi:10.1393/ncr/i2009-10048-0}}.

\bibitem{Majumder:2010qh}
A.~Majumder, M.~Van~Leeuwen, {The Theory and Phenomenology of Perturbative QCD
  Based Jet Quenching}, Prog. Part. Nucl. Phys. 66 (2011) 41--92.
\newblock \href {http://arxiv.org/abs/1002.2206} {\path{arXiv:1002.2206}},
  \href {https://doi.org/10.1016/j.ppnp.2010.09.001}
  {\path{doi:10.1016/j.ppnp.2010.09.001}}.

\bibitem{ATLAS:2012sjl}
G.~Aad, et~al., {Determination of the strange quark density of the proton from
  ATLAS measurements of the $W \to \ell \nu$ and $Z \to \ell\ell$ cross
  sections}, Phys. Rev. Lett. 109 (2012) 012001.
\newblock \href {http://arxiv.org/abs/1203.4051} {\path{arXiv:1203.4051}},
  \href {https://doi.org/10.1103/PhysRevLett.109.012001}
  {\path{doi:10.1103/PhysRevLett.109.012001}}.

\bibitem{ATLAS:2016nqi}
M.~Aaboud, et~al., {Precision measurement and interpretation of inclusive $W^+$
  , $W^-$ and $Z/\gamma ^*$ production cross sections with the ATLAS detector},
  Eur. Phys. J. C 77~(6) (2017) 367.
\newblock \href {http://arxiv.org/abs/1612.03016} {\path{arXiv:1612.03016}},
  \href {https://doi.org/10.1140/epjc/s10052-017-4911-9}
  {\path{doi:10.1140/epjc/s10052-017-4911-9}}.

\bibitem{NuSea:1998kqi}
E.~A. Hawker, et~al., {Measurement of the light anti-quark flavor asymmetry in
  the nucleon sea}, Phys. Rev. Lett. 80 (1998) 3715--3718.
\newblock \href {http://arxiv.org/abs/hep-ex/9803011}
  {\path{arXiv:hep-ex/9803011}}, \href
  {https://doi.org/10.1103/PhysRevLett.80.3715}
  {\path{doi:10.1103/PhysRevLett.80.3715}}.

\bibitem{NuSea:2001idv}
R.~S. Towell, et~al., {Improved measurement of the anti-d / anti-u asymmetry in
  the nucleon sea}, Phys. Rev. D 64 (2001) 052002.
\newblock \href {http://arxiv.org/abs/hep-ex/0103030}
  {\path{arXiv:hep-ex/0103030}}, \href
  {https://doi.org/10.1103/PhysRevD.64.052002}
  {\path{doi:10.1103/PhysRevD.64.052002}}.

\bibitem{Alekhin:2008mb}
S.~Alekhin, S.~A. Kulagin, R.~Petti, {Determination of Strange Sea
  Distributions from Neutrino-Nucleon Deep Inelastic Scattering}, Phys. Lett. B
  675 (2009) 433--440.
\newblock \href {http://arxiv.org/abs/0812.4448} {\path{arXiv:0812.4448}},
  \href {https://doi.org/10.1016/j.physletb.2009.04.033}
  {\path{doi:10.1016/j.physletb.2009.04.033}}.

\bibitem{Alekhin:2014sya}
S.~Alekhin, J.~Blumlein, L.~Caminada, K.~Lipka, K.~Lohwasser, S.~Moch,
  R.~Petti, R.~Placakyte, {Determination of Strange Sea Quark Distributions
  from Fixed-target and Collider Data}, Phys. Rev. D 91~(9) (2015) 094002.
\newblock \href {http://arxiv.org/abs/1404.6469} {\path{arXiv:1404.6469}},
  \href {https://doi.org/10.1103/PhysRevD.91.094002}
  {\path{doi:10.1103/PhysRevD.91.094002}}.

\bibitem{Alekhin:2017olj}
S.~Alekhin, J.~Bl\"umlein, S.~Moch, {Strange sea determination from collider
  data}, Phys. Lett. B 777 (2018) 134--140.
\newblock \href {http://arxiv.org/abs/1708.01067} {\path{arXiv:1708.01067}},
  \href {https://doi.org/10.1016/j.physletb.2017.12.024}
  {\path{doi:10.1016/j.physletb.2017.12.024}}.

\bibitem{Alekhin:2017kpj}
S.~Alekhin, J.~Bl\"umlein, S.~Moch, R.~Placakyte, {Parton distribution
  functions, $\alpha_s$, and heavy-quark masses for LHC Run II}, Phys. Rev. D
  96~(1) (2017) 014011.
\newblock \href {http://arxiv.org/abs/1701.05838} {\path{arXiv:1701.05838}},
  \href {https://doi.org/10.1103/PhysRevD.96.014011}
  {\path{doi:10.1103/PhysRevD.96.014011}}.

\bibitem{Cooper-Sarkar:2018ufj}
A.~M. Cooper-Sarkar, K.~Wichmann, {QCD analysis of the ATLAS and CMS $W^{\pm}$
  and $Z$ cross-section measurements and implications for the strange sea
  density}, Phys. Rev. D 98~(1) (2018) 014027.
\newblock \href {http://arxiv.org/abs/1803.00968} {\path{arXiv:1803.00968}},
  \href {https://doi.org/10.1103/PhysRevD.98.014027}
  {\path{doi:10.1103/PhysRevD.98.014027}}.

\bibitem{HERMES:2008pug}
A.~Airapetian, et~al., {Measurement of Parton Distributions of Strange Quarks
  in the Nucleon from Charged-Kaon Production in Deep-Inelastic Scattering on
  the Deuteron}, Phys. Lett. B 666 (2008) 446--450.
\newblock \href {http://arxiv.org/abs/0803.2993} {\path{arXiv:0803.2993}},
  \href {https://doi.org/10.1016/j.physletb.2008.07.090}
  {\path{doi:10.1016/j.physletb.2008.07.090}}.

\bibitem{HERMES:2013ztj}
A.~Airapetian, et~al., {Reevaluation of the parton distribution of strange
  quarks in the nucleon}, Phys. Rev. D 89~(9) (2014) 097101.
\newblock \href {http://arxiv.org/abs/1312.7028} {\path{arXiv:1312.7028}},
  \href {https://doi.org/10.1103/PhysRevD.89.097101}
  {\path{doi:10.1103/PhysRevD.89.097101}}.

\bibitem{Leader:2010rb}
E.~Leader, A.~V. Sidorov, D.~B. Stamenov, {Determination of Polarized PDFs from
  a QCD Analysis of Inclusive and Semi-inclusive Deep Inelastic Scattering
  Data}, Phys. Rev. D 82 (2010) 114018.
\newblock \href {http://arxiv.org/abs/1010.0574} {\path{arXiv:1010.0574}},
  \href {https://doi.org/10.1103/PhysRevD.82.114018}
  {\path{doi:10.1103/PhysRevD.82.114018}}.

\bibitem{Leader:2011tm}
E.~Leader, A.~V. Sidorov, D.~B. Stamenov, {A Possible Resolution of the Strange
  Quark Polarization Puzzle ?}, Phys. Rev. D 84 (2011) 014002.
\newblock \href {http://arxiv.org/abs/1103.5979} {\path{arXiv:1103.5979}},
  \href {https://doi.org/10.1103/PhysRevD.84.014002}
  {\path{doi:10.1103/PhysRevD.84.014002}}.

\bibitem{Sato:2016wqj}
N.~Sato, J.~J. Ethier, W.~Melnitchouk, M.~Hirai, S.~Kumano, A.~Accardi, {First
  Monte Carlo analysis of fragmentation functions from single-inclusive $e^+
  e^-$ annihilation}, Phys. Rev. D 94~(11) (2016) 114004.
\newblock \href {http://arxiv.org/abs/1609.00899} {\path{arXiv:1609.00899}},
  \href {https://doi.org/10.1103/PhysRevD.94.114004}
  {\path{doi:10.1103/PhysRevD.94.114004}}.

\bibitem{Aschenauer:2015rna}
E.~C. Aschenauer, H.~E. Jackson, S.~Joosten, K.~Rith, G.~Schnell, C.~Van~Hulse,
  {Reply to \textquotedblleft{}Comment on \textquoteleft{}Reevaluation of the
  parton distribution of strange quarks in the
  nucleon\textquoteright{}\textquotedblright{}}, Phys. Rev. D 92~(9) (2015)
  098102.
\newblock \href {http://arxiv.org/abs/1508.04020} {\path{arXiv:1508.04020}},
  \href {https://doi.org/10.1103/PhysRevD.92.098102}
  {\path{doi:10.1103/PhysRevD.92.098102}}.

\bibitem{Borsa:2017vwy}
I.~Borsa, R.~Sassot, M.~Stratmann, {Probing the Sea Quark Content of the Proton
  with One-Particle-Inclusive Processes}, Phys. Rev. D 96~(9) (2017) 094020.
\newblock \href {http://arxiv.org/abs/1708.01630} {\path{arXiv:1708.01630}},
  \href {https://doi.org/10.1103/PhysRevD.96.094020}
  {\path{doi:10.1103/PhysRevD.96.094020}}.

\bibitem{Sato:2019yez}
N.~Sato, C.~Andres, J.~J. Ethier, W.~Melnitchouk, {Strange quark suppression
  from a simultaneous Monte Carlo analysis of parton distributions and
  fragmentation functions}, Phys. Rev. D 101~(7) (2020) 074020.
\newblock \href {http://arxiv.org/abs/1905.03788} {\path{arXiv:1905.03788}},
  \href {https://doi.org/10.1103/PhysRevD.101.074020}
  {\path{doi:10.1103/PhysRevD.101.074020}}.

\bibitem{Brady:2011uy}
L.~T. Brady, A.~Accardi, T.~J. Hobbs, W.~Melnitchouk, {Next-to leading order
  analysis of target mass corrections to structure functions and asymmetries},
  Phys. Rev. D 84 (2011) 074008, [Erratum: Phys.Rev.D 85, 039902 (2012)].
\newblock \href {http://arxiv.org/abs/1108.4734} {\path{arXiv:1108.4734}},
  \href {https://doi.org/10.1103/PhysRevD.84.074008}
  {\path{doi:10.1103/PhysRevD.84.074008}}.

\bibitem{Hobbs:2008mm}
T.~Hobbs, W.~Melnitchouk, {Finite-Q**2 corrections to parity-violating DIS},
  Phys. Rev. D 77 (2008) 114023.
\newblock \href {http://arxiv.org/abs/0801.4791} {\path{arXiv:0801.4791}},
  \href {https://doi.org/10.1103/PhysRevD.77.114023}
  {\path{doi:10.1103/PhysRevD.77.114023}}.

\bibitem{Hou:2022onq}
T.-J. Hou, H.-W. Lin, M.~Yan, C.~P. Yuan, {Impact of Lattice Strangeness
  Asymmetry Data in the CTEQ-TEA Global Analysis} (11 2022).
\newblock \href {http://arxiv.org/abs/2211.11064} {\path{arXiv:2211.11064}}.

\bibitem{Liu:2020rvc}
T.~Liu, W.~Melnitchouk, J.-W. Qiu, N.~Sato, {Factorized approach to radiative
  corrections for inelastic lepton-hadron collisions}, Phys. Rev. D 104~(9)
  (2021) 094033.
\newblock \href {http://arxiv.org/abs/2008.02895} {\path{arXiv:2008.02895}},
  \href {https://doi.org/10.1103/PhysRevD.104.094033}
  {\path{doi:10.1103/PhysRevD.104.094033}}.

\bibitem{Cocuzza:2022jye}
C.~Cocuzza, W.~Melnitchouk, A.~Metz, N.~Sato, {Polarized antimatter in the
  proton from a global QCD analysis}, Phys. Rev. D 106~(3) (2022) L031502.
\newblock \href {http://arxiv.org/abs/2202.03372} {\path{arXiv:2202.03372}},
  \href {https://doi.org/10.1103/PhysRevD.106.L031502}
  {\path{doi:10.1103/PhysRevD.106.L031502}}.

\bibitem{Ball:2016neh}
R.~D. Ball, V.~Bertone, M.~Bonvini, S.~Carrazza, S.~Forte, A.~Guffanti, N.~P.
  Hartland, J.~Rojo, L.~Rottoli, {A Determination of the Charm Content of the
  Proton}, Eur. Phys. J. C 76~(11) (2016) 647.
\newblock \href {http://arxiv.org/abs/1605.06515} {\path{arXiv:1605.06515}},
  \href {https://doi.org/10.1140/epjc/s10052-016-4469-y}
  {\path{doi:10.1140/epjc/s10052-016-4469-y}}.

\bibitem{Ball:2022qks}
R.~D. Ball, A.~Candido, J.~Cruz-Martinez, S.~Forte, T.~Giani, F.~Hekhorn,
  K.~Kudashkin, G.~Magni, J.~Rojo, {Evidence for intrinsic charm quarks in the
  proton}, Nature 608~(7923) (2022) 483--487.
\newblock \href {http://arxiv.org/abs/2208.08372} {\path{arXiv:2208.08372}},
  \href {https://doi.org/10.1038/s41586-022-04998-2}
  {\path{doi:10.1038/s41586-022-04998-2}}.

\bibitem{NNPDF:2021njg}
R.~D. Ball, et~al., {The path to proton structure at 1\% accuracy}, Eur. Phys.
  J. C 82~(5) (2022) 428.
\newblock \href {http://arxiv.org/abs/2109.02653} {\path{arXiv:2109.02653}},
  \href {https://doi.org/10.1140/epjc/s10052-022-10328-7}
  {\path{doi:10.1140/epjc/s10052-022-10328-7}}.

\bibitem{Hobbs:2013bia}
T.~J. Hobbs, J.~T. Londergan, W.~Melnitchouk, {Phenomenology of nonperturbative
  charm in the nucleon}, Phys. Rev. D 89~(7) (2014) 074008.
\newblock \href {http://arxiv.org/abs/1311.1578} {\path{arXiv:1311.1578}},
  \href {https://doi.org/10.1103/PhysRevD.89.074008}
  {\path{doi:10.1103/PhysRevD.89.074008}}.

\bibitem{Guzzi:2022rca}
M.~Guzzi, T.~J. Hobbs, K.~Xie, J.~Huston, P.~Nadolsky, C.~P. Yuan, {The
  persistent nonperturbative charm enigma} (11 2022).
\newblock \href {http://arxiv.org/abs/2211.01387} {\path{arXiv:2211.01387}}.

\bibitem{Kelsey:2021gpk}
M.~Kelsey, R.~Cruz-Torres, X.~Dong, Y.~Ji, S.~Radhakrishnan, E.~Sichtermann,
  {Constraints on gluon distribution functions in the nucleon and nucleus from
  open charm hadron production at the Electron-Ion Collider}, Phys. Rev. D
  104~(5) (2021) 054002.
\newblock \href {http://arxiv.org/abs/2107.05632} {\path{arXiv:2107.05632}},
  \href {https://doi.org/10.1103/PhysRevD.104.054002}
  {\path{doi:10.1103/PhysRevD.104.054002}}.

\bibitem{AbdulKhalek:2021gbh}
R.~Abdul~Khalek, et~al., {Science Requirements and Detector Concepts for the
  Electron-Ion Collider}: {EIC Yellow Report}, Nucl. Phys. A 1026 (2022)
  122447.
\newblock \href {http://arxiv.org/abs/2103.05419} {\path{arXiv:2103.05419}},
  \href {https://doi.org/10.1016/j.nuclphysa.2022.122447}
  {\path{doi:10.1016/j.nuclphysa.2022.122447}}.

\bibitem{Gao:2021fle}
J.~Gao, T.~J. Hobbs, P.~M. Nadolsky, C.~Sun, C.~P. Yuan, {General heavy-flavor
  mass scheme for charged-current DIS at NNLO and beyond}, Phys. Rev. D 105~(1)
  (2022) L011503.
\newblock \href {http://arxiv.org/abs/2107.00460} {\path{arXiv:2107.00460}},
  \href {https://doi.org/10.1103/PhysRevD.105.L011503}
  {\path{doi:10.1103/PhysRevD.105.L011503}}.

\bibitem{Gauld:2015yia}
R.~Gauld, J.~Rojo, L.~Rottoli, J.~Talbert, {Charm production in the forward
  region: constraints on the small-x gluon and backgrounds for neutrino
  astronomy}, JHEP 11 (2015) 009.
\newblock \href {http://arxiv.org/abs/1506.08025} {\path{arXiv:1506.08025}},
  \href {https://doi.org/10.1007/JHEP11(2015)009}
  {\path{doi:10.1007/JHEP11(2015)009}}.

\bibitem{Gao:2017yyd}
J.~Gao, L.~Harland-Lang, J.~Rojo, {The Structure of the Proton in the LHC
  Precision Era}, Phys. Rept. 742 (2018) 1--121.
\newblock \href {http://arxiv.org/abs/1709.04922} {\path{arXiv:1709.04922}},
  \href {https://doi.org/10.1016/j.physrep.2018.03.002}
  {\path{doi:10.1016/j.physrep.2018.03.002}}.

\bibitem{PDF4LHCWorkingGroup:2022cjn}
R.~D. Ball, et~al., {The PDF4LHC21 combination of global PDF fits for the LHC
  Run III}, J. Phys. G 49~(8) (2022) 080501.
\newblock \href {http://arxiv.org/abs/2203.05506} {\path{arXiv:2203.05506}},
  \href {https://doi.org/10.1088/1361-6471/ac7216}
  {\path{doi:10.1088/1361-6471/ac7216}}.

\bibitem{Hou:2019efy}
T.-J. Hou, et~al., {New CTEQ global analysis of quantum chromodynamics with
  high-precision data from the LHC}, Phys. Rev. D 103~(1) (2021) 014013.
\newblock \href {http://arxiv.org/abs/1912.10053} {\path{arXiv:1912.10053}},
  \href {https://doi.org/10.1103/PhysRevD.103.014013}
  {\path{doi:10.1103/PhysRevD.103.014013}}.

\bibitem{Bailey:2020ooq}
S.~Bailey, T.~Cridge, L.~A. Harland-Lang, A.~D. Martin, R.~S. Thorne, {Parton
  distributions from LHC, HERA, Tevatron and fixed target data: MSHT20 PDFs},
  Eur. Phys. J. C 81~(4) (2021) 341.
\newblock \href {http://arxiv.org/abs/2012.04684} {\path{arXiv:2012.04684}},
  \href {https://doi.org/10.1140/epjc/s10052-021-09057-0}
  {\path{doi:10.1140/epjc/s10052-021-09057-0}}.

\bibitem{Collins:1977iv}
J.~C. Collins, D.~E. Soper, {Angular Distribution of Dileptons in High-Energy
  Hadron Collisions}, Phys. Rev. D 16 (1977) 2219.
\newblock \href {https://doi.org/10.1103/PhysRevD.16.2219}
  {\path{doi:10.1103/PhysRevD.16.2219}}.

\bibitem{Ball:2022qtp}
R.~D. Ball, A.~Candido, S.~Forte, F.~Hekhorn, E.~R. Nocera, J.~Rojo, C.~Schwan,
  {Parton distributions and new physics searches: the Drell\textendash{}Yan
  forward\textendash{}backward asymmetry as a case study}, Eur. Phys. J. C
  82~(12) (2022) 1160.
\newblock \href {http://arxiv.org/abs/2209.08115} {\path{arXiv:2209.08115}},
  \href {https://doi.org/10.1140/epjc/s10052-022-11133-y}
  {\path{doi:10.1140/epjc/s10052-022-11133-y}}.

\bibitem{Greljo:2021kvv}
A.~Greljo, S.~Iranipour, Z.~Kassabov, M.~Madigan, J.~Moore, J.~Rojo, M.~Ubiali,
  C.~Voisey, {Parton distributions in the SMEFT from high-energy Drell-Yan
  tails}, JHEP 07 (2021) 122.
\newblock \href {http://arxiv.org/abs/2104.02723} {\path{arXiv:2104.02723}},
  \href {https://doi.org/10.1007/JHEP07(2021)122}
  {\path{doi:10.1007/JHEP07(2021)122}}.

\bibitem{Gao:2022srd}
J.~Gao, M.~Gao, T.~J. Hobbs, D.~Liu, X.~Shen, {Simultaneous CTEQ-TEA extraction
  of PDFs and SMEFT parameters from jet and $ t\overline{t} $ data}, JHEP 05
  (2023) 003.
\newblock \href {http://arxiv.org/abs/2211.01094} {\path{arXiv:2211.01094}},
  \href {https://doi.org/10.1007/JHEP05(2023)003}
  {\path{doi:10.1007/JHEP05(2023)003}}.

\bibitem{Ruso:2022qes}
L.~A. Ruso, et~al., {Theoretical tools for neutrino scattering: interplay
  between lattice QCD, EFTs, nuclear physics, phenomenology, and neutrino event
  generators} (3 2022).
\newblock \href {http://arxiv.org/abs/2203.09030} {\path{arXiv:2203.09030}}.

\bibitem{AdSCFT}
T.~Liu, R.~S. Sufian, G.~F. de~T\'eramond, H.~G. Dosch, S.~J. Brodsky, A.~Deur,
  \href{https://link.aps.org/doi/10.1103/PhysRevLett.124.082003}{Unified
  description of polarized and unpolarized quark distributions in the proton},
  Phys. Rev. Lett. 124 (2020) 082003.
\newblock \href {https://doi.org/10.1103/PhysRevLett.124.082003}
  {\path{doi:10.1103/PhysRevLett.124.082003}}.
\newline\urlprefix\url{https://link.aps.org/doi/10.1103/PhysRevLett.124.082003}

\bibitem{Deur:2016tte}
A.~Deur, S.~J. Brodsky, G.~F. de~Teramond, {The QCD Running Coupling}, Nucl.
  Phys. 90 (2016) 1.
\newblock \href {http://arxiv.org/abs/1604.08082} {\path{arXiv:1604.08082}},
  \href {https://doi.org/10.1016/j.ppnp.2016.04.003}
  {\path{doi:10.1016/j.ppnp.2016.04.003}}.

\bibitem{ParticleDataGroup:2020ssz}
P.~A. Zyla, et~al., {Review of Particle Physics}, PTEP 2020~(8) (2020) 083C01.
\newblock \href {https://doi.org/10.1093/ptep/ptaa104}
  {\path{doi:10.1093/ptep/ptaa104}}.

\bibitem{dEnterria:2022hzv}
D.~d'Enterria, et~al., {The strong coupling constant: State of the art and the
  decade ahead} (3 2022).
\newblock \href {http://arxiv.org/abs/2203.08271} {\path{arXiv:2203.08271}}.

\bibitem{Bjorken:1966jh}
J.~D. Bjorken, {Applications of the Chiral U(6) x (6) Algebra of Current
  Densities}, Phys. Rev. 148 (1966) 1467--1478.
\newblock \href {https://doi.org/10.1103/PhysRev.148.1467}
  {\path{doi:10.1103/PhysRev.148.1467}}.

\bibitem{EG12}
S.~Kuhn, et~al., The Longitudinal Spin Structure of the Nucleon
  \href{https://misportal.jlab.org/mis/physics/experiments/viewProposal.cfm?paperId=688}{Jlab
  experiment E12-06-109 ''}. (2006).

\bibitem{Kataev:1994gd}
A.~L. Kataev, {The Ellis-Jaffe sum rule: The Estimates of the next to
  next-to-leading order QCD corrections}, Phys. Rev. D 50 (1994) R5469--R5472.
\newblock \href {http://arxiv.org/abs/hep-ph/9408248}
  {\path{arXiv:hep-ph/9408248}}, \href
  {https://doi.org/10.1103/PhysRevD.50.R5469}
  {\path{doi:10.1103/PhysRevD.50.R5469}}.

\bibitem{Bjorken_a5}
A.~L. Kataev, private communication in
  \href{https://www.slac.stanford.edu/exp/e154/incerti_thesis.pdf}{S. Incerti,
  Ph. D dissertation ``Mesure de la fonction de structure polaris\'ee $g_1^n$
  du neutron par l'experience e154 au slac''}. (Jan. 1998).

\bibitem{Deur:2014vea}
A.~Deur, Y.~Prok, V.~Burkert, D.~Crabb, F.~X. Girod, K.~A. Griffioen, N.~Guler,
  S.~E. Kuhn, N.~Kvaltine, {High precision determination of the $Q^2$ evolution
  of the Bjorken Sum}, Phys. Rev. D 90~(1) (2014) 012009.
\newblock \href {http://arxiv.org/abs/1405.7854} {\path{arXiv:1405.7854}},
  \href {https://doi.org/10.1103/PhysRevD.90.012009}
  {\path{doi:10.1103/PhysRevD.90.012009}}.

\bibitem{Kniehl:2006bg}
B.~A. Kniehl, A.~V. Kotikov, A.~I. Onishchenko, O.~L. Veretin, {Strong-coupling
  constant with flavor thresholds at five loops in the anti-MS scheme}, Phys.
  Rev. Lett. 97 (2006) 042001.
\newblock \href {http://arxiv.org/abs/hep-ph/0607202}
  {\path{arXiv:hep-ph/0607202}}, \href
  {https://doi.org/10.1103/PhysRevLett.97.042001}
  {\path{doi:10.1103/PhysRevLett.97.042001}}.

\bibitem{Deur:2022msf}
A.~Deur, V.~Burkert, J.~P. Chen, W.~Korsch, {Experimental determination of the
  QCD effective charge $\alpha_{g_1}(Q)$}, Particles 5 (2022) 171.
\newblock \href {http://arxiv.org/abs/2205.01169} {\path{arXiv:2205.01169}},
  \href {https://doi.org/10.3390/particles5020015}
  {\path{doi:10.3390/particles5020015}}.

\bibitem{Brodsky:2010ur}
S.~J. Brodsky, G.~F. de~Teramond, A.~Deur, {Nonperturbative QCD Coupling and
  its $\beta$-function from Light-Front Holography}, Phys. Rev. D 81 (2010)
  096010.
\newblock \href {http://arxiv.org/abs/1002.3948} {\path{arXiv:1002.3948}},
  \href {https://doi.org/10.1103/PhysRevD.81.096010}
  {\path{doi:10.1103/PhysRevD.81.096010}}.

\bibitem{Cui:2019dwv}
Z.-F. Cui, J.-L. Zhang, D.~Binosi, F.~de~Soto, C.~Mezrag, J.~Papavassiliou,
  C.~D. Roberts, J.~Rodr\'\i{}guez-Quintero, J.~Segovia, S.~Zafeiropoulos,
  {Effective charge from lattice QCD}, Chin. Phys. C 44~(8) (2020) 083102.
\newblock \href {http://arxiv.org/abs/1912.08232} {\path{arXiv:1912.08232}},
  \href {https://doi.org/10.1088/1674-1137/44/8/083102}
  {\path{doi:10.1088/1674-1137/44/8/083102}}.

\bibitem{Barry:2023qqh}
P.~C. Barry, L.~Gamberg, W.~Melnitchouk, E.~Moffat, D.~Pitonyak, A.~Prokudin,
  N.~Sato, {Tomography of pions and protons via transverse momentum dependent
  distributions} (2 2023).
\newblock \href {http://arxiv.org/abs/2302.01192} {\path{arXiv:2302.01192}}.

\bibitem{Cao:2021aci}
N.~Y. Cao, P.~C. Barry, N.~Sato, W.~Melnitchouk, {Towards the three-dimensional
  parton structure of the pion: Integrating transverse momentum data into
  global QCD analysis}, Phys. Rev. D 103~(11) (2021) 114014.
\newblock \href {http://arxiv.org/abs/2103.02159} {\path{arXiv:2103.02159}},
  \href {https://doi.org/10.1103/PhysRevD.103.114014}
  {\path{doi:10.1103/PhysRevD.103.114014}}.

\bibitem{PionTDISProposal}
C.~E. Keppel, et~al., {C12-15-006 JLab} experiment: Measurement of tagged deep
  inelastic scattering (2015).

\bibitem{KaonTDISProposal}
K.~Park, et~al., {C12-15-006A JLab} run group: Measurement of kaon structure
  through tagged deep inelastic scattering (2017).

\bibitem{NA10:1985ibr}
B.~Betev, et~al., {Differential Cross-section of High Mass Muon Pairs Produced
  by a 194-{GeV}/$c \pi^-$ Beam on a Tungsten Target}, Z. Phys. C 28 (1985) 9.
\newblock \href {https://doi.org/10.1007/BF01550243}
  {\path{doi:10.1007/BF01550243}}.

\bibitem{Conway:1989fs}
J.~S. Conway, et~al., {Experimental Study of Muon Pairs Produced by 252-GeV
  Pions on Tungsten}, Phys. Rev. D 39 (1989) 92--122.
\newblock \href {https://doi.org/10.1103/PhysRevD.39.92}
  {\path{doi:10.1103/PhysRevD.39.92}}.

\bibitem{H1:2010hym}
F.~D. Aaron, et~al., {Measurement of Leading Neutron Production in
  Deep-Inelastic Scattering at HERA}, Eur. Phys. J. C 68 (2010) 381--399.
\newblock \href {http://arxiv.org/abs/1001.0532} {\path{arXiv:1001.0532}},
  \href {https://doi.org/10.1140/epjc/s10052-010-1369-4}
  {\path{doi:10.1140/epjc/s10052-010-1369-4}}.

\bibitem{ZEUS:2002gig}
S.~Chekanov, et~al., {Leading neutron production in e+ p collisions at HERA},
  Nucl. Phys. B 637 (2002) 3--56.
\newblock \href {http://arxiv.org/abs/hep-ex/0205076}
  {\path{arXiv:hep-ex/0205076}}, \href
  {https://doi.org/10.1016/S0550-3213(02)00439-X}
  {\path{doi:10.1016/S0550-3213(02)00439-X}}.

\bibitem{Arrington:2021biu}
J.~Arrington, et~al., {Revealing the Structure of Light Pseudoscalar Mesons at
  the Electron-Ion Collider}, J. Phys. G 48 (2021) 075106.

\bibitem{Bacchetta:2006tn}
A.~Bacchetta, M.~Diehl, K.~Goeke, A.~Metz, P.~J. Mulders, M.~Schlegel,
  {Semi-inclusive deep inelastic scattering at small transverse momentum}, JHEP
  02 (2007) 093.
\newblock \href {http://arxiv.org/abs/hep-ph/0611265}
  {\path{arXiv:hep-ph/0611265}}, \href
  {https://doi.org/10.1088/1126-6708/2007/02/093}
  {\path{doi:10.1088/1126-6708/2007/02/093}}.

\bibitem{Gonzalez-Hernandez:2018ipj}
J.~Gonzalez-Hernandez, T.~Rogers, N.~Sato, B.~Wang, {Challenges with Large
  Transverse Momentum in Semi-Inclusive Deeply Inelastic Scattering}, Phys.
  Rev. D 98~(11) (2018) 114005.
\newblock \href {http://arxiv.org/abs/1808.04396} {\path{arXiv:1808.04396}},
  \href {https://doi.org/10.1103/PhysRevD.98.114005}
  {\path{doi:10.1103/PhysRevD.98.114005}}.

\bibitem{Wang:2019bvb}
B.~Wang, J.~O. Gonzalez-Hernandez, T.~C. Rogers, N.~Sato, {Large Transverse
  Momentum in Semi-Inclusive Deeply Inelastic Scattering Beyond Lowest Order},
  Phys. Rev. D 99~(9) (2019) 094029.
\newblock \href {http://arxiv.org/abs/1903.01529} {\path{arXiv:1903.01529}},
  \href {https://doi.org/10.1103/PhysRevD.99.094029}
  {\path{doi:10.1103/PhysRevD.99.094029}}.

\bibitem{Boglione:2016bph}
M.~Boglione, J.~Collins, L.~Gamberg, J.~O. Gonzalez-Hernandez, T.~C. Rogers,
  N.~Sato, {Kinematics of Current Region Fragmentation in Semi-Inclusive Deeply
  Inelastic Scattering}, Phys. Lett. B 766 (2017) 245--253.
\newblock \href {http://arxiv.org/abs/1611.10329} {\path{arXiv:1611.10329}},
  \href {https://doi.org/10.1016/j.physletb.2017.01.021}
  {\path{doi:10.1016/j.physletb.2017.01.021}}.

\bibitem{Collins:2016hqq}
J.~Collins, L.~Gamberg, A.~Prokudin, T.~C. Rogers, N.~Sato, B.~Wang, {Relating
  Transverse Momentum Dependent and Collinear Factorization Theorems in a
  Generalized Formalism}, Phys. Rev. D 94~(3) (2016) 034014.
\newblock \href {http://arxiv.org/abs/1605.00671} {\path{arXiv:1605.00671}},
  \href {https://doi.org/10.1103/PhysRevD.94.034014}
  {\path{doi:10.1103/PhysRevD.94.034014}}.

\bibitem{Boglione:2019nwk}
M.~Boglione, A.~Dotson, L.~Gamberg, S.~Gordon, J.~Gonzalez-Hernandez,
  A.~Prokudin, T.~Rogers, N.~Sato, {Mapping the Kinematical Regimes of
  Semi-Inclusive Deep Inelastic Scattering}, JHEP 10 (2019) 122.
\newblock \href {http://arxiv.org/abs/1904.12882} {\path{arXiv:1904.12882}},
  \href {https://doi.org/10.1007/JHEP10(2019)122}
  {\path{doi:10.1007/JHEP10(2019)122}}.

\bibitem{CLAS:2010fns}
H.~Avakian, et~al., {Measurement of Single and Double Spin Asymmetries in Deep
  Inelastic Pion Electroproduction with a Longitudinally Polarized Target},
  Phys. Rev. Lett. 105 (2010) 262002.
\newblock \href {http://arxiv.org/abs/1003.4549} {\path{arXiv:1003.4549}},
  \href {https://doi.org/10.1103/PhysRevLett.105.262002}
  {\path{doi:10.1103/PhysRevLett.105.262002}}.

\bibitem{CLAS:2017yrm}
S.~Jawalkar, et~al., {Semi-Inclusive $\pi_0$ target and beam-target asymmetries
  from 6 GeV electron scattering with CLAS}, Phys. Lett. B 782 (2018) 662--667.
\newblock \href {http://arxiv.org/abs/1709.10054} {\path{arXiv:1709.10054}},
  \href {https://doi.org/10.1016/j.physletb.2018.06.014}
  {\path{doi:10.1016/j.physletb.2018.06.014}}.

\bibitem{Musch:2010ka}
B.~U. Musch, P.~Hagler, J.~W. Negele, A.~Schafer, {Exploring quark transverse
  momentum distributions with lattice QCD}, Phys. Rev. D 83 (2011) 094507.
\newblock \href {http://arxiv.org/abs/1011.1213} {\path{arXiv:1011.1213}},
  \href {https://doi.org/10.1103/PhysRevD.83.094507}
  {\path{doi:10.1103/PhysRevD.83.094507}}.

\bibitem{Avakian:2019uzf}
H.~Avakian, {Hadronization of quarks and correlated di-hadron production in
  hard scattering}, PoS DIS2019 (2019) 265.
\newblock \href {https://doi.org/10.22323/1.352.0265}
  {\path{doi:10.22323/1.352.0265}}.

\bibitem{Bebek:1975pn}
C.~J. Bebek, C.~N. Brown, M.~Herzlinger, S.~D. Holmes, C.~A. Lichtenstein,
  F.~M. Pipkin, S.~Raither, L.~K. Sisterson, {Scaling Behavior of Inclusive
  Pion Electroproduction}, Phys. Rev. Lett. 34 (1975) 759.
\newblock \href {https://doi.org/10.1103/PhysRevLett.34.759}
  {\path{doi:10.1103/PhysRevLett.34.759}}.

\bibitem{Bebek:1976wv}
C.~J. Bebek, A.~Browman, C.~N. Brown, K.~M. Hanson, R.~V. Kline, D.~Larson,
  F.~M. Pipkin, S.~W. Raither, A.~Silverman, L.~K. Sisterson, {Charged Pion
  Electroproduction from Protons Up to Q**2 = 9.5-GeV**2}, Phys. Rev. Lett. 37
  (1976) 1525--1528.
\newblock \href {https://doi.org/10.1103/PhysRevLett.37.1525}
  {\path{doi:10.1103/PhysRevLett.37.1525}}.

\bibitem{Bebek:1977pf}
C.~J. Bebek, C.~N. Brown, M.~S. Herzlinger, S.~D. Holmes, C.~A. Lichtenstein,
  F.~M. Pipkin, S.~W. Raither, L.~K. Sisterson, {Inclusive Charged Pion
  Electroproduction}, Phys. Rev. D 15 (1977) 3085.
\newblock \href {https://doi.org/10.1103/PhysRevD.15.3085}
  {\path{doi:10.1103/PhysRevD.15.3085}}.

\bibitem{Bacchetta:2008xw}
A.~Bacchetta, D.~Boer, M.~Diehl, P.~J. Mulders, {Matches and mismatches in the
  descriptions of semi-inclusive processes at low and high transverse
  momentum}, JHEP 08 (2008) 023.
\newblock \href {http://arxiv.org/abs/0803.0227} {\path{arXiv:0803.0227}},
  \href {https://doi.org/10.1088/1126-6708/2008/08/023}
  {\path{doi:10.1088/1126-6708/2008/08/023}}.

\bibitem{Anselmino:2005nn}
M.~Anselmino, M.~Boglione, U.~D'Alesio, A.~Kotzinian, F.~Murgia, A.~Prokudin,
  {The Role of Cahn and sivers effects in deep inelastic scattering}, Phys.
  Rev. D 71 (2005) 074006.
\newblock \href {http://arxiv.org/abs/hep-ph/0501196}
  {\path{arXiv:hep-ph/0501196}}, \href
  {https://doi.org/10.1103/PhysRevD.71.074006}
  {\path{doi:10.1103/PhysRevD.71.074006}}.

\bibitem{Bacchetta:2022awv}
A.~Bacchetta, V.~Bertone, C.~Bissolotti, G.~Bozzi, M.~Cerutti, F.~Piacenza,
  M.~Radici, A.~Signori, {Unpolarized transverse momentum distributions from a
  global fit of Drell-Yan and semi-inclusive deep-inelastic scattering data},
  JHEP 10 (2022) 127.
\newblock \href {http://arxiv.org/abs/2206.07598} {\path{arXiv:2206.07598}},
  \href {https://doi.org/10.1007/JHEP10(2022)127}
  {\path{doi:10.1007/JHEP10(2022)127}}.

\bibitem{COMPASS:2014kcy}
C.~Adolph, et~al., {Measurement of azimuthal hadron asymmetries in
  semi-inclusive deep inelastic scattering off unpolarised nucleons}, Nucl.
  Phys. B 886 (2014) 1046--1077.
\newblock \href {http://arxiv.org/abs/1401.6284} {\path{arXiv:1401.6284}},
  \href {https://doi.org/10.1016/j.nuclphysb.2014.07.019}
  {\path{doi:10.1016/j.nuclphysb.2014.07.019}}.

\bibitem{Moretti:2021naj}
A.~Moretti, {TMD observables in unpolarised Semi-Inclusive DIS at COMPASS},
  SciPost Phys. Proc. 8 (2022) 144.
\newblock \href {http://arxiv.org/abs/2107.10740} {\path{arXiv:2107.10740}},
  \href {https://doi.org/10.21468/SciPostPhysProc.8.144}
  {\path{doi:10.21468/SciPostPhysProc.8.144}}.

\bibitem{Airapetian:2012yg}
A.~Airapetian, et~al., {Azimuthal distributions of charged hadrons, pions, and
  kaons produced in deep-inelastic scattering off unpolarized protons and
  deuterons}, Phys. Rev. D 87~(1) (2013) 012010.
\newblock \href {http://arxiv.org/abs/1204.4161} {\path{arXiv:1204.4161}},
  \href {https://doi.org/10.1103/PhysRevD.87.012010}
  {\path{doi:10.1103/PhysRevD.87.012010}}.

\bibitem{Osipenko:2008aa}
M.~Osipenko, et~al., {Measurement of unpolarized semi-inclusive pi+
  electroproduction off the proton}, Phys. Rev. D 80 (2009) 032004.
\newblock \href {http://arxiv.org/abs/0809.1153} {\path{arXiv:0809.1153}},
  \href {https://doi.org/10.1103/PhysRevD.80.032004}
  {\path{doi:10.1103/PhysRevD.80.032004}}.

\bibitem{Diehl:2021rnj}
S.~Diehl, et~al., {First multidimensional, high precision measurements of
  semi-inclusive $\pi^+$ beam single spin asymmetries from the proton over a
  wide range of kinematics} (1 2021).
\newblock \href {http://arxiv.org/abs/2101.03544} {\path{arXiv:2101.03544}}.

\bibitem{Collins:1992kk}
J.~C. Collins, {Fragmentation of transversely polarized quarks probed in
  transverse momentum distributions}, Nucl. Phys. B 396 (1993) 161--182.
\newblock \href {http://arxiv.org/abs/hep-ph/9208213}
  {\path{arXiv:hep-ph/9208213}}, \href
  {https://doi.org/10.1016/0550-3213(93)90262-N}
  {\path{doi:10.1016/0550-3213(93)90262-N}}.

\bibitem{Kerbizi:2018qpp}
A.~Kerbizi, X.~Artru, Z.~Belghobsi, F.~Bradamante, A.~Martin, {Recursive model
  for the fragmentation of polarized quarks}, Phys. Rev. D 97~(7) (2018)
  074010.
\newblock \href {http://arxiv.org/abs/1802.00962} {\path{arXiv:1802.00962}},
  \href {https://doi.org/10.1103/PhysRevD.97.074010}
  {\path{doi:10.1103/PhysRevD.97.074010}}.

\bibitem{Matevosyan:2016fwi}
H.~H. Matevosyan, A.~Kotzinian, A.~W. Thomas, {Monte Carlo Implementation of
  Polarized Hadronization}, Phys. Rev. D 95~(1) (2017) 014021.
\newblock \href {http://arxiv.org/abs/1610.05624} {\path{arXiv:1610.05624}},
  \href {https://doi.org/10.1103/PhysRevD.95.014021}
  {\path{doi:10.1103/PhysRevD.95.014021}}.

\bibitem{Kerbizi:2021pzn}
A.~Kerbizi, L.~L\"onnblad, {StringSpinner - adding spin to the PYTHIA string
  fragmentation}, Comput. Phys. Commun. 272 (2022) 108234.
\newblock \href {http://arxiv.org/abs/2105.09730} {\path{arXiv:2105.09730}},
  \href {https://doi.org/10.1016/j.cpc.2021.108234}
  {\path{doi:10.1016/j.cpc.2021.108234}}.

\bibitem{Sjostrand:2014zea}
T.~Sj\"ostrand, S.~Ask, J.~R. Christiansen, R.~Corke, N.~Desai, P.~Ilten,
  S.~Mrenna, S.~Prestel, C.~O. Rasmussen, P.~Z. Skands, {An introduction to
  PYTHIA 8.2}, Comput. Phys. Commun. 191 (2015) 159--177.
\newblock \href {http://arxiv.org/abs/1410.3012} {\path{arXiv:1410.3012}},
  \href {https://doi.org/10.1016/j.cpc.2015.01.024}
  {\path{doi:10.1016/j.cpc.2015.01.024}}.

\bibitem{Hayward:2021psm}
T.~B. Hayward, et~al., {Observation of Beam Spin Asymmetries in the Process
  $ep\rightarrow{e}^{'}{\pi}^{+}{\pi}^{-}X$ with CLAS12}, Phys. Rev. Lett. 126
  (2021) 152501.
\newblock \href {http://arxiv.org/abs/2101.04842} {\path{arXiv:2101.04842}},
  \href {https://doi.org/10.1103/PhysRevLett.126.152501}
  {\path{doi:10.1103/PhysRevLett.126.152501}}.

\bibitem{Kerbizi:2021gos}
A.~Kerbizi, X.~Artru, A.~Martin, {Production of vector mesons in the
  String+${}^3P_0$ model of polarized quark fragmentation}, Phys. Rev. D
  104~(11) (2021) 114038.
\newblock \href {http://arxiv.org/abs/2109.06124} {\path{arXiv:2109.06124}},
  \href {https://doi.org/10.1103/PhysRevD.104.114038}
  {\path{doi:10.1103/PhysRevD.104.114038}}.

\bibitem{Bacchetta:2000jk}
A.~Bacchetta, P.~J. Mulders, {Deep inelastic leptoproduction of spin-one
  hadrons}, Phys. Rev. D 62 (2000) 114004.
\newblock \href {http://arxiv.org/abs/hep-ph/0007120}
  {\path{arXiv:hep-ph/0007120}}, \href
  {https://doi.org/10.1103/PhysRevD.62.114004}
  {\path{doi:10.1103/PhysRevD.62.114004}}.

\bibitem{COMPASS:2022jth}
T.~C. Collaboration, {Collins and Sivers transverse-spin asymmetries in
  inclusive muoproduction of $\rho^0$ mesons}, CERN-EP-2022-234 (10 2022).
\newblock \href {http://arxiv.org/abs/2211.00093} {\path{arXiv:2211.00093}}.

\bibitem{COMPASS:2014bze}
C.~Adolph, et~al., {Collins and Sivers asymmetries in muonproduction of pions
  and kaons off transversely polarised protons}, Phys. Lett. B 744 (2015)
  250--259.
\newblock \href {http://arxiv.org/abs/1408.4405} {\path{arXiv:1408.4405}},
  \href {https://doi.org/10.1016/j.physletb.2015.03.056}
  {\path{doi:10.1016/j.physletb.2015.03.056}}.

\bibitem{Trentadue:1993ka}
L.~Trentadue, G.~Veneziano, {Fracture functions: An Improved description of
  inclusive hard processes in QCD}, Phys. Lett. B 323 (1994) 201--211.
\newblock \href {https://doi.org/10.1016/0370-2693(94)90292-5}
  {\path{doi:10.1016/0370-2693(94)90292-5}}.

\bibitem{Anselmino:2011ss}
M.~Anselmino, V.~Barone, A.~Kotzinian, {SIDIS in the target fragmentation
  region: Polarized and transverse momentum dependent fracture functions},
  Phys. Lett. B 699 (2011) 108--118.
\newblock \href {http://arxiv.org/abs/1102.4214} {\path{arXiv:1102.4214}},
  \href {https://doi.org/10.1016/j.physletb.2011.03.067}
  {\path{doi:10.1016/j.physletb.2011.03.067}}.

\bibitem{CLAS:2022sqt}
H.~Avakian, et~al., {Observation of Correlations between Spin and Transverse
  Momenta in Back-to-Back Dihadron Production at CLAS12}, Phys. Rev. Lett.
  130~(2) (2023) 022501.
\newblock \href {http://arxiv.org/abs/2208.05086} {\path{arXiv:2208.05086}},
  \href {https://doi.org/10.1103/PhysRevLett.130.022501}
  {\path{doi:10.1103/PhysRevLett.130.022501}}.

\bibitem{Schweitzer:2012hh}
P.~Schweitzer, M.~Strikman, C.~Weiss, {Intrinsic transverse momentum and parton
  correlations from dynamical chiral symmetry breaking}, JHEP 01 (2013) 163.
\newblock \href {http://arxiv.org/abs/1210.1267} {\path{arXiv:1210.1267}},
  \href {https://doi.org/10.1007/JHEP01(2013)163}
  {\path{doi:10.1007/JHEP01(2013)163}}.

\bibitem{Sargsian:2005rm}
M.~Sargsian, M.~Strikman, {Model independent method for determination of the
  DIS structure of free neutron}, Phys. Lett. B 639 (2006) 223--231.
\newblock \href {http://arxiv.org/abs/hep-ph/0511054}
  {\path{arXiv:hep-ph/0511054}}, \href
  {https://doi.org/10.1016/j.physletb.2006.05.091}
  {\path{doi:10.1016/j.physletb.2006.05.091}}.

\bibitem{Cosyn:2020kwu}
W.~Cosyn, C.~Weiss, {Polarized electron-deuteron deep-inelastic scattering with
  spectator nucleon tagging}, Phys. Rev. C 102 (2020) 065204.
\newblock \href {http://arxiv.org/abs/2006.03033} {\path{arXiv:2006.03033}},
  \href {https://doi.org/10.1103/PhysRevC.102.065204}
  {\path{doi:10.1103/PhysRevC.102.065204}}.

\bibitem{Bonus12:2006}
S.~Bueltmann, M.~Christy, H.~Fenker, K.~Griffioen, C.~Keppel, S.~Kuhn,
  W.~Melnitchouk, V.~s. Tvaskis,
  \href{http://www.jlab.org/exp_prog/12GEV_EXP/E1206113.html}{{The Structure of
  the Free Neutron at Large x-Bjorken;
  \url{http://www.jlab.org/exp_prog/12GEV_EXP/E1206113.html}}}JLab Experiment
  E1206113 (2006).
\newline\urlprefix\url{http://www.jlab.org/exp_prog/12GEV_EXP/E1206113.html}

\bibitem{Armstrong:2017wfw}
W.~Armstrong, et~al., {Partonic Structure of Light Nuclei} (2017).
\newblock \href {http://arxiv.org/abs/1708.00888} {\path{arXiv:1708.00888}}.

\bibitem{Londergan:1996vf}
J.~T. Londergan, A.~Pang, A.~W. Thomas, {Probing charge symmetry violating
  quark distributions in semiinclusive leptoproduction of hadrons}, Phys. Rev.
  D 54 (1996) 3154--3161.
\newblock \href {http://arxiv.org/abs/hep-ph/9604446}
  {\path{arXiv:hep-ph/9604446}}, \href
  {https://doi.org/10.1103/PhysRevD.54.3154}
  {\path{doi:10.1103/PhysRevD.54.3154}}.

\bibitem{Martin:2003sk}
A.~D. Martin, R.~G. Roberts, W.~J. Stirling, R.~S. Thorne, {Uncertainties of
  predictions from parton distributions. 2. Theoretical errors}, Eur. Phys. J.
  C 35 (2004) 325--348.
\newblock \href {http://arxiv.org/abs/hep-ph/0308087}
  {\path{arXiv:hep-ph/0308087}}, \href
  {https://doi.org/10.1140/epjc/s2004-01825-2}
  {\path{doi:10.1140/epjc/s2004-01825-2}}.

\bibitem{deFlorian:2007aj}
D.~de~Florian, R.~Sassot, M.~Stratmann, {Global analysis of fragmentation
  functions for pions and kaons and their uncertainties}, Phys. Rev. D 75
  (2007) 114010.
\newblock \href {http://arxiv.org/abs/hep-ph/0703242}
  {\path{arXiv:hep-ph/0703242}}, \href
  {https://doi.org/10.1103/PhysRevD.75.114010}
  {\path{doi:10.1103/PhysRevD.75.114010}}.

\bibitem{Scimemi:2019cmh}
I.~Scimemi, A.~Vladimirov, {Non-perturbative structure of semi-inclusive
  deep-inelastic and Drell-Yan scattering at small transverse momentum}, JHEP
  06 (2020) 137.
\newblock \href {http://arxiv.org/abs/1912.06532} {\path{arXiv:1912.06532}},
  \href {https://doi.org/10.1007/JHEP06(2020)137}
  {\path{doi:10.1007/JHEP06(2020)137}}.

\bibitem{Bury:2022czx}
M.~Bury, F.~Hautmann, S.~Leal-Gomez, I.~Scimemi, A.~Vladimirov, P.~Zurita, {PDF
  bias and flavor dependence in TMD distributions}, JHEP 10 (2022) 118.
\newblock \href {http://arxiv.org/abs/2201.07114} {\path{arXiv:2201.07114}},
  \href {https://doi.org/10.1007/JHEP10(2022)118}
  {\path{doi:10.1007/JHEP10(2022)118}}.

\bibitem{BermudezMartinez:2022ctj}
A.~Bermudez~Martinez, A.~Vladimirov, {Determination of the Collins-Soper kernel
  from cross-sections ratios}, Phys. Rev. D 106~(9) (2022) L091501.
\newblock \href {http://arxiv.org/abs/2206.01105} {\path{arXiv:2206.01105}},
  \href {https://doi.org/10.1103/PhysRevD.106.L091501}
  {\path{doi:10.1103/PhysRevD.106.L091501}}.

\bibitem{Boglione:2014oea}
M.~Boglione, J.~O. Gonzalez~Hernandez, S.~Melis, A.~Prokudin, {A study on the
  interplay between perturbative QCD and CSS/TMD formalism in SIDIS processes},
  JHEP 02 (2015) 095.
\newblock \href {http://arxiv.org/abs/1412.1383} {\path{arXiv:1412.1383}},
  \href {https://doi.org/10.1007/JHEP02(2015)095}
  {\path{doi:10.1007/JHEP02(2015)095}}.

\bibitem{Yoon:2017qzo}
B.~Yoon, M.~Engelhardt, R.~Gupta, T.~Bhattacharya, J.~R. Green, B.~U. Musch,
  J.~W. Negele, A.~V. Pochinsky, A.~Sch\"afer, S.~N. Syritsyn, {Nucleon
  Transverse Momentum-dependent Parton Distributions in Lattice QCD:
  Renormalization Patterns and Discretization Effects}, Phys. Rev. D 96~(9)
  (2017) 094508.
\newblock \href {http://arxiv.org/abs/1706.03406} {\path{arXiv:1706.03406}},
  \href {https://doi.org/10.1103/PhysRevD.96.094508}
  {\path{doi:10.1103/PhysRevD.96.094508}}.

\bibitem{Ji:2013dva}
X.~Ji, {Parton Physics on a Euclidean Lattice}, Phys. Rev. Lett. 110 (2013)
  262002.
\newblock \href {http://arxiv.org/abs/1305.1539} {\path{arXiv:1305.1539}},
  \href {https://doi.org/10.1103/PhysRevLett.110.262002}
  {\path{doi:10.1103/PhysRevLett.110.262002}}.

\bibitem{Ji:2020ect}
X.~Ji, Y.-S. Liu, Y.~Liu, J.-H. Zhang, Y.~Zhao, {Large-momentum effective
  theory}, Rev. Mod. Phys. 93~(3) (2021) 035005.
\newblock \href {http://arxiv.org/abs/2004.03543} {\path{arXiv:2004.03543}},
  \href {https://doi.org/10.1103/RevModPhys.93.035005}
  {\path{doi:10.1103/RevModPhys.93.035005}}.

\bibitem{Ebert:2018gzl}
M.~A. Ebert, I.~W. Stewart, Y.~Zhao, {Determining the Nonperturbative
  Collins-Soper Kernel From Lattice QCD}, Phys. Rev. D 99~(3) (2019) 034505.
\newblock \href {http://arxiv.org/abs/1811.00026} {\path{arXiv:1811.00026}},
  \href {https://doi.org/10.1103/PhysRevD.99.034505}
  {\path{doi:10.1103/PhysRevD.99.034505}}.

\bibitem{Ji:2019ewn}
X.~Ji, Y.~Liu, Y.-S. Liu, {Transverse-momentum-dependent parton distribution
  functions from large-momentum effective theory}, Phys. Lett. B 811 (2020)
  135946.
\newblock \href {http://arxiv.org/abs/1911.03840} {\path{arXiv:1911.03840}},
  \href {https://doi.org/10.1016/j.physletb.2020.135946}
  {\path{doi:10.1016/j.physletb.2020.135946}}.

\bibitem{Shanahan:2020zxr}
P.~Shanahan, M.~Wagman, Y.~Zhao, {Collins-Soper kernel for TMD evolution from
  lattice QCD}, Phys. Rev. D 102~(1) (2020) 014511.
\newblock \href {http://arxiv.org/abs/2003.06063} {\path{arXiv:2003.06063}},
  \href {https://doi.org/10.1103/PhysRevD.102.014511}
  {\path{doi:10.1103/PhysRevD.102.014511}}.

\bibitem{LatticeParton:2020uhz}
Q.-A. Zhang, et~al., {Lattice-QCD Calculations of TMD Soft Function Through
  Large-Momentum Effective Theory}, Phys. Rev. Lett. 125~(19) (2020) 192001.
\newblock \href {http://arxiv.org/abs/2005.14572} {\path{arXiv:2005.14572}},
  \href {https://doi.org/10.22323/1.396.0477} {\path{doi:10.22323/1.396.0477}}.

\bibitem{Schlemmer:2021aij}
M.~Schlemmer, A.~Vladimirov, C.~Zimmermann, M.~Engelhardt, A.~Sch\"afer,
  {Determination of the Collins-Soper Kernel from Lattice QCD}, JHEP 08 (2021)
  004.
\newblock \href {http://arxiv.org/abs/2103.16991} {\path{arXiv:2103.16991}},
  \href {https://doi.org/10.1007/JHEP08(2021)004}
  {\path{doi:10.1007/JHEP08(2021)004}}.

\bibitem{LPC:2022ibr}
M.-H. Chu, et~al., {Nonperturbative determination of the Collins-Soper kernel
  from quasitransverse-momentum-dependent wave functions}, Phys. Rev. D 106~(3)
  (2022) 034509.
\newblock \href {http://arxiv.org/abs/2204.00200} {\path{arXiv:2204.00200}},
  \href {https://doi.org/10.1103/PhysRevD.106.034509}
  {\path{doi:10.1103/PhysRevD.106.034509}}.

\bibitem{Shu:2023cot}
H.-T. Shu, M.~Schlemmer, T.~Sizmann, A.~Vladimirov, L.~Walter, M.~Engelhardt,
  A.~Sch\"afer, Y.-B. Yang, {Universality of the Collins-Soper kernel in
  lattice calculations} (2 2023).
\newblock \href {http://arxiv.org/abs/2302.06502} {\path{arXiv:2302.06502}}.

\bibitem{LPC:2022zci}
J.-C. He, M.-H. Chu, J.~Hua, X.~Ji, A.~Sch\"afer, Y.~Su, W.~Wang, Y.~Yang,
  J.-H. Zhang, Q.-A. Zhang, {Unpolarized Transverse-Momentum-Dependent Parton
  Distributions of the Nucleon from Lattice QCD} (11 2022).
\newblock \href {http://arxiv.org/abs/2211.02340} {\path{arXiv:2211.02340}}.

\bibitem{Zhou:2022wzm}
Y.~Zhou, N.~Sato, W.~Melnitchouk, {How well do we know the gluon polarization
  in the proton?}, Phys. Rev. D 105~(7) (2022) 074022.
\newblock \href {http://arxiv.org/abs/2201.02075} {\path{arXiv:2201.02075}},
  \href {https://doi.org/10.1103/PhysRevD.105.074022}
  {\path{doi:10.1103/PhysRevD.105.074022}}.

\bibitem{Alberg:2017ijg}
M.~Alberg, G.~A. Miller, {Chiral Light Front Perturbation Theory and the Flavor
  Dependence of the Light-Quark Nucleon Sea}, Phys. Rev. C 100~(3) (2019)
  035205.
\newblock \href {http://arxiv.org/abs/1712.05814} {\path{arXiv:1712.05814}},
  \href {https://doi.org/10.1103/PhysRevC.100.035205}
  {\path{doi:10.1103/PhysRevC.100.035205}}.

\bibitem{NewMuon:1991hlj}
P.~Amaudruz, et~al., {The Gottfried sum from the ratio F2(n) / F2(p)}, Phys.
  Rev. Lett. 66 (1991) 2712--2715.
\newblock \href {https://doi.org/10.1103/PhysRevLett.66.2712}
  {\path{doi:10.1103/PhysRevLett.66.2712}}.

\bibitem{Garvey:2001yq}
G.~T. Garvey, J.-C. Peng, {Flavor asymmetry of light quarks in the nucleon
  sea}, Prog. Part. Nucl. Phys. 47 (2001) 203--243.
\newblock \href {http://arxiv.org/abs/nucl-ex/0109010}
  {\path{arXiv:nucl-ex/0109010}}, \href
  {https://doi.org/10.1016/S0146-6410(01)00155-7}
  {\path{doi:10.1016/S0146-6410(01)00155-7}}.

\bibitem{Nagai:2017dhp}
K.~Nagai, {Measurement of Antiquark Flavor Asymmetry in the Proton by the
  Drell\textendash{}Yan Experiment SeaQuest at Fermilab}, JPS Conf. Proc. 13
  (2017) 020051.
\newblock \href {https://doi.org/10.7566/JPSCP.13.020051}
  {\path{doi:10.7566/JPSCP.13.020051}}.

\bibitem{Aidala:2012mv}
C.~A. Aidala, S.~D. Bass, D.~Hasch, G.~K. Mallot, {The Spin Structure of the
  Nucleon}, Rev. Mod. Phys. 85 (2013) 655--691.
\newblock \href {http://arxiv.org/abs/1209.2803} {\path{arXiv:1209.2803}},
  \href {https://doi.org/10.1103/RevModPhys.85.655}
  {\path{doi:10.1103/RevModPhys.85.655}}.

\bibitem{STAR:2014wox}
L.~Adamczyk, et~al., {Precision Measurement of the Longitudinal Double-spin
  Asymmetry for Inclusive Jet Production in Polarized Proton Collisions at
  $\sqrt{s}=200$ GeV}, Phys. Rev. Lett. 115~(9) (2015) 092002.
\newblock \href {http://arxiv.org/abs/1405.5134} {\path{arXiv:1405.5134}},
  \href {https://doi.org/10.1103/PhysRevLett.115.092002}
  {\path{doi:10.1103/PhysRevLett.115.092002}}.

\bibitem{STAR:2019yqm}
J.~Adam, et~al., {Longitudinal double-spin asymmetry for inclusive jet and
  dijet production in pp collisions at $\sqrt{s} = 510$ GeV}, Phys. Rev. D
  100~(5) (2019) 052005.
\newblock \href {http://arxiv.org/abs/1906.02740} {\path{arXiv:1906.02740}},
  \href {https://doi.org/10.1103/PhysRevD.100.052005}
  {\path{doi:10.1103/PhysRevD.100.052005}}.

\bibitem{STAR:2021mfd}
M.~S. Abdallah, et~al., {Longitudinal double-spin asymmetry for inclusive jet
  and dijet production in polarized proton collisions at $\sqrt{s}=200$ GeV},
  Phys. Rev. D 103~(9) (2021) L091103.
\newblock \href {http://arxiv.org/abs/2103.05571} {\path{arXiv:2103.05571}},
  \href {https://doi.org/10.1103/PhysRevD.103.L091103}
  {\path{doi:10.1103/PhysRevD.103.L091103}}.

\bibitem{STAR:2021mqa}
M.~S. Abdallah, et~al., {Longitudinal double-spin asymmetry for inclusive jet
  and dijet production in polarized proton collisions at $\sqrt{s}=510$ GeV},
  Phys. Rev. D 105~(9) (2022) 092011.
\newblock \href {http://arxiv.org/abs/2110.11020} {\path{arXiv:2110.11020}},
  \href {https://doi.org/10.1103/PhysRevD.105.092011}
  {\path{doi:10.1103/PhysRevD.105.092011}}.

\bibitem{PHENIX:2010aru}
A.~Adare, et~al., {Event Structure and Double Helicity Asymmetry in Jet
  Production from Polarized $p+p$ Collisions at $\sqrt{s} =
  200$\textasciitilde{}GeV}, Phys. Rev. D 84 (2011) 012006.
\newblock \href {http://arxiv.org/abs/1009.4921} {\path{arXiv:1009.4921}},
  \href {https://doi.org/10.1103/PhysRevD.84.012006}
  {\path{doi:10.1103/PhysRevD.84.012006}}.

\bibitem{Bass:2009ed}
S.~D. Bass, A.~W. Thomas, {The Nucleon's octet axial-charge g(A)**(8) with
  chiral corrections}, Phys. Lett. B 684 (2010) 216--220.
\newblock \href {http://arxiv.org/abs/0912.1765} {\path{arXiv:0912.1765}},
  \href {https://doi.org/10.1016/j.physletb.2010.01.008}
  {\path{doi:10.1016/j.physletb.2010.01.008}}.

\bibitem{Ethier:2017zbq}
J.~J. Ethier, N.~Sato, W.~Melnitchouk, {First simultaneous extraction of
  spin-dependent parton distributions and fragmentation functions from a global
  QCD analysis}, Phys. Rev. Lett. 119~(13) (2017) 132001.
\newblock \href {http://arxiv.org/abs/1705.05889} {\path{arXiv:1705.05889}},
  \href {https://doi.org/10.1103/PhysRevLett.119.132001}
  {\path{doi:10.1103/PhysRevLett.119.132001}}.

\bibitem{Candido:2020yat}
A.~Candido, S.~Forte, F.~Hekhorn, {Can $ \overline{\mathrm{MS}} $ parton
  distributions be negative?}, JHEP 11 (2020) 129.
\newblock \href {http://arxiv.org/abs/2006.07377} {\path{arXiv:2006.07377}},
  \href {https://doi.org/10.1007/JHEP11(2020)129}
  {\path{doi:10.1007/JHEP11(2020)129}}.

\bibitem{Collins:2021vke}
J.~Collins, T.~C. Rogers, N.~Sato, {Positivity and renormalization of parton
  densities}, Phys. Rev. D 105~(7) (2022) 076010.
\newblock \href {http://arxiv.org/abs/2111.01170} {\path{arXiv:2111.01170}},
  \href {https://doi.org/10.1103/PhysRevD.105.076010}
  {\path{doi:10.1103/PhysRevD.105.076010}}.

\bibitem{Jager:2003ch}
B.~Jager, M.~Stratmann, S.~Kretzer, W.~Vogelsang, {QCD hard scattering and the
  sign of the spin asymmetry A**pi(LL)}, Phys. Rev. Lett. 92 (2004) 121803.
\newblock \href {http://arxiv.org/abs/hep-ph/0310197}
  {\path{arXiv:hep-ph/0310197}}, \href
  {https://doi.org/10.1103/PhysRevLett.92.121803}
  {\path{doi:10.1103/PhysRevLett.92.121803}}.

\bibitem{PHENIX:2015fxo}
A.~Adare, et~al., {Inclusive cross section and double-helicity asymmetry for
  $\pi^{0}$ production at midrapidity in $p+p$ collisions at $\sqrt{s}=510$
  GeV}, Phys. Rev. D 93~(1) (2016) 011501.
\newblock \href {http://arxiv.org/abs/1510.02317} {\path{arXiv:1510.02317}},
  \href {https://doi.org/10.1103/PhysRevD.93.011501}
  {\path{doi:10.1103/PhysRevD.93.011501}}.

\bibitem{PHENIX:2014axc}
A.~Adare, et~al., {Charged-pion cross sections and double-helicity asymmetries
  in polarized p+p collisions at $\sqrt{s}$=200 GeV}, Phys. Rev. D 91~(3)
  (2015) 032001.
\newblock \href {http://arxiv.org/abs/1409.1907} {\path{arXiv:1409.1907}},
  \href {https://doi.org/10.1103/PhysRevD.91.032001}
  {\path{doi:10.1103/PhysRevD.91.032001}}.

\bibitem{PHENIX:2020trf}
U.~A. Acharya, et~al., {Measurement of charged pion double spin asymmetries at
  midrapidity in longitudinally polarized $p+p$ collisions at $\sqrt {s}$ = 510
  GeV}, Phys. Rev. D 102~(3) (2020) 032001.
\newblock \href {http://arxiv.org/abs/2004.02681} {\path{arXiv:2004.02681}},
  \href {https://doi.org/10.1103/PhysRevD.102.032001}
  {\path{doi:10.1103/PhysRevD.102.032001}}.

\bibitem{Whitehill:2022mpq}
R.~M. Whitehill, Y.~Zhou, N.~Sato, W.~Melnitchouk, {Accessing gluon
  polarization with high-PT hadrons in SIDIS}, Phys. Rev. D 107~(3) (2023)
  034033.
\newblock \href {http://arxiv.org/abs/2210.12295} {\path{arXiv:2210.12295}},
  \href {https://doi.org/10.1103/PhysRevD.107.034033}
  {\path{doi:10.1103/PhysRevD.107.034033}}.

\bibitem{Polyakov:2018zvc}
M.~V. Polyakov, P.~Schweitzer, {Forces inside hadrons: pressure, surface
  tension, mechanical radius, and all that}, Int. J. Mod. Phys. A 33~(26)
  (2018) 1830025.
\newblock \href {http://arxiv.org/abs/1805.06596} {\path{arXiv:1805.06596}},
  \href {https://doi.org/10.1142/S0217751X18300259}
  {\path{doi:10.1142/S0217751X18300259}}.

\bibitem{Lorce:2018egm}
C.~Lorc\'e, H.~Moutarde, A.~P. Trawi\'nski, {Revisiting the mechanical
  properties of the nucleon}, Eur. Phys. J. C 79~(1) (2019) 89.
\newblock \href {http://arxiv.org/abs/1810.09837} {\path{arXiv:1810.09837}},
  \href {https://doi.org/10.1140/epjc/s10052-019-6572-3}
  {\path{doi:10.1140/epjc/s10052-019-6572-3}}.

\bibitem{Lorce:2017xzd}
C.~Lorc\'e, {On the hadron mass decomposition}, Eur. Phys. J. C 78~(2) (2018)
  120.
\newblock \href {http://arxiv.org/abs/1706.05853} {\path{arXiv:1706.05853}},
  \href {https://doi.org/10.1140/epjc/s10052-018-5561-2}
  {\path{doi:10.1140/epjc/s10052-018-5561-2}}.

\bibitem{Hatta:2018sqd}
Y.~Hatta, A.~Rajan, K.~Tanaka, {Quark and gluon contributions to the QCD trace
  anomaly}, JHEP 12 (2018) 008.
\newblock \href {http://arxiv.org/abs/1810.05116} {\path{arXiv:1810.05116}},
  \href {https://doi.org/10.1007/JHEP12(2018)008}
  {\path{doi:10.1007/JHEP12(2018)008}}.

\bibitem{Metz:2020vxd}
A.~Metz, B.~Pasquini, S.~Rodini, {Revisiting the proton mass decomposition},
  Phys. Rev. D 102~(11) (2021) 114042.
\newblock \href {http://arxiv.org/abs/2006.11171} {\path{arXiv:2006.11171}},
  \href {https://doi.org/10.1103/PhysRevD.102.114042}
  {\path{doi:10.1103/PhysRevD.102.114042}}.

\bibitem{Goeke:2001tz}
K.~Goeke, M.~V. Polyakov, M.~Vanderhaeghen, {Hard exclusive reactions and the
  structure of hadrons}, Prog. Part. Nucl. Phys. 47 (2001) 401--515.
\newblock \href {http://arxiv.org/abs/hep-ph/0106012}
  {\path{arXiv:hep-ph/0106012}}, \href
  {https://doi.org/10.1016/S0146-6410(01)00158-2}
  {\path{doi:10.1016/S0146-6410(01)00158-2}}.

\bibitem{Diehl:2003ny}
M.~Diehl, {Generalized parton distributions}, Phys. Rept. 388 (2003) 41--277.
\newblock \href {http://arxiv.org/abs/hep-ph/0307382}
  {\path{arXiv:hep-ph/0307382}}, \href
  {https://doi.org/10.1016/j.physrep.2003.08.002}
  {\path{doi:10.1016/j.physrep.2003.08.002}}.

\bibitem{Belitsky:2005qn}
A.~V. Belitsky, A.~V. Radyushkin, {Unraveling hadron structure with generalized
  parton distributions}, Phys. Rept. 418 (2005) 1--387.
\newblock \href {http://arxiv.org/abs/hep-ph/0504030}
  {\path{arXiv:hep-ph/0504030}}, \href
  {https://doi.org/10.1016/j.physrep.2005.06.002}
  {\path{doi:10.1016/j.physrep.2005.06.002}}.

\bibitem{Carman:2023zke}
D.~S. Carman, R.~W. Gothe, V.~I. Mokeev, C.~D. Roberts, {Nucleon Resonance
  Electroexcitation Amplitudes and Emergent Hadron Mass}, Particles 6~(1)
  (2023) 416.

\bibitem{Kharzeev:1998bz}
D.~Kharzeev, H.~Satz, A.~Syamtomov, G.~Zinovjev, {$J/\psi$ photoproduction and
  the gluon structure of the nucleon}, Eur. Phys. J. C 9 (1999) 459--462.
\newblock \href {http://arxiv.org/abs/hep-ph/9901375}
  {\path{arXiv:hep-ph/9901375}}, \href {https://doi.org/10.1007/s100529900047}
  {\path{doi:10.1007/s100529900047}}.

\bibitem{Gryniuk:2016mpk}
O.~Gryniuk, M.~Vanderhaeghen, {Accessing the real part of the forward
  $J/\psi$-p scattering amplitude from $J/\psi$ photoproduction on protons
  around \ threshold}, Phys. Rev. D 94~(7) (2016) 074001.
\newblock \href {http://arxiv.org/abs/1608.08205} {\path{arXiv:1608.08205}},
  \href {https://doi.org/10.1103/PhysRevD.94.074001}
  {\path{doi:10.1103/PhysRevD.94.074001}}.

\bibitem{Mamo:2019mka}
K.~A. Mamo, I.~Zahed, {Diffractive photoproduction of $J/\psi$ and $\Upsilon$
  using holographic QCD: gravitational form factors and GPD of gluons in the
  proton}, Phys. Rev. D 101~(8) (2020) 086003.
\newblock \href {http://arxiv.org/abs/1910.04707} {\path{arXiv:1910.04707}},
  \href {https://doi.org/10.1103/PhysRevD.101.086003}
  {\path{doi:10.1103/PhysRevD.101.086003}}.

\bibitem{Mamo:2022eui}
K.~A. Mamo, I.~Zahed, {J/\ensuremath{\psi} near threshold in holographic QCD: A
  and D gravitational form factors}, Phys. Rev. D 106~(8) (2022) 086004.
\newblock \href {http://arxiv.org/abs/2204.08857} {\path{arXiv:2204.08857}},
  \href {https://doi.org/10.1103/PhysRevD.106.086004}
  {\path{doi:10.1103/PhysRevD.106.086004}}.

\bibitem{Guo:2021ibg}
Y.~Guo, X.~Ji, Y.~Liu, {QCD Analysis of Near-Threshold Photon-Proton Production
  of Heavy Quarkonium}, Phys. Rev. D 103~(9) (2021) 096010.
\newblock \href {http://arxiv.org/abs/2103.11506} {\path{arXiv:2103.11506}},
  \href {https://doi.org/10.1103/PhysRevD.103.096010}
  {\path{doi:10.1103/PhysRevD.103.096010}}.

\bibitem{Duran:2022xag}
B.~Duran, et~al., {Determining the gluonic gravitational form factors of the
  proton}, Nature 615~(7954) (2023) 813--816.
\newblock \href {http://arxiv.org/abs/2207.05212} {\path{arXiv:2207.05212}},
  \href {https://doi.org/10.1038/s41586-023-05730-4}
  {\path{doi:10.1038/s41586-023-05730-4}}.

\bibitem{Pefkou:2021fni}
D.~A. Pefkou, D.~C. Hackett, P.~E. Shanahan, {Gluon gravitational structure of
  hadrons of different spin}, Phys. Rev. D 105~(5) (2022) 054509.
\newblock \href {http://arxiv.org/abs/2107.10368} {\path{arXiv:2107.10368}},
  \href {https://doi.org/10.1103/PhysRevD.105.054509}
  {\path{doi:10.1103/PhysRevD.105.054509}}.

\bibitem{Chen:2014psa}
J.~P. Chen, H.~Gao, T.~K. Hemmick, Z.~E. Meziani, P.~A. Souder, {A White Paper
  on SoLID (Solenoidal Large Intensity Device)} (9 2014).
\newblock \href {http://arxiv.org/abs/1409.7741} {\path{arXiv:1409.7741}}.

\bibitem{Polyakov:1999gs}
M.~V. Polyakov, C.~Weiss, {Skewed and double distributions in pion and
  nucleon}, Phys. Rev. D 60 (1999) 114017.
\newblock \href {http://arxiv.org/abs/hep-ph/9902451}
  {\path{arXiv:hep-ph/9902451}}, \href
  {https://doi.org/10.1103/PhysRevD.60.114017}
  {\path{doi:10.1103/PhysRevD.60.114017}}.

\bibitem{Polyakov:2002yz}
M.~V. Polyakov, {Generalized parton distributions and strong forces inside
  nucleons and nuclei}, Phys. Lett. B 555 (2003) 57--62.
\newblock \href {http://arxiv.org/abs/hep-ph/0210165}
  {\path{arXiv:hep-ph/0210165}}, \href
  {https://doi.org/10.1016/S0370-2693(03)00036-4}
  {\path{doi:10.1016/S0370-2693(03)00036-4}}.

\bibitem{Burkert:2018bqq}
V.~D. Burkert, L.~Elouadrhiri, F.~X. Girod, {The pressure distribution inside
  the proton}, Nature 557~(7705) (2018) 396--399.
\newblock \href {https://doi.org/10.1038/s41586-018-0060-z}
  {\path{doi:10.1038/s41586-018-0060-z}}.

\bibitem{Burkert:2023wzr}
V.~D. Burkert, L.~Elouadrhiri, F.~X. Girod, C.~Lorc\'e, P.~Schweitzer, P.~E.
  Shanahan, {Colloquium: Gravitational Form Factors of the Proton} (3 2023).
\newblock \href {http://arxiv.org/abs/2303.08347} {\path{arXiv:2303.08347}}.

\bibitem{Kumericki:2019ddg}
K.~Kumeri\v{c}ki, {Measurability of pressure inside the proton}, Nature
  570~(7759) (2019) E1--E2.
\newblock \href {https://doi.org/10.1038/s41586-019-1211-6}
  {\path{doi:10.1038/s41586-019-1211-6}}.

\bibitem{Moutarde:2019tqa}
H.~Moutarde, P.~Sznajder, J.~Wagner, {Unbiased determination of DVCS Compton
  Form Factors}, Eur. Phys. J. C 79~(7) (2019) 614.
\newblock \href {http://arxiv.org/abs/1905.02089} {\path{arXiv:1905.02089}},
  \href {https://doi.org/10.1140/epjc/s10052-019-7117-5}
  {\path{doi:10.1140/epjc/s10052-019-7117-5}}.

\bibitem{CLAS:2022syx}
G.~Christiaens, et~al., {First CLAS12 measurement of DVCS beam-spin asymmetries
  in the extended valence region} (11 2022).
\newblock \href {http://arxiv.org/abs/2211.11274} {\path{arXiv:2211.11274}}.

\bibitem{Ji:1996ek}
X.-D. Ji, {Gauge-Invariant Decomposition of Nucleon Spin}, Phys. Rev. Lett. 78
  (1997) 610--613.
\newblock \href {http://arxiv.org/abs/hep-ph/9603249}
  {\path{arXiv:hep-ph/9603249}}, \href
  {https://doi.org/10.1103/PhysRevLett.78.610}
  {\path{doi:10.1103/PhysRevLett.78.610}}.

\bibitem{Radyushkin:1997ki}
A.~Radyushkin, {Nonforward parton distributions}, Phys. Rev. D 56 (1997)
  5524--5557.
\newblock \href {http://arxiv.org/abs/hep-ph/9704207}
  {\path{arXiv:hep-ph/9704207}}, \href
  {https://doi.org/10.1103/PhysRevD.56.5524}
  {\path{doi:10.1103/PhysRevD.56.5524}}.

\bibitem{Kumericki:2016ehc}
K.~Kumericki, S.~Liuti, H.~Moutarde, {GPD phenomenology and DVCS fitting}:
  {Entering the high-precision era}, Eur. Phys. J. A 52~(6) (2016) 157.
\newblock \href {http://arxiv.org/abs/1602.02763} {\path{arXiv:1602.02763}},
  \href {https://doi.org/10.1140/epja/i2016-16157-3}
  {\path{doi:10.1140/epja/i2016-16157-3}}.

\bibitem{Soper:1976jc}
D.~E. Soper, {The Parton Model and the Bethe-Salpeter Wave Function}, Phys.
  Rev. D 15 (1977) 1141.
\newblock \href {https://doi.org/10.1103/PhysRevD.15.1141}
  {\path{doi:10.1103/PhysRevD.15.1141}}.

\bibitem{Burkardt:2000za}
M.~Burkardt, {Impact parameter dependent parton distributions and off forward
  parton distributions for zeta ---\ensuremath{>} 0}, Phys. Rev. D 62 (2000)
  071503, [Erratum: Phys.Rev.D 66, 119903 (2002)].
\newblock \href {http://arxiv.org/abs/hep-ph/0005108}
  {\path{arXiv:hep-ph/0005108}}, \href
  {https://doi.org/10.1103/PhysRevD.62.071503}
  {\path{doi:10.1103/PhysRevD.62.071503}}.

\bibitem{Belitsky:2001ns}
A.~V. Belitsky, D.~Mueller, A.~Kirchner, {Theory of deeply virtual Compton
  scattering on the nucleon}, Nucl. Phys. B 629 (2002) 323--392.
\newblock \href {http://arxiv.org/abs/hep-ph/0112108}
  {\path{arXiv:hep-ph/0112108}}, \href
  {https://doi.org/10.1016/S0550-3213(02)00144-X}
  {\path{doi:10.1016/S0550-3213(02)00144-X}}.

\bibitem{Belitsky:2010jw}
A.~V. Belitsky, D.~Mueller, {Exclusive electroproduction revisited: treating
  kinematical effects}, Phys. Rev. D 82 (2010) 074010.
\newblock \href {http://arxiv.org/abs/1005.5209} {\path{arXiv:1005.5209}},
  \href {https://doi.org/10.1103/PhysRevD.82.074010}
  {\path{doi:10.1103/PhysRevD.82.074010}}.

\bibitem{Kriesten:2020apm}
B.~Kriesten, S.~Liuti, A.~Meyer, {Novel Rosenbluth extraction framework for
  Compton form factors from deeply virtual exclusive experiments}, Phys. Lett.
  B 829 (2022) 137051.
\newblock \href {http://arxiv.org/abs/2011.04484} {\path{arXiv:2011.04484}},
  \href {https://doi.org/10.1016/j.physletb.2022.137051}
  {\path{doi:10.1016/j.physletb.2022.137051}}.

\bibitem{Kriesten:2020wcx}
B.~Kriesten, S.~Liuti, {Theory of deeply virtual Compton scattering off the
  unpolarized proton}, Phys. Rev. D 105~(1) (2022) 016015.
\newblock \href {http://arxiv.org/abs/2004.08890} {\path{arXiv:2004.08890}},
  \href {https://doi.org/10.1103/PhysRevD.105.016015}
  {\path{doi:10.1103/PhysRevD.105.016015}}.

\bibitem{Kriesten:2019jep}
B.~Kriesten, S.~Liuti, L.~Calero-Diaz, D.~Keller, A.~Meyer, G.~R. Goldstein,
  J.~Osvaldo Gonzalez-Hernandez, {Extraction of generalized parton distribution
  observables from deeply virtual electron proton scattering experiments},
  Phys. Rev. D 101~(5) (2020) 054021.
\newblock \href {http://arxiv.org/abs/1903.05742} {\path{arXiv:1903.05742}},
  \href {https://doi.org/10.1103/PhysRevD.101.054021}
  {\path{doi:10.1103/PhysRevD.101.054021}}.

\bibitem{Kriesten:2021sqc}
B.~Kriesten, P.~Velie, E.~Yeats, F.~Y. Lopez, S.~Liuti, {Parametrization of
  quark and gluon generalized parton distributions in a dynamical framework},
  Phys. Rev. D 105~(5) (2022) 056022.
\newblock \href {http://arxiv.org/abs/2101.01826} {\path{arXiv:2101.01826}},
  \href {https://doi.org/10.1103/PhysRevD.105.056022}
  {\path{doi:10.1103/PhysRevD.105.056022}}.

\bibitem{Kumericki:2019mgk}
K.~Kumeri\v{c}ki, {Extraction of DVCS form factors with uncertainties}, in:
  {Probing Nucleons and Nuclei in High Energy Collisions}: {Dedicated to the
  Physics of the Electron Ion Collider}, 2020, pp. 25--29.
\newblock \href {http://arxiv.org/abs/1910.04806} {\path{arXiv:1910.04806}},
  \href {https://doi.org/10.1142/9789811214950_0005}
  {\path{doi:10.1142/9789811214950_0005}}.

\bibitem{Grigsby:2020auv}
J.~Grigsby, B.~Kriesten, J.~Hoskins, S.~Liuti, P.~Alonzi, M.~Burkardt, {Deep
  learning analysis of deeply virtual exclusive photoproduction}, Phys. Rev. D
  104~(1) (2021) 016001.
\newblock \href {http://arxiv.org/abs/2012.04801} {\path{arXiv:2012.04801}},
  \href {https://doi.org/10.1103/PhysRevD.104.016001}
  {\path{doi:10.1103/PhysRevD.104.016001}}.

\bibitem{Cuic:2020iwt}
M.~\v{C}ui\'c, K.~Kumeri\v{c}ki, A.~Sch\"afer, {Separation of Quark Flavors
  Using Deeply Virtual Compton Scattering Data}, Phys. Rev. Lett. 125~(23)
  (2020) 232005.
\newblock \href {http://arxiv.org/abs/2007.00029} {\path{arXiv:2007.00029}},
  \href {https://doi.org/10.1103/PhysRevLett.125.232005}
  {\path{doi:10.1103/PhysRevLett.125.232005}}.

\bibitem{Almaeen:2022imx}
M.~Almaeen, J.~Grigsby, J.~Hoskins, B.~Kriesten, Y.~Li, H.-W. Lin, S.~Liuti,
  {Benchmarks for a Global Extraction of Information from Deeply Virtual
  Exclusive Scattering} (7 2022).
\newblock \href {http://arxiv.org/abs/2207.10766} {\path{arXiv:2207.10766}}.

\bibitem{Guidal:2002kt}
M.~Guidal, M.~Vanderhaeghen, {Double deeply virtual Compton scattering off the
  nucleon}, Phys. Rev. Lett. 90 (2003) 012001.
\newblock \href {http://arxiv.org/abs/hep-ph/0208275}
  {\path{arXiv:hep-ph/0208275}}, \href
  {https://doi.org/10.1103/PhysRevLett.90.012001}
  {\path{doi:10.1103/PhysRevLett.90.012001}}.

\bibitem{Belitsky:2002tf}
A.~V. Belitsky, D.~Mueller, {Exclusive electroproduction of lepton pairs as a
  probe of nucleon structure}, Phys. Rev. Lett. 90 (2003) 022001.
\newblock \href {http://arxiv.org/abs/hep-ph/0210313}
  {\path{arXiv:hep-ph/0210313}}, \href
  {https://doi.org/10.1103/PhysRevLett.90.022001}
  {\path{doi:10.1103/PhysRevLett.90.022001}}.

\bibitem{Zhao:2021zsm}
S.~Zhao, A.~Camsonne, D.~Marchand, M.~Mazouz, N.~Sparveris, S.~Stepanyan,
  E.~Voutier, Z.~W. Zhao, {Double deeply virtual Compton scattering with
  positron beams at SoLID}, Eur. Phys. J. A 57~(7) (2021) 240.
\newblock \href {http://arxiv.org/abs/2103.12773} {\path{arXiv:2103.12773}},
  \href {https://doi.org/10.1140/epja/s10050-021-00551-3}
  {\path{doi:10.1140/epja/s10050-021-00551-3}}.

\bibitem{Ivanov:2002jj}
D.~Y. Ivanov, B.~Pire, L.~Szymanowski, O.~V. Teryaev, {Probing chiral odd GPD's
  in diffractive electroproduction of two vector mesons}, Phys. Lett. B 550
  (2002) 65--76.
\newblock \href {http://arxiv.org/abs/hep-ph/0209300}
  {\path{arXiv:hep-ph/0209300}}, \href
  {https://doi.org/10.1016/S0370-2693(02)02856-3}
  {\path{doi:10.1016/S0370-2693(02)02856-3}}.

\bibitem{Boussarie:2016qop}
R.~Boussarie, B.~Pire, L.~Szymanowski, S.~Wallon, {Exclusive photoproduction of
  a $\gamma\,\rho$ pair with a large invariant mass}, JHEP 02 (2017) 054,
  [Erratum: JHEP 10, 029 (2018)].
\newblock \href {http://arxiv.org/abs/hep-ph/1609.03830}
  {\path{arXiv:hep-ph/1609.03830}}, \href
  {https://doi.org/10.1007/JHEP02(2017)054}
  {\path{doi:10.1007/JHEP02(2017)054}}.

\bibitem{Duplancic:2023kwe}
G.~Duplan\v{c}i\'c, S.~Nabeebaccus, K.~Passek-Kumeri\v{c}ki, B.~Pire,
  L.~Szymanowski, S.~Wallon, {Probing chiral-even and chiral-odd leading twist
  quark generalised parton distributions through the exclusive photoproduction
  of a $ \gamma \rho $ pair} (2 2023).
\newblock \href {http://arxiv.org/abs/2302.12026} {\path{arXiv:2302.12026}}.

\bibitem{Pedrak:2017cpp}
A.~Pedrak, B.~Pire, L.~Szymanowski, J.~Wagner, {Hard photoproduction of a
  diphoton with a large invariant mass}, Phys. Rev. D 96~(7) (2017) 074008,
  [Erratum: Phys.Rev.D 100, 039901 (2019)].
\newblock \href {http://arxiv.org/abs/hep-ph/1708.01043}
  {\path{arXiv:hep-ph/1708.01043}}, \href
  {https://doi.org/10.1103/PhysRevD.96.074008}
  {\path{doi:10.1103/PhysRevD.96.074008}}.

\bibitem{Grocholski:2022rqj}
O.~Grocholski, B.~Pire, P.~Sznajder, L.~Szymanowski, J.~Wagner, {Phenomenology
  of diphoton photoproduction at next-to-leading order}, Phys. Rev. D 105~(9)
  (2022) 094025.
\newblock \href {http://arxiv.org/abs/2204.00396} {\path{arXiv:2204.00396}},
  \href {https://doi.org/10.1103/PhysRevD.105.094025}
  {\path{doi:10.1103/PhysRevD.105.094025}}.

\bibitem{Grocholski:2021man}
O.~Grocholski, B.~Pire, P.~Sznajder, L.~Szymanowski, J.~Wagner, {Collinear
  factorization of diphoton photoproduction at next to leading order}, Phys.
  Rev. D 104~(11) (2021) 114006.
\newblock \href {http://arxiv.org/abs/2110.00048} {\path{arXiv:2110.00048}},
  \href {https://doi.org/10.1103/PhysRevD.104.114006}
  {\path{doi:10.1103/PhysRevD.104.114006}}.

\bibitem{Qiu:2022pla}
J.-W. Qiu, Z.~Yu, {Single diffractive hard exclusive processes for the study of
  generalized parton distributions}, Phys. Rev. D 107~(1) (2023) 014007.
\newblock \href {http://arxiv.org/abs/2210.07995} {\path{arXiv:2210.07995}},
  \href {https://doi.org/10.1103/PhysRevD.107.014007}
  {\path{doi:10.1103/PhysRevD.107.014007}}.

\bibitem{Goloskokov_2014}
S.~V. Goloskokov, P.~Kroll,
  \href{https://doi.org/10.1140%2Fepja%2Fi2014-14146-2}{The pion pole in hard
  exclusive vector-meson leptoproduction}, The European Physical Journal A
  50~(9) (sep 2014).
\newblock \href {https://doi.org/10.1140/epja/i2014-14146-2}
  {\path{doi:10.1140/epja/i2014-14146-2}}.
\newline\urlprefix\url{https://doi.org/10.1140%2Fepja%2Fi2014-14146-2}

\bibitem{Mezrag:2013mya}
C.~Mezrag, H.~Moutarde, F.~Sabati\'e, {Test of two new parametrizations of the
  generalized parton distribution H}, Phys. Rev. D 88~(1) (2013) 014001.
\newblock \href {http://arxiv.org/abs/1304.7645} {\path{arXiv:1304.7645}},
  \href {https://doi.org/10.1103/PhysRevD.88.014001}
  {\path{doi:10.1103/PhysRevD.88.014001}}.

\bibitem{Pedrak:2020mfm}
A.~Pedrak, B.~Pire, L.~Szymanowski, J.~Wagner, {Electroproduction of a large
  invariant mass photon pair}, Phys. Rev. D 101~(11) (2020) 114027.
\newblock \href {http://arxiv.org/abs/hep-ph/2003.03263}
  {\path{arXiv:hep-ph/2003.03263}}, \href
  {https://doi.org/10.1103/PhysRevD.101.114027}
  {\path{doi:10.1103/PhysRevD.101.114027}}.

\bibitem{Berthou:2015oaw}
B.~Berthou, et~al., {PARTONS: PARtonic Tomography Of Nucleon Software}: {A
  computing framework for the phenomenology of Generalized Parton
  Distributions}, Eur. Phys. J. C 78~(6) (2018) 478.
\newblock \href {http://arxiv.org/abs/hep-ph/1512.06174}
  {\path{arXiv:hep-ph/1512.06174}}, \href
  {https://doi.org/10.1140/epjc/s10052-018-5948-0}
  {\path{doi:10.1140/epjc/s10052-018-5948-0}}.

\bibitem{Aschenauer:2022aeb}
E.~C. Aschenauer, V.~Batozskaya, S.~Fazio, K.~Gates, H.~Moutarde, D.~Sokhan,
  H.~Spiesberger, P.~Sznajder, K.~Tezgin, {EpIC: novel Monte Carlo generator
  for exclusive processes}, Eur. Phys. J. C 82~(9) (2022) 819.
\newblock \href {http://arxiv.org/abs/2205.01762} {\path{arXiv:2205.01762}},
  \href {https://doi.org/{10.1140/epjc/s10052-022-10651-z}}
  {\path{doi:{10.1140/epjc/s10052-022-10651-z}}}.

\bibitem{Ji:1994av}
X.-D. Ji, {A QCD analysis of the mass structure of the nucleon}, Phys.Rev.Lett.
  74 (1995) 1071--1074.
\newblock \href {http://arxiv.org/abs/hep-ph/9410274}
  {\path{arXiv:hep-ph/9410274}}, \href
  {https://doi.org/10.1103/PhysRevLett.74.1071}
  {\path{doi:10.1103/PhysRevLett.74.1071}}.

\bibitem{Ji:1995sv}
X.-D. Ji, {Breakup of hadron masses and energy - momentum tensor of QCD}, Phys.
  Rev. D 52 (1995) 271--281.
\newblock \href {http://arxiv.org/abs/hep-ph/9502213}
  {\path{arXiv:hep-ph/9502213}}, \href
  {https://doi.org/10.1103/PhysRevD.52.271}
  {\path{doi:10.1103/PhysRevD.52.271}}.

\bibitem{Qiu:2022bpq}
J.-W. Qiu, Z.~Yu, {Exclusive production of a pair of high transverse momentum
  photons in pion-nucleon collisions for extracting generalized parton
  distributions}, JHEP 08 (2022) 103.
\newblock \href {http://arxiv.org/abs/2205.07846} {\path{arXiv:2205.07846}},
  \href {https://doi.org/10.1007/JHEP08(2022)103}
  {\path{doi:10.1007/JHEP08(2022)103}}.

\bibitem{Qiu:2023mrm}
J.-W. Qiu, Z.~Yu, {Extraction of the $x$-dependence of generalized parton
  distributions from exclusive photoproduction} (5 2023).
\newblock \href {http://arxiv.org/abs/2305.15397} {\path{arXiv:2305.15397}}.

\bibitem{Goloskokov:2005sd}
S.~V. Goloskokov, P.~Kroll, {Vector meson electroproduction at small Bjorken-x
  and generalized parton distributions}, Eur. Phys. J. C 42 (2005) 281--301.
\newblock \href {http://arxiv.org/abs/hep-ph/0501242}
  {\path{arXiv:hep-ph/0501242}}, \href
  {https://doi.org/10.1140/epjc/s2005-02298-5}
  {\path{doi:10.1140/epjc/s2005-02298-5}}.

\bibitem{Goloskokov:2007nt}
S.~Goloskokov, P.~Kroll, {The Role of the quark and gluon GPDs in hard
  vector-meson electroproduction}, Eur. Phys. J. C 53 (2008) 367--384.
\newblock \href {http://arxiv.org/abs/0708.3569} {\path{arXiv:0708.3569}},
  \href {https://doi.org/10.1140/epjc/s10052-007-0466-5}
  {\path{doi:10.1140/epjc/s10052-007-0466-5}}.

\bibitem{Goloskokov:2009ia}
S.~V. Goloskokov, P.~Kroll, {An Attempt to understand exclusive pi+
  electroproduction}, Eur. Phys. J. C 65 (2010) 137--151.
\newblock \href {http://arxiv.org/abs/0906.0460} {\path{arXiv:0906.0460}},
  \href {https://doi.org/10.1140/epjc/s10052-009-1178-9}
  {\path{doi:10.1140/epjc/s10052-009-1178-9}}.

\bibitem{Kroll:2012sm}
P.~Kroll, H.~Moutarde, F.~Sabatie, {From hard exclusive meson electroproduction
  to deeply virtual Compton scattering}, Eur. Phys. J. C 73~(1) (2013) 2278.
\newblock \href {http://arxiv.org/abs/1210.6975} {\path{arXiv:1210.6975}},
  \href {https://doi.org/10.1140/epjc/s10052-013-2278-0}
  {\path{doi:10.1140/epjc/s10052-013-2278-0}}.

\bibitem{Bertone:2021yyz}
V.~Bertone, H.~Dutrieux, C.~Mezrag, H.~Moutarde, P.~Sznajder, {Deconvolution
  problem of deeply virtual Compton scattering}, Phys. Rev. D 103~(11) (2021)
  114019.
\newblock \href {http://arxiv.org/abs/2104.03836} {\path{arXiv:2104.03836}},
  \href {https://doi.org/10.1103/PhysRevD.103.114019}
  {\path{doi:10.1103/PhysRevD.103.114019}}.

\bibitem{Moffat:2023svr}
E.~Moffat, A.~Freese, I.~Clo\"et, T.~Donohoe, L.~Gamberg, W.~Melnitchouk,
  A.~Metz, A.~Prokudin, N.~Sato, {Shedding light on shadow generalized parton
  distributions} (3 2023).
\newblock \href {http://arxiv.org/abs/2303.12006} {\path{arXiv:2303.12006}}.

\bibitem{Kroll:2022roq}
P.~Kroll, K.~Passek-Kumeri\v{c}ki, {Transition GPDs and exclusive
  electroproduction of $\pi$-$\Delta(1232)$ final states}, Phys. Rev. D 107~(5)
  (2023) 054009.
\newblock \href {http://arxiv.org/abs/2211.09474} {\path{arXiv:2211.09474}},
  \href {https://doi.org/10.1103/PhysRevD.107.054009}
  {\path{doi:10.1103/PhysRevD.107.054009}}.

\bibitem{Guichon:2003ah}
P.~A.~M. Guichon, L.~Moss\'e, M.~Vanderhaeghen, {Pion production in deeply
  virtual Compton scattering}, Phys. Rev. D 68 (2003) 034018.
\newblock \href {http://arxiv.org/abs/hep-ph/0305231}
  {\path{arXiv:hep-ph/0305231}}, \href
  {https://doi.org/10.1103/PhysRevD.68.034018}
  {\path{doi:10.1103/PhysRevD.68.034018}}.

\bibitem{Semenov-Tian-Shansky:2023bsy}
K.~M. Semenov-Tian-Shansky, M.~Vanderhaeghen, {Deeply-Virtual Compton Process
  $e^- N \to e^- \gamma \pi N$ to Study Nucleon to Resonance Transitions -
  arXiv:2303.00119 [hep-ph]} (2023).
\newblock \href {http://arxiv.org/abs/2303.00119} {\path{arXiv:2303.00119}}.

\bibitem{Jones:1972ky}
H.~F. Jones, M.~D. Scadron, {Multipole $\gamma N$--$\Delta$ form factors and
  resonant photoproduction and electroproduction}, Annals Phys. 81 (1973)
  1--14.
\newblock \href {https://doi.org/10.1016/0003-4916(73)90476-4}
  {\path{doi:10.1016/0003-4916(73)90476-4}}.

\bibitem{Adler:1968tw}
S.~L. Adler, {Photoproduction, electroproduction and weak single pion
  production in the (3,3) resonance region}, Annals Phys. 50 (1968) 189--311.
\newblock \href {https://doi.org/10.1016/0003-4916(68)90278-9}
  {\path{doi:10.1016/0003-4916(68)90278-9}}.

\bibitem{Adler:1975mt}
S.~L. Adler, {Application of Current Algebra Techniques to Soft Pion Production
  by the Weak Neutral Current: V,a Case}, Phys. Rev. D 12 (1975) 2644.
\newblock \href {https://doi.org/10.1103/PhysRevD.12.2644}
  {\path{doi:10.1103/PhysRevD.12.2644}}.

\bibitem{Kim:2022bwn}
J.-Y. Kim, {Parametrization of transition energy-momentum tensor form factors},
  Phys. Lett. B 834 (2022) 137442.
\newblock \href {http://arxiv.org/abs/2206.10202} {\path{arXiv:2206.10202}},
  \href {https://doi.org/10.1016/j.physletb.2022.137442}
  {\path{doi:10.1016/j.physletb.2022.137442}}.

\bibitem{Kim:2023xvw}
J.-Y. Kim, H.-Y. Won, J.~L. Goity, C.~Weiss, {QCD angular momentum in $N
  \rightarrow \Delta$ transitions} (4 2023).
\newblock \href {http://arxiv.org/abs/2304.08575} {\path{arXiv:2304.08575}}.

\bibitem{Pascalutsa:2006ne}
V.~Pascalutsa, M.~Vanderhaeghen, {New large-N(c) relations among the nucleon
  and nucleon-to-Delta GPDs} (11 2006).
\newblock \href {http://arxiv.org/abs/hep-ph/0611050}
  {\path{arXiv:hep-ph/0611050}}.

\bibitem{Schweitzer:2016jmd}
P.~Schweitzer, C.~Weiss, {Spin-flavor structure of chiral-odd generalized
  parton distributions in the large- N$_c$ limit}, Phys. Rev. C 94~(4) (2016)
  045202.
\newblock \href {http://arxiv.org/abs/1606.08388} {\path{arXiv:1606.08388}},
  \href {https://doi.org/10.1103/PhysRevC.94.045202}
  {\path{doi:10.1103/PhysRevC.94.045202}}.

\bibitem{CLAS:2023akb}
S.~Diehl, et~al., {First measurement of hard exclusive $\pi^- \Delta^{++}$
  electroproduction beam-spin asymmetries off the proton} (3 2023).
\newblock \href {http://arxiv.org/abs/2303.11762} {\path{arXiv:2303.11762}}.

\bibitem{CLAS:2022iqy}
S.~Diehl, et~al., {A multidimensional study of the structure function ratio
  \ensuremath{\sigma}LT'/\ensuremath{\sigma}0 from hard exclusive
  \ensuremath{\pi}+ electro-production off protons in the GPD regime}, Phys.
  Lett. B 839 (2023) 137761.
\newblock \href {http://arxiv.org/abs/2210.14557} {\path{arXiv:2210.14557}},
  \href {https://doi.org/10.1016/j.physletb.2023.137761}
  {\path{doi:10.1016/j.physletb.2023.137761}}.

\bibitem{Kim23}
A.~Kim, S.~Diehl, K.~J. et~al. (CLAS~Collaboration), to be submitted to Phys.
  Lett. B (2023).

\bibitem{Gayoso:2021rzj}
C.~A. Gayoso, et~al., {Progress and opportunities in backward angle (u-channel)
  physics}, Eur. Phys. J. A 57~(12) (2021) 342.
\newblock \href {http://arxiv.org/abs/2107.06748} {\path{arXiv:2107.06748}},
  \href {https://doi.org/10.1140/epja/s10050-021-00625-2}
  {\path{doi:10.1140/epja/s10050-021-00625-2}}.

\bibitem{Frankfurt:1999fp}
L.~L. Frankfurt, P.~V. Pobylitsa, M.~V. Polyakov, M.~Strikman, {Hard exclusive
  pseudoscalar meson electroproduction and spin structure of a nucleon}, Phys.
  Rev. D 60 (1999) 014010.
\newblock \href {http://arxiv.org/abs/hep-ph/9901429}
  {\path{arXiv:hep-ph/9901429}}, \href
  {https://doi.org/10.1103/PhysRevD.60.014010}
  {\path{doi:10.1103/PhysRevD.60.014010}}.

\bibitem{Pire:2004ie}
B.~Pire, L.~Szymanowski, {Hadron annihilation into two photons and backward VCS
  in the scaling regime of QCD}, Phys. Rev. D 71 (2005) 111501.
\newblock \href {http://arxiv.org/abs/hep-ph/0411387}
  {\path{arXiv:hep-ph/0411387}}, \href
  {https://doi.org/10.1103/PhysRevD.71.111501}
  {\path{doi:10.1103/PhysRevD.71.111501}}.

\bibitem{Pire:2021hbl}
B.~Pire, K.~Semenov-Tian-Shansky, L.~Szymanowski, {Transition distribution
  amplitudes and hard exclusive reactions with baryon number transfer}, Phys.
  Rept. 940 (2021) 1--121.
\newblock \href {http://arxiv.org/abs/2103.01079} {\path{arXiv:2103.01079}},
  \href {https://doi.org/10.1016/j.physrep.2021.09.002}
  {\path{doi:10.1016/j.physrep.2021.09.002}}.

\bibitem{CLAS:2017rgp}
K.~Park, et~al., {Hard exclusive pion electroproduction at backward angles with
  CLAS}, Phys. Lett. B 780 (2018) 340--345.
\newblock \href {http://arxiv.org/abs/1711.08486} {\path{arXiv:1711.08486}},
  \href {https://doi.org/10.1016/j.physletb.2018.03.026}
  {\path{doi:10.1016/j.physletb.2018.03.026}}.

\bibitem{JeffersonLabFp:2019gpp}
W.~B. Li, et~al., {Unique Access to $u$-Channel Physics: Exclusive
  Backward-Angle Omega Meson Electroproduction}, Phys. Rev. Lett. 123~(18)
  (2019) 182501.
\newblock \href {http://arxiv.org/abs/1910.00464} {\path{arXiv:1910.00464}},
  \href {https://doi.org/10.1103/PhysRevLett.123.182501}
  {\path{doi:10.1103/PhysRevLett.123.182501}}.

\bibitem{CLAS:2020yqf}
S.~Diehl, et~al., {Extraction of Beam-Spin Asymmetries from the Hard Exclusive
  $\pi^+$ Channel off Protons in a Wide Range of Kinematics}, Phys. Rev. Lett.
  125~(18) (2020) 182001.
\newblock \href {http://arxiv.org/abs/2007.15677} {\path{arXiv:2007.15677}},
  \href {https://doi.org/10.1103/PhysRevLett.125.182001}
  {\path{doi:10.1103/PhysRevLett.125.182001}}.

\bibitem{Li:2020nsk}
W.~B. Li, et~al., {Backward-angle Exclusive pi0 Production above the Resonance
  Region} (8 2020).
\newblock \href {http://arxiv.org/abs/2008.10768} {\path{arXiv:2008.10768}}.

\bibitem{Pire:2011xv}
B.~Pire, K.~Semenov-Tian-Shansky, L.~Szymanowski, {$\pi$ N transition
  distribution amplitudes: their symmetries and constraints from chiral
  dynamics}, Phys. Rev. D 84 (2011) 074014.
\newblock \href {https://doi.org/10.1103/PhysRevD.84.074014}
  {\path{doi:10.1103/PhysRevD.84.074014}}.

\bibitem{Lansberg:2006uh}
J.~P. Lansberg, B.~Pire, L.~Szymanowski, {Backward DVCS and Proton to Photon
  Transition Distribution Amplitudes}, Nucl. Phys. A 782 (2007) 16--23.
\newblock \href {http://arxiv.org/abs/hep-ph/0607130}
  {\path{arXiv:hep-ph/0607130}}, \href
  {https://doi.org/10.1016/j.nuclphysa.2006.10.014}
  {\path{doi:10.1016/j.nuclphysa.2006.10.014}}.

\bibitem{Pire:2022fbi}
B.~Pire, K.~M. Semenov-Tian-Shansky, A.~A. Shaikhutdinova, L.~Szymanowski,
  {Backward timelike Compton scattering to decipher the photon content of the
  nucleon}, Eur. Phys. J. C 82~(7) (2022) 656.
\newblock \href {http://arxiv.org/abs/2201.12853} {\path{arXiv:2201.12853}},
  \href {https://doi.org/10.1140/epjc/s10052-022-10587-4}
  {\path{doi:10.1140/epjc/s10052-022-10587-4}}.

\bibitem{Pire:2022kwu}
B.~Pire, K.~M. Semenov-Tian-Shansky, A.~A. Shaikhutdinova, L.~Szymanowski,
  {Pion and photon beam initiated backward charmonium or lepton pair
  production} (12 2022).
\newblock \href {http://arxiv.org/abs/2212.07688} {\path{arXiv:2212.07688}}.

\bibitem{GlueX:2023pev}
S.~Adhikari, et~al., {Measurement of the J/$\psi $ photoproduction cross
  section over the full near-threshold kinematic region} (4 2023).
\newblock \href {http://arxiv.org/abs/2304.03845} {\path{arXiv:2304.03845}}.

\bibitem{Jain:2022xzo}
P.~Jain, B.~Pire, J.~P. Ralston, {The Status and Future of Color Transparency
  and Nuclear Filtering}, MDPI Physics 4~(2) (2022) 578--589.
\newblock \href {http://arxiv.org/abs/2203.02579} {\path{arXiv:2203.02579}},
  \href {https://doi.org/10.3390/physics4020038}
  {\path{doi:10.3390/physics4020038}}.

\bibitem{Huber:2022wns}
G.~M. Huber, W.~B. Li, W.~Cosyn, B.~Pire, {u-Channel Color Transparency
  Observables}, MDPI Physics 4~(2) (2022) 451--461.
\newblock \href {http://arxiv.org/abs/2202.04470} {\path{arXiv:2202.04470}},
  \href {https://doi.org/10.3390/physics4020030}
  {\path{doi:10.3390/physics4020030}}.

\bibitem{E12-09-011}
T.~Horn, G.~M. Huber, P.~Markowitz, et~al., {Studies of the L/T Separated Kaon
  Electroproduction Cross Sections from 5-11~GeV}, jefferson Lab 12 GeV
  Experiment E12-09-011.

\bibitem{E12-19-006}
G.~M. Huber, D.~Gaskell, T.~Horn, et~al.,
  \href{https://www.jlab.org/exp_prog/proposals/19/E12-19-006.pdf}{{Measurement
  of the Charged Pion Form Factor to High $Q^{2}$ and Scaling Study of the
  L/T-Separated Pion Electroproduction Cross Section at 11~GeV}}, jefferson Lab
  12 GeV Experiment E12-19-006 (2019).
\newline\urlprefix\url{https://www.jlab.org/exp_prog/proposals/19/E12-19-006.pdf}

\bibitem{JeffersonLabHallA:1999epl}
M.~K. Jones, et~al., $g_{Ep}/g_{Mp}$ ratio by polarization transfer in $\vec e
  p \to e \vec p$, Phys. Rev. Lett. 84 (2000) 1398--1402.
\newblock \href {http://arxiv.org/abs/nucl-ex/9910005}
  {\path{arXiv:nucl-ex/9910005}}, \href
  {https://doi.org/10.1103/PhysRevLett.84.1398}
  {\path{doi:10.1103/PhysRevLett.84.1398}}.

\bibitem{JeffersonLabHallA:2001qqe}
O.~Gayou, et~al., {Measurement of $G_{Ep}/G_{Mp}$ in $\vec e p \to e \vec p$ to
  $Q^2 = 5.6$-GeV$^2$}, Phys. Rev. Lett. 88 (2002) 092301.
\newblock \href {http://arxiv.org/abs/nucl-ex/0111010}
  {\path{arXiv:nucl-ex/0111010}}, \href
  {https://doi.org/10.1103/PhysRevLett.88.092301}
  {\path{doi:10.1103/PhysRevLett.88.092301}}.

\bibitem{Puckett:2010ac}
A.~J.~R. Puckett, et~al., {Recoil Polarization Measurements of the Proton
  Electromagnetic Form Factor Ratio to $Q^2$ = 8.5 GeV$^2$}, Phys. Rev. Lett.
  104 (2010) 242301.
\newblock \href {http://arxiv.org/abs/1005.3419} {\path{arXiv:1005.3419}},
  \href {https://doi.org/10.1103/PhysRevLett.104.242301}
  {\path{doi:10.1103/PhysRevLett.104.242301}}.

\bibitem{perdrisat_bonner}
{American Physical Society 2017 Bonner Prize in Nuclear Physics Recipient
  Charles F. Perdrisat (College of William and Mary)},
  \href{https://www.aps.org/programs/honors/prizes/prizerecipient.cfm?last_nm=F&first_nm=C&year=2017}{[Webpage]}.

\bibitem{Barabanov:2020jvn}
M.~Y. Barabanov, et~al., {Diquark correlations in hadron physics: Origin,
  impact and evidence}, Prog. Part. Nucl. Phys. 116 (2021) 103835.
\newblock \href {http://arxiv.org/abs/2008.07630} {\path{arXiv:2008.07630}},
  \href {https://doi.org/10.1016/j.ppnp.2020.103835}
  {\path{doi:10.1016/j.ppnp.2020.103835}}.

\bibitem{Gross:2022hyw}
F.~Gross, et~al., {50 Years of Quantum Chromodynamics} (12 2022).
\newblock \href {http://arxiv.org/abs/2212.11107} {\path{arXiv:2212.11107}}.

\bibitem{Schmookler:2022gxw}
B.~Schmookler, A.~Pierre-Louis, A.~Deshpande, D.~Higinbotham, E.~Long, A.~J.~R.
  Puckett, {High $Q^2$ electron-proton elastic scattering at the future
  Electron-Ion Collider} (7 2022).
\newblock \href {http://arxiv.org/abs/2207.04378} {\path{arXiv:2207.04378}}.

\bibitem{Englert:2014zpa}
F.~Englert, {Nobel Lecture: The BEH Mechanism and its Scalar Boson}, Rev. Mod.
  Phys. 86 (2014) 843.

\bibitem{Higgs:2014aqa}
P.~W. Higgs, {Nobel Lecture: Evading the Goldstone theorem}, Rev. Mod. Phys. 86
  (2014) 851.

\bibitem{Binosi:2022djx}
D.~Binosi, {Emergent Hadron Mass in Strong Dynamics}, Few Body Syst. 63~(2)
  (2022) 42.

\bibitem{Ferreira:2023fva}
M.~N. Ferreira, J.~Papavassiliou, {Gauge Sector Dynamics in QCD}, Particles
  6~(1) (2023) 312.

\bibitem{Ding:2022ows}
M.~Ding, C.~D. Roberts, S.~M. Schmidt, {Emergence of Hadron Mass and
  Structure}, Particles 6~(1) (2023) 57.

\bibitem{Barabanov_2021}
M.~Y. Barabanov, et~al., {Diquark Correlations in Hadron Physics: Origin,
  Impact and Evidence}, Progress in Particle and Nuclear Physics 116 (2021)
  103835.

\bibitem{Proceedings:2020fyd}
S.~J. Brodsky, et~al., {Strong QCD from Hadron Structure Experiments}: {Newport
  News, VA, USA, November 4-8, 2019}, Int. J. Mod. Phys. E 29~(08) (2020)
  2030006.

\bibitem{Burkert:2017djo}
V.~D. Burkert, C.~D. Roberts, {Colloquium : Roper Resonance: Toward a Solution
  to the Fifty Year Puzzle}, Rev. Mod. Phys. 91~(1) (2019) 011003.

\bibitem{Flambaum:2005kc}
V.~V. Flambaum, et~al., {Sigma Terms of Light-Quark Hadrons}, Few Body Syst. 38
  (2006) 31.

\bibitem{RuizdeElvira:2017stg}
J.~Ruiz~de Elvira, M.~Hoferichter, B.~Kubis, U.-G. Mei\ss{}ner, {Extracting the
  $\sigma$-Term from Low-Energy Pion-Nucleon Scattering}, J. Phys. G 45~(2)
  (2018) 024001.

\bibitem{Aoki:2019cca}
S.~Aoki, et~al., {FLAG Review 2019}, Eur. Phys. J. C 80 (2020) 113.

\bibitem{Aguilar:2019teb}
A.~C. Aguilar, et~al., {Pion and Kaon Structure at the Electron-Ion Collider},
  Eur. Phys. J. A 55 (2019) 190.

\bibitem{Anderle:2021wcy}
D.~P. Anderle, et~al., {Electron-Ion Collider in China}, Front. Phys. (Beijing)
  16~(6) (2021) 64701.

\bibitem{Schwinger:1962tp}
J.~S. Schwinger, {Gauge Invariance and Mass. 2.}, Phys. Rev. 128 (1962) 2425.

\bibitem{Cornwall:1981zr}
J.~M. Cornwall, {Dynamical Mass Generation in Continuum QCD}, Phys. Rev. D 26
  (1982) 1453.

\bibitem{Mandula:1987rh}
J.~Mandula, M.~Ogilvie, {The Gluon Is Massive: A Lattice Calculation of the
  Gluon Propagator in the Landau Gauge}, Phys. Lett. B 185 (1987) 127.

\bibitem{Roberts:2021nhw}
C.~D. Roberts, D.~G. Richards, T.~Horn, L.~Chang, {Insights into the Emergence
  of Mass from Studies of Pion and Kaon Structure}, Prog. Part. Nucl. Phys. 120
  (2021) 103883.

\bibitem{Suzuki:2009nj}
N.~Suzuki, B.~Julia-Diaz, H.~Kamano, T.~S.~H. Lee, A.~Matsuyama, T.~Sato,
  {Disentangling the Dynamical Origin of P-11 Nucleon Resonances}, Phys. Rev.
  Lett. 104 (2010) 042302.

\bibitem{Giannini:2015zia}
M.~M. Giannini, E.~Santopinto, {The Hypercentral Constituent Quark Model and
  its Application to Baryon Properties}, Chin. J. Phys. 53 (2015) 020301.

\bibitem{Qin:2020rad}
S.-X. Qin, C.~D. Roberts, {Impressions of the Continuum Bound State Problem in
  QCD}, Chin. Phys. Lett. 37~(12) (2020) 121201.

\bibitem{Gao:2017mmp}
F.~Gao, L.~Chang, Y.-X. Liu, C.~D. Roberts, P.~C. Tandy, {Exposing Strangeness:
  Projections for Kaon Electromagnetic Form Factors}, Phys. Rev. D 96~(3)
  (2017) 034024.

\bibitem{Xu:2018cor}
S.-S. Xu, Z.-F. Cui, L.~Chang, J.~Papavassiliou, C.~D. Roberts, H.-S. Zong,
  {New Perspective on Hybrid Mesons}, Eur. Phys. J. A (Lett.) 55 (2019) 113.

\bibitem{Wang:2018kto}
Q.-W. Wang, S.-X. Qin, C.~D. Roberts, S.~M. Schmidt, {Proton Tensor Charges
  from a Poincar{\'e}-Covariant Faddeev Equation}, Phys. Rev. D 98 (2018)
  054019.

\bibitem{Chen:2018rwz}
M.~Chen, M.~Ding, L.~Chang, C.~D. Roberts, {Mass-Dependence of Pseudoscalar
  Meson Elastic Form Factors}, Phys. Rev. D 98 (2018) 091505(R).

\bibitem{Binosi:2018rht}
D.~Binosi, L.~Chang, M.~Ding, F.~Gao, J.~Papavassiliou, C.~D. Roberts,
  {Distribution Amplitudes of Heavy-Light Mesons}, Phys. Lett. B 790 (2019)
  257.

\bibitem{Chen:2019fzn}
C.~Chen, G.~I. Krein, C.~D. Roberts, S.~M. Schmidt, J.~Segovia, {Spectrum and
  Structure of Octet and Decuplet Baryons and Their Positive-Parity
  Excitations}, Phys. Rev. D 100 (2019) 054009.

\bibitem{Qin:2019hgk}
S.-X. Qin, C.~D. Roberts, S.~M. Schmidt, {Spectrum of Light- and
  Heavy-Baryons}, Few Body Syst. 60 (2019) 26.

\bibitem{Lu:2019bjs}
Y.~Lu, C.~Chen, Z.-F. Cui, C.~D. Roberts, S.~M. Schmidt, J.~Segovia, H.~S.
  Zong, {Transition Form Factors: $\gamma^* + p \to \Delta(1232)$,
  $\Delta(1600)$}, Phys. Rev. D 100~(3) (2019) 034001.

\bibitem{Souza:2019ylx}
E.~V. Souza, M.~Narciso~Ferreira, A.~C. Aguilar, J.~Papavassiliou, C.~D.
  Roberts, S.-S. Xu, {Pseudoscalar Glueball Mass: A Window on Three-Gluon
  Interactions}, Eur. Phys. J. A (Lett.) 56 (2020) 25.

\bibitem{Raya:2021zrz}
K.~Raya, Z.-F. Cui, L.~Chang, J.-M. Morgado, C.~D. Roberts,
  J.~Rodr{\'{\i}}guez-Quintero, {Revealing Pion and Kaon Structure via
  Generalised Parton Distributions}, Chin. Phys. C 46~(26) (2022) 013105.

\bibitem{Cui:2021mom}
Z.~F. Cui, M.~Ding, J.~M. Morgado, K.~Raya, D.~Binosi, L.~Chang,
  J.~Papavassiliou, C.~D. Roberts, J.~Rodr\'\i{}guez-Quintero, S.~M. Schmidt,
  {Concerning Pion Parton Distributions}, Eur. Phys. J. A 58~(1) (2022) 10.

\bibitem{Liu:2022nku}
L.~Liu, C.~Chen, C.~D. Roberts, {Wave functions of
  $(I,J^P)=(\tfrac{1}{2},\tfrac{3}{2}^{\mp})$ baryons}, Phys. Rev. D 107~(1)
  (2023) 014002.

\bibitem{Gao:2017uox}
F.~Gao, S.-X. Qin, C.~D. Roberts, J.~Rodr{\'{\i}}guez-Quintero, {Locating the
  Gribov Horizon}, Phys. Rev. D 97 (2018) 034010.

\bibitem{Oliveira:2018lln}
O.~Oliveira, P.~J. Silva, J.-I. Skullerud, A.~Sternbeck, {Quark Propagator with
  Two Flavors of $O(a)$-Improved Wilson Fermions}, Phys. Rev. D 99~(9) (2019)
  094506.

\bibitem{Binosi:2019ecz}
D.~Binosi, R.-A. Tripolt, {Spectral Functions of Confined Particles}, Phys.
  Lett. B 801 (2020) 135171.

\bibitem{Boito:2022rad}
D.~Boito, A.~Cucchieri, C.~Y. London, T.~Mendes, {Probing the Singularities of
  the Landau-Gauge Gluon and Ghost Propagators with Rational Approximants},
  JHEP 02 (2023) 144.

\bibitem{Burkert:2022ioj}
V.~D. Burkert, {Nucleon Resonances and Transition Form
  Factors\,--\,arXiv:2212.08980 [hep-ph]} (2022).

\bibitem{Aznauryan:2011qj}
I.~G. Aznauryan, V.~D. Burkert, {Electroexcitation of Nucleon Resonances},
  Prog. Part. Nucl. Phys. 67 (2012) 1--54.

\bibitem{Mokeev:2022xfo}
V.~I. Mokeev, D.~S. Carman, {Photo- and Electrocouplings of Nucleon
  Resonances}, Few Body Syst. 63~(3) (2022) 59.

\bibitem{Segovia:2015hra}
J.~Segovia, B.~El-Bennich, E.~Rojas, I.~C. Cloet, C.~D. Roberts, S.-S. Xu,
  H.-S. Zong, {Completing the Picture of the Roper Resonance}, Phys. Rev. Lett.
  115~(17) (2015) 171801.

\bibitem{Wilson:2011aa}
D.~J. Wilson, I.~C. Cloet, L.~Chang, C.~D. Roberts, {Nucleon and Roper
  Electromagnetic Elastic and Transition Form Factors}, Phys. Rev. C 85 (2012)
  025205.

\bibitem{Cui:2020rmu}
Z.-F. Cui, C.~Chen, D.~Binosi, F.~de~Soto, C.~D. Roberts,
  J.~Rodr{\'{\i}}guez-Quintero, S.~M. Schmidt, J.~Segovia, {Nucleon Elastic
  Form Factors at Accessible Large Spacelike Momenta}, Phys. Rev. D 102 (2020)
  014043.

\bibitem{Chen:2022odn}
C.~Chen, C.~D. Roberts, {Nucleon Axial Form Factor at Large Momentum
  Transfers}, Eur. Phys. J. A 58~(10) (2022) 206.

\bibitem{Ding:2018xwy}
M.~Ding, K.~Raya, A.~Bashir, D.~Binosi, L.~Chang, M.~Chen, C.~D. Roberts,
  {$\gamma^\ast \gamma \to \eta, \eta^\prime$ Transition Form Factors}, Phys.
  Rev. D 99 (2019) 014014.

\bibitem{Egiyan_2003}
K.~S. Egiyan, et~al., Observation of nuclear scaling in the $a(e,e^{'})$
  reaction at ${x}_{B}>1$, Phys. Rev. C 68 (2003) 014313.
\newblock \href {https://doi.org/10.1103/PhysRevC.68.014313}
  {\path{doi:10.1103/PhysRevC.68.014313}}.

\bibitem{Egiyan_2006}
K.~S. Egiyan, et~al., Measurement of two- and three-nucleon short-range
  correlation probabilities in nuclei, Phys. Rev. Lett. 96 (2006) 082501.
\newblock \href {https://doi.org/10.1103/PhysRevLett.96.082501}
  {\path{doi:10.1103/PhysRevLett.96.082501}}.

\bibitem{Fomin_2012}
N.~Fomin, et~al., {New Measurements of High-Momentum Nucleons and Short-Range
  Structures in Nuclei}, Phys. Rev. Lett. 108 (2012) 092502.
\newblock \href {https://doi.org/10.1103/PhysRevLett.108.092502}
  {\path{doi:10.1103/PhysRevLett.108.092502}}.

\bibitem{Frankfurt:1993sp}
L.~L. Frankfurt, M.~I. Strikman, D.~B. Day, M.~Sargsian, {Evidence for short
  range correlations from high Q**2 (e, e-prime) reactions}, Phys. Rev. C 48
  (1993) 2451--2461.
\newblock \href {https://doi.org/10.1103/PhysRevC.48.2451}
  {\path{doi:10.1103/PhysRevC.48.2451}}.

\bibitem{FRANKFURT:2008}
L.~Frankfurt, M.~Sargsian, M.~Strikman, Recent observation of short range
  nucleon correlations in nuclei and their implications for the structure of
  nuclei and neutron stars, Int. J. Mod. Phys. A 23~(20) (2008) 2991--3055.
\newblock \href {https://doi.org/10.1142/s0217751x08041207}
  {\path{doi:10.1142/s0217751x08041207}}.

\bibitem{Piazetzky_2006}
E.~Piasetzky, M.~Sargsian, L.~Frankfurt, M.~Strikman, J.~W. Watson, Evidence
  for strong dominance of proton-neutron correlations in nuclei, Phys. Rev.
  Lett. 97 (2006) 162504.
\newblock \href {https://doi.org/10.1103/PhysRevLett.97.162504}
  {\path{doi:10.1103/PhysRevLett.97.162504}}.

\bibitem{Subedi_2008}
R.~Subedi, et~al., Probing cold dense nuclear matter, Science 320~(5882) (2008)
  1476--1478.
\newblock \href
  {http://arxiv.org/abs/https://science.sciencemag.org/content/320/5882/1476.full.pdf}
  {\path{arXiv:https://science.sciencemag.org/content/320/5882/1476.full.pdf}},
  \href {https://doi.org/10.1126/science.1156675}
  {\path{doi:10.1126/science.1156675}}.

\bibitem{Duer_2019}
M.~Duer, et~al., {Direct Observation of Proton-Neutron Short-Range Correlation
  Dominance in Heavy Nuclei}, Phys. Rev. Lett. 122 (2019) 172502.
\newblock \href {https://doi.org/10.1103/PhysRevLett.122.172502}
  {\path{doi:10.1103/PhysRevLett.122.172502}}.

\bibitem{Sargsian:2005}
M.~Sargsian, T.~Abrahamyan, M.~Strikman, L.~Frankfurt, Exclusive
  electrodisintegration of $^{3}\mathrm{He}$ at high ${Q}^{2}$.~ii.~decay
  function formalism, Phys. Rev. C 71~(4) (2005).
\newblock \href {https://doi.org/10.1103/physrevc.71.044615}
  {\path{doi:10.1103/physrevc.71.044615}}.

\bibitem{Schiavilla:2006xx}
R.~Schiavilla, R.~B. Wiringa, S.~C. Pieper, J.~Carlson, {Tensor Forces and the
  Ground-State Structure of Nuclei}, Phys. Rev. Lett. 98 (2007) 132501.
\newblock \href {http://arxiv.org/abs/nucl-th/0611037}
  {\path{arXiv:nucl-th/0611037}}, \href
  {https://doi.org/10.1103/PhysRevLett.98.132501}
  {\path{doi:10.1103/PhysRevLett.98.132501}}.

\bibitem{Sargsian:2012sm}
M.~M. Sargsian, {New properties of the high-momentum distribution of nucleons
  in asymmetric nuclei}, Phys. Rev. C 89~(3) (2014) 034305.
\newblock \href {http://arxiv.org/abs/1210.3280} {\path{arXiv:1210.3280}},
  \href {https://doi.org/10.1103/PhysRevC.89.034305}
  {\path{doi:10.1103/PhysRevC.89.034305}}.

\bibitem{Hen_2014}
O.~Hen, et~al., Momentum sharing in imbalanced fermi systems, Science
  346~(6209) (2014) 614--617.
\newblock \href
  {http://arxiv.org/abs/https://science.sciencemag.org/content/346/6209/614.full.pdf}
  {\path{arXiv:https://science.sciencemag.org/content/346/6209/614.full.pdf}},
  \href {https://doi.org/10.1126/science.1256785}
  {\path{doi:10.1126/science.1256785}}.

\bibitem{Duer_2018}
M.~Duer, et~al., Probing the high-momentum protons and neutrons in neutron-rich
  nuclei, Nature 560 (2018) 617--621,
  \url{https://doi.org/10.1038/s41586-018-0400-z}.

\bibitem{Jastrow:1951}
R.~Jastrow, On the nucleon-nucleon interaction, Phys. Rev. 81 (1951) 165--170.
\newblock \href {https://doi.org/10.1103/PhysRev.81.165}
  {\path{doi:10.1103/PhysRev.81.165}}.

\bibitem{Wiringa:1994wb}
R.~B. Wiringa, V.~G.~J. Stoks, R.~Schiavilla, {An Accurate nucleon-nucleon
  potential with charge independence breaking}, Phys. Rev. C 51 (1995) 38--51.
\newblock \href {http://arxiv.org/abs/nucl-th/9408016}
  {\path{arXiv:nucl-th/9408016}}, \href
  {https://doi.org/10.1103/PhysRevC.51.38} {\path{doi:10.1103/PhysRevC.51.38}}.

\bibitem{Epelbaum:2008ga}
E.~Epelbaum, H.-W. Hammer, U.-G. Meissner, {Modern Theory of Nuclear Forces},
  Rev. Mod. Phys. 81 (2009) 1773--1825.
\newblock \href {http://arxiv.org/abs/0811.1338} {\path{arXiv:0811.1338}},
  \href {https://doi.org/10.1103/RevModPhys.81.1773}
  {\path{doi:10.1103/RevModPhys.81.1773}}.

\bibitem{Harvey:1981}
M.~Harvey, Effective nuclear forces in the quark model with delta and
  hidden-color channel coupling, Nucl. Phys. A 352~(3) (1981) 326--342.
\newblock \href {https://doi.org/https://doi.org/10.1016/0375-9474(81)90413-9}
  {\path{doi:https://doi.org/10.1016/0375-9474(81)90413-9}}.

\bibitem{Brodsky:1986}
C.~Ji, S.~Brodsky, Quantum-chromodynamic evolution of six-quark states, Phys.
  Rev. D 34 (1986) 1460--1473.
\newblock \href {https://doi.org/10.1103/PhysRevD.34.1460}
  {\path{doi:10.1103/PhysRevD.34.1460}}.

\bibitem{Frankfurt:1981}
L.~Frankfurt, M.~Strikman, High-energy phenomena, short-range nuclear structure
  and {QCD}, Phys. Rept. 76~(4) (1981) 215--347.
\newblock \href {https://doi.org/https://doi.org/10.1016/0370-1573(81)90129-0}
  {\path{doi:https://doi.org/10.1016/0370-1573(81)90129-0}}.

\bibitem{Miller:2014}
G.~Miller, Pionic and hidden-color, six-quark contributions to the deuteron
  ${b}_{1}$ structure function, Phys. Rev. C 89 (2014) 045203.
\newblock \href {https://doi.org/10.1103/PhysRevC.89.045203}
  {\path{doi:10.1103/PhysRevC.89.045203}}.

\bibitem{West:2020tyo}
J.~Rittenhouse~West, {Diquark induced short-range nucleon-nucleon correlations
  \& the EMC effect}, Nucl. Phys. A 1029 (2023) 122563.
\newblock \href {http://arxiv.org/abs/2009.06968} {\path{arXiv:2009.06968}},
  \href {https://doi.org/10.1016/j.nuclphysa.2022.122563}
  {\path{doi:10.1016/j.nuclphysa.2022.122563}}.

\bibitem{West:2020rlk}
J.~Rittenhouse~West, S.~J. Brodsky, G.~F. de~Teramond, A.~S. Goldhaber,
  I.~Schmidt, {QCD hidden-color hexadiquark in the core of nuclei}, Nucl. Phys.
  A 1007 (2021) 122134.
\newblock \href {http://arxiv.org/abs/2004.14659} {\path{arXiv:2004.14659}},
  \href {https://doi.org/10.1016/j.nuclphysa.2020.122134}
  {\path{doi:10.1016/j.nuclphysa.2020.122134}}.

\bibitem{Frankfurt:1988}
L.~Frankfurt, M.~Strikman, Hard nuclear processes and microscopic nuclear
  structure, Phys. Rept. 160~(5) (1988) 235--427.
\newblock \href {https://doi.org/10.1016/0370-1573(88)90179-2}
  {\path{doi:10.1016/0370-1573(88)90179-2}}.

\bibitem{Sargsian:2007gd}
M.~M. Sargsian, {Superfast quarks in the nuclear medium}, Nucl. Phys. A 782
  (2007) 199--206.
\newblock \href {https://doi.org/10.1016/j.nuclphysa.2006.10.057}
  {\path{doi:10.1016/j.nuclphysa.2006.10.057}}.

\bibitem{Freese:2015}
A.~Freese, M.~Sargsian, M.~Strikman, Probing superfast quarks in nuclei through
  dijet production at the {LHC}, Eur. Phys. J. C 75~(11) (nov 2015).
\newblock \href {https://doi.org/10.1140/epjc/s10052-015-3755-4}
  {\path{doi:10.1140/epjc/s10052-015-3755-4}}.

\bibitem{Freese:2019}
A.~Freese, W.~Cosyn, M.~Sargsian, {QCD} evolution of superfast quarks, Phys.
  Rev. D 99 (2019) 114019.
\newblock \href {https://doi.org/10.1103/PhysRevD.99.114019}
  {\path{doi:10.1103/PhysRevD.99.114019}}.

\bibitem{Fomin:2010}
N.~Fomin, et~al., Scaling of the ${F}_{2}$ structure function in nuclei and
  quark distributions at $x > 1$, Phys. Rev. Lett. 105 (2010) 212502.
\newblock \href {https://doi.org/10.1103/PhysRevLett.105.212502}
  {\path{doi:10.1103/PhysRevLett.105.212502}}.

\bibitem{E12-06-105}
J.~Arrington, D.~Day, N.~Fomin, P.~Solvignon,
  \href{https://www.jlab.org/exp\_prog/proposals/06/PR12-06-105.pdf}{{E12-06-105:
  Inclusive Scattering from Nuclei at $x > 1$ in the quasielastic and deeply
  inelastic regimes}} (2006).
\newline\urlprefix\url{https://www.jlab.org/exp\_prog/proposals/06/PR12-06-105.pdf}

\bibitem{Yero:2020}
C.~Yero, et~al., Probing the deuteron at very large internal momenta, Phys.
  Rev. Lett. 125 (2020) 262501.
\newblock \href {https://doi.org/10.1103/PhysRevLett.125.262501}
  {\path{doi:10.1103/PhysRevLett.125.262501}}.

\bibitem{Sargsian:2019}
M.~Sargsian, D.~Day, L.~Frankfurt, M.~Strikman, Searching for three-nucleon
  short-range correlations, Phys. Rev. C 100~(4) (2019).
\newblock \href {https://doi.org/10.1103/physrevc.100.044320}
  {\path{doi:10.1103/physrevc.100.044320}}.

\bibitem{Day:2023}
D.~Day, L.~Frankfurt, M.~Sargsian, M.~Strikman, Toward observation of
  three-nucleon short-range correlations in high-${Q}^{2}
  a(e,{e}^{\ensuremath{'}})x$ reactions, Phys. Rev. C 107 (2023) 014319.
\newblock \href {https://doi.org/10.1103/PhysRevC.107.014319}
  {\path{doi:10.1103/PhysRevC.107.014319}}.

\bibitem{Aubert:1983}
J.~J. Aubert, et~al., {The ratio of the nucleon structure functions $F2_n$ for
  iron and deuterium}, Phys. Lett. B 123 (1983) 275--278.
\newblock \href {https://doi.org/10.1016/0370-2693(83)90437-9}
  {\path{doi:10.1016/0370-2693(83)90437-9}}.

\bibitem{Seely:2009}
J.~Seely, et~al., {New measurements of the {EMC} effect in very light nuclei},
  Phys. Rev. Lett. 103 (2009) 202301.
\newblock \href {http://arxiv.org/abs/0904.4448} {\path{arXiv:0904.4448}},
  \href {https://doi.org/10.1103/PhysRevLett.103.202301}
  {\path{doi:10.1103/PhysRevLett.103.202301}}.

\bibitem{Gomez:2011}
L.~B. Weinstein, E.~Piasetzky, D.~W. Higinbotham, J.~Gomez, O.~Hen, R.~Shneor,
  {Short Range Correlations and the {EMC} Effect}, Phys. Rev. Lett. 106 (2011)
  052301.
\newblock \href {http://arxiv.org/abs/1009.5666} {\path{arXiv:1009.5666}},
  \href {https://doi.org/10.1103/PhysRevLett.106.052301}
  {\path{doi:10.1103/PhysRevLett.106.052301}}.

\bibitem{Schmookler:2019}
B.~Schmookler, et~al., {Modified structure of protons and neutrons in
  correlated pairs}, Nature 566~(7744) (2019) 354--358.
\newblock \href {http://arxiv.org/abs/2004.12065} {\path{arXiv:2004.12065}},
  \href {https://doi.org/10.1038/s41586-019-0925-9}
  {\path{doi:10.1038/s41586-019-0925-9}}.

\bibitem{Brodsky:2014yha}
S.~J. Brodsky, G.~F. de~Teramond, H.~G. Dosch, J.~Erlich, {Light-Front
  Holographic QCD and Emerging Confinement}, Phys. Rept. 584 (2015) 1--105.
\newblock \href {http://arxiv.org/abs/1407.8131} {\path{arXiv:1407.8131}},
  \href {https://doi.org/10.1016/j.physrep.2015.05.001}
  {\path{doi:10.1016/j.physrep.2015.05.001}}.

\bibitem{Kim:2022lng}
D.~N. Kim, G.~A. Miller, {Light-front holography model of the EMC effect},
  Phys. Rev. C 106~(5) (2022) 055202.
\newblock \href {http://arxiv.org/abs/2209.13753} {\path{arXiv:2209.13753}},
  \href {https://doi.org/10.1103/PhysRevC.106.055202}
  {\path{doi:10.1103/PhysRevC.106.055202}}.

\bibitem{Frankfurt:2011cs}
L.~Frankfurt, V.~Guzey, M.~Strikman, {Leading Twist Nuclear Shadowing Phenomena
  in Hard Processes with Nuclei}, Phys. Rept. 512 (2012) 255--393.
\newblock \href {http://arxiv.org/abs/1106.2091} {\path{arXiv:1106.2091}},
  \href {https://doi.org/10.1016/j.physrep.2011.12.002}
  {\path{doi:10.1016/j.physrep.2011.12.002}}.

\bibitem{Miller:2001yf}
G.~A. Miller, {Revealing nuclear pions using electron scattering}, Phys. Rev. C
  64 (2001) 022201.
\newblock \href {http://arxiv.org/abs/nucl-th/0104025}
  {\path{arXiv:nucl-th/0104025}}, \href
  {https://doi.org/10.1103/PhysRevC.64.022201}
  {\path{doi:10.1103/PhysRevC.64.022201}}.

\bibitem{Alde:1990im}
D.~M. Alde, et~al., {Nuclear dependence of dimuon production at 800-GeV.
  FNAL-772 experiment}, Phys. Rev. Lett. 64 (1990) 2479--2482.
\newblock \href {https://doi.org/10.1103/PhysRevLett.64.2479}
  {\path{doi:10.1103/PhysRevLett.64.2479}}.

\bibitem{Alvioli:2022tij}
M.~Alvioli, M.~Strikman, {Hunting for an EMC-like effect for antiquarks} (10
  2022).
\newblock \href {http://arxiv.org/abs/2210.12597} {\path{arXiv:2210.12597}}.

\bibitem{Kotko:2017oxg}
P.~Kotko, K.~Kutak, S.~Sapeta, A.~M. Stasto, M.~Strikman, {Estimating nonlinear
  effects in forward dijet production in ultra-peripheral heavy ion collisions
  at the LHC}, Eur. Phys. J. C 77~(5) (2017) 353.
\newblock \href {http://arxiv.org/abs/1702.03063} {\path{arXiv:1702.03063}},
  \href {https://doi.org/10.1140/epjc/s10052-017-4906-6}
  {\path{doi:10.1140/epjc/s10052-017-4906-6}}.

\bibitem{Frankfurt:1994hf}
L.~L. Frankfurt, G.~A. Miller, M.~Strikman, {The Geometrical color optics of
  coherent high-energy processes}, Ann. Rev. Nucl. Part. Sci. 44 (1994)
  501--560.
\newblock \href {http://arxiv.org/abs/hep-ph/9407274}
  {\path{arXiv:hep-ph/9407274}}, \href
  {https://doi.org/10.1146/annurev.ns.44.120194.002441}
  {\path{doi:10.1146/annurev.ns.44.120194.002441}}.

\bibitem{E791:2000kym}
E.~M. Aitala, et~al., {Observation of color transparency in diffractive
  dissociation of pions}, Phys. Rev. Lett. 86 (2001) 4773--4777.
\newblock \href {http://arxiv.org/abs/hep-ex/0010044}
  {\path{arXiv:hep-ex/0010044}}, \href
  {https://doi.org/10.1103/PhysRevLett.86.4773}
  {\path{doi:10.1103/PhysRevLett.86.4773}}.

\bibitem{Clasie:2007aa}
B.~Clasie, et~al., {Measurement of nuclear transparency for the A(e, e-prime'
  pi+) reaction}, Phys. Rev. Lett. 99 (2007) 242502.
\newblock \href {http://arxiv.org/abs/0707.1481} {\path{arXiv:0707.1481}},
  \href {https://doi.org/10.1103/PhysRevLett.99.242502}
  {\path{doi:10.1103/PhysRevLett.99.242502}}.

\bibitem{CLAS:2012tlh}
L.~El~Fassi, et~al., {Evidence for the onset of color transparency in $\rho^0$
  electroproduction off nuclei}, Phys. Lett. B 712 (2012) 326--330.
\newblock \href {http://arxiv.org/abs/1201.2735} {\path{arXiv:1201.2735}},
  \href {https://doi.org/10.1016/j.physletb.2012.05.019}
  {\path{doi:10.1016/j.physletb.2012.05.019}}.

\bibitem{Elfassi:2022}
L.~El~{F}assi, {Chasing QCD Signatures in Nuclei Using Color Coherence
  Phenomena}, Physics 4~(3) (2022) 970--980.
\newblock \href {https://doi.org/10.3390/physics4030064}
  {\path{doi:10.3390/physics4030064}}.

\bibitem{HallC:2020ijh}
D.~Bhetuwal, et~al., {Ruling out Color Transparency in Quasielastic
  $^{12}$C(e,e'p) up to $Q^2$ of 14.2 (GeV/c)$^2$}, Phys. Rev. Lett. 126~(8)
  (2021) 082301.
\newblock \href {http://arxiv.org/abs/2011.00703} {\path{arXiv:2011.00703}},
  \href {https://doi.org/10.1103/PhysRevLett.126.082301}
  {\path{doi:10.1103/PhysRevLett.126.082301}}.

\bibitem{Caplow-Munro:2021xwi}
O.~Caplow-Munro, G.~A. Miller, {Color transparency and the proton form factor:
  Evidence for the Feynman mechanism}, Phys. Rev. C 104~(1) (2021) L012201.
\newblock \href {http://arxiv.org/abs/2104.11168} {\path{arXiv:2104.11168}},
  \href {https://doi.org/10.1103/PhysRevC.104.L012201}
  {\path{doi:10.1103/PhysRevC.104.L012201}}.

\bibitem{Egiian:1994ey}
K.~Egiian, L.~Frankfurt, W.~R. Greenberg, G.~A. Miller, M.~Sargsian,
  M.~Strikman, {Searching for color coherent effects at intermediate Q**2 via
  double scattering processes}, Nucl. Phys. A 580 (1994) 365--382.
\newblock \href {http://arxiv.org/abs/nucl-th/9401002}
  {\path{arXiv:nucl-th/9401002}}, \href
  {https://doi.org/10.1016/0375-9474(94)90903-2}
  {\path{doi:10.1016/0375-9474(94)90903-2}}.

\bibitem{Frankfurt:1994kt}
L.~L. Frankfurt, W.~R. Greenberg, G.~A. Miller, M.~M. Sargsian, M.~I. Strikman,
  {Color transparency effects in electron deuteron interactions at intermediate
  Q**2}, Z. Phys. A 352 (1995) 97--113.
\newblock \href {http://arxiv.org/abs/nucl-th/9501009}
  {\path{arXiv:nucl-th/9501009}}, \href {https://doi.org/10.1007/BF01292764}
  {\path{doi:10.1007/BF01292764}}.

\bibitem{Frankfurt:1994kk}
L.~L. Frankfurt, W.~R. Greenberg, G.~A. Miller, M.~M. Sargsian, M.~I. Strikman,
  {Color transparency and the vanishing deuterium shadow}, Phys. Lett. B 369
  (1996) 201--206.
\newblock \href {http://arxiv.org/abs/nucl-th/9412033}
  {\path{arXiv:nucl-th/9412033}}, \href
  {https://doi.org/10.1016/0370-2693(95)01558-2}
  {\path{doi:10.1016/0370-2693(95)01558-2}}.

\bibitem{GrossWilczek73}
D.~J. Gross, F.~Wilczek, Ultraviolet behavior of non-abelian gauge theories,
  Phys. Rev. Lett. 30 (1973) 1343--1346.
\newblock \href {https://doi.org/10.1103/PhysRevLett.30.1343}
  {\path{doi:10.1103/PhysRevLett.30.1343}}.

\bibitem{Dokshitzer:2003bt}
Y.~L. Dokshitzer, {QCD phenomenology}, 2003, pp. 1--33.
\newblock \href {http://arxiv.org/abs/hep-ph/0306287}
  {\path{arXiv:hep-ph/0306287}}.

\bibitem{Andersson:1979ue}
B.~Andersson, G.~Gustafson, C.~Peterson, {Quark Jet Fragmentation}, Phys.
  Scripta 19 (1979) 184--190.
\newblock \href {https://doi.org/10.1088/0031-8949/19/2/015}
  {\path{doi:10.1088/0031-8949/19/2/015}}.

\bibitem{Andersson:1983ia}
B.~Andersson, G.~Gustafson, G.~Ingelman, T.~Sjostrand, {Parton Fragmentation
  and String Dynamics}, Phys. Rept. 97 (1983) 31--145.
\newblock \href {https://doi.org/10.1016/0370-1573(83)90080-7}
  {\path{doi:10.1016/0370-1573(83)90080-7}}.

\bibitem{Osborne:1978ai}
L.~S. Osborne, C.~Bolon, R.~L. Lanza, D.~Luckey, D.~G. Roth, J.~F. Martin,
  G.~J. Feldman, M.~E.~B. Franklin, G.~Hanson, M.~L. Perl, {Electroproduction
  of Hadrons From Nuclei}, Phys. Rev. Lett. 40 (1978) 1624.
\newblock \href {https://doi.org/10.1103/PhysRevLett.40.1624}
  {\path{doi:10.1103/PhysRevLett.40.1624}}.

\bibitem{EuropeanMuon:1991jmx}
J.~Ashman, et~al., {Comparison of forward hadrons produced in muon interactions
  on nuclear targets and deuterium}, Z. Phys. C 52 (1991) 1--12.
\newblock \href {https://doi.org/10.1007/BF01412322}
  {\path{doi:10.1007/BF01412322}}.

\bibitem{EuropeanMuon:1984inj}
A.~Arvidson, et~al., {Hadron production in 200-GeV $\mu$ - copper and $\mu$ -
  carbon deep inelastic interactions}, Nucl. Phys. B 246 (1984) 381--407.
\newblock \href {https://doi.org/10.1016/0550-3213(84)90045-2}
  {\path{doi:10.1016/0550-3213(84)90045-2}}.

\bibitem{Artru:1974hr}
X.~Artru, G.~Mennessier, {String model and multiproduction}, Nucl. Phys. B 70
  (1974) 93--115.
\newblock \href {https://doi.org/10.1016/0550-3213(74)90360-5}
  {\path{doi:10.1016/0550-3213(74)90360-5}}.

\bibitem{Shuryak:1978ij}
E.~V. Shuryak, {Quark-Gluon Plasma and Hadronic Production of Leptons, Photons
  and Psions}, Phys. Lett. B 78 (1978) 150.
\newblock \href {https://doi.org/10.1016/0370-2693(78)90370-2}
  {\path{doi:10.1016/0370-2693(78)90370-2}}.

\bibitem{Wang:2003aw}
X.-N. Wang, {Why the observed jet quenching at RHIC is due to parton energy
  loss}, Phys. Lett. B 579 (2004) 299--308.
\newblock \href {https://doi.org/10.1016/j.physletb.2003.11.011}
  {\path{doi:10.1016/j.physletb.2003.11.011}}.

\bibitem{Kopeliovich:2003py}
B.~Z. Kopeliovich, J.~Nemchik, E.~Predazzi, A.~Hayashigaki, {Nuclear
  hadronization: Within or without?}, Nucl. Phys. A 740 (2004) 211--245.
\newblock \href {https://doi.org/10.1016/j.nuclphysa.2004.04.110}
  {\path{doi:10.1016/j.nuclphysa.2004.04.110}}.

\bibitem{BROOKS2021136171}
{W.~K.~Brooks and J.~A.~López}, {Estimating the color lifetime of energetic
  quarks}, Phys. Lett. B 816 (2021) 136171.
\newblock \href {https://doi.org/10.1016/j.physletb.2021.136171}
  {\path{doi:10.1016/j.physletb.2021.136171}}.

\bibitem{Brodsky:1987xw}
S.~J. Brodsky, G.~F. de~Teramond, {Spin Correlations, QCD Color Transparency
  and Heavy Quark Thresholds in Proton Proton Scattering}, Phys. Rev. Lett. 60
  (1988) 1924.
\newblock \href {https://doi.org/10.1103/PhysRevLett.60.1924}
  {\path{doi:10.1103/PhysRevLett.60.1924}}.

\bibitem{Sargsian:2003}
M.~Sargsian, et~al., Hadrons in the nuclear medium, J. Phys. G: Nucl. Part.
  Phys. 29~(3) (2003) R1--R45.
\newblock \href {https://doi.org/10.1088/0954-3899/29/3/201}
  {\path{doi:10.1088/0954-3899/29/3/201}}.

\bibitem{Frankfurt:1993es}
L.~Frankfurt, G.~A. Miller, M.~Strikman, {Precocious dominance of point - like
  configurations in hadronic form-factors}, Nucl. Phys. A 555 (1993) 752--764.
\newblock \href {https://doi.org/10.1016/0375-9474(93)90504-Q}
  {\path{doi:10.1016/0375-9474(93)90504-Q}}.

\bibitem{ARRINGTON:2012}
J.~Arrington, D.~Higinbotham, G.~Rosner, M.~Sargsian, Hard probes of
  short-range nucleon–nucleon correlations, Prog. Part. Nucl. Phys. 67~(4)
  (2012) 898--938.
\newblock \href {https://doi.org/https://doi.org/10.1016/j.ppnp.2012.04.002}
  {\path{doi:https://doi.org/10.1016/j.ppnp.2012.04.002}}.

\bibitem{Arrington:2022sov}
J.~Arrington, N.~Fomin, A.~Schmidt, {Progress in understanding short-range
  structure in nuclei: an experimental perspective}, Ann. Rev. Nucl. Part. Sci.
  (2022) 307\href {http://arxiv.org/abs/2203.02608} {\path{arXiv:2203.02608}}.

\bibitem{Arrington:2003qt}
J.~Arrington, {Do ordinary nuclei contain exotic states of matter?}, Acta Phys.
  Hung. A 21 (2004) 295.
\newblock \href {http://arxiv.org/abs/hep-ph/0304213}
  {\path{arXiv:hep-ph/0304213}}, \href
  {https://doi.org/10.1556/APH.21.2004.2-4.30}
  {\path{doi:10.1556/APH.21.2004.2-4.30}}.

\bibitem{Mulders:1983au}
P.~J. Mulders, A.~W. Thomas, {The 'Six Quark' Component in the Deuteron From a
  Comparison of Electron and Neutrino / Anti-neutrinos Structure Functions},
  Phys. Rev. Lett. 52 (1984) 1199.
\newblock \href {https://doi.org/10.1103/PhysRevLett.52.1199}
  {\path{doi:10.1103/PhysRevLett.52.1199}}.

\bibitem{Hen:2016}
O.~Hen, G.~Miller, E.~Piasetzky, L.~Weinstein, {Nucleon-Nucleon Correlations,
  Short-lived Excitations, and the Quarks Within}, Rev. Mod. Phys. 89~(4)
  (2017) 045002.
\newblock \href {https://doi.org/10.1103/RevModPhys.89.045002}
  {\path{doi:10.1103/RevModPhys.89.045002}}.

\bibitem{Niculescu:2000tk}
I.~Niculescu, et~al., {Experimental verification of quark hadron duality},
  Phys. Rev. Lett. 85 (2000) 1186--1189.
\newblock \href {https://doi.org/10.1103/PhysRevLett.85.1186}
  {\path{doi:10.1103/PhysRevLett.85.1186}}.

\bibitem{Niculescu:2015wka}
I.~Niculescu, et~al., {Direct observation of quark-hadron duality in the free
  neutron $F_2$ structure function}, Phys. Rev. C 91~(5) (2015) 055206.
\newblock \href {http://arxiv.org/abs/1501.02203} {\path{arXiv:1501.02203}},
  \href {https://doi.org/10.1103/PhysRevC.91.055206}
  {\path{doi:10.1103/PhysRevC.91.055206}}.

\bibitem{Arrington:2003nt}
J.~Arrington, R.~Ent, C.~E. Keppel, J.~Mammei, I.~Niculescu, {Low Q scaling,
  duality, and the EMC effect}, Phys. Rev. C 73 (2006) 035205.
\newblock \href {http://arxiv.org/abs/nucl-ex/0307012}
  {\path{arXiv:nucl-ex/0307012}}, \href
  {https://doi.org/10.1103/PhysRevC.73.035205}
  {\path{doi:10.1103/PhysRevC.73.035205}}.

\bibitem{Boeglin:2015}
W.~Boeglin, M.~Sargsian, {Modern Studies of the Deuteron: from the Lab Frame to
  the Light Front}, Int. J. Mod. Phys. E 24~(03) (2015) 1530003.
\newblock \href {http://arxiv.org/abs/1501.05377} {\path{arXiv:1501.05377}},
  \href {https://doi.org/10.1142/S0218301315300039}
  {\path{doi:10.1142/S0218301315300039}}.

\bibitem{Boeglin_2011}
W.~U. Boeglin, et~al., Probing the high momentum component of the deuteron at
  high ${Q}^{2}$, Phys. Rev. Lett. 107 (2011) 262501.
\newblock \href {https://doi.org/10.1103/PhysRevLett.107.262501}
  {\path{doi:10.1103/PhysRevLett.107.262501}}.

\bibitem{Sargsian_2010}
M.~M. Sargsian, Large ${Q}^{2}$ electrodisintegration of the deuteron in the
  virtual nucleon approximation, Phys. Rev. C 82 (2010) 014612.
\newblock \href {https://doi.org/10.1103/PhysRevC.82.014612}
  {\path{doi:10.1103/PhysRevC.82.014612}}.

\bibitem{Laget_2005}
J.~Laget, The electro-disintegration of few body systems revisited, Physics
  Letters B 609~(1) (2005) 49--56.
\newblock \href
  {https://doi.org/https://doi.org/10.1016/j.physletb.2005.01.046}
  {\path{doi:https://doi.org/10.1016/j.physletb.2005.01.046}}.

\bibitem{Orden_2014}
W.~P. Ford, S.~Jeschonnek, J.~W. Van~Orden, Momentum distributions for
  $^{2}\mathrm{H}(e,{e}^{\ensuremath{'}}p)$, Phys. Rev. C 90 (2014) 064006.
\newblock \href {https://doi.org/10.1103/PhysRevC.90.064006}
  {\path{doi:10.1103/PhysRevC.90.064006}}.

\bibitem{Vera_2021}
F.~Vera, \href{https://arxiv.org/abs/2108.11502}{Probing the structure of
  deuteron at very short distances} (2021).
\newblock \href {https://doi.org/10.48550/ARXIV.2108.11502}
  {\path{doi:10.48550/ARXIV.2108.11502}}.
\newline\urlprefix\url{https://arxiv.org/abs/2108.11502}

\bibitem{Sargsian_2022}
M.~M. Sargsian, F.~Vera, {New Structure in the Deuteron}, Phys. Rev. Lett.
  130~(11) (2023) 112502.
\newblock \href {http://arxiv.org/abs/2208.00501} {\path{arXiv:2208.00501}},
  \href {https://doi.org/10.1103/PhysRevLett.130.112502}
  {\path{doi:10.1103/PhysRevLett.130.112502}}.

\bibitem{Yero2023}
C.~Yero, Deuteron disintegration at large missing momenta (January 2023).

\bibitem{CiofidegliAtti:2015lcu}
C.~Ciofi~degli Atti, {In-medium short-range dynamics of nucleons: Recent
  theoretical and experimental advances}, Phys. Rept. 590 (2015) 1--85.
\newblock \href {https://doi.org/10.1016/j.physrep.2015.06.002}
  {\path{doi:10.1016/j.physrep.2015.06.002}}.

\bibitem{Heiselberg:2000dn}
H.~Heiselberg, V.~Pandharipande, {Recent progress in neutron star theory}, Ann.
  Rev. Nucl. Part. Sci. 50 (2000) 481--524.
\newblock \href {http://arxiv.org/abs/astro-ph/0003276}
  {\path{arXiv:astro-ph/0003276}}, \href
  {https://doi.org/10.1146/annurev.nucl.50.1.481}
  {\path{doi:10.1146/annurev.nucl.50.1.481}}.

\bibitem{Sargsian:2005ASF}
M.~Sargsian, T.~Abrahamyan, M.~Strikman, L.~Frankfurt, Exclusive
  electrodisintegration of $^{3}\mathrm{He}$ at high ${Q}^{2}$. i. generalized
  eikonal approximation, Phys. Rev. C 71 (2005) 044614.
\newblock \href {https://doi.org/10.1103/PhysRevC.71.044614}
  {\path{doi:10.1103/PhysRevC.71.044614}}.

\bibitem{Fomin:2017}
N.~Fomin, D.~Higinbotham, M.~Sargsian, P.~Solvignon, New results on short-range
  correlations in nuclei, Ann. Rev. Nucl. Part. Sci. 67~(1) (2017) 129--159.
\newblock \href {https://doi.org/10.1146/annurev-nucl-102115-044939}
  {\path{doi:10.1146/annurev-nucl-102115-044939}}.

\bibitem{Day:1979bx}
D.~Day, J.~S. Mccarthy, I.~Sick, R.~G. Arnold, B.~T. Chertok, S.~Rock, Z.~M.
  Szalata, F.~Martin, B.~A. Mecking, G.~Tamas, {INCLUSIVE ELECTRON SCATTERING
  FROM HE-3}, Phys. Rev. Lett. 43 (1979) 1143.
\newblock \href {https://doi.org/10.1103/PhysRevLett.43.1143}
  {\path{doi:10.1103/PhysRevLett.43.1143}}.

\bibitem{Rock:1981aa}
S.~Rock, R.~G. Arnold, B.~T. Chertok, Z.~M. Szalata, D.~Day, J.~S. McCarthy,
  F.~Martin, B.~A. Mecking, I.~Sick, G.~Tamas, {Inelastic Electron Scattering
  From $^{3}$He and $^{4}$He in the Threshold Region at High Momentum
  Transfer}, Phys. Rev. C 26 (1982) 1592.
\newblock \href {https://doi.org/10.1103/PhysRevC.26.1592}
  {\path{doi:10.1103/PhysRevC.26.1592}}.

\bibitem{Geesaman:1995}
D.~F. Geesaman, K.~Saito, A.~W. Thomas, {The nuclear {EMC} effect}, Ann. Rev.
  Nucl. Part. Sci. 45 (1995) 337--390.
\newblock \href {https://doi.org/10.1146/annurev.ns.45.120195.002005}
  {\path{doi:10.1146/annurev.ns.45.120195.002005}}.

\bibitem{Norton:2003cb}
P.~R. Norton, {The EMC effect}, Rept. Prog. Phys. 66 (2003) 1253--1297.
\newblock \href {https://doi.org/10.1088/0034-4885/66/8/201}
  {\path{doi:10.1088/0034-4885/66/8/201}}.

\bibitem{Malace:2014uea}
S.~Malace, D.~Gaskell, D.~W. Higinbotham, I.~Cloet, {The Challenge of the EMC
  Effect: existing data and future directions}, Int. J. Mod. Phys. E 23~(08)
  (2014) 1430013.
\newblock \href {http://arxiv.org/abs/1405.1270} {\path{arXiv:1405.1270}},
  \href {https://doi.org/10.1142/S0218301314300136}
  {\path{doi:10.1142/S0218301314300136}}.

\bibitem{CLAS:2011qvj}
N.~Baillie, et~al., {Measurement of the neutron F2 structure function via
  spectator tagging with CLAS}, Phys. Rev. Lett. 108 (2012) 142001, [Erratum:
  Phys.Rev.Lett. 108, 199902 (2012)].
\newblock \href {http://arxiv.org/abs/1110.2770} {\path{arXiv:1110.2770}},
  \href {https://doi.org/10.1103/PhysRevLett.108.142001}
  {\path{doi:10.1103/PhysRevLett.108.142001}}.

\bibitem{Tkachenko:2014}
S.~Tkachenko, et~al., {Measurement of the structure function of the nearly free
  neutron using spectator tagging in inelastic $^2$H(e, e'p)X scattering with
  CLAS}, Phys. Rev. C 89 (2014) 045206, [Addendum: Phys.Rev.C 90, 059901
  (2014)].
\newblock \href {http://arxiv.org/abs/1402.2477} {\path{arXiv:1402.2477}},
  \href {https://doi.org/10.1103/PhysRevC.89.045206}
  {\path{doi:10.1103/PhysRevC.89.045206}}.

\bibitem{Griffioen:2015hxa}
K.~A. Griffioen, et~al., {Measurement of the EMC Effect in the Deuteron}, Phys.
  Rev. C 92~(1) (2015) 015211.
\newblock \href {http://arxiv.org/abs/1506.00871} {\path{arXiv:1506.00871}},
  \href {https://doi.org/10.1103/PhysRevC.92.015211}
  {\path{doi:10.1103/PhysRevC.92.015211}}.

\bibitem{CLAS:2005ekq}
A.~V. Klimenko, et~al., {Electron scattering from high-momentum neutrons in
  deuterium}, Phys. Rev. C 73 (2006) 035212.
\newblock \href {http://arxiv.org/abs/nucl-ex/0510032}
  {\path{arXiv:nucl-ex/0510032}}, \href
  {https://doi.org/10.1103/PhysRevC.73.035212}
  {\path{doi:10.1103/PhysRevC.73.035212}}.

\bibitem{Lovato:2016gkq}
A.~Lovato, S.~Gandolfi, J.~Carlson, S.~C. Pieper, R.~Schiavilla,
  {Electromagnetic response of $^{12}$C: A first-principles calculation}, Phys.
  Rev. Lett. 117~(8) (2016) 082501.
\newblock \href {http://arxiv.org/abs/1605.00248} {\path{arXiv:1605.00248}},
  \href {https://doi.org/10.1103/PhysRevLett.117.082501}
  {\path{doi:10.1103/PhysRevLett.117.082501}}.

\bibitem{Cloet:2015tha}
I.~C. Clo\"et, W.~Bentz, A.~W. Thomas, {Relativistic and Nuclear Medium Effects
  on the Coulomb Sum Rule}, Phys. Rev. Lett. 116~(3) (2016) 032701.
\newblock \href {http://arxiv.org/abs/1506.05875} {\path{arXiv:1506.05875}},
  \href {https://doi.org/10.1103/PhysRevLett.116.032701}
  {\path{doi:10.1103/PhysRevLett.116.032701}}.

\bibitem{Rvachev_2005}
M.~M. Rvachev, et~al., Quasielastic
  $^{3}\mathrm{He}(e,{e}^{\ensuremath{'}}p)^{2}\mathrm{H}$ reaction at
  ${Q}^{2}=1.5\text{ }\text{ }{\mathrm{gev}}^{2}$ for recoil momenta up to
  $1\text{ }\text{ }\mathrm{GeV}/c$, Phys. Rev. Lett. 94 (2005) 192302.
\newblock \href {https://doi.org/10.1103/PhysRevLett.94.192302}
  {\path{doi:10.1103/PhysRevLett.94.192302}}.

\bibitem{Hu:2006fy}
B.~Hu, et~al., {Polarization transfer in the H-2(polarized-e, e-prime
  polarized-p) n reaction up to Q**2 = 1.61-(GeV/c)**2}, Phys. Rev. C 73 (2006)
  064004.
\newblock \href {http://arxiv.org/abs/nucl-ex/0601025}
  {\path{arXiv:nucl-ex/0601025}}, \href
  {https://doi.org/10.1103/PhysRevC.73.064004}
  {\path{doi:10.1103/PhysRevC.73.064004}}.

\bibitem{Malace:2010ft}
S.~P. Malace, et~al., {A precise extraction of the induced polarization in the
  4He(e,e'p)3H reaction}, Phys. Rev. Lett. 106 (2011) 052501.
\newblock \href {http://arxiv.org/abs/1011.4483} {\path{arXiv:1011.4483}},
  \href {https://doi.org/10.1103/PhysRevLett.106.052501}
  {\path{doi:10.1103/PhysRevLett.106.052501}}.

\bibitem{Ford:2014lra}
W.~P. Ford, R.~Schiavilla, J.~W. Van~Orden, {The $^3$He$(e,e^\prime p)^2$H and
  $^4$He$(e,e^\prime p)^3$H reactions at high momentum transfer}, Phys. Rev. C
  89~(3) (2014) 034004.
\newblock \href {http://arxiv.org/abs/1401.4399} {\path{arXiv:1401.4399}},
  \href {https://doi.org/10.1103/PhysRevC.89.034004}
  {\path{doi:10.1103/PhysRevC.89.034004}}.

\bibitem{Dupre:2021}
R.~Dupr\'e, et~al., {Measurement of deeply virtual Compton scattering off
  $^{4}\mathrm{He}$ with the CEBAF Large Acceptance Spectrometer at Jefferson
  Lab}, Phys. Rev. C 104~(2) (2021) 025203.
\newblock \href {http://arxiv.org/abs/2102.07419} {\path{arXiv:2102.07419}},
  \href {https://doi.org/10.1103/PhysRevC.104.025203}
  {\path{doi:10.1103/PhysRevC.104.025203}}.

\bibitem{Guzey:2008fe}
V.~Guzey, A.~W. Thomas, K.~Tsushima, {Medium modifications of the bound nucleon
  GPDs and incoherent DVCS on nuclear targets}, Phys. Lett. B 673 (2009) 9--14.
\newblock \href {http://arxiv.org/abs/0806.3288} {\path{arXiv:0806.3288}},
  \href {https://doi.org/10.1016/j.physletb.2009.01.064}
  {\path{doi:10.1016/j.physletb.2009.01.064}}.

\bibitem{Liuti:2005qj}
S.~Liuti, S.~K. Taneja, {Nuclear medium modifications of hadrons from
  generalized parton distributions}, Phys. Rev. C 72 (2005) 034902.
\newblock \href {http://arxiv.org/abs/hep-ph/0504027}
  {\path{arXiv:hep-ph/0504027}}, \href
  {https://doi.org/10.1103/PhysRevC.72.034902}
  {\path{doi:10.1103/PhysRevC.72.034902}}.

\bibitem{Guidal:2013rya}
M.~Guidal, H.~Moutarde, M.~Vanderhaeghen, {Generalized Parton Distributions in
  the valence region from Deeply Virtual Compton Scattering}, Rept. Prog. Phys.
  76 (2013) 066202.
\newblock \href {http://arxiv.org/abs/1303.6600} {\path{arXiv:1303.6600}},
  \href {https://doi.org/10.1088/0034-4885/76/6/066202}
  {\path{doi:10.1088/0034-4885/76/6/066202}}.

\bibitem{Armstrong:2017}
W.~Armstrong, et~al., Spectator-tagged deeply virtual compton scattering on
  light nuclei, arXiv:1708.00835 (2017).
\newblock \href {https://doi.org/10.48550/ARXIV.1708.00835}
  {\path{doi:10.48550/ARXIV.1708.00835}}.

\bibitem{PhysRevC.98.015203}
S.~Fucini, S.~Scopetta, M.~Viviani, Coherent deeply virtual compton scattering
  off $^{4}\mathrm{He}$, Phys. Rev. C 98 (2018) 015203.
\newblock \href {https://doi.org/10.1103/PhysRevC.98.015203}
  {\path{doi:10.1103/PhysRevC.98.015203}}.

\bibitem{Zurita:2021kli}
P.~Zurita, {Medium modified Fragmentation Functions with open source xFitter}
  (1 2021).
\newblock \href {http://arxiv.org/abs/2101.01088} {\path{arXiv:2101.01088}}.

\bibitem{Eskola:2021nhw}
K.~Eskola, P.~Paakkinen, H.~Paukkunen, C.~Salgado, {EPPS21: a global QCD
  analysis of nuclear PDFs}, Eur. Phys. J. C 82~(5) (2022) 413.
\newblock \href {http://arxiv.org/abs/2112.12462} {\path{arXiv:2112.12462}},
  \href {https://doi.org/10.1140/epjc/s10052-022-10359-0}
  {\path{doi:10.1140/epjc/s10052-022-10359-0}}.

\bibitem{PhysRevD.100.096015}
M.~Walt, I.~Helenius, W.~Vogelsang, Open-source qcd analysis of nuclear parton
  distribution functions at nlo and nnlo, Phys. Rev. D 100 (2019) 096015.
\newblock \href {https://doi.org/10.1103/PhysRevD.100.096015}
  {\path{doi:10.1103/PhysRevD.100.096015}}.

\bibitem{pEMC_proposal}
W.~Brooks, S.~Kuhn, et~al., The {EMC} {E}ffect in {S}pin {S}tructure
  {F}unctions,
  \href{https://www.jlab.org/exp_prog/proposals/14/PR12-14-001.pdf}{CLAS12
  E12-14-00 Experiment (Run Group G)} (2014).

\bibitem{pEMC_update}
W.~Brooks, S.~Kuhn, et~al., The {EMC} {E}ffect in {S}pin {S}tructure
  {F}unctions,
  \href{https://www.jlab.org/exp_prog/proposals/20/Jeopardy/Run\%20Group\%20G\_Update.pdf}{CLAS12
  Run Group G Jeopardy Update} (2020).

\bibitem{PhysRevD.70.116003}
S.~J. Brodsky, I.~Schmidt, J.-J. Yang, Nuclear antishadowing in neutrino deep
  inelastic scattering, Phys. Rev. D 70 (2004) 116003.
\newblock \href {https://doi.org/10.1103/PhysRevD.70.116003}
  {\path{doi:10.1103/PhysRevD.70.116003}}.

\bibitem{PhysRevC.61.014002}
V.~Guzey, M.~Strikman, Nuclear effects in ${g}_{1A}{(x,Q}^{2})$ at small x in
  deep inelastic scattering on ${}^{7}\mathrm{Li}$ and ${}^{3}\mathrm{He}$,
  Phys. Rev. C 61 (1999) 014002.
\newblock \href {https://doi.org/10.1103/PhysRevC.61.014002}
  {\path{doi:10.1103/PhysRevC.61.014002}}.

\bibitem{PhysRevC.95.055208}
L.~Frankfurt, V.~Guzey, M.~Strikman, Dynamical model of antishadowing of the
  nuclear gluon distribution, Phys. Rev. C 95 (2017) 055208.
\newblock \href {https://doi.org/10.1103/PhysRevC.95.055208}
  {\path{doi:10.1103/PhysRevC.95.055208}}.

\bibitem{Cloet:2006}
I.~Clo\"et, W.~Bentz, A.~Thomas, {EMC} and polarized {EMC} effects in nuclei,
  Phy. Lett. B 642~(3) (2006) 210--217.
\newblock \href
  {https://doi.org/https://doi.org/10.1016/j.physletb.2006.08.076}
  {\path{doi:https://doi.org/10.1016/j.physletb.2006.08.076}}.

\bibitem{Miller:2005}
J.~Smith, G.~Miller, Polarized quark distributions in nuclear matter, Phys.
  Rev. C 72 (2005) 022203.
\newblock \href {https://doi.org/10.1103/PhysRevC.72.022203}
  {\path{doi:10.1103/PhysRevC.72.022203}}.

\bibitem{Fanchiotti:2014}
H.~Fanchiotti, C.~A. Garc\'\i{}a-Canal, T.~Tarutina, V.~Vento, {Medium Effects
  in DIS from Polarized Nuclear Targets}, Eur. Phys. J. A 50 (2014) 116.
\newblock \href {http://arxiv.org/abs/1404.3047} {\path{arXiv:1404.3047}},
  \href {https://doi.org/10.1140/epja/i2014-14116-8}
  {\path{doi:10.1140/epja/i2014-14116-8}}.

\bibitem{cloet:2005}
I.~Clo\"et, W.~Bentz, A.~Thomas, Spin-dependent structure functions in nuclear
  matter and the polarized emc effect, Phys. Rev. Lett. 95 (2005) 052302.
\newblock \href {https://doi.org/10.1103/PhysRevLett.95.052302}
  {\path{doi:10.1103/PhysRevLett.95.052302}}.

\bibitem{Brodsky:1988xz}
S.~J. Brodsky, A.~H. Mueller, {Using Nuclei to Probe Hadronization in QCD},
  Phys. Lett. B 206 (1988) 685--690.
\newblock \href {https://doi.org/10.1016/0370-2693(88)90719-8}
  {\path{doi:10.1016/0370-2693(88)90719-8}}.

\bibitem{Brodsky:1994kf}
S.~J. Brodsky, L.~Frankfurt, J.~F. Gunion, A.~H. Mueller, M.~Strikman,
  {Diffractive leptoproduction of vector mesons in {QCD}}, Phys. Rev. D 50
  (1994) 3134--3144.
\newblock \href {http://arxiv.org/abs/hep-ph/9402283}
  {\path{arXiv:hep-ph/9402283}}, \href
  {https://doi.org/10.1103/PhysRevD.50.3134}
  {\path{doi:10.1103/PhysRevD.50.3134}}.

\bibitem{Dutta:2012ii}
D.~Dutta, K.~Hafidi, M.~Strikman, {Color Transparency: past, present and
  future}, Prog. Part. Nucl. Phys. 69 (2013) 1--27.
\newblock \href {http://arxiv.org/abs/1211.2826} {\path{arXiv:1211.2826}},
  \href {https://doi.org/10.1016/j.ppnp.2012.11.001}
  {\path{doi:10.1016/j.ppnp.2012.11.001}}.

\bibitem{Brodsky:2022bum}
S.~J. Brodsky, G.~F. de~Teramond, {Onset of Color Transparency in Holographic
  Light-Front {QCD}}, MDPI Physics 4~(2) (2022) 633--646.
\newblock \href {http://arxiv.org/abs/2202.13283} {\path{arXiv:2202.13283}},
  \href {https://doi.org/10.3390/physics4020042}
  {\path{doi:10.3390/physics4020042}}.

\bibitem{Clasie:2007}
B.~Clasie, et~al., {Measurement of Nuclear Transparency for the
  $A(e,{e}'{\pi}^{+})$ Reaction}, Phys. Rev. Lett. 99 (2007) 242502.
\newblock \href {http://arxiv.org/abs/0707.1481} {\path{arXiv:0707.1481}},
  \href {https://doi.org/10.1103/PhysRevLett.99.242502}
  {\path{doi:10.1103/PhysRevLett.99.242502}}.

\bibitem{ElFassi:2012}
L.~El~Fassi, et~al., {Evidence for the onset of color transparency in $\rho^0$
  electroproduction off nuclei}, Phys. Lett. B 712 (2012) 326--330.
\newblock \href {http://arxiv.org/abs/1201.2735} {\path{arXiv:1201.2735}},
  \href {https://doi.org/10.1016/j.physletb.2012.05.019}
  {\path{doi:10.1016/j.physletb.2012.05.019}}.

\bibitem{Frankfurt:2008pz}
L.~Frankfurt, G.~A. Miller, M.~Strikman, {Color Transparency in Semi-Inclusive
  Electroproduction of rho Mesons}, Phys. Rev. C 78 (2008) 015208.
\newblock \href {http://arxiv.org/abs/0803.4012} {\path{arXiv:0803.4012}},
  \href {https://doi.org/10.1103/PhysRevC.78.015208}
  {\path{doi:10.1103/PhysRevC.78.015208}}.

\bibitem{Gallmeister:2010wn}
K.~Gallmeister, M.~Kaskulov, U.~Mosel, {Color transparency in hadronic
  attenuation of $\rho^0$ mesons}, Phys. Rev. C 83 (2011) 015201.
\newblock \href {http://arxiv.org/abs/1007.1141} {\path{arXiv:1007.1141}},
  \href {https://doi.org/10.1103/PhysRevC.83.015201}
  {\path{doi:10.1103/PhysRevC.83.015201}}.

\bibitem{Cosyn:2013qe}
W.~Cosyn, J.~Ryckebusch, {Nuclear \ensuremath{\rho} meson transparency in a
  relativistic Glauber model}, Phys. Rev. C 87~(6) (2013) 064608.
\newblock \href {http://arxiv.org/abs/1301.1904} {\path{arXiv:1301.1904}},
  \href {https://doi.org/10.1103/PhysRevC.87.064608}
  {\path{doi:10.1103/PhysRevC.87.064608}}.

\bibitem{Carroll:1988}
A.~S. Carroll, et~al., {Nuclear Transparency to Large Angle $p p$ Elastic
  Scattering}, Phys. Rev. Lett. 61 (1988) 1698--1701.
\newblock \href {https://doi.org/10.1103/PhysRevLett.61.1698}
  {\path{doi:10.1103/PhysRevLett.61.1698}}.

\bibitem{Mardor:1998zf}
I.~Mardor, et~al., {Nuclear transparency in large momentum transfer
  quasielastic scattering}, Phys. Rev. Lett. 81 (1998) 5085--5088.
\newblock \href {https://doi.org/10.1103/PhysRevLett.81.5085}
  {\path{doi:10.1103/PhysRevLett.81.5085}}.

\bibitem{Leksanov:2001}
A.~Leksanov, et~al., {Energy dependence of nuclear transparency in C$(p, 2p)$
  scattering}, Phys. Rev. Lett. 87 (2001) 212301.
\newblock \href {http://arxiv.org/abs/hep-ex/0104039}
  {\path{arXiv:hep-ex/0104039}}, \href
  {https://doi.org/10.1103/PhysRevLett.87.212301}
  {\path{doi:10.1103/PhysRevLett.87.212301}}.

\bibitem{Aclander:2004zm}
J.~Aclander, et~al., {Nuclear transparency in
  ${90}_{\mathrm{c.m.}}^{\ifmmode^\circ\else\textdegree\fi{}}$ quasielastic
  $A(p,2p)$ reactions}, Phys. Rev. C 70 (2004) 015208.
\newblock \href {http://arxiv.org/abs/nucl-ex/0405025}
  {\path{arXiv:nucl-ex/0405025}}, \href
  {https://doi.org/10.1103/PhysRevC.70.015208}
  {\path{doi:10.1103/PhysRevC.70.015208}}.

\bibitem{Makins:1994mm}
N.~Makins, et~al., {Momentum transfer dependence of nuclear transparency from
  the quasielastic $^{12}\mathrm{C}$(e,e'p) reaction}, Phys. Rev. Lett. 72
  (1994) 1986--1989.
\newblock \href {https://doi.org/10.1103/PhysRevLett.72.1986}
  {\path{doi:10.1103/PhysRevLett.72.1986}}.

\bibitem{ONeill:1994znv}
T.~G. O'Neill, et~al., {$A$-dependence of nuclear transparency in quasielastic
  $A(e, e' p)$ at high $Q^2$}, Phys. Lett. B 351 (1995) 87--92.
\newblock \href {http://arxiv.org/abs/hep-ph/9408260}
  {\path{arXiv:hep-ph/9408260}}, \href
  {https://doi.org/10.1016/0370-2693(95)00362-O}
  {\path{doi:10.1016/0370-2693(95)00362-O}}.

\bibitem{Abbott:1997}
D.~Abbott, et~al., {Quasifree $(e,e'p)$ reactions and proton propagation in
  nuclei}, Phys. Rev. Lett. 80 (1998) 5072--5076.
\newblock \href {https://doi.org/10.1103/PhysRevLett.80.5072}
  {\path{doi:10.1103/PhysRevLett.80.5072}}.

\bibitem{Garrow:2001}
K.~Garrow, et~al., {Nuclear transparency from quasielastic
  ${A(e,e}^{\ensuremath{'}}p)$ reactions up to
  ${Q}^{2}=8.1(\mathrm{GeV}{/c)}^{2}$}, Phys. Rev. C 66 (2002) 044613.
\newblock \href {http://arxiv.org/abs/hep-ex/0109027}
  {\path{arXiv:hep-ex/0109027}}, \href
  {https://doi.org/10.1103/PhysRevC.66.044613}
  {\path{doi:10.1103/PhysRevC.66.044613}}.

\bibitem{Dutta:2021}
D.~Bhetuwal, J.~Matter, H.~Szumila-Vance, M.~L. Kabir, D.~Dutta, R.~Ent,
  et~al., Ruling out color transparency in quasielastic
  $^{12}\mathrm{C}(\mathrm{e},{e}^{\ensuremath{'}}\mathrm{p})$ up to ${Q}^{2}$
  of $14.2\text{ }\text{ }(\mathrm{GeV}/\mathrm{c}{)}^{2}$, Phys. Rev. Lett.
  126 (2021) 082301.
\newblock \href {https://doi.org/10.1103/PhysRevLett.126.082301}
  {\path{doi:10.1103/PhysRevLett.126.082301}}.

\bibitem{physics4040092}
S.~Li, C.~Yero, J.~R. West, C.~Bennett, W.~Cosyn, D.~Higinbotham, M.~Sargsian,
  H.~Szumila-Vance, Searching for an enhanced signal of the onset of color
  transparency in baryons with d(e,e'p)n scattering, Physics 4~(4) (2022)
  1426--1439.
\newblock \href {https://doi.org/10.3390/physics4040092}
  {\path{doi:10.3390/physics4040092}}.

\bibitem{Field:1976ve}
R.~D. Field, R.~P. Feynman, {Quark Elastic Scattering as a Source of High
  Transverse Momentum Mesons}, Phys. Rev. D 15 (1977) 2590--2616.
\newblock \href {https://doi.org/10.1103/PhysRevD.15.2590}
  {\path{doi:10.1103/PhysRevD.15.2590}}.

\bibitem{aubert1983ratio}
J.-J. Aubert, G.~Bassompierre, K.~Becks, C.~Best, E.~B{\"o}hm, X.~de~Bouard,
  F.~Brasse, C.~Broll, S.~Brown, J.~Carr, et~al., The ratio of the nucleon
  structure functions {$F_2^N$} for iron and deuterium, Physics Letters B
  123~(3-4) (1983) 275--278.

\bibitem{adcox2005formation}
K.~Adcox, S.~Adler, S.~Afanasiev, C.~Aidala, N.~Ajitanand, Y.~Akiba,
  A.~Al-Jamel, J.~Alexander, R.~Amirikas, K.~Aoki, et~al., Formation of dense
  partonic matter in relativistic nucleus--nucleus collisions at rhic:
  experimental evaluation by the phenix collaboration, Nuclear Physics A
  757~(1-2) (2005) 184--283.

\bibitem{STAR:2005gfr}
J.~Adams, et~al., {Experimental and theoretical challenges in the search for
  the quark gluon plasma: The STAR Collaboration's critical assessment of the
  evidence from RHIC collisions}, Nucl. Phys. A 757 (2005) 102--183.
\newblock \href {http://arxiv.org/abs/nucl-ex/0501009}
  {\path{arXiv:nucl-ex/0501009}}, \href
  {https://doi.org/10.1016/j.nuclphysa.2005.03.085}
  {\path{doi:10.1016/j.nuclphysa.2005.03.085}}.

\bibitem{HERMES:2010}
A.~Airapetian, et~al., Transverse momentum broadening of hadrons produced in
  semi-inclusive deep-inelastic scattering on nuclei, Phys. Lett. B 684 (2010)
  114--118.
\newblock \href {https://doi.org/10.1016/j.physletb.2010.01.020}
  {\path{doi:10.1016/j.physletb.2010.01.020}}.

\bibitem{HERMES:2011}
A.~Airapetian, et~al., {Multidimensional Study of Hadronization in Nuclei},
  Eur. Phys. J. A 47 (2011) 113.
\newblock \href {http://arxiv.org/abs/1107.3496} {\path{arXiv:1107.3496}},
  \href {https://doi.org/10.1140/epja/i2011-11113-5}
  {\path{doi:10.1140/epja/i2011-11113-5}}.

\bibitem{HERMES:2007plz}
A.~Airapetian, et~al., {Hadronization in semi-inclusive deep-inelastic
  scattering on nuclei}, Nucl. Phys. B 780 (2007) 1--27.
\newblock \href {http://arxiv.org/abs/0704.3270} {\path{arXiv:0704.3270}},
  \href {https://doi.org/10.1016/j.nuclphysb.2007.06.004}
  {\path{doi:10.1016/j.nuclphysb.2007.06.004}}.

\bibitem{PhysRevLett.96.162301}
A.~Airapetian, et~al., Double-hadron leptoproduction in the nuclear medium,
  Phys. Rev. Lett. 96 (2006) 162301.
\newblock \href {https://doi.org/10.1103/PhysRevLett.96.162301}
  {\path{doi:10.1103/PhysRevLett.96.162301}}.

\bibitem{Paul:2022}
S.~J. Paul, S.~Mor\'an, M.~Arratia, A.~El~Alaoui, H.~Hakobyan, W.~Brooks,
  et~al., Observation of azimuth-dependent suppression of hadron pairs in
  electron scattering off nuclei, Phys. Rev. Lett. 129 (2022) 182501.
\newblock \href {https://doi.org/10.1103/PhysRevLett.129.182501}
  {\path{doi:10.1103/PhysRevLett.129.182501}}.

\bibitem{HERMES:2001}
A.~Airapetian, et~al., {Hadron formation in deep inelastic positron scattering
  in a nuclear environment}, Eur. Phys.~J. C 20 (2001) 479--486.
\newblock \href {https://doi.org/10.1007/s100520100697}
  {\path{doi:10.1007/s100520100697}}.

\bibitem{Moran:2022}
S.~Mor\'an, R.~Dupr\'e, H.~Hakobyan, M.~Arratia, W.~K. Brooks, A.~B\'orquez,
  A.~El~Alaoui, L.~El~Fassi, K.~Hafidi, R.~Mendez, T.~Mineeva, S.~J. Paul,
  et~al., Measurement of charged-pion production in deep-inelastic scattering
  off nuclei with the {CLAS} detector, Phys. Rev. C 105 (2022) 015201.
\newblock \href {https://doi.org/10.1103/PhysRevC.105.015201}
  {\path{doi:10.1103/PhysRevC.105.015201}}.

\bibitem{HERMES:2003}
A.~Airapetian, et~al., {Quark fragmentation to $\pi^\pm$, $\pi^0$, $K^\pm$, $p$
  and $\bar p$ in the nuclear environment}, Phys.~Lett.~B 577 (2003) 37--46.
\newblock \href {https://doi.org/10.1016/j.physletb.2003.10.026}
  {\path{doi:10.1016/j.physletb.2003.10.026}}.

\bibitem{HERMES:2007}
A.~Airapetian, et~al., {Hadronization in semi-inclusive deep-inelastic
  scattering on nuclei}, Nucl. Phys.~B 780 (2007) 1--27.
\newblock \href {https://doi.org/10.1016/j.nuclphysb.2007.06.004}
  {\path{doi:10.1016/j.nuclphysb.2007.06.004}}.

\bibitem{PhysRevLett.130.142301}
T.~Chetry, L.~El~Fassi, et~al., First measurement of
  $\mathrm{\ensuremath{\Lambda}}$ electroproduction off nuclei in the current
  and target fragmentation regions, Phys. Rev. Lett. 130 (2023) 142301.
\newblock \href {https://doi.org/10.1103/PhysRevLett.130.142301}
  {\path{doi:10.1103/PhysRevLett.130.142301}}.

\bibitem{EIC_whitepaper}
A.~Accardi, et~al., Electron-{I}on {C}ollider: The next {QCD} frontier, The
  European Physical Journal A 52~(9) (2016) 268.
\newblock \href {https://doi.org/10.1140/epja/i2016-16268-9}
  {\path{doi:10.1140/epja/i2016-16268-9}}.

\bibitem{2017489}
V.~Khachatryan, et~al., {Coherent $\ensuremath{J/\psi}$ photoproduction in
  ultra-peripheral PbPb collisions at $\sqrt{s_{NN}}=2.76$ TeV with the CMS
  experiment}, Physics Letters B 772 (2017) 489--511.

\bibitem{20131273}
B.~Abelev, et~al., {Coherent $\ensuremath{J/\psi}$ photoproduction in
  ultra-peripheral Pb--Pb collisions at $\sqrt{s_{NN}}=2.76 TeV$}, Physics
  Letters B 718~(4) (2013) 1273--1283.
\newblock \href
  {https://doi.org/https://doi.org/10.1016/j.physletb.2012.11.059}
  {\path{doi:https://doi.org/10.1016/j.physletb.2012.11.059}}.

\bibitem{2021136280}
S.~Acharya, et~al., {First measurement of the $|t|$-dependence of coherent
  $\ensuremath{J/\psi}$ photonuclear production}, Physics Letters B 817 (2021)
  136280.
\newblock \href
  {https://doi.org/https://doi.org/10.1016/j.physletb.2021.136280}
  {\path{doi:https://doi.org/10.1016/j.physletb.2021.136280}}.

\bibitem{ALICE_jpsi_EPJC}
S.~Acharya, et~al., {Coherent $\ensuremath{J/\psi}$ and $\psi'$ photoproduction
  at midrapidity in ultra-peripheral Pb--Pb collisions at $\sqrt{s_{NN}}=5.02$
  TeV}, The European Physical Journal C 81~(8) (2021) 712.
\newblock \href {https://doi.org/10.1140/epjc/s10052-021-09437-6}
  {\path{doi:10.1140/epjc/s10052-021-09437-6}}.

\bibitem{PhysRevC.105.L032201}
R.~Aaij, et~al., {$\ensuremath{J/\psi}$ photoproduction in Pb-Pb peripheral
  collisions at $\sqrt{{s}_{NN}}=5$ TeV}, Phys. Rev. C 105 (2022) L032201.
\newblock \href {https://doi.org/10.1103/PhysRevC.105.L032201}
  {\path{doi:10.1103/PhysRevC.105.L032201}}.

\bibitem{PhysRevLett.128.122303}
M.~S. Abdallah, et~al., Probing the gluonic structure of the deuteron with
  $\ensuremath{J/\psi}$ photoproduction in $\mathit{d}+\mathrm{Au}$
  ultraperipheral collisions, Phys. Rev. Lett. 128 (2022) 122303.
\newblock \href {https://doi.org/10.1103/PhysRevLett.128.122303}
  {\path{doi:10.1103/PhysRevLett.128.122303}}.

\bibitem{PhysRevLett.123.072001}
A.~Ali, et~al., {First Measurement of Near-Threshold $\ensuremath{J/\psi}$
  Exclusive Photoproduction off the Proton}, Phys. Rev. Lett. 123 (2019)
  072001.
\newblock \href {https://doi.org/10.1103/PhysRevLett.123.072001}
  {\path{doi:10.1103/PhysRevLett.123.072001}}.

\bibitem{PhysRevC.87.024913}
T.~Toll, T.~Ullrich, Exclusive diffractive processes in electron-ion
  collisions, Phys. Rev. C 87 (2013) 024913.
\newblock \href {https://doi.org/10.1103/PhysRevC.87.024913}
  {\path{doi:10.1103/PhysRevC.87.024913}}.

\bibitem{GlueX:2020dvv}
O.~Hen, et~al., {Studying Short-Range Correlations with Real Photon Beams at
  GlueX} (9 2020).
\newblock \href {http://arxiv.org/abs/2009.09617} {\path{arXiv:2009.09617}}.

\bibitem{PhysRevLett.35.483}
U.~Camerini, J.~G. Learned, R.~Prepost, C.~M. Spencer, D.~E. Wiser, W.~W. Ash,
  R.~L. Anderson, D.~M. Ritson, D.~J. Sherden, C.~K. Sinclair, {Photoproduction
  of the $\ensuremath{\psi}$ Particles}, Phys. Rev. Lett. 35 (1975) 483--486.
\newblock \href {https://doi.org/10.1103/PhysRevLett.35.483}
  {\path{doi:10.1103/PhysRevLett.35.483}}.

\bibitem{PhysRevC.89.024305}
R.~B. Wiringa, R.~Schiavilla, S.~C. Pieper, J.~Carlson, {Nucleon and
  nucleon-pair momentum distributions in $A\ensuremath{\le}12$ nuclei}, Phys.
  Rev. C 89 (2014) 024305.
\newblock \href {https://doi.org/10.1103/PhysRevC.89.024305}
  {\path{doi:10.1103/PhysRevC.89.024305}}.

\bibitem{Gan:2020aco}
L.~Gan, B.~Kubis, E.~Passemar, S.~Tulin, {Precision tests of fundamental
  physics with \ensuremath{\eta} and \ensuremath{\eta}' mesons}, Phys. Rept.
  945 (2022) 1--105.
\newblock \href {http://arxiv.org/abs/2007.00664} {\path{arXiv:2007.00664}},
  \href {https://doi.org/10.1016/j.physrep.2021.11.001}
  {\path{doi:10.1016/j.physrep.2021.11.001}}.

\bibitem{Adler:1969gk}
S.~L. Adler, {Axial vector vertex in spinor electrodynamics}, Phys. Rev. 177
  (1969) 2426--2438.
\newblock \href {https://doi.org/10.1103/PhysRev.177.2426}
  {\path{doi:10.1103/PhysRev.177.2426}}.

\bibitem{Bell:1969ts}
J.~S. Bell, R.~Jackiw, {A PCAC puzzle: $\pi^0 \to \gamma \gamma$ in the
  $\sigma$ model}, Nuovo Cim. A 60 (1969) 47--61.
\newblock \href {https://doi.org/10.1007/BF02823296}
  {\path{doi:10.1007/BF02823296}}.

\bibitem{tHooft:1976rip}
G.~'t~Hooft, {Symmetry Breaking Through Bell-Jackiw Anomalies}, Phys. Rev.
  Lett. 37 (1976) 8--11.
\newblock \href {https://doi.org/10.1103/PhysRevLett.37.8}
  {\path{doi:10.1103/PhysRevLett.37.8}}.

\bibitem{Witten:1979vv}
E.~Witten, {Current Algebra Theorems for the U(1) Goldstone Boson}, Nucl. Phys.
  B 156 (1979) 269--283.
\newblock \href {https://doi.org/10.1016/0550-3213(79)90031-2}
  {\path{doi:10.1016/0550-3213(79)90031-2}}.

\bibitem{Gell-Mann:1968hlm}
M.~Gell-Mann, R.~J. Oakes, B.~Renner, {Behavior of current divergences under
  SU(3) x SU(3)}, Phys. Rev. 175 (1968) 2195--2199.
\newblock \href {https://doi.org/10.1103/PhysRev.175.2195}
  {\path{doi:10.1103/PhysRev.175.2195}}.

\bibitem{Bell:1968wta}
J.~S. Bell, D.~G. Sutherland, {Current algebra and eta ---\ensuremath{>} 3 pi},
  Nucl. Phys. B 4 (1968) 315--325.
\newblock \href {https://doi.org/10.1016/0550-3213(68)90316-7}
  {\path{doi:10.1016/0550-3213(68)90316-7}}.

\bibitem{Sutherland:1966zz}
D.~G. Sutherland, {Current algebra and the decay $\eta \to 3\pi$}, Phys. Lett.
  23 (1966) 384--385.
\newblock \href {https://doi.org/10.1016/0031-9163(66)90477-X}
  {\path{doi:10.1016/0031-9163(66)90477-X}}.

\bibitem{Kuzmin:1985mm}
V.~A. Kuzmin, V.~A. Rubakov, M.~E. Shaposhnikov, {On the Anomalous Electroweak
  Baryon Number Nonconservation in the Early Universe}, Phys. Lett. B 155
  (1985) 36.
\newblock \href {https://doi.org/10.1016/0370-2693(85)91028-7}
  {\path{doi:10.1016/0370-2693(85)91028-7}}.

\bibitem{Aoyama:2020ynm}
T.~Aoyama, et~al., {The anomalous magnetic moment of the muon in the Standard
  Model}, Phys. Rept. 887 (2020) 1--166.
\newblock \href {http://arxiv.org/abs/2006.04822} {\path{arXiv:2006.04822}},
  \href {https://doi.org/10.1016/j.physrep.2020.07.006}
  {\path{doi:10.1016/j.physrep.2020.07.006}}.

\bibitem{Hoferichter:2018dmo}
M.~Hoferichter, B.-L. Hoid, B.~Kubis, S.~Leupold, S.~P. Schneider, {Pion-pole
  contribution to hadronic light-by-light scattering in the anomalous magnetic
  moment of the muon}, Phys. Rev. Lett. 121~(11) (2018) 112002.
\newblock \href {http://arxiv.org/abs/1805.01471} {\path{arXiv:1805.01471}},
  \href {https://doi.org/10.1103/PhysRevLett.121.112002}
  {\path{doi:10.1103/PhysRevLett.121.112002}}.

\bibitem{Hoferichter:2018kwz}
M.~Hoferichter, B.-L. Hoid, B.~Kubis, S.~Leupold, S.~P. Schneider, {Dispersion
  relation for hadronic light-by-light scattering: pion pole}, JHEP 10 (2018)
  141.
\newblock \href {http://arxiv.org/abs/1808.04823} {\path{arXiv:1808.04823}},
  \href {https://doi.org/10.1007/JHEP10(2018)141}
  {\path{doi:10.1007/JHEP10(2018)141}}.

\bibitem{PrimEx-II:2020jwd}
I.~Larin, et~al., {Precision measurement of the neutral pion lifetime}, Science
  368~(6490) (2020) 506--509.
\newblock \href {https://doi.org/10.1126/science.aay6641}
  {\path{doi:10.1126/science.aay6641}}.

\bibitem{Gerardin:2019vio}
A.~G\'erardin, H.~B. Meyer, A.~Nyffeler, {Lattice calculation of the pion
  transition form factor with $N_f=2+1$ Wilson quarks}, Phys. Rev. D 100~(3)
  (2019) 034520.
\newblock \href {http://arxiv.org/abs/1903.09471} {\path{arXiv:1903.09471}},
  \href {https://doi.org/10.1103/PhysRevD.100.034520}
  {\path{doi:10.1103/PhysRevD.100.034520}}.

\bibitem{Goity:2002nn}
J.~L. Goity, A.~M. Bernstein, B.~R. Holstein, {The Decay pi0 ---\ensuremath{>}
  gamma gamma to next to leading order in chiral perturbation theory}, Phys.
  Rev. D 66 (2002) 076014.
\newblock \href {http://arxiv.org/abs/hep-ph/0206007}
  {\path{arXiv:hep-ph/0206007}}, \href
  {https://doi.org/10.1103/PhysRevD.66.076014}
  {\path{doi:10.1103/PhysRevD.66.076014}}.

\bibitem{Ananthanarayan:2002kj}
B.~Ananthanarayan, B.~Moussallam, {Electromagnetic corrections in the anomaly
  sector}, JHEP 05 (2002) 052.
\newblock \href {http://arxiv.org/abs/hep-ph/0205232}
  {\path{arXiv:hep-ph/0205232}}, \href
  {https://doi.org/10.1088/1126-6708/2002/05/052}
  {\path{doi:10.1088/1126-6708/2002/05/052}}.

\bibitem{Kampf:2009tk}
K.~Kampf, B.~Moussallam, {Chiral expansions of the pi0 lifetime}, Phys. Rev. D
  79 (2009) 076005.
\newblock \href {http://arxiv.org/abs/0901.4688} {\path{arXiv:0901.4688}},
  \href {https://doi.org/10.1103/PhysRevD.79.076005}
  {\path{doi:10.1103/PhysRevD.79.076005}}.

\bibitem{Burri:2022gdg}
S.~A. Burri, et~al., {Pseudoscalar-pole contributions to the muon $g-2$ at the
  physical point}, PoS LATTICE2022 (2023) 306.
\newblock \href {http://arxiv.org/abs/2212.10300} {\path{arXiv:2212.10300}},
  \href {https://doi.org/10.22323/1.430.0306} {\path{doi:10.22323/1.430.0306}}.

\bibitem{Verplanke:2022eto}
A.~G\'erardin, J.~N. Guenther, L.~Varnhorst, W.~E.~A. Verplanke, {Pseudoscalar
  transition form factors and the hadronic light-by-light contribution to the
  muon g-2}, PoS LATTICE2022 (2023) 332.
\newblock \href {http://arxiv.org/abs/2211.04159} {\path{arXiv:2211.04159}},
  \href {https://doi.org/10.22323/1.430.0332} {\path{doi:10.22323/1.430.0332}}.

\bibitem{Hoferichter:2014vra}
M.~Hoferichter, B.~Kubis, S.~Leupold, F.~Niecknig, S.~P. Schneider, {Dispersive
  analysis of the pion transition form factor}, Eur. Phys. J. C 74 (2014) 3180.
\newblock \href {http://arxiv.org/abs/1410.4691} {\path{arXiv:1410.4691}},
  \href {https://doi.org/10.1140/epjc/s10052-014-3180-0}
  {\path{doi:10.1140/epjc/s10052-014-3180-0}}.

\bibitem{Hanhart:2013vba}
C.~Hanhart, A.~Kup\'s\'c, U.-G. Mei\ss{}ner, F.~Stollenwerk, A.~Wirzba,
  {Dispersive analysis for $\eta\to \gamma\gamma^*$}, Eur. Phys. J. C 73~(12)
  (2013) 2668, [Erratum: Eur. Phys. J. C \textbf{75}, 242 (2015)].
\newblock \href {http://arxiv.org/abs/1307.5654} {\path{arXiv:1307.5654}},
  \href {https://doi.org/10.1140/epjc/s10052-013-2668-3}
  {\path{doi:10.1140/epjc/s10052-013-2668-3}}.

\bibitem{Kubis:2015sga}
B.~Kubis, J.~Plenter, {Anomalous decay and scattering processes of the $\eta $
  meson}, Eur. Phys. J. C 75~(6) (2015) 283.
\newblock \href {http://arxiv.org/abs/1504.02588} {\path{arXiv:1504.02588}},
  \href {https://doi.org/10.1140/epjc/s10052-015-3495-5}
  {\path{doi:10.1140/epjc/s10052-015-3495-5}}.

\bibitem{Holz:2015tcg}
S.~Holz, J.~Plenter, C.-W. Xiao, T.~Dato, C.~Hanhart, B.~Kubis, U.-G.
  Mei\ss{}ner, A.~Wirzba, {Towards an improved understanding of $\eta \to
  \gamma^* \gamma^*$}, Eur. Phys. J. C 81~(11) (2021) 1002.
\newblock \href {http://arxiv.org/abs/1509.02194} {\path{arXiv:1509.02194}},
  \href {https://doi.org/10.1140/epjc/s10052-021-09661-0}
  {\path{doi:10.1140/epjc/s10052-021-09661-0}}.

\bibitem{Holz:2022hwz}
S.~Holz, C.~Hanhart, M.~Hoferichter, B.~Kubis, {A dispersive analysis of $\eta
  '\rightarrow \pi ^+\pi ^-\gamma $ and $\eta '\rightarrow \ell ^+\ell ^-\gamma
  $}, Eur. Phys. J. C 82~(5) (2022) 434, [Addendum: Eur. Phys. J. C
  \textbf{82}, 1159 (2022)].
\newblock \href {http://arxiv.org/abs/2202.05846} {\path{arXiv:2202.05846}},
  \href {https://doi.org/10.1140/epjc/s10052-022-10247-7}
  {\path{doi:10.1140/epjc/s10052-022-10247-7}}.

\bibitem{Masjuan:2017tvw}
P.~Masjuan, P.~S\'anchez-Puertas, {Pseudoscalar-pole contribution to the
  $(g_{\mu}-2)$: a rational approach}, Phys. Rev. D 95~(5) (2017) 054026.
\newblock \href {http://arxiv.org/abs/1701.05829} {\path{arXiv:1701.05829}},
  \href {https://doi.org/10.1103/PhysRevD.95.054026}
  {\path{doi:10.1103/PhysRevD.95.054026}}.

\bibitem{Escribano:2015yup}
R.~Escribano, S.~Gonz\`alez-Sol\'\i{}s, P.~Masjuan, P.~S\'anchez-Puertas,
  {$\eta$' transition form factor from space- and timelike experimental data},
  Phys. Rev. D 94~(5) (2016) 054033.
\newblock \href {http://arxiv.org/abs/1512.07520} {\path{arXiv:1512.07520}},
  \href {https://doi.org/10.1103/PhysRevD.94.054033}
  {\path{doi:10.1103/PhysRevD.94.054033}}.

\bibitem{Alexandrou:2022qyf}
C.~Alexandrou, et~al., {The $\eta \rightarrow \gamma^* \gamma^*$ transition
  form factor and the hadronic light-by-light $\eta$-pole contribution to the
  muon $g-2$ from lattice QCD} (12 2022).
\newblock \href {http://arxiv.org/abs/2212.06704} {\path{arXiv:2212.06704}}.

\bibitem{Browman:1974sj}
A.~Browman, J.~DeWire, B.~Gittelman, K.~M. Hanson, E.~Loh, R.~Lewis, {The
  Radiative Width of the eta Meson}, Phys. Rev. Lett. 32 (1974) 1067.
\newblock \href {https://doi.org/10.1103/PhysRevLett.32.1067}
  {\path{doi:10.1103/PhysRevLett.32.1067}}.

\bibitem{Primakoff:1951iae}
H.~Primakoff, {Photoproduction of neutral mesons in nuclear electric fields and
  the mean life of the neutral meson}, Phys. Rev. 81 (1951) 899.
\newblock \href {https://doi.org/10.1103/PhysRev.81.899}
  {\path{doi:10.1103/PhysRev.81.899}}.

\bibitem{Gan:2014pna}
L.~Gan, {Test of fundamental symmetries via the Primakoff effect}, EPJ Web
  Conf. 73 (2014) 07004.
\newblock \href {https://doi.org/10.1051/epjconf/20147307004}
  {\path{doi:10.1051/epjconf/20147307004}}.

\bibitem{Bernstein:2011bx}
A.~M. Bernstein, B.~R. Holstein, {Neutral Pion Lifetime Measurements and the
  QCD Chiral Anomaly}, Rev. Mod. Phys. 85 (2013) 49.
\newblock \href {http://arxiv.org/abs/1112.4809} {\path{arXiv:1112.4809}},
  \href {https://doi.org/10.1103/RevModPhys.85.49}
  {\path{doi:10.1103/RevModPhys.85.49}}.

\bibitem{ParticleDataGroup:2018ovx}
M.~Tanabashi, et~al., {Review of Particle Physics}, Phys. Rev. D 98~(3) (2018)
  030001.
\newblock \href {https://doi.org/10.1103/PhysRevD.98.030001}
  {\path{doi:10.1103/PhysRevD.98.030001}}.

\bibitem{Ioffe:2007eg}
B.~L. Ioffe, A.~G. Oganesian, {Axial anomaly and the precise value of the pi0
  ---\ensuremath{>} 2 gamma decay width}, Phys. Lett. B 647 (2007) 389--393.
\newblock \href {http://arxiv.org/abs/hep-ph/0701077}
  {\path{arXiv:hep-ph/0701077}}, \href
  {https://doi.org/10.1016/j.physletb.2007.02.021}
  {\path{doi:10.1016/j.physletb.2007.02.021}}.

\bibitem{Kastner:2008ch}
A.~Kastner, H.~Neufeld, {The K(l3) scalar form factors in the standard model},
  Eur. Phys. J. C 57 (2008) 541--556.
\newblock \href {http://arxiv.org/abs/0805.2222} {\path{arXiv:0805.2222}},
  \href {https://doi.org/10.1140/epjc/s10052-008-0703-6}
  {\path{doi:10.1140/epjc/s10052-008-0703-6}}.

\bibitem{Giusti:2017dmp}
D.~Giusti, V.~Lubicz, C.~Tarantino, G.~Martinelli, F.~Sanfilippo, S.~Simula,
  N.~Tantalo, {Leading isospin-breaking corrections to pion, kaon and
  charmed-meson masses with Twisted-Mass fermions}, Phys. Rev. D 95~(11) (2017)
  114504.
\newblock \href {http://arxiv.org/abs/1704.06561} {\path{arXiv:1704.06561}},
  \href {https://doi.org/10.1103/PhysRevD.95.114504}
  {\path{doi:10.1103/PhysRevD.95.114504}}.

\bibitem{PrimExeta:2010}
A.~Gasparian, L.~Gan, et~al., A precision measurement of the $\eta$ radiative
  decay width via the primakoff effect,
  \text{https://www.jlab.org/exp$_-$prog/proposals/10/PR12-10-011.pdf}.

\bibitem{Leutwyler:1996np}
H.~Leutwyler, {Implications of eta eta-prime mixing for the decay eta
  ---\ensuremath{>} 3 pi}, Phys. Lett. B 374 (1996) 181--185.
\newblock \href {http://arxiv.org/abs/hep-ph/9601236}
  {\path{arXiv:hep-ph/9601236}}, \href
  {https://doi.org/10.1016/0370-2693(96)00167-0}
  {\path{doi:10.1016/0370-2693(96)00167-0}}.

\bibitem{Essig:2013lka}
R.~Essig, et~al., {Working Group Report: New Light Weakly Coupled Particles},
  in: {Community Summer Study 2013}: {Snowmass on the Mississippi}, 2013.
\newblock \href {http://arxiv.org/abs/1311.0029} {\path{arXiv:1311.0029}}.

\bibitem{Alexander:2016aln}
J.~Alexander, et~al., {Dark Sectors 2016 Workshop: Community Report}, 2016.
\newblock \href {http://arxiv.org/abs/1608.08632} {\path{arXiv:1608.08632}}.

\bibitem{Battaglieri:2017aum}
M.~Battaglieri, et~al., {US Cosmic Visions: New Ideas in Dark Matter 2017:
  Community Report}, in: {U.S. Cosmic Visions: New Ideas in Dark Matter}, 2017.
\newblock \href {http://arxiv.org/abs/1707.04591} {\path{arXiv:1707.04591}}.

\bibitem{Arkani-Hamed:2008hhe}
N.~Arkani-Hamed, D.~P. Finkbeiner, T.~R. Slatyer, N.~Weiner, {A Theory of Dark
  Matter}, Phys. Rev. D 79 (2009) 015014.
\newblock \href {http://arxiv.org/abs/0810.0713} {\path{arXiv:0810.0713}},
  \href {https://doi.org/10.1103/PhysRevD.79.015014}
  {\path{doi:10.1103/PhysRevD.79.015014}}.

\bibitem{Pospelov:2008jd}
M.~Pospelov, A.~Ritz, {Astrophysical Signatures of Secluded Dark Matter}, Phys.
  Lett. B 671 (2009) 391--397.
\newblock \href {http://arxiv.org/abs/0810.1502} {\path{arXiv:0810.1502}},
  \href {https://doi.org/10.1016/j.physletb.2008.12.012}
  {\path{doi:10.1016/j.physletb.2008.12.012}}.

\bibitem{Liu:2018qgl}
Y.-S. Liu, I.~C. Clo\"et, G.~A. Miller, {Eta Decay and Muonic Puzzles}, Nucl.
  Phys. B (2019) 114638.
\newblock \href {http://arxiv.org/abs/1805.01028} {\path{arXiv:1805.01028}},
  \href {https://doi.org/10.1016/j.nuclphysb.2019.114638}
  {\path{doi:10.1016/j.nuclphysb.2019.114638}}.

\bibitem{Fayet:2007ua}
P.~Fayet, {U-boson production in e+ e- annihilations, psi and Upsilon decays,
  and Light Dark Matter}, Phys. Rev. D 75 (2007) 115017.
\newblock \href {http://arxiv.org/abs/hep-ph/0702176}
  {\path{arXiv:hep-ph/0702176}}, \href
  {https://doi.org/10.1103/PhysRevD.75.115017}
  {\path{doi:10.1103/PhysRevD.75.115017}}.

\bibitem{Pospelov:2008zw}
M.~Pospelov, {Secluded U(1) below the weak scale}, Phys. Rev. D 80 (2009)
  095002.
\newblock \href {http://arxiv.org/abs/0811.1030} {\path{arXiv:0811.1030}},
  \href {https://doi.org/10.1103/PhysRevD.80.095002}
  {\path{doi:10.1103/PhysRevD.80.095002}}.

\bibitem{Krasznahorkay:2015iga}
A.~J. Krasznahorkay, et~al., {Observation of Anomalous Internal Pair Creation
  in Be8 : A Possible Indication of a Light, Neutral Boson}, Phys. Rev. Lett.
  116~(4) (2016) 042501.
\newblock \href {http://arxiv.org/abs/1504.01527} {\path{arXiv:1504.01527}},
  \href {https://doi.org/10.1103/PhysRevLett.116.042501}
  {\path{doi:10.1103/PhysRevLett.116.042501}}.

\bibitem{Feng:2016jff}
J.~L. Feng, B.~Fornal, I.~Galon, S.~Gardner, J.~Smolinsky, T.~M.~P. Tait,
  P.~Tanedo, {Protophobic Fifth-Force Interpretation of the Observed Anomaly in
  $^8$Be Nuclear Transitions}, Phys. Rev. Lett. 117~(7) (2016) 071803.
\newblock \href {http://arxiv.org/abs/1604.07411} {\path{arXiv:1604.07411}},
  \href {https://doi.org/10.1103/PhysRevLett.117.071803}
  {\path{doi:10.1103/PhysRevLett.117.071803}}.

\bibitem{Tulin:2017ara}
S.~Tulin, H.-B. Yu, {Dark Matter Self-interactions and Small Scale Structure},
  Phys. Rept. 730 (2018) 1--57.
\newblock \href {http://arxiv.org/abs/1705.02358} {\path{arXiv:1705.02358}},
  \href {https://doi.org/10.1016/j.physrep.2017.11.004}
  {\path{doi:10.1016/j.physrep.2017.11.004}}.

\bibitem{Tulin:2012wi}
S.~Tulin, H.-B. Yu, K.~M. Zurek, {Resonant Dark Forces and Small Scale
  Structure}, Phys. Rev. Lett. 110~(11) (2013) 111301.
\newblock \href {http://arxiv.org/abs/1210.0900} {\path{arXiv:1210.0900}},
  \href {https://doi.org/10.1103/PhysRevLett.110.111301}
  {\path{doi:10.1103/PhysRevLett.110.111301}}.

\bibitem{Tulin:2013teo}
S.~Tulin, H.-B. Yu, K.~M. Zurek, {Beyond Collisionless Dark Matter: Particle
  Physics Dynamics for Dark Matter Halo Structure}, Phys. Rev. D 87~(11) (2013)
  115007.
\newblock \href {http://arxiv.org/abs/1302.3898} {\path{arXiv:1302.3898}},
  \href {https://doi.org/10.1103/PhysRevD.87.115007}
  {\path{doi:10.1103/PhysRevD.87.115007}}.

\bibitem{Aloni:2018vki}
D.~Aloni, Y.~Soreq, M.~Williams, {Coupling QCD-Scale Axionlike Particles to
  Gluons}, Phys. Rev. Lett. 123~(3) (2019) 031803.
\newblock \href {http://arxiv.org/abs/1811.03474} {\path{arXiv:1811.03474}},
  \href {https://doi.org/10.1103/PhysRevLett.123.031803}
  {\path{doi:10.1103/PhysRevLett.123.031803}}.

\bibitem{Aloni:2019ruo}
D.~Aloni, C.~Fanelli, Y.~Soreq, M.~Williams, {Photoproduction of Axionlike
  Particles}, Phys. Rev. Lett. 123~(7) (2019) 071801.
\newblock \href {http://arxiv.org/abs/1903.03586} {\path{arXiv:1903.03586}},
  \href {https://doi.org/10.1103/PhysRevLett.123.071801}
  {\path{doi:10.1103/PhysRevLett.123.071801}}.

\bibitem{Dolan:2017osp}
M.~J. Dolan, T.~Ferber, C.~Hearty, F.~Kahlhoefer, K.~Schmidt-Hoberg, {Revised
  constraints and Belle II sensitivity for visible and invisible axion-like
  particles}, JHEP 12 (2017) 094, [Erratum: JHEP 03, 190 (2021)].
\newblock \href {http://arxiv.org/abs/1709.00009} {\path{arXiv:1709.00009}},
  \href {https://doi.org/10.1007/JHEP12(2017)094}
  {\path{doi:10.1007/JHEP12(2017)094}}.

\bibitem{Bjorken:1988as}
J.~D. Bjorken, S.~Ecklund, W.~R. Nelson, A.~Abashian, C.~Church, B.~Lu, L.~W.
  Mo, T.~A. Nunamaker, P.~Rassmann, {Search for Neutral Metastable Penetrating
  Particles Produced in the SLAC Beam Dump}, Phys. Rev. D 38 (1988) 3375.
\newblock \href {https://doi.org/10.1103/PhysRevD.38.3375}
  {\path{doi:10.1103/PhysRevD.38.3375}}.

\bibitem{OPAL:2002vhf}
G.~Abbiendi, et~al., {Multiphoton production in e+ e- collisions at s**(1/2) =
  181-GeV to 209-GeV}, Eur. Phys. J. C 26 (2003) 331--344.
\newblock \href {http://arxiv.org/abs/hep-ex/0210016}
  {\path{arXiv:hep-ex/0210016}}, \href
  {https://doi.org/10.1140/epjc/s2002-01074-5}
  {\path{doi:10.1140/epjc/s2002-01074-5}}.

\bibitem{Knapen:2016moh}
S.~Knapen, T.~Lin, H.~K. Lou, T.~Melia, {Searching for Axionlike Particles with
  Ultraperipheral Heavy-Ion Collisions}, Phys. Rev. Lett. 118~(17) (2017)
  171801.
\newblock \href {http://arxiv.org/abs/1607.06083} {\path{arXiv:1607.06083}},
  \href {https://doi.org/10.1103/PhysRevLett.118.171801}
  {\path{doi:10.1103/PhysRevLett.118.171801}}.

\bibitem{Blumlein:1991xh}
J.~Blumlein, et~al., {Limits on the mass of light (pseudo)scalar particles from
  Bethe-Heitler e+ e- and mu+ mu- pair production in a proton - iron beam dump
  experiment}, Int. J. Mod. Phys. A 7 (1992) 3835--3850.
\newblock \href {https://doi.org/10.1142/S0217751X9200171X}
  {\path{doi:10.1142/S0217751X9200171X}}.

\bibitem{PVDIS_PAC}
P.~A. Souder, P.~E. Reimer, X.~Zheng, {Precision Measurement of
  Parity-violation in Deep Inelastic Scattering Over a Broad Kinematic Range,
  Jefferson Lab Experiment E12-10-007, 2010 with 2022 update}.

\bibitem{COHERENT:2021xmm}
D.~Akimov, et~al., {Measurement of the Coherent Elastic Neutrino-Nucleus
  Scattering Cross Section on CsI by COHERENT}, Phys. Rev. Lett. 129~(8) (2022)
  081801.
\newblock \href {http://arxiv.org/abs/2110.07730} {\path{arXiv:2110.07730}},
  \href {https://doi.org/10.1103/PhysRevLett.129.081801}
  {\path{doi:10.1103/PhysRevLett.129.081801}}.

\bibitem{BDX:2016akw}
M.~Battaglieri, et~al., {Dark Matter Search in a Beam-Dump eXperiment (BDX) at
  Jefferson Lab} (7 2016).
\newblock \href {http://arxiv.org/abs/1607.01390} {\path{arXiv:1607.01390}}.

\bibitem{Marsicano:2018vin}
L.~Marsicano, M.~Battaglieri, A.~Celentano, R.~De~Vita, Y.-M. Zhong, {Probing
  Leptophilic Dark Sectors at Electron Beam-Dump Facilities}, Phys. Rev. D
  98~(11) (2018) 115022.
\newblock \href {http://arxiv.org/abs/1812.03829} {\path{arXiv:1812.03829}},
  \href {https://doi.org/10.1103/PhysRevD.98.115022}
  {\path{doi:10.1103/PhysRevD.98.115022}}.

\bibitem{Battaglieri:2022dcy}
M.~Battaglieri, et~al., {Dark matter search with the BDX-MINI experiment},
  Phys. Rev. D 106~(7) (2022) 072011.
\newblock \href {http://arxiv.org/abs/2208.01387} {\path{arXiv:2208.01387}},
  \href {https://doi.org/10.1103/PhysRevD.106.072011}
  {\path{doi:10.1103/PhysRevD.106.072011}}.

\bibitem{Bartnik:2020pos}
A.~Bartnik, et~al., {CBETA: First Multipass Superconducting Linear Accelerator
  with Energy Recovery}, Phys. Rev. Lett. 125~(4) (2020) 044803.
\newblock \href {https://doi.org/10.1103/PhysRevLett.125.044803}
  {\path{doi:10.1103/PhysRevLett.125.044803}}.

\bibitem{Bogacz:2021}
S.~Bogacz, et~al., {20-24 GeV FFA CEBAF Energy Upgrade}, Proc. IPAC’21,
  Campinas, Brazil, May 2021 (2023) 715–718\href
  {https://doi.org/10.18429/JACoW-IPAC2021-MOPAB216}
  {\path{doi:10.18429/JACoW-IPAC2021-MOPAB216}}.

\bibitem{Brooks:2022}
S.~Brooks, S.~Bogacz, {Permanent Magnets forthe CEBAF 24GeV Upgrade}, Proc.
  IPAC’22, Bangkok, Thailand, Jun. 2022 (2022) 2792–2795\href
  {https://doi.org/10.18429/JACoW-IPAC2022-THPOTK011}
  {\path{doi:10.18429/JACoW-IPAC2022-THPOTK011}}.

\bibitem{Brooks:2023}
S.~Brooks, et~al., {Open-Midplane Gradient Permanent Magnet with 1.53 T Peak
  Field}, Proc. IPAC’23, Venice, Italy, May 2023 (2023).

\end{thebibliography}
\end{document}